%% file: Thesis.tex
\documentclass[12pt]{book} 
\usepackage{graphicx} 

\usepackage{appendix}

\usepackage{setspace}

\usepackage{multicol}

\usepackage{fix-cm}
\usepackage{xcolor}
\usepackage{multicol}
\usepackage[bf,rm,medium,compact]{titlesec}
\definecolor{gray75}{gray}{0.75}

\titleformat{\chapter}[display]{\filright\Huge\bfseries}{\fontsize{100}{100}\selectfont\textcolor{gray75}\thechapter}{1ex}{}[]%

\usepackage{hyperref}
\hypersetup{
    colorlinks,
    allcolors = blue,
    linktoc=page 
    }

\usepackage{url} 

\usepackage[defernumbers=true,
style=nature,
eprint=true,doi=true]{biblatex} 

\makeatletter
\newcommand{\makeAlph}[1]{\@Roman{#1}}
\makeatother
\DeclareFieldFormat{labelnumber}{\ifkeyword{own}
{\makeAlph{#1}}
{#1}} 

\title{PhD Thesis}
\author{SANCHEZ GARRIDO Adrian}
\date{\today}

\usepackage{dsfont}
\usepackage{amsmath}
\usepackage{enumitem}
\usepackage{mathabx} 

\usepackage{titlesec}

\titleformat{\paragraph}
{\normalfont\normalsize\bfseries}{\theparagraph}{1em}{}
\titlespacing*{\paragraph}
{0pt}{3.25ex plus 1ex minus .2ex}{1.5ex plus .2ex}

\usepackage{geometry} 
\usepackage{subcaption}
\usepackage{color}
\usepackage{braket}
\usepackage{relsize}
\usepackage{float}
\usepackage[normalem]{ulem} 
\newcommand{\be}{\begin{equation}}
\newcommand{\ee}{\end{equation}}
\newcommand{\bea}{\begin{eqnarray}}
\newcommand{\eea}{\end{eqnarray}}
\newcommand{\tr}{\mathrm{Tr}}

\usepackage{soul}
\usepackage{afterpage} 
\usepackage{emptypage} 

\newcommand\blankpage{%
    \null
    \thispagestyle{empty}%
    \newpage}

\begin{document}

\begin{onehalfspacing}
\pagestyle{empty}
\input{content/TitlePage.tex}

\end{onehalfspacing}


\input{content/Copyright}


\input{content/committee}

\cleardoublepage


\begin{flushright}
    \thispagestyle{empty}
    \vspace*{\fill}
    \textit{A la librería Albatros, lugar de resistencia}
    \vspace*{\fill}
\end{flushright}
\cleardoublepage 

\pagenumbering{roman}
\pagestyle{plain}

\input{content/Acknowledgements}

\cleardoublepage

\input{content/AbstractFrench}
\input{content/AbstractEnglish}
\cleardoublepage



\tableofcontents
\afterpage{\blankpage}
\cleardoublepage

\pagenumbering{arabic}

\input{content/Introduction}


\part{Fundamentals}
\label{part:part1}
\input{content/Chapter01}

\input{content/Chapter02}


\part{Contributions}
\label{part:part2}

\input{content/Chapter03}

\input{content/Chapter04}
\input{content/Chapter05}
\afterpage{\blankpage} 


\input{content/Discussion}
\afterpage{\blankpage} 


\appendix

\input{content/AppxCh01}
\input{content/AppxCh03}
\input{content/AppxCh04}
\input{content/AppxCh05}


\linespread{1}
\printbibliography[keyword=own,
resetnumbers=true,
title = {Publications},
heading=bibintoc]


\linespread{1}
\printbibliography[notkeyword=own,
resetnumbers=true,
title={Bibliography},
heading=bibintoc]


\end{document}

%% file: content/TitlePage.tex
\begin{titlepage}
\begin{center}

\begin{minipage}[t]{.49\textwidth}
\flushleft
UNIVERSITÉ DE GENÈVE \\
Département de physique théorique
\end{minipage}
\begin{minipage}[t]{.49\textwidth}
\flushright
FACULTÉ DES SCIENCES \\
Professeur Julian Sonner
\end{minipage}
    
\vspace{.5cm}

\noindent\rule{\textwidth}{.5pt}

\vspace{1.8cm}

{\bfseries\LARGE  On Krylov Complexity}
\vspace{1.8cm}

\vspace{0.7cm}
{\large THÈSE}
\vspace{.5cm}

\noindent
Présentée à la Faculté des sciences de l'Université de Genève\\
Pour obtenir le grade de Docteur ès sciences, mention physique\\

\vspace{1.0cm}

{\large Par}
\vspace{.5cm}

{\large Adrián SÁNCHEZ GARRIDO}

\vspace{.5cm}

{\large de}

\vspace{0.5cm}

{\large Grenade (Espagne)}


\vspace{1.0cm}
{\large Thèse Nº 5820}
\vfill

{\large GENÈVE}

{\large Centre d'impression de l’Université de Genève}

{\large 2024}

\end{center}
\end{titlepage}

%% file: content/Copyright.tex
\thispagestyle{empty}


~ \vfill
\noindent \copyright ~  { \small The author. This work is licensed under a Creative Commons Attribution (CC BY 4.0).} \url{https://creativecommons.org/licenses/by/4.0} .

%% file: content/committee.tex
\chapter*{\rm\bfseries Comité de thèse}
\label{ch:committee}

\thispagestyle{empty}

\noindent
\textbf{Prof. Julian Sonner}\\
\textbf{Directeur de thèse}\\
Départment de physique théorique\\
Université de Genève\\
Genève, Suisse \\

\noindent
\textbf{Prof. Marcos Mariño Beiras}\\
Départment de physique théorique et Section de mathématiques\\
Université de Genève\\
Genève, Suisse\\

\noindent
\textbf{Prof. Mark Mezei}\\
Mathematical Institute\\
University of Oxford\\
Oxford, England\\

%% file: content/Acknowledgements.tex
\chapter*{\rm\bfseries Acknowledgements}
\label{ch:acknowledgement}

\begin{flushright}
\noindent\rule{7.5cm}{.5pt} \\
    \textit{Gracias a la vida, que me ha dado tanto}\\
    Violeta Parra \\
    \noindent\rule{7.5cm}{.5pt}
\end{flushright}

I am indebted, in the first place, to my supervisor Julian Sonner, for his always wise and helpful insights, both on the scientific and on the personal level. Culminating this Thesis would have been an impossible task without his orientation. Over the years, during work meetings, lunch breaks, academic trips and different leisure activities with varying degrees of (in-)formality, we gradually built a relationship of mutual understanding and appreciation that will surely endure the passing of time. I wish to also thank our long-term collaborators Eliezer Rabinovici and Ruth Shir, as working with them has been a true pleasure. Despite not being my official supervisors, they have made an essential contribution to my training as a young scientist. In particular, I will treasure as a precious memory the countless hours of one-on-one technical work with Ruth, whom I finally met in person after three years of virtual calls, which included a global pandemic. Many other people have contributed in one way or another to the achievement of the results of this PhD research, and I therefore wish to thank D. Abanin, T. Anous, V. Balasubramanian, J. Barbón, A. Belin, B. Craps, A. Dymarsky, J. Erdmenger, D. Galante, J. Kames-King, S. Shashi and G. Di Ubaldo for stimulating conversations and correspondence. I must not forget to acknowledge the staff from the HPC cluster at the University of Geneva, which was used for the heavy numerics of my research projects. On this note, the assistance of J. Rougemont probably spared me a significant amount of hours of hopeless fights against my computer.

I must also include in these acknowledgements a special mention to all the members of my Solvay class, who participated together with me in the Solvay doctoral school in 2019, which meant for me an absolute phase transition in my mindset as a researcher. It is always a pleasure to encounter my former classmates in different scientific events all around the world. The Solvay school is a very valuable tool for training young theoretical physicists and I am very grateful to its organizers and to all the involved professors for their massive work. I hope it will continue to exist for many years.

In Geneva I have been a member of a particularly welcoming and charming group, whose stimulating and at the same time relaxed atmosphere has been a great host for these years of PhD. I am therefore thankful to Alex, Arunabha, Ben, Bharath, Carl, Claudia, Davide, Igor, Manuel, Manus, Marco, Max, Michael, Paulo, Pietro, Pranjal, Rahel, Rea, Sergio, Sonia and Tomás (note that some names on this list are doubly degenerate). We created some nice memories together, inside and outside of the office, which will stay with me. The coffee breaks with J. P. Eckmann, where we discussed about science from all sorts of free-style perspectives, are also something that I will miss. 

Apart from my PhD degree, I also studied my Master and one of my Bachelor years in Geneva. I have been around for a while and this has allowed me to come across with professors that have become an inspiration to me: T. Giamarchi, M. Maggiore, M. Mariño, M. Pohl, F. Riva and P. Wittwer are therefore partly responsible of the achievement of this Thesis. This responsibility is nevertheless shared with great classmates and friends I met along the way, such as Perrine, Giulia, Matzi, Laurent, Amélie, Jérémie, Yannick, Rebecka, Florian, Flan, Thomas, Mick, Ameek, Varun, Karthik, Jon, Miguel...

Beyond physics, in Geneva I have met people who have become very important to me and whose friendship has been a pillar of my well-being. I am talking about Francesco, Imola, Ahmad and Pramod\footnote{Christelle Perrier, the directress of the Centre Universitaire Protestant 1, where we once shared an apartment, is to be blamed for putting all these characters together. I am not sure if she knew what she was doing, but we are all very thankful to her.}. I would also like to thank Elena for her good friendship and support, together with Aki, Aurélien, Bea, Bibi, Ece, Miquel and Zhimei. Thanks for the good times together! At some point during my years in this city I had the fortunate idea of joining the Emmet university theater group, where I met very special people whose visions about art and society, especially those coming from Martin, Xi, Inès, Ciro and Sanam, have never ceased to inspire me.

\begin{center}
\vspace{1.1cm}
 {\large  * $\qquad$ * $\qquad$  *}
\vspace{1.1cm}
\end{center}

Para continuar, necesariamente tengo que cambiar de idioma. El camino hasta aquí ha sido largo y empieza en un colegio llamado Alquería en el que aprendí a pensar y a sentir, para a continuación pasar por el instituto Fray Luis de Granada, en el que profesores como M. Ángel, Sebastián, J. Ignacio, Vicente, M. Jesús y, especialmente, Fran, prendieron en mí una llama por la ciencia en general y por la física en particular que en algún momento el astrónomo Emilio Alfaro avivó. El Proyecto de Iniciación a la Investigación en Astrofísica en Secundaria (PIIAS), realizado en el Instituto de Astrofísica de Andalucía (en años sucesivos se amplió a múltiples disciplinas en diversos centros de la Universidad de Granada y del CSIC) y en el que Emilio fue mi primer mentor, fue lo que terminó de decantar mi vocación por la física, de modo que siempre estaré en deuda con Javier Cáceres, que impulsó aquel inmenso proyecto con su habitual entusiasmo por la ciencia.

En 2013 el camino alcanzó, al fin, la Universidad de Granada, donde aprendí de profesores que me abrieron nuevos horizontes. Algunos de ellos son F. Del Águila, J. A. Aguilar, F. Aguirre, M. Barros, E. Battaner, J. L. Bernier, D. Blanco, A. Bueno, M. Cabrera, M. Á. Cabrerizo, J. Callejas, M. C. Carrión, Y. Castro, J. S. Dehesa, Á. Delgado, E. Florido, M. J. Gálvez, C. García, J. F. Gómez (Kiko), R. González, J. I. Illana, B. Janssen, J. Marro, L. L. Salcedo, S. Verley, A. Zurita y mi tutor del trabajo de fin de grado, Manolo Pérez-Victoria, que tan sabiamente me introdujo en la quinta dimensión. Quisiera mencionar separadamente al profesor Roque Hidalgo Álvarez, una persona a la que admiro tanto por su rigor científico como por su compromiso y lucha constante en pos de una sociedad mejor\footnote{Tampoco olvido cómo jamás le tembló el pulso a la hora de suspender informes de laboratorio en los que no apreciara un mínimo nivel de madurez y honestidad científica por parte de sus autores.}. En la Universidad (y fuera de ella) me rodeé de personas con intereses, convicciones y pasiones que me nutrieron diariamente; es por eso que recuerdo con mucho cariño a Jesús, Chulani, María, Lucía, Pablo Torres, Pablo Rodríguez, Taro, Ana, Orestis, Chemi, Juan, Olalla, Fran, Blanca, José, Rubén, Antonio, Iván y un largo etcétera. De esta época saqué amistades sólidas y duraderas, como la de Lourdes, que ya es prácticamente familia. Fueron tiempos de nuevos descubrimientos, crecimiento personal, búsquedas artísticas y luchas apasionadas, y por ello no quiero dejar de recordar a mis compañeras de teatro de la escuela Remiendo, así como a mis apreciadas camaradas de la UJCE y de la CSE.

Fuera de la academia y del mundo que gira en torno a ella hay muchas personas responsables de que haya llegado hasta este punto en condiciones físicas y psíquicas medianamente aceptables. Esta tesis es de mis amigos de toda la vida Ana, Chuso, Javi, Luis, Sánchez y Quero, que siempre están ahí, de mi cuñada Marce y mi sobrinita Sophie, de Cata po, de mi viejo y reaparecido amigo Héctor, de Ana, Mercedes y Patri, que fueron mi día a día aquel primer año en Ginebra, de Ángela, Paty y Blanca, y de Irene, a quien veo poco pero quiero mucho. Y es también de mi familia toda; es de mi madre que, no contenta con darme la vida una vez, me la dio una segunda; es de mi hermano; y es de mi padre, que me vio doctorarme.

Los agradecimientos que siguen son exclusivamente culpa de una sola persona: Rodrigo Díaz Pino, dueño de Albatros, la librería latinoamericana plantada en medio de Ginebra gracias a la que me he reencontrado a mí mismo y he conocido a una serie de personas que van a hacer que me cueste irme. Rodrigo es uno de los imprescindibles de Bertolt Brecht, cuya labor por crear redes de apoyo, solidaridad y amistad, y por mantener viva la cultura latina en el contexto del exilio en Suiza, es tan importante que nunca recibirá suficientes muestras de cariño y agradecimiento.
A través de Albatros he encontrado grandes maestros de la música, como son mis queridos Sergio Valdeos, Cali Flores, Carla Derpic y Gabo Guzmán (cuya última lección aún no terminé de entender). Y por Albatros encontré también unos amigos que me han cambiado la vida y merecen por tanto una dedicatoria. Hay muchos, como mis camaradas Javi, Mariana y Susana, así como toda la plétora de artistas, activistas, e intelectuales que desfilan por la librería: Raúl, Juan, Gladys, Alois, Nathalie, Yves, Peter, Pedro, Lola, Víctor, Hortensia, Ana, Elena, Charito, Carla Claros, Emma, Mariel, mi compay Pablo, Nico, Alonzo, Xihuitl, Fulvia, Ileana, Yves Cerf, Jorge, Lakshmi, Héctor, Miriam, Arlene, Silvia, Patricia... Entre tanta gente, no quiero dejar de mencionar a Karen, André, Isma, Carla, Rous, y a Mar, que me has enseñado tanto.

Tengo, en resumen, mucho que agradecer. Quién le iba a decir a aquel chaval que entraba con las manos en los bolsillos en aquella primera clase de álgebra lineal de la carrera que algún día sería doctor...

%% file: content/AbstractFrench.tex
\chapter*{\rm\bfseries Résumé}
\label{ch:abstratfr}

La dualité holographique constitue l’un des paradigmes de la physique théorique des hautes energies depuis la proposition de la correspondance AdS/CFT pendant les dernières années du siècle passé. Cette correspondance, démontrée dans des cas spécifiques et conjecturée comme une propriété générique de la gravité, postule l’équivalence mathématique entre des théories quantiques de la gravité formulées sur des variétés $d$-dimensionnelles (à l’origine, des variétés se comportant asymptotiquement comme l’espace-temps anti-de Sitter) et des théories quantiques en $d-1$ dimensions avec symétrie conforme, mais sans gravité. Cette dualité permet d’étudier certaines propriétés de la gravité, qui autrement seraient inaccessibles ou difficiles à analyser, à partir des calculs sur la théorie duale correspondante. Un \textit{dictionnaire holographique} reliant des différents observables et propriétés entre les pairs de théories duales à été progressivement construit dans le cadre de cette correspondance.

Un élément de ce dictionnaire qui n’est pas encore pleinement établi aujourd’hui est la complexité quantique. Selon des conjectures récentes, dans une théorie avec un dual gravitationnel et pour un état quantique dont la géométrie associée est décrite par un trou noir, cette quantité devrait capturer la taille du trou de ver derrière l’horizon du trou noir. Celle-ci est une information qui est, selon la théorie classique de la Relativité Générale, inaccessible pour un observateur à l’extérieur du trou noir, ce qui rend la question sur la complexité quantique dans le cadre de la dualité holographique un sujet attirant.

Cependant, il existe plusieurs définitions de complexité quantique, notamment proposées dans des contextes de la computation quantique, qui ne sont pas bien adaptées pour une éventuelle application dans le domaine de la gravité quantique et de l’holographie. Conventionnellement, la complexité quantique est entendue comme une mesure du nombre minimal de portes logiques quantiques nécessaires pour reproduire un état quantique donné à partir d’un état de référence. Autrement dit, elle est donnée par la taille du circuit quantique minimale requis pour reproduire l’état en question. Cette notion nécessite de la définition ad-hoc des portes logiques de base, ainsi que d’un état de référence et, incidemment, d’un paramètre extrinsèque de tolérance assurant que la taille du circuit optimal soit finie. Tous ces choix humains, qui ne viennent pas des propriétés intrinsèques du système quantique en considération, ont de la peine à trouver une interprétation dans une théorie gravitationnelle putativement duale.

Pour cette raison, une nouvelle notion de complexité quantique, appelée complexité de Krylov, a été récemment proposée. Cette notion est entièrement définie donnés seulement un état initial et un générateur de l’évolution temporelle, c’est-à dire un Hamiltonian, puisqu’elle quantifie la complexité de l’évolution temporelle de l’état comme l’espérance quantique de sa position par rapport à une certaine base de l’espace de Hilbert, la base de Krylov, qui est par construction adaptée à une telle évolution, car elle est le résultat de la procedure de Gram-Schmidt appliquée aux puissances successives du générateur des translations temporelles agissant sur la condition initiale. Cette base est générée par un algorithme historiquement bien connu dans le domaine des méthodes numériques, ainsi que dans le champ de la physique de la matière condensée: L'algorithme de Lanczos.

Cette thèse explore l’applicabilité de la complexité de Krylov dans le domaine de l’holographie, une question qui passe obligatoirement par l’examination de son comportement dans des systèmes quantiques chaotiques en contraposition à ceux intégrables, car il est connu que les trous noirs sont des systèmes chaotiques. Le manuscrit donne une introduction extensive à l’algorithme de Lanczos, ainsi qu'une vue d’ensemble sur la recherche sur la complexité de Krylov dans la communauté des hautes énergies. Ensuite, les publications \cite{I,II,III,IV}, sur lesquelles ce travail est basé, sont présentées. Ces articles constituent des contributions de relevance dans le domaine de recherche décrit: Les projets \cite{I,II,III} on développé des méthodes numériques efficaces pour le calcul de la complexité de Krylov pour des systèmes chaotiques et intégrables de taille finie mais grande, donnant accès à des échelles temporelles exponentiellement longues par rapport à l’entropie du système. Les résultats suggèrent que la complexité de Krylov peut distinguer les systèmes chaotiques des systèmes intégrables, ce qui es satisfaisant pour une éventuelle application en holographie. Avec cette motivation, l’article \cite{IV} présente la première construction explicite d’une équivalence mathématique entre la complexité de Krylov et la longueur du trou de ver dans un example concret de l’holographie en basses dimensions: Celui donné par le système SYK et la théorie JT de la gravité en deux dimensions.


%% file: content/AbstractEnglish.tex
\chapter*{\rm\bfseries Abstract}
\label{ch:abstraten}

This Thesis explores the notion of Krylov complexity as a probe of quantum chaos and as a candidate for holographic complexity. The first Part is devoted to presenting the fundamental notions required to conduct research in this area. Namely, an extensive introduction to the Lanczos algorithm, its properties and associated algebraic structures, as well as technical details related to its practical implementation, is given. Subsequently, an overview of the seminal references and the main debates regarding Krylov complexity and its relation to chaos and holography is provided. The text throughout this first Part combines review material with original analyses which either intend to contextualize, compare and criticize results in the literature, or are the fruit of the investigations leading to the publications \cite{I,II,III,IV} on which this Thesis is based. These research projects are the subject of the second Part of the manuscript. In them, methods for the efficient implementation of the Lanczos algorithm in finite many-body systems were developed, allowing to compute numerically the Krylov complexity of models like SYK or the XXZ spin chain up to time scales exponentially large in system size. It was observed that the operator Krylov complexity profile in SYK, a paradigmatic low-dimensional chaotic system with a holographic dual, agrees with holographic expectations \cite{I}, while in the case of integrable models like XXZ complexity is affected by a novel localization effect in the so-called Krylov space which hinders its growth \cite{II,III}. Finally, an exact, analytical, correspondence between the Krylov complexity of the infinite-temperature thermofield double state in the low-energy regime of the double-scaled SYK model and bulk length in the theory of JT gravity was established in \cite{IV}. 

%% file: content/Introduction.tex
\chapter*{\rm\bfseries Introduction}\label{ch:introduction}
\addcontentsline{toc}{part}{\protect\numberline{}Introduction}%


In the year 1931, the Soviet mathematician and engineer Aleksey N. Krylov published a seminal article \cite{Krylov:1931} in which he proposed an efficient method for constructing numerically the secular equation associated to a given matrix. His method was based on the iterative action of the matrix of study on a fixed reference vector. Krylov's proposal was at the time the most efficient algorithm for the construction of the characteristic polynomial of a matrix and, in fact, it gave rise to a whole branch of numerical techniques still used nowadays which are loosely referred to as \textit{Krylov methods} \cite{Parlett,KrylovBook_Vorst,KrylovBook_Liesen}. The common denominator of all these methods is the fact that they study the restriction of a linear operator in a certain vector space generated by successive applications of the latter on a given, fixed vector. Spaces constructed in this fashion are dubbed Krylov subspaces, and they are extremely useful for the numerical eigenvalue problem, as well as for the computation of the exponential of linear operators, being therefore suitable for the resolution of linear matrix differential equations, which are ubiquitous in mathematics, physics and engineering (see e.g. \cite{Parlett,DE_Caesa,DE_Teschl,GalindoI,GalindoII}). The fact that this problem attracted the interest of A. N. Krylov, a major figure of naval engineering in his country at the time \cite{KrylovBiography_Shtraykh,KrylovWorks_Smirnov}, requires no explanation.

About twenty years later, while working for the National Bureau of Standards, the Martian\footnote{During the first half of the 20th Century a generation of Hungarian scientists, mostly of Jewish origin, migrated to the United States and eventually became prominent figures in physics, mathematics and engineering with a major role in the rapid scientific development of their host country. They were jocularly referred to as ``the Martians'', a term they proudly adopted, due to their common origin, superior wit, and the fact that they were able to communicate among themselves in a language that was barely understandable for any English speaker or, in fact, for almost any non-Hungarian speaker. Cornelius Lanczos is considered to be a part of this group, together with figures of relevance such as Theodore von Kármán, Leo Szilard, Eugene Wigner, John von Neumann, and Edward Teller \cite{MartianBook_Marx,MartianBook_Hargittai,MartianBook_Marton,LanczosBiography_Gellai}.} mathematician and physicist Cornelius Lanczos was also confronted with the eigenvalue problem. In his article \cite{Lanczos:1950zz}, he analyzed numerical methods for the diagonalization of matrices based precisely on the iterative action of the target matrix on a probe vector. In particular, aiming for the minimization of numerical errors, he proposed an algorithm that constructs efficiently an orthonormal\footnote{In Lanczos' original article \cite{Lanczos:1950zz}, the basis constructed is only orthogonal, but it can immediately be made orthonormal by normalizing its elements, which is the common practice nowadays.} basis for the Hilbert space spanned by the vectors iteratively constructed. Krylov's work, published in Russian, appeared in the News of the Academy of Sciences of the USSR, and Lanczos claimed to be unaware of it \cite{Lanczos:1950zz}. However, the motivations and methods that both authors describe are so intimately related that nowadays Lanczos' proposal is commonly known as the \textit{Lanczos algorithm}\footnote{An alternative name for this algorithm, popularly used in the condensed matter physics community, is \textit{the recursion method} \cite{ViswanathMuller}.}, while the orthonormal basis that such a method outputs has been dubbed the \textit{Krylov basis} \cite{Parlett}.

Little did Krylov, nor Lanczos, know that, almost a century after the first of the two articles appeared, their work would be used to describe the quantum properties of black holes\footnote{The first black hole metric ever found, which in fact constituted the first exact solution (other than the trivial flat spacetime) of Einstein's field equations, is the Schwarzschild solution \cite{Schwarzschild:1916uq,Schwarzschild:1916ae}, which appeared in 1916, very shortly after Einstein's publication of General Relativity \cite{Einstein:1916vd} and 15 years before Krylov's article.}\textsuperscript{,}\footnote{Lanczos himself produced several fruitful results in General Relativity \cite{Lanczos:1924zz,Lanczos:1932zz,Lanczos:1938sf,Lanczos:1942zz,Lanczos:1949zz,Lanczos:1957zz,Lanczos:1962zz,Lanczos:1963zz,Lanczos:1975su}.}. In order to elaborate on this claim, we first need to contextualize it within the main areas of research in modern-day high-energy theoretical physics. In the last years of the 20th Century, the equivalence between the theory of $\mathcal{N}=4$ super-Yang Mills in four spacetime dimensions and type IIB string theory defined on an AdS$_5\times$S$^5$ manifold was established by \cite{Maldacena:1997re}. This was the first concrete realization of what is nowadays known as the AdS/CFT correspondence\footnote{This correspondence had already been anticipated by the study on the relation between the asymptotic symmetries of AdS$_3$ and the two-dimensional conformal group in \cite{Brown:1986nw}.} \cite{Aharony:1999ti}, a conjectured duality between quantum field theories with conformal symmetry in flat space and quantum gravity defined in an asymptotically anti-de Sitter space with one extra spatial dimension. This correspondence is also commonly referred to as gauge/gravity duality or, simply, the holographic principle, reflecting the aim for generalizing it to instances in which the gravitational side of the duality is not an asymptotically anti-de Sitter spacetime \cite{Hull:1998vg,Strominger:2001pn,Witten:2001kn,Klemm:2001ea,Balasubramanian:2002zh,Anninos:2017hhn,Anninos:2018svg,Anninos:2021ihe,Anninos:2022ujl,Galante:2022nhj,Galante:2023uyf,Strominger:2017zoo,Donnay:2020guq,Pasterski:2021rjz,Pasterski:2021raf,Mizera:2022sln,Gonzo:2022tjm,Sonner:2018rmt}. In this framework, a \textit{holographic dictionary} \cite{Maldacena:1997re,Witten:1998qj,Gubser:1998bc,Heemskerk:2009pn,Penedones:2010ue,Fitzpatrick:2011ia,Harlow:2011ke,Aharony:1999ti,Sonner:2018rmt} has been progressively constructed, relating various properties and calculable observables on the field-theory side to their bulk, gravitational counterpart. The mapping between the spectrum of scaling dimensions in the conformal field theory and the mass spectrum in the gravity side, or the correspondence between bulk gauge symmetries and global symmetries of the field theory, are examples of items of the holographic dictionary. A more conceptually profound item of this dictionary is the one relating the ultra-violet cutoff in the field theory to the bulk radial direction, which implies that renormalization-group flows can be understood as probing deeper and deeper regions of the bulk; this has been used to derive the Callan-Symanzik equation and even the Weyl anomaly from semiclassical computations in gravity \cite{Henningson:1998gx,deBoer:1999tgo,Skenderis:2002wp}. 

The use of the term \textit{holographic} principle is not only due to the fact that the gravitational theory has an extra dimension compared to its field theory dual: Additionally, there is a sense in which the conformal field theory can be thought of as being defined on the boundary of the manifold in which the gravitational theory lies\footnote{Note, however, that the two theories do not coexist: On the one hand there is the gravitational theory defined on the bulk of the anti-de Sitter spacetime, and on the other hand there is the conformal field theory defined on the boundary manifold; these two are, according to the AdS/CFT correspondence, equivalent and interchangeable descriptions.} \cite{Witten:1998qj,Harlow:2011ke}. In this spirit, in gravitational scenarios that involve black holes, it is intriguing to wonder how the boundary theory encodes the properties of the black hole interior, which are classically inaccessible to a bulk observer outside of the event horizon. In particular, a very characteristic feature associated to the black hole interior is the growth of the \textit{Einstein-Rosen bridge} \cite{Einstein:1935tc}, conventionally denoted as ERB, which is the wormhole joining the two otherwise disconnected geometries outside the black hole horizon. From the gravity perspective, in the framework of eternal two-sided anti-de Sitter black holes, observables like the length of a spacelike geodesic with anchoring points on both asymptotic boundaries grow as a function of boundary time as a consequence of the growth of the ERB, and they do so following a very specific profile \cite{Shenker:2013pqa}. The analysis of such a profile and its response to perturbations led to the proposal \cite{Susskind:2014rva,Brown:2015bva,Brown:2015lvg} that this phenomenon should be captured by some notion of quantum complexity \cite{nielsen_chuang_2010} on the boundary theory. In particular, the known notions of quantum complexity respond satisfactorily to the so-called \textit{switchback effect} which, in gravity, consists on the disruption of the geometry due to an in-falling high-energy shock wave sourced on the boundary: This shock wave has the effect of pushing the event horizon towards the future, yielding an increase of the wormhole length. As a function of the boundary time at which the shock is sourced, the profile of this increment transitions from exponential growth before the scrambling time (logarithmic in the number of degrees of freedom or, equivalently, in the black hole area \cite{Bekenstein:1973ur}) to linear increase (see \cite{Shenker:2013pqa} for a concrete computation in three-dimensional gravity). The boundary description of this phenomenon consists on the perturbation of an initial state (originally dual to the unperturbed black hole geometry) by an operator inserted in the past which acts as the source of the in-falling shock wave. For a notion of quantum complexity to be \textit{holographic}, it should follow, as a function of the operator insertion time, the same profile that has been described above \cite{Stanford:2014jda,Ben-Ami:2016qex,Chapman:2016hwi,Carmi:2017jqz}.
The holographic complexity proposals refer to bulk gravitational observables such as the volume of spacelike hypersurfaces \cite{Susskind:2014rva} or the value of the action of a Wheeler-de Witt patch \cite{Brown:2015bva,Brown:2015lvg}; however, the seemingly intrinsic ambiguities in the definition of quantum complexity are important obstacles to the establishment of a precise item in the holographic dictionary \cite{Belin:2021bga,Belin:2022xmt}. Let us elaborate on this below.

The ordinary and most widely known notion of quantum complexity is \textit{circuit complexity} \cite{nielsen_chuang_2010}. Given a quantum state belonging to a certain Hilbert space, its circuit complexity is defined as the minimal number of unitary gates that are required to reconstruct such a state out of a starting reference state. For operators, circuit complexity can be defined analogously as the minimal number of gates that need to be conjugated with some starting operator in order to reproduce the target one.
This definition arises from the framework of quantum computation and, as an input, it requires the choice of both the reference state and the fixed set of unitary gates used as building blocks for constructing arbitrary quantum circuits. Furthermore, for a fixed and finite set of unitary gates, nothing guarantees in principle that any arbitrary state of the Hilbert space can be reconstructed out of the same fixed reference state by concatenating a finite number of gates, even if the Hilbert space has a finite dimension: In order to ensure finiteness of circuit complexity, it is thus required to introduce a notion of distance in the manifold defined by the Hilbert space \cite{doi:10.1126/science.1121541,Susskind:2018pmk} and a \textit{tolerance parameter}: A state shall be considered as ``reproduced'' if the distance between the constructed state and the target is smaller than the given tolerance threshold. These features are potentially unsatisfactory for an application in gauge/gravity duality: the extrinsic tolerance parameter is not a property of the boundary theory, but rather a human input used to define the complexity notion\footnote{The tolerance parameter cannot be related to an ultra-violet cutoff whatsoever since, as explained, it is required even for systems with finite-dimensional Hilbert spaces.}, and as such it will not be associated to any intrinsic features of the bulk gravitational theory. The situation becomes even more delicate if one considers that all known bounds on circuit complexity become divergent in the limit in which the extrinsic tolerance parameter is sent to zero (see for instance \cite{Susskind:2018pmk}). Similar considerations on arbitrariness may be made regarding the choice of the set of unitary gates.

Building up on the original works by Krylov and Lanczos, a new notion of quantum complexity has recently been proposed, dubbed \textit{Krylov complexity}, or K-complexity \cite{Parker:2018yvk}. Originally defined as an instance of operator complexity, it quantifies the complexity of a given time-evolving operator as its position expectation value with respect to an orthonormal basis of operator space built out by iteratively orthonormalizing nested commutators of the system's Hamiltonian with the initial operator. Such an orthogonalization procedure is precisely achieved by the Lanczos algorithm, which provides a basis of operator space that is, by construction, explored gradually by the evolving observable. This definition of complexity can be immediately extended to the case of states evolving in the Schrödinger picture \cite{Balasubramanian:2022tpr}. The potential advantage of Krylov complexity with respect to other complexity notions resides in the fact that it only requires of an initial condition and the generator of time evolution in order to be well-defined\footnote{In the case of operator Krylov complexity, an inner product in the Hilbert space of operators needs to be additionally specified. It may be the one naturally inherited from the inner product in the space of states, or some other physically sensible choice, such as a thermal correlator \cite{Parker:2018yvk,ViswanathMuller}. This provides a very rich structure relating operator Krylov complexity to two-point functions, the knowledge of whose properties may thus be used to shed light on features of this new notion of quantum complexity.} and finite. It is therefore free from ad-hoc, extrinsic elements, an attractive feature that makes it a promising candidate for holographic complexity. 

As presented, K-complexity describes the progressive exploration of the available Hilbert space by a certain operator (or state) evolving in time. Each successive element of the Krylov basis is more ``complicated'' than the previous one in the sense that it involves one more power of the time translation generator acting on the initial condition. Even though the Hamiltonian is not a unitary operator in general, it is seen to morally play a qualitatively similar role to that of the basic quantum gates in the case of circuit complexity, with the difference that, in the current case, the ``fundamental gate'' is fully dictated by the choice of the system. The fact that the meaning of \textit{complexity} is determined by the system's Hamiltonian, together with the initial condition for time evolution, is at the same time a technical advantage of Krylov complexity and an unusual feature that makes it not immediately analogous to circuit complexity, where an operator or a state can be assigned a degree of complexity without the need to specify the Hamiltonian that generates their evolution\footnote{Nonetheless, the choice of sensible unitary gates in circuit complexity is often influenced by the properties of the Hamiltonian. For instance, local unitary gates will be well suited for tracking the time evolution generated by a local Hamiltonian. From this perspective, Krylov complexity is not so different from other complexity notions. We will see in this Thesis that, in fact, it automatizes the procedure of the optimal choice of system-dependent gates \cite{Balasubramanian:2022tpr}.}. It is therefore necessary to investigate to what extent K-complexity can be taken as an actual instance of quantum complexity.

An important feature to which a good notion of quantum complexity should be sensitive is the chaotic or integrable nature of a system. Roughly speaking, a chaotic system is an efficient information scrambler \cite{Maldacena:2015waa,Roberts:2014isa,Stanford:2014jda}, which implies that local perturbations can rapidly spread in space: In the case of circuit complexity, given a local Hamiltonian and a fixed set of local quantum gates, this would, in turn, imply a fast increase in the complexity of an initially local operator \cite{Susskind:2018pmk}. Probing quantum chaos was in fact the original motivation of the authors of \cite{Parker:2018yvk}, where Krylov complexity was introduced\footnote{Historically, the recursion method had already been used extensively to study operator evolution in many-body systems \cite{ViswanathMuller}.}. They established a rigorous bound, for systems in the thermodynamic limit and at infinite temperature, according to which K-complexity of operators can grow at most exponentially in time, and they went further and proposed a \textit{universal operator growth hypothesis} according to which such a bound is saturated by maximally chaotic systems. In the same article, they demonstrated that K-complexity is a representative of a formal class of observables that can be thought of as \textit{generalized complexities}, which happen to be bounded from above by the Krylov complexity. Familiar notions of complexity like operator size \cite{Roberts:2014isa}, or probes of quantum chaos like out-of-time-ordered correlation functions \cite{Maldacena:2015waa}, belong to this class, supporting the authors' hypothesis that K-complexity is a useful probe of chaos. The sensitivity to quantum chaos is, furthermore, also of interest for applications to holography: It has been proposed that the spectrum of quantum gravity should be, above the black hole threshold, that of a chaotic system \cite{Anous:2019yku,Schlenker:2022dyo}, and semiclassical considerations indicate that black holes are the fastest information scramblers in nature \cite{Sekino:2008he}. Therefore, for consistency, a notion of quantum complexity probing properties of the black hole interior should be sensitive to this.

Its suggested usefulness as a probe of quantum chaos, together with its potential applicability in holography to diagnose properties of black holes such as the wormhole growth, have made Krylov complexity an attractive object of study for both the condensed matter and the high-energy theory communities\footnote{Section \ref{sect:reseachKC} will give a list of the most prominent areas of research on Krylov complexity, mostly focusing on those related to topics of interest for the high-energy community, together with the main references in each case.}. The area is, nevertheless, not exempt of debate: Counterexamples to the universal operator growth hypothesis at finite temperature, a regime in which the hypothesis is conjectured but not proved, have appeared in the literature, in particular in the framework of quantum field theory \cite{Dymarsky:2021bjq}, suggesting that in this realm the usefulness of K-complexity as a probe of chaos or even as a holographic complexity might be limited; additionally, a concrete item of the holographic dictionary relating Krylov complexity to the wormhole length in arbitrary space-time dimensions has not been established yet. The motivation behind the work presented in this Thesis has been to shed light on these two open questions, as well as the investigation of the universal operator growth hypothesis in systems away from the thermodynamic limit: Black holes are systems with a finite number of degrees of freedom \cite{Bekenstein:1973ur}, and hence, entropy-dependent time scales leaving a imprint on the K-complexity profile provide valuable information about the bulk. In particular, the late-time saturation value of complexity is only accessible for finite systems, and from the bulk point of view it probes an intrinsically quantum phenomenon, namely the fact that the expectation value of the observable capturing the wormhole growth fails to continue increasing because the system is exploring linear recombinations of microstates previously visited during its time evolution, due to the finite dimensionality of the Hilbert space: This is the regime in which the fully classical and geometrical description of gravity ceases to apply.

The publications supporting this Thesis \cite{I,II,III,IV} can be divided into two closely related lines of research: Articles \cite{I,II,III} study operator Krylov complexity in chaotic and integrable systems with finite size, while the work \cite{IV} presents the first explicit construction in the literature relating quantitatively K-complexity of a specific state (the \textit{thermofield double state} \cite{Maldacena:2001kr}) to bulk wormhole length in an instance of low-dimensional holography. Inspired by this, and by all the considerations manifested throughout the Introduction, the structure of the Thesis is the following:

\begin{itemize}
    \item Part \ref{part:part1} provides the fundamental notions and tools for conducting research in the area of Krylov complexity. It combines the review of seminal references with the exposition of original analyses. 
    \begin{itemize}
         \item Chapter \ref{ch:chapter01_Lanczos} gives a historical introduction to Krylov methods and the Lanczos algorithm by presenting the main aspects of the original works by A. Krylov and C. Lanczos. In particular, the work by the former author \cite{Krylov:1931} has only been published in Russian and there exists a review of it in German in the Zentralblatt (see the details specified in the Bibliography section for \cite{Krylov:1931}): This Thesis provides the first review in English of Krylov's main results. In addition, the Chapter contains all the technical details necessary to tackle the Lanczos algorithm and presents some further analysis on the algebraic structures associated to it, which is a result of the practical experience gained by implementing this algorithm numerically for the contributions \cite{I,II,III} and by studying Krylov complexity from a more analytical point of view in the project leading to \cite{IV}.

        \item Chapter \ref{ch:chapter02_KC} proceeds then to introduce Krylov complexity as a probe of quantum chaos and a potentially suitable candidate for being the complexity item in the holographic dictionary. Seminal results are reviewed and, once again, combined with some original analyses attempting to contextualize and compare them, addressing possible points of conflict that suggest putative lines of future research. In particular, a comparative discussion of two relevant theorems regarding K-complexity \cite{Parker:2018yvk,Balasubramanian:2022tpr} is given, and some attention is devoted to the ongoing debate on whether K-complexity is adequately sensitive to quantum chaos, since this matter is one of the points addressed in contributions \cite{I,II,III}. A discussion on the switchback effect from the perspective of Krylov complexity is also given, and the Chapter is closed by a compilation of the main lines of research currently active in the area of K-complexity.
    \end{itemize}
    \item Part \ref{part:part2} presents partial or total reproductions of the publications \cite{I,II,III,IV} contributing to this Thesis.
    \begin{itemize}
        \item Chapter \ref{ch:chapter03_SYK} presents the results of publication \cite{I}. This work studied K-complexity in the complex \textit{Sachdev-Ye-Kitaev (SYK) model} \cite{Kitaev:2015,Sachdev:2015efa,Sachdev:1992fk,Maldacena:2016hyu}, a one-dimensional system of spinless fermions with all-to-all random interactions known to be dual to two-dimensional gravity, which also provides a paradigmatic example of a quantum-chaotic system \cite{Maldacena:2016hyu,Sonner:2017hxc,Altland:2021rqn,Garcia-Garcia:2016mno,Garcia-Garcia:2017pzl}. The goal of this project consisted on studying SYK at finite size in order to compute numerically the K-complexity profile for a typical operator up to time scales exponentially large in system size, in order to verify its agreement with holographic expectations. Along the way, an upper bound on operator K-complexity for finite systems was proved from algebraic considerations. The main technical achievement of this project is the numerical implementation of the Lanczos algorithm for a many-body system with a finite but large Hilbert space dimension, for which two re-orthogonalization algorithms (\textit{Full Orthogonalization} and \textit{Partial Re-Orthogonalization}) needed to be applied in order to cure the intrinsic numerical instabilities of the Lanczos algorithm. This article was the first to present the full sequence of Lanczos coefficients for a finite many-body system: These are the orthogonalization coefficients output by the Lanczos algorithm, and they are required in order to compute the K-complexity profile up to exponentially late times, giving access to the regime of complexity saturation.

        \item Chapter \ref{ch:chapter04_Integrable} merges the results of \cite{II,III}. These works seek to identify the imprints of integrability, as opposed to quantum chaos, in the K-complexity profile of systems away from the thermodynamic limit. It is proposed that, as a consequence of the structure of the spectrum of an integrable system, a localization effect hinders the propagation of an evolving operator through its Krylov basis, eventually yielding a smaller late-time K-complexity saturation value as opposed to that in a quantum chaotic model. The work \cite{II} proposes and explores this \textit{Krylov localization} phenomenon in the context of the XXZ spin chain \cite{Heisenberg:1928mqa}, while \cite{III} supports this proposal further by studying K-complexity in a system whose Hamiltonian is given by that of the XXZ model, perturbed by an integrability-breaking defect, which can be used to interpolate between integrable and chaotic regimes: It was observed that, the stronger the defect is, the weaker the Krylov localization phenomenon becomes, yielding a larger K-complexity saturation value at late times, confirming that this indicator is sensitive to the chaotic or integrable character of the system.

        \item Finally, Chapter \ref{ch:chapter05_DSSYK} is devoted to the results published in \cite{IV}. Using the duality between the low-energy regime of the SYK model and the two-dimensional theory of Jackiw-Teitelboim (JT) gravity \cite{Jackiw:1984je,Teitelboim:1983ux}, which is analytically tractable thanks to the protocol of the \textit{double-scaling limit} \cite{Berkooz:2018jqr,Lin:2022rbf}, the article \cite{IV} proves the exact correspondence between the K-complexity of the infinite-temperature thermofield double state of a double-scaled SYK system near its ground state and the regularized two-sided bulk length in its corresponding dual geometry described by JT gravity. This is not only the first time an exact analytical matching of this sort is demonstrated for K-complexity but, as a matter of fact, it also constitutes the first explicit construction of a quantitative matching between bulk and boundary observables for \textit{any} notion of quantum complexity, going beyond the qualitative comparison of the profile of the corresponding quantities as a function of time.
        One may regard this Chapter as the culmination of the Thesis: While the preceding investigations on finite systems, quantum chaos and integrability serve as an exploration of the potential adequacy of Krylov complexity for holographic applications, with promising conclusions, this last result succeeds in constructing rigorously an holographic bulk dual. 
    \end{itemize}
\end{itemize}

%% file: content/Chapter01.tex
\chapter{\rm\bfseries The Lanczos algorithm}\label{ch:chapter01_Lanczos}

The present Chapter serves as a modern introduction to the main tool supporting the framework of Krylov complexity, namely the Lanczos algorithm, which we shall eventually use to unveil a very specific structure of the Hilbert space of a time-evolving state or operator, with respect to which this useful notion of complexity with applications to quantum gravity, holography and black holes, as well as quantum chaos and many-body physics, can be defined.

The method originated from the works by A. Krylov \cite{Krylov:1931} and C. Lanczos \cite{Lanczos:1950zz}, who studied independently, and within a time lapse of about twenty years, efficient algorithms to solve the numerical eigenvalue problem \cite{Parlett,KrylovBook_Vorst,KrylovBook_Liesen} based on the iterative exploration of the available Hilbert space by the means of the nested application on a certain probe vector of the operator whose spectrum is sought. 

The methods by Krylov and Lanczos are generic tools for the resolution of the eigenvalue problem of some operator in an unspecified Hilbert space. For the research presented in this Thesis, the operator of interest will be the time-evolution generator in a quantum-mechanical system, either for the space of states (i.e. the Hamiltonian), or for operator space (i.e. the Liouvillian). Nevertheless, for the sake of generality and in order to illustrate the wide applicability of the methods, this Chapter will provide a purely algebraic discussion of the Lanczos algorithm, applied to some hermitian operator $H$ acting over some Hilbert space $\mathcal{H}$. The original developments by Krylov and Lanczos shall be presented using modern notation and concepts related to Hilbert spaces, adapted to their eventual implementation in the realm of quantum mechanics.
The exposition will be accompanied by a further critical analysis of both works, which expands on the original articles \cite{Krylov:1931,Lanczos:1950zz} and is a result of the investigations that lead to the numerical implementations in publications \cite{I,II,III} and the analytical developments achieved in \cite{IV}. In summary, the Chapter can be thought of as a compilation of the necessary knowledge and technical requirements related to the Lanczos algorithm which serve as a basis to approach the field of Krylov complexity, stressing on the aspects that were more relevant for the research projects that compose this Thesis.

\section{Krylov's method for the efficient construction of the secular equation}\label{sect:Krylov}

Krylov's article \cite{Krylov:1931} approached the eigenvalue problem from an original perspective, seeking an efficient algorithm to compute the coefficients of the characteristic polynomial associated to a given matrix, whose zeroes provide the spectrum of the latter. This work was published exclusively in Russian, in the News of the Academy of Sciences of the USSR and, shortly after its appearance, a review article by N. N. Luzin \cite{Luzin:1931} was published in the same journal. There exist short reports on both papers, written in German\footnote{I would like to thank Rahel L. Baumgartner for her help translating these documents.}, in the Zentralblatt (see the Bibliography items for \cite{Krylov:1931,Luzin:1931}). Therefore, besides the historical contextualization of the methodology of interest in this Thesis, the current section aims to provide a brief summary in English of the main aspects of Krylov's work.

The review that will be presented below, however, does not follow literally the steps detailed in the seminal article \cite{Krylov:1931}. Instead, it focuses on the elements that are more relevant to the current research in Krylov complexity, and it does so using the more modern language and notation of Hilbert spaces, which is not employed in the original reference\footnote{In fact, the notion of abstract Hilbert spaces started to be generalized and axiomatized by John von Neumann, David Hilbert and others for applications to quantum mechanics precisely around the year 1930 \cite{Hilbert:1928,Neumann:1930,Jordan:1933vh}, simultaneously to the publication of Krylov's work.}. Modern-day terminology associated to the algebraic structure underlying Krylov's method \cite{Parlett,ViswanathMuller,Parker:2018yvk,I} will also be used for the sake of pedagogy, even though it was not originally introduced in \cite{Krylov:1931}.

\subsection{Formulation of the method}

Let us consider a hermitian operator $H$ acting over a $D$-dimensional Hilbert space\footnote{Krylov's original article considers real vector spaces and does not necessarily restrict the problem to symmetric matrices. Nevertheless, our interest in this problem comes from the framework of quantum mechanics, and hence we shall focus on hermitian operators over complex Hilbert spaces.} $\mathcal{H}$. 
The characteristic polynomial associated to it is defined as
\begin{equation}
    \label{Sect_Krylov_charac_polyn_def}
    S(\lambda) := \text{det}\left(H-\lambda \mathds{1}\right)~,
\end{equation}
and it is a polynomial of degree $D$ in $\lambda$. Its zeroes provide the eigenvalues of $H$. Otherwise stated, the expression 
\begin{equation}
    \label{Sect_Krylov_Secular_eqn_def}
    S(\lambda) = 0
\end{equation}
is the secular equation associated to the operator $H$.

Because of the determinant in \eqref{Sect_Krylov_charac_polyn_def}, the coefficients of the polynomial $S(\lambda)$ are basis-independent, but we can still give their expression using coordinates over a certain basis,
\begin{equation}
    \label{Sect_Krylov_Arbitrary_Basis}
    \left\{ |\mathcal{B}_n\rangle  \right\}_{n=1}^D~,
\end{equation}
with respect to which the matrix elements of $H$ are denoted as
\begin{equation}
    \label{Sect_Krylov_H_matrix elements}
    H_{mn}:=\langle \mathcal{B}_m |H|\mathcal{B}_n\rangle~.
\end{equation}
With this, we may express the characteristic polynomial \eqref{Sect_Krylov_charac_polyn_def} as:
\begin{equation}
    \label{Sect_Krylov_charac_polyn_coeffs}
    S(\lambda) = \begin{vmatrix}
        H_{11}-\lambda & H_{12}&\dots&H_{1D} \\
        H_{21} & H_{22}-\lambda &\dots & H_{2D} \\
        \vdots & \vdots &\ddots &\vdots \\
        H_{D1} & H_{D2} &\dots & H_{DD}-\lambda
    \end{vmatrix} =: \sum_{k=0}^{D} S_k\,\lambda^k~,
\end{equation}
where the last identity serves as a definition for the coefficients $S_k$. 
Determining them order by order for arbitrary $D$ is, \textit{a priori}, a computationally costly task, because of the way in which $\lambda$ enters in the determinant.
It is possible to anticipate that $S_D=(-1)^D$; however, the calculation of the rest of the coefficients $\left\{ S_k \right\}_{k=0}^{D-1}$ for an arbitrary Hilbert space dimension $D$ would either require some symbolic manipulation of the determinant, or the numerical implementation of some recursive relation. Alternatively, A. Krylov proposed to rewrite the determinant in \eqref{Sect_Krylov_charac_polyn_coeffs} in a way such that the different powers of $\lambda$ appear isolated in a single column, allowing for the systematic computation of $S_k$ by developing the new determinant expression along such a column. Let us elaborate on this below.

We consider a certain probe vector $|\Omega\rangle \in \mathcal{H}$, with which the Hilbert space will be iteratively explored, and we consider the following set of vectors:
\begin{equation}
    \label{Sect_Krylov_Set_Vectors_Powers_H}
    \left\{|\Omega^{(k)}\rangle := H^k|\Omega\rangle\right\}_{k\geq 0}~.
\end{equation}
In coordinates over the arbitrary basis \eqref{Sect_Krylov_Arbitrary_Basis}, we can decompose the probe vector as
\begin{equation}
    \label{Sect_Krylov_Probe_coords}
    |\Omega\rangle = \sum_{n=1}^D \Omega_n |\mathcal{B}_n\rangle
\end{equation}
and, likewise, there is a decomposition for every element in \eqref{Sect_Krylov_Set_Vectors_Powers_H}:
\begin{equation}
    \label{Sect_Krylov_Coords_All_vecs_of_set}
    |\Omega^{(k)}\rangle = \sum_{n=1}^D \Omega_n^{(k)}|\mathcal{B}_n\rangle~,
\end{equation}
where all $\Omega_n^{(k)}$ with $k>0$ can be determined systematically out of $\Omega_n\equiv \Omega_n^{(0)}$ and $H_{mn}$ via the definition of $|\Omega^{(k)}\rangle$ in \eqref{Sect_Krylov_Set_Vectors_Powers_H}. 
Not being specific about whether some subset of \eqref{Sect_Krylov_Set_Vectors_Powers_H} forms a basis of the Hilbert space $\mathcal{H}$ (or of a subspace of it), an aspect on which we shall elaborate later, we can consider the projection the vectors $|\Omega^{(k)}\rangle$ over some new vector $|\psi\rangle\in \mathcal H$, not necessarily equal to $|\Omega\rangle$:
\begin{equation}
    \label{Sect_Krylov_Projections}
    x^{(k)}:=\langle \psi | \Omega^{(k)}\rangle=\langle \psi | H^k | \Omega \rangle~,
\end{equation}
for\footnote{Originally \cite{Krylov:1931}, Krylov did not consider a fixed vector $|\psi\rangle$ but a one-parameter family of them given by the solution $|\psi(t)\rangle$ of a differential equation of the form $\frac{d}{dt}|\psi(t)\rangle = H|\psi(t)\rangle$ seeded by some initial condition $|\psi\rangle = |\psi(0)\rangle$. With this, the quantities defined in \eqref{Sect_Krylov_Projections} become $t$-dependent, denoted $x^{(k)}(t)$, and are in fact given by the $k$-th derivative of $x^{(0)}(t)=\langle \psi(t)|\Omega \rangle\equiv x(t)$. However, this additional structure is not required for our discussion.} $k\geq 0$.
For simplicity, we shall assume that $|\psi\rangle$ is normalized, so that the quantities \eqref{Sect_Krylov_Projections} can indeed be called projections.
Again, using the decomposition of $|\psi\rangle$ in terms of the basis \eqref{Sect_Krylov_Arbitrary_Basis},
\begin{equation}
    \label{Sect_Krylov_psi_coords}
    |\psi\rangle = \sum_{n=1}^D \psi_n |\mathcal{B}_n\rangle~,
\end{equation}
we can rewrite \eqref{Sect_Krylov_Projections} as:
\begin{equation}
    \label{Sect_Krylov_Projections_coords}
    x^{(k)}=\sum_{n=1}^D \psi_n^{*}\, \Omega_n^{(k)} ~.
\end{equation}
Note that, for every $k$, the identity \eqref{Sect_Krylov_Projections_coords} holds regardless of the choice of basis, since $x^{(k)}$ is an inner product. We can now consider the expressions with $0\leq k \leq D$ and regard them as an inhomogeneous system of $D+1$ equations for $D$ unknowns $\left\{\psi_n^{*}\right\}_{n=1}^D$. Such a system is at risk of being over-constrained, but it is possible to find a condition for it to yield a non-trivial solution: We can introduce the auxiliary variable $\psi_0^{*}$ and pose the system
\begin{equation}
    \label{Sect_Krylov_system_aux}
    0=-\psi_0^{*} \, x^{(k)} \,+\sum_{n=1}^D \psi_n^{*}\,\Omega_n^{(k)}~,\qquad \text{for }k=0,...,D~.
\end{equation}
The equations in \eqref{Sect_Krylov_system_aux} form an homogeneous system of $D+1$ equations for $D+1$ unknowns $\left\{\psi_n^{*}\right\}_{n=0}^k$. It either admits only the trivial solution or a family of solutions that spans a vector space. A solution of \eqref{Sect_Krylov_system_aux} with $\psi_0=1$ yields automatically a solution of \eqref{Sect_Krylov_Projections_coords} and, therefore, a necessary condition for \eqref{Sect_Krylov_Projections_coords} to admit a non-trivial solution is the vanishing of the following determinant:
\begin{equation}
    \label{Sect_Krylov_det_x_k}
    \begin{vmatrix}    
    x^{(0)} & \Omega_1&\dots &\Omega_D\\
    x^{(1)} & \Omega_1^{(1)}&\dots &\Omega_D^{(1)}\\
    \vdots & \vdots & \ddots & \vdots \\
    x^{(D)} & \Omega_1^{(D)}&\dots & \Omega_D^{(D)}
    \end{vmatrix} =0~.
\end{equation}
Note that \eqref{Sect_Krylov_det_x_k} may be regarded as the condition that the vectors $\left\{|\Omega^{(k)}\rangle \right\}_{k=0}^D$ are not all linearly independent:
If some subset $\left\{|\Omega^{(k)}\rangle \right\}_{k\in I}$, where $I$ denotes a collection of non-negative integer values for $k$, satisfies a linear relation, then the values of the projections $\left\{x^{(k)}\right\}_{k\in I}$ will satisfy the same linear relation, and so will the corresponding lines in the determinant \eqref{Sect_Krylov_det_x_k}, making it vanish. Such a linear relation must compulsorily exist within $\left\{|\Omega^{(k)}\rangle \right\}_{k=0}^D\in \mathcal{H}$ because $\dim \mathcal{H}=D$.

The determinant in \eqref{Sect_Krylov_det_x_k} takes a very suggestive form if the vector $|\psi\rangle$ onto which the elements of \eqref{Sect_Krylov_Set_Vectors_Powers_H} are projected is an eigenvector of $H$, verifying
\begin{equation}
    \label{Sect_Krylov_Psi_eigenstate}
    H|\psi\rangle = \lambda |\psi\rangle~,
\end{equation}
where $\lambda$ belongs to the spectrum of $H$, i.e. $\lambda\in\sigma(H)$. Given an eigenbasis of $H$, denoted by
$\left\{|\lambda_n\rangle\right\}_{n=1}^D$, we have that $|\psi\rangle$ could either be equal to an element of the basis or, in case some eigenvalues are degenerate, to some linear combination of the basis elements that span the degenerate eigenspace. This is why we kept the notation $|\psi\rangle$ instead of using something like ``$|\lambda\rangle$'', keeping in mind that $|\psi\rangle$ does satisfy \eqref{Sect_Krylov_Psi_eigenstate}. Applying this property in \eqref{Sect_Krylov_Projections} we find that
\begin{equation}
    \label{Sect_Krylov_Projections_lambda}
    x^{(k)}=\lambda^k \langle \psi | \Omega\rangle = \lambda^k\,x^{(0)}
\end{equation}
and, inserting this in the determinant \eqref{Sect_Krylov_det_x_k}, we reach:
\begin{equation}
    \label{Sect_Krylov_det_lambda_k_overlap}
    \langle\psi  |\Omega\rangle\, \begin{vmatrix}    
    1 & \Omega_1&\dots &\Omega_D\\
    \lambda &\Omega_1^{(1)}&\dots &\Omega_D^{(1)}\\
    \vdots & \vdots & \ddots & \vdots \\
    \lambda^D & \Omega_1^{(D)}&\dots & \Omega_D^{(D)}
    \end{vmatrix} =0~.
\end{equation}
Assuming that $\langle \psi  | \Omega\rangle\neq 0$, i.e. that our probe vector $|\Omega\rangle$ has a non-vanishing overlap with the eigenvector of interest $|\psi\rangle$, \eqref{Sect_Krylov_det_lambda_k_overlap} implies:
\begin{equation}
    \label{Sect_Krylov_det_lambda_k}
    \begin{vmatrix}    
    1 & \Omega_1&\dots &\Omega_D\\
    \lambda & \Omega_1^{(1)}&\dots &\Omega_D^{(1)}\\
    \vdots & \vdots & \ddots & \vdots \\
    \lambda^D & \Omega_1^{(D)}&\dots & \Omega_D^{(D)}
    \end{vmatrix} =0~.
\end{equation}

If the left-hand side of \eqref{Sect_Krylov_det_lambda_k} is not identically zero for all values of $\lambda$, then the above expression
will be a polynomial equation of degree $D$ in $\lambda$, whose solutions will therefore yield the eigenvalues of $H$. It is therefore the secular equation for this linear operator and, furthermore, we can define the polynomial
\begin{equation}
    \label{Sect_Krylov_charac_polyn_up_to_constant}
    P(\lambda):=\begin{vmatrix}    
    1 & \Omega_1&\dots &\Omega_D\\
    \lambda & \Omega_1^{(1)}&\dots &\Omega_D^{(1)}\\
    \vdots & \vdots & \ddots & \vdots \\
    \lambda^D & \Omega_1^{(D)}&\dots & \Omega_D^{(D)}
    \end{vmatrix} =:\sum_{n=0}^D P_k\,\lambda^k
\end{equation}
which, up to a multiplicative constant (which is irrelevant for the secular equation) is equal to the characteristic polynomial $S(\lambda)$ defined in \eqref{Sect_Krylov_charac_polyn_def}. As announced, and as claimed by Krylov, the advantage of rewriting the characteristic polynomial as \eqref{Sect_Krylov_charac_polyn_up_to_constant} resides in the fact that its coefficients $P_k$ can be computed systematically by developing the determinant along its first column. 
Explicitly, they are given by
\begin{equation}
    \label{Sect_Krylov_Coefficients_Krylov_charac_polyn}
    P_k = (-1)^k \begin{vmatrix}
        \Omega_1 & \Omega_2 & \dots &\Omega_D \\
        \vdots & \vdots & \ddots & \vdots \\
        \widecheck{\Omega_1^{(k)}} & \widecheck{\Omega_2^{(k)}} & \dots & \widecheck{\Omega_D^{(k)} }\\
        \vdots & \vdots & \ddots & \vdots \\
        \Omega_1^{(D)} & \Omega_2^{(D)} & \dots & \Omega_D^{(D)}
    \end{vmatrix}=:(-1)^{k} \mathcal{D}_\Omega^{(k)}~,
\end{equation}
where the symbol `` $\widecheck{\quad}$ '' indicates that the marked elements are not present in the determinant: $\mathcal{D}_\Omega^{(k)}$ is the determinant of the $D$-dimensional matrix built, line by line, out of the coordinates of the vectors $\left\{|\Omega^{(l)}\rangle\right\}_{l=0}^D \setminus \left\{|\Omega^{(k)}\rangle\right\}$, and it may be regarded as the determinant testing whether they are linearly independent.

Furthermore, having accepted that $P(\lambda)\propto S(\lambda)$, the proportionality constant relating both polynomials can be computed by comparing their leading coefficients, $S_D=(-1)^D$ and $P_D=(-1)^D \mathcal{D}_\Omega ^{(D)}$,
yielding:
\begin{equation}    \label{Sect_Krylov_charac_polyns_proportionality_constant}
     \frac{P(\lambda)}{S(\lambda)} = \frac{P_D}{S_D} = \mathcal{D}_\Omega^{(D)}~. 
\end{equation}

Finally, Krylov's method for the efficient numerical construction of the characteristic polynomial associated to a hermitian operator $H$ may be formulated algorithmically in terms of the following steps:

\begin{enumerate}
\setcounter{enumi}{-1}
    \item Choose\footnote{For the sake of the numerical implementation, it is understood that the basis \eqref{Sect_Krylov_Arbitrary_Basis} is the computational basis, and hence the choice of $|\Omega\rangle$ is equivalent to the choice of the coordinates $\left\{\Omega_n\right\}_{n=1}^D$. Likewise, it is understood that $H$ is given through its coordinates $\left\{H_{mn}\right\}_{m,n=1}^D$ over such a basis.} a probe vector $|\Omega\rangle$.
    \item Build the vectors $|\Omega^{(k)}\rangle=H^k|\Omega\rangle$ for $0\leq k \leq D$ via the iterative application of $H$ on $|\Omega\rangle$.
    \item With the vectors built in the previous step, compute the determinants $\mathcal{D}_\Omega^{(k)}$ for $0\leq k \leq D$ and use them to obtain the coefficients $P_k$ of the characteristic polynomial, following \eqref{Sect_Krylov_Coefficients_Krylov_charac_polyn}.
\end{enumerate}

Even though expressed in modern terms, these are the main ingredients of the mathematical arguments presented in \cite{Krylov:1931} and reviewed in \cite{Luzin:1931}. These works suffer from some caveats that have already been announced. Namely, they do not specify under which conditions the objects $|\Omega^{(k)}\rangle$ in the set \eqref{Sect_Krylov_Set_Vectors_Powers_H} span the full Hilbert space $\mathcal{H}$ or, alternatively, what is the subspace that they span. If the eigenvector $|\psi\rangle$ associated to some eigenvalue $\lambda$ does not belong to such a subspace, we have that $\langle \psi |\Omega \rangle = 0$ and therefore the vanishing of \eqref{Sect_Krylov_det_lambda_k} does not necessarily follow from \eqref{Sect_Krylov_det_lambda_k_overlap}, invalidating the subsequent argument on the construction of the secular equation. Likewise, as we shall see below, the determinant in \eqref{Sect_Krylov_det_lambda_k} vanishes identically for any value of $\lambda$ if there are degeneracies in the spectrum of $H$ and if the associated degenerate eigenspaces are contained in the space spanned by \eqref{Sect_Krylov_Set_Vectors_Powers_H}.

\subsection{A critique of Krylov's method}\label{sect:Krylov_critique}

We shall now provide a more complete analysis of the algebraic structures at hand that parallels the reasoning presented in \cite{I} in the context of operator Krylov complexity, the details of which can be found in Chapter \ref{ch:chapter03_SYK}. The starting point is to choose a convenient basis in terms of which the manipulations in the previous section can be rewritten and shown to take a suggestive form. This can be done since all the relevant expressions are basis-independent, even though they have been written in terms of coordinates. The natural basis to use for our purposes is the eigenbasis of $H$, that is
\begin{equation}
    \label{Sect_Krylov_basis_is_eigenbasis}
    |\mathcal{B}_n\rangle = |\lambda_n\rangle~,
\end{equation}
for $n=1,\, ...,\,D$, in terms of which the coordinates of the probe state $|\Omega\rangle$ introduced in \eqref{Sect_Krylov_Probe_coords} become
\begin{equation}
    \label{Sect_Krylov_probe_coords_eigenbasis}
    \Omega_n = \langle \lambda_n|\Omega\rangle~.
\end{equation}
Furthermore, we can see that the coordinates \eqref{Sect_Krylov_Coords_All_vecs_of_set} of the elements in the set \eqref{Sect_Krylov_Set_Vectors_Powers_H} can be easily determined in terms of the $\Omega_n$:
\begin{equation}
    \label{Sect_Krylov_set_vectors_eigenbasis}
    |\Omega\rangle = \sum_{n=1}^D \Omega_n|\lambda_n\rangle \quad \Longrightarrow \quad H^k|\Omega \rangle = \sum_{n=1}^D \lambda_n^k\,\Omega_n|\lambda_n\rangle~,
\end{equation}
for any $k\geq 0$. That is:
\begin{equation}
    \label{Sect_Krylov_set_vectors_eigenbasis_coords}
    \Omega_n^{(k)}= \lambda_n^k\,\Omega_n\,,\quad \forall\; k\geq 0~. 
\end{equation}
We note that, according to \eqref{Sect_Krylov_set_vectors_eigenbasis}, each application of $H$ has the effect of amplifying the projection of $|\Omega\rangle$ over the eigenspaces with which it overlaps. Hence, the infinite set \eqref{Sect_Krylov_Set_Vectors_Powers_H} spans a $K$-dimensional subspace of $\mathcal H$, 
where $K$ is the number of eigenspaces of $H$ on which $|\Omega\rangle$ has a non-vanishing projection, and the first $K$ elements of the set suffice to form a basis of such a subspace, as we may prove in this section.
Following \cite{ViswanathMuller,Parker:2018yvk,I}, the space
\begin{equation}
    \label{Sect_Krylov_Krylov_Space_Def}
    \mathcal{H}_{|\Omega\rangle} := \text{span}\left\{H^k|\Omega\rangle\right\}_{k\geq 0}
\end{equation}
shall be dubbed the \textit{Krylov space} associated to the probe vector $|\Omega\rangle$, and its dimension $K$ shall be referred to as the \textit{Krylov dimension} \cite{I}. Since $K$ is given by the number of eigenspaces of $H$ with which the probe overlaps, it immediately follows that
\begin{equation}
    \label{Sect_Krylov_bound_generic}
    1\leq K\leq D
\end{equation}
for any non-null seed. As a corollary, we may note that, if $D$ is finite, even though the set \eqref{Sect_Krylov_Set_Vectors_Powers_H} has a countably infinite cardinality, its linear span must have a finite dimension because it is always contained in $\mathcal{H}$.

Using \eqref{Sect_Krylov_set_vectors_eigenbasis_coords}, we can rewrite the polynomial $P(\lambda)$ defined in \eqref{Sect_Krylov_charac_polyn_up_to_constant} as 
\begin{equation}
    \label{Sect_Krylov_polynomial_Vdm}
    P(\lambda) = \prod_{n=1}^D\bigg( \langle  \lambda_n| \Omega\rangle \bigg) \, \begin{vmatrix}
        1 & 1 & \dots & 1 \\
        \lambda & \lambda_1 & \dots & \lambda_D \\
        \vdots & \vdots & \ddots & \vdots \\
        \lambda^D & \lambda_1^D & \dots & \lambda_D^D
    \end{vmatrix}~.
\end{equation}
The determinant above is a Vandermonde determinant, which can be developed to yield:
\begin{equation}
    \label{Sect_Krylov_polynomial_Vdm_developed}
    P(\lambda) = \prod_{n=1}^D\bigg( \langle \lambda_n | \Omega\rangle \bigg) \, \prod_{\substack{n,m=0 \\ n>m}}^D\left(\lambda_n-\lambda_m\right)~,
\end{equation}
where, for notational simplicity, we have made the identification $\lambda_0\equiv \lambda$. If the spectrum is non-degenerate and all the overlaps $\langle\lambda_n|\Omega\rangle$ are non-zero, expression \eqref{Sect_Krylov_polynomial_Vdm_developed} proves that the zeroes of $P(\lambda)$ are the eigenvalues of $H$, and thus it is equal (up to a constant) to the characteristic polynomial. At this stage, two comments are in order:

\begin{itemize}
    \item[\textit{i)}] If $\langle\lambda_n|\Omega\rangle=0$ for some $n\in\{1,\,...,\,D\}$, then $P(\lambda)=0$ for any value of $\lambda$, becoming a meaningless object. For it to yield a non-trivial polynomial, one would have to repeat the steps leading to its construction excluding from the Hilbert space those eigenspaces over which $|\Omega\rangle$ does not have a projection; this would yield a polynomial of lower degree whose zeroes would give (assuming the absence of degeneracies) the remaining eigenvalues whose associated eigenspaces $|\Omega\rangle$ is able to probe.
    \item[\textit{ii)}] If there is a degeneracy in the spectrum of $H$, e.g. $\lambda_n=\lambda_m$ for some $m\neq n$ such that $n,m\in\{1,\, ...,\, D\}$, then the Vandermonde determinant in \eqref{Sect_Krylov_polynomial_Vdm_developed} is identically zero and therefore $P(\lambda)$ will be trivial again. In this case, one would need to repeat the steps excluding from the Hilbert space all but one direction from the degenerate subspace.
    Assuming there are no other degeneracies in the spectrum of $H$, the resulting polynomial would have a lower degree and non-degenerate roots giving the eigenvalues that $|\Omega\rangle$ probes, but nothing in the structure of $P(\lambda)$ would signal which one was the originally degenerate eigenvalue.
\end{itemize}

Of course, the actions described in (\textit{i}) and (\textit{ii}) are virtually impossible to carry out if one does not know \textit{a priori} the spectrum of $H$ and the corresponding eigenvectors. In practice, a possible way out consists on blindly trying different choices for the probe vector $|\Omega\rangle$ and combining the information obtained from each one. Originally, Krylov's article \cite{Krylov:1931} assumed the absence of degeneracies in the spectrum of $H$ and proposed to use as a probe a vector whose coordinates in some arbitrary basis take the simple form $\Omega_n=\delta_{n,1}$, with the expectation that this should be enough to probe all eigenspaces of $H$ if the basis is completely uncorrelated from the eigenbasis. 

The last discussion, together with the fact that it is able to give information on the eigenvalues but not on the eigenvectors, are the main limitations of Krylov's method in the form in which it has been described.
The work by Cornelius Lanczos \cite{Lanczos:1950zz}, which will be reviewed next, improved on these issues by proposing systematic procedure to obtain the full spectrum of eigenvalues and their associated eigenvectors even in the case of an operator $H$ with a degenerate spectrum.

\section{The Lanczos algorithm}\label{sect:Lanczos}

In his paper \cite{Lanczos:1950zz}, the Hungarian mathematician and physicist C. Lanczos reviewed different algorithms for the numerical resolution of first-order linear matrix differential equations, namely the Liouville-Neumann expansion and the Schmidt series \cite{Whittaker_Watson_1996} (see \cite{Lanczos:1950zz} for a brief summary of both methods) and, building upon them, he proposed a method, which he dubbed the S-expansion, for the numerical diagonalization of matrices, which exploits the use of the vectors $|\Omega^{(k)}\rangle = H^k|\Omega\rangle$ built out of a probe $|\Omega\rangle$ in order to generate the characteristic polynomial of $H$ and to express the eigenvectors in coordinates over a (non-orthogonal) basis built out of the maximal linearly independent subset of \eqref{Sect_Krylov_Set_Vectors_Powers_H}, $\left\{|\Omega^{(k)}\rangle\right\}_{k=0}^{K-1}$. In the very same article, Lanczos already noticed that a method like this suffers from important numerical errors when implemented in a machine using finite-precision arithmetics, since large eigenvalues are strongly amplified in the construction of $|\Omega^{(k)}\rangle=H^k|\Omega\rangle$, overshadowing the smaller ones, which eventually become hard to determine numerically with good accuracy. In order to improve on this, he proposed yet another iterative method, still based on the repeated action of the matrix of study on a probe vector, which is nowadays known as the \textit{Lanczos algorithm} or the \textit{recursion method} \cite{ViswanathMuller}. We will directly present this procedure, as its derivation is self-contained and it constitutes the main building block of our eventual study of Krylov complexity. Just like the preceding one, this section and the subsequent ones will use modern tools and notation related to Hilbert spaces in order to review and expand on Lanczos' proposal.

We consider again a hermitian operator $H$ acting on a $D$-dimensional Hilbert space $\mathcal{H}$ from which we take a probe vector $|\Omega\rangle$. As announced, numerical methods based on the construction with finite-precision arithmetics of a basis $\left\{H^k |\Omega\rangle\right\}_{k=0}^{K-1}$ will suffer from numerical errors that may result in an inaccurate prediction of the eigenvalues of $H$ with smaller absolute value. To understand this, we may observe expression \eqref{Sect_Krylov_set_vectors_eigenbasis}, which shows that successive applications of $H$ over the probe vector amplify more significantly its projection over the eigenspaces corresponding to the largest eigenvalues, relatively washing away the contribution from the smaller ones. In order to cure this, Lanczos proposed an iteration method that constructs progressively a basis $\left\{|K_n\rangle\right\}_{n=0}^{K-1}$ for the Krylov space\footnote{Lanczos' work was independent from Krylov's, and hence the term \textit{Krylov space} is not originally used in \cite{Lanczos:1950zz}.} of $|\Omega\rangle$, $\mathcal{H}_{|\Omega\rangle}$ defined in \eqref{Sect_Krylov_Krylov_Space_Def}: This basis, commonly referred to as the \textit{Krylov basis} \cite{ViswanathMuller,Parker:2018yvk,Parlett}, is generated sequentially, acting in each step with $H$ over the previously constructed vector, but subtracting from it a certain linear combination of the preceding basis elements demanding that the result has minimal norm: this allows for all the eigenvectors of the matrix $H$ contained in the Krylov space to alternate dominance in the different basis elements, avoiding the direction associated to leading eigenvalue from outstanding over the rest. As we shall review, the basis output by this optimization process is orthonormal, and the Lanczos algorithm turns out to be a refined version of the Gram-Schmidt process particularized to the objects $H^k|\Omega\rangle$. Let us derive it step by step:

\begin{enumerate}
\setcounter{enumi}{-1}
    \item The initial basis element is assigned to be the seed vector itself:
    \begin{equation}
        \label{Sect_Lanczos_Zeroth_Krylov_element}
        |K_0\rangle = |\Omega\rangle~.
    \end{equation}
    For simplicity, we assume the seed vector to be normalized to unity, and hence so is the Krylov element $|K_0\rangle$. In what follows, shall use the terms ``seed'' and ``probe'' interchangeably, since the vector $|\Omega\rangle$ is indeed the seed of the algorithm, but it may also be regarded as the probe with which the Hilbert space $\mathcal{H}$ is explored, as it determines the dimension of the Krylov space \eqref{Sect_Krylov_Krylov_Space_Def}, depending on which a bigger or smaller portion of the spectrum of $H$ will be accessible. We already gave some hints about this in section \ref{sect:Krylov_critique}, and in section \ref{sect:eigenspace_representatives} we will provide a more complete argument.
    \item We start building a non-normalized version of the next basis element by applying $H$ to the previous one and subtracting from it a vector in the direction of the latter:
    \begin{equation}
        \label{Sect_Lanczos_A1}
        |A_1\rangle = H|K_0\rangle-x|K_0\rangle~,
    \end{equation}
    for some c-number $x$, which shall be determined from the requirement that $|A_1\rangle$ has minimal norm, as announced:
    \begin{equation}
        \label{Sect_Lanczos_Norm_A1}
        f(x):= {\lVert |A_1\rangle \rVert}^2 = {\lVert H|K_0\rangle  \rVert}^2 +x^2 -2x\,\langle K_0 | H | K_0\rangle~.
    \end{equation}
    We note that $f(x)$ is a convex parabola, hence admitting an absolute minimum, which is what we seek, located at
    \begin{equation}
        \label{Sect_Lanczos_a0_min}
        x=\langle K_0 | H | K_0 \rangle =: a_0~.
    \end{equation}
    The coefficient $a_0$ is the first item of one of the two sequences of \textit{Lanczos coefficients} that we shall encounter in this process. As a consequence of this optimization, the vector $|A_1\rangle$ is orthogonal to $|K_0\rangle$:
    \begin{equation}
        \label{Sect_Lanczos_orthogonality_first_Lanczos_step}
        \langle K_0 | A_1\rangle = \langle K_0 | \Big( H|K_0\rangle - \langle K_0 | H |K_0\rangle\,|K_0\rangle \Big) = 0~.
    \end{equation}

    We can now proceed to normalize $|A_1\rangle$:
    \begin{equation}
        \label{Sect_Lanczos_b1_def}
        b_1:=\lVert |A_1\rangle \rVert = \sqrt{f(a_0)}~,
    \end{equation}
    where $b_1>0$ by definition. The normalized basis element is:
    \begin{equation}
        \label{Sect_Lanczos_K1_def}
        |K_1\rangle = \frac{1}{b_1}|A_1\rangle = \frac{1}{b_1}\left(H-a_0\right)|K_0\rangle
    \end{equation}

    \item For the next basis element, we again apply $H$ on the last element and subtract some linear combination of the two already constructed ones:
    \begin{equation}
        \label{Sect_Lanczos_A2_def}
        |A_2\rangle = H |K_1\rangle-x |K_1\rangle - y |K_0\rangle~,
    \end{equation}
    where again the value of the unknown variables $x$ and $y$ shall be fixed by minimizing the norm of $|A_2\rangle$:
    \begin{equation}
        \label{Sect_Lanczos_Norm_A2}
        F(x,y):= {\lVert |A_2\rangle \rVert}^2= {\lVert H |K_1\rangle\rVert}^2 +x^2-2x\,\langle K_1| H |K_1\rangle +y^2 -2y\,\text{Re}\Big(\langle K_1| H |K_0\rangle\Big)~.
    \end{equation}
    Thanks to the fact that $|K_0\rangle$ and $|K_1\rangle$ are orthogonal, the function $F(x,y)$ contains no mixing term proportional to $xy$: It is a sum of two convex parabolas of a single variable and as such it has no saddle point but an absolute minimum, which can be determined taking the corresponding partial derivatives:
    \begin{equation}
        \label{Sect_Lanczos_Derivs_Fxy}
        \begin{split}
            &\partial_x F(x,y)=0\quad \Longleftrightarrow \quad x = \langle K_1 | H | K_1 \rangle =:a_1 ~,\\
            & \partial_y F(x,y)=0 \quad\Longleftrightarrow \quad y = \text{Re}\,\langle K_1 | H | K_0 \rangle~.
        \end{split}
    \end{equation}
    The optimal value for $y$ can be simplified further making use of the knowledge of the already constructed basis elements $|K_0\rangle$ and $|K_1\rangle$: Recalling \eqref{Sect_Lanczos_K1_def}, we have that 
    \begin{equation}
        \label{Sect_Lanczos_b1_developped}
        \langle K_1 | H | K_0 \rangle =  \langle K_1 | \Big( b_1 |K_1\rangle + a_0 | K_0 \rangle \Big)=b_1~,
    \end{equation}
    where we have used the fact that $\{ |K_0\rangle, |K_1\rangle \}$ is an orthonormal set, once $|K_1\rangle$ is constructed using the right coefficients, explicitly written in \eqref{Sect_Lanczos_K1_def}. Summarizing, the square of the norm of the vector $|A_2\rangle$ constructed in \eqref{Sect_Lanczos_A2_def} is given by $F(x,y)$, defined in \eqref{Sect_Lanczos_Norm_A2}, and it is minimal when $(x,y)=(a_1,b_1)$, where $a_1$ is given in \eqref{Sect_Lanczos_Derivs_Fxy} and $b_1$ already appeared in \eqref{Sect_Lanczos_b1_def}.

    Just like in step 1, we can now verify that this optimization grants orthogonality of $|A_2\rangle$ with $|K_0\rangle$ and $|K_1\rangle$. On the one hand we have:
    \begin{equation}
        \label{Sect_Lanczos_Orthog_K1_A2}
        \langle K_1 | A_2 \rangle = \langle K_1 |(H-a_1)|K_1\rangle - b_1 \langle K_1|K_0\rangle = 0~,
    \end{equation}
    where we have used the definition of $a_1$ and the fact that $\{|K_0\rangle,|K_1\rangle\}$ is an orthonormal set. On the other hand:
    \begin{equation}
        \label{Sect_Lanczos_orthog_K0_A2}
        \langle K_0 |A_2\rangle = \langle K_0 |\Big\{ (H-a_1)|K_1\rangle -b_1 |K_0\rangle \Big\}=0~,
    \end{equation}
    where we have used, again, orthonormality of $\{|K_0\rangle,|K_1\rangle\}$, together with \eqref{Sect_Lanczos_b1_developped}.
    We are now in position to normalize $|A_2\rangle$. Defining a new Lanczos coefficient,
    \begin{equation}
        \label{Sect_Lanczos_b2_def}
        b_2:= \lVert|A_2\rangle \rVert = \sqrt{F(a_1,b_1)}~,
    \end{equation}
    where $b_2>0$ by definition, the new normalized Krylov element reads:
    \begin{equation}
        \label{Sect_Lanczos_K2_def}
        |K_2\rangle = \frac{1}{b_2}|A_2\rangle= \frac{1}{b_2}\left(H-a_1\right)|K_1\rangle - b_1|K_0\rangle~.
    \end{equation}

    At this point, we could directly move on to the induction step in order to establish the generic recursion for all basis elements $|K_n\rangle$, but for the sake of pedagogy we shall perform an extra step explicitly, in order to illustrate the fact that no more than two lists of coefficients $a_n$ and $b_n$ is required, in contrast to the upper-triangular matrix of orthogonalization coefficients that the regular Gram-Schmidt procedure outputs.
    \item Naively, one would pose
    \begin{equation}
        \label{Sect_Lanczos_A3_naive}
        |A_3\rangle =H|K_2\rangle - x|K_2\rangle - y |K_1\rangle -z |K_0\rangle~,
    \end{equation}
    admitting in principle a subtraction term for all of the previous Krylov elements. We may now observe that the minimization of the norm of $|A_3\rangle$ is equivalent to making this vector orthogonal to $\left\{|K_n\rangle\right\}_{n=0}^2$, since such vectors form an orthonormal basis of a subspace of $\mathcal{H}$ over which $|A_3\rangle$ may have a non-vanishing projection and, therefore, minimizing the norm of the linear combination \eqref{Sect_Lanczos_A3_naive} amounts to making it orthogonal to this subspace. We will provide a formal proof of this statement shortly, but for now let us show explicitly that the norm optimization process does impose $z=0$. The square of the norm of \eqref{Sect_Lanczos_A3_naive} is given by:
    \begin{equation}
        \label{Sect_Lanczos_norm_A3_G_function}
        \begin{split}
            &G(x,y,z):= {\lVert |A_3\rangle \rVert}^2 = {\lVert H |K_2\rangle \rVert}^2 + x^2 + y ^2 + z^2  \\
            & -2x\, \langle K_2 | H | K_2\rangle - 2y\, \text{Re}\,\langle K_2 | H | K_1\rangle - 2z \,\text{Re}\,\langle K_2 | H | K_0 \rangle~.
        \end{split}
    \end{equation}
    Using \eqref{Sect_Lanczos_K2_def}, we have that
    \begin{equation}
        \label{Sect_Lanczos_cross_term_A3_2_1}
        \langle K_2 | H | K_1\rangle = b_2~,
    \end{equation}
    and likewise, with \eqref{Sect_Lanczos_K1_def}, we find 
    \begin{equation}
        \label{Sect_Lanczos_cross_term_A3_2_0}
        \langle K_2 | H | K_0\rangle = 0~.
    \end{equation}
    Importantly, the proofs of both \eqref{Sect_Lanczos_cross_term_A3_2_1} and \eqref{Sect_Lanczos_cross_term_A3_2_0} involve the use of the fact that the set $\left\{|K_n\rangle\right\}_{n=0}^2$ is orthonormal. Note that \eqref{Sect_Lanczos_cross_term_A3_2_0} tells us that the term $H|K_2\rangle$ in \eqref{Sect_Lanczos_A3_naive} is automatically orthogonal to $|K_0\rangle$. Together with the results \eqref{Sect_Lanczos_cross_term_A3_2_1} and \eqref{Sect_Lanczos_cross_term_A3_2_0}, we \textit{define}
    \begin{equation}
        \label{Sect_Lanczos_a2_definition_for_G}
        a_2:=\langle K_2 | H | K_2\rangle~,
    \end{equation}
    so that the function $G$ in \eqref{Sect_Lanczos_norm_A3_G_function} now takes a simpler form:
    \begin{equation}
        \label{Sect_Lanczos_function_G_simple_form}
        G(x,y,z) = {\lVert H|K_2\rangle\rVert}^2 + x^2-2x a_2 + y^2-2y b_2 + z^2~,
    \end{equation}
    with which one can immediately find that the minimum of $G(x,y,z)$ is located at $(x,y,z) = (a_2,b_2,0)$. With this value of the coefficients, the norm of $|A_3\rangle$ is minimal and given by
    \begin{equation}
        \label{Sect_Lanczos_norm_A3_b3_def}
        b_3:= \lVert |A_3\rangle \rVert = \sqrt{G(a_2,b_2,0)}~,
    \end{equation}
    where once again $b_3>0$ by definition. The new, normalized Krylov basis element is therefore 
    \begin{equation}
        \label{Sect_Lanczos_K3_normalized}
        |K_3\rangle = \frac{1}{b_3}|A_3\rangle = \frac{1}{b_3}\left(H-a_2\right)|K_2\rangle- \frac{b_2}{b_3}|K_1\rangle~.
    \end{equation}
    Again, it can be checked that $|K_3\rangle$ is orthogonal to $\left\{|K_n\rangle\right\}_{n=0}^2$.

    \item[\underline{Induction step:}] We can now show by induction that the recursion
    \begin{equation}
        \label{Sect_Lanczos_recursion_proposal}
        |A_{n}\rangle = \left(H-a_{n-1}\right)|K_{n-1}\rangle - b_{n-1} |K_{n-2}\rangle
    \end{equation}
    for $n \geq 2$ is the output of the norm optimization process, yielding an orthonormal basis, when
    \begin{equation}
        \label{Sect_Krylov_a_b_proposal}
        a_n = \langle K_n | H | K_n \rangle~,\quad b_n = \sqrt{\langle A_n|A_n\rangle}~,
    \end{equation}
    and where the normalized Krylov elements satisfy 
    \begin{equation}
        \label{Sect_Lanczos_Kn_proposal}
        |K_n\rangle = \frac{1}{b_n}|A_n\rangle = \frac{1}{b_n}\left(H-a_{n-1}\right)|K_{n-1}\rangle -\frac{b_{n-1}}{b_n}|K_{n-2}\rangle~.
    \end{equation}
    Since \eqref{Sect_Lanczos_recursion_proposal} is a recursion that involves the last as well as the second-to-last items, a proof by induction requires of the verification of the proposed relation for two seeds, which are accounted for by the previous steps explicitly performed. 
    The remaining induction step consists on assuming that the basis constructed as \eqref{Sect_Lanczos_Kn_proposal} is optimal (and hence orthonormal) for $\left\{|K_m\rangle\right\}_{m=0}^n$ and showing that the next item $|K_{n+1}\rangle$ satisfies the same recursion if the requirement of minimal norm of the non-normalized vector $|A_{n+1}\rangle$ is made. We may start by proving in full generality that the requirement of minimal norm is equivalent to making the new non-normalized element $|A_{n+1}\rangle$ orthogonal to all the previous ones: We consider the action of $H$ on $|K_n\rangle$, which can be decomposed as
    \begin{equation}
        \label{Sect_Lanczos_H_Kn_decomposition}
        H|K_n\rangle = |V_{n+1}\rangle + |W_{n+1}\rangle~,
    \end{equation}
    where 
    \begin{equation}
        \label{Sect_Lanczos_H_Kn_decomposition_terms}
        |V_{n+1}\rangle \, \bot\, \text{span}\{|K_m\rangle\}_{m=0}^n~,\qquad |W_{n+1}\rangle \in \text{span}\{|K_m\rangle\}_{m=0}^n~. 
    \end{equation}
    The element $|A_{n+1}\rangle$ is defined by subtracting from $H|K_n\rangle$ some vector $|\xi_n\rangle\in\text{span}\{|K_m\rangle\}_{m=0}^n$,
\begin{equation}
    \label{Sect_Lanczos_An1_decomp}
    |A_{n+1}\rangle = H|K_n\rangle - |\xi_n\rangle
\end{equation}
    such that the result has minimal norm. Combining \eqref{Sect_Lanczos_An1_decomp} with \eqref{Sect_Lanczos_H_Kn_decomposition} and using Pythagoras' theorem we obtain that the square of the norm of $|A_{n+1}\rangle$ is given by
    \begin{equation}
        \label{Sect_Lanczos_An1_norm_decomp}
    \lVert |A_{n+1}\rangle \rVert ^2 = \lVert |V_{n+1}\rangle \rVert ^2 + \lVert |W_{n+1}\rangle-|\xi_n\rangle \rVert ^2~.
    \end{equation}
    This quantity is the sum of two terms that are greater than or equal to zero. It follows immediately that the optimal choice for the subtraction vector $|\xi_n\rangle$ is the one that makes
    \begin{equation}
        \label{Sect_Lanczos_Optimization_orthog_decomp}
        \lVert |W_{n+1}\rangle-|\xi_n\rangle \rVert ^2 =0 \quad \Longleftrightarrow\quad |\xi_n\rangle = |W_{n+1}\rangle
    \end{equation}
    and, as a consequence, the resulting $|A_{n+1}\rangle$ is orthogonal to all $|K_m\rangle$ with $m=0,...,n$.

    In order to build such an $|A_{n+1}\rangle$ from the starting object $H|K_n\rangle$, we begin by noting that the latter is already orthogonal to $\left\{|K_m\rangle\right\}_{m=0}^{n-2}$:
    \begin{equation}
        \label{Sect_Lanczos_HKn_orthog_n2}
        \langle K_m|H|K_n\rangle = \Big(b_{m+1}\langle K_{m+1}| + a_m \langle K_m| + b_m \langle K_{m-1}|\Big)|K_n\rangle~.
    \end{equation}
    According to the assumption of the induction step, $\left\{|K_m\rangle\right\}_{m=0}^n$ is an orthonormal set, and therefore \eqref{Sect_Lanczos_HKn_orthog_n2} is zero if $m+1\leq n-1$, i.e. if $m\leq n-2$. This proves that 
    \begin{equation}
        \label{Sect_Lanczos_Space_up_to_HKn}
        \begin{split}
            & \text{span}\Big( \left\{|K_m\rangle\right\}_{m=0}^n\,\bigcup\, \{H|K_n\rangle\} \Big) \\
            &= \text{span}\{|K_m\rangle\}_{m=0}^{n-2}\,\bigobot \, \text{span}\Big\{|K_{n-1}\rangle,\,|K_n\rangle,\,H|K_n\rangle\Big\}~,
        \end{split}
    \end{equation}
    where the operation in the second line is an orthogonal direct sum.
    Therefore, in order to yield a new vector perpendicular to all previous ones, $H|K_n\rangle$ only needs to be orthogonalized against $|K_{n-1}\rangle$ and $|K_n\rangle$:
    \begin{equation}
        \label{Sect_Lanczos_An1_conclusion}
        |A_{n+1}\rangle = (H-x)|K_n\rangle - y |K_{n-1}\rangle~.
    \end{equation}
    Either minimizing the norm of \eqref{Sect_Lanczos_An1_conclusion} or explicitly demanding orthogonality of $|A_{n+1}\rangle$ to $|K_n\rangle$ and $|K_{n-1}\rangle$ yields
    \begin{equation}
        \label{Sect_Lanczos_Lanczos_coeffs}
        x= a_n = \langle K_n| H |K_n\rangle,\qquad y= b_n= \lVert |A_n\rangle \rVert~.
    \end{equation}
    As usual, the last task is to normalize the vector:
    \begin{equation}
        \label{Sect_Lanczos_Kn1_def_induction}
        b_{n+1}:= \lVert |A_{n+1}\rangle \rVert,\quad\Longrightarrow\quad |K_{n+1}\rangle = \frac{1}{b_{n+1}}|A_{n+1}\rangle~.
    \end{equation}
    Combining \eqref{Sect_Lanczos_An1_conclusion} with \eqref{Sect_Lanczos_Lanczos_coeffs} and \eqref{Sect_Lanczos_Kn1_def_induction} we find that $|K_{n+1}\rangle$ does satisfy the same recursion as $|K_n\rangle$, given in \eqref{Sect_Lanczos_Kn_proposal}. This concludes the proof by induction.
\end{enumerate}

For the sake of clarity, let us write concisely the steps of the \textit{Lanczos algorithm} that has just been derived, in a way that is directly implementable in numerical calculations.

\begin{itemize}
    \item[\underline{$n=0$:}] 
    \begin{itemize}
        \item[$\bullet$] Set $|K_0\rangle = |\Omega\rangle$ (normalized by assumption).
        \item[$\bullet$] Assign $a_0=\langle K_0 | H | K_0\rangle$.
    \end{itemize}
    \item[\underline{$n=1$:}] 
    \begin{itemize}
        \item[$\bullet$] Set $|A_1\rangle = (H-a_0)|K_0\rangle$.
        \item[$\bullet$] \textbf{If} $\sqrt{\langle A_1 | A_1\rangle}=0$ \textbf{break}. \textbf{Otherwise} continue.
        \item[$\bullet$] Assign $b_1 = \sqrt{\langle A_1 | A_1\rangle}$.
        \item[$\bullet$] Set $|K_1\rangle = \frac{1}{b_1}|A_1\rangle$.
        \item[$\bullet$] Assign $a_1 = \langle K_1 | H | K_1 \rangle$.
    \end{itemize}
    \item[\underline{$n\geq 2$:}]
        \begin{itemize}
            \item[$\bullet$] Set $|A_n\rangle = (H-a_{n-1})|K_{n-1}\rangle-b_{n-1}|K_{n-2}\rangle$.
            \item[$\bullet$] \textbf{If} $\sqrt{\langle A_n|A_n\rangle}=0$ \textbf{break}. \textbf{Otherwise} continue.
            \item[$\bullet$] Assign $b_n = \sqrt{\langle A_n|A_n\rangle}$
            \item[$\bullet$] Set $|K_n\rangle=\frac{1}{b_n}|A_n\rangle$.
            \item[$\bullet$] Assign $a_{n} = \langle K_{n} | H | K_{n} \rangle$.
            \item[$\bullet$] Increase $n\longmapsto n+1$ and \textbf{repeat} the current step.
        \end{itemize}
\end{itemize}

In practical implementations, checking whether a variable is zero amounts to testing if it is smaller than some customizable numerical threshold, typically taken to be equal to the working precision, or the square root of it in a less conservative approach. In this sense, testing whether the norm of a vector is zero, or whether the square of this quantity is null, are nonequivalent operations.

As explained in section \ref{sect:Krylov_critique}, the Krylov space 
\begin{equation}
    \label{Sect_Lanczos_Krylov_space}
    \mathcal{H}_{|\Omega\rangle}=\text{span}\left\{H^k|\Omega\rangle\right\}_{k\geq 0}
\end{equation}
is a $K$-dimensional subspace of the full Hilbert space $\mathcal{H}$, where $K$ is the number of eigenspaces of $H$ over which the probe state $|\Omega\rangle$ has a non-vanishing projection, satisfying $1\leq K \leq D$ as already noted by Lanczos \cite{Lanczos:1950zz}.
Importantly, in the cases in which the Krylov dimension is finite, since in every step the Lanczos algorithm constructs a vector that is orthogonal to all previous ones, the only possibility in the $n=K$ step is to stumble on the null vector, since all the directions of Krylov space have been exhausted at that point:
\begin{equation}
    \label{Sect_Lanczos_termination}
    |A_K\rangle = 0 \quad \Longleftrightarrow \quad b_K=0~.
\end{equation}
This justifies the termination criterion included in the algorithmic description above. The output of the recursion method is an orthonormal \textit{Krylov basis} $\left\{|K_n\rangle\right\}_{n=0}^{K-1}$ and two sequences of the nowadays \cite{ViswanathMuller,Parlett} called \textit{Lanczos coefficients}, $\left\{ a_n \right\}_{n=0}^{K-1}$ and $\left\{b_n\right\}_{n=1}^{K-1}$. Altogether, these three sets suffice for the full resolution of the eigenvalue problem restricted to the eigenspaces of $H$ that the seed $|\Omega\rangle$ probes (i.e. those that belong to the Krylov space). 

\section{The structure of Krylov space}\label{sect:eigenspace_representatives}

For completeness, let us now be more specific about the eigenvectors of $H$ that are \textit{represented} in the Krylov space of $|\Omega\rangle$. This discussion follows closely the one in \cite{I} on operator Krylov complexity, which will be presented in Chapter \ref{ch:chapter03_SYK}, but its fundamental aspects may already be generalized at this point. Both the Krylov basis $\left\{|K_n\rangle\right\}_{n=0}^{K-1}$ and the infinite set $\left\{|\Omega^{(k)}\rangle = H^k|\Omega\rangle\right\}_{k\geq 0}$ span the same Krylov space, $\mathcal{H}_{|\Omega\rangle}$, and in order to elucidate its structure it turns out to be more useful, in a first approach, to consider the elements in the latter set. As already noted in \eqref{Sect_Krylov_set_vectors_eigenbasis}, the vectors $|\Omega^{(k)}\rangle$ take a suggestive form, which we shall repeat here, when expressed in terms of the eigenbasis of $H$, denoted\footnote{The nature of the Lanczos algorithm presented in this section suggests the use of a computational indexing convention, starting from $0$ rather than from $1$, which differs from the more mathematically formal convention employed throughout section \ref{sect:Krylov} on Krylov's work. We shall stick to the computational convention for the rest of this Thesis.} $\left\{|\lambda_n\rangle\right\}_{n=0}^{D-1}$:
\begin{equation}
    \label{Sect_Lanczos_Hk_Omega_restatement}
    |\Omega^{(k)}\rangle = H^k|\Omega\rangle = \sum_{n=0}^{D-1} \lambda_n^k\,\Omega_n |\lambda_n\rangle~,
\end{equation}
where $\left\{\Omega_n\right\}_{n=0}^{D-1}$ are the coordinates of the seed $|\Omega\rangle$ over the eigenbasis, given explicitly by
\begin{equation}
    \label{Sect_Lanczos_coords_seed_explicit}
    \Omega_n = \langle \lambda_n|\Omega\rangle~.
\end{equation}

We immediately note in \eqref{Sect_Lanczos_Hk_Omega_restatement} that if $\Omega_n=0$ for some value of $n$, then all vectors $|\Omega^{(k)}\rangle$ will have zero projection over the eigen-direction $|\lambda_n\rangle$. Furthermore, if several eigenvectors correspond to the same, degenerate eigenvalue, they will all be multiplied by the same factor $\lambda_n^k$, hence only picking one particular direction from the degenerate eigenspace. In order to be more specific about this, let us consider the $K$ eigenspaces over which $|\Omega\rangle$ has a non-zero projection and ignore the rest, since they will not contribute to \eqref{Sect_Lanczos_Hk_Omega_restatement}. When we write the spectrum of $H$ as $\sigma(H)=\left\{\lambda_n\right\}_{n=0}^{D-1}$ we are implicitly counting eigenvalues taking their multiplicity into account, i.e. there might be some values $m\neq n$ for which $\lambda_n=\lambda_m$, and in this case the eigenstates $|\lambda_n\rangle$ and $|\lambda_m\rangle$ represent two orthogonal directions within the degenerate eigenspace. In the generic case, the subset of $\sigma(H)$ that is \textit{represented} in Krylov space (i.e. whose eigenspaces overlap with $|\Omega\rangle$) may be structured as follows:
\begin{equation}
    \centering
    \label{Sect_Lanczos_Krylov_space_spectrum}
    \sigma(H) \ni \bigcup_{j=0}^{K-1}\left\{\lambda_n\right\}_{n\in I_j}~,
\end{equation}
where the family of sets $\left\{I_j\right\}_{j=0}^{K-1}$ is \textit{almost} a partition of $\left\{0,\dots,D-1\right\}$, in the following sense:
\begin{equation}
    \label{Sect_Lanczos_almost_partitions_spectrum}
    \begin{split}
        &I_j\subseteq \left\{0,\dots,D-1\right\}~,\qquad\forall\,j=0,...,K-1~, \\
        &I_j\bigcap I_l = \emptyset~, \qquad \forall\, j\neq l\; \text{such that}\; j,l\,\in\,\left\{0,\dots,K-1\right\}~,\\
        &\bigcup_{j=0}^{K-1}I_j~ \subseteq ~\left\{0,\dots,D-1\right\}~.
    \end{split}
\end{equation}
Rigorously speaking, the absence of a strict equality sign in the last line of \eqref{Sect_Lanczos_almost_partitions_spectrum} is what prevents us from calling the family of sets $I_j$ a partition of $\left\{0,\dots,D-1\right\}$. It is designed to account only for the eigenvalues that are represented in Krylov space, in a way such that they are grouped according to degeneracies, that is:
\begin{equation}
    \label{Sect_Lanczos_Krylov_space_evals_def}
    \lambda_n=\lambda_m=:\widetilde{\lambda}_j\qquad\forall\,n,m\in I_j \qquad \text{for fixed }j\in\left\{0,\dots,K-1\right\}~, 
\end{equation}
and $\widetilde{\lambda}_j \neq \widetilde{\lambda}_l$ whenever $j\neq l$.
With this, we may rewrite \eqref{Sect_Lanczos_Hk_Omega_restatement} as:
\begin{equation}
    \label{Sect_Lanczos_Omega_k_representatives}
    |\Omega^{(k)}\rangle = H^k|\Omega\rangle = \sum_{j=0}^{K-1}\widetilde{\lambda}_j^k\sum_{n\in I_j}\Omega_n|\lambda_n\rangle~, 
\end{equation}
where \eqref{Sect_Lanczos_Krylov_space_evals_def} was used in order to bring the eigenvalue outside of the sum over $n$. We see that in all vectors $|\Omega^{(k)}\rangle$, each eigenspace contributes with the same direction, picked by the projection of $|\Omega\rangle$ over it. Making a slight abuse of notation, we define the normalized elements:
\begin{equation}
    \label{Sect_Lanczos_eigenspace_representative_def}
    |\widetilde{\lambda}_j\rangle := \frac{1}{\widetilde{\Omega}_j}\sum_{n\in I_j}\Omega_n |\lambda_n\rangle~,\qquad \text{where}\qquad\widetilde{\Omega}_j:=\sqrt{\sum_{n\in I_j}\Omega_n^2}=\lVert P_j|\Omega\rangle\rVert~,
\end{equation}
where $P_j$ is the projector onto the eigenspace corresponding to the eigenvalue $\widetilde{\lambda}_j$. The vectors $|\widetilde{\lambda}_j\rangle$ are eigenvectors of $H$ with eigenvalue $\widetilde{\lambda}_j$, they are orthonormal for $j=0,\dots,K-1$, and with them we can write
\begin{equation}
    \label{Sect_Lanczos_Omega_k_eigenspace_representatives}
    H^k|\Omega\rangle = \sum_{j=0}^{K-1}\widetilde{\lambda}_j^k \,\widetilde{\Omega}_j|\widetilde{\lambda}_j\rangle~,
\end{equation}
where we note that 
\begin{equation}
    \label{Sect_Lanczos_projection_Omega_representative}
    \widetilde{\Omega}_j=\langle \widetilde{\lambda}_j|\Omega\rangle~,
\end{equation}
which justifies the choice of the notation for the normalization factor introduced in \eqref{Sect_Lanczos_eigenspace_representative_def}.

In view of \eqref{Sect_Lanczos_Omega_k_eigenspace_representatives}, we have three equivalent ways to define the Krylov space of $|\Omega\rangle$:
\begin{equation}
    \label{Sect_Lanczos_Krylov_space_span_representatives}
    \mathcal{H}_{|\Omega\rangle}=\text{span}\left\{H^k|\Omega\rangle\right\}_{k\geq 0}=\text{span}\left\{|\widetilde{\lambda}_j\rangle\right\}_{j=0}^{K-1}=\text{span}\left\{|K_n\rangle\right\}_{n=0}^{K-1}~.
\end{equation}
Because of this, we shall refer to the vectors $|\widetilde{\lambda}_j\rangle$ as the \textit{eigenspace representatives}, since they give the particular direction of each eigenspace of $H$ that contributes to Krylov space. Along the same lines, the subset of eigenvalues
\begin{equation}
    \label{Sect_Lanczos_KrylovSpectrum}
    \widetilde{\sigma}(H):=\left\{\widetilde{\lambda}_j\right\}_{j=0}^{K-1}\in\sigma(H)
\end{equation}
shall be referred to as the \textit{Krylov-space spectrum}. By construction, the action of $H$ over the Krylov space its closed and, as we shall see in section \ref{sect:Lanczos_exact_spectrum}, the knowledge of the Krylov basis and the Lanczos coefficients will allow to reconstruct the spectrum of the restriction of $H$ over $\mathcal{H}_{|\Omega\rangle}$, given by $\widetilde{\sigma}(H)$, as well as the associated eigenvectors, given by the eigenspace representatives $\left\{|\widetilde{\lambda}_j\rangle\right\}_{j=0}^{K-1}$. 

As an important corollary, we note that the restriction of $H$ over the Krylov space always features a non-degenerate spectrum, since each potentially degenerate eigenspace contributes with only one direction to the Krylov space, given by its eigenspace representative.

Once the structure of Krylov space has become neater, we can study the restriction of $H$ over this subspace of $\mathcal{H}$. 
Even though it was not originally noted by Lanczos in \cite{Lanczos:1950zz}, one can verify \cite{Parlett} that the restriction of $H$ over Krylov space, which we shall sometimes denote by $T$, takes the form of a tridiagonal matrix $(H_{mn})_{m,n=0}^{K-1}$ when expressed in coordinates over the Krylov basis, where the non-zero diagonals are given by the Lanczos coefficients. This can be seen using the recursion \eqref{Sect_Lanczos_Kn_proposal}, which implies:
\begin{equation}
    \label{Sect_Lanczos_matrix_elements_H}
    \begin{split}
        &H_{m,0}= b_1\delta_{m,1} +a_0\delta_{m,0}~, \\
        & H_{mn} = b_{n+1}\delta_{m,n+1} + a_n \delta_{m,n} + b_n \delta_{m,n-1}~,\qquad n=1,...,K-2~, \\
        &H_{m,K-1} = a_{K-1}\delta_{m,K-1} + b_{K-1}\delta_{m,K-2}~,
    \end{split}
\end{equation}
where $H_{mn}\equiv \langle K_m | H | K_n \rangle$ and $m= 0,\, ..., \, K-1$ in all cases. In matrix form, this gives:
\begin{equation}
    \label{Sect_Lanczos_Tridiag_H_matrix}
    T \overset{*}{=} (H_{mn}) = \begin{pmatrix}
        a_0 & b_1 & 0  &\dots &0 & 0 & 0 \\
        b_1 & a_1 & b_2 & \dots & 0 & 0 & 0 \\
        0 & b_2 & a_2 & \dots & 0 & 0 & 0 \\
        \vdots & \vdots & \vdots & \ddots & \vdots & \vdots & \vdots \\
        0 & 0 & 0 & \dots & a_{K-3} & b_{K-2} & 0 \\
        0 & 0 & 0 & \dots & b_{K-2} & a_{K-2} & b_{K-1} \\
        0 & 0 & 0 & \dots & 0 & b_{K-1} & a_{K-1}
    \end{pmatrix}~,
\end{equation}
where the asterisk above the equality sign denotes that the right-hand side is an expression in coordinates over some chosen basis (in this case, the Krylov basis).
Hence, the Lanczos algorithm reduces the eigenvalue problem of an arbitrary hermitian operator $H$ to the diagonalization of a tridiagonal matrix of the form \eqref{Sect_Lanczos_Tridiag_H_matrix} \cite{Parlett,KrylovBook_Liesen,KrylovBook_Vorst}.

\section{Krylov polynomials}

As noted in \cite{Lanczos:1950zz}, the recursion between the Krylov elements $|K_n\rangle$ \eqref{Sect_Lanczos_Kn_proposal} defining the Lanczos algorithm can be turned into a recursion constructing a family of orthogonal polynomials, which we shall dub \textit{Krylov polynomials}. These analytical objects will turn out to be very intimately related to the spectrum of $H$ and, whenever such a spectrum is continuous, to the eigenvalue density\footnote{In quantum-mechanical applications, this is often called the density of states. But in order to keep this algebraic discussion as generic as possible we shall refrain from using that terminology in the present Chapter.}.

The recursion \eqref{Sect_Lanczos_Kn_proposal} can be unwrapped in order to express an arbitrary Krylov element $|K_n\rangle$ in terms of the seed $|\Omega\rangle=|K_0\rangle$, yielding:
\begin{equation}
    \label{Sect_Lanczos_recursion_gives_polynomial}
    b_{n}|K_n\rangle = p_n (H) |\Omega\rangle~.
\end{equation}
That is, the $n$-th Krylov basis element is given by the action on the probe vector of an operator which is a polynomial in $H$ of degree $n$, whose coefficients are dictated by the Lanczos coefficients. The recursion between the Krylov elements immediately implies an analogous recursion between the corresponding polynomials:
\begin{equation}
    \label{Sect_Lanczos_recursion_polynomials}
    \begin{split}
        & p_0(\lambda) = 1~, \\
        & p_1(\lambda) = \lambda-a_0~, \\
        & p_n (\lambda) = \frac{\lambda-a_{n-1}}{b_{n-1}}p_{n-1}(\lambda) - \frac{b_{n-1}}{b_{n-2}} p_{n-2} (\lambda)~,\qquad  2 \leq n \leq K~,
    \end{split}
\end{equation}
where we use the convention $b_0\equiv 1$ to ensure well-definedness\footnote{Note that $b_0$ is a ``fictitious'' Lanczos coefficient in the sense that it does not appear in the tridiagonal form of the restriction of $H$ over the Krylov basis \eqref{Sect_Lanczos_Tridiag_H_matrix}; its value may therefore be chosen at will in order to make sense of the recursive relation one wishes to write. As an example, the recursion \eqref{Sect_Lanczos_recursion_proposal} can be extended to $n=1$ by setting $b_0\equiv 1$ together with $|K_{-1}\rangle \equiv 0$.} of the step $n=2$. We shall refer to $\left\{p_n (\lambda)\right\}_{n=0}^{K-1}$ as the \textit{Krylov polynomials}.

The recursion \eqref{Sect_Lanczos_recursion_polynomials} allows to extend the definition of the Krylov polynomials to an extra, \textit{last} Krylov polynomial $p_K(\lambda)$ which is of high importance: The termination condition $|A_K\rangle = 0$ discussed in \eqref{Sect_Lanczos_termination} translates into the statement that 
\begin{equation}
    \label{Sect_Lanczos_p_K_H_zero}
    p_K(H)=0
\end{equation}
\textit{as an operator restricted over Krylov space}. In order to show this, 
we note that the termination condition $|A_K\rangle = 0$ is equivalent to:
\begin{equation}
    \label{Sect_Lanczos_termination_polyn_spectral_decomp}
    p_K(H)|\Omega\rangle = 0 \quad \Longleftrightarrow \quad \sum_{n=0}^{K-1} p_n(\widetilde{\lambda_n})\,\widetilde{\Omega}_n |\widetilde{\lambda}_n\rangle = 0~,
\end{equation}
which immediately implies 
\begin{equation}
    \label{Sect_Lanczos_p_K_0_evals}
    p_K(\widetilde{\lambda}_n)=0\qquad\forall\;n=0,\,1,\,...,\,K-1
\end{equation}
because of orthogonality of the vectors $\left\{|\widetilde{\lambda}_n\rangle\right\}_{n=0}^{K-1}$ and since all $\widetilde{\Omega}_n\neq 0$ by the definition of eigenspace representative. In other words, if we regard $p_K$ as a polynomial of a real variable $\lambda$, namely $p_K(\lambda)$, equation \eqref{Sect_Lanczos_p_K_0_evals} is telling us that its zeroes are given by the eigenvalues of $H$ that lie in Krylov space. Hence, $p_K(\lambda)$ is, up to a multiplicative constant, the characteristic polynomial of the restriction of $H$ to such a subspace, while
\begin{equation}
    \label{Sect_Lanczos_secular_equation_p_K}
    p_K(\lambda) = 0
\end{equation}
is the associated secular equation.

The Krylov polynomials $\left\{p_n(\lambda)\right\}_{n=0}^{K-1}$ form a family of orthogonal polynomials defined over $\widetilde{\sigma}(H)$. This is a consequence of the fact that the Krylov elements $\left\{|K_n\rangle\right\}_{n=0}^{K-1}$ form an orthonormal basis of Krylov space. Using \eqref{Sect_Lanczos_recursion_gives_polynomial}, together with the fact that
\begin{equation}
    \label{Sect_Lanczos_Krylov_Elements_Orthonormal}
    \langle K_m | K_n\rangle = \delta_{mn}~,
\end{equation}
we can prove orthogonality of the polynomials as follows:
\begin{equation}
\label{Sect_Lanczos_polyns_orthogonal_discrete}
    b_n^2\,\delta_{mn}=\langle \Omega | p_m (H) p_n(H)|\Omega\rangle = \sum_{j=0}^{K-1} {|\widetilde{\Omega}_j|}^2 p_m(\widetilde{\lambda}_j) p_n(\widetilde{\lambda}_j)~, 
\end{equation}
where in the last step we used the resolution of the identity over Krylov space, that is:
\begin{equation}
    \label{Sect_Lanczos_resolution_identity_Krylov_space_discrete}
    \sum_{j=0}^{K-1}|\widetilde{\lambda}_j\rangle \langle \widetilde{\lambda}_j | = \widetilde{\mathds{1}}~,
\end{equation}
where $\widetilde{\mathds{1}}$ denotes the identity over Krylov space. Equation \eqref{Sect_Lanczos_polyns_orthogonal_discrete} states that the normalized Krylov polynomials
\begin{equation}
   \label{Sect_Lanczos_normalized_Krylov_polynomials_def}
   q_n(\lambda):=\frac{p_n(\lambda)}{b_n}
\end{equation}
are orthonormal with respect to the discrete measure $\widetilde{\Omega}_j$. Note that we have explicitly excluded the last Krylov polynomial, $p_K$, from the relation \eqref{Sect_Lanczos_polyns_orthogonal_discrete} because it is identically zero when evaluated over $\widetilde{\sigma}(H)$.

Additionally, the Krylov polynomials $\left\{ p_n(\lambda) \right\}_{n=0}^{K-1}$ form a complete basis for the space of polynomials of degree $K-1$ over $\widetilde{\sigma}(H)$, as a completeness relation can be deduced from that of the Krylov elements,
\begin{equation}
    \label{Sect_Lanczos_Completeness_Krylov_basis}
    \sum_{n=0}^{K-1} |K_n\rangle \langle K_n | = \widetilde{\mathds{1}}~.
\end{equation}
Acting with $\langle \widetilde{\lambda}_i |$ from the left and with $|\widetilde{\lambda}_j\rangle$ from the right, which is equivalent to computing the $(ij)$ matrix element of the operator in \eqref{Sect_Lanczos_Completeness_Krylov_basis}, we find
\begin{equation}
    \label{Sect_Lanczos_Krylov_polynomials_completeness_discrete}
    \sum_{n=0}^{K-1} \frac{p_n(\widetilde{\lambda}_i) p_n(\widetilde{\lambda}_j)}{b_n^2} =\frac{\delta_{ij}}{{|\widetilde{\Omega}_i|}^2}~.
\end{equation}

\subsection{The case of a continuous spectrum}\label{sect:KryPolCont}

So far we have implicitly assumed that the spectrum represented in Krylov space, $\widetilde{\sigma}(H)$, is discrete. Let us now provide, for reference, the relations satisfied by the family of Krylov polynomials in the case of a continuous Krylov-space spectrum.

To start, we note that the resolution of the identity \eqref{Sect_Lanczos_resolution_identity_Krylov_space_discrete} should be replaced by
\begin{equation}
    \label{Sect_Lanczos_resolution_identity_Krylov_space_cont}
    \int_{\widetilde{\sigma}(H)} d\lambda\, \widetilde{\rho}(\lambda) |\lambda\rangle \langle \lambda | = \mathds{1}_K~,
\end{equation}
where $\widetilde{\rho}(\lambda)$ is not the eigenvalue density $\rho(\lambda)$ for the full spectrum $\sigma(H)$ of the operator $H$, but that for the subset of $\sigma(H)$ represented in Krylov space, i.e. $\widetilde{\sigma}(H)$. For finite systems whose spectrum becomes continuous in the $D\to +\infty$ limit (whenever such a limit can be defined), the measure may be expressed as
\begin{equation}
    \label{Sect_Lanczos_Krylov_density_of_states_limit}
    \widetilde{\rho}(\lambda) = \lim_{D\to + \infty} \sum_{j=0}^{K(D)-1} \delta\Big( \lambda - \widetilde{\lambda}_j \Big)\qquad \text{in the sense of distributions.}
\end{equation}
Note that we have additionally assumed that the Krylov dimension has some dependence on $D$, $K(D)$. In order for \eqref{Sect_Lanczos_Krylov_density_of_states_limit} to yield a continuous function of $\lambda$ it is necessary that $K(D)\overset{D\to +\infty}{\longrightarrow} +\infty$, which depends additionally on the choice of probe vector $|\Omega\rangle$. Using \eqref{Sect_Lanczos_resolution_identity_Krylov_space_cont} and the fact that Krylov elements are orthonormal, the orthogonality relation for the Krylov polynomials becomes
\begin{equation}
    \label{Sect_Lanczos_polyns_orthogonal_cont}
    b_n^2 \delta_{mn} = \int_{\widetilde{\sigma}(H)}d\lambda\,\widetilde{\rho}(\lambda) {|\widetilde{\Omega}(\lambda)|}^2 p_m(\lambda) p_n(\lambda)~,
\end{equation}
where $\widetilde{\Omega}(\lambda)$ is the continuous version of \eqref{Sect_Lanczos_projection_Omega_representative}. Note that \eqref{Sect_Lanczos_polyns_orthogonal_discrete} may be recovered from \eqref{Sect_Lanczos_polyns_orthogonal_cont} by replacing $\widetilde{\rho}(\lambda)$ and $\widetilde{\Omega}(\lambda)$ by the corresponding finite sums of delta functions.
Equation \eqref{Sect_Lanczos_polyns_orthogonal_cont} states that the normalized Krylov polynomials $q_n(\lambda)$ defined in \eqref{Sect_Lanczos_normalized_Krylov_polynomials_def} are orthonormal with respect to the probe-vector-dependent measure $\widetilde{\rho}(\lambda) {|\widetilde{\Omega}(\lambda)|}^2$. Note, however, that the fact that we are still using discrete labels for the Krylov polynomials, $p_n(\lambda)$, and Krylov elements, $|K_n\rangle$, means that we have implicitly assumed that the Krylov space has a countably infinite dimension\footnote{For an infinite-dimensional operator $H$ that can be reached as the $D\to +\infty$ limit of some family of finite, $D$-dimensional operators, the resulting eigenvalue density can indeed be continuous while having an underlying Hilbert space with a countably infinite dimension if, for example, the spectrum is bounded (from above and from below) by the same constants for any $D$, because in such a case the eigenvalues can accumulate inside the finite interval that contains them (except for special situations in which, for example, a finite fraction of them converges to a fixed point). In other words, the mean level spacing tends to zero. This is the case of the Hamiltonian of the double-scaled SYK model, studied in Chapter \ref{ch:chapter05_DSSYK}, if it is adequately normalized.}: We did this because the Lanczos recursion \eqref{Sect_Lanczos_Kn_proposal} and the very nature of the Lanczos algorithm, constructed as a concatenation of iterative steps, seems to be intrinsically designed to yield a discrete Krylov basis (finite or infinite). Formal extensions of the Lanczos algorithm to the case of systems with an uncountably-infinite Hilbert space have not been rigorously studied yet, and the question appears to be relevant for the realm of quantum field theory and quantum gravity \cite{Witten:2021jzq}.

Obtaining the continuous-spectrum version of the completeness relation \eqref{Sect_Lanczos_Krylov_polynomials_completeness_discrete} involves taking some necessary subtleties into account. In particular, the normalization of the Krylov-space eigenvectors consistent with \eqref{Sect_Lanczos_resolution_identity_Krylov_space_cont} is
\begin{equation}
    \label{Sect_Lanczos_eigenstate_normalization_cont}
    \langle \lambda | \lambda^\prime\rangle = \frac{\delta(\lambda-\lambda^\prime)}{\widetilde{\rho}(\lambda)}~,\qquad \lambda,\,\lambda^\prime\,\in\,\widetilde{\sigma}(H)~.
\end{equation}
This can be checked, for example, posing $\langle \lambda | \lambda^\prime\rangle = \frac{\delta(\lambda-\lambda^\prime)}{f(\lambda)}$ for some unknown function $f$ and inserting the resolution of the identity \eqref{Sect_Lanczos_resolution_identity_Krylov_space_cont} inside the bra-ket of the left-hand side, which will impose $f(\lambda)=\widetilde{\rho}(\lambda)$. Taking this into account, acting on \eqref{Sect_Lanczos_Completeness_Krylov_basis} with $\langle \lambda |$ from the left and with $|\lambda^\prime\rangle$ from the right yields:
\begin{equation}
    \label{Sect_Lanczos_Krylov_polynomials_completeness_cont}
    \sum_{n=0}^{+\infty} \frac{p_n(\lambda) p_n(\lambda^\prime)}{b_n^2} =\frac{\delta(\lambda-\lambda^\prime)}{{|\widetilde{\Omega}(\lambda)|}^2\,\widetilde{\rho}(\lambda)}~,\qquad\;\lambda,\,\lambda^\prime\,\in\,\widetilde{\sigma}(H)~.
\end{equation}

Observing \eqref{Sect_Lanczos_polyns_orthogonal_cont} and \eqref{Sect_Lanczos_Krylov_polynomials_completeness_cont}, one might conclude that the correct redefinition of the Krylov polynomials in order to yield simultaneously canonical orthonormality and completeness relations with respect to the measure $\widetilde{\rho}(\lambda)$ should be 
\begin{equation}
    \label{Sect_Lanczos_canonical_Krylov_polynomials}
    r_n(\lambda):= \frac{\widetilde{\Omega}(\lambda) p_n(\lambda)}{b_n}~.
\end{equation}
Nevertheless, for the present discussion, we shall choose not to absorb the projection of the probe vector over the Krylov-space eigenvector, $\widetilde{\Omega}(\lambda)=\langle \widetilde{\lambda}|\Omega\rangle$, into the polynomial\footnote{Note that, strictly speaking, $\{r_n(\lambda)\}_{n\geq 0}$ need not be a family of polynomials anymore, as this depends on the seed profile $\widetilde{\Omega}(\lambda)$. They will nevertheless constitute some family of orthonormal functions defined over the domain $\widetilde{\sigma}(H)$.}, as the original setup of the problem at hand is to use the Lanczos algorithm to determine the eigenvalues and eigenvectors of the operator $H$. The Lanczos recursion allows to construct the Lanczos coefficients as well as the Krylov elements and polynomials, but $\widetilde{\Omega}(\lambda)$ remains unknown so far because eigenvectors have not been computed yet. An alternative perspective consists on considering that the normalized polynomials $q_n(\lambda)$ defined in \eqref{Sect_Lanczos_normalized_Krylov_polynomials_def} satisfy orthonormality and completeness relations with respect to the seed-dependent measure ${|\widetilde{\Omega}(\lambda)|}^2\,\widetilde{\rho}(\lambda)$.

\section{Exact solution of the eigenvalue problem with the Lanczos algorithm}\label{sect:Lanczos_exact_spectrum}

We are now in conditions to explain how the knowledge of the Lanczos coefficients, together with the Krylov elements and polynomials, is sufficient to obtain a formal solution to the eigenvalue problem, which can be systematically implemented in a computer.

As argued previously, equation \eqref{Sect_Lanczos_secular_equation_p_K} gives the eigenvalues of the operator $H$ that are represented in Krylov space. Once the eigenvalues have been determined, and making use of the Krylov polynomials, it is also possible to compute the associated eigenvectors in closed form. We may start by considering the combination of \eqref{Sect_Lanczos_recursion_gives_polynomial} and \eqref{Sect_Lanczos_normalized_Krylov_polynomials_def}:
\begin{equation}
    \label{Sect_Lanczos_Krylov_element_qn}
    |K_n\rangle = q_n(H)|\Omega\rangle~,
\end{equation}
whose projection over a given eigenspace representative $|\widetilde{\lambda}_j\rangle$ is thus given by:
\begin{equation}
    \label{Sect_Lanczos_Krylov_eigenstate_basis_discrete}
    \langle {\widetilde \lambda}_j | K _n \rangle = \widetilde{\Omega}_j\,q_n(\widetilde{\lambda}_j)~.
\end{equation}
Using this, together with the resolution of the identity granted by the Krylov basis \eqref{Sect_Lanczos_Completeness_Krylov_basis}, we immediately obtain the corresponding Krylov-space eigenvector in coordinates over the Krylov basis:
\begin{equation}
    \label{Sect_Lanczos_evects_in_Krylov_basis_discrete}
    |\widetilde{\lambda}_j\rangle =\sum_{n=0}^{K-1} |K_n\rangle \langle K_n|\widetilde{\lambda}_j\rangle=\widetilde{\Omega}_j^* \sum_{n=0}^{K-1}q_n(\widetilde{\lambda}_j)|K_n\rangle~.
\end{equation}
Note that the Krylov-space eigenvectors are correctly normalized:
\begin{equation}
    \label{Sect_Lanczos_evec_normalized_discrete}
    \langle \widetilde{\lambda}_i | \widetilde{\lambda}_j\rangle = \widetilde{\Omega}_i\widetilde{\Omega}_j^*\sum_{n=0}^{+\infty}q_n(\widetilde{\lambda}_i)q_n(\widetilde{\lambda}_j)=\delta_{ij}~,
\end{equation}
where in the first equality we used orthonormality of the Krylov basis, while in the second we invoked the completeness relation of the Krylov polynomials \eqref{Sect_Lanczos_Krylov_polynomials_completeness_discrete}.

Computationally, when solving the eigenvalue problem for the operator $H$ using the probe $|\Omega\rangle$, the Lanczos algorithm outputs the Krylov basis elements and the Lanczos coefficients from which the Krylov polynomials are built, but the overlaps of the seed with each Krylov-space eigenvector, $\widetilde{\Omega}_j$, remain unknown until the eigenvectors are explicitly constructed. Hence, for the practical task of determining $|\widetilde{\lambda}_j\rangle$ using \eqref{Sect_Lanczos_evects_in_Krylov_basis_discrete}, one may proceed by constructing a non-normalized version of the eigenvector,
\begin{equation}
    \label{Sect_Lanczos_evects_in_Krylov_basis_discrete_unnormalized}
    \sum_{n=0}^{K-1}q_n(\widetilde{\lambda}_j)|K_n\rangle~,
\end{equation}
and ultimately it may be divided by its norm in order to obtain a unit vector. Such a normalization factor will give the overlap $\widetilde{\Omega}_j$, according to \eqref{Sect_Lanczos_evec_normalized_discrete}, up to a global phase which can be chosen arbitrarily and independently for each eigenvector, as this does not affect orthogonality of the eigenbasis. 

\subsection{Continuous case}

For the sake of completeness, let us provide the corresponding expressions for the case of a continuous Krylov-space spectrum. The Krylov-space eigenvectors are given, in terms of the Krylov basis elements, by\footnote{For expressions in the continuum case, for the sake of notational simplicity we have avoided the use of the tilde symbol, always assuming implicitly that the domain of the continuous variable $\lambda$ is the Krylov-space spectrum $\widetilde{\sigma}(H)$. Accordingly, the vectors $|\lambda\rangle$ should be understood as the eigenspace representatives picked by the seed $|\Omega\rangle$, $\widetilde{\Omega}(\lambda)$ being the overlap between the former and the latter.}
\begin{equation}
    \label{Sect_Lanczos_eigenvectors_Krylov_basis_cont}
    |\lambda\rangle = \widetilde{\Omega}(\lambda)^*\sum_{n\geq 0} q_n(\lambda) |K_n\rangle~,
\end{equation}
which can be checked to be consistent with the normalization of the eigenvectors in the continuous case \eqref{Sect_Lanczos_eigenstate_normalization_cont}:
\begin{equation}
    \label{Sect_Lanczos_evecs_normalization_check}
    \langle \lambda | \lambda^\prime \rangle = \widetilde{\Omega}(\lambda) \widetilde{\Omega}(\lambda^\prime)^* \sum_{n\geq 0} q_n(\lambda) q_n(\lambda^\prime)=\frac{\delta(\lambda-\lambda^\prime)}{\widetilde{\rho} (\lambda)}~,
\end{equation}
where, once again, in the first equality we used the fact that the Krylov basis is orthonormal, while in the second we applied the completeness relation \eqref{Sect_Lanczos_Krylov_polynomials_completeness_cont}.

\subsection{Probing the full Hilbert space}

Up to now we have carefully kept track of the Krylov dimension $K$, keeping in mind that $K\leq D$ and that, due to the specific choice of the probe vector $|\Omega\rangle$, or to the presence of degeneracies in the spectrum of $H$, it can happen to be strictly smaller than the full Hilbert space dimension $D$. In such a case, similarly to Krylov (see section \ref{sect:Krylov}), Lanczos proposed in \cite{Lanczos:1950zz} to combine the information coming from different seeds, say $|\Omega\rangle$ and $|\Omega^\prime\rangle$. However, the Lanczos algorithm does not trivialize if there is a degeneracy in the spectrum (the Krylov dimension gets reduced, but it is still able to compute the corresponding Krylov basis and Lanczos coefficients), and it additionally allows to obtain the eigenvectors, so we are in a much better position as compared to section \ref{sect:Krylov_critique}. We can distinguish the two following cases:

\begin{itemize}
    \item[\textit{i})] If the application of the Lanczos algorithm to $|\Omega^\prime\rangle$ yields a new eigenvalue that had not been obtained from the analysis of $|\Omega\rangle$, then we have successfully found another eigenspace of $H$. We may append the corresponding eigenvector to the eigenbasis constructed out of $|\Omega\rangle$.
    \item[\textit{ii})] If $|\Omega^\prime\rangle$ yields the same eigenvalues, then it probes the same eigenspaces as $|\Omega\rangle$. However, there might be a degenerate eigenspace, and this can be checked by comparing the eigenvectors. In particular, let $|\widetilde{\lambda}\rangle$ be the (normalized) eigenvector associated to the eigenvalue $\widetilde{\lambda}$ obtained from the analysis of $|\Omega\rangle$, and let $|\widetilde{\lambda}^\prime\rangle$ be the (also normalized) eigenvector corresponding to the \textit{same} eigenvalue $\widetilde{\lambda}$ obtained from $|\Omega^\prime \rangle$. If 
    \begin{equation}
        \label{Sect_Lanczos_different_probes_degeneracy_alert}
        \langle \widetilde{\lambda} |\widetilde{\lambda}^\prime\rangle \neq 1~,
    \end{equation}
    then the two eigenvectors are different, signaling a degenerate eigenspace. We may then append to the list of eigenvectors obtained from $|\Omega\rangle$ the component of $|\widetilde{\lambda}^\prime\rangle$ orthogonal to $|\widetilde{\lambda}\rangle$, adequately normalized.
\end{itemize}

We can go even further and claim that there exists a systematic procedure that guarantees the extraction of the full spectrum of $H$, together with its complete basis of eigenvectors, regardless of whether there are degeneracies, by the means of a finite number of applications of the Lanczos algorithm. Such a procedure consists on using a complete, orthonormal basis of probe vectors $\left\{|\Omega_l\rangle\right\}_{l=0}^{D-1}$, applying the Lanczos algorithm to every single one of them and comparing the resulting information according to cases (\textit{i}) and (\textit{ii}) above. The end result is granted to yield the complete basis of eigenvectors of $H$. In other words, let $\mathcal{H}_{|\Omega_l\rangle}\equiv \mathcal{H}_l$ be the Krylov space of the probe vector $|\Omega_l\rangle$; then it holds that
\begin{equation}
    \label{Sect_Lanczos_Krylov_spaces_basis_probes_full_space}
    \bigoplus_{l=0}^{D-1}\mathcal{H}_l = \mathcal{H}~,
\end{equation}
i.e. combining all the Krylov spaces always builds up the full Hilbert space $\mathcal{H}$. This statement is not made in the article by Lanczos \cite{Lanczos:1950zz}, and we may prove it as follows: We begin by noting that, in order to obtain the Krylov dimension $K_l$ of each Krylov space $\mathcal{H}_l$, it suffices to decompose formally the probe vector $|\Omega_l\rangle$ in terms of the eigenbasis of $H$:
\begin{equation}
    \label{Sect_Lanczos_probe_basis_decomposition_eigenbasis}
    |\Omega_l\rangle = \sum_{j=0}^{D-1} c_{lj} |\lambda_j\rangle~,\qquad \forall \; l=0,\,...,\,D-1~.  
\end{equation}
Since both $\left\{|\Omega_l\rangle\right\}_{l=0}^{D-1}$ and $\left\{|\lambda_l\rangle\right\}_{l=0}^{D-1}$ are orthonormal bases of the full Hilbert space $\mathcal{H}$, it follows that the matrix of coefficients $(c_{lj})_{l,j=0}^{D-1}$ is unitary and, in particular, its determinant is non-null. We may write it explicitly as
\begin{equation}
    \label{Sect_Lanczos_probe_basis_eigenbasis_matrix}
    (c_{lj})\equiv \begin{pmatrix}
        c_{00} & c_{01} & \dots & c_{0,D-1} \\
        c_{10} & c_{11} & \dots & c_{1,D-1} \\
        \vdots & \vdots  & \ddots & \vdots \\
        c_{D-1,0} & c_{D-1, 1} & \dots & c_{D-1,D-1}
    \end{pmatrix}~.
\end{equation}
We note that each row of \eqref{Sect_Lanczos_probe_basis_eigenbasis_matrix} gives the coordinates of a fixed probe vector over the eigenbasis of $H$. In order to prove \eqref{Sect_Lanczos_Krylov_spaces_basis_probes_full_space}, we only need to show that every eigenvector $|\lambda_j\rangle$ has a non-zero overlap with at least one probe vector $|\Omega_l\rangle$, so that it will be represented in its Krylov space $\mathcal{H}_l$. We can prove this by contradiction: If a given eigenvector was absent from all Krylov spaces $\mathcal{H}_l$ with $l=0,\,...,\,D-1$, then a full column of \eqref{Sect_Lanczos_probe_basis_eigenbasis_matrix} would need to be zero, but this would imply that the determinant of the matrix $(c_{lj})$ is zero, which would be in contradiction with the fact that it is unitary. This concludes the proof.

We have therefore shown that applying the Lanczos algorithm to \textit{any} set of probe vectors is granted to yield eventually all eigenvectors and eigenvalues (accounting for potential degeneracies) of the hermitian operator $H$, as long as such a set, which can be chosen by convenience of the user in a numerical application\footnote{The most immediate example of this is the computational basis, i.e. the basis in which each vector is numerically represented by a $1$ in a given position, and $0$ in all the rest.}, forms an orthonormal basis of the full Hilbert space $\mathcal{H}$. In a very generic situation one might get lucky and succeed in finding the full spectrum after having used only a few probe vectors but, in any case, it is granted that such a task will be accomplished after the use of at most $D$ probes. 

Krylov had also proposed in \cite{Krylov:1931} to combine different probe vectors in case the polynomial $P(\lambda)$ developed in \eqref{Sect_Krylov_polynomial_Vdm_developed} happens to be zero due to the potential presence of degeneracies in the spectrum, or to the fact that the probe vector does not overlap with a certain eigenspace. However, in his case the situation is worse because of the impossibility of obtaining the eigenvectors through his argument. Specific instances of operator $H$ and smart choices of probe vectors to combine might yield in some cases the full spectrum of the operator, but there is no general result similar to \eqref{Sect_Lanczos_Krylov_spaces_basis_probes_full_space} ensuring that there is always a finite protocol that will eventually lead to the solution of the eigenvalue problem. We can illustrate this with two examples:

\begin{enumerate}
    \item The simplest example is that of a completely degenerate spectrum. Let us compare the outcome of applying Krylov's method (i.e. constructing Krylov's version of the characteristic polynomial, $P(\lambda)$ as given in \eqref{Sect_Krylov_charac_polyn_up_to_constant}) to that of applying the Lanczos algorithm, in both cases sweeping through an orthonormal basis of probe vectors.
    \begin{itemize}
        \item[\textit{i})] Krylov's method: No matter which basis of probe vectors is chosen, each of its probe vectors will always yield an identically vanishing $P(\lambda)$, either because it does not overlap with all eigenvectors or, in case it does, because the degeneracies make the determinant vanish. The fact that the only tool in this case is the determinant constructed out of applying iteratively $H$ to the probe vectors is rather limiting.
        \item[\textit{ii})] Lanczos' method: Regardless of the particular choice of probe basis, the application of the Lanczos algorithm to its elements will in all cases yield a $1$-dimensional Krylov space, granting an eigenvalue and its associated eigenvector (equal to the probe itself). Comparing the information coming from all probe vectors it will be possible to deduce that such an eigenvalue is $D$ times degenerate and to construct the eigenbasis (in fact, having noted that the eigenvalue is maximally degenerate, any orthonormal basis of $\mathcal{H}$ is an eigenbasis).
        \end{itemize}

        The present example is, however, too obvious. In fact, a maximally degenerate spectrum implies that the operator $H$ is a multiple of the identity, where the multiplicative coefficient gives the eigenvalue, namely
        \begin{equation}
            \label{Sect_Lanczos_maximally_degenerate_operator}
            H=\lambda \, \mathds{1} ~,
        \end{equation}
        and hence its diagonalization is immediate and no algorithm needs to be performed. Below we shall present a more sophisticated case that illustrates the differences between Krylov's and Lanczos' methods.
    \item Let us now consider an arbitrary hermitian operator $H$ (with or without degeneracies in its spectrum) and a probe basis whose elements happen to be rotations of eigenvectors two by two (we consider for simplicity that the Hilbert space dimension is even, $D=2L$). More specifically, the eigenbasis can be written as
    \begin{equation}
        \label{Sect_Lanczos_smart_example_eigenbasis}
        \mathcal{B}=\bigcup_{i=0}^{L-1} \mathcal{B}^{(i)}~,\qquad \mathcal{B}^{(i)}=\left\{|\lambda_0^{(i)}\rangle,\;|\lambda_1^{(i)}\rangle\right\}~,
    \end{equation}
    while the probe basis admits an analogue structure:
    \begin{equation}
        \label{Sect_Lanczos_smart_example_probe_basis}
        \mathcal{B}_{\Omega}=\bigcup_{i=0}^{L-1} \mathcal{B}_{\Omega}^{(i)}~,\qquad \mathcal{B}_{\Omega}^{(i)}=\left\{|\Omega_0^{(i)}\rangle,\;|\Omega_1^{(i)}\rangle\right\}~.
    \end{equation}
    As announced above, the probe basis happens to verify the following property:
    \begin{equation}
        \label{Sect_Lanczos_smart_example_probe_basis_rotations_eigenbasis}
        |\Omega_k^{(i)}\rangle = \sum_{l=0}^1 U_{kl}^{(i)} |\lambda_l^{(i)}\rangle~,\qquad k=0,1~,
    \end{equation}
    where, for each fixed $i=0,\,\dots,\,L-1$, $\Big(U_{kl}^{(i)}\Big)_{k,l=0}^1$ is a two-by-two unitary matrix. With this choice of probe basis, we can now compare again what would be the outcomes of Krylov and Lanczos:
    \begin{itemize}
        \item[\textit{i})] Krylov's method: All elements of the probe basis would yield $P(\lambda)=0$, $\forall \lambda \in \mathds{R}$, because they all leave out eigenspaces of $H$. After sweeping through the full basis we would have achieved nothing. 
        \item[\textit{ii})] Lanczos' method: The application of the Lanczos algorithm to each probe vector would always yield either a one- or a two-dimensional Krylov space (depending on potential degeneracies in the spectrum of $H$). Combining eigenvalues and eigenvectors computed out of each probe vector it would be eventually possible to obtain the full spectrum and eigenbasis, and to identify any degeneracies.
    \end{itemize}
\end{enumerate}

The last example illustrates that Lanczos' method is guaranteed to work for \textit{any} choice of orthonormal probe basis, thanks to the result \eqref{Sect_Lanczos_Krylov_spaces_basis_probes_full_space} proved in this Thesis, while Krylov's method need not succeed in finding the full spectrum given a fixed probe basis and operator $H$. Since in a practical problem the spectrum of $H$ is not known \textit{a priori}, it is not possible to anticipate which probe basis may be the optimal one, and hence the fact that any such basis will achieve a successful outcome puts the usefulness of Lanczos' method on a more solid footing. On the other hand, there is no systematic recipe for finding the spectrum of any hermitian operator by building Krylov's version of the characteristic polynomial \eqref{Sect_Krylov_charac_polyn_up_to_constant}; still, for applications of interest at the time of the publication of \cite{Krylov:1931} (complicated operators with a generically non-degenerate spectrum and where a non-fine tuned probe vector is likely to have some overlap with all eigenvectors), Krylov's method would most of the times be successful and it was already received as a breakthrough in efficiency when it was formulated \cite{Luzin:1931}. Here we have only proposed a comparative analysis of both methods for the sake of pedagogy. Lanczos' work was independent from Krylov's but, in the frame of this exposition, the developments by the latter should be seen as a first theoretical step in the understanding of the construction and structure of Krylov space, eventually leading to the formulation of the Lanczos algorithm, rather than as a competing method.

\section{Symmetries and the Lanczos algorithm}\label{sect:Lanczos_alg_syms}

The Lanczos algorithm features some properties under transformations of either the operator $H$ or the seed vector $|\Omega\rangle$ that might shed some light on the structure of the Hilbert space or, in particular, the Krylov space probed by the seed. This discussion was not originally present in Lanczos' article \cite{Lanczos:1950zz} but it will reveal itself useful for a deep understanding of the recursion method, in particular whenever it is applied in the realm of quantum mechanics. Let us describe the most immediate transformations that may come to mind:

\begin{enumerate}
    \item \textbf{Unitary symmetries:} Let $U$ be a \textit{unitary symmetry} of $H$ verifying, by definition, 
    \begin{equation}
        \label{Sect_Lanczos_Unitary_symmetry}
        U^\dagger H U = H\qquad \text{and} \qquad U^\dagger U = U U^\dagger = \mathds{1}~,
    \end{equation}
    which in turn implies $[H,U]=0$. If $U$ belongs to a Lie group, it may be expressed as $U=e^{i \sum_{l=1}^r \alpha_l J_l }$, where $r$ is the dimension of the group manifold, $\left\{\alpha_l\right\}_{l=1}^r$ are some group parameters (i.e. coordinates covering a patch of the group manifold connected to the identity), and $\left\{J_l\right\}_{l=1}^r$ are the (hermitian) generators of its algebra, which also commute with $H$. Changing the seed vector by a symmetry transformation, namely
    \begin{equation}
        \label{Sect_Lanczos_seed_unitary_change}
        |\Omega\rangle \longmapsto |\overline{\Omega}\rangle = U |\Omega\rangle~,
    \end{equation}
    can be shown inductively to lead to a new Krylov basis $\left\{|\overline{K}_n\rangle\right\}_{n=0}^{\overline{K}-1}$ of length $\overline{K}$, and new sequences of Lanczos coefficients $\left\{\overline{a}_n\right\}_{n=0}^{\overline{K}-1}$ and $\left\{\overline{b}_n\right\}_{n=1}^{\overline{K}-1}$, satisfying:
    \begin{equation}
        \label{Sect_Lanczos_Effect_of_Unitary_transf}
        \begin{split}
            &\overline{K}=K~,\\
            &\overline{a}_n = a_n\qquad\forall\;n=0,\,\dots,\,K-1~,\\
            &\overline{b_n} = b_n\qquad \forall\;n=1,\,\dots,\,K-1~,\\
            &|\overline{K}_n\rangle = U |K_n\rangle\qquad\forall\;n=0,\,\dots,\,K-1~.
        \end{split}
    \end{equation}
    That is to say, the Krylov dimension remains the same, the Lanczos coefficients are invariant, and the Krylov basis gets unitarily transformed. The fact that $\overline{K}=K$ does not necessarily mean that the Krylov spaces of $|\overline{\Omega}\rangle$ and $|\Omega\rangle$ are the same: In each case, they consist of the span of the eigenspace representatives picked by the seed vector; since $[H,U]=0$, the action of $U$ on $|\Omega\rangle$ to generate $|\overline{\Omega}\rangle$ does not mix eigenspace representatives together, but it might still transform non trivially those representatives lying in degenerate eigenspaces of $H$, whose dimension is bigger than $1$: In such a case, $\mathcal{H}_{|\overline{\Omega}\rangle}$ will be a different subspace of $\mathcal{H}$ from $\mathcal{H}_{|\Omega\rangle}$; both Krylov spaces will probe the same eigenspaces of $H$ (and will therefore give access to the same subset of the spectrum of $H$), but through potentially different eigenspace representatives. If $H$ does not have any degenerate eigenspace, then $\mathcal{H}_{|\overline{\Omega}\rangle} = \mathcal{H}_{|\Omega\rangle}$. Alternatively, if $U$ is a function of $H$, $U\equiv f(H)$, then it will not change the direction of any eigenspace representatives, as it will just amount to multiplying them by a factor $f(\widetilde{\lambda})$, where $\widetilde{\lambda}$ is the eigenvalue associated to the given eigenspace, not changing the direction of the Hilbert space such a representative spans: In this simple case\footnote{From the linear algebra perspective this is a relatively simple statement, but it is still relevant to point it out at this stage because of subsequent applications of the Lanczos algorithm in the realm of quantum mechanics, where $H$ shall be the Hamiltonian or, more generally, the time evolution generator for the Hilbert space of study: A corollary of this property states the fact that a time translation of the seed state $|\Omega\rangle \mapsto e^{-itH}|\Omega\rangle$ does not modify its Lanczos coefficients or its Krylov space, only changing its Krylov basis by the corresponding time-translated version. Quantities computed solely out of the Lanczos coefficients
    will therefore be invariant under time translations of the seed state.
    This will be the case of the Krylov complexity profile as a function of time, as can be deduced from its definition in section \ref{sect:KCdef}. This will not necessarily mean that Krylov complexity of a given, fixed seed is invariant under time translations (i.e. constant): Rather, the profiles followed by the K-complexities of the evolution of two initial conditions related by a time translation will be identical.}, the Krylov spaces of both seeds are also identical, despite the fact that the Krylov bases are a unitary transformation of each other. 

    In view of the polynomial recursion \eqref{Sect_Lanczos_recursion_polynomials}, since the Lanczos coefficients in \eqref{Sect_Lanczos_Effect_of_Unitary_transf} are invariant, so will be the Krylov polynomials, i.e.
    \begin{equation}
        \label{Sect_Lanczos_polyns_unitary_invariant}
        p_n(\lambda)\longmapsto \overline{p}_n(\lambda) = p_n(\lambda)
    \end{equation}
    and, therefore, their roots are the same for any $n=0,\,\dots,\, K$. This is in agreement with the fact that both seeds $|\Omega\rangle$ and $|\overline{\Omega}\rangle$ in \eqref{Sect_Lanczos_seed_unitary_change} probe the same eigenspaces of $H$ (even though they do so through potentially different eigenspace representatives).
    \item \textbf{Translations of $H$:} Instead of changing the seed, one may also envision certain modifications of the Hermitian operator of study while keeping $|\overline{\Omega}\rangle = |\Omega\rangle$. Let us consider a shift,
    \begin{equation}
        \label{Sect_Lanczos_shift_H}
        H\longmapsto\overline{H}=H-\alpha\mathds{1},\qquad \alpha \in \mathds{R}~.
    \end{equation}
    In quantum mechanics, constant energy shifts are known not to have an impact on observable quantities. It is therefore insightful to analyse the effect of such a transformation on the Lanczos algorithm. It can be shown inductively to only shift the $a$-coefficients, leaving everything else invariant:
    \begin{equation}
        \label{Sect_Lanczos_Effect_of_shift}
        \begin{split}
            &\overline{K}=K~,\\
            &\overline{a}_n = a_n-\alpha \qquad\forall\;n=0,\,\dots,\,K-1~,\\
            &\overline{b_n} = b_n\qquad \forall\;n=1,\,\dots,\,K-1~,\\
            &|\overline{K}_n\rangle = |K_n\rangle\qquad\forall\;n=0,\,\dots,\,K-1~.
        \end{split}
    \end{equation}
    As a corollary, since the Krylov basis remains invariant, so does the Krylov space. The effect of \eqref{Sect_Lanczos_Effect_of_shift} on the Krylov polynomials is a shift in their domain, namely:
    \begin{equation}
        \label{Sect_Lanczos_effect_shift_polynomials}
        \overline{p}_n(\lambda) = p_n(\lambda + \alpha)~.
    \end{equation}
    Denoting the roots of $p_n$ (resp., $\overline{p}_n$) as $\widetilde{\lambda}_j^{(n)}$ (resp. $\overline{\widetilde{\lambda}}_j^{(n)}$)\footnote{The author apologises for the convoluted notation. The tilde is still kept because these roots are related to the spectrum of the \textit{restriction} of $H$ to Krylov space, $\widetilde{\lambda}_j = \widetilde{\lambda}_j^{(K)}$ for $j=0,\dots, K-1$, which need not constitute the full spectrum of the hermitian operator under consideration, but potentially only a subset of it.}, where $j=0,\dots , n-1$ for fixed $n$, \eqref{Sect_Lanczos_effect_shift_polynomials} implies that the roots are accordingly shifted:
    \begin{equation}
        \label{Sect_Lanczos_effect_shift_polynomials_roots}
        \overline{\widetilde{\lambda}}_j^{(n)} = \widetilde{\lambda}_j^{(n)} - \alpha~.
    \end{equation}
    In particular, the eigenvalues of $H$ represented in Krylov space, given by the roots of the $K$-th Krylov polynomial, are shifted in the same way as the operator itself is in \eqref{Sect_Lanczos_shift_H}, consistently.
    \item \textbf{Rescalings of $H$:} Let us also consider the effect of a rescaling in $H$,
    \begin{equation}
        \label{Sect_rescaling_H}
        H\longmapsto \overline{H}=\kappa H~,\qquad\kappa\in \mathds{R},
    \end{equation}
    while keeping the seed fixed. This transformation still does not change the Krylov basis (nor the Krylov space), but rescales accordingly the Lanczos coefficients:
    \begin{equation}
        \label{Sect_Lanczos_Effect_of_rescaling}
        \begin{split}
            &\overline{K}=K~,\\
            &\overline{a}_n = \kappa a_n \qquad\forall\;n=0,\,\dots,\,K-1~,\\
            &\overline{b_n} = \kappa b_n\qquad \forall\;n=1,\,\dots,\,K-1~,\\
            &|\overline{K}_n\rangle = |K_n\rangle\qquad\forall\;n=0,\,\dots,\,K-1~.
        \end{split}
    \end{equation}
     When \eqref{Sect_Lanczos_Effect_of_rescaling} is plugged into \eqref{Sect_Lanczos_recursion_polynomials}, the Krylov polynomials get accordingly rescaled:
    \begin{equation}
        \label{Sect_Lanczos_effect_rescaling_polynomials}
        \overline{p}_n(\lambda) = p_n\left(\frac{\lambda}{\kappa}\right)~,
    \end{equation}
    hence also rescaling their roots as
    \begin{equation}
        \label{Sect_Lanczos_effect_rescaling_polynomial_roots}
        \overline{\widetilde{\lambda}}_j^{(n)} = \kappa \widetilde{\lambda}_j^{(n)}~,
    \end{equation}
    consistently with the fact that $\overline{H}=\kappa H$.
    \item \textbf{Sign flip of $H$:} Finally, let us study the effect of changing the sign of the operator:
    \begin{equation}
        \label{Sect_Lanczos_sign_flip_H}
        H\longmapsto \overline{H} = -H~.
    \end{equation}
    Inductively, the effect of \eqref{Sect_Lanczos_sign_flip_H} is shown to be:
    \begin{equation}
        \label{Sect_Lanczos_Effect_of_sign_flip}
        \begin{split}
            &\overline{K}=K~,\\
            &\overline{a}_n = - a_n \qquad\forall\;n=0,\,\dots,\,K-1~,\\
            &\overline{b_n} =  b_n\qquad \forall\;n=1,\,\dots,\,K-1~,\\
            &|\overline{K}_n\rangle = (-1)^n |K_n\rangle\qquad\forall\;n=0,\,\dots,\,K-1~, \\
            & \overline{p}_n(\lambda) = (-1)^n p_n(-\lambda)\qquad\forall\;n=0,\,\dots,\,K~.
        \end{split}
    \end{equation}
    And the last line of \eqref{Sect_Lanczos_Effect_of_sign_flip} implies, consistently, that the roots of the Krylov polynomials change sign, namely
    \begin{equation}
        \label{Sect_Lanczos_Effect_of_sign_flip_roots}
        \overline{\widetilde{\lambda}}_j^{(n)} = - \widetilde{\lambda}_j^{(n)}~,
    \end{equation}
    for $j=0,\dots,n-1$ and $n=0,\dots, K$.
 \end{enumerate}

The roots of the last Krylov polynomial $p_K(\lambda)$ are the exact eigenvalues of the restriction of $H$ to Krylov space. Therefore, the roots of $p_K$ should transform in the same way as the eigenvalues of $H$ do under the corresponding transformations. This is confirmed by equations \eqref{Sect_Lanczos_polyns_unitary_invariant}, \eqref{Sect_Lanczos_effect_shift_polynomials_roots}, \eqref{Sect_Lanczos_effect_rescaling_polynomial_roots} and \eqref{Sect_Lanczos_Effect_of_sign_flip_roots}. Furthermore, these equations also show that the roots of the lower-degree Krylov polynomials ($p_n$ with $n<K$) still transform in the same way: This is insightful because, as we shall argue in section \ref{sect:Lanczos_approximation_method}, such roots can be understood as numerical approximations to the eigenvalues of $H$ obtained via successive truncations of Krylov space. 

The behavior of Lanczos coefficients and Krylov elements under the transformations discussed here will also turn out to be useful in order to understand the mechanism through which the Lanczos algorithm resolves progressively the spectrum when it is used as an approximation method, as we shall discuss in section \ref{sect:Lanczos_generic_behavior}.

\section{The Lanczos algorithm as an approximation method}\label{sect:Lanczos_approximation_method}

So far, we have presented the Lanczos algorithm and argued how it may be used as an efficient method to solve exactly the eigenvalue problem of a certain hermitian operator $H$, which has the additional advantage of reducing potential numerical errors due to the amplification of the contribution of eigenspaces corresponding to large eigenvalues, an effect that is avoided thanks to the iterative orthogonalization of each new Krylov element against the previous ones in the Lanczos algorithm. However, this algorithm can be also used as an approximation method: Running only a subset of the Lanczos steps allows to calculate a numerical approximation to a subset of the spectrum of $H$ in the way that will be explained in this section.

Let us suppose that we run the Lanczos algorithm formulated at the end of section \ref{sect:Lanczos} up to the step $n_*-1$, building successfully the first $n_*$ Krylov elements, $\left\{|K_m\rangle\right\}_{m=0}^{n_*-1}$, together with the Lanczos coefficients $\left\{a_m\right\}_{m=0}^{n_*-1}$ and $\left\{b_m\right\}_{m=1}^{n_*-1}$. The latter are sufficient, according to the polynomial recursion \eqref{Sect_Lanczos_recursion_polynomials}, to build the $n_*$-th (non-normalized) Krylov polynomial $p_{n_*}(\lambda)$. In the same way the roots of the last Krylov polynomial $p_K(\lambda)$ yield the $K$ eigenvalues represented in Krylov space, as discussed around \eqref{Sect_Lanczos_p_K_0_evals}, one may attempt to use the $n_*$ roots of $p_{n_*}(\lambda)$ as an approximation to $n_*$ of those eigenvalues.
There is a sense in which this is correct and, in order to illustrate it, let us for now consider an ideal case in which the Krylov-space spectrum features a very pronounced hierarchy:
\begin{equation}
    \label{Sect_Lanczos_spectrum_hierarchy}
     \widetilde{\lambda}_0  \gg  \widetilde{\lambda}_1  \gg \dots \gg  \widetilde{\lambda}_{K-1} >0 ~.
\end{equation}
Note that the requirement that all eigenvalues are positive is not invariant under shifts of $H$, despite the fact that the Krylov elements are expected to be invariant under such a shift according to \eqref{Sect_Lanczos_Effect_of_shift}. It is also not invariant under a sign flip of $H$, while the Krylov elements only change by a phase in such a case, as stated in \eqref{Sect_Lanczos_Effect_of_sign_flip}. It will nevertheless be pedagogical, at this stage, to particularize our discussion to the spectrum \eqref{Sect_Lanczos_spectrum_hierarchy}. The treatment of the most generic case will be postponed to section \ref{sect:Lanczos_generic_behavior}.
A hierarchy like \eqref{Sect_Lanczos_spectrum_hierarchy} can be achieved, for example, if there is a parametric separation between the eigenvalues, e.g.
\begin{equation}
    \label{Sect_Lanczos_eigenvalues_parametric_separation}
    \widetilde{\lambda}_j = \lambda^{K-j}~,\qquad \text{for }j=0,\,...\,K-1
\end{equation}
and where $\lambda>1$ is some constant setting the scale.

In the process of the implementation of the Lanczos algorithm, each application of the operator $H$ on a given vector amplifies the projection of such a vector over the eigenspace corresponding to the largest eigenvalue and, in order to avoid this, each Lanczos step effectively subtracts the eigenspace (or eigenspaces) contributing the most to the norm of the previous Krylov element. As a result of this, if the spectrum ideally features a very pronounced eigenvalue hierarchy, the \textit{Krylov subspace} spanned by $\left\{|K_m\rangle\right\}_{m=0}^{n_*-1}$ effectively probes the eigenspaces of the $n_*$ largest eigenvalues for which, therefore, the roots of $p_{n_*}(\lambda)$ give a numerical approximation. Let us be more precise about this: 
We shall probe Krylov space with a seed whose profile with respect to the eigenspace representatives is given by
\begin{equation}
    \label{Sect_Lanczos_probe_vector_profile_Krylov_space}
    |\Omega\rangle =\sum_{j=0}^{K-1} \widetilde{\Omega}_j |\widetilde{\lambda}_j\rangle~.
\end{equation}
Note that all $\widetilde{\Omega}_j\neq 0$ because we are already focusing on Krylov space. We also assume \eqref{Sect_Lanczos_probe_vector_profile_Krylov_space} to be normalized, so that
\begin{equation}
\label{Sect_Lanczos_probe_normalized_Krylov_space}
\sum_{j=0}^{K-1} {\lvert \widetilde{\Omega}_j \rvert}^2=1~.
\end{equation}
This vector already gives the zeroth Krylov element, $|K_0\rangle = |\Omega\rangle$. Let us apply explicitly the first Lanczos step to it. We begin by computing the coefficient $a_0$:
\begin{equation}
    \label{Sect_Lanczos_a0_explicit}
    a_0 = \langle K_0 | H | K_0 \rangle = \sum_{j=0}^{K-1} {\lvert \widetilde{\Omega}_j \rvert}^2 \, \widetilde{\lambda}_j~.
\end{equation}
That is: $a_0$ is the weighted average of all the eigenvalues represented in Krylov space. If we do not assume any special ``fine-tuning'' of the coefficients $\widetilde{\Omega}_j$ that may compete with the hierarchy \eqref{Sect_Lanczos_spectrum_hierarchy}, then it is natural to expect that $a_0$ will be numerically close to the largest eigenvalue $\widetilde{\lambda}_0$. In the particular case of a parametric scale separation between eigenvalues \eqref{Sect_Lanczos_eigenvalues_parametric_separation}, as long as the coefficients $\widetilde{\Omega}_j$ do not feature a similar or competing parametric scaling with $j$ (which is usually not the case because the probe vector is chosen independently from the spectrum), it is safe to claim that $a_0$ and $\widetilde{\lambda}_0$ are of the same order, $a_0\sim \widetilde{\lambda_0}\sim \lambda^K$. Hence, $a_0$ is already a good approximation to the largest eigenvalue and, in fact, the first Krylov polynomial, written in \eqref{Sect_Lanczos_recursion_polynomials}, is precisely $p_1(\lambda)=\lambda-a_0$, which already justifies that the root of $p_1(\lambda)$ approximates the largest eigenvalue.

Furthermore, we can complete the Lanczos step and construct the first, non-normalized Krylov element:
\begin{equation}
    \label{Sect_Lanczos_A1_explicit}
    |A_1\rangle = (H-a_0)|K_0\rangle = \sum_{j=0}^{K-1} \widetilde{\Omega}_j \Big( \widetilde{\lambda}_j - a_0 \Big) |\widetilde{\lambda}_j\rangle~.
\end{equation}
Now, since $a_0\sim\widetilde{\lambda}_0$, the subtraction inside the bracket in the last term of \eqref{Sect_Lanczos_A1_explicit} has the effect of practically cancelling out the contribution from $j=0$, so that we may approximate
\begin{equation}
    \label{Sect_Lanczos_A1_direction_removed}
    |A_1\rangle \approx \sum_{j=1}^{K-1} \widetilde{\Omega}_j \Big( \widetilde{\lambda}_j - a_0 \Big) |\widetilde{\lambda}_j\rangle~,
\end{equation}
and in this way this Krylov element does no longer have a contribution from the eigenspace of the largest eigenvalue. 
This consideration will be quantitatively more accurate the more peaked the seed vector is around $|\widetilde{\lambda}_0\rangle$, as this will make $a_0$ numerically closer to $\widetilde{\lambda}_0$ according to the weighted average \eqref{Sect_Lanczos_a0_explicit}.
A new application of $H$ will now amplify prominently the direction $|\widetilde{\lambda}_1\rangle$, but at the same time the coefficient $a_1$ will be numerically close to $\widetilde{\lambda}_1$, and therefore the projection of $H|K_1\rangle$ over $|\widetilde{\lambda}_1\rangle$ will be effectively removed in the $n=2$ Lanczos step\footnote{If we had kept track of the contribution along the direction of $|\widetilde{\lambda}_0\rangle$, which had been neglected in \eqref{Sect_Lanczos_A1_direction_removed}, it would have equally gotten amplified by the action of $H$; however, it would eventually be effectively removed by the corresponding subtractions of the Lanczos step that ensure orthogonality of the new Krylov elements with respect to the previously-constructed ones. Thus, for the sake of this qualitative discussion, we may directly not consider this direction at all.}, as a result of which the Krylov element $|K_2\rangle$ will effectively not probe the eigenspaces $\left\{|\widetilde{\lambda}_j\rangle\right\}_{j=0}^1$. In general, this may serve as a qualitative argument for the statement
\begin{equation}
    \label{Sect_Lanczos_Krylov_elements_dont_probe_certain_eigenspaces}
    \langle \widetilde{\lambda}_j | A_n\rangle = \text{\textit{small}}\qquad \text{for } j=0,\,...,\,n-1~.
\end{equation}
This was Lanczos' original motivation for developing the recursion method, as announced. The above expression may now be compared to equation \eqref{Sect_Lanczos_Krylov_eigenstate_basis_discrete}, whose version in terms of the non-normalized Krylov elements and polynomials we provide here:
\begin{equation}
    \label{Sect_Lanczos_Krylov_eigenstate_basis_discrete_unnormalized_version}
    \langle \widetilde{\lambda}_j|A_n\rangle = \widetilde{\Omega}_j p_n(\widetilde{\lambda}_j)~.
\end{equation}
From \eqref{Sect_Lanczos_Krylov_elements_dont_probe_certain_eigenspaces} and \eqref{Sect_Lanczos_Krylov_eigenstate_basis_discrete_unnormalized_version}, assuming the coordinates $\widetilde{\Omega}_j$ do not feature a pronounced hierarchy competing with that in the spectrum, we conclude that 
\begin{equation}
    \label{Sect_Lanczos_krylov_polynomials_almost_zero}
    p_n(\widetilde{\lambda}_j)=\text{\textit{small}}\qquad \text{for } j=0,\,...,\,n-1~.
\end{equation}
Since $p_n(\lambda)$ is a polynomial of degree $n$, which diverges when $\lambda\to\pm\infty$, one may consider \eqref{Sect_Lanczos_krylov_polynomials_almost_zero} as an indication that its roots are ``near'' $\left\{\widetilde{\lambda}_j\right\}_{j=0}^{n-1}$. It is in this sense that we can claim that the roots of $p_n(\lambda)$ give a numerical approximation to the $n$ largest eigenvalues of $n$, at least in the case of the positive spectrum with a well-defined hierarchy \eqref{Sect_Lanczos_spectrum_hierarchy} that we took in order to simplify the discussion, and given a seed whose profile over the eigenbasis is not in competition with such a hierarchy. For the sake of generality, however, we have not been quantitatively precise about the meaning of ``approximating'' or about the notions of ``smallness'' or ``being close to a number''. The case of a parametric hierarchy between the eigenvalues \eqref{Sect_Lanczos_eigenvalues_parametric_separation} provides nevertheless a clear example than can be tested numerically, as will be done in section \ref{sect:Lanczos_Numerics}. In such a case, the suppression of objects like \eqref{Sect_Lanczos_Krylov_elements_dont_probe_certain_eigenspaces} may be understood as a suppression in terms of powers of the scale $\lambda$ controlling the parametric separation in \eqref{Sect_Lanczos_eigenvalues_parametric_separation}.

\subsection{Krylov space truncations}\label{sect:Lanczos_Krylov_truncations}

The approximation scheme being considered here amounts to \textit{truncating} Krylov space to a dimension $n_{*}<K$. Recalling the discussion around \eqref{Sect_Lanczos_secular_equation_p_K}, we note that finding the numerical approximation to $n_{*}$ of the eigenvalues of $H$ by solving the equation
\begin{equation}
    \label{Sect_Lanczos_truncated_secular_eq}
    p_{n_{*}}(\lambda)=0
\end{equation}
amounts to having effectively pretended that $|A_{n_{*}}\rangle\equiv0$, i.e. placing the termination of the Lanczos algorithm at step $n_{*}$ rather than at step $K$, providing a truncated Krylov basis $\left\{|K_n\rangle\right\}_{n=0}^{n_{*}-1}$ that spans a truncated Krylov subspace:
\begin{equation}
    \label{Sect_Lanczos_truncated_Krylov_subspace}
    \mathcal{H}_{|\Omega\rangle}^{(n_{*})}:=\text{span}\left\{|K_n\rangle\right\}_{n=0}^{n_{*}-1}< \mathcal{H}_{|\Omega\rangle}~.
\end{equation}
The ``fake'' termination $|A_{n_{*}}\rangle\equiv0$ is equivalent to pretending that the action of $H$ over the truncated Krylov subspace is closed, admitting a well-defined restriction which could be diagonalized within this subspace. The fact that the roots of \eqref{Sect_Lanczos_truncated_secular_eq} are not exactly a subset of the eigenvalues of $H$, but a numerical approximation to them as argued below equation \eqref{Sect_Lanczos_krylov_polynomials_almost_zero}, is a reflection of the fact that there does not exist such a well-defined restriction of $H$ over $\mathcal{H}_{|\Omega\rangle}^{(n_{*})}$. Nevertheless, in practical situations with a limited computational power available, where only a subset of the spectrum is required, or in which the Krylov dimension might be infinite (which is often the case for differential operators\footnote{The Lanczos algorithm can be applied to a differential operator either by turning it into a finite matrix through some regularization scheme, or by iteratively applying it analytically to some seed function for a finite collection of Lanczos steps. This may result useful e.g. for computing the exponential of such an operator, an object which is often required for the solution of linear differential equations. An immediate example is the computation of the time evolution operator in quantum mechanics, whenever the Hamiltonian is given in the form of a differential operator.}) the use of this truncation protocol is very useful and often unavoidable.

The solutions of \eqref{Sect_Lanczos_truncated_secular_eq} give $n_{*}$ approximations to Krylov-space eigenvalues of $H$, which we shall denote $\widetilde{\lambda}_j^{(n_{*})}$ with $j=0,\dots,n_{*}-1$. Using them, together with the truncated Krylov basis that spans \eqref{Sect_Lanczos_truncated_Krylov_subspace} and its associated Krylov polynomials, we can equally build the corresponding approximate eigenvectors by the means of the truncated version of expression \eqref{Sect_Lanczos_evects_in_Krylov_basis_discrete}:
\begin{equation}
    \label{Sect_Lanczos_evec_approx_truncated}
    |\widetilde{\lambda}_j^{(n_{*})}\rangle = \widetilde{\Omega}_j^{*}\sum_{n=0}^{n_{*}-1}q_n\Big( \widetilde{\lambda}_j^{(n_{*})} \Big) |K_n\rangle~.
\end{equation}

This approximation scheme will be illustrated in the numerical examples of section \ref{sect:Lanczos_Numerics} for three models with different seed vectors and hermitian operators $H$. We note that, if $n_{*}=K$, then the truncated Krylov subspace becomes the full Krylov space, and hence the approximate eigenvalues and eigenvectors become exact. Namely, for all $j=0,\dots,K-1$ we have:
\begin{equation}
    \label{Sect_Lanczos_truncated_becomes_exact}
    \widetilde{\lambda}_j^{(K)}=\widetilde{\lambda}_j~,\qquad |\widetilde{\lambda}_j^{(K)}\rangle = |\widetilde{\lambda}_j\rangle~.
\end{equation}

\subsection{Successive characteristic polynomials}

In order to understand more visually this approximation, in which successive truncations of Krylov space yield numerical approximations to the largest eigenvalues of the spectrum \eqref{Sect_Lanczos_spectrum_hierarchy}, we can show explicitly that the last Krylov polynomial $p_K(\lambda)$ is proportional to the characteristic polynomial of the tridiagonal matrix \eqref{Sect_Lanczos_Tridiag_H_matrix} and that, likewise, each Krylov polynomial $p_n(\lambda)$ is proportional to the $n$-dimensional truncation of such a matrix, denoted $T_n$, built out of the Lanczos coefficients $\left\{a_m\right\}_{m=0}^{n-1}$ and $\left\{b_m\right\}_{m=1}^{n-1}$. In order to prove this, we will establish a recursive relation between the aforementioned characteristic polynomials and show that it maps to the Krylov polynomial recursion \eqref{Sect_Lanczos_recursion_polynomials} by rescaling the polynomials with the adequate non-vanishing constant. Consider the following characteristic polynomial:
\begin{equation}
    \label{Sect_Lanczos_charac_polynomial_truncated_tridiagonal}
    \begin{split}
        & \qquad \qquad \qquad \qquad\quad R_n(\lambda):= \det\left( \lambda \mathds{1}_n - T_n \right) \\
        & = \begin{vmatrix}
        \lambda -a_0 & -b_1 & 0 & \dots & 0 & 0 & 0 \\
        -b_1 & \lambda - a_1 & -b_2 & \dots & 0 & 0 & 0 \\
        0 & -b_2 & \lambda - a_2 & \dots & 0 & 0 & 0 \\
        \vdots & \vdots & \vdots & \ddots & \vdots & \vdots & \vdots \\
        0 & 0 & 0 & \dots & \lambda - a_{n-3} & - b_{n-2} & 0 \\
        0 & 0 & 0 & \dots & - b_{n-2} & \lambda - a_{n-2} & - b_{n-1} \\
        0 & 0 & 0 & \dots & 0 & - b_{n-1} & \lambda - a_{n-1}
    \end{vmatrix} ~.
    \end{split}
\end{equation}
Note that, for the present discussion, we have chosen a convention to define the characteristic polynomial that differs from that in \eqref{Sect_Krylov_charac_polyn_def} by a power of $-1$.
This determinant will yield a polynomial of degree $n$, and we may develop it along its rightmost column, obtaining:
\begin{equation}
    \label{Sect_Lanczos_charac_polynomial_recursion_step_1}
    R_n(\lambda) = (\lambda - a_{n-1})R_{n-1}(\lambda) + b_{n-1} X_{n-2}(\lambda)~,
\end{equation}
where $X_{n-2}(\lambda)$ is an intermediary polynomial of degree $n-2$ given by the determinant
\begin{equation}
    \label{Sect_Lanczos_charac_polynomial_recursion_intermediary_det}
    X_{n-2}(\lambda) := \begin{vmatrix}
         \lambda -a_0 & -b_1 & 0 & \dots & 0 & 0  \\
        -b_1 & \lambda - a_1 & -b_2 & \dots & 0 & 0  \\
        0 & -b_2 & \lambda - a_2 & \dots & 0 & 0  \\
        \vdots & \vdots & \vdots & \ddots & \vdots & \vdots  \\
        0 & 0 & 0 & \dots & \lambda - a_{n-3} & - b_{n-2}  \\
        
        0 & 0 & 0 & \dots & 0 & - b_{n-1} 
    \end{vmatrix}~,
\end{equation}
which can be immediately developed along its last line, yielding
\begin{equation}
    \label{Sect_Lanczos_aux_det_developed}
    X_{n-2}(\lambda) = -b_{n-1} R_{n-2}(\lambda)~.
\end{equation}
Plugging \eqref{Sect_Lanczos_aux_det_developed} into \eqref{Sect_Lanczos_charac_polynomial_recursion_step_1} yields a recursion between the characteristic polynomials:
\begin{equation}
    \label{Sect_Lanczos_charac_polynomial_recursion_final_form}
    R_n(\lambda) = (\lambda - a_{n-1}) R_{n-1} (\lambda) - b_{n-1}^2 R_{n-2} (\lambda)~,
\end{equation}
whose seeds are $R_0(\lambda)=1$ and $R_1(\lambda) = \lambda - a_0$. We now note that, using the convention $b_0\equiv 1$ and posing
\begin{equation}
    \label{Sect_Lanczos_relation_Krylov_polynomials_and_charac_polyns}
    \begin{split}
        & R_0(\lambda) = p_0(\lambda) = 1~, \\
        & R_n(\lambda) = \Big( \prod_{m=0}^{n-1} b_m \Big) p_n (\lambda)~,\qquad 1\leq n \leq K~,
    \end{split}
\end{equation}
the recursion \eqref{Sect_Lanczos_charac_polynomial_recursion_final_form} gets mapped to \eqref{Sect_Lanczos_recursion_polynomials}, confirming that the Krylov polynomials are the characteristic polynomials of the successive tridiagonal matrices up to a non-zero proportionality constant\footnote{The $b$-coefficients are never zero because they are originally given by the norms of the non-normalized Krylov elements in the Lanczos algorithm. As discussed at the end of section \ref{sect:Lanczos}, if one of those norms is zero, the algorithm terminates and such a coefficient ($b_K=0$) is not stored.}. The fact that their zeroes give numerical approximations to the eigenvalues of $H$ probed by these successive Krylov space truncations is now intuitively easier to understand.

\subsection{The \textit{reversed} Lanczos algorithm}

The recursion \eqref{Sect_Lanczos_charac_polynomial_recursion_final_form} was derived by the means of developing the determinant \eqref{Sect_Lanczos_charac_polynomial_truncated_tridiagonal} along its rightmost column. This immediately suggests the possibility of proceeding differently, namely, developing the determinant for the full characteristic polynomial $R_K(\lambda)$ along its leftmost column, yielding a new polynomial recursion whose final element is still $R_K(\lambda)$, just like with \eqref{Sect_Lanczos_charac_polynomial_recursion_final_form}, but whose intermediary polynomials are different from the Krylov polynomials $p_n(\lambda)$, still admitting an interpretation in the framework of the Lanczos algorithm, as we shall see below.

We start by considering the following pattern of truncation of Krylov space: Instead of keeping the first $n$ Krylov elements, we shall keep the subset $\left\{ |K_m\rangle\right\}_{m=n}^{K-1}$. The approximate restriction of $H$ over this Krylov subspace\footnote{Along the lines of the discussion in section \ref{sect:Lanczos_Krylov_truncations}, it is \textit{approximate} in the sense that the action of $H$ is not really closed in this subspace, but we explicitly neglect the subspace of the image that does not belong to the subspace spanned by the truncated basis.} is denoted as $\overline{T}_{K-n}$ and, in coordinates over the above-mentioned restricted Krylov basis, it takes the following form:
\begin{equation}
    \label{Sect_Lanczos_tridiag_reverse_truncation}
    \overline{T}_{K-n}\overset{*}{=} \begin{pmatrix}
        a_n & b_{n+1} & 0  &\dots &0 & 0 & 0 \\
        b_{n+1} & a_{n+1} & b_{n+2} & \dots & 0 & 0 & 0 \\
        0 & b_{n+2} & a_{n+2} & \dots & 0 & 0 & 0 \\
        \vdots & \vdots & \vdots & \ddots & \vdots & \vdots & \vdots \\
        0 & 0 & 0 & \dots & a_{K-3} & b_{K-2} & 0 \\
        0 & 0 & 0 & \dots & b_{K-2} & a_{K-2} & b_{K-1} \\
        0 & 0 & 0 & \dots & 0 & b_{K-1} & a_{K-1}
    \end{pmatrix}~,
\end{equation}
where the star over the equality sign denotes that the right-hand side is an expression in coordinates over a particular basis (in this case, the truncated Krylov basis described in the preceding paragraph).
We define the characteristic polynomial of this matrix as
\begin{equation}
    \label{Sect_Lanczos_reverse_charac_polyn}
    \overline{R}_{K-n} (\lambda) := \det (\lambda \mathds{1} - \overline{T}_{K-n})~.
\end{equation}
Notationally, both for the matrix and the polynomial, the indexing $K-n$ has been preferred because in this way the subindex indicates their dimension and degree, respectively. Proceeding similarly to the steps that led to \eqref{Sect_Lanczos_charac_polynomial_recursion_final_form}, but this time developing the determinant along its leftmost column, one reaches the recursion
\begin{equation}
    \label{Sect_Lanczos_reverse_charac_polyns_recursion_messy_indexing}
    \overline{R}_{K-n} (\lambda) = (\lambda - a_n) \overline{R}_{K-(n+1)}(\lambda) -b_{n+1}^2 \overline{R}_{K-(n+2)}(\lambda)~.
\end{equation}
For $n=K-1,\dots, 0$. Note that, for $n=0$, the truncated operator becomes the full original hermitian operator restricted to Krylov space $T_K=T$, where we recall that $T$ is given in coordinates over the full Krylov basis in \eqref{Sect_Lanczos_Tridiag_H_matrix}, and hence we consistently have that
\begin{equation}
    \label{Sect_Lanczos_Reversed_charac_polyn_vs_actual_polyn}
    \overline{R}_K(\lambda)=R_K(\lambda) = \det(\lambda \mathds{1}-T)~,
\end{equation}
where the determinant should be understood to be restricted to Krylov space $\mathcal{H}_{|\Omega\rangle}$. 

The recursion above uses the polynomials at levels $n+1$ and $n+2$ to determine that at level $n$, so the seeds required are the polynomials corresponding to $n=K$ and $n=K-1$, which are given by
\begin{equation}
    \label{Sect_Lanczos_charac_polyn_recursion_seeds}
    \overline{R}_0(\lambda)=1,\qquad \overline{R}_1(\lambda)=\lambda - a_{K-1}~.
\end{equation}
In order to make the recursion \eqref{Sect_Lanczos_reverse_charac_polyns_recursion_messy_indexing} more readable, and to make contact with the discussion on Krylov polynomials that we shall shortly take, it is convenient to relabel the indices as $K-n\longmapsto n$, resulting in
\begin{equation}
    \label{Sect_Lanczos_reverse_charac_polyns_recursion_good_indexing}
    \overline{R}_n (\lambda) = (\lambda - a_{K-n}) \overline{R}_{n-1}(\lambda) - b_{K-(n-1)}^2\overline{R}_{n-2}(\lambda)~,
\end{equation}
whose seed is still \eqref{Sect_Lanczos_charac_polyn_recursion_seeds}. The connection of these successive truncations and their corresponding characteristic polynomials to the Lanczos algorithm is immediate after noticing that they correspond to ordered truncations of Krylov space starting from the last (rather than the zeroth) Krylov element. As we shall now show, the polynomials $\overline{R}_n(\lambda)$ are related to the Krylov polynomials obtained applying the Lanczos algorithm to the last Krylov element $|K_{K-1}\rangle$, used now as the new seed of the iterative process. This protocol gives a new Krylov basis $\left\{|\overline{K}_n\rangle\right\}_{n=0}^{K-1}$ of the same length as the original one, together with new sets of Lanczos coefficients $\left\{\overline{a}_n\right\}_{n=0}^{K-1}$ and $\left\{\overline{b}_n\right\}_{n=1}^{K-1}$. The new seed gives directly the zeroth element of the basis,
\begin{equation}
    \label{Sect_Lanczos_reversed_basis_K0}
    |\overline{K}_0\rangle = |K_{K-1}\rangle~,
\end{equation}
from which $\overline{a}_0$ can be immediately computed:
\begin{equation}
   \label{Sect_Lanczos_reversed_a0}
   \overline{a}_0=\langle \overline{K}_0|H|\overline{K}_0\rangle = \langle K_{K-1}|H|K_{K-1}\rangle = a_{K-1}~.
\end{equation}
The first Lanczos step can now be applied:
\begin{equation}
    \label{Sect_Lanczos_reversed_basis_A1}
    |\overline{A}_1\rangle = (H-\overline{a}_0)|\overline{K}_0\rangle = (H-a_{K-1})|K_{K-1}\rangle = b_{K-1}|K_{K-2}\rangle~,
\end{equation}
where in the last step we made use of the termination condition \eqref{Sect_Lanczos_termination}
\begin{equation}
    \label{Sect_Lanczos_termination_revisited}
    |A_K\rangle = (H-a_{K-1})|K_{K-1}\rangle - b_{K-1}|K_{K-2}\rangle = 0~.
\end{equation}
The first new Lanczos coefficient is given by the norm of $|\overline{A}_1\rangle$,
\begin{equation}
    \label{Sect_Lanczos_reversed_b1}
    \overline{b_1}=b_{K-1}~,
\end{equation}
and the next Krylov element is the normalized version of $|\overline{A}_1\rangle$, i.e.
\begin{equation}
    \label{Sect_Lanczos_reversed_basis_K1}
    |\overline{K}_1\rangle = |K_{K-2}\rangle~.
\end{equation}
To conclude this Lanczos step, we can compute the next $a$-coefficient:
\begin{equation}
    \label{Sect_Lanczos_reversed_a1}
    \overline{a}_1 = \langle \overline{K}_1|H|\overline{K}_1\rangle = \langle K_{K-2}|H|K_{K-2}\rangle = a_{K-2}~.
\end{equation}

At this point, we are in conditions to propose the following relations:
\begin{equation}
    \label{Sect_Lanczos_reversed_vs_original}
    \begin{split}
        & \overline{a}_n = a_{K-1-n}~,\qquad n=0,\,1\,,\dots,\,K-1~, \\
        & \overline{b}_n = b_{K-n}~,\qquad n = 1,\,2,\,,\dots ,\, K-1~, \\
        &|\overline{K}_n\rangle = |K_{K-1-n}\rangle~,\qquad n= 0,\,1,\,\dots,\,K-1~.
    \end{split}
\end{equation}
This proposal may be proved by induction. The starting seeds (steps $n=0,\,1$) have already been provided above, and it is only left to show the induction step. We assume that \eqref{Sect_Lanczos_reversed_vs_original} holds up to the step $n$, and take a step further:
\begin{equation}
    \label{Sect_Lanczos_reversed_induction_step}
    \begin{split}
        & |\overline{A}_{n+1}\rangle = (H-\overline{a}_n)|\overline{K}_n\rangle - \overline{b}_n|\overline{K}_{n-1}\rangle = (H-a_{K-1-n})|K_{K-1-n}\rangle - b_{K-n}|K_{K-n}\rangle  \\
        & = b_{K-n}|K_{K-n}\rangle + b_{K-1-n}|K_{K-n-2}\rangle - b_{K-n}|K_{K-n}\rangle = b_{K-1-n}|K_{K-n-2}\rangle~,
    \end{split}
\end{equation}
where we made use of the generic Lanczos step \eqref{Sect_Lanczos_Kn_proposal}.
The norm of $|\overline{A}_{n+1}\rangle$ gives the new $b$-coefficient,
\begin{equation}
    \label{Sect_Lanczos_reversed_b_coeffs_induction_step}
    \overline{b}_{n+1} = b_{K-(n+1)}~,
\end{equation}
and its normalized version yields the new Krylov element,
\begin{equation}
    \label{Sect_Lanczos_reversed_basis_induction_step}
    |\overline{K}_{n+1}\rangle = |K_{K-1-(n+1)}\rangle~.
\end{equation}
Finally, the next $a$-coefficient reads:
\begin{equation}
    \label{Sect_Lanczos_reversed_a_coeffs_induction_step}
    \overline{a}_{n+1} = \langle \overline{K}_{n+1}|H|\overline{K}_{n+1}\rangle = a_{K-1-(n+1)}~,
\end{equation}
where \eqref{Sect_Lanczos_reversed_basis_induction_step} and the definition of $a_{K-1-(n+1)}$ have been used. Expressions \eqref{Sect_Lanczos_reversed_b_coeffs_induction_step}, \eqref{Sect_Lanczos_reversed_basis_induction_step} and \eqref{Sect_Lanczos_reversed_a_coeffs_induction_step} are consistent with \eqref{Sect_Lanczos_reversed_vs_original}, which is therefore proved. The Krylov basis associated to $|K_{K-1}\rangle$ simply sweeps backwards the original Krylov basis, and as such we shall refer to it as the \textit{reversed} Krylov basis. Likewise, its associated $\overline{a}_n$ and $\overline{b}_n$ sequences shall be dubbed the reversed Lanczos coefficients.

The termination of the reversed basis can be verified considering the $n=K$ Lanczos step:
\begin{equation}
    \label{Sect_Lanczos_reversed_basis_termination}
    \begin{split}
        & |\overline{A}_K\rangle = (H-\overline{a}_{K-1})|\overline{K}_{K-1}\rangle - \overline{b}_{K-1}|\overline{K}_{K-2}\rangle \\
        & = (H-a_0)|K_0\rangle - b_1|K_1\rangle = b_1|K_1\rangle - b_1|K_1\rangle = 0~,
    \end{split}
\end{equation}
where, in the second-to-last equality, the $n=1$ Lanczos step of the original basis, \eqref{Sect_Lanczos_K1_def}, which has one subtraction term less than the generic steps with $n\geq 2$, has been applied.

With this, it is equally possible to build a family of \textit{reversed} Krylov polynomials, denoted $\overline{p}_n(\lambda)$ with $n=0,\,\dots,\,K$, by noting that 
\begin{equation}
    \label{Sect_Lanczos_reversed_Krylov_polynomials_def}
    |\overline{A}_n\rangle = \overline{p}_n(H)|\overline{K}_0\rangle = \overline{p}_n(H)|K_{K-1}\rangle~.
\end{equation}
Using the Lanczos recursion satisfied by the reversed Krylov basis, given in the first equality of \eqref{Sect_Lanczos_reversed_induction_step}, one can show that the reversed polynomials satisfy a relation which is analogous to that in \eqref{Sect_Lanczos_recursion_polynomials}:

\begin{equation}
    \label{Sect_Lanczos_recursion_reversed_polynomials}
    \begin{split}
        & \overline{p}_0(\lambda) = 1~, \\
        & \overline{p}_1(\lambda) = \lambda-\overline{a}_0=\lambda - a_{K-1}~, \\
        & \overline{p}_n (\lambda) = \frac{\lambda-\overline{a}_{n-1}}{\overline{b}_{n-1}}\overline{p}_{n-1}(\lambda) - \frac{\overline{b}_{n-1}}{\overline{b}_{n-2}} \overline{p}_{n-2} (\lambda) \\
        &\qquad\;\,= \frac{\lambda - a_{K-n}}{b_{K-(n-1)}}\overline{p}_{n-1}(\lambda)-\frac{b_{K-(n-1)}}{b_{K-(n-2)}}\overline{p}_{n-2}(\lambda)~,\qquad n\geq 2~.
    \end{split}
\end{equation}
Now, in the same way that recursions \eqref{Sect_Lanczos_recursion_polynomials} and \eqref{Sect_Lanczos_charac_polynomial_recursion_final_form} are mapped to each other through \eqref{Sect_Lanczos_relation_Krylov_polynomials_and_charac_polyns}, the recursions defining the reversed characteristic polynomials \eqref{Sect_Lanczos_reverse_charac_polyns_recursion_good_indexing} and the reversed Krylov polynomials \eqref{Sect_Lanczos_recursion_reversed_polynomials} are mapped to each other through\footnote{This time, we prefer not to use the convention ``$\overline{b}_0\equiv1$'', since a putative extension of the relation between original and reversed Lanczos coefficients given in \eqref{Sect_Lanczos_reversed_vs_original} to $n=0$ would suggest to associate $\overline{b}_0\equiv b_K=0$, which might induce to confusion.}:
\begin{equation}
    \label{Sect_Lanczos_relation_reversed_polyns_Krylov_charac}
    \begin{split}
        &\overline{R}_0(\lambda) = \overline{p}_0(\lambda)=1~,\qquad \overline{R}_1(\lambda) = \overline{p}_1(\lambda) = \lambda - \overline{a}_0=\lambda - a_{K-1}~, \\
        &\overline{R}_n (\lambda) = \Big(\prod_{m=1}^{n-1} \overline{b}_m\Big)\,\overline{p}_n (\lambda) = \Big(\prod_{m=1}^{n-1} b_{K-m}\Big)\,\overline{p}_n(\lambda)~,\qquad 2\leq n \leq K~.
    \end{split}
\end{equation}
In particular, setting $n=K$, we expect to obtain the full characteristic polynomial of the restriction of $H$ to the Krylov space of the original seed $|\Omega\rangle$, as stated in \eqref{Sect_Lanczos_Reversed_charac_polyn_vs_actual_polyn}. Combining this with the recursion \eqref{Sect_Lanczos_relation_reversed_polyns_Krylov_charac} yields
\begin{equation}
    \label{Sect_Lanczos_full_charac_polyn_from_reversed_check}
    \Big(\prod_{m=1}^{K-1} b_{K-m}\Big)\,\overline{p}_K(\lambda) = \Big(\prod_{m=1}^{K-1} b_{m}\Big)\,p_K(\lambda)~.
\end{equation}
The products on sides of the equality involve the full sequence of $b$-coefficients, which are all non-vanishing. This implies
\begin{equation}
    \label{Sect_Lanczos_last_Krylov_polyn_vs_last_reversed_polyn}
    \overline{p}_K(\lambda) = p_K(\lambda)~.
\end{equation}

The interesting feature of these reversed families of Krylov basis elements and polynomials is that, since they sweep backwards their original counterparts, the vector $|\overline{K}_n\rangle$ will be mostly supported over the eigenspaces over which $|K_{K-1-n}\rangle$ is supported, which in the case of the discussion at the beginning of section \ref{sect:Lanczos_approximation_method} would be those corresponding to the $n$ \textit{smallest} eigenvalues of $H$ represented in Krylov space. Therefore, the roots of the reversed Krylov polynomials $\overline{p}_n(\lambda)$ will, this time, give numerical approximations to increasingly larger eigenvalues, as we shall verify in the numerical illustration presented in section \ref{sect:Lanczos_Numerics}.

This \textit{reversed} Lanczos algorithm remains, nevertheless, an academic exercise, and cannot be used as a resource-saving protocol to approximate the smallest eigenvalues of the spectrum (instead of the largest), as in order to construct this reversed Krylov basis one needs to have constructed in the first place the full original Krylov basis of the seed $|\Omega\rangle$, since the seed of the former is the last element of the latter. Rather, the discussion serves as an illustration of the fact that, once the Krylov space associated to a given vector $|\Omega\rangle$ has been constructed, it is not possible to exit it by applying the Lanczos procedure to any vector contained within that Krylov space, because the action of $H$ on such a space is closed, by construction. As an alternative example, one may attempt to seed the Lanczos algorithm with any other element of the Krylov basis, namely $|\overline{K}_0\rangle = |K_{n_0}\rangle$ for some $n_0\in [0,K-1]\bigcap \mathds{Z}$: In such a case, analytical expressions rapidly become cumbersome, but one can verify that $|\overline{K}_n\rangle \in \text{span}\left\{|K_m\rangle\right\}_{m = n_0-n}^{n_0 + n}$, i.e. the new Krylov basis explores symmetrically the original Krylov basis around the new seed \footnote{Modulo eventual \textit{bounces} of the terms in the linear combination if it reaches the left or right edges of the original Krylov basis, that is, if $m$ in a term $|K_m\rangle$ contributing to $|\overline{K}_n\rangle$ reaches $0$ (\textit{left edge}) or $K-1$ (\textit{right edge}): When this occurs, a new application of $H$ to the conflictive term will yield terms proportional to $|K_0\rangle$ and $|K_1\rangle$ (resp. $|K_{K-1}\rangle$ and $|K_{K-2}\rangle$), which is what we mean by a \textit{bounce} of the support of the linear combination at the edge of Krylov space.}.

\subsection{Generic behavior of the Lanczos algorithm}\label{sect:Lanczos_generic_behavior}

The discussion below the example spectrum \eqref{Sect_Lanczos_spectrum_hierarchy} stated that, for such a particular instance of a positive spectrum with a marked hierarchy between the eigenvalues, the profile of the vectors $|K_n\rangle$ in the eigenbasis of $H$ is such that the first Krylov elements are peaked near eigenvectors corresponding to the largest eigenvalues. For a generic, potentially not positive-definite spectrum, inspection of expressions like \eqref{Sect_Lanczos_A1_explicit} might appear to suggest that the relevant eigenvalues for the initial Krylov elements are those with the largest \textit{absolute value}, as the phase of the coordinates of a vector over an orthonormal basis is not relevant when it comes to evaluating the contribution of a given direction to the total norm. This, however, cannot be a generic statement about the behavior of the Lanczos algorithm for any hermitian operator $H$ with an arbitrary spectrum, as can be seen from the symmetry properties of the recursion method discussed in section \ref{sect:Lanczos_alg_syms}: The Krylov elements are invariant under constant shifts of $H$ \eqref{Sect_Lanczos_shift_H}, which also shift the spectrum but leave invariant the eigenvectors; therefore, the profile of $|K_n\rangle$ in the eigenbasis is invariant under constant shifts of the spectrum, but the identification of the eigenvalue with the largest absolute value is not a shift-invariant statement, so this cannot be a valid criterion to identify the dominant directions of a Krylov element.

As an example of the above discussion, we can take the spectrum in \eqref{Sect_Lanczos_spectrum_hierarchy} and a seed vector which is, for simplicity, initially peaked near the direction with the largest eigenvalue $\widetilde{\lambda}_0$. The discussion at the beginning of section \ref{sect:Lanczos_approximation_method} implied that the first Lanczos step would effectively remove the direction of the largest eigenvalue, $|\widetilde{\lambda}_0\rangle$, so that $|K_1\rangle$ is peaked near the direction of $|\widetilde{\lambda}_1\rangle$, the second largest eigenvalue, and so forth. We may now perform a shift of the full spectrum by the value of the originally largest eigenvalue, i.e. a shift \eqref{Sect_Lanczos_shift_H} with $\alpha = \widetilde{\lambda}_0$:
\begin{equation}
    \label{Sect_Lanczos_shifted_spectrum_example}
    \overline{\widetilde{\lambda}}_j = \widetilde{\lambda}_j - \widetilde{\lambda}_0~, 
\end{equation}
which results in the sign-flipped hierarchy
\begin{equation}
\label{Sect_Lanczos_hierarchy_sign_flipped}
    0=\overline{\widetilde{\lambda}}_0 \gg \overline{\widetilde{\lambda}}_1 \gg \dots \gg \overline{\widetilde{\lambda}}_{K-1}
\end{equation}
and, therefore, in the shifted spectrum the eigenvalue with the largest absolute value is $\overline{\widetilde{\lambda}}_{K-1}$.
However, since both the Krylov elements and the eigenvectors are invariant under this shift, it will still hold that $|\overline{K}_0\rangle$ is peaked near $|\overline{\widetilde{\lambda}}_0\rangle$, $|\overline{K}_1\rangle$ near $|\overline{\widetilde{\lambda}}_1\rangle$, etc, in apparent contradiction with our earlier qualitative considerations.

The way out from this apparent paradox is to rephrase the practical effect of a Lanczos step in terms that are explicitly invariant under sign flips and constant shifts or rescalings of $H$, and which reduce to the identification of the largest eigenvalues in the particular case of \eqref{Sect_Lanczos_spectrum_hierarchy}. For this, we consider a generic Lanczos step with $n\geq 2$, which we restate here:
\begin{equation}
    \label{Sect_Lanczos_generic_step_restatement}
    |A_n\rangle = (H-a_{n-1})|K_{n-1}\rangle - b_{n-1}|K_{n-2}\rangle~.
\end{equation}
The new, non-normalized Krylov element is constructed in \eqref{Sect_Lanczos_generic_step_restatement} by modifying $H|K_{n-1}\rangle$ with two subtraction terms, whose effect we may describe qualitatively:
\begin{itemize}
    \item The coefficient $a_{n-1}$ is given by 
    \begin{equation}
        \label{Sect_Lanczos_a_coeff_restatement}
        a_{n-1}=\langle K_{n-1}| H | K_{n-1}\rangle~,
    \end{equation}
    and as such it gives the (weighted) average value of the eigenvalues effectively probed by $|K_{n-1}\rangle$. Thus, the effect of the term $-a_{n-1}|K_{n-1}\rangle$ can be analyzed from the inspection of
    \begin{equation}
        \label{Sect_Krylov_effect_first_subtraction}
        (H-a_{n-1})|K_{n-1}\rangle = \sum_{j=0}^{K-1} \mathcal{K}_j^{(n-1)} (\widetilde{\lambda}_j - a_0)|\widetilde{\lambda}_j\rangle~,
    \end{equation}
    where $\mathcal{K}_j^{(n-1)}$ are the coordinates of $|K_{n-1}\rangle$ over the eigenbasis. Considering only those directions $|\widetilde{\lambda}_j\rangle$ that contain most of the norm of this Krylov element, the eigenvalues that will get more affected by the subtraction $\widetilde{\lambda}_j - a_0$ will be those belonging to regions of the spectrum where the eigenvalues \textit{accumulate}, as $a_0$ will be numerically close to them. Consequently, this subtraction will substantially reduce the projection of the vector over the corresponding eigenspaces. 
    \item Broadly speaking, the subtraction term $-b_{n-1}|K_{n-2}\rangle$ in \eqref{Sect_Lanczos_generic_step_restatement} serves the purpose of avoiding the previous Krylov element from reemerging when $H$ is applied to $|K_{n-1}\rangle$, in order to ensure orthogonality of the Krylov basis. For instance, in the step $n=1$, given in \eqref{Sect_Lanczos_K1_def}, a term of this sort is not required because there is no $|K_{-1}\rangle\equiv 0$ term requiring to be subtracted.
\end{itemize}

Putting these two considerations together, the more correct (yet qualitative) description of the net effect of the subtraction terms in the $n$-th Lanczos step can now be stated. The projections more drastically reduced are those along the directions of eigenvectors verifying that:
\begin{itemize}
    \item[\textit{i})] $|K_{n-1}\rangle$ has a more significant overlap with them.
    \item[\textit{ii})] Their associated eigenvalues belong to a region of the spectrum effectively probed by $|K_{n-1}\rangle$ where the eigenvalues accumulate more.
\end{itemize}
Furthermore, directions belonging to the linear span of the previous Krylov elements do not reemerge in subsequent Lanczos steps.

Consequently, isolated eigenvalues are the first to be resolved efficiently in the scheme of successive approximations using progressive Krylov space truncations, as long as the seed state overlapped significantly with their eigenspace, and small eigenvalue differences take longer to be resolved and require more numerical accuracy. This statement is adequately invariant under shifts or rescalings of $H$ in the sense that eigenvalues belonging to densely or sparsely populated regions of the spectrum (relative to the total spectral width, or to the average eigenvalue spacing) are mapped under a shift or a rescaling to eigenvalues in the corresponding regions of the new spectrum (defined relatively with respect to the new spectral width or spacing). The statement is also invariant under a sign flip in the spectrum because this transformation does not change the distance between eigenvalues. For instance, in the shifted, negative semi-definite spectrum \eqref{Sect_Lanczos_shifted_spectrum_example}, the hierarchy of absolute values is reversed with respect to that in \eqref{Sect_Lanczos_spectrum_hierarchy}, as can be seen in \eqref{Sect_Lanczos_hierarchy_sign_flipped}, but it still holds that, in order, the more isolated eigenvalues are those with $j=0,\,1,\,\dots,\,K-1$, both for $\widetilde{\lambda}_j$ and for $\overline{\widetilde{\lambda}}_j$.

On can think of the Lanczos algorithm as an approximation method that \textit{resolves eigenvalue differences}. Given the mean eigenvalue spacing $\Delta$ in the spectrum of $H$, we can define the normalized distances between consecutive eigenvalues,
\begin{equation}
    \label{Sect_Lanczos_normalized_level_spacing}
    s_n:=\frac{\lvert \lambda_n - \lambda_{n+1}\rvert}{\Delta}~,
\end{equation}
where we have implicitly assumed that the eigenvalues $\lambda_n$ are sorted so that $\lambda_0\geq \lambda_1\geq \dots\geq \lambda_{D-1}$. The quantity $s_n$ is explicitly invariant under both shifts and rescalings of $H$, as well as a sign flip of the spectrum. With this, we can generically say that, assuming the use of a seed vector $|\Omega\rangle$ probing all eigenspaces democratically, it will take fewer Lanczos steps to resolve eigenvalue pairs $\left(\lambda_n,\lambda_{n+1}\right)$ with a larger value of $s_n$, while those featuring a smaller normalized difference will require finer approximations where the Krylov truncation dimension $n_{*}$ is closer to $K$.

All these qualitative claims will be illustrated in next section, combining different spectra and seed vectors with different starting profiles over the eigenbasis.

For the eventual applications to physics, and particularly to gravity, of interest in this Thesis, approximation methods based on the Lanczos algorithm will not play a central role. We will in general assume the capability of building the full Krylov basis and sequences of Lanczos coefficients (either numerically or analytically) and use this information to arrange the Hilbert space in a manner that will be useful for studying systematically the time evolution of a given initial state or operator, as will be described in detail in \ref{sect:KCdef}. Nevertheless, in the cases in which $H$ is the time evolution generator (either the Hamiltonian for state evolution, or the Liouvillian\footnote{The Liouvillian, defined in equation \eqref{Sect_KCdef_Liouvillian_def}, is an operator acting over the space of operators, consisting on the adjoint action of the Hamiltonian. This is the correct time evolution generator for operators in the Heisenberg picture} for operator evolution), the analysis provided throughout the present Chapter may still be valuable in the sense that it sheds light on how successive Krylov elements probe the different gravity microstates in the available Hilbert space. The previous conclusion stating that the last Krylov elements probe finer and finer level differences in the spectrum will, for instance, turn out to be insightful in the context of state and operator dynamics in quantum-chaotic systems.

\section{Numerical examples}\label{sect:Lanczos_Numerics}

In this section we shall provide three numerical examples of the application of the Lanczos algorithm, which serve to illustrate how Krylov elements probe the different eigenvectors of $H$ depending on the seed vector and on the structure of the spectrum itself, reinforcing the discussions in section \ref{sect:Lanczos_approximation_method}, especially those in its subsection \ref{sect:Lanczos_generic_behavior}. The three toy models considered are:

\begin{enumerate}
    \item An operator $H$ whose spectrum features the exponential hierarchy \eqref{Sect_Lanczos_eigenvalues_parametric_separation} with $\lambda=2$ and a seed vector $|\Omega\rangle$ which is exponentially peaked on the largest eigenvalue, reinforcing such a hierarchy:
    \begin{equation}
    \label{Sect_Lanczos_Numerics_seed_exp}
    |\Omega\rangle = \frac{1}{N(\lambda)}\sum_{j=0}^{K-1} \lambda^{-j} |\widetilde{\lambda}_j\rangle~,
\end{equation}
where the factor $N(\lambda)$ ensures normalization and is given by
\begin{equation}
    \label{Sect_Lanczos_numerics_exp_seed_normalization}
    N(\lambda) = \sqrt{\sum_{j=0}^{K-1}\lambda^{-2j}}~.
\end{equation}
A simpler version of this model was presented in the original work by Lanczos \cite{Lanczos:1950zz} in order to illustrate the exchange in dominance of the different eigenvectors contributing to the successive Krylov elements. 

It is worth to note that, in the context of quantum mechanics and holography, a state with an exponential profile over the energy basis like \eqref{Sect_Lanczos_Numerics_seed_exp} is reminiscent of a (non-factorizable) version of the thermofield double state \cite{Maldacena:2001kr}, which is the gravity microstate describing a two-sided black hole, and can be also seen as the purification of the thermal state. However, for $\lambda>1$, the vector \eqref{Sect_Lanczos_Numerics_seed_exp} is peaked near the largest\footnote{In this section we use an indexing convention such that $\widetilde{\lambda}_0$ is the largest, rather than the smallest, eigenvalue. This choice was sensible for the discussion of interest, as in the examples considered $\widetilde{\lambda}_0$ was amongst the first eigenvalues to be reproduced in the scheme of successive approximations, being numerically comparable to the first few Lanczos coefficients, as Figures \ref{fig:Sect_Lanczos_Numerics_exp_seed_Lanczos_coeffs_vs_evals}, \ref{fig:Sect_Lanczos_Numerics_flat_seed_Lanczos_coeffs_vs_evals} and \ref{fig:Sect_Lanczos_Numerics_flat_seed_quasideg_Lanczos_coeffs_vs_evals} will illustrate. In quantum-mechanical problems it is customary to reserve the smallest index for the ground state of the Hamiltonian, and we shall stick to that convention in the subsequent Chapters of this Thesis.} eigenvalue $\widetilde{\lambda}_0$, while the actual thermofield double state at finite temperature is always peaked on the ground state.

\item The same operator $H$ as in the previous case, but probed with a different seed vector, this time with a strictly flat profile over the eigenbasis:
\begin{equation}
    \label{Sect_Lanczos_Numerics_seed_flat}
    |\Omega\rangle = \frac{1}{\sqrt{K}}\sum_{j=0}^{K-1} |\widetilde{\lambda}_j\rangle~.
\end{equation}
This might be thought of as a more realistic example, in which the seed $|\Omega\rangle$ is not necessarily correlated with the eigenvectors of $H$ and thus probes all of them equally\footnote{A vector whose coordinates over the eigenbasis are randomly drawn from independent and identical distributions might be a better generalization of this, but for the purpose of the intended analysis the simpler instance with constant coefficients will be insightful enough.}.

The vector \eqref{Sect_Lanczos_Numerics_seed_flat} can be obtained as the $\lambda\to 1$ limit of \eqref{Sect_Lanczos_Numerics_seed_exp}, and it can be thought of as a toy version of the infinite-temperature thermofield double state.

\item In this case, the seed kept a flat profile \eqref{Sect_Lanczos_Numerics_seed_flat}, but the spectrum of $H$ in \eqref{Sect_Lanczos_eigenvalues_parametric_separation} was tweaked to feature a \textit{quasi-degeneracy} near the largest eigenvalue:
\begin{equation}
    \label{Sect_Lanczos_Numerics_spectrum_quasideg}
    \begin{split}
        &\widetilde{\lambda}_0 = \widetilde{\lambda}_1 + \lambda ~,\\
        &\widetilde{\lambda}_j = \lambda^{K-j}~,\qquad 1\leq j \leq K-1~,
    \end{split}
\end{equation}
still with $\lambda=2$.
\end{enumerate}

Since all the seed vectors \eqref{Sect_Lanczos_Numerics_seed_exp} and \eqref{Sect_Lanczos_Numerics_seed_flat} have been written directly in coordinates over the Krylov eigenspace representatives (projections over other potential eigenspaces of $H$ being null by definition of an eigenspace representative), constructing numerically these vectors amounts to initializing column vectors in a space of dimension $D\equiv K$. In all the analyses the Krylov dimension was taken to be $K=20$ and we worked in coordinates over the eigenbasis of $H$, so that the coordinates of $|\Omega\rangle$ could be directly read off from \eqref{Sect_Lanczos_Numerics_seed_exp} and \eqref{Sect_Lanczos_Numerics_seed_flat}, and $H$ took the form of a diagonal matrix with entries given by the eigenvalues. This fully defines the seed and operator in a given coordinate system, and no more information is required in order to run the Lanczos algorithm.

Models 1 and 2 serve the purpose of illustrating the discussion at the beginning of section \ref{sect:Lanczos_approximation_method} on a positive-definite spectrum with a parametrically controlled hierarchy, where the successive Krylov space truncations serve to gradually approximate numerically the eigenvalues of $H$ from the largest to eventually the smallest. On the other hand, model 3 is an example of the more general behavior announced in section \ref{sect:Lanczos_generic_behavior}: the largest eigenvalues of the spectrum are very close to each other (much closer than the average eigenvalue spacing), and consequently it takes various Lanczos steps to resolve them accurately, despite being the biggest: This supports the claim that the relevant feature determining how fast eigenvalues are resolved is not their absolute value, but whether they are isolated or not (compared to the rest of the eigenvalues probed by the seed).

For each model, a numerical analysis demonstrating the various aspects of the Lanczos algorithm presented throughout this Chapter was performed. Specifically, the steps of the analysis were the following:

\begin{enumerate}
    \item Application of the Lanczos algorithm for the construction of all the Krylov elements $|K_n\rangle$ and Lanczos coefficients $a_n$, $b_n$. In order to estimate quantitatively what eigenvectors are probed, and to what extent, by each Krylov element, the following indicators describing the structure of a vector $|\chi\rangle$ in coordinates over a given basis (in this case, the basis of eigenspace representatives in Krylov space) were used:
    \begin{itemize}
    \item Average position of a given vector $|\chi\rangle$ on the basis:
    \begin{equation}
    \label{Sect_Lanczos_Numerics_position_ebasis_def}
        P(|\chi\rangle) := \sum_{j=0}^{K-1} j\, {\lvert \langle \widetilde{\lambda}_j |\chi\rangle \rvert}^2~.
    \end{equation}
    \item Participation ratio (PR): It estimates the effective number of basis elements contributing to the norm of a certain vector $|\chi\rangle$, hence its name. It requires of the construction of an intermediary quantity, called the Inverse Participation Ratio (IPR):
    \begin{equation}
        \label{Sect_Lanczos_Numerics_IPR_def}
        IPR(|\chi\rangle) :=  \sum_{j=0}^{K-1} {\lvert \langle \widetilde{\lambda}_j |\chi\rangle \rvert}^4~,
    \end{equation}
    from which the PR can be computed as
    \begin{equation}
        \label{Sect_Lanczos_Numerics_PR_def}
        PR(|\chi\rangle) = \frac{1}{IPR(|\chi\rangle)}~.  
    \end{equation}
    If $|\chi\rangle$ has unit norm, this quantity satisfies $0\leq PR(|\chi\rangle) \leq K$ \cite{Kramer_1993}.
\end{itemize}
\item Using the Lanczos coefficients computed and the polynomial recursion \eqref{Sect_Lanczos_recursion_polynomials}, the Krylov polynomials $\left\{p_n(\lambda)\right\}_{n=0}^K$ were constructed explicitly. Next, the roots of each Krylov polynomial were numerically determined; these give the successive numerical approximations to the eigenvalues of $H$ that one obtains from the progressive Krylov space truncations of dimension $n_{*}$, verifying $1\leq n_{*}\leq K$. For each truncated Krylov subspace of dimension $n_*$, the roots of the polynomial $p_{n_*}(\lambda)$ yield $n_*$ approximate eigenvalues, $\left\{\widetilde{\lambda}^{(n_{*})}_j\right\}_{j=0}^{n_{*}-1}$. When $n_{*}=K$, the truncated Krylov subspace becomes equal to the full Krylov space, $p_K(\lambda)$ is proportional to the characteristic polynomial of $H$, and its roots were verified to be equal (within the working numerical precision) to the exact eigenvalues of $H$.
\item Within each Krylov space truncation, the approximate eigenvectors associated to the approximate eigenvalues obtained in the previous point were computed via \eqref{Sect_Lanczos_evec_approx_truncated},
where \textit{a-priori} knowledge of $\widetilde{\Omega}_j^*$ is not required\footnote{Even if in these specific toy models we did have such a knowledge, since the seed vector was directly constructed in coordinates over the eigenbasis, which give the coefficients $\widetilde{\Omega}_j$. In a practical situation, such coordinates are not known \textit{a priori} because the goal of the problem is precisely to construct the eigenbasis.}, since it amounts to a normalization constant of the vector. These approximate eigenvectors were compared to the exact ones computing their position \eqref{Sect_Lanczos_Numerics_position_ebasis_def} and participation ratio \eqref{Sect_Lanczos_Numerics_PR_def} over the eigenbasis. Since all computations were performed in coordinates over the latter basis, exact eigenvectors take the form of Kronecker deltas, and it was verified that approximate eigenvectors reduced to this shape in the last truncation $n_{*}=K$.
\item As an ``academic exercise'', we constructed the reversed Krylov polynomials $\overline{p}_n(\lambda)$ implementing the recursion \eqref{Sect_Lanczos_recursion_reversed_polynomials}. Computing their roots, we obtained approximations to the spectrum within the successive Krylov space truncations of dimension $n_{*}$, spanned by $\left\{|\overline{K}_n\rangle = |K_{K-1-n}\rangle \right\}_{n=0}^{n_{*}-1}$. We verified that they reconstructed the spectrum of $H$ in the opposite order to the ordinary Krylov space truncations in step 2, even though with a slightly slower convergence.
\item For the sake of completeness, the approximate eigenvectors corresponding to the approximate eigenvalues in the previous step were constructed analogously to the analysis in step 3 and compared to the exact eigenvectors, which are still recovered when $n_{*}=K$.
\end{enumerate}

The method by Lanczos proposes to solve the eigenvalue problem restricted to Krylov space in terms of the basis $\left\{|K_n\rangle\right\}_{n=0}^{K-1}$ rather than the (non orthonormal) basis $\left\{H^n|\Omega\rangle\right\}_{n=0}^{K-1}$ because the construction of the latter is numerically unreliable due to the uncontrolled amplification of the largest eigenvalues every time $H$ acts on the vector. However, the algorithm is still not exempt from numerical instabilities inherent to its recursive nature: Finite-precision errors may build up accross the successive Lanczos steps, resulting in the eventual loss of orthogonality of the constructed Krylov elements. There exist re-orthogonalization algorithms \cite{Parlett} that may be used to fix these issues, which were studied in detail and implemented for the first time in the context of Krylov complexity for quantum many-body systems in \cite{I,II,III}, works in which these algorithms were combined with the use of high-performance computing facilities that allowed the exploration of large Hilbert spaces using double-floating-point precision. However, a deep discussion on re-orthogonalization techniques was not required for the toy models that shall be presented here: In this case, the original Lanczos algorithm was run using the software Mathematica \cite{Mathematica} at a sufficiently high numerical precision, whose important memory and time requirements are at the origin of the smallness of the Krylov dimension $K=20$ chosen. Nevertheless, with this value of $K$, the parametric separation in the spectrum \eqref{Sect_Lanczos_eigenvalues_parametric_separation} is sufficiently pronounced for the results to be instructive, even if, as announced above, the scale parameter was set to $\lambda = 2$: The ratio between the largest and the smallest eigenvalue of the spectrum in models 1 and 2 is $\lambda^{K-1}=2^{19}=524288$, and the use of high-precision arithmetic is therefore required in order to prevent the eigenvector of the largest eigenvalue from being spuriously amplified after it had been effectively subtracted in the initial Lanczos steps.

The numerical results are depicted in Figures 
\ref{fig:Sect_Lanczos_Numerics_exp_seed_Krylov_elements} to \ref{fig:Sect_Lanczos_Numerics_flat_seed_quasideg_EvalEvec_Convergence_Reversed}, which have been collected in Appendix \ref{ch:AppxCh01}. We shall discuss them in detail for each model next.

\subsection{Model 1. Exponentially peaked seed vector}

Applying the Lanczos algorithm to the seed \eqref{Sect_Lanczos_Numerics_seed_exp} yields a Krylov basis whose properties with respect to the eigenbasis can be analyzed by the means of the functions \eqref{Sect_Lanczos_Numerics_position_ebasis_def} and \eqref{Sect_Lanczos_Numerics_PR_def} defined previously. The results, depicted in Figure \ref{fig:Sect_Lanczos_Numerics_exp_seed_Krylov_elements}, confirm the expectation that each Lanczos step effectively removes from the linear combination the eigenspace corresponding to the largest eigenvalue dominating in the previous iteration. The effect is mainly due to the fact that the seed vector is already very localized near $|\widetilde{\lambda}_0\rangle$ and, as discussed in section \ref{sect:Lanczos_approximation_method}, each Lanczos step removes the dominant direction in the previous Krylov element.


Figure \ref{fig:Sect_Lanczos_Numerics_exp_seed_Krylov_elements} appears to suggest that the Krylov basis elements are already \textit{almost} eigenstates of $H$, since they are very localized near the corresponding eigenvectors. This observation is in qualitative agreement with the tridiagonal (hence, \textit{almost} diagonal) form of $H$ when it is written in coordinates over the Krylov basis \eqref{Sect_Lanczos_Tridiag_H_matrix}. In fact, since the Lanczos coefficients are the items of such a matrix representation, they already provide an order-of-magnitude estimate of the eigenvalues of $H$, as illustrated in Figure \ref{fig:Sect_Lanczos_Numerics_exp_seed_Lanczos_coeffs_vs_evals}.


In Figure \ref{fig:Sect_Lanczos_Numerics_exp_seed_EvalEvec_Convergence} we can observe the successive approximations to the spectrum of $H$ and to the corresponding eigenvectors obtained in the scheme of progressive truncations of Krylov space, as a function of the dimension of the Krylov subspace considered, $n_{*}$. When $n_{*}=1$ only one eigenvalue is obtained, numerically close to the largest eigenvalue of $H$. Progressively, smaller and smaller eigenvalues start to appear as $n_{*}$ is increased. Consistently, the approximate eigenvectors gradually converge to Kronecker deltas in the eigenbasis, centered in the position of the exact eigenvector they correspond to, and with a participation ratio equal to $1$ when $n_{*}=K$.

Finally, Figure \ref{fig:Sect_Lanczos_Numerics_exp_seed_EvalEvec_Convergence_Reversed} depicts the result of successive approximations to the spectrum obtained from the family of reversed Krylov polynomials. Correctly, they start approximating the smaller eigenvalues and their corresponding eigenvectors, and yield the exact spectrum when the truncation is maximal, $n_{*}=K$.


\subsection{Model 2. Flat seed vector} 

The result of the application of the Lanczos algorithm to the seed \eqref{Sect_Lanczos_Numerics_seed_flat} with an operator $H$ whose spectrum features the exponential hierarchy \eqref{Sect_Lanczos_eigenvalues_parametric_separation} also illustrates the exchange of dominance of the different eigenvectors, in a setup in which the seed is not correlated with the spectrum in the sense that its profile over the eigenbasis does not reinforce or counter the spectral hierarchy. The initial Krylov element $|K_0\rangle$, coinciding with the seed vector, is a flat linear superposition of all eigenvectors and, in agreement with \eqref{Sect_Lanczos_a0_explicit}, $a_0$ is the arithmetic average of the eigenvalues of $H$, slightly numerically biased towards lower eigenvalues because, as a result of the exponential hierarchy, they are closer to each other than the larger ones, which are more isolated. Consequently, the subtraction $\widetilde{\lambda}_j-a_0$ in \eqref{Sect_Lanczos_A1_explicit} is maximal for $j=0$, implying that the Krylov element $|K_1\rangle$ is sharply peaked near $|\widetilde{\lambda}_0\rangle$. Similarly, subsequent Krylov elements are peaked around the eigenvectors corresponding to smaller and smaller eigenvalues, as illustrated in Figure \ref{fig:Sect_Lanczos_Numerics_flat_seed_Krylov_elements}. Figure \ref{fig:Sect_Lanczos_Numerics_flat_seed_Lanczos_coeffs_vs_evals} depicts the comparison between the Lanczos coefficients and the eigenvalues of $H$, featuring a somewhat significant discrepancy between $a_0$ and $\widetilde{\lambda}_0$ due to the fact that the former is the arithmetic spectral average. Figure \ref{fig:Sect_Lanczos_Numerics_flat_seed_EvalEvec_Convergence} depicts the convergence of the approximate eigenvalues and eigenvectors as a function of the Krylov space truncation: In each Krylov subspace, the lowest approximate eigenvalue is close to the arithmetic average of the eigenvalues that are left out of the truncated subspace, as one can confirm by observing that the corresponding approximate eigenvector receives contributions from the exact eigenvectors of all such low-lying eigenvalues. Finally, Figure \ref{fig:Sect_Lanczos_Numerics_flat_seed_EvalEvec_Convergence_Reversed} illustrates that the reconstruction of the spectrum of $H$ in the reverse order is still achieved by the use of the reversed Krylov polynomials.

\subsection{Model 3. Flat seed vector, spectrum with a quasi-degeneracy}

Probing the spectrum \eqref{Sect_Lanczos_Numerics_spectrum_quasideg} with a flat seed \eqref{Sect_Lanczos_Numerics_seed_flat} through the Lanczos algorithm illustrates the fact that the recursion method takes longer to resolve eigenvalues in areas where they accumulate, regardless of whether their absolute value is large or small, since this is not a translationally-invariant feature, as argued in section \ref{sect:Lanczos_generic_behavior}. Figure \ref{fig:Sect_Lanczos_Numerics_flat_seed_quasideg_Krylov_elements} depicts the structure of the Krylov elements: While $|K_0\rangle=|\Omega\rangle$ is a flat superposition of all exact eigenvectors, $|K_1\rangle$ is a linear combination that involves both $|\widetilde{\lambda}_0\rangle$ and $|\widetilde{\lambda}_1\rangle$, the eigenvectors of the quasi-degenerate eigenvalues, not being able to resolve them; eventually, $|K_7\rangle$ is also some linear combination of $|\widetilde{\lambda}_0\rangle$ and $|\widetilde{\lambda}_1\rangle$, so that the truncations with dimension $n_{*}\geq 8$ start to show two approximate eigenvalues very close to each other of the order of $\widetilde{\lambda}_0$, having therefore effectively resolved the quasi-degeneracy. This can be observed in Figure \ref{fig:Sect_Lanczos_Numerics_flat_seed_quasideg_EvalEvec_Convergence}. For completeness, Figures \ref{fig:Sect_Lanczos_Numerics_flat_seed_quasideg_Lanczos_coeffs_vs_evals} and \ref{fig:Sect_Lanczos_Numerics_flat_seed_quasideg_EvalEvec_Convergence_Reversed} depict, respectively, the comparison between Lanczos coefficients and eigenvalues, and the convergence results upon applying the reversed Lanczos algorithm.

\section{Combinatorial relations and Lanczos coefficients}\label{sect:relations_Lan_Moments}

There exists \cite{ViswanathMuller,SANCHEZDEHESA1978275,Parker:2018yvk} a bijective map between the set of Lanczos coefficients $a_n,b_n$ and the sequence \textit{moments} of the operator $H$, defined as:
\begin{equation}
    \label{Sect_Lanczos_moments_of_H}
    M_k:=\langle \Omega | H^k | \Omega \rangle~.
\end{equation}
The correspondence is of a combinatorial nature, which allows for some analytical estimates and bounds, as we shall see for instance in section \ref{sect:KC_chaos_holog}, but which can also be systematically implemented on a computer. Let us study each direction of the correspondence separately.

\subsection{From Lanczos coefficients to moments}\label{subsect:From_Lanczos_to_Moments}

This direction may appear to be the least interesting one, as in practice one can often easily compute the moments $M_k$ and needs the Lanczos coefficients. However, in section \ref{sect:KC_chaos_holog} it will reveal itself as a useful tool to establish bounds on the growth of the $b_n$ coefficients as a function of $n$.

The cleanest way to understand how to obtain moments from Lanczos coefficients \cite{ViswanathMuller,Parker:2018yvk,Barbon:2019wsy} is to consider the tridiagonal form of $H$ in coordinates over the Krylov basis, given in \eqref{Sect_Lanczos_Tridiag_H_matrix}. Such an expression can be written algebraically as
\begin{equation}
    \label{Sect_Lanczos_H_tridiagonal_algebraic_expression}
    H = \widehat{a} + \widehat{a}^\dagger + \widehat{n}~,
\end{equation}
where, borrowing the conventional notation from the simple quantum harmonic oscillator, we have defined the \textit{ladder-like} operators\footnote{Notationally, the hat in the definition of these operators is used to distinguish them from the $a$-coefficients and the index $n$, symbols which have been used previously in the text.} $\widehat{a},~\widehat{a}^\dagger$ and the \textit{number-like} operator $\widehat{n}$:
\begin{equation}
    \label{Sect_Lanczos_ladder_ops}
    \widehat{a} := \sum_{n=0}^{K-2} b_{n+1}\, |K_n\rangle\langle K_{n+1}|~,\qquad \widehat{n} := \sum_{n=0}^{K-1} a_n\, |K_n\rangle \langle K_n |~.
\end{equation}
These operators are reminiscent of the ladder and number operators of the harmonic oscillator because $\widehat{a}$ and $\widehat{a}^\dagger$ decrease (resp. increase) the position in the Krylov basis by one unit, while $\widehat{n}$ preserves it. However, the algebra of the operators at hand is determined by the specific value of the Lanczos coefficients, and in fact they need not close an algebra in general.

Now, using that $|\Omega\rangle = |K_0\rangle$, we can rewrite the moments \eqref{Sect_Lanczos_moments_of_H} in a more suggestive manner:
\begin{equation}
    \label{Sect_Lanczos_Moments_ladder_ops}
    M_k = \langle K_0| {\Big( \widehat{a} + \widehat{a}^\dagger + \widehat{n} \Big)}^k | K_0\rangle~.
\end{equation}
Expanding the power inside of the bra-ket we will obtain a sum of monomials of degree $k$, consisting of powers of $\widehat{a},~\widehat{a}^\dagger$ and $\widehat{n}$, sandwiched by $\langle K_0 |$ and $|K_0\rangle$. It is possible to understand intuitively how to compute in a diagrammatic fashion the expectation value of each monomial in the zeroth state\footnote{Since we are borrowing terminology from the quantum harmonic oscillator, we shall also refer to vectors as states, and to diagonal matrix elements of operators as expectation values in the corresponding state. But we shall keep in mind that the algebraic discussions throughout the present Chapter are applicable to generic Hilbert spaces, potentially outside the realm of quantum mechanics.} of the Krylov basis: Starting from $|K_0\rangle$, we apply in an ordered fashion the operators composing a given monomial, which have the following effect:
\begin{itemize}
    \item $\widehat{a}$ sends $|K_n\rangle\longmapsto b_n~ |K_{n-1}\rangle$.
    \item $\widehat{a}^\dagger$ sends $|K_n\rangle\longmapsto b_{n+1}~ |K_{n+1}\rangle$.
    \item $\widehat{n}$ performs the operation $|K_n\rangle\longmapsto a_n~|K_{n}\rangle$.
\end{itemize}
Lastly, after having applied the full monomial to $|K_0\rangle$, the inner product between the resulting state and $\langle K_0|$ is computed in order to obtain the sought expectation value; if the former state is not itself proportional to $|K_0\rangle$, the result will be zero. Additionally, the action of the monomial on $|K_0\rangle$ will vanish if in an intermediary step of the sequence of operations we encounter either $\widehat{a}|K_0\rangle = 0$ or $\widehat{a}^\dagger|K_{K-1}\rangle = 0$.
Conceiving the positions over the Krylov basis as levels in a vertical direction, and the power $k$ (giving the degree of the monomial) as a total number of horizontal steps, the above protocol can be represented by \textit{Motzkin paths} \cite{MotzkinRef}: These are paths of $k$ steps that depict concatenations of positions $\left\{P_l=(l,n_l)\right\}_{l=0}^k$ in a two-dimensional grid starting from the point $P_0=(0,0)$ and ending in $P_k=(k,0)$ in which $n_{l+1}$ is either $n_l+1$, $n_l-1$ or $n_l$, with the condition that $n_l\geq 0$ (i.e. no path goes below the zero level). The contribution of a full Motzkin path, giving the expectation value of a given monomial of ladder operators, can be evaluated assigning to it a product of Lanczos coefficients, where the index of each coefficient is dictated by each one of the steps $P_l\longmapsto P_{l+1}$ according to a rule inherited from the sequence of operations described above:
\begin{itemize}
    \item A \textit{descending} step $(l,n_l)\longmapsto (l+1,n_l-1)$ is evaluated as $b_{n_l}$.
    \item An \textit{ascending} step $(l,n_l)\longmapsto (l+1,n_l+1)$ is evaluated as $b_{n_l+1}$.
    \item A \textit{horizontal} step $(l,n_l)\longmapsto (l+1,n_l)$ evaluates to $a_{n_l}$.
\end{itemize}
As announced, for finite $K$, the paths need to be satisfy the additional requirement of not reaching higher than the level $K-1$, i.e. $n_l\leq K-1$. These are called \textit{restricted} Motzkin paths. In the discussion that follows we will focus on unrestricted Motzkin paths, simply referred to as Motzkin paths.

These combinatorial objects admit a rather detailed analytical treatment, presented in \cite{MotzkinRef}. In particular, the number of (unrestricted) Motzkin paths of length $k$ is given by the \textit{Motzkin number} \cite{MotzkinNumbers}, $m_k$:
\begin{equation}
    \label{Sect_Lanczos_Motzkin_number}
    m_k=\sum_{l=0}^{\lfloor k/2 \rfloor}\binom{k}{2l} C_l~,
\end{equation}
where $C_l$ denotes the Catalan numbers,
\begin{equation}
    \label{Sect_Lanczos_Catalan_numbers}
    C_l=\frac{1}{l+1}\binom{2l}{l}~,
\end{equation}
which in turn count the number of \textit{Dyck}\footnote{Note the ``y''.} \textit{paths} \cite{Stanley_2015} of length $2l$. A Dyck path of length $2l$ is a Motzkin path of the same length with no horizontal steps (which is the reason why they can only have an even length). For future reference, we may give the asymptotic behavior of both numbers for large length \cite{Stanley_2015,BERNHART199973}:
\begin{equation}
    \label{Sect_Lanczos_Motzkin_Dyck_asympt}
    m_k\sim \frac{1}{2\sqrt{\pi}}{\left(\frac{3}{k}\right)}^{3/2}3^k~,\qquad C_k\sim \frac{2^{2k}}{k^{3/2}\sqrt{\pi}}~,
\end{equation}
for large $k$.

\begin{figure}[t]
    \centering
    \includegraphics[width=6cm]{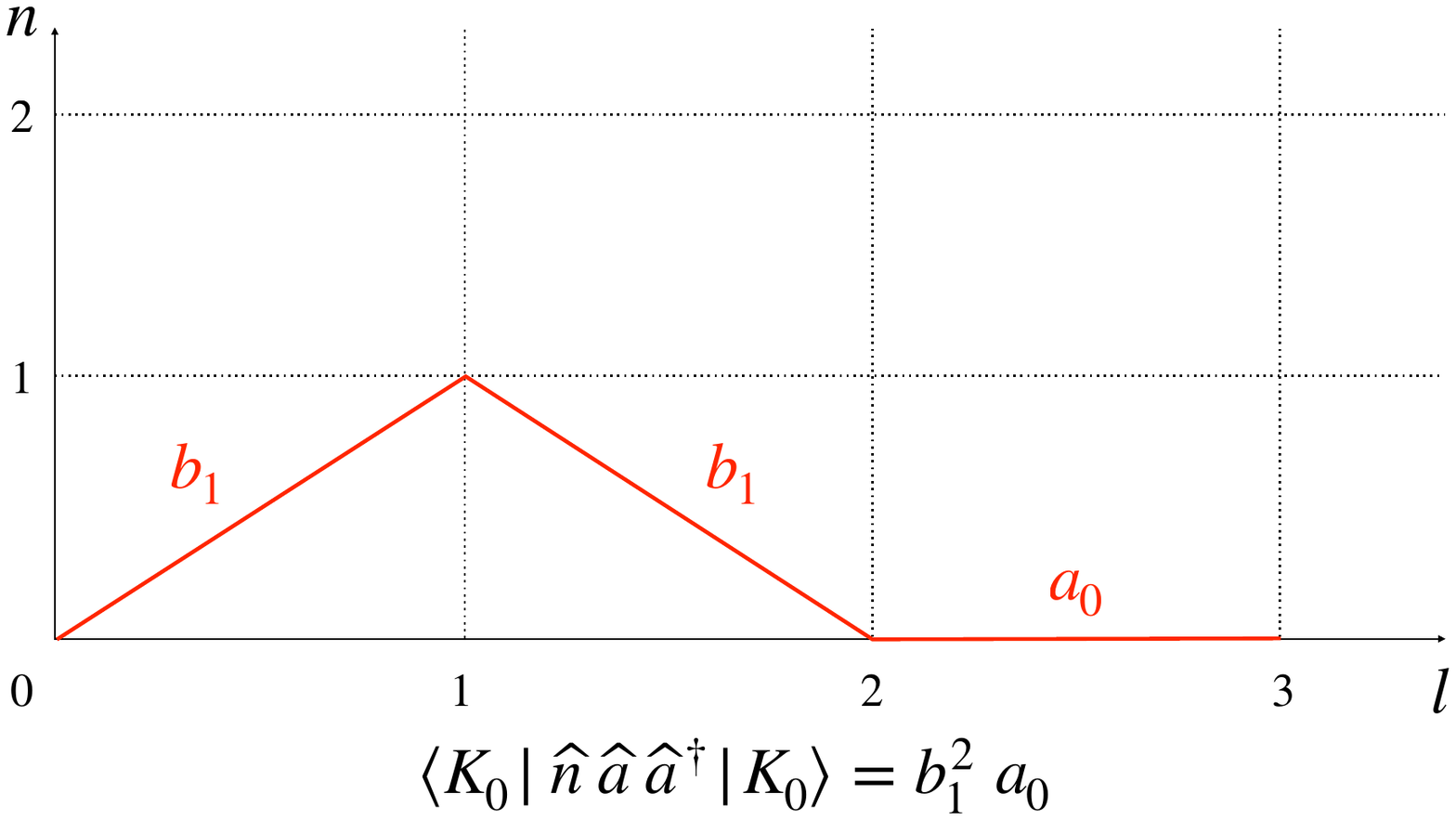}  \includegraphics[width=6cm]{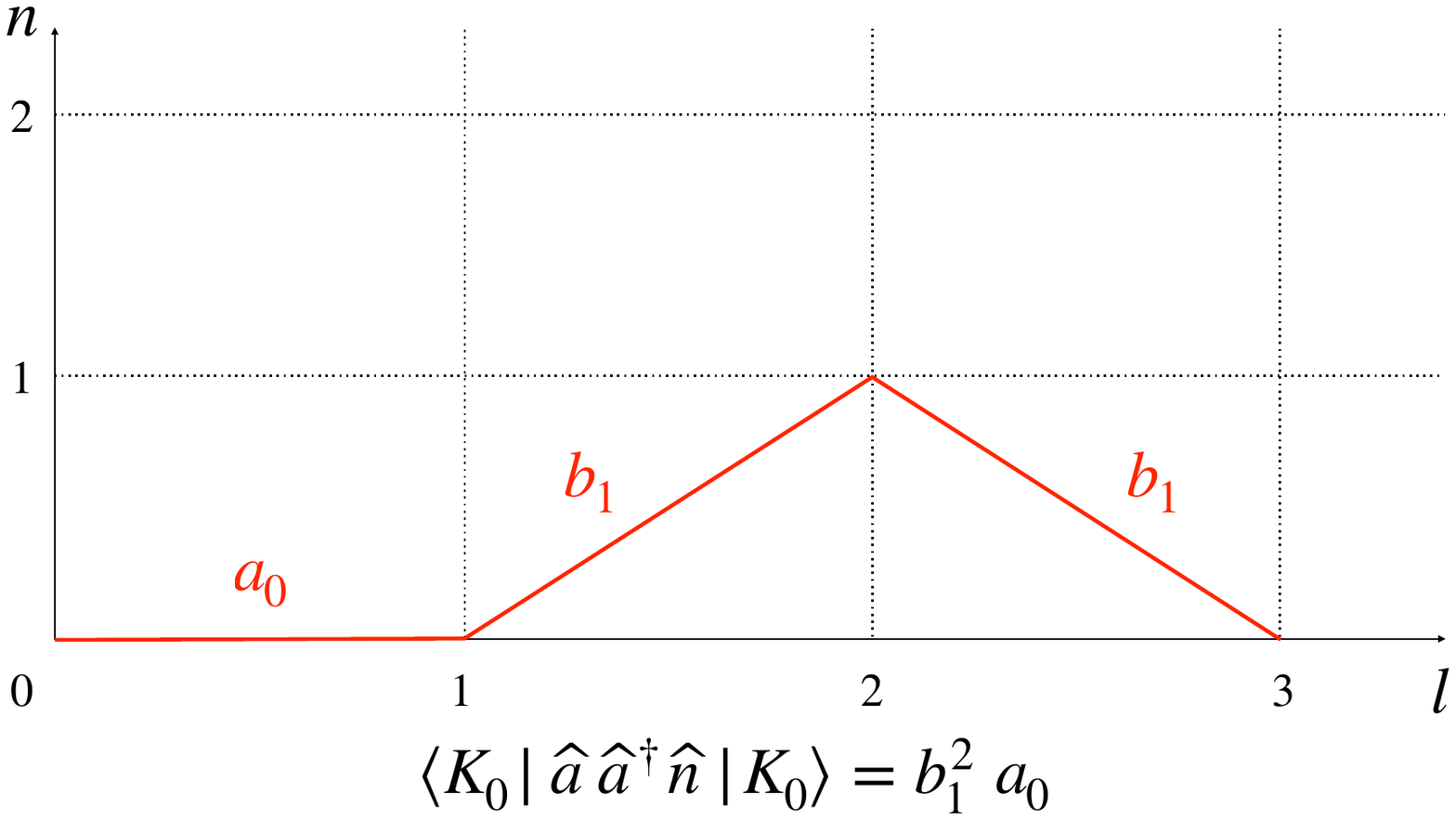} \\ 
     \includegraphics[width=6cm]{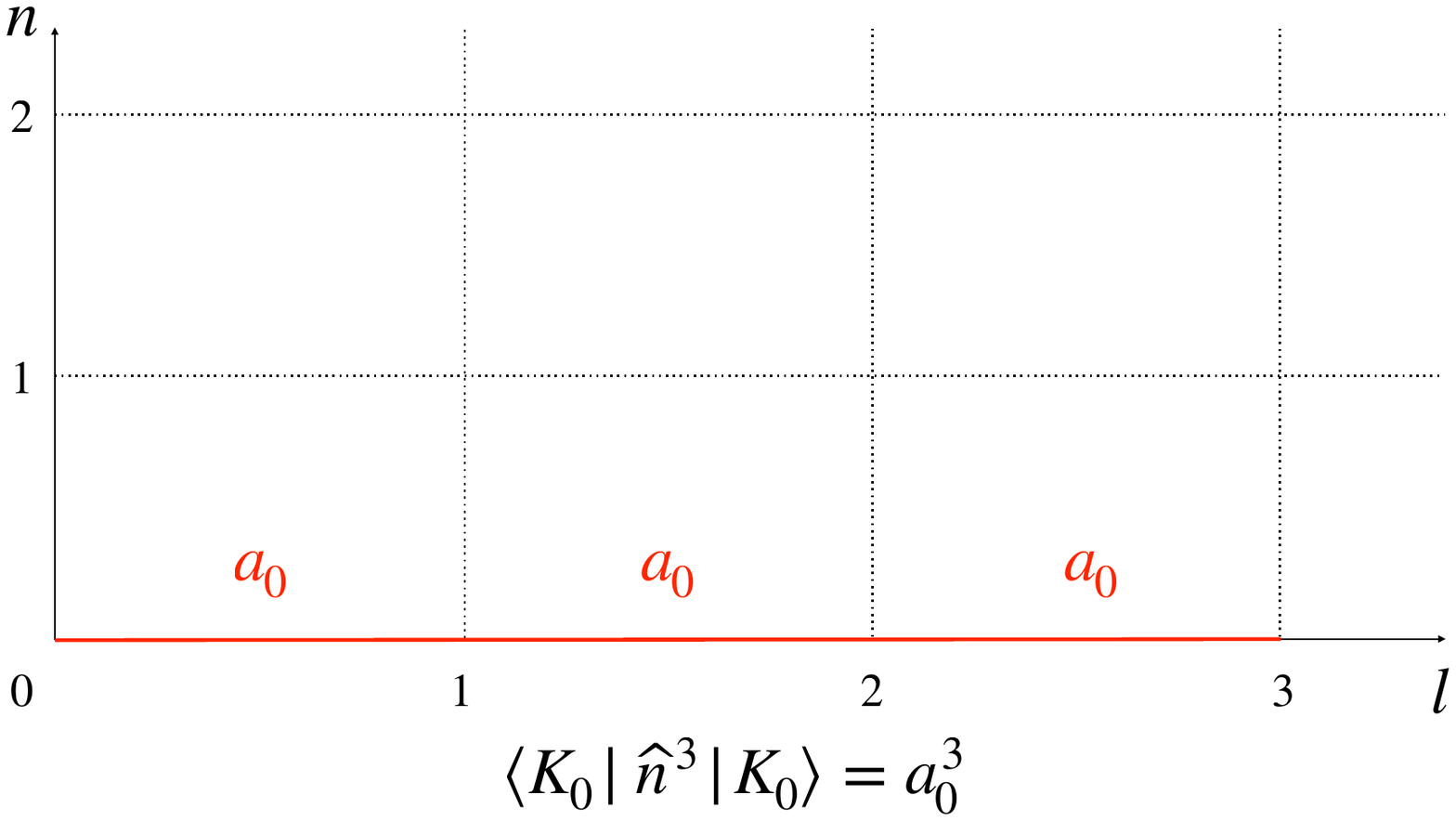}  \includegraphics[width=6cm]{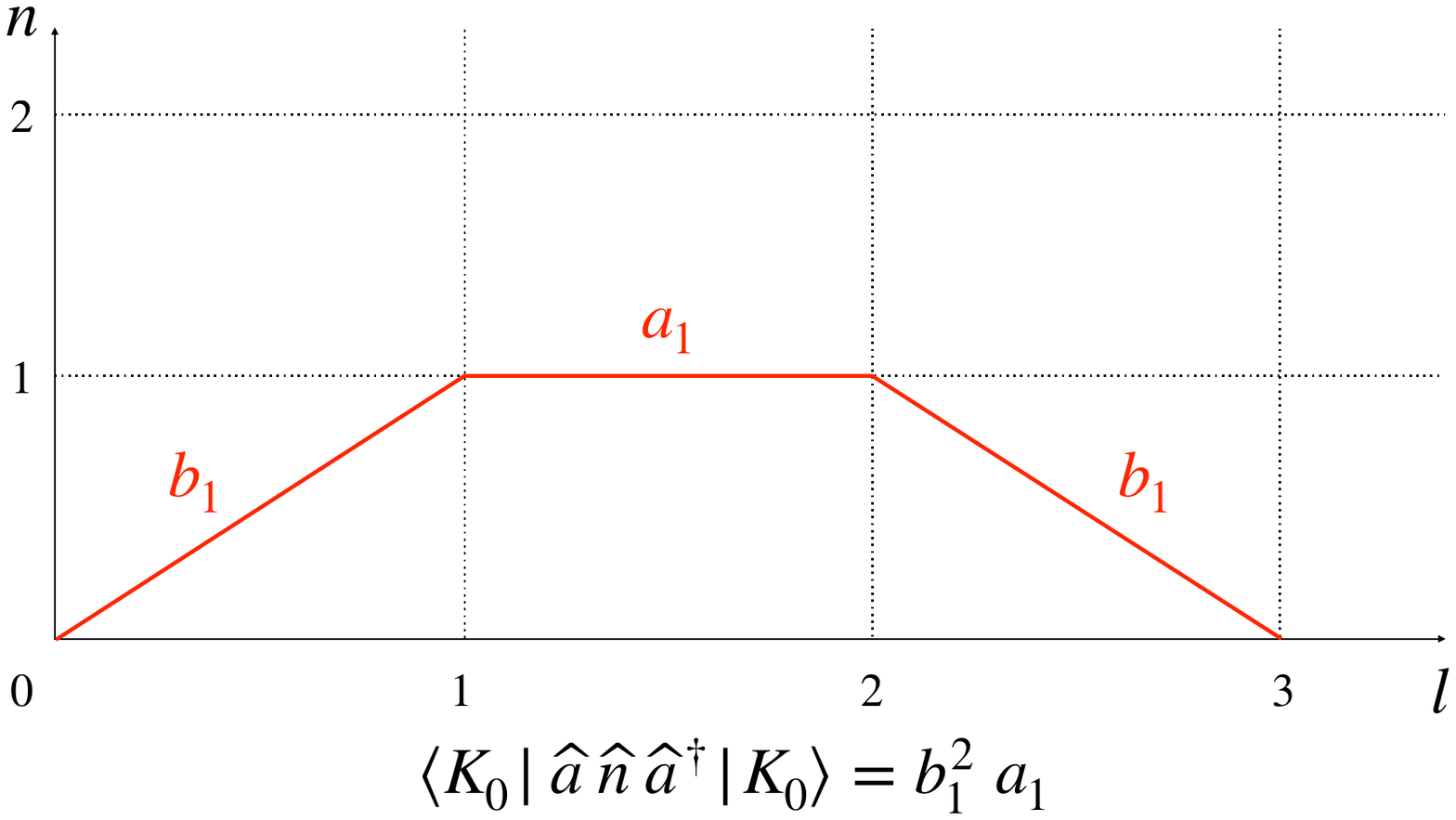}
    \caption{The four nonequivalent Motzkin paths contributing to $M_3$, with their corresponding evaluation. Note that the third Motzkin number given by \eqref{Sect_Lanczos_Motzkin_number} is indeed $m_3=4$.}
    \label{fig:MotzkinPaths_M3}
\end{figure}

As an example, let us show the explicit computation of the third moment $M_3$, assuming no restriction on the height of the Motzkin path (or, equivalently, assuming $K\geq2$), depicted diagrammatically in Figure \ref{fig:MotzkinPaths_M3}:
\begin{equation}
    \label{Sect_Lanczos_moment_M3_explicit}
    \begin{split}
        M_3 &= \langle K_0 | \widehat{n}^3 |K_0\rangle + \langle K_0 | \widehat{a} \widehat{n} \widehat{a}^\dagger |K_0\rangle + \langle K_0 | \widehat{n} \widehat{a} \widehat{a}^\dagger |K_0\rangle + \langle K_0 |  \widehat{a} \widehat{a}^\dagger \widehat{n} |K_0\rangle \\
        &= a_0^3 + b_1^2 a_1 + 2b_1^2 a_0~.
    \end{split}
\end{equation}

At this point, we note that all odd moments $M_{2l+1}$ are given by Motzkin paths of odd length, which must necessarily involve at least a horizontal step, since the number of ascending steps must equal that of descending steps. As each horizontal step is evaluated by an $a$-coefficient, this immediately implies the following corollary:
\begin{equation}
    \label{Sect_Lanczos_corollary_zero_a_coeffs}
    M_{2m+1}=0\qquad \forall ~ m=0,\dots,n \qquad \Longleftrightarrow\qquad a_m=0\qquad \forall~m=0,\dots,n~.
\end{equation}
In particular, all $a$-coefficients will vanish if and only if all odd moments are null\footnote{This result is hard to prove inductively using the Lanczos recursion \eqref{Sect_Lanczos_Kn_proposal}, while it becomes self-evident when analyzed from the optics of Motzkin paths.}. In this case, only even moments $M_{2n}$ will remain being generically non-null, and they will be exclusively given by Dyck paths of length $2n$. This result applies in cases of particular relevance for the subjects of study of this Thesis, such as the Lanczos coefficients associated to quantum-mechanical observables evolving in the Heisenberg picture, as will be detailed in Chapter \ref{ch:chapter02_KC}, or the coefficients of the infinite-temperature thermofield double state in the disorder-averaged and double-scaled SYK model, analyzed in detail in Chapter \ref{ch:chapter05_DSSYK} following publication \cite{IV}.

\subsection{From moments to Lanczos coefficients}\label{sect:Moments_to_Lanczos}

In most practical situations, the moments $M_n$ will either be available or efficiently calculable, and the task will be to compute the Lanczos coefficients out of them. In such a case, knowing that the relation between moments and Lanczos coefficients is bijective, one may exploit the analytical results in the previous section and guess Ansätze that may give at least bounds on the profile of the Lanczos coefficients that are compatible with the given moments. Sometimes this is sufficient to give the asymptotic behavior of $a_n$ and $b_n$ at large $n$, as illustrated in \cite{Parker:2018yvk,Barbon:2019wsy} and reviewed in section \ref{sect:KC_chaos_holog}. Alternatively, one can implement the direct transformation from moments to Lanczos coefficients \cite{ViswanathMuller}. The main objects involved in this transformation are Hankel matrices of moments, defined as
\begin{equation}
    \label{Sect_Lanczos_Hankel_matrices}
    H_n:=\begin{pmatrix}
        M_0 & M_1 & \dots & M_n \\
        M_1 & M_2 & \dots & M_{n+1} \\
        \vdots & \vdots & \ddots & \vdots \\
        M_n & M_{n+1} & \dots & M_{2n}
    \end{pmatrix}~,
\end{equation}
whose determinant we denote as 
\begin{equation}
    \label{Sect_Lanczos_Hankel_det_def}
    D_n:=\det H_n~.
\end{equation}
The Lanczos $a_n$ and $b_n$ coefficients can be written in terms of these determinants as \cite{SANCHEZDEHESA1978275}:
\begin{equation}
    \label{Sect_Lanczos_Lanczos_coeffs_from_moments_most_general}
    \begin{split}
        &a_n = \frac{\mathcal{C}_{n}^{n,~n-1}}{D_{n-1}}-\frac{\mathcal{C}_{n+1}^{n+1,n}}{D_{n}}~,\\
        & b_n^2 = \frac{D_{n-2} D_n}{D_{n-1}^2}~, 
    \end{split}
\end{equation}
where $\mathcal{C}_n^{i,j}$ denotes the cofactor associated to the $(i,j)$ element\footnote{Note that $H_n$ is a $(n+1)\times (n+1)$ matrix, and we use an indexing convention such that matrix elements $(H_n)_{ij}$ are labelled by $i,j=0,\dots,n$.} of the matrix $H_n$ given in \eqref{Sect_Lanczos_Hankel_matrices}.

Operationally, the computational implementation of \eqref{Sect_Lanczos_Lanczos_coeffs_from_moments_most_general} can be made by the means of a recursive algorithm \cite{ViswanathMuller}. If we have access to the moments $\left\{M_n\right\}_{n=0}^{2n_{*}+1}$, we can compute $\left\{a_n\right\}_{n=0}^{n_{*}}$ and $\left\{b_n\right\}_{n=1}^{n_{*}}$. For this, we initialize the following quantities:
\begin{equation}
    \label{Sect_Lanczos_Lanczos_coeffs_from_moments_most_general_recursive_algorithm_seed}
    M_k^{(0)}=(-1)^k M_k~,\qquad L_k^{(0)}=(-1)^{k+1}M_{k+1}~,\qquad k = 0,1,\dots,2n_{*}~.
\end{equation}
Next, we can construct the successive lists $\left\{M^{(n)}_k\right\}_{k=n}^{2n_{*}-n+1}$ and $\left\{L^{(n)}_k\right\}_{k=n}^{2n_{*}-n}$, for $n=1,\dots,n_{*}$, through the recursion:
\begin{equation}
\label{Sect_Lanczos_Lanczos_coeffs_from_moments_most_general_recursive_algorithm_recursion}
    \begin{split}
        &M_k^{(n)}=L_{k}^{(n-1)} - \frac{L_{n-1}^{(n-1)}}{M_{n-1}^{(n-1)}}M_k^{(n-1)}~,\\
        & L_k^{(n)} = \frac{M_{k+1}^{(n)}}{M_n^{(n)}}-\frac{M_k^{(n-1)}}{M_{n-1}^{(n-1)}}~.
    \end{split}
\end{equation}
In practice, this recursion may be implemented by nesting two loops, an outer one for $n=0,\dots, n_{*}$ and an inner one for\footnote{Note that the range of $k$ in $L_k^{(n)}$ is $k=n,\dots,2n_{*}-n$. The overlooking of this fact, together with a wrong range for $n$ in the outer loop, constitute typos in the reference \cite{ViswanathMuller}.} $k=n,n+1,\dots,2n_{*}-n+1$. With this, the Lanczos coefficients can be evaluated from:
\begin{equation}
    \label{Sect_Lanczos_Lanczos_coeffs_from_moments_most_general_recursive_algorithm_Lanczos_coeffs}
    b_n^2 = M_n^{(n)}~,\qquad a_n = -L_n^{(n)}~,
\end{equation}
for $n=0,\dots,n_{*}$ in both cases. We note that the range of $k$ for both $M^{(n)}_k$ and $L^{(n)}_k$ at fixed $n$ is empty whenever $n\geq n_{*}+1$, consistent with the fact that no more Lanczos coefficients $a_n$ or $b_n$ with $n>n_{*}$ can be computed out of the given finite collection of moments $\left\{M_n\right\}_{n=0}^{2n_{*}+1}$. See \cite{ViswanathMuller} for more details on this recursive algorithm and for the symbolic expression of the first few Lanczos coefficients in terms of moments.

As argued around \eqref{Sect_Lanczos_corollary_zero_a_coeffs}, whenever odd moments are zero all the $a_n$ coefficients vanish. In this case the transformation from moments to Lanczos coefficients simplifies slightly: There are only $b_n$ coefficients, still given by the second line of \eqref{Sect_Lanczos_Lanczos_coeffs_from_moments_most_general}, and in order to evaluate the determinants involved, one can implement another recursive algorithm \cite{ViswanathMuller}, consisting on a simplified version of \eqref{Sect_Lanczos_Lanczos_coeffs_from_moments_most_general_recursive_algorithm_recursion}. Assuming access to the even moments $\left\{M_{2n}\right\}_{n=0}^{n_{*}}$, it is possible to compute $\left\{b_n\right\}_{n=1}^{n_{*}}$ by initializing the list $M^{(0)}_{2k}=M_{2k}$ for $k=0,\dots,n_{*}$ and progressively generating the successive lists $\left\{M^{(n)}_{2k}\right\}_{k=n}^{n_{*}}$, with $n=1,\dots,n_{*}$, by the means of the recursion:
\begin{equation}
    \label{Sect_Lanczos_Lanczos_coeffs_from_moments_aNull_recursive_algorithm}
    M^{(n)}_{2k} = \frac{M^{(n-1)}_{2k}}{b_{n-1}^2}-\frac{M^{(n-2)}_{2k-2}}{b_{n-2}^2}~,
\end{equation}
with which the Lanczos coefficients can be evaluated from
\begin{equation}
    \label{Sect_Lanczos_Lanczos_coeffs_from_moments_aNull_recursive_algorithm_coeffs}
    b_n^2 = M^{(n)}_{2n}~.
\end{equation}
For consistency, we note that the recursion \eqref{Sect_Lanczos_Lanczos_coeffs_from_moments_aNull_recursive_algorithm} needs to be complemented with the boundary conditions
\begin{equation}
\label{Sect_Lanczos_Lanczos_coeffs_from_moments_aNull_recursive_algorithm_bdry_condits}
b_0\equiv b_{-1} \equiv 1~,\qquad M^{(-1)}_{2k} = 0 \quad \text{for any }k~.
\end{equation}

There exist \cite{ViswanathMuller} recursive computational algorithms similar to \eqref{Sect_Lanczos_Lanczos_coeffs_from_moments_most_general_recursive_algorithm_recursion} and \eqref{Sect_Lanczos_Lanczos_coeffs_from_moments_aNull_recursive_algorithm} for the evaluation of the moments out of the Lanczos coefficients (i.e. the reverse transformation, which would belong to section \ref{subsect:From_Lanczos_to_Moments}), but we shall not consider them here, as the problems of interest in this Thesis involve the computation of Lanczos coefficients in contexts in which the moments are known or easily computable. For the reverse direction, we shall only consider the combinatorial relations described in section \ref{subsect:From_Lanczos_to_Moments} because they allow for some analytical grasp.

As a matter of fact, in computational implementations, not even the recursions \eqref{Sect_Lanczos_Lanczos_coeffs_from_moments_most_general_recursive_algorithm_recursion} and \eqref{Sect_Lanczos_Lanczos_coeffs_from_moments_aNull_recursive_algorithm} shall be extensively used in our case, because they require high precision arithmetics in order to compute reliably more than a few Lanczos coefficients, the reason for this being that the moments in \eqref{Sect_Lanczos_moments_of_H} grow roughly exponentially in $k$, while the Lanczos coefficients are of the order of the spectral width; hence, important numerical cancellations between large numbers need to operate in order to yield smaller numbers (the Lanczos coefficients) in the afore-mentioned recursions, making them difficult to work with in practice. Due to this, in the case of the systems studied numerically in Chapters \ref{ch:chapter03_SYK} and \ref{ch:chapter04_Integrable}, the preferable method was a direct implementation of the Lanczos algorithm, which itself also suffers from some numerical instabilities that are nevertheless controllable, as we shall review in the next section.

\section{Numerical instabilities of the Lanczos algorithm} \label{sect:reorthog_algorithms}

As formulated in section \ref{sect:Lanczos}, the Lanczos algorithm suffers from severe numerical instabilities due to its inherent recursive nature: In a given Lanczos step $n$, the Krylov element $|K_n\rangle$ is built out of $|K_{n-1}\rangle$ and $|K_{n-2}\rangle$, as illustrated in \eqref{Sect_Lanczos_Kn_proposal}, which are themselves the outcomes of the two previous Lanczos steps, as a consequence of which numerical errors due to the implementation of the algorithm with finite-precision arithmetic accumulate through the Lanczos steps and result in the eventual loss of orthogonality in the numerically constructed Krylov basis.

In his seminal article \cite{Lanczos:1950zz}, Lanczos himself proposed to combine the recursion method with full Gram-Schmidt (GS) steps in order to maintain orthogonality of the basis. Nowadays, in the literature on numerical methods there exist more systematic algorithms for curing this numerical instability (see e.g. \cite{Parlett,PRO,SO}). The numerical works of \cite{I,II,III}, presented in Chapters \ref{ch:chapter03_SYK} and \ref{ch:chapter04_Integrable} make use of the \textit{Full Orthogonalization} (FO) and \textit{Partial Re-Orthogonalization} (PRO) algorithms, which we shall describe generically in this section. Such algorithms may allow to compute the full sequences of Lanczos coefficients and Krylov elements at fixed, double-floating point precision for reasonably large matrices, but they come at the cost requiring more execution time and more memory in order to store all the Krylov basis elements\footnote{From the point of view of the numerical eigenvalue problem, the Krylov elements needed to be stored anyway because they are used to obtain the eigenvectors. However, in the physical applications of the Lanczos algorithm that we will introduce in next Chapter (in particular, the computation of Krylov complexity), only the Lanczos coefficients are needed, and hence this memory requirement may be regarded as a drawback of the re-orthogonalization algorithms.}.

In what follows we will give a description of the FO and PRO algorithms, providing their systematic formulation which can be readily implemented on a computer. Within the framework of those presentations, objects like $a_n$, $b_n$, $|A_n\rangle$ and $|K_n\rangle$ will stand for the \textit{numerically computed} coefficients or vectors, rather than their exact counterparts.

\subsection{Full Orthogonalization (FO)}

This algorithm \cite{Parlett} systematizes the proposal by Lanczos \cite{Lanczos:1950zz} and amounts to replacing each Lanczos step by a full GS step. That is, in every step the operator $H$ is applied to the last Krylov element built, and the result is explicitly orthogonalized against \textit{all} of the previous basis elements computed, eventually evaluating the Lanczos coefficients according to their original definition. Conventionally \cite{Parlett}, the Gram-Schmidt orthogonalization is performed explicitly twice in order to reduce errors.

Altogether, the FO algorithm may be formulated as follows:
\begin{itemize}
    \item[\underline{$n=0$:}]
    \begin{itemize}
        \item[$\bullet$] Set $|K_0\rangle = |\Omega\rangle$ (normalized by assumption).
        \item[$\bullet$] Assign $a_0 = \langle K_0 |H|K_0\rangle$.
    \end{itemize}
    \item[\underline{$n\geq 1$}:] 
    \begin{itemize}
        \item[$\bullet$] Set $|A_n\rangle = H|K_{n-1}\rangle$
        \item[$\bullet$] GS step: $|A_n\rangle \longmapsto |A_n\rangle - \sum_{m=0}^{n-1} \langle K_m|A_n\rangle \, |K_m\rangle$.
        \item[$\bullet$] \textbf{Repeat} the previous subtraction.
        \item[$\bullet$] \textbf{If} $\sqrt{\langle A_n|A_n\rangle} = 0$ \textbf{break}. \textbf{Otherwise} continue.
        \item[$\bullet$] Assign $b_n = \sqrt{\langle A_n|A_n\rangle}$.
        \item[$\bullet$] Set $|K_n\rangle = \frac{1}{b_n}|A_n\rangle$.
        \item[$\bullet$] Assign $a_n = \langle K_n | H | K_n\rangle$.
        \item[$\bullet$] Increase $n\longmapsto n+1$ and \textbf{repeat} the current step.
    \end{itemize}
\end{itemize}

We remind that, just like for the original Lanczos algorithm formulated in section \ref{sect:Lanczos}, the termination condition checking whether the norm of $|A_n\rangle$ is null is computationally implemented by comparing such a quantity to a numerical threshold of choice, generically related to the working precision.

\subsection{Partial Re-Orthogonalization (PRO)}

The FO algorithm is somewhat inefficient because the Gram-Schmidt steps increase significantly the execution time of the recursion: In every FO step one needs to compute $n$ inner products and perform $2n$ subtractions, as opposed to the computation of two coefficients and two subtractions per step in the original Lanczos algorithm. Furthermore, it is an overkill method because it always performs the explicit re-orthogonalization even in steps in which it might not be required. Partial Re-Orthogonalization \cite{PRO,Parlett,SO} improves on this by running a new recursion, parallel to the Lanczos recursion, which keeps track of how numerical errors build up in the Krylov basis, and performs a full orthogonalization step only when such errors cross a customizable threshold. This ensures orthogonality of the output Krylov basis up to a controllable parameter.

We start by writing a generic Lanczos step where, as announced, the objects written represent the numerically constructed vectors and coefficients; because of this, we need to add an extra vector in the relation between the computed Krylov elements that accounts for the accumulated numerical error, which deviates them from their exact counterparts and makes them fail to fulfill exactly \eqref{Sect_Lanczos_Kn_proposal}:
\begin{equation}
    \label{Sect_Lanczos_recursion_with_error}
    b_n |K_n\rangle = (H-a_{n-1}) |K_{n-1}\rangle -b_{n-1}|K_{n-2}\rangle + |\xi_{n-1}\rangle~,
\end{equation}
where $|\xi_{n-1}\rangle$ is the promised extra vector accounting for numerical error. Equation \eqref{Sect_Lanczos_recursion_with_error} can be turned into a recursion that estimates the buildup of the spurious overlap between different Krylov elements (which should be orthogonal). Defining the inner product 
\begin{equation}
    \label{Sect_Lanczos_PRO_W_def}
    W_{mn}:= \langle K_m | K_n\rangle 
\end{equation}
and acting with $\langle K_m|$ on \eqref{Sect_Lanczos_recursion_with_error} yields
\begin{equation}
    \label{Sect_Lanczos_recursion_errors_raw}
    b_n W_{mn} = T_{m, n-1}-a_{n-1} W_{m,n-1} - b_{n-1} W_{m, n-2} +\langle K_m | \xi_{n-1}\rangle~, 
\end{equation}
where $T_{mn}:= \langle K_m | K_n\rangle$ is the matrix element of the restriction of $H$ to Krylov space. It may not necessarily be real, symmetric and tridiagonal as expected from \eqref{Sect_Lanczos_Tridiag_H_matrix} because of the accumulation of numerical errors. We may now relabel the indices in \eqref{Sect_Lanczos_recursion_errors_raw} as $m \leftrightarrow n-1$, obtaining the following equivalent expression:
\begin{equation}
    \label{Sect_Lanczos_recursion_errors_indices_relabeled}
    b_{m+1} W_{n-1,m+1} = T_{n-1,m}-a_m W_{n-1,m} - b_{m} W_{n-1, m-1}+ \langle K_{n-1} | \xi_m\rangle~. 
\end{equation}
Subtracting \eqref{Sect_Lanczos_recursion_errors_indices_relabeled} from \eqref{Sect_Lanczos_recursion_errors_raw} and using that $T_{mn}=T_{nm}$ up to numerical errors that we loosely absorb into the contributions involving the vectors $|\xi_n\rangle$, we find a recursion between the overlaps:
\begin{equation}
    \label{Sect_Lanczos_recursion_errors_indices_final}
    \begin{split}
        W_{mn} = &\frac{1}{b_n}\Big\{ b_{m+1} W^{*}_{m+1,n-1} + a_m W^{*}_{m,n-1} + b_m W^{*}_{m-1,n-1} - a_{n-1} W_{m,n-1} -b_{n-1}W_{m,n-2}  \Big. \\
        & \Big. + \langle K_m |\xi_{n-1}\rangle - \langle K_{n-1} | \xi_m \rangle \Big\}\equiv \frac{1}{b_n}\Big\{ \widetilde{W}_{mn} + \mathcal{E}_{m,n-1}  \Big\}~,
    \end{split}
\end{equation}
where in the last step we have recast the recursion into a term $\widetilde{W}_{mn}$ involving the computable overlaps, given in the first line of the expression \eqref{Sect_Lanczos_recursion_errors_indices_final}, plus an error contribution $\mathcal{E}_{m,n-1}$ that can be read from the second line and which is in principle unknown. Given a working machine precision $\varepsilon_M$, it may be estimated as a quantity of order $\varepsilon_M \lVert H \rVert$, where $\lVert H \rVert$ is the norm\footnote{It should be thought of as the norm in operator space inherited from the norm in the departing Hilbert space $\mathcal{H}$.} of the operator $H$:
\begin{equation}
    \label{Sect_Lanczos_PRO_error_estimate}
    \mathcal{E}_{m,n-1}:=\langle K_m |\xi_{n-1}\rangle - \langle K_{n-1} | \xi_m \rangle \approx \frac{\widetilde{W}_{mn}}{\lvert \widetilde{W}_{mn} \rvert}\cdot 2 \varepsilon_M \lVert H \rVert~.
\end{equation}
Since $\mathcal{E}_{m,n-1}$ is in general a complex number, it needs to be given a phase, and \eqref{Sect_Lanczos_PRO_error_estimate} proposes the conservative estimate of giving it the same phase as that of $\widetilde{W}_{mn}$, so that its contribution is maximally disruptive. Along the same lines, the factor of $2$ is introduced because of the fact that the error contribution has been defined as the sum of two terms, each of which are modelled to be of the same order.

The recursion \eqref{Sect_Lanczos_recursion_errors_indices_final} may be exploited to keep track of how spurious overlaps between Krylov elements grow in the successive Lanczos steps. Specifically, PRO proposes a modified Lanczos algorithm in which at every step $n$ one computes not only the Krylov element $|K_n\rangle$, but also estimates the spurious overlap with previously computed Krylov elements, $W_{mn}$ with $m<n$ by the means of \eqref{Sect_Lanczos_recursion_errors_indices_final}.

Graphically, the algorithm computes iteratively the columns of the hermitian matrix
\begin{equation}
    \label{Sect_Lanczos_PRO_W_matrix}
    \Big(W_{mn}\Big) = \begin{pmatrix}
       W_{00} & W_{01}& W_{02} & \dots & W_{0,K-1} \\
       \cdot & W_{11}& W_{12} & \dots & W_{1,K-1} \\
       \cdot  & \cdot & W_{22} & \dots & W_{2,K-1} \\
       \vdots & \vdots & \vdots &  \ddots & \vdots \\
       \cdot & \cdot & \cdot &\dots & W_{K-1,K-1}
    \end{pmatrix}~.
\end{equation}
The matrix elements below the main diagonal can in principle be determined by hermiticity; however, within a fixed column (i.e. fixed Lanczos step $n$), they cannot be predicted in terms of the elements above the diagonal of the previous columns (i.e. previous Lanczos steps), as they correspond to overlaps $W_{mn}=\langle K_m | K_n\rangle$ with $m>n$, that is, overlaps with Krylov elements that have not been generated yet. The PRO algorithm uses, at a given Lanczos step $n$, the recursion \eqref{Sect_Lanczos_recursion_errors_indices_final} to estimate the spurious overlaps with the previous basis elements, $W_{mn}$ with $m\leq n$ assuming knowledge of all $\left\{W_{ij},\; i=0,\dots,j,\; \forall j = 0,\dots, n-1\right\}$. It should be noted, however, that the presence of the term $W^{*}_{m+1,n-1}$ in the cited expression prevents it from being able to predict both $W_{n-1,n}$ and $W_{nn}$ in terms of the above-mentioned set of inner products. The lack of predictability of both quantities is not a problem because in the Lanczos step \eqref{Sect_Lanczos_recursion_proposal} $|A_n\rangle$ is explicitly orthogonalized against $|K_{n-1}\rangle$ and subsequently renormalized so that $|K_n\rangle$ has unit norm, and hence one can assign $W_{n-1,n}=\varepsilon_M$ and $W_{nn}=1$. 

After computing all $W_{mn}$ with $m<n$, given a certain numerical threshold $\varepsilon_T\geq \varepsilon_M$ of choice, the PRO algorithm will implement a full Gram-Schmidt step whenever any of these quantities surpasses the threshold, granting orthogonality of the output Krylov basis up to $\varepsilon_T$.

All in all, the Lanczos algorithm incorporating PRO, sometimes denoted as \textit{LanPRO} \cite{Parlett}, may be formulated as follows:
\begin{itemize}
    \item[\underline{$n=0$:}]
    \begin{itemize}
        \item[$\bullet$] Set $|K_0\rangle = |\Omega\rangle$ (normalized by assumption).
        \item[$\bullet$] Assign $W_{00}=1$.
        \item[$\bullet$] Assign $a_0=\langle K_0| H | K_0\rangle$.
    \end{itemize}
    \item[\underline{$n=1$:}]
    \begin{itemize}
        \item[$\bullet$] Lanczos step: $|A_1\rangle = (H-a_0)|K_0\rangle$.
        \item[$\bullet$] Assign $b_1=\sqrt{\langle A_1 | A_1\rangle}$. \textbf{If} $b_1<\varepsilon_T$, \textbf{break}. \textbf{Otherwise} continue.
        \item[$\bullet$] Set $|K_1\rangle = \frac{1}{b_1}|A_1\rangle$.
        \item[$\bullet$] Assign $W_{01} = \varepsilon_M$ and $W_{11}=1$.
        \item[$\bullet$] Assign $a_1 = \langle K_1 | H | K_1 \rangle$.
    \end{itemize}
    \item[\underline{$n\geq 2$:}] External \textbf{loop} for $n$:
    \begin{itemize}
        \item[$\bullet$] Lanczos step: $|A_n\rangle = (H-a_{n-1})|K_{n-1}\rangle - b_{n-1}|K_{n-2}\rangle$.
        \item[$\bullet$] Compute \textit{a-priori b-coefficient:} $b_n = \sqrt{\langle A_n|A_n\rangle}$. \textbf{If} $b_n<\varepsilon_T$ \textbf{break}. \textbf{Otherwise} continue.
        \item[$\bullet$] Assign $W_{n-1,n}=\varepsilon_M$ and $W_{nn}=1$.
        \item[$\bullet$] Internal \textbf{loop} for $m=0,\dots,n-2$:
        \begin{itemize}
            \item[$-$] Assign, in order: 
            \begin{equation}
            \label{Sect_Lanczos_PRO_error_recursion_algorithm_formulation}
            \begin{split}
                & \widetilde{W}_{mn} = b_{m+1} W^{*}_{m+1,n-1} + a_m W^{*}_{m,n-1} + b_m W^{*}_{m-1,n-1} - a_{n-1} W_{m,n-1} -b_{n-1}W_{m,n-2}~,\\
                &\mathcal{E}_{m,n-1} = \frac{\widetilde{W}_{mn}}{\lvert \widetilde{W}_{mn} \rvert}\cdot 2 \varepsilon_M \lVert H \rVert~, \\
                & W_{mn} = \frac{1}{b_n}\Big\{ \widetilde{W}_{mn} + \mathcal{E}_{m,n-1} \Big\}~.
            \end{split}
            \end{equation}
        \end{itemize}
        \item[$\bullet$] \textbf{If} there is some $m\leq n-2$ such that $W_{mn}>\varepsilon_T$, \textbf{do}:
        \begin{itemize}
            \item[$-$] GS for $|A_{n-1}\rangle \longmapsto |A_{n-1}\rangle - \sum_{l=0}^{n-2} \langle K_l|A_{n-1}\rangle ~|K_l\rangle$.
            \item[$-$] \textbf{Repeat} the previous subtraction.
            \item[$-$] Recompute $b_{n-1}$. \textbf{If} $b_{n-1}< \varepsilon_T$ \textbf{break}. \textbf{Otherwise} recompute $|K_{n-1}\rangle =\frac{1}{b_{n-1}}|A_{n-1}\rangle $, and recompute $a_{n-1}=\langle K_{n-1}| H | K_{n-1}\rangle$.
            \item[$-$] GS for $|A_{n}\rangle \longmapsto |A_{n}\rangle - \sum_{l=0}^{n-1} \langle K_l|A_{n}\rangle ~ |K_l\rangle$.
            \item[$-$] \textbf{Repeat} the previous subtraction.
            \item[$-$] Recompute $b_{n}$. \textbf{If} $b_{n}< \varepsilon_T$ \textbf{break}. \textbf{Otherwise} compute $|K_{n}\rangle =\frac{1}{b_{n}}|A_{n}\rangle $, and assign $a_{n}=\langle K_{n}| H | K_{n}\rangle$.
            \item[$-$] Assign $W_{l,n-1} = \delta_{l,n-1} + (1-\delta_{l,n-1})\varepsilon_M$ for all $l=0,\dots,n-1$.
            \item[$-$] Assign $W_{l,n} = \delta_{l,n} + (1-\delta_{l,n})\varepsilon_M$ for all $l=0,\dots,n$.
        \end{itemize}
        \item[$\bullet$] \textbf{Otherwise do:}
        \begin{itemize}
            \item[$-$] Set $|K_n\rangle = \frac{1}{b_n}|A_n\rangle$.
            \item[$-$] Assign $a_n = \langle K_n|H|K_n\rangle$.
        \end{itemize}
    \end{itemize}

\end{itemize}

Let us make some clarifications at this point:

\begin{itemize}
    \item The numerical threshold for deciding whether a quantity is effectively zero, $\varepsilon_T$, is a parameter of choice. It may be taken to be e.g. $\varepsilon_T = \sqrt{\varepsilon_M}>\varepsilon_M$ (assuming $\varepsilon_M<1$). The interpretation of the interplay between $\varepsilon_M$ and $\varepsilon_T$ is that spurious overlaps are initialized to be of the order of the working precision, $\varepsilon_M$, and iteratively implementing the recursion \eqref{Sect_Lanczos_recursion_errors_indices_final} they are allowed to gradually grow up to $\varepsilon_T$, when the GS step will be enforced. That is, orthogonality of the output Krylov basis is granted up to errors of order $\varepsilon_T$.
    \item Since the spurious overlaps are initialized to quantities of order $\varepsilon_M$ and grow from there, in cases in which the hermitian operator $H$ is bounded so that its norm is of order one, the error term $\mathcal{E}_{m,n-1}$ in \eqref{Sect_Lanczos_PRO_error_recursion_algorithm_formulation} can be directly neglected as it will be at least of the order of the rest of the terms in the recursion for $\widetilde{W}_{mn}$, whose numerical implementation will anyway generate automatically new errors of order $\varepsilon_M$ themselves. This simplifying choice was made in the implementations of LanPRO for the projects \cite{I,II,III}.
    \item We note from \eqref{Sect_Lanczos_PRO_error_recursion_algorithm_formulation} that, at a given step $n$, the overlaps $\left\{W_{mn}\right\}_{m=0}^{n-2}$ are determined solely out of $\left\{W_{m,n-1}\right\}_{m=0}^{n-1}$ and $\left\{W_{m,n-2}\right\}_{m=0}^{n-2}$. That is, for the determination of the items above the diagonal of a given column of the matrix \eqref{Sect_Lanczos_PRO_W_matrix} only the items (above the diagonal) of the two previous columns are required. The preceding ones do not need to be saved in the local memory. Likewise, because of this reason, whenever the threshold $\varepsilon_T$ is crossed the explicit GS step is performed both for $|A_{n-1}\rangle$ and $|A_n\rangle$, since in this way both lists $\left\{W_{m,n-1}\right\}_{m=0}^{n-1}$ and $\left\{W_{m,n}\right\}_{m=0}^{n}$ can be re-initialized to orthogonality up to machine precision, so that the error recursion starts over in the next step $n+1$.
\end{itemize}

%% file: content/Chapter02.tex
\chapter{\rm\bfseries Krylov complexity}
\label{ch:chapter02_KC}

Chapter \ref{ch:chapter01_Lanczos} introduced the Lanczos algorithm and the algebraic structure of Krylov space in an intentionally generic manner. There, $H$ denoted some \textit{hermitian} operator acting over some Hilbert space, probed with the combination of the latter and a seed vector $|\Omega\rangle$. These tools were originally developed as an efficient approach to the numerical eigenvalue problem, see \cite{Parlett,KrylovBook_Vorst,KrylovBook_Liesen} for a review, but they have eventually found wide application in physics for tasks that go far beyond the computation of the spectrum of some operator of interest.

An area in which the recursion method has provided new insights and has even become a line of research of its own is many-body physics. For the analysis of the dynamics of quantum systems, one can take the hermitian operator $H$ to be the generator of time evolution: Either the quantum Hamiltonian (denoted as $H_{\text{q}}$ only for the sake of this discussion) for the evolution of states in the Schrödinger picture, or the Liouvillian if one focuses on operator evolution in the Heisenberg picture, in which case the discussions in the previous Chapter carry over upon the replacement $H\mapsto \mathcal{L}\equiv [H_{\text{q}},\cdot]$ and $|\Omega\rangle \longmapsto \mathcal{O}$, where $\mathcal{O}$ is the operator whose evolution is of interest (at the end of the day, operator space is a perfectly valid Hilbert space of which $\mathcal{O}$ is an element and on which $\mathcal{L}$ acts as a hermitian operator). The application of the Lanczos algorithm to the initial condition using the corresponding generator of time translations yields, as we will see, a basis of the Hilbert space adapted to time evolution and a set of Lanczos coefficients that governs the dynamics in Krylov space. Crucially, these coefficients are intimately related to observables such as the two-point function in the case of operator evolution and its various cousins obtained through Laplace and Fourier transforms, whose analytical structures encode information on the system's spectrum. These quantities can be reconstructed from the Lanczos coefficients, and importantly they may also be approximated out of finite truncations of such sequences. The reader interested in applications of the recursion method to many-body physics as an efficient numerical (and analytical) method for analyzing dynamical and spectral properties is encouraged to consult the detailed review \cite{ViswanathMuller}\footnote{This review also contains some illustrations of the application of the Lanczos algorithm to classical evolution, where the Hilbert space is the space of functions over phase space, and the time evolution generator is the action of the Poisson bracket involving the classical Hamiltonian $H_{\text{cl}}$, where the role of the generic $H$ in the discussion from Chapter \ref{ch:chapter01_Lanczos} would be played by $H\longmapsto\left\{H_{\text{cl}},\cdot\right\}$.}.

Building up on these applications of the recursion method in the context of condensed matter, the article \cite{Parker:2018yvk} proposed to use the Lanczos algorithm to study operator growth and quantum chaos. The notion of \textit{Krylov complexity} (or \textit{K-complexity}) was presented in that work as the position expectation value of the evolving observable $\mathcal{O}(t)$ with respect to its own Krylov basis. A higher value of K-complexity denotes the fact that the operator is further within the time evolution that can be generated by the system's Liouvillian. At early times (that is before the scrambling time scale\footnote{See \cite{Altland:2020ccq} for a summary of quantum and classical chaotic time scales.}), for systems with $k$-local Hamiltonians, K-complexity is comparable to size complexity, defined as the spatial extent over which the operator has a non-trivial support, but it eventually surpasses it because the Krylov basis is a complete basis of Krylov space, and therefore K-complexity still has scope of growth once size complexity is saturated in a finite system \cite{Barbon:2019wsy}. A faster propagation through the Krylov basis denotes that the time evolution generator is an efficient scrambler, and \cite{Parker:2018yvk} proposes to use this as a diagnostic of quantum chaos. In particular, since the Krylov space dynamics are controlled by the Lanczos coefficients, the \textit{universal operator growth hypothesis} presented in that work proposes that maximally chaotic systems in the thermodynamic limit will exhibit a maximal growth of the $b_n$-sequence for simple operators (proved to be linear at infinite temperature, and conjectured to be so at finite temperature), implying the fastest possible Krylov complexity growth (namely, exponential). Additional mathematical relations to other notions of complexity presented by the authors and which we shall review in this Chapter put on more solid footing the use of K-complexity as a probe of quantum chaos.

Bringing this notion of complexity to the context of holography and black hole physics as a plausible candidate for being the item in the holographic dictionary describing the size of the ERB from the boundary perspective was proposed in \cite{Barbon:2019wsy}. The authors of this work complemented the study of \cite{Parker:2018yvk} by considering systems away from the thermodynamic limit. This allowed them to make some predictions on the expected behavior of K-complexity in the post-scrambling regime, which was found to be linear for thermalizing systems\footnote{We define a thermalizing system as a system in which typical (local or non-extensive) operators satisfy the eigenstate thermalization hypothesis (ETH), which is generally fulfilled in quantum chaotic systems \cite{PhysRevA.30.504,PhysRevA.43.2046,Srednicki_1999,Srednicki:1994mfb,DAlessio:2015qtq}.} in agreement with gravitational expectations for holographic complexities \cite{Stanford:2014jda,Ben-Ami:2016qex,Chapman:2016hwi,Carmi:2017jqz}. The immediate next step was to verify numerically this K-complexity profile for a specific instance of a finite system with a holographic gravitational dual, in order to also access the eventual complexity saturation at late times: This was done for the first time in \cite{I} studying the complex SYK model, in a work to which we shall devote Chapter \ref{ch:chapter03_SYK}.

Complementing the work by \cite{Parker:2018yvk} and given that, as explained, the Lanczos algorithm can be applied to any hermitian operator and probe vector belonging to any Hilbert space, the authors of \cite{Balasubramanian:2022tpr} propose to use Krylov complexity to describe the evolution (in the Scrödinger picture) of states (rather than observables) in a system's Hilbert space with respect to their Krylov basis. In the case of the evolution of the thermofield double state (TFD) \cite{Maldacena:2001kr}, a state relevant for studies in holography, the Lanczos coefficients generating this evolution happen to be in correspondence to the moments of the spectral form factor \cite{Altland:2020ccq,HaakeBook,efetov_1996,stöckmann_1999}, which itself also contains imprints of the chaotic nature of systems, as it is related to the two-point spectral correlations. The first analytical study of the state Krylov complexity of the TFD in a low-dimensional instance of holography, which resulted the establishment of an exact correspondence between K-complexity and wormhole length, was carried out in the project \cite{IV} on the double-scaled SYK system, which will be treated in Chapter \ref{ch:chapter05_DSSYK}.

The area of Krylov complexity is very young and, as such, not exempt from debate. For instance, it is not immediate how to generalize the Lanczos algorithm to systems whose Hilbert space has a more intricate structure as compared to quantum mechanical models whose space consists of a discrete tensor product. This is the case of quantum field theory (QFT) in general and conformal field theory (CFT) in particular. In this context, the works \cite{Dymarsky:2021bjq,Avdoshkin:2022xuw} pointed out that the universal operator growth hypothesis is not immediately correct at finite temperature in QFT. We will review in this Chapter the interesting caveats raised by the authors of these works, on which publications \cite{I,II,III} also intend to shed light. In addition, providing support for the holographic interpretation of Krylov complexity requires the explicit computation of such a quantity in concrete models, which are often low-dimensional, and an eventual generalization to high dimensions constitutes an open area of research.

To summarize, the driving forces of the projects supporting this Thesis are the use of K-complexity as a probe of quantum chaos and as an item in the holographic dictionary, two proposals that are intimately related. Hence, the present Chapter intends to provide the essential definitions and constructions required for this task, taken from the seminal references in the area and combined with some original analyses that will hopefully enrich them. Because of this, the selection of topics is, even though relatively basic and generic, oriented towards the above-mentioned goals. Nevertheless, the recent years have seen an explosion of articles on K-complexity: As a guide for the reader, a final section is devoted to list the main lines of research in the area up to this Thesis' submission date, together with the relevant references.

\section{Definition of Krylov complexity}\label{sect:KCdef}

In the spirit of the wide applicability of the Lanczos algorithm that Chapter \ref{ch:chapter01_Lanczos} emphasized, this section shall present separately the definition of Krylov complexity for operators and states. The algebraic structures underlying both constructions will be completely analogous but we shall see that, in a concrete physical application, each one gives access to different types of observables and, consequently, to different features of the system studied, therefore constituting complementary approaches. 

\subsection{Operator K-complexity}\label{sect:KC_op}

The notion of Krylov complexity was introduced by Parker et al. in \cite{Parker:2018yvk}, an article that proposed to study operator growth building up on the applications of the recursion method in the context of many-body physics. Applying the Lanczos algorithm to an initial observable (seen as an element of the operator Hilbert space) and the system's Liouvillian (which is a hermitian operator acting on operator Hilbert space\footnote{In the condensed matter literature, in order to avoid confusion, the term \textit{superoperator} is often used to refer to an operator that acts over operator space. Similarly, elements of operator space are often dubbed \textit{superstates} in this context.}) results in an ordered Krylov basis that is well adapted to time evolution. Because of the importance of this construction, even though it is a particularization of the general algorithm described in section \ref{sect:Lanczos}, let us define explicitly the main objects and tools that arise in the framework of operator Krylov complexity.

Given a Hilbert space $\mathcal{H}$ of dimension $D$, we consider the space of linear operators acting on it, $\widehat{\mathcal{H}}$, whose dimension is $D^2$. The time evolution of an operator $\mathcal{O}\in \widehat{\mathcal{H}}$ in the Heisenberg picture is given by\footnote{Throughout this Thesis we shall use natural units, setting $\hbar \equiv c \equiv 1$.}:
\begin{equation}
    \label{Sect_KCdef_Ot_BCH}
    \mathcal{O}(t)=e^{itH} \mathcal{O} e^{-itH}=\sum_{n=0}^{+\infty}\frac{{(it)}^{n}}{n!}\overbrace{\Big[H,\big[\dots,}^{(n)}\mathcal{O}\big]\Big]~,
\end{equation}
where the last step makes use of the Baker-Campbell-Hausdorff (BCH) formula. For notational convenience, we shall label elements of operator space with smooth kets, i.e. $\mathcal{O}$ es represented by $|\mathcal{O})$. Likewise, we will make use of the Liouvillian superoperator $\mathcal{L}$ acting over $\widehat{\mathcal{H}}$, defined through
\begin{equation}
    \label{Sect_KCdef_Liouvillian_def}
    \mathcal{L}|\mathcal{O}) = \Big| [H,\mathcal{O}] \Big)
\end{equation}
for any $|\mathcal{O})\in\widehat{\mathcal{H}}$. With this, expression \eqref{Sect_KCdef_Ot_BCH} can be written more compactly as:
\begin{equation}
    \label{Sect_KCdef_Ot_notation}
    \big| \mathcal{O}(t)\big) = e^{it\mathcal{L}}|\mathcal{O})=\sum_{n=0}^{+\infty}\frac{{(it)}^n}{n!}\mathcal{L}^n|\mathcal{O})~,
\end{equation}
i.e. the nested commutators of the Hamiltonian with the operator are conveniently re-expressed as powers of the Liouvillian acting on the initial condition. From \eqref{Sect_KCdef_Ot_notation}, the definition of the Krylov space associated to the observable $\mathcal{O}$ is direct:
\begin{equation}
    \label{Sect_KCdef_OperatorKrylovSpace_def}
    \widehat{\mathcal{H}}_{\mathcal{O}}:=\text{span}\left\{\mathcal{L}^n |\mathcal{O})\right\}_{n\geq 0}~.
\end{equation}
By construction, \eqref{Sect_KCdef_OperatorKrylovSpace_def} is a subspace of $\widehat{\mathcal{H}}$, and therefore the Krylov space dimension, denoted $K:=\dim \widehat{\mathcal{H}}$, must satisfy $K\leq D^2$. In \cite{I} a slightly stronger bound which is useful in numerical applications was proved to be
\begin{equation}
    \label{Sect_KCdef_Kbound_operators}
    1\leq K \leq D^2-D+1~,
\end{equation}
for any non-vanishing operator $\mathcal{O}$. However, the basic notions on K-complexity that will be reported in this Chapter will mostly refer, unless otherwise stated, to systems in the thermodynamic limit with an infinite Krylov dimension for the operator of interest. The proof of \eqref{Sect_KCdef_Kbound_operators} will be presented in Chapter \ref{ch:chapter03_SYK}.

Also by construction, the Krylov space \eqref{Sect_KCdef_OperatorKrylovSpace_def} is the smallest subspace of $\widehat{\mathcal{H}}$ that contains the time-evolved \textit{superstate} $\big|\mathcal{O}(t)\big)$ for any time\footnote{Strictly speaking, given that $t$ appears in the coefficients of the infinite linear combination \eqref{Sect_KCdef_Ot_notation}, one could say that $\widehat{\mathcal{H}}_{\mathcal{O}}$ is the minimal subspace of operator space that contains $\big|\mathcal{O}(t)\big)$ for any $t\in \mathds{C}$, as long as the Taylor expansion in \eqref{Sect_KCdef_Ot_notation} is convergent.}. The elements $\left\{\mathcal{L}^n |\mathcal{O})\right\}_{n\geq 0}$ are gradually explored by the time-evolving observable, so one might think that they are adequate for a systematic study of the progression of time evolution. However, this set does not necessarily constitute an orthonormal basis for Krylov space (and, furthermore, in the cases in which $K$ is finite it is an overcomplete basis). The Lanczos algorithm is able to provide an orthonormal basis (upon the specification of an inner product in $\widehat{\mathcal{H}}$) that still retains some notion of ordering adapted to time evolution, in the sense that the $n$-th Krylov element will consist of a linear combination of up to $n$ powers of the Liouvillian acting on the initial condition or, equivalently, up to $n$ nested commutators of $H$ with $O$.

As announced, the application of the Lanczos algorithm for the construction of an orthonormal Krylov basis requires the specification of an inner product in operator space. Mathematically, the most immediate choice is the Frobenius inner product, as this is the one that is naturally inherited from the inner product in the space of states:
\begin{equation}
    \label{Sect_KCdef_inner_product_infinite_temp}
    \left( \mathcal{A}|\mathcal{B} \right) := \frac{1}{D}\text{Tr}\left[\mathcal{A}^\dagger \mathcal{B}\right]
\end{equation}
for any two operators $\mathcal{A},\mathcal{B}\in\widehat{\mathcal{H}}$. Note that the normalization factor $\frac{1}{D}$ ensures that the unit operator has unit norm. However, one might choose other scalar products of some physical relevance \cite{ViswanathMuller,Parker:2018yvk}, as is the case of the thermal inner product with inverse temperature\footnote{Throughout this Thesis we shall also use the convention $k_B\equiv 1$ for the Boltzmann constant, implying that temperatures have units of energy.} $\beta=\frac{1}{T}$. There are several ways in which this inner product may be defined; the conventional thermal product in quantum mechanics is just
\begin{equation}
    \label{Sect_KCdef_thermal_inner_prod_conventional}
    \left( \mathcal{A} | \mathcal{B} \right)_\beta := \frac{1}{Z(\beta)}\text{Tr}\left[ e^{-\beta H} \mathcal{A}^\dagger \mathcal{B}  \right]~,
\end{equation}
where $Z(\beta) = \text{Tr}\left[e^{-\beta H}\right]$ is the thermal partition function. In linear-response theory, it is more customary \cite{ViswanathMuller,Parker:2018yvk} to use the so-called \textit{standard} thermal inner product:
\begin{equation}
    \label{Sect_KCdef_thermal_inner_prod_standard}
    \left(\mathcal{A} | \mathcal{B}\right)_{\beta}^S := \frac{1}{2Z(\beta)} \text{Tr}\left[ e^{-\beta H} \mathcal{A}^\dagger \mathcal{B} + \mathcal{A}^\dagger e^{-\beta H} \mathcal{B} \right]~.
\end{equation} 
On the other hand, in applications to quantum field theory in which correlation functions display contact singularities it is often preferred to use a regularized thermal inner product, dubbed the \textit{Wightman inner product} (see e.g. \cite{Parker:2018yvk}). This product splits evenly the Boltzmann factor amongst the operator insertions, which is equivalent to analytically continuing one of them to imaginary times:
\begin{equation}
    \label{Sect_KCdef_Wightman_inner_prod}
    \left(\mathcal{A}|\mathcal{B}\right)_{\beta}^W := \frac{1}{Z(\beta)} \text{Tr}\left[ e^{-\beta H /2} \mathcal{A}^\dagger e^{-\beta H /2} \mathcal{B} \right] = \left( \mathcal{A} \Bigg| \mathcal{B} \left(\frac{i\beta}{2} \right) \right)_{\beta}~. 
\end{equation}

The three options above are particular cases of a more generic thermal inner product \cite{ViswanathMuller}, which can be defined by the means of some function $g(\tau)$ on the thermal circle:
\begin{equation}
    \label{Sect_KCdef_thermal_inner_product_g}
    \left(\mathcal{A}|\mathcal{B}\right)_{\beta}^g := \frac{1}{Z(\beta)} \int_0^\beta d\tau g(\tau) \text{Tr}\left[ e^{-(\beta - \tau)H} \mathcal{A}^\dagger e^{-\tau H} \mathcal{B} \right] = \int_0^\beta d\tau g(\tau) \left( \mathcal{A} \Big| \mathcal{B} \left( i \tau \right) \right)_{\beta}~.
\end{equation}
This is an inner product as long as $g(\tau)$ is a positive-semidefinite function on the thermal circle $[0,\beta]$ which integrates to some constant in such a domain; the constant is chosen to be $1$ so that the unit operator has unit norm.
We note that \eqref{Sect_KCdef_thermal_inner_prod_conventional}, \eqref{Sect_KCdef_thermal_inner_prod_standard} and \eqref{Sect_KCdef_Wightman_inner_prod} can be retrieved from \eqref{Sect_KCdef_thermal_inner_product_g} by setting $g(\tau) = \delta(\tau)$, $g(\tau) = \frac{1}{2}\left( \delta(\tau) + \delta(\tau - \beta) \right)$ and $g(\tau) = \delta(\tau-\beta / 2)$, respectively. Likewise, the Frobenius product \eqref{Sect_KCdef_inner_product_infinite_temp} can be interpreted as an infinite-temperature scalar product, reachable in the $\beta=0$ limit of any of the thermal products defined above. With respect of any member of the family of inner products \eqref{Sect_KCdef_thermal_inner_product_g}, the Liouvillian $\mathcal{L}$ is a hermitian superoperator acting over operator space.

For a hermitian initial operator $\mathcal{O} = \mathcal{O}^\dagger$ (i.e. an observable), the Lanczos algorithm formulated in section \ref{sect:Lanczos} can be applied using the Liouvillian $\mathcal{L}$ in order to generate an orthogonal Krylov basis $\left\{|\mathcal{O}_n )\right\}_{n\geq 0}$ for Krylov space \eqref{Sect_KCdef_OperatorKrylovSpace_def}, together with a set of Lanczos coefficients $\left\{a_n\right\}_{n\geq 0}$ and $\left\{b_n\right\}_{n\geq 1}$. Given the generic thermal inner product \eqref{Sect_KCdef_thermal_inner_product_g}, if the $g$-function additionally satisfies $g(\tau) = g(\beta - \tau)$, i.e. it is even in the thermal circle\footnote{This requirement is often demanded in linear-response theory because it allows correlation functions, defined as the overlap $\left(\mathcal{O} | \mathcal{O}(t)\right)_{\beta}^g$, to satisfy the usual fluctuation-dissipation theorem \cite{ViswanathMuller}.}, which is the case of \eqref{Sect_KCdef_thermal_inner_prod_standard} and \eqref{Sect_KCdef_Wightman_inner_prod} but not of \eqref{Sect_KCdef_thermal_inner_prod_conventional}, then the algorithm simplifies and it is possible to prove inductively the two following results:
\begin{itemize}
    \item[\textit{i})] The Krylov elements have alternating hermiticity properties: $\mathcal{O}_n^\dagger = (-1)^n \mathcal{O}_n$. Equivalently, $i^n \mathcal{O}_n$ is hermitian for any $n\geq 0$.
    \item[\textit{ii})] $a_n=0$ for all $n\geq 0$.
\end{itemize}

Hence, in this case the Lanczos algorithm applied to operators only yields one sequence of $b$-coefficients. Krylov elements $\mathcal{O}_n$ with even (resp. odd) index $n$ are a linear combination of terms containing an even (resp. odd) number of nested commutators of $H$ with $\mathcal{O}$.

For now we shall focus on the infinite-temperature inner product \eqref{Sect_KCdef_inner_product_infinite_temp}, which probes the full Hilbert space of the system democratically. Let us write the explicit formulation of the Lanczos algorithm for this case\footnote{It can be generalized to the family of inner products \eqref{Sect_KCdef_thermal_inner_product_g} by replacing the occurrences of the $\beta=0$ scalar product by the corresponding one, and including $a$-coefficients according to the prescription in section \ref{sect:Lanczos} whenever $g(\tau)$ is not even.}:

\begin{itemize}
    \item[\underline{$n=0$:}]
    \begin{itemize}
        \item[$\bullet$] $|\mathcal{O}_0) = |\mathcal{O})$ (normalized by assumption).
    \end{itemize}
    \item[\underline{$n=1$:}]
    \begin{itemize}
        \item[$\bullet$] $|\mathcal{A}_1) = \mathcal{L}|\mathcal{O}_0)$.
        \item[$\bullet$] $b_1 = \sqrt{\left(\mathcal{A}_1|\mathcal{A}_1\right)}$.
        \item[$\bullet$] $|\mathcal{O}_1) = \frac{1}{b_1}|\mathcal{A}_1)$.
    \end{itemize}
    \item[\underline{$n\geq 2$}:] 
    \begin{itemize}
        \item[$\bullet$] $|\mathcal{A}_n) = \mathcal{L}|\mathcal{O}_{n-1})-b_{n-1}|\mathcal{O}_{n-2})$.
        \item[$\bullet$] $b_n = \sqrt{\left( \mathcal{A}_n | \mathcal{A}_n \right)}$
        \item[$\bullet$]$|\mathcal{O}_n) = \frac{1}{b_n}|\mathcal{A}_n)$.
    \end{itemize}
\end{itemize}
The formulation above emphasizes on the exact analytical definition of the objects involved. For the specific numerical implementation of the algorithm, see Chapter \ref{ch:chapter03_SYK}. For now, it suffices to say that, in cases with a finite-dimensional Krylov space, the algorithm shall terminate by hitting a zero at $b_K=0$. We note that the Krylov elements satisfy:
\begin{equation}
    \label{Sect_KCdef_KrylovElementsRelation}
    \begin{split}
        & b_1|\mathcal{O}_1) = \mathcal{L}|\mathcal{O}_0)~, \\
        & b_n|\mathcal{O}_n) = \mathcal{L}|\mathcal{O}_{n-1})-b_{n-1}|\mathcal{O}_{n-2})~,\qquad 2\leq n \leq K-1~, \\
        & 0 = \mathcal{L}|\mathcal{O}_{K-1})-b_{K-1}|\mathcal{O}_{K-2})~,
    \end{split}
\end{equation}
where the last line only applies if $K$ is finite.

These Krylov elements are probed gradually by the time-evolving observable $\left| \mathcal{O}(t) \right)$. It is thus useful to decompose it in terms of them:
\begin{equation}
    \label{Sect_KCdef_operator_wavefn_decomp}
    \big| \mathcal{O}(t) \big)= e^{it \mathcal{L}} |\mathcal{O}) = \sum_{n=0}^{K-1} i^n \varphi_n(t) |\mathcal{O}_n)~,
\end{equation}
where $\varphi_n(t) := i^{-n}\left( \mathcal{O}_n | \mathcal{O} (t) \right)$ is dubbed the \textit{operator wave function}, which is real\footnote{This definition of the wave function modding out some power of the imaginary unit $i$ is useful for numerical applications, as it allows to store the values of this function in exclusively real variables, granting some memory optimization.} thanks to the alternating hermiticity properties of the Krylov elements discussed previously. We may think of the Krylov basis as some chain (the \textit{Krylov chain}) through which the evolving operator is spreading according to the wave function $\varphi_n(t)$, where $n$ plays the role of the position coordinate on the chain. This perspective can be pushed further by noting the tridiagonal form taken by the Liouvillian in coordinates over the Krylov basis, which can be deduced from \eqref{Sect_KCdef_KrylovElementsRelation}:
\begin{equation}
    \label{Sect_KCdef_Liouvillian_tridiag}
    \mathcal{L} = \sum_{n=0}^{K-2} b_{n+1} \Big( |\mathcal{O}_n)(\mathcal{O}_{n+1}| + |\mathcal{O}_{n+1})(\mathcal{O}_{n}| \Big)~.
\end{equation}
We observe that \eqref{Sect_KCdef_Liouvillian_tridiag}, which generates the time evolution of the initial condition $|\mathcal{O}) = |\mathcal{O}_0)$ takes the form of the Hamiltonian of a hopping model \cite{Anderson_AbsDiff} where the Lanczos coefficients play the role of hopping amplitudes on the Krylov chain. In other words, the Lanczos construction allows to map the time evolution of a certain operator to a one-dimensional quantum-mechanical hopping problem \cite{Parker:2018yvk,ViswanathMuller}. The projects \cite{II,III}, presented in Chapter \ref{ch:chapter04_Integrable}, exploit this fact and study the imprints of the system's properties on the hopping amplitudes governing Krylov space propagation. As we have already announced, the article \cite{Parker:2018yvk} proposed to use the position on the Krylov chain as a notion of (operator) complexity, dubbed \textit{Krylov complexity} or K-complexity. Considering the position (super-)operator over the Krylov basis, which may itself be referred to as the K-complexity (super-)operator,
\begin{equation}
    \label{Sect_KCdef_KC_operator}
    \widehat{\mathcal{C}_K}:=\sum_{n=0}^{K-1} n~ |\mathcal{O}_n)(\mathcal{O}_n|~,
\end{equation}
K-complexity is defined to be the expectation value of such a superoperator for the time-evolving element $\big| \mathcal{O}(t) \big)$:
\begin{equation}
    \label{Sect_KCdef_KC_def}
    C_K(t):= \Big( \mathcal{O}(t) \Big| \widehat{\mathcal{C}_K} \Big| \mathcal{O}(t) \Big) = \sum_{n=0}^{K-1} n~ \varphi_n(t)^2~,
\end{equation}
where the last step made use of \eqref{Sect_KCdef_operator_wavefn_decomp}, together with the fact that the operator wave function is always real. By construction of the Krylov basis, the operator wave function is initially localized at $n=0$, i.e. $\varphi_n(0) = \delta_{n0}$, and from then on it spreads over the Krylov chain according to the Schrödinger equation dictated by the Liouvillian \eqref{Sect_KCdef_Liouvillian_tridiag}, which takes the form of a recursive equation:
\begin{equation}
    \label{Sect_KCdef_recursion_phin}
    \dot{\varphi}_n (t) = b_n\varphi_{n-1}(t) - b_{n+1}\varphi_{n+1}(t)~,
\end{equation}
whose initial condition is, as announced, $\varphi_{n}(0) = \delta_{n0}$ and where we need some boundary conditions to account for the \textit{edges} of Krylov space, namely $\varphi_{-1}\equiv\varphi_{K}\equiv 0$. Equation \eqref{Sect_KCdef_recursion_phin} manifests that the knowledge of the Lanczos coefficients uniquely determines the operator wave function $\varphi_n(t)$, and is therefore sufficient for the eventual computation of K-complexity \eqref{Sect_KCdef_KC_def}. We note that the time-evolution (super-) operator $e^{it\mathcal{L}}$ is unitary since $\mathcal{L}$ is hermitian, as has been discussed. Therefore, the norm of the operator remains constant as a function of (real) time\footnote{Note that imaginary-time evolution does not preserve the norm. $\Big|\mathcal{O}(i\tau)\Big)$ still belongs to Krylov space by construction, but it would need to be renormalized to unity in order to compute its K-complexity for some arbitrary $\tau\in [0,\beta]$.}:
\begin{equation}
    \label{Sect_KCdef_norm_op_constant_unitarity}
    \Big( \mathcal{O}(t) \Big| \mathcal{O}(t) \Big) = \sum_{n=0}^{K-1} \varphi_n(t)^2=1~,\qquad\forall t\in \mathds{R}~.
\end{equation}

The larger K-complexity is, the further $\Big| \mathcal{O}(t) \Big)$ is within the time evolution that can be generated by $\mathcal{L}$, i.e. one requires more applications of the Liovillian in order to reconstruct $\Big| \mathcal{O}(t) \Big)$ out of the initial condition $|\mathcal{O})$. In this sense, this notion may be understood as a generalization of circuit complexity that does not require any human input designating what the reference state or the set of unitary gates are, and which is well-defined and finite without introducing a tolerance parameter. In fact, despite not being frequently mentioned, the choice of unitary gates in the context of circuit complexity does have some relation with the Hamiltonian generating time evolution: The complexity computed out of $k$-local gates will only give a sensible result for the study of the time evolution of an initially localized state or operator if the Hamiltonian is itself $k$-local and if the reference state (or operator) is not highly non-local; similarly, size-complexity \cite{Roberts:2014isa} of an initially localized operator will grow progressively in time only if the Hamiltonian is built with $k$-local interactions. At the end of the day, the gates out of which complexity is computed are often related to the building blocks of the Hamiltonian generating time evolution, and K-complexity generalizes this fact: The reference state is nothing but the initial condition of time evolution, and ``the only unitary gate'' is the time evolution operator itself. $C_K(t)$ is therefore fully specified once the initial condition, the time evolution generator, and and inner product over the Hilbert space of interest, are given. This comes at the cost of being a very specific construction in the sense that the basis adapted to the time evolution of a given initial condition $|\mathcal{O)}$ might be completely unrelated to the Krylov basis of a different initial condition $|\mathcal{O}^\prime)$.

An object which will be useful to characterize the exploration of the Krylov chain by the time-evolving operator is the Shannon entropy of its operator wave function with respect to the Krylov basis, introduced in \cite{Barbon:2019wsy} and dubbed \textit{Krylov entropy} or K-entropy:
\begin{equation}
    \label{Sect_KCdef_KS_def}
    S_K(t) = - \sum_{n=0}^{K-1} \varphi_n(t)^2~\log \varphi_n(t)^2~.
\end{equation}

At this point, we recall that section \ref{sect:relations_Lan_Moments} illustrated how the Lanczos coefficients are in correspondence to the moments of the generator. In the case of operator K-complexity, the relevant moments are
\begin{equation}
    \label{Sect_KCdef_operator_moments}
    \mu_{2n}:= \left( \mathcal{O} | \mathcal{L}^{2n} | \mathcal{O} \right)~,
\end{equation}
where we have explicitly omitted odd moments, which vanish identically for any hermitian operator $\mathcal{O}$ given an inner product belonging to the family \eqref{Sect_KCdef_thermal_inner_product_g} that has an even $g$ function, consistently with the fact that the $a$-coefficients vanish in this case, as argued in the discussion leading to \eqref{Sect_KCdef_KrylovElementsRelation}. We thus have the following bijective relation:
\begin{equation}
    \label{Sect_KCdef_bijection_b_mu}
    \left\{ b_n \right\}_{n=1}^{K-1} \longleftrightarrow \left\{\mu_{2n}\right\}_{n=1}^{K-1}~,
\end{equation}
where we note that $\mu_0=1$ identically because of the requirement of normalization of the initial condition. In this physical context, the relation \eqref{Sect_KCdef_bijection_b_mu} is in fact extremely useful because the moments $\mu_{2n}$ are very intimately related to an observable of importance: The operator's two point function or auto-correlation function, given by
\begin{equation}
    \label{Sect_KCdef_twoPt_def}
    C(t):= \left( \mathcal{O}| \mathcal{O}(t) \right)~.
\end{equation}
Note that, since the initial condition gives the zeroth Krylov element, the quantity \eqref{Sect_KCdef_twoPt_def} is equal to the zeroth component of the operator wave function, $C(t) = \varphi_0(t)$. For hermitian operators, the auto-correlation function is even in the time domain, and one can see that the coefficients of its Taylor series around $t=0$ are precisely the moments $\mu_{2n}$:
\begin{equation}
    \label{Sect_KCdef_moments_from_TwoPt}
    C(t) = \left(\mathcal{O} | e^{it\mathcal{L}} | \mathcal{O}\right) = \sum_{n=0}^{+\infty} \frac{{(it)}^{2n}}{(2n)!}\mu_{2n}~.
\end{equation}
Operationally, when one does not have direct access to the concrete algebraic description of the underlying Hilbert space of the system in order to apply explicitly the Lanczos algorithm, or if such an application is costly and limited (e.g. in the case of field theory, or systems of infinite size), it may be preferred to compute analytically the two-point function \eqref{Sect_KCdef_twoPt_def}, obtain the moments from it and, numerically, transform them into Lanczos coefficients using the algorithm described in section \ref{sect:Moments_to_Lanczos}, at the cost of coping with its high-precision requirements.

Additionally, the fact that the moments $\mu_{2n}$ are the Taylor coefficients of the operator's two-point function allows for some further analytical understanding of the behavior of the Lanczos sequence by extrapolating features of the sequence of moments that can in turn be obtained from known results on the two-point function or its various cousins. This was exploited in \cite{Parker:2018yvk} in order to propose the \textit{operator growth hypothesis}, which is the central contribution of the article together with the introduction of the notion of K-complexity. We shall review this hypothesis in section \ref{sect:KC_chaos_holog}, but for now let us just present the aforementioned cousins:
\begin{itemize}
    \item The \textit{spectral function}: It is the Fourier transform of the two-point function \eqref{Sect_KCdef_twoPt_def},
    \begin{equation}
        \label{Sect_KCdef_spectral_funct}
        \Phi(\omega):=\int_{\mathds{R}} dt~ C(t) ~ e^{-i\omega t}~,
    \end{equation}
    from which the moments $\mu_{2n}$ can be computed as
    \begin{equation}
        \label{Sect_KCdef_moments_from_spectral}
        \mu_{2n} = \frac{1}{2\pi}\int_{\mathds{R}}d\omega~\omega^{2n}~\Phi(\omega)~.
    \end{equation}
    \item The \textit{Green function}, given by
    \begin{equation}
        \label{Sect_KCdef_GreenFunction_def}
        G(z) := \left( \mathcal{O} \Big| \frac{1}{z-\mathcal{L}} \Big| \mathcal{O} \right)~.
    \end{equation}
    This function has singularities along the real axis, such that the discontinuity in the vertical direction gives a version of the density of states of the Liouvillian, weighted by the operator $\mathcal{O}$ due to the fact that \eqref{Sect_KCdef_GreenFunction_def} is an expectation value evaluated in the state $|\mathcal{O})$ rather than a full trace over $\widehat{\mathcal{H}}$. We will see that such a weighted density is nothing but the spectral function $\Phi(\omega)$. The Green function $G(z)$ admits different integral representations depending on whether its argument belongs to the upper or to the lower half of the complex plane:
    \begin{equation}
        \label{Sect_KCdef_GreenFunction_Integral_rep}
        G(z^{\pm}) = \mp i\int_{\mathds{R}^{\mp}} dt ~ C(t) ~ e^{-itz^\pm}~,
    \end{equation}
    where $z^+$ (resp. $z^{-}$) denotes that the imaginary part of $z$ is positive (resp. negative). From \eqref{Sect_KCdef_GreenFunction_Integral_rep} and \eqref{Sect_KCdef_spectral_funct} we obtain, as promised, the spectral function in terms of the jump across the real-axis discontinuity of $G(z)$:
    \begin{equation}
        \label{Sect_KCdef_Phi_from_G}
        \Phi(\omega) = -i \lim_{\varepsilon\to 0^+}\Big( G\left( \omega - i\varepsilon \right) - G\left( \omega + i\varepsilon \right) \Big)
    \end{equation}
    for real $\omega$.

    Importantly, the Green function $G(z)$ admits a continued fraction expansion in terms of the Lanczos coefficients $b_n$. This can be seen considering the operator wave functions $\varphi_n(t)$ defined in \eqref{Sect_KCdef_operator_wavefn_decomp}: Defining some ``generalized'' Green functions $G_n(z)$ as
    \begin{equation}
        \label{Sect_KCdef_Gn}
        G_n(z^{\pm}) = \mp i\int_{\mathds{R}^{\mp}} dt ~ \varphi_n(t) ~ e^{-itz^\pm} = i^{-n} \left(\mathcal{O}_n \Big | \frac{1}{z^\pm-\mathcal{L}} \Big| \mathcal{O}\right)~,
    \end{equation}
    we may turn the differential recursion \eqref{Sect_KCdef_recursion_phin} for $\varphi_n(t)$ into an algebraic recurrence relation for $G_n(z)$:
    \begin{equation}
        \label{Sect_KCdef_recursion_Gn}
        iz G_n(z) - i \delta_{n0} = b_n G_{n-1}(z) - b_{n+1}G_{n+1}(z)~,
    \end{equation}
    supplemented with the boundary conditions $G_{-1}\equiv G_K \equiv 0$. Iterating this recursion, it is possible to obtain the promised continued-fraction expansion of $G_0(z)=G(z)$:
    \begin{equation}
        \label{Sect_KCdef_Gz_continued_fraction}
        G(z) = \frac{1}{z-\frac{b_1^2}{z - \frac{b_2^2}{z-\dots}}}~.
    \end{equation}
    This last equation illustrates how the Lanczos coefficients are intimately related to the spectral function $\Phi(\omega)$, as they are directly responsible for the singularity structure of the Green function through \eqref{Sect_KCdef_Gz_continued_fraction}. The reader interested in the use of truncated sequences of Lanczos coefficients in order to reconstruct numerically the spectral function is referred to the review \cite{ViswanathMuller}, where the $b$-sequence is treated as a ``genetic code of spectral densities'' since it encodes information on the features of $\Phi(\omega)$. For example, a $b_n$ sequence that asymptotes to a constant implies a spectral density of bounded support, while indefinitely growing Lanczos coefficients correspond to a $\Phi(\omega)$ whose domain is unbounded. For a compilation of properties and relations, see \cite{ViswanathMuller}, and for a rigorous mathematical analysis of some particular cases see \cite{Magnus}.
\end{itemize}

\subsection{State K-complexity}\label{sect:stateKC}

Chronologically, operator Krylov complexity was introduced before state Krylov complexity because the relation between the Lanczos coefficients and the various observables related to the two-point function make the recursion method a very useful tool to study operator growth. However, articles \cite{Barbon:2019wsy,I} proposed to consider Krylov complexity as an adequate candidate for holographic complexity, a task for which the K-complexity of a time-evolving state in the system's Hilbert space may be equally insightful, as shall be discussed in section \ref{sect:KC_chaos_holog}. The work \cite{Balasubramanian:2022tpr} was the first to study analytically Krylov complexity of states for some symmetry-dominated systems, as well as numerically in random matrix theory. For reference, we shall give in this section the relevant definitions regarding state K-complexity.

Given a time-evolving state in the Schrödinger picture,
\begin{equation}
    \label{Sect_KCdef_state_t}
    |\phi(t)\rangle = e^{-itH} |\phi\rangle~,
\end{equation}
the Lanczos algorithm can be applied just as it was described\footnote{Throughout Chapter \ref{ch:chapter01_Lanczos}, the probe vector was generically denoted by $|\Omega\rangle$, representing the seed element in whichever Hilbert space where one envisions to apply the Lanczos algorithm. Here, we prefer the use of $|\phi\rangle$ because the symbol $|\Omega\rangle$ is frequently reserved to the vacuum state in theories where it can be defined, whereas we intend to refer to an arbitrary state in the system's Hilbert space.} in section \ref{sect:Lanczos}. The reader is referred to that section for the details but, schematically, let us simply recall that the algorithm produces an orthonormal Krylov basis $\left\{|K_n\rangle \right\}_{n=0}^{K-1}$ and, this time, generically two sets of Lanczos coefficients, $\left\{a_n\right\}_{n=0}^{K-1}$ and $\left\{b_n\right\}_{n=1}^{K-1}$, all these objects being related by the recursion
\begin{equation}
    \label{Sect_KCdef_Lanczos_recursion_states}
    b_n |K_n\rangle = \left( H-a_{n-1} \right)|K_{n-1}\rangle - b_{n-1}|K_{n-2}\rangle~,
\end{equation}
where it is understood that $|K_{-1}\rangle \equiv |K_K\rangle \equiv 0$, as usual. Expressed in coordinates over the Krylov basis, the Hamiltonian takes, once again, the form of a tight-binding Hamiltonian \cite{Anderson_AbsDiff} where the $b$-coefficients play the role of hopping amplitudes, while the $a$-coefficients act as local potential energies on each site of the Kryvlov chain. Explicitly, we have:
\begin{equation}
    \label{Sect_KCdef_Htridiag}
    H = \sum_{n=0}^{K-1} a_n |K_n\rangle\langle K_n | + \sum_{n=0}^{K-2}b_{n+1}\Big( |K_n\rangle \langle K_{n+1}| + |K_{n+1}\rangle \langle K_{n}| \Big)~.
\end{equation}

The time-evolving state may be decomposed in coordinates over the Krylov basis as
\begin{equation}
    \label{Sect_KCdef_state_wavefn_decomp}
    |\phi(t)\rangle = e^{-itH}|\phi\rangle = \sum_{n=0}^{K-1}\phi_n(t)|K_n\rangle~,
\end{equation}
where this time the Krylov space wave function $\phi_n(t):=\langle K_n|\phi(t)\rangle$ has been defined without modding out a a power of $i$ as in \eqref{Sect_KCdef_operator_wavefn_decomp} because the function would still be generically complex-valued even after such a redefinition. The Schrödinger equation dictated by the Hamiltonian \eqref{Sect_KCdef_Htridiag} imposes a differential recurrence relation for the wave function which has a diagonal term controlled by the $a$-coefficients:
\begin{equation}
    \label{Sect_KCdef_recursion_phin_states}
    i\dot{\phi}_n(t) = a_n\phi_n(t) + b_{n+1}\phi_{n+1}(t) + b_n\phi_{n-1}(t)~,
\end{equation}
with the boundary conditions $\phi_{-1}\equiv \phi_K \equiv 0$ and the initial condition $\phi_n(0)=\delta_{n0}$, since the initial state is itself the zeroth Krylov element\footnote{As a consistency check for \eqref{Sect_KCdef_recursion_phin_states}, one may redefine $\phi_n(t)=i^n\varphi_n(t)$: Substituting this in the recursion and setting $a_n=0$ recovers \eqref{Sect_KCdef_recursion_phin} but with the opposite sign on the left-hand side. This is perfectly consistent with the fact that \eqref{Sect_KCdef_recursion_phin_states} was derived for a state evolving in the Schrödinger picture, where the time evolution operator is $e^{-itH}$, while \eqref{Sect_KCdef_recursion_phin} was obtained for operator evolution in the Heisenberg picture, where the time-translation (super-)operator is $e^{it\mathcal{L}}$.}.

Noting that unitarity preserves the norm of the Krylov space wave function,
\begin{equation}
    \label{Sect_KCdef_wavefn_state_normalized}
    \sum_{n=0}^{K-1} {|\phi_n(t)|}^2=1~,\qquad\forall t\in\mathds{R}~,
\end{equation}
the state K-complexity and K-entropy can now be defined, analogously to \eqref{Sect_KCdef_KC_def} and \eqref{Sect_KCdef_KS_def}, as the expectation value of position on the Krylov chain and as the Shannon entropy of the wave function, respectively:
\begin{equation}
    \label{Sect_KCdef_KC_state_def}
    \begin{split}
        &C_K(t) := \sum_{n=0}^{K-1}n~{|\phi_n(t)|}^2~, \\
        &S_K(t) := - \sum_{n=0}^{K-1} {|\phi_n(t)|}^2 \log {|\phi_n(t)|}^2~.
    \end{split}
\end{equation}

Since the $a$-coefficients do not necessarily vanish for states, the moments to which the Lanczos coefficients are in correspondence can have indistinctly even or odd index:
\begin{equation}
    \label{Sect_KCdef_moments_states}
    M_n := \langle \phi | H^n |\phi\rangle = \langle K_0 | H^n | K_0\rangle ~. 
\end{equation}
These are, in turn, the Taylor series coefficients of the survival amplitude of the evolving state defined as:
\begin{equation}
    \label{Sect_KCdef_survival_amplitude}
    S(t):=\langle \phi | \phi(t)\rangle = \sum_{n=0}^{+\infty} \frac{{(-it)}^n}{n!}M_n~,
\end{equation}
as one can see considering \eqref{Sect_KCdef_state_t} and Taylor-expanding the time evolution operator. Depending on the state $|\phi\rangle$ of interest, the survival amplitude might be given by some relevant quantity with well-known properties. As an example, the authors of \cite{Balasubramanian:2022tpr}, motivated by the usefulness of Krylov complexity in holography, take the seed state to be the thermofield double state, as we shall review in section \ref{sect:KC_chaos_holog}, in which case the survival amplitude is given by the analytic continuation of the system's partition function. Even more interestingly, in the latter case the survival \textit{probability} equals the spectral form factor \cite{Altland:2020ccq,HaakeBook,efetov_1996,stöckmann_1999}, which is itself a well-studied object sensitive to the quantum-chaotic properties of systems.

Let us conclude this section by stating a useful property of K-complexity, applicable to both state and operator complexity. Observing the most generic form of the recursion relation \eqref{Sect_KCdef_recursion_phin_states} for the Krylov space wave function we may note that, since the Lanczos coefficients are always real by construction, complex conjugating the full expression yields the same new recursion as sending $t\mapsto-t$, as long as $t\in\mathds{R}$, leaving the initial and boundary conditions unchanged. Consequently, we have the relation
\begin{equation}
    \label{Sect_KCdef_phin_conjugate_minust}
    \phi_n^{*}(t)=\phi_n(-t)\quad\text{for }t\in\mathds{R}~.
\end{equation}
We may now plug this result into the definition of Krylov complexity \eqref{Sect_KCdef_KC_state_def}, reaching:
\begin{equation}
    \label{Sect_KCdef_KCeven}
    C_K(t) = \sum_{n=0}^{K-1} n~\phi_n(t) \phi_n(-t)\quad\text{for }t\in\mathds{R}~.
\end{equation}
Since the right-hand side of expression \eqref{Sect_KCdef_KCeven} is manifestly invariant under the replacement $t\mapsto -t$, we conclude that Krylov complexity is always an even function of time, regardless of the details of the initial state, i.e.:
\begin{equation}
    \label{Sect_KCdef_KCeven_statement}
    C_K(t) = C_K(-t)~,
\end{equation}
always for $t\in\mathds{R}$. In particular, this property applies to operator Krylov complexity; we recall that the operator recursion \eqref{Sect_KCdef_recursion_phin} is, modulo a wave function redefinition and an extra sign on the left-hand side coming from the difference between the Heisenberg and the Schrödinger pictures, a particular case of the more generic recursion \eqref{Sect_KCdef_recursion_phin_states} that applies whenever the $a$-coefficients vanish. Finally, let us note that using \eqref{Sect_KCdef_phin_conjugate_minust} it is immediate to show that K-entropy, as defined in \eqref{Sect_KCdef_KC_state_def}, is also an even function of time since, just like K-complexity, it only depends on the modulus squared of the Krylov space wave function.

\section{Two theorems}\label{sect:Two_theorems}

This section will present two theorems that appeared in \cite{Parker:2018yvk} and \cite{Balasubramanian:2022tpr}. These independent results (the first one in the frame of operator Krylov complexity, and the second in that of state K-complexity) put the notion of K-complexity on a more solid footing by relating it to other generic complexity notions. The theorem in \cite{Parker:2018yvk} shows that $C_K(t)$ is an upper bound on a class of generalized complexities to which familiar notions like size complexity \cite{Roberts:2014isa} or out-of-time-ordered correlators (OTOCs) \cite{Maldacena:2015waa} belong. On the other hand, \cite{Balasubramanian:2022tpr} argues that the Krylov basis is an optimal choice of basis for defining a complexity in the sense that it minimizes the local rate of growth near $t=0$. The statements and proofs of both theorems will be presented, together with an eventual comparative discussion of the results which will provide some clarifying remarks.

\subsection{Q-theorem}\label{sect:QTheorem}

This first theorem, presented in \cite{Parker:2018yvk}, constructs a class of generalized complexities, dubbed \textit{Q-complexities} (hence the name we are using to refer to this result), where the \textit{Q} comes from the French word \textit{quelconque}, and shows that K-complexity is, up to a multiplicative factor, an upper bound of the family. The theorem was originally formulated for operator complexities but, as we shall see below, it can be immediately extrapolated to any other Hilbert space, and in particular it applies to state complexity as well. 

Let us start by introducing the family of Q-complexities for a given time-evolving (super-)state $\left|\mathcal{O}(t)\right) = e^{it\mathcal{L}}|\mathcal{O})\in \widehat{\mathcal{H}}$, where each member is defined as the expectation value of some hermitian (super-)operator $\mathcal{Q}$ in the state $|\mathcal{O}(t))$, i.e.
\begin{equation}
    \label{Sect_theoremQ_QC_def}
    Q(t):=\Big( \mathcal{O}(t) \Big| \mathcal{Q} \Big| \mathcal{O}(t) \Big)~.
\end{equation}
For it to define a Q-complexity, the operator $\mathcal{Q}$ must fulfill the following properties:
\begin{enumerate}
    \item The operator is positive-semidefinite, so that $Q(t)$ is never negative. Explicitly:
    \begin{equation}
        \label{Sect_theoremQ_Assumptions_Q_positive_semidef}
        \mathcal{Q} = \sum_{a\geq 0} q_a | q_a ) (q_a|~,\qquad q_a\geq 0\quad\forall a\geq 0~,
    \end{equation}
    where $q_a$ denote the Q-complexity eigenvalues and $|q_a)$ are the associated eigenstates. Note that, for the sake of generality, we have not been specific about whether the space has finite of infinite dimension. Additionally, note that the set $\left\{q_a\right\}_{a\geq 0}$ need not be sorted.
     \item There exists a constant $M>0$ such that the two following properties hold simultaneously:   
    \begin{enumerate}
        \item[\textit{i})] The action of the time-evolution generator (the Liouvillian, in the current formulation) on a Q-complexity eigenstate cannot increase arbitrarily the Q-complexity:
        \begin{equation}
            \label{Sect_theoremQ_Assumptions_L_on_q}
            \left(q_a | \mathcal{L} | q_b\right) = 0\qquad\text{if } |q_a-q_b|>M~.
        \end{equation}
        \item[\textit{ii})] The initial condition does not feature an arbitrarily high Q-complexity:
        \begin{equation}
            \label{Sect_theoremQ_Assumptions_initial_compl}
            \left(q|\mathcal{O}\right)=0\qquad\text{if } q>M~.
        \end{equation}
    \end{enumerate}
\end{enumerate}

We will later provide some examples of Q-complexities (specifically, operator size and out-of-time-ordered correlation functions), but for now it shall suffice to say that K-complexity is an element of this family, where the $\mathcal{Q}$ operator is given by the K-complexity operator $\widehat{\mathcal{C}_K}$ defined in \eqref{Sect_KCdef_KC_operator}, for which the $q$-values are simply $q_n=n$ and the complexity eigenstates are the Krylov elements $|\mathcal{O}_n)$. K-complexity is, in fact, a special representative of the family of Q-complexities because the eigenvalues are sorted and, by construction, the initial state is the zeroth element of the Krylov basis and the action of $\mathcal{L}$ on a basis element changes its K-complexity by at most one unit; therefore, the constant $M$ in \eqref{Sect_theoremQ_Assumptions_L_on_q} and \eqref{Sect_theoremQ_Assumptions_initial_compl} can be chosen to be $M=1$. The eigenbasis of the K-complexity operator is, by construction, adapted to $\mathcal{L}$ and $\mathcal{O}$ so that $C_K(t)$ increases progressively due to the hopping between nearest neighbours on the Krylov chain. From this point of view, the eigenbasis of any other $\mathcal{Q}$ operator may be regarded as a dilated version of the Krylov basis, in the sense that a given state $|q)$ may receive contributions from several Krylov elements, so that larger jumps between different $|q)$ states may occur during time evolution. We shall see that this implies that $Q(t)$ is, at most, a multiple of $C_K(t)$, where the multiplicative coefficient is controlled by the constant $M$.

The Q-theorem states that
\begin{equation}
    \label{Sect_theoremQ_statement}
    Q(t)\leq M ~ C_K(t)~.
\end{equation}
In order to prove it, let us begin by bounding the subset of $q$-eigenstates with which a given Krylov element overlaps. We propose that 
\begin{equation}
    \label{Sect_theoremQ_overlap_q_Ln_O}
    \left(q | \mathcal{L}^n| \mathcal{O}\right)=0\qquad \text{if } q>M(n+1)~.
\end{equation}
This may be shown inductively. For $n=0$, \eqref{Sect_theoremQ_overlap_q_Ln_O} reduces to the assumption \eqref{Sect_theoremQ_Assumptions_initial_compl}, and it implies that $|\mathcal{O})$ belongs to the span of $ |q_a)$ with $a$ such that $q_a\leq M$, that is:
\begin{equation}
    \label{Sect_theoremQ_O_in_Qbasis}
    |\mathcal{O}) = \sum_{\substack{a~\text{s.t.} \\ q_a\leq M}} c_a |q_a)~,
\end{equation}
for some coefficients $c_a$. Now, for the $n=1$ case of \eqref{Sect_theoremQ_overlap_q_Ln_O} we have:
\begin{equation}
    \label{Sect_theoremQ_overlap_q_L_O}
    \left( q | \mathcal{L}| \mathcal{O} \right)= \sum_{\substack{a~\text{s.t.} \\ q_a\leq M}} c_a \left( q | \mathcal{L}| q_a \right)~.
\end{equation}
Applying \eqref{Sect_theoremQ_Assumptions_L_on_q} to the term of the sum with the highest value of $q_a$ we reach that 
\begin{equation}
    \label{Sect_theoremQ_overlap_q_L_O_v2}
    \left( q | \mathcal{L}| \mathcal{O} \right)= 0\qquad\text{if }q> 2M~.
\end{equation}
The induction step is completely analogous to the $n=1$ step that we have just performed explicitly, resulting in \eqref{Sect_theoremQ_overlap_q_Ln_O}. Since each Krylov element $|\mathcal{O}_n)$ is a linear combination of $\mathcal{L}^m|\mathcal{O})$ with $m=0,\dots,n$, it immediately follows that
\begin{equation}
    \label{Sect_theoremQ_overlap_qbasis_Kbasis}
    \left( q | \mathcal{O}_n \right)=0\qquad \text{if } q>M(n+1).
\end{equation}
Equation \eqref{Sect_theoremQ_overlap_qbasis_Kbasis} may be understood as follows: For a fixed value of $q$, the Krylov elements $|\mathcal{O}_n)$ have zero overlap with $|q)$ (i.e. the former \textit{do not reach that far in the $q$-basis}) if $n$ is strictly smaller than some critical value $n (q)$. The critical value suggested by \eqref{Sect_theoremQ_overlap_qbasis_Kbasis} is
\begin{equation}
    \label{Sect_theoremQ_critical_nq}
    n(q) = \frac{q}{M}-1~.
\end{equation}

The expectation value of $\mathcal{Q}$ in an arbitrary state $|\mathcal{W})\in \widehat{\mathcal{H}}$ is
\begin{equation}
    \label{Sect_theoremQ_ExpQ_W}
    \left( \mathcal{W} | \mathcal{Q} |\mathcal{W}\right) = \sum_{a\geq 0} q_a {\lvert \left(q_a | \mathcal{W}\right)\rvert}^2 \equiv \int_{\mathds{R}^{+}}dq~ q~ w(q)~,
\end{equation}
where in the last step we have introduced the probability distribution function (pdf) $w(q)$, which may be thought of as a sum of delta functions in order to match the discrete sum prior to the last equality sign. We may now invoke a useful relation between the expectation value of a positive-semidefinite variable $q$ with a pdf $w(q)$ and the integral of the (reciprocal of the) cumulative distribution function (cdf), $W(q)$, defined as
\begin{equation}
    \label{Sect_theoremQ_Wq}
    W(q):=\int_{q}^{+\infty}ds~w(s)~.
\end{equation}
The announced property is:
\begin{equation}
    \label{Sect_theoremQ_ExpValue_vs_cumulative}
    \int_{\mathds{R}^+}dq~q~w(q) = \int_{\mathds{R}^+}dq~W(q)~,
\end{equation}
relating the expectation value of $q$ to the sum of the cumulative probabilities of large values of the variable\footnote{Identity \eqref{Sect_theoremQ_ExpValue_vs_cumulative} may be proved as follows:
\begin{equation*}
    \int_{\mathds{R}^+}dq~W(q) = \int_{\mathds{R}^+}dq\int_{q}^{+\infty}ds~w(s) = \int_{\mathds{R}^+} ds \int_0^s dq ~w(s) = \int_{\mathds{R}^+} ds ~ s ~w(s)~,
\end{equation*} 
and \eqref{Sect_theoremQ_ExpValue_vs_cumulative} is proved. Note that the argument did not involve potentially ill-behaved tools such as derivatives and integration by parts, only an exchange of the integration symbols, which is acceptable as long as Fubini's theorem is applicable. In particular, this proof allows to use the identity \eqref{Sect_theoremQ_ExpValue_vs_cumulative} for the distributions $w(q)$ and $W(q)$ at hand even if they are sums of delta functions, reflecting their original discrete nature.}. The distribution $W(q)$ may be expressed in terms of projectors to eigenspaces of $\mathcal{Q}$ with eigenvalues larger than or equal to $q$. For simplicity, we give their discrete version:
\begin{equation}
    \label{Sect_theoremQ_projectors_large_values}
    \begin{split}
        & P_q^{Q} := \sum_{\substack{a~\text{s.t.} \\ q_a \geq q}} |q_a) ( q_a |~, \\
        & P_n^{K} := \sum_{m\geq n} |\mathcal{O}_m) ( \mathcal{O}_m |~,
    \end{split}
\end{equation}
where in the second line we have additionally defined the analogous object for the Krylov basis, which shall be used later. $W(q)$ is given in terms of $P_q^Q$ by
\begin{equation}
    \label{Sect_theoremQ_Wq_projector}
    W(q) = \left(\mathcal{W}|P_q^{Q}|\mathcal{W}\right)~,
\end{equation}
and combining \eqref{Sect_theoremQ_ExpQ_W} with \eqref{Sect_theoremQ_ExpValue_vs_cumulative} and \eqref{Sect_theoremQ_Wq_projector} we reach:
\begin{equation}
    \label{Sect_theoremQ_ExpQ_projector}
    \left( \mathcal{W}|\mathcal{Q}|\mathcal{W} \right) = \int_{\mathds{R}^+}dq ~\left(\mathcal{W}|P_q^{Q}|\mathcal{W}\right)~.
\end{equation}
Having expressed the expectation value of $\mathcal{Q}$ in the state $|\mathcal{W})$ in terms of the projector over large spectral values of $q$ is useful because it allows to employ inequalities like \eqref{Sect_theoremQ_overlap_qbasis_Kbasis}, which state to what extent Krylov elements probe high $Q$-complexity values. In order to proceed further, we consider the identity
\begin{equation}
    \label{Sect_theoremQ_projectors_product_zero}
    P_q^Q (1-P^K_{n(q)})= \sum_{\substack{a~\text{s.t.} \\ q_a \geq q}} \sum_{m<n(q)}|q_a) \left(q_a | \mathcal{O}_m\right) ( \mathcal{O}_m | = 0~.
\end{equation}
Technically, \eqref{Sect_theoremQ_projectors_product_zero} vanishes because, at fixed $q$, for any $m<n(q)$ the overlap $\left(q_a | \mathcal{O}_m\right)$ is zero by definition of $n(q)$, as discussed below equation \eqref{Sect_theoremQ_overlap_qbasis_Kbasis}. Equivalently, one may understand directly the vanishing of the product $P_q^Q (1-P^K_{n(q)})$ because the term in brackets is the projector onto the subspace of Krylov elements $|\mathcal{O}_m)$ with $m<n(q)$ which, by definition of $n(q)$, is orthogonal to the span of the elements $|q_a)$ with $q_a\geq q$, onto which the first term of the product projects. Expression \eqref{Sect_theoremQ_projectors_product_zero} implies that the span of $\left\{|q_a)\right\}_{q_a\geq q}$ is entirely contained in the span of $\left\{|\mathcal{O}_m)\right\}_{m\geq n(q)}$, and therefore we have the identities:
\begin{equation}
    \label{Sect_theoremQ_projectors_identity}
    P_q^Q P_{n(q)}^K = P_q^Q \qquad \Longleftrightarrow \qquad P_{n(q)}^K P_q^Q = P_q^Q~,
\end{equation}
where the second expression is the hermitian adjoint of the first. With this, we may manipulate the expectation value of the projector in \eqref{Sect_theoremQ_ExpQ_projector} as follows:
\begin{equation}
\label{Sect_theoremQ_projectors_Expect_ineq}
    \left(\mathcal{W} \left| P_q^Q \right| \mathcal{W}\right) = \left(\mathcal{W} \left| P_{n(q)}^K P_q^Q P_{n(q)}^K\right| \mathcal{W}\right) \leq \left(\mathcal{W} \Big| {\left( P_{n(q)}^K \right)}^2 \Big| \mathcal{W}\right) = \left(\mathcal{W} \left| P_{n(q)}^K \right| \mathcal{W}\right)~.
\end{equation}
For the first step we applied iteratively the identities \eqref{Sect_theoremQ_projectors_identity}, and the inequality simply used the fact that $P_q^Q$ is a projector, so its presence in the expectation value can only have the effect of reducing the norm contributions. The last equality uses the fact that projectors are idempotent. Plugging \eqref{Sect_theoremQ_projectors_Expect_ineq} into \eqref{Sect_theoremQ_ExpQ_projector} we find:
\begin{equation}
    \label{Sect_theoremQ_inequality_ExpVals_W}
    \begin{split}
        &\left(\mathcal{W} | \mathcal{Q} |\mathcal{W} \right) = \int_{\mathds{R}^+} dq~ \left( \mathcal{W} \left| P_q^Q \right| \mathcal{W} \right) \leq \int_{\mathds{R}^+} dq ~ \left( \mathcal{W} \left| P_{n(q)}^K \right| \mathcal{W} \right) \\
        & = \int_{n(0)}^{n(+\infty)} \frac{dn}{\left.n^\prime \left(q\right)\right|_{q=q(n)}} \left( \mathcal{W} \left| P_{n}^K \right| \mathcal{W} \right)~,
    \end{split}
\end{equation}
where in the last step we have noted explicitly the change of variables $q\mapsto n(q)$. For $n(q) = \frac{q}{M}-1$, as given in \eqref{Sect_theoremQ_critical_nq}, the Jacobian is constant and equal to $M$, and the integration domain becomes $n\in[-1,+\infty[$. However, the spectrum of the K-complexity operator has no support over $n<0$, and therefore the final integral is equal to an integral over $n\in\mathds{R}^+$. This gives:
\begin{equation}
    \label{Sect_theoremQ_inequality_ExpVals_W_v2}
    \left(\mathcal{W} | \mathcal{Q} |\mathcal{W} \right) \leq M \int_{\mathds{R}^+} dn ~ \left( \mathcal{W} \left| P_{n}^K \right| \mathcal{W} \right) = M~\left(\mathcal{W} \left| \widehat{\mathcal{C}_K}\right|\mathcal{W} \right)~.
\end{equation}
As stated, expression \eqref{Sect_theoremQ_inequality_ExpVals_W_v2} holds for any $|\mathcal{W})\in \widehat{\mathcal{H}}$. Our interest, nevertheless, is for the case $|\mathcal{W})=\left| \mathcal{O}(t) \right)$, in which case relation \eqref{Sect_theoremQ_inequality_ExpVals_W_v2} is the announced upper bound on the family of K-complexities:
\begin{equation}
    \label{Sect_theoremQ_Bound_on_QC_proved}
    Q(t)\leq M~C_K(t)~,
\end{equation}
which holds for all times, precisely because \eqref{Sect_theoremQ_inequality_ExpVals_W_v2} holds for any element of the Hilbert space $\widehat{\mathcal{H}}$. This concludes the proof\footnote{The original formulation of the theorem in \cite{Parker:2018yvk} poses a more conservative bound, $Q(t)\leq 2 M C_K(t)$, which resulted from the use of a different critical function $n(q)=\frac{q}{2M}$. Such a choice was not judged necessary for the proof presented in this text.} of the Q-theorem \eqref{Sect_theoremQ_statement}.

\subsubsection{Some examples of Q-complexities}

The Q-theorem \eqref{Sect_theoremQ_statement} puts K-complexity in a prominent position within the family of Q-complexities. Given this, showing that some members of this family are familiar notions of quantum complexity that have been widely studied in the context of operator growth, quantum chaos and holography serves to provide a more solid justification to the study of Krylov complexity.

\paragraph*{Size complexity}

Operator size \cite{Roberts:2014isa} measures the spatial extent over which an operator has a non-trivial support. For the sake of illustration, we shall consider a system of $N$ qudits, whose Hilbert space is the tensor product of the $d$-dimensional spaces of each individual qudit, therefore having dimension $D=d^N$, which implies that the dimension of operator space is $\dim \widehat{\mathcal{H}}=D^2 = d^{2N}$. The tensor product structure of the total Hilbert space allows to define the size of an operator $\mathcal{O}$ as the number of sites (i.e. the number of qudits) on which the operator acts non-trivially (i.e. differently from the identity over that qudit space). If $\mathcal{O}$ is simply a tensor product of operators, the definition just given applies immediately, but if it is a linear combination of several operators consisting on tensor products themselves, then the notion of size needs to be modified to be the weighted average of the size of each individual tensor product contributing to $\mathcal{O}$. Phrased in this manner, showing that operator size is a $Q$-complexity is rather straightforward: We just need to construct a basis of operator space consisting of tensor-product operators of fixed size. For a given size $s$, there are $h(s)$ linearly independent operator chains, where
\begin{equation}
    \label{Sect_theoremQ_sizeComp_lin_indep_sizes}
    h(s)=\binom{N}{s}{(d^2-1)}^s~,
\end{equation}
where $d^2-1$ is the dimension of the subspace of the operator space of a single qudit that excludes the identity. We note that, consistently,
\begin{equation}
    \label{Sect_theoremQ_sizeComp_sum_sizes_dim}
    \sum_{s=0}^N h(s)=d^{2N} = \dim\widehat{\mathcal{H}}~.
\end{equation}
Thus, one may form an orthonormal basis of operator space constructed out of fixed-size chains $|s,j)$, with $j=1,\dots,h(s)$ and $s=0,\dots,N$, which can be used to build the $\mathcal{Q}$ operator for size complexity through its spectral resolution:
\begin{equation}
    \label{Sect_theoremQ_sizeComp_Qop}
    \mathcal{Q} = \sum_{s=0}^N s \sum_{j=1}^{h(s)}|s,j) (s,j|~.
\end{equation}
The complexity eigenvalues are given by $s=0,\dots,N$, which are $h(s)$ times degenerate. They are positive-semidefinite and therefore the assumption \eqref{Sect_theoremQ_Assumptions_Q_positive_semidef} is satisfied. Furthermore, given a $k$-local Hamiltonian and an initial operator $\mathcal{O}$ of size $r$, assumptions \eqref{Sect_theoremQ_Assumptions_L_on_q} and \eqref{Sect_theoremQ_Assumptions_initial_compl} are satisfied with $M=\max(k,r)$. Thus, \eqref{Sect_theoremQ_sizeComp_Qop} defines correctly a $Q$-complexity. The size of the time-evolving operator $|\mathcal{O}(t))$ may be computed as the expectation value of $\mathcal{Q}$ in \eqref{Sect_theoremQ_sizeComp_Qop} and is subject to the bound \eqref{Sect_theoremQ_statement}.

Given a system in the thermodynamic limit with a $k$-local Hamiltonian and an initial operator of small size, 
each power of the Liouvillian in the Taylor series of $e^{it\mathcal{L}}|\mathcal{O})$ has the effect of increasing the size of the operator by a finite amount, and likewise it increases its projection over the corresponding Krylov elements. As a result, for these large-$N$ systems, K-complexity and size complexity may be expected to be qualitatively similar \cite{Roberts:2014isa,I}, at least for chaotic systems, where both quantities may grow in time as fast as allowed (see section \ref{sect:KC_chaos_holog}). However, for finite size systems, the degeneracy in the size eigenvalues forbids size complexity from growing further than $N$, the total length of the qudit chain in the example above, while the Krylov complexity eigenvalues are completely non-degenerate and increasing, as a result of which K-complexity has a further scope of growth once size complexity has saturated. In the case of chaotic systems, size complexity will typically grow exponentially \cite{Roberts:2014isa} up to the scrambling time \cite{Altland:2020ccq} $t_s\sim \log N$, after which it saturates, while K-complexity may continue growing up until the Heisenberg time $t\sim e^N$, as argued in \cite{Barbon:2019wsy} and shown numerically for the first time in publication \cite{I}, detailed in Chapter \ref{ch:chapter03_SYK}.

\paragraph*{Out-of-time-ordered correlation functions}

OTOCs are a standard probe of chaos (see e.g. \cite{Maldacena:2015waa,Sonner:2017hxc,Anous:2019yku}), as they measure the capacity of a system to scramble information efficiently. Let us consider two localized observables in a system (morally, one could think of the example qudit system in the previous discussion), $\mathcal{V}$ and $\mathcal{O}$. Assuming that they are localized on different sites, their commutator vanishes, $[\mathcal{V},\mathcal{O}]=0$, implying that the measurements of both observables are completely uncorrelated, they do not affect each other. If we now evolve the operator $\mathcal{O}(t)$ with the system's Hamiltonian, it may start to grow, potentially developing some non-trivial support over the sites on which $\mathcal{V}$ is localized. As a consequence, a measurement of $\mathcal{V}$ at time zero and a measurement of $\mathcal{O}$ at time $t$ (i.e. a measurement of $\mathcal{O}(t)$) will be correlated, and this will be reflected in the non-vanishing of the commutator $[\mathcal{V},\mathcal{O}(t)]$. In order to measure the growth of such a quantity, the OTOC is defined to be (the square of) its norm:
\begin{equation}
    \label{Sect_theoremQ_OTOC_def}
    \mathcal{F}(t):= {\Big\lVert [\mathcal{V},\mathcal{O}(t)] \Big\rVert}^2 = \Big( [\mathcal{V},\mathcal{O}(t)] \Big| [\mathcal{V},\mathcal{O}(t)] \Big)~.
\end{equation}
Given the inner product \eqref{Sect_KCdef_thermal_inner_prod_conventional}, expression \eqref{Sect_theoremQ_OTOC_def} takes the familiar form
\begin{equation}
    \label{Sect_theoremQ_OTOC_familiarform}
    \mathcal{F}(t) = \frac{-1}{Z(\beta)}\text{Tr}\left\{ e^{-\beta H} {\left[ \mathcal{V},\mathcal{O}(t) \right]}^2 \right\}~,
\end{equation}
whose non-trivial \cite{Maldacena:2015waa} connected components are of the form $\langle \mathcal{V} \mathcal{O}(t) \mathcal{V} \mathcal{O}(t) \rangle_{\beta}$, hence the name \textit{out-of-time-ordered} correlation function. The relevance of OTOCs for the study of quantum chaos has peaked since \cite{Maldacena:2015waa} showed that these functions grow at most exponentially, $\mathcal{F}(t)\sim e^{\lambda_L t}$, with an exponent that is bounded from above by\footnote{Strictly speaking, the so-called \textit{MSS bound} \eqref{Sect_theoremQ_OTOC_MSS_bound} was proved in \cite{Maldacena:2015waa} for a regulated version of the thermal inner product. As of today, applying the same bound to OTOCs \eqref{Sect_theoremQ_OTOC_familiarform} computed with the conventional thermal inner product remains a conjecture.}:
\begin{equation}
    \label{Sect_theoremQ_OTOC_MSS_bound}
    \lambda_L\leq \frac{2\pi}{\beta}~.
\end{equation}
Proving that OTOCs belong to the family of $Q$-complexities, bounded from above by (a multiple of) K-complexity, is therefore an important result of \cite{Parker:2018yvk} which suggests the usefulness of the latter as a probe of quantum chaos. Let us consider the superoperator defined by the adjoint action of $\mathcal{V}$, $\mathcal{L}_{\mathcal{V}}\equiv [\mathcal{V},\cdot]$, verifying
\begin{equation}
    \label{Sect_theoremQ_OTOC_Lv}
    \mathcal{L}_{\mathcal{V}}|\mathcal{W}) = \Big| \left[\mathcal{V},\mathcal{W}\right] \Big)
\end{equation}
for any $|\mathcal{W})\in\widehat{\mathcal{H}}$. With respect to the inner products \eqref{Sect_KCdef_thermal_inner_product_g}, this superoperator is hermitian, $\mathcal{L}_{\mathcal{V}}=\mathcal{L}_{\mathcal{V}}^\dagger$, and as such it satisfies
\begin{equation}
    \label{Sect_theoremQ_OTOC_Lv_dagger}
    (\mathcal{W}|\mathcal{L}_{\mathcal{V}} = \Big( \left[\mathcal{V},\mathcal{W}\right] \Big|~,
\end{equation}
for any $(\mathcal{W}|$ belonging to the dual of operator space, $\widehat{\mathcal{H}}^{*}$. With this, we can define the $\mathcal{Q}$ operator relevant to the OTOC \eqref{Sect_theoremQ_OTOC_def} as:
\begin{equation}
    \label{Sect_theoremQ_OTOC_Qop}
    \mathcal{Q} = \mathcal{L}_{\mathcal{V}}^2~.
\end{equation}
And one can verify that
\begin{equation}
    \label{Sect_theoremQ_OTOC_Ft_Q}
    \mathcal{F}(t) = \Big( \mathcal{O}(t)\Big| \mathcal{Q} \Big| \mathcal{O}(t) \Big) = \Big( \mathcal{O}(t)\Big| \mathcal{L}_{\mathcal{V}}^2\Big| \mathcal{O}(t) \Big)
\end{equation}
Because of the square, \eqref{Sect_theoremQ_OTOC_Qop} is positive-semidefinite, automatically fulfilling \eqref{Sect_theoremQ_Assumptions_Q_positive_semidef}. Even though we do not have the explicit eigenbasis of $\mathcal{L}_{\mathcal{V}}^2$, given a $k$-local Hamiltonian, assumptions \eqref{Sect_theoremQ_Assumptions_L_on_q} and \eqref{Sect_theoremQ_Assumptions_initial_compl} are also satisfied thanks to the locality of $\mathcal{V}$ and $\mathcal{O}$. Even though OTOCs are often thought of as four-point functions (which they are by definition), they are also a perfectly valid $Q$-complexity. Their relation to K-complexity and the use of the latter to improve the MSS bound \eqref{Sect_theoremQ_OTOC_MSS_bound} will be the subject of section \ref{sect:KC_chaos}.

\subsection{Optimal basis theorem}\label{sect:Btheorem}

Let us now proceed to present the second theorem on Krylov complexity, which appeared in \cite{Balasubramanian:2022tpr}. It was originally formulated for state complexities defined in the Hilbert space $\mathcal{H}$ of some physical system but, just like the Q-theorem, it is applicable to any Hilbert space (such as operator space) for which there may be a time evolution generator that can be used to define the Krylov complexity of a certain time-evolving element of the space. The optimal basis theorem proceeds very similarly to the Q-theorem: It establishes a class of generalized complexities or \textit{cost functions}, of which Krylov complexity (or a very closely related notion) is a representative, and performs an optimization procedure that will make the latter stand out from the rest. This optimization, however, is of a different nature from that in section \ref{sect:QTheorem}, and as a result it will yield that K-complexity bounds any other element of the family from below, but for a finite interval of time that will be dependent on the latter. Let us present the theorem in its original formulation.

Given a time-evolving state $|\phi(t)\rangle$, written in \eqref{Sect_KCdef_state_t}, in the system's $D$-dimensional Hilbert space $\mathcal{H}$, and given some orthonormal basis for the space,
\begin{equation}
    \label{Sect_theoremB_basis}
    \mathcal{B}:=\left\{|\mathcal{B}_n\rangle\right\}_{n=0}^{D-1}~,
\end{equation}
one can define a generalized complexity or cost function associated to the basis $\mathcal{B}$ as:
\begin{equation}
    \label{Sect_theoremB_costfunct_def}
    C_{\mathcal{B}}(t) := \sum_{n=0}^{D-1} c_n {\left| \langle \mathcal{B}_n|\phi(t) \rangle \right|}^2 \equiv \sum_{n=0}^{D-1} c_n ~p_{\mathcal{B}}(n,t)~,  
\end{equation}
where the complexity eigenvalues $c_n$ form a sorted, strictly increasing sequence $0\leq c_0 < c_1 < \dots < c_{D-1}$. 
The motivation of the argument is to consider these coefficients as a \textit{fixed} collection of values and to find a basis that is optimal in some sense. In order to clarify the optimization process, let us consider the Taylor series of the cost function \eqref{Sect_theoremB_costfunct_def} around $t=0$:
\begin{equation}
    \label{Sect_theoremB_cost_Taylor}
    C_{\mathcal{B}}(t) = \sum_{m=0}^{+\infty}\frac{t^m}{m!}C_{\mathcal{B}}^{(m)}~,
\end{equation}
where
\begin{equation}
    \label{Sect_theoremB_cost_Taylor_coeff}
    C_{\mathcal{B}}^{(m)} := \sum_{n=0}^{D-1} c_n p_{\mathcal{B}}^{(m)}(n)~,\qquad p_{\mathcal{B}}^{(m)}(n):=\left. \frac{d^m}{dt^m} p_{\mathcal{B}}(n,t) \right|_{t=0}~.
\end{equation}

We now introduce a relational notion between sets of cost function Taylor coefficients: Given two bases $\mathcal{B}$ and $\mathcal{B}^\prime$ defining the corresponding cost functions, we say that $\left\{C_{\mathcal{B}}^{(m)}\right\}_{m\geq 0} < \left\{C_{\mathcal{B}^\prime}^{(m)}\right\}_{m\geq 0} $ if there exists some $M\in \mathds{N}_0$ such that $C_{\mathcal{B}}^{(m)} = C_{\mathcal{B}^\prime}^{(m)} $ for $m<M$ and $C_{\mathcal{B}}^{(M)} < C_{\mathcal{B}^\prime}^{(M)} $. In words, one set is said to be \textit{smaller} than the other if the first differing Taylor coefficient is smaller in the first set. This implies that there exists some time $t_{*}$ such that $\mathcal{C}_{\mathcal{B}}(t)< \mathcal{C}_{\mathcal{B}^\prime}(t)$ for $t\in[ 0,t_{*} [$. The particular value of $t_{*}$ is strongly dependent on the subsequent Taylor coefficients of both series.

With this, we can now state the optimal basis theorem: For a fixed set of complexity eigenvalues $c_n$, and given some orthonormal basis $\mathcal{B}$, the Krylov basis (dubbed $\mathcal{K}$) is always optimal for defining the cost function \eqref{Sect_theoremB_costfunct_def} in the sense that 
\begin{equation}
    \label{Sect_theoremB_statement}
    \left\{C_{\mathcal{K}}^{(m)}\right\}_{m\geq 0} < \left\{C_{\mathcal{B}}^{(m)}\right\}_{m\geq 0} \qquad \text{for any }\mathcal{B}~.
\end{equation}
Again, this implies that, for any basis $\mathcal{B}$, there will always exist some time interval around $t=0$ such that the cost function $C_{\mathcal{B}}(t)$ is larger than K-complexity\footnote{Strictly speaking, $C_{\mathcal{K}}(t)$ matches our definition of Krylov complexity \eqref{Sect_KCdef_KC_def} if the fixed complexity eigenvalues are $c_n=n$. Nevertheless, for any other choice of increasing $c_n$, the Krylov basis remains being optimal. This is why, within this section, we shall denote the cost function defined through the Krylov basis as $C_{\mathcal{K}}(t)$ instead of $C_K(t)$, even though we may still loosely refer to it as K-complexity.}. 

In order to prove this theorem, we shall compute explicitly the moments of the cost function of $\mathcal{B}$ and compare them to those of the cost function associated to the Krylov basis. The only property of the Krylov basis that will enter in the argument is the fact that it is an orthonormal basis $\left\{|K_n\rangle\right\}_{n=0}^{K-1}$ where the $n$-th element is a linear combination of $\left\{H^m|\phi\rangle\right\}_{m=0}^{n}$; it can be inductively shown that this requirement defines uniquely the basis (up to global phases), and hence the details of the Lanczos algorithm will not be necessary.

The probability moments $p_{\mathcal{B}}^{(m)}(n)$ appearing in \eqref{Sect_theoremB_cost_Taylor_coeff} can be developed by using the definition \eqref{Sect_theoremB_costfunct_def}, together with the explicit expression of the time-evolving state \eqref{Sect_KCdef_state_t}. We find:
\begin{equation}
    \label{Sect_theoremB_probability_moments}
    p_{\mathcal{B}}^{(m)}(n) = i^m \sum_{k=0}^m (-1)^k \langle \phi | H^{m-k} |\mathcal{B}_n\rangle \langle \mathcal{B}_n | H^k | \phi\rangle~.
\end{equation}

To show \eqref{Sect_theoremB_statement}, we may now proceed in an inductive fashion: We consider a basis $\mathcal{B}$ such that its first $N$ elements are Krylov elements, i.e. $|\mathcal{B}_n\rangle = |K_n\rangle$ for $n=0,\dots,N-1$, and we shall analyse the difference between having the next basis element given by the next Krylov element, i.e. $|\mathcal{B}_N\rangle = |K_N\rangle$, or by something else. The conclusion will be that the former option minimises the cost function Taylor coefficients, and proceeding iteratively one eventually constructs the full Krylov basis\footnote{If the Krylov dimension $K$ is smaller than the full Hilbert space dimension $D$, then any basis whose first $K$ elements are the Krylov elements is optimal for \eqref{Sect_theoremB_costfunct_def}, the subsequent basis elements being irrelevant because they lie outside of the Krylov space and are never probed by $|\phi(t)\rangle$.}. As a seed for the inductive argument, we note that the zeroth Krylov element is the initial condition itself, $|K_0\rangle = |\phi(0)\rangle=|\phi\rangle$, and therefore $C_{\mathcal{K}}^{(0)}=C_{\mathcal{K}}(0) = c_0$; if $|\mathcal{B}_0\rangle$ differs from $|\phi\rangle$ by more than a global phase, then $C_{\mathcal{B}}^{(0)}=C_{\mathcal{B}}(0)$ will be some linear combination of $\left\{c_n\right\}_{n=0}^{D-1}$ (where the coefficients of such a linear combination add up to one because of normalization of the initial condition and orthonormality of the basis $\mathcal{B}$), probing non-trivially complexity eigenvalues higher than $c_0$. Since the $c_n$ are sorted and increasing, it follows that $C_{\mathcal{K}}^{(0)}<C_{\mathcal{B}}^{(0)}$, establishing \eqref{Sect_theoremB_statement} for this particular choice of $\mathcal{B}$ (which is in fact a class of bases, as we have not been specific about the subsequent basis elements $\left\{|\mathcal{B}_n\rangle\right\}_{n=1}^{D-1}$).

For a basis $\mathcal{B}$ that coincides by assumption with the Krylov basis in its $N$ first elements, we can decompose the contributions to the coefficients \eqref{Sect_theoremB_cost_Taylor_coeff} into those coming from the Krylov basis, and those coming from the remaining unspecified part of $\mathcal{B}$:
\begin{equation}
    \label{Sect_theoremB_Taylor_coeffs_contribs}
    C_{\mathcal{B}}^{(m)} = \sum_{n=0}^{N-1} c_n ~p_{\mathcal{K}}^{(m)}(n) + \sum_{n\geq N} c_n~ p_{\mathcal{B}}^{(m)}(n)~.
\end{equation}

For the contributions with $n\geq N$, we can observe \eqref{Sect_theoremB_probability_moments} and note that, by assumption, the elements $|\mathcal{B}_n\rangle$ with $n\geq N$ are orthogonal to $\left\{|K_n\rangle\right\}_{n<N}$ and, therefore, at least one of the two bra-kets inside of the sum will vanish for $m<2N$, yielding a vanishing result for the probability moment. That is:
\begin{equation}
    \label{Sect_theoremB_lemma1}
    p_{\mathcal{B}}^{(m)}(n) = 0\qquad\text{for }n\geq N\text{ and }m<2N~.
\end{equation}
Plugging \eqref{Sect_theoremB_lemma1} into \eqref{Sect_theoremB_Taylor_coeffs_contribs} shows that 
\begin{equation}
    \label{Sect_theoremB_TaylorCoeffs_equal_up_to_2N}
    C_{\mathcal{B}}^{(m)} = C_{\mathcal{K}}^{(m)}\qquad\text{for }m<2N~.
\end{equation}
Hence, the potentially differing Taylor coefficient is $C_{\mathcal{B}}^{(2N)}$. Once again, using \eqref{Sect_theoremB_probability_moments} and recalling that $\mathcal{B}$ coincides with the Krylov basis in its first $N$ elements, we have that
\begin{equation}
    \label{Sect_theoremB_2N_coeff}
    C_{\mathcal{B}}^{(2N)}= \sum_{n=0}^{N-1}c_n~p_{\mathcal{K}}^{(2N)}(n) + \binom{2N}{N}\sum_{n\geq N} c_n ~\langle\phi | H^N |\mathcal{B}_n\rangle\langle \mathcal{B}_n | H^N | \phi\rangle~.
\end{equation}
At this point, it is useful to perform the orthogonal decomposition 
\begin{equation}
    \label{Sect_theoremB_HN_orthogonal_decomp}
    H^N|\phi\rangle = |\chi_N\rangle + |\xi_{N-1}\rangle~,
\end{equation}
where
\begin{equation}
    \label{Sect_theoremB_xiN1_belongs}
    |\xi_{N-1}\rangle \in \text{span}\left\{|K_n\rangle\right\}_{n=0}^{N-1} = \text{span}\left\{|\mathcal{B}_n\rangle\right\}_{n=0}^{N-1}
\end{equation}
and, by definition of the Krylov basis, $|\chi_N\rangle \propto |K_N\rangle$. Its full norm must therefore be contained in the span of all the $|\mathcal{B}_n\rangle$ with $n\geq N$, in particular:
\begin{equation}
    \label{Sect_theoremB_chiN_norm}
    \langle \chi_N|\chi_N\rangle = \sum_{n\geq N} {\lvert \langle \mathcal{B}_n|\chi_N\rangle \rvert}^2.
\end{equation}
We therefore have that \eqref{Sect_theoremB_2N_coeff} equals to:
\begin{equation}
    \label{Sect_theoremB_final_steps}
    \begin{split}
        C_{\mathcal{B}}^{(2N)} &= \sum_{n=0}^{N-1} c_n~p_{\mathcal{K}}^{(2N)}(n) + \binom{2N}{N}\sum_{n\geq N} c_n~\langle \chi_N|\mathcal{B}_n\rangle \langle \mathcal{B}_n | \chi_N \rangle \\
        & \geq \sum_{n=0}^{N-1} c_n~p_{\mathcal{K}}^{(2N)}(n)+ c_N ~\binom{2N}{N}\sum_{n\geq N} \langle \chi_N|\mathcal{B}_n\rangle \langle \mathcal{B}_n | \chi_N \rangle \\
        &=\sum_{n=0}^{N-1} c_n~p_{\mathcal{K}}^{(2N)}(n)+ c_N ~\binom{2N}{N} \langle \chi_N|\chi_N\rangle = C_{\mathcal{K}}^{(2N)}~,
    \end{split}
\end{equation}
where the inequality has made use of the fact that $\left\{c_n\right\}_{n\geq 0}$ is an increasing sequence. Equality holds either if $|\chi_N\rangle = 0$, in which case the Krylov basis terminates and the basis $\mathcal{B}$ is already optimal (and equal to the $\mathcal{K}$ basis), or if $|\mathcal{B}_N\rangle = |K_N\rangle$, as in this case $|\chi_N\rangle$ does not have overlap with the elements $|\mathcal{B}_n\rangle$ with $n>N$. This shows that the optimal Taylor coefficient is obtained if $|\mathcal{B}_{N}\rangle$ coincides with the next Krylov element. As announced, the iterative application of this argument makes us conclude that the Krylov basis optimizes the Taylor series coefficients of the cost function, proving the theorem \eqref{Sect_theoremB_statement}.

The authors of \cite{Balasubramanian:2022tpr} present this theorem as a manner to obtain Krylov complexity as an output of an optimization process. The caveats of this optimization will nevertheless be discussed in section \ref{sect:Theorems_comparison}.

\subsection{Discussion and comparison of the theorems}\label{sect:Theorems_comparison}

The two theorems that have been discussed provide a solid motivation for the study of Krylov complexity: The Q-theorem shows that it appears in the upper bound of a large family of generalized complexities to which familiar notions like operator size or OTOCs belong, while the optimal basis theorem proves that the Krylov basis can be obtained as an output of an optimization process that, roughly speaking, seeks the smallest growth rate of complexity in the vicinity of $t=0$. 

However, at first glance, it might appear that the two theorems are in tension, since K-complexity is simultaneously appearing as an upper and as a lower bound on the generalized complexities. Several subtleties need to be discussed in order to understand how these results can hold simultaneously. First, we have already pointed out that, even though the theorems were originally formulated for different Hilbert spaces (one dealt with operator complexity, and the other addressed state complexity), nothing in their proofs prevents us from extending them to other Hilbert spaces. In particular, one may attempt to apply both theorems simultaneously in the same space, but the first caveat in such a situation is to note that their hypotheses are not exactly identical:

\begin{itemize}
    \item The complexity eigenvalues in the Q-theorem are positive-semidefinite and not necessarily sorted, according to \eqref{Sect_theoremQ_Assumptions_Q_positive_semidef}, while those in the basis theorem, introduced in \eqref{Sect_theoremB_costfunct_def}, are positive-semidefinite and sorted. This is not a major obstacle: The Q-theorem can be reformulated in terms of a sorted sequence of $q$-values by re-arranging the eigenbasis of the $\mathcal{Q}$ operator. Additionally, the $q$-values in \eqref{Sect_theoremQ_Assumptions_Q_positive_semidef} are allowed to be degenerate, while the sequence $c_n$ of cost-function coefficients was originally \cite{Balasubramanian:2022tpr} defined to be strictly increasing. Nevertheless, the proof of the basis theorem can be extended to the case where the $c_n$ feature degeneracies, since at the end of the day the inequality reached in \eqref{Sect_theoremB_final_steps} between the Taylor series coefficients is not strict.
    \item The Q-theorem has two additional assumptions, \eqref{Sect_theoremQ_Assumptions_L_on_q} and \eqref{Sect_theoremQ_Assumptions_initial_compl}, which makes it restrict to bases for which the complexity only increases gradually upon the action of the time-evolution generator, and such that the initial condition does not have an arbitrarily large complexity. Hence, having accepted the previous point, it follows that all Q-complexities define acceptable cost functions for the basis theorem, but the reverse is not true.
\end{itemize}

Taking the above considerations into account, we deduce that there exist \textit{some} generalized complexities that fulfill simultaneously the assumptions of both theorems, and therefore both bounds apply simultaneously for them. However, the bounds are of a different nature:

\begin{itemize}
    \item The Q-complexity bound \eqref{Sect_theoremQ_statement} holds at arbitrary times and for any Q-complexity, since it was derived as a bound on the expectation value of the $\mathcal{Q}$ superoperator evaluated on an arbitrary state $|\mathcal{W})$, which was eventually particularized to be $\big|\mathcal{O}(t)\big)$. 
    \item The bound in the time domain that one can deduce from the basis theorem operates differently: Given a cost function $C_{\mathcal{B}}(t)$ for a certain basis $\mathcal{B}$, the fact that its Taylor coefficients are bigger (in the sense defined in section \ref{sect:Btheorem}) than those of $C_{\mathcal{K}}(t)$ implies that there exists some interval of time around $t=0$ for which $C_{\mathcal{K}}(t)\leq C_{\mathcal{B}}(t)$, but the size of that interval is completely dependent on $\mathcal{B}$. In other words, there does not exist a finite time range for which Krylov complexity is smaller or equal than \textit{any} cost function\footnote{As an example, for any time $t_0>0$, one can always construct a basis $\mathcal{B}$ such that $|\mathcal{B}_0\rangle = |\phi(t_0)\rangle$, and as a result $C_{\mathcal{B}}(t)$ will be minimal at $t=t_0$. In particular, $t_0$ can be arbitrarily close to $t=0$.}. The bound of this theorem acts \textit{pairwise}, in the sense that for any basis there is always an interval of time for which the Krylov basis gives a smaller cost function, but such an interval is not universal\footnote{The authors of \cite{Balasubramanian:2022tpr} refer to this statement as a \textit{functional optimization}, as opposed to an optimization in time.}. The fact that the $c_n$ values are left \textit{fixed} in the optimization argument reflects the aforementioned \textit{pairwise} nature of the latter.
\end{itemize}

To summarise, there may exist a generalized complexity $C_{\mathcal{B}}(t)$ for which both theorems apply simultaneously, suggesting the hierarchy:
\begin{equation}
    \label{Sect_Theorems_Comparison_hierarchy}
    \text{ `` }C_{\mathcal{K}}(t) \leq C_{\mathcal{B}}(t) \leq M C_K(t)~\text{.'' }
\end{equation}
However, the first inequality holds for some $\mathcal{B}$-dependent time interval around $t=0$, while the second one holds at arbitrary times and for some $\mathcal{B}$-dependent constant $M$. Additionally, for the \textit{actual} K-complexity $C_K(t)$ to enter in the leftmost term of \eqref{Sect_Theorems_Comparison_hierarchy}, the complexity eigenvalues defining $C_{\mathcal{B}}(t)$ must be given by $c_n = n$, since they are left fixed in the optimal basis theorem, which is the one dictating the leftmost inequality\footnote{Otherwise, the leftmost term of \eqref{Sect_Theorems_Comparison_hierarchy} is given by a \textit{version} of Krylov complexity constructed with the same complexity eigenbasis (i.e. the Krylov basis), but whose complexity eigenvalues are set to $c_n$ instead of $n$. Such a version was denoted as $C_\mathcal{K}(t)$, instead of $C_K(t)$, in section \ref{sect:Btheorem}. Nevertheless, in such a situation the rightmost term of \eqref{Sect_Theorems_Comparison_hierarchy} will still be $MC_K(t)$ (i.e. $M$ times the \textit{actual} Krylov complexity) because that inequality comes from the Q-theorem, whose proof does not require the set of complexity eigenvalues to be fixed.}.

\section{K-complexity, chaos and holography}\label{sect:KC_chaos_holog}

In this section we shall introduce the remaining ingredients motivating and contextualizing the research conducted for the contributions \cite{I,II,III,IV}.

\subsection{K-complexity and chaos}\label{sect:KC_chaos}

Krylov complexity was introduced in \cite{Parker:2018yvk} because of its usefulness for the study of operator growth. Not only did the authors put this notion in relation to other complexities through their Q-theorem, but they also proposed the so-called \textit{universal operator growth hypothesis} according to which the Lanczos coefficients (and therefore K-complexity) grow as fast as possible in maximally chaotic systems.

Before presenting the hypothesis in more detail, we shall elaborate on what we mean by a quantum chaotic system. Classically, a system is said to be chaotic if it features ergodic trajectories in phase space, as opposed to classically integrable systems where orbits are confined to the Cartesian product of invariant tori defined by the action-angle variables \cite{Arnold:1989who}. Typically, trajectories in a classically chaotic system with slightly different initial conditions will feature exponential departure as a function of time \cite{ClassicalChaosBook}, where the exponent controlling the separation is the (classical) Lyapunov exponent. Extrapolating this notion of classical chaos to the quantum realm is not entirely obvious, as the lack of periodicity in the phase space orbits of classically chaotic systems does not allow for a semi-classical quantization scheme \textit{à la} Bohr-Sommerfeld \cite{stöckmann_1999}, and therefore alternative approaches to quantum chaos have been developed historically. Inspired by the study of heavy atomic nuclei spectra (see \cite{Bohigas_survey_paper,Bohigas_survey_proceedings} for a wide survey), E. Wigner proposed to model the spectral correlations in such systems by those of random matrices: This boosted the study of random matrix theory in the context of quantum chaos, pioneered by Wigner himself, as well as by F. Dyson and M. Mehta \cite{HaakeBook,stöckmann_1999,MehtaBook}. Subsequently, based on numerical evidences, the \textit{BGS conjecture} by O. Bohigas, M. J. Giannoni, and C. Schmit \cite{BGSpaper} advocated for the use of random matrix theory as a paradigm for the study of quantum chaos: They proposed that classically chaotic systems feature, when quantized, eigenvalue level repulsion, which is the imprint of random-matrix spectral correlations, even if the system has an order-one number of degrees of freedom (which was not the case of the nuclei of heavy atoms). On the other hand, the \textit{Berry-Tabor conjecture} \cite{BerryTabor} proposes that in integrable systems the amount of symmetry is such that energy levels are effectively uncorrelated and the level spacing probability distribution becomes Poissonian. The use of the spectral statistics as a diagnose for quantum chaos or quantum integrability is nowadays paradigmatic, see \cite{HaakeBook} for an extensive review.

There exists, however, a notion of quantum chaos that is in many cases related to the one above, but potentially not mathematically equivalent, and it is that of operator growth and out-of-time-ordered correlation functions. Roughly speaking, chaotic systems may be regarded as efficient information scramblers that rapidly spread initially localized perturbations, reflected in a fast operator growth \cite{Roberts:2014isa} and in the corresponding exponential increase of OTOCs \cite{Maldacena:2015waa}. Often, many-body systems exhibiting spectral chaos also feature scrambling dynamics in their four-point functions, as is the case of the \textit{Sachdev-Ye-Kitaev (SYK) model} \cite{Sachdev:1992fk,Sachdev:2015efa,Kitaev:2015}, a paradigmatic instance of a quantum-chaotic model in $0+1$ dimensions with a gravitational dual. The article \cite{Parker:2018yvk} proposing the universal operator growth hypothesis for K-complexity focuses on this notion of chaos. 

The starting point for the formulation of the hypothesis is to exploit the combinatorial relations between the two-point function and the Lanczos coefficients stated in section \ref{sect:KC_op} in order to translate the behavior of the high-frequency tails of the spectral function $\Phi(\omega)$ into the asymptotic behavior of the Lanczos coefficients $b_n$ as a function of $n$ (so far we assume the system to be in the thermodynamic limit, allowing for the Krylov dimension of the operator to be infinite). In \cite{Magnus} it is shown that the two following behaviours are equivalent:
\begin{equation}
    \label{Sect_KC_chaos_equivalence_behaviours}
    b_n\sim bn^{\delta} \qquad\Longleftrightarrow \qquad \Phi(\omega) \sim  e^{-{\left| \frac{\omega}{\omega_0} \right|}^{1/\delta}}~,
\end{equation}
where $b$ and $\omega_0$ are some dimensionful constants and $\delta > 0$. This can be combined with the bound on the spectral function proved in \cite{PhysRevLett.115.256803}, according to which the slowest possible decay rate of the spectral function at large frequencies is the exponential, giving $\delta\leq 1$. More specifically:
\begin{equation}
    \label{Sect_KC_chaos_bound_Lanczos}
    b_n\lesssim \alpha n\qquad\Longleftrightarrow \qquad \Phi(\omega)\lesssim e^{-\frac{|\omega|}{\omega_0}}~.
\end{equation}
Note that, because of its importance, for the case of an asymptotically linearly growing Lanczos sequence ($\delta=1$) the constant $b$ in \eqref{Sect_KC_chaos_equivalence_behaviours} has been relabeled as $\alpha$. In systems with one spatial dimension, the bound on $b_n$ acquires a logarithmic correction \cite{Parker:2018yvk}.

The translation between the behaviors \eqref{Sect_KC_chaos_bound_Lanczos} is relatively simple given the exponential behavior of the spectral function: Since it comes from the Fourier transform of the two-point function \eqref{Sect_KCdef_spectral_funct}, this exponential tail implies that the latter has a singularity along the imaginary axis at $t = \pm i\omega_0^{-1}$, which in turn enforces a certain asymptotic behavior of the moments $\mu_{2n}$ such that the convergence radius of the Taylor series of the two-point function does not exceed the distance between such a singularity and the origin. Finally, the relation between moments and Lanczos coefficients through Dyck paths reviewed in section \ref{subsect:From_Lanczos_to_Moments} can be used to show that this is compatible with the linear growth of the Lanczos coefficients. The details are worked out in \cite{Parker:2018yvk}, where the argument here described allows to additionally show that 
\begin{equation}
    \label{Sect_KC_chaos_omega0_alpha}
    \omega_0 = \frac{2}{\pi}\alpha~.
\end{equation}

The proof of the theorem bounding the tails of the spectral function by an exponential decay involves \cite{Parker:2018yvk} the study of the asymptotic behavior of the moments $\mu_{2n} = \left(\mathcal{O}\left|\mathcal{L}^{2n}\right|\mathcal{O}\right)$, which probes the growth of the operator upon successive applications of the Liouvillian, leading to the proposal that the bound \eqref{Sect_KC_chaos_bound_Lanczos} is saturated by maximally chaotic systems (in the sense of scrambling). This is the statement of the universal\footnote{The slope $\alpha$, or equivalently $\omega_0$, in the bound \eqref{Sect_KC_chaos_bound_Lanczos} is system-dependent \cite{PhysRevLett.115.256803}, but the fact that the asymptotic growth of the Lanczos coefficients is at most linear is not: This is the universal content of the hypothesis.} operator growth hypothesis.

An analysis of the recurrence relation for the operator wave functions \eqref{Sect_KCdef_recursion_phin} shows that a linearly increasing sequence of Lanczos coefficients $b_n\sim \alpha n$ implies an exponential growth of Krylov complexity, $C_K(t)\sim e^{2\alpha t}$. This is verified for some analytical toy models in \cite{Parker:2018yvk}, but it may also be derived from a continuous approximation of the discrete recurrence relation\footnote{See Appendix \ref{Appx:Cont_approx} for a detailed analysis of such a continuous approximation.}. The result is very useful in connection to OTOCs: As shown in \cite{Maldacena:2015waa}, the latter grow at most exponentially with a quantum Lyapunov exponent $\lambda_L$ subject to the universal bound \eqref{Sect_theoremQ_OTOC_MSS_bound} and, at the same time, they are subject to the Q-complexity bound \eqref{Sect_theoremQ_statement}, implying that their growth rate cannot exceed that of K-complexity. This provides a new bound for the Lyapunov exponent in terms of the Krylov complexity exponent: For maximally chaotic systems, both the OTOC and $C_K(t)$ grow exponentially, with exponents satisfying the relation
\begin{equation}
    \label{Sect_KC_chaos_bound_Lyapunov}
    \lambda_L\leq 2\alpha~.
\end{equation}

A clarification is in order at this point: Both the Lanczos coefficients and the two-point function require the specification of an inner product for their computation.
Despite not having explicitly mentioned it, the bound \eqref{Sect_KC_chaos_bound_Lanczos} on the tails of the spectral function is proved only at infinite-temperature $\beta=0$, and hence the corresponding inner product \eqref{Sect_KCdef_inner_product_infinite_temp} has been assumed throughout. In the case of infinite temperature, the MSS bound \eqref{Sect_theoremQ_OTOC_MSS_bound} becomes meaningless, while the bound \eqref{Sect_KC_chaos_bound_Lyapunov} remains finite, even though it is system-dependent due to the appearance of $\alpha$. This is an illustration of the power of K-complexity as a probe of chaos.

For finite temperatures $\beta>0$, a choice of thermal inner product \eqref{Sect_KCdef_thermal_inner_product_g} needs to be made. The MSS bound was proved in \cite{Maldacena:2015waa} for a specific choice of regulated thermal inner product that spread evenly the Boltzmann factor across the operator insertions. Hence, the authors of \cite{Parker:2018yvk} propose that the corresponding choice for the computation of K-complexity in order to make contact with such a regulated OTOC is the Wightman inner product \eqref{Sect_KCdef_Wightman_inner_prod}, which leads them to \textit{conjecture} that
\begin{equation}
    \label{Sect_KC_chaos_bound_Lyapunov_T}
    \lambda_L^{(T)}\leq 2\alpha_T^{(W)}~,
\end{equation}
where the temperature dependence and the inner product choice have been stressed for clarity. This conjecture is tested in \cite{Parker:2018yvk} numerically for the SYK model at various temperatures, and it is found to be much tighter than the MSS bound in such a system.

The universal operator growth hypothesis triggered a full field of research on the use of K-complexity as a probe of quantum chaos. A plethora of articles (see section \ref{sect:reseachKC}) have been produced in this context, including contributions \cite{I,II,III}. Nonetheless, the field is not exempt of debate: The universal operator growth hypothesis has been strongly contested for finite temperature where, as has been discussed, it is still conjectural. Let us give a brief overview of such a debate, as some valuable lessons can be extracted from it and it has influenced the research line of the contributions leading to this Thesis.

The main criticism to the operator growth hypothesis expressed as \eqref{Sect_KC_chaos_bound_Lyapunov_T} comes from the realm of quantum field theory and CFT \cite{Dymarsky:2021bjq}. In cases in which the operators of interest have positive energy dimension, contact singularities at $t=0$ are expected \cite{DiFrancesco:1997nk} in the conventional thermal two-point function, defined as
\begin{equation}
    \label{Sect_KC_chaos_conventional_thermal_TwoPt}
    C_{\beta}(t) := \big( \mathcal{O} | \mathcal{O}(t) \big)_{\beta} = \frac{1}{Z(\beta)}\text{Tr}\left[ e^{-\beta H} \mathcal{O} \mathcal{O}(t) \right]~,
\end{equation}
and due to this non-analyticity, no moments $\mu_{2n}$ can be computed. In order to compute K-complexity in these situations, one is forced to employ a regulated version of $C_{\beta}(t)$, namely the Wightman correlator, obtained from the corresponding Wightman inner product \eqref{Sect_KCdef_Wightman_inner_prod}:
\begin{equation}
    \label{Sect_KC_chaos_Wightman_Two_Pt}
    C_{\beta}^{W}(t) := \big( \mathcal{O} | \mathcal{O}(t) \big)_{\beta}^W = C_{\beta}\left(t+\frac{i\beta}{2}\right)~.
\end{equation}
The argument in \cite{Dymarsky:2021bjq} proceeds as follows: If $C_{\beta}(t)$ is singular at $t=0$, this implies immediately that $C_{\beta}^W(t)$ is singular at $t = -\frac{i\beta}{2}$. This is sufficient to dictate the asymptotic behavior of the moments $\mu_{2n}$ that leads to an asymptotically linear growth of the Lanczos coefficients with slope $\alpha = \pi / \beta$, yielding an exponential growth for K-complexity $C_K(t)\sim e^{2\alpha t} = e^{\frac{2\pi t}{\beta}}$. In other words, the choice of the regularization scheme for the two-point function automatically enforces exponential growth of Krylov complexity, regardless of the details of the system. The authors of \cite{Dymarsky:2021bjq} compute explicitly Lanczos coefficients\footnote{In order to compute analytically the Lanczos coefficients out of the moments in certain specific cases, the authors of \cite{Dymarsky:2021bjq} use a connection, previously pointed out in \cite{Dymarsky:2019elm}, between the recursion method defining the Lanczos coefficients and the recursion defining integrable Toda hierarchies, for which there exist families of known solutions.} in two-dimensional CFT, as well as free and generalized free theories \cite{Alday:2020eua} in $d$ spacetime dimensions, confirming their expectation. The fact that for all these evidently non-chaotic systems the Lanczos coefficients still feature linear growth because of the regularization scheme chosen for the two-point function is presented in the article as a counterexample to the universal operator growth hypothesis. Furthermore, given that in the case at hand $\alpha = \pi / \beta$, the bound \eqref{Sect_KC_chaos_bound_Lyapunov_T} reduces to the MSS bound \eqref{Sect_theoremQ_OTOC_MSS_bound}.

This criticism can be addressed from various perspectives, which have led to interesting avenues of research. Let us note some pertinent observations:
\begin{itemize}
    \item The slope of the Lanczos coefficients is dictated by the singularity of the two-point function that is closest to the origin of the complex plane for $t$. The fact that the Wightman regularization scheme imposes a singularity at $t=-\frac{i\beta}{2}$ does not necessarily rule out the existence of other system-specific singularities closer to the origin which would then dominate the slope of the Lanczos coefficients. Nevertheless, the point that this $\beta$-dependent singularity needs to be there regardless of the nature of the system remains valid and is sufficient to ensure exponential growth of Krylov complexity.
    \item The conflictive singularity that appears to make K-complexity featureless (or, in a sense, universal) in the context of field theory is a result of the inner product choice. Hence, an immediate alternative that one may consider is to choose a different scalar product. This approach was taken, in the context of two-dimensional conformal field theory with a large central charge, by \cite{Kundu:2023hbk}: The article proposes to take as an inner product the expectation value in a primary state of the theory, that should be thought of as acting as a thermal bath. With this, invoking the operator-state correspondence, the auto-correlation function of a certain light operator becomes a four-point function, carrying dynamical information. It is found in \cite{Kundu:2023hbk} that the K-complexity of a light primary operator defined with this inner product behaves either oscillatory or exponentially depending on whether the ``thermal bath'' is generated by a light or a heavy primary, respectively. Heavy primaries are those whose scaling dimension is above the black hole threshold, $\Delta_{BH}=\frac{c}{12}$. It is in fact expected that the spectrum of the CFT (and of its gravitational dual) is chaotic above such a threshold \cite{Anous:2019yku,Schlenker:2022dyo}, and hence this result illustrates that K-complexity is sensitive to chaos\footnote{In the article \cite{Kundu:2023hbk}, the two-point function still needs to be regulated in order to avoid singularities at $t=0$, but the authors show that the regulator does not determine the asymptotic behavior of K-complexity.}. Holographically, the dual interpretation of the black hole threshold in the CFT$_2$ spectrum is the Hawking-Page transition point in the AdS$_3$ bulk \cite{Hawking:1982dh}: Above this point, the bulk features a black hole, which is expected to be a chaotic system \cite{Sekino:2008he,Schlenker:2022dyo,Anous:2019yku}, while below it the emergent geometry corresponding to the light states consists on empty AdS$_3$ with a conical defect. The fact that K-complexity is found to oscillate below the threshold and to grow exponentially above it is consistent with the bulk picture.
    \item Lastly, we may note that up to now all the results on K-complexity reviewed in the current section have dealt with systems in the thermodynamic limit. Even the formulation of the universal operator growth hypothesis is itself adapted to infinite systems, since it addresses the asymptotic behavior of the Lanczos coefficients $b_n$, hence implicitly assuming that the sequence does not terminate. This leaves scope for the study of finite-size (and potentially non-perturbative) effects in Krylov complexity, which may as well retain some imprints of the chaotic or integrable nature of systems. Such imprints might potentially cure the limitations of the universal operator growth hypothesis pointed out previously, being invisible in the thermodynamic limit. Studying (mostly numerically) the behavior of the Lanczos coefficients towards the edge of the Krylov chain and the associated Krylov space dynamics that their features imply was one of the main motivations of the projects \cite{I,II,III}, together with the applicability of Krylov complexity to holography that will be discussed in the next section, even though it has already been suggested in this text.
\end{itemize}

\subsection{K-complexity and holography}\label{sect:KC_holog}

As mentioned in the Introduction of this Thesis and at the beginning of the present Chapter, it has been proposed that the growth of the Einstein-Rosen bridge in a two-sided\footnote{See \cite{DeBoer:2019yoe} for a study of one-sided black holes where the Tomita-Takesaki formalism allows for the partial reconstruction of some ``left side'', which probably does not asymptote to a left boundary because of an unavoidable cutoff in space whose details are currently not well understood.} black hole should be captured by a notion of quantum complexity in the boundary theory. To present day, such a proposal rests mostly on conjectural grounds. The size of the wormhole throat, computed as the length of space-like geodesics with anchoring points on opposite boundaries, or as related quantities such as bulk volumes \cite{Susskind:2014rva} or actions defined on a Wheeler-DeWitt patch \cite{Brown:2015bva}, follows a very specific profile as a function of time and, most importantly, satisfies the so-called \textit{switchback effect}, according to which an initially small perturbation (a shock wave sent from the boundary) disrupts the geometry exponentially fast, as we shall describe more precisely below. The observation that quantum complexity, understood as circuit complexity, follows a similar profile in time and is equally sensitive to initially small perturbations led to the holographic complexity proposal, which suffers from numerous ambiguities \cite{Belin:2021bga,Belin:2022xmt} both on the boundary and on the bulk side: On the boundary, extrinsic tolerance parameters and the choice of sets of unitary gates or cost functions in the case of circuit complexity make it unlikely to be an item of the holographic dictionary, since such choices are not features of the theory; on the other hand, from the bulk perspective, the authors of \cite{Belin:2021bga,Belin:2022xmt} show that it is possible to build families of observables defined in codimension-zero and one hypersurfaces whose growth as a function of time in a black hole background is qualitatively the same and controlled by the size of the wormhole throat.

Before moving on to analyzing why Krylov complexity is a good candidate to be the complexity item in the holographic dictionary, let us give some further details on the gravitational scenarios that one expects to capture via a complexity computation on the boundary. This will allow to identify the objects to focus on and the relevant time scales of the problem. In this discussion we shall use as a guiding example the results in \cite{Shenker:2013pqa} on the length of space-like geodesics in a BTZ black hole geometry in AdS$_3$ \cite{Banados:1992wn}.

\begin{figure}[t]
    \centering
    \includegraphics[width=10cm]{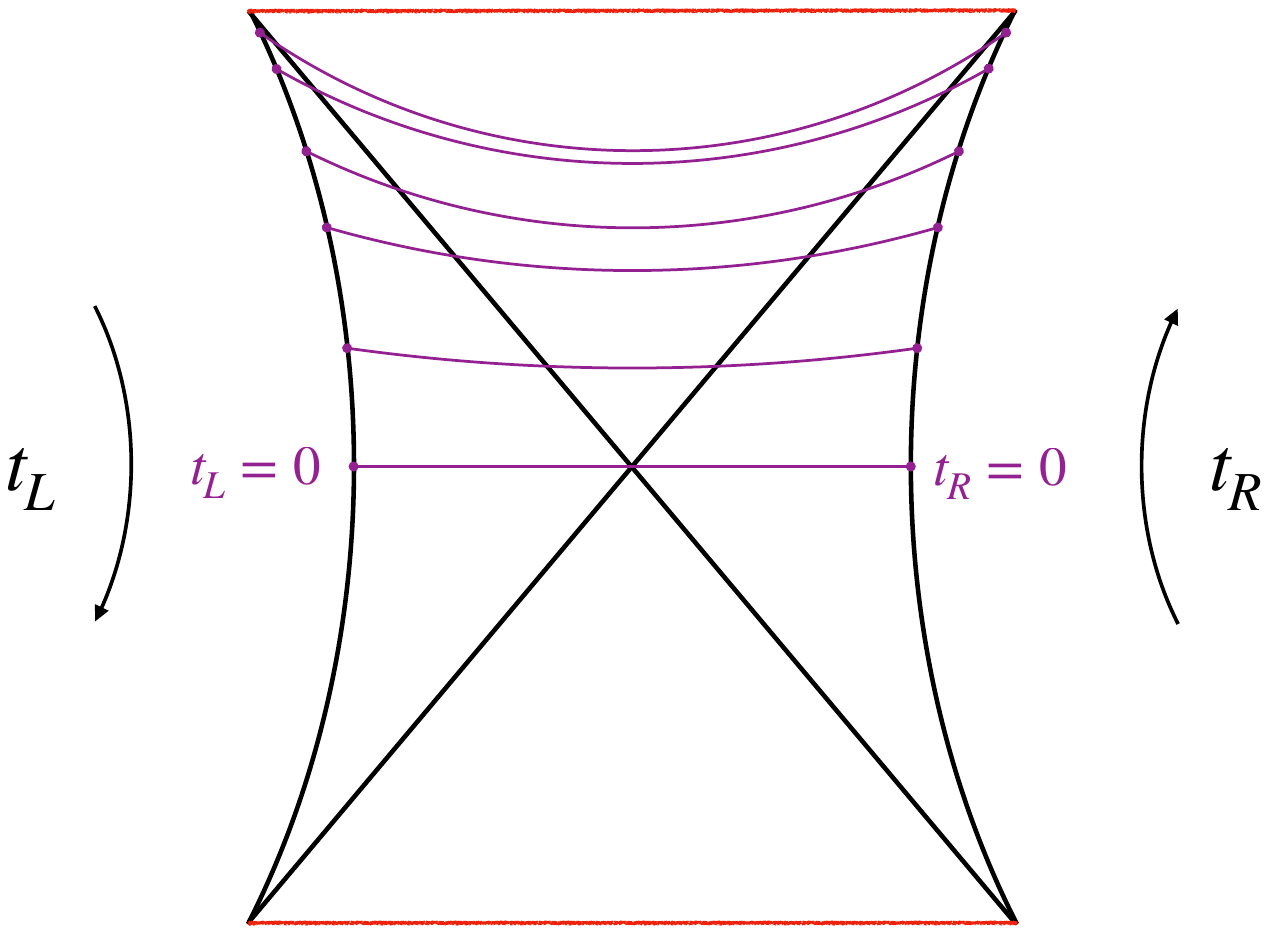}
    \caption{Schematic depiction of the universal extension of a two-sided black hole. Red lines represent the black hole and white hole singularities, and the black diagonal lines are the black hole and white hole horizons. Spacelike geodesics anchored at a regulated boundary at times $t_R=- t_L =t$ have been depicted in purple. The wormhole length is given by the length of the portion of each geodesic contained behind the horizon, and it can be measured by the difference between the geodesic length at time $t$ and the one at $t=0$, since the latter features no wormhole at all.}
    \label{fig:BH_geodesics_unperturbed}
\end{figure}

The most immediate gravitational construction where one can study the growth of the wormhole throat is that of an unperturbed two-sided black hole. Figure \ref{fig:BH_geodesics_unperturbed} depicts such a background, plus a family of space-like geodesics joining the left and right boundaries\footnote{In a specific computation, the anchoring points of the geodesics should be placed on a regulated boundary, i.e. at some finite distance in the radial direction, since otherwise the length would be divergent.} anchored at coincident times on the left and the right side. The difference between the length at a time $t$ and that at time $t=0$ is associated to the growth of the wormhole since, as one can see in the diagram, there is a space-time region building up behind the horizon for $t>0$ that cannot be accessed with the ordinary Schwarzschild coordinates. In \cite{Shenker:2013pqa} the result for the geodesic length (upon subtracting constant terms so that it vanishes at $t=0$) in the aforementioned BTZ background is:
\begin{equation}
    \label{Sect_KC_holog_length_BTZ_unperturbed}
    \frac{l(t)}{l_{AdS}} = 2\log\left\{ \cosh\left( \frac{R}{l_{AdS}^2} t \right) \right\}~,
\end{equation}
where $l_{AdS}$ and $R$ denote the AdS length and the black hole horizon radius, respectively. The horizon radius $R$ is related to the black hole's inverse temperature,
\begin{equation}
    \label{Sect_KC_holog_BTZ_R_beta}
    R = \frac{2\pi l_{AdS}^2}{\beta}~,
\end{equation}
and by inspection of \eqref{Sect_KC_holog_length_BTZ_unperturbed} we note that $l(t)$ features early quadratic growth before the thermal, or Rindler time \cite{Maldacena:2013xja}, 
\begin{equation}
    \label{Sect_KC_holog_Rindler_time}
    t_\beta:=\frac{\beta}{2\pi}~,
\end{equation}
after which it exhibits linear growth. If we want to analyze the growth of the ERB from the boundary perspective by computing the complexity of the microstate dual to the emergent bulk geometry, we need to decide which time-evolving state we should compute the complexity of. The state dual to the $t=0$ slice depicted in Figure \ref{fig:BH_geodesics_unperturbed} is, as argued in \cite{Maldacena:2001kr}, the \textit{thermofield double state}, which is a highly entangled state belonging to the tensor product of the left and right Hilbert spaces and is equal to the purification of the thermal density matrix:
\begin{equation}
    \label{Sect_KC_holog_TFD}
    |TFD\rangle := \frac{1}{\sqrt{Z(\beta)}} \sum_{n} e^{-\frac{\beta E_n}{2}} \left( \mathcal{T} |E_n\rangle \right) \otimes |E_n\rangle~,
\end{equation}
where $\left\{E_n\right\}$ denote the proper energies of the Hamiltonian of a single side, and $\mathcal{T}$ stands for the time-reversal operator\footnote{The path-integral construction of the TFD state requires of an anti-unitary operator in its definition, see for instance \cite{Maldacena:2001kr,Cottrell:2018ash}. It may be taken to be some more generic operator, such as the CPT operator, but here for simplicity we have chosen time reversal which, roughly speaking, enforces the fact that asymptotic boundary times run differently in the left and right boundaries, as depicted schematically in Figure \ref{fig:BH_geodesics_unperturbed}.}. 
Since the anchoring points of a given geodesic run vertically upwards on both boundaries, i.e. $t_R=-t_L=t$, the state whose complexity one should consider is the evolution of the TFD state generated by both the left and right Hamiltonian, with the same sign:
\begin{equation}
    \label{Sect_KC_holog_TFD_t}
    |\Psi(t)\rangle = e^{-it(H_L + H_R)}|TFD\rangle~,
\end{equation}
where $H_L\equiv \left(\mathcal{T}H\mathcal{T}^\dagger\right)\otimes \mathds{1}$ and $H_R\equiv \mathds{1}\otimes H$, the left and right spaces being identical copies of each other\footnote{By convention, all left operators are constructed as the insertion in the left Hilbert space of time-reversed versions of the ``original'' operators in the Hilbert space that has been doubled in this two-sided construction. That is, for any operator $\mathcal{A}\in \widehat{\mathcal{H}}$, we define $\mathcal{A}_L:= \left(\mathcal{T}\mathcal{A}\mathcal{T}^\dagger\right)\otimes \mathds{1}$. Importantly, the left time-evolution operator (which makes left time increase \textit{downwards} on the left side) satisfies $U_L(t) = \left(\mathcal{T}U(t)\mathcal{T}^\dagger\right)\otimes\mathds{1}=\left(\mathcal{T}e^{-i t H}\mathcal{T}^\dagger\right)\otimes\mathds{1} = e^{it H_L}$, which explains why the relative plus sign in equation \eqref{Sect_KC_holog_TFD_t} is the correct one for the protocol described in Figure \ref{fig:BH_geodesics_unperturbed}, where both left and right times are required to increase \textit{upwards}.}.

\begin{figure}
    \centering
    \includegraphics[width=10cm]{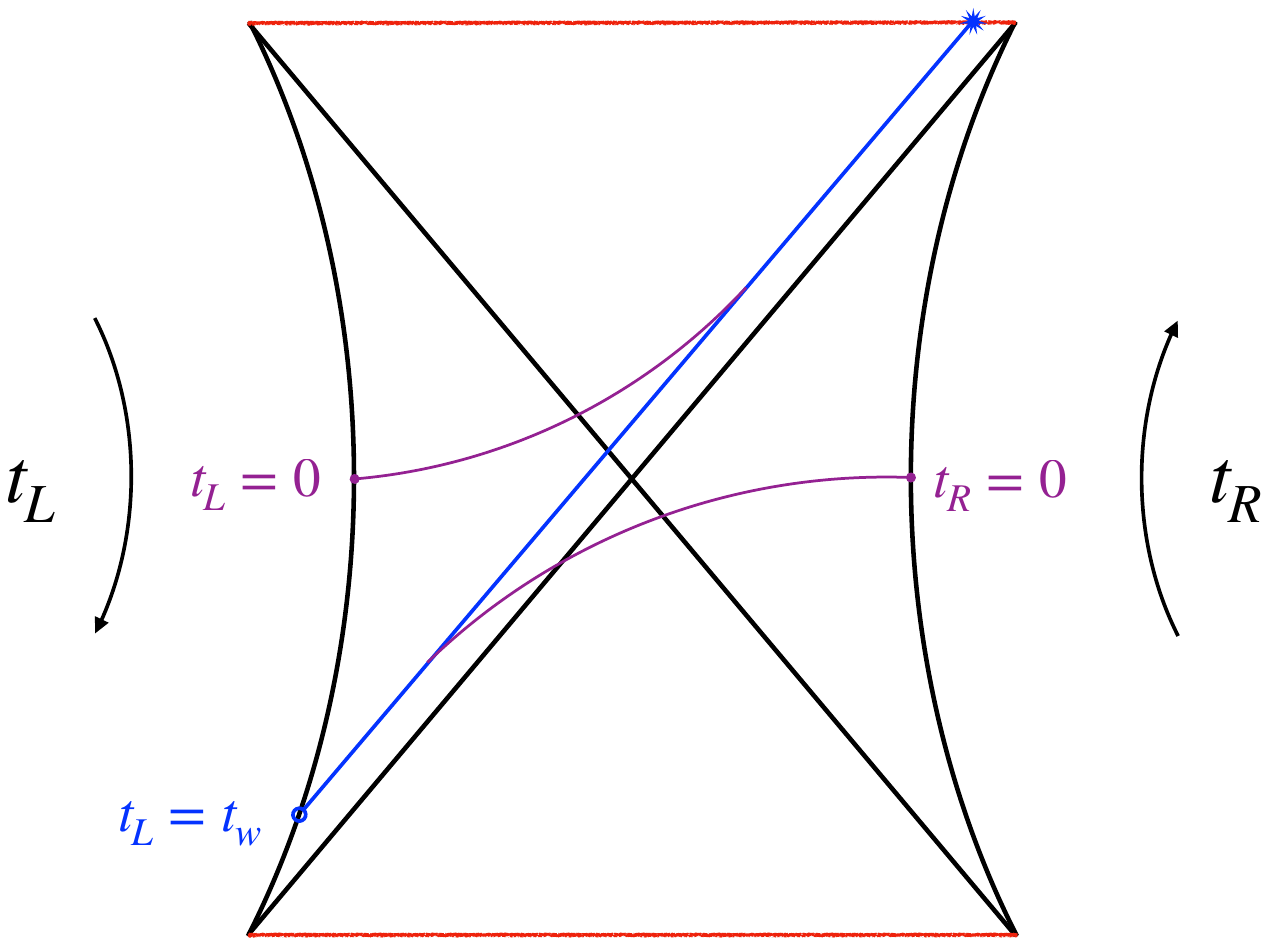}
    \caption{Schematic depiction of a two-sided eternal black hole perturbed by an almost null shock wave sourced at a point on the left (regulated) boundary at time $t_L=t_w>0$ in the past. Its effect \cite{Shenker:2013pqa,Vaidya:1943,Vaidya:1953,DRAY1985173} is to produce a shift in the corresponding null coordinate. The resulting geometry is not discontinuous \cite{DRAY1985173}; rather, the diagram manifests a coordinate singularity due to the gluing of different patches on the left and right of the shock wave, which were adapted to the unperturbed geometry. Roughly speaking, the effect of the shock wave is to ``push'' the singularity towards the future for the spacetime on the right side of the shock. The purple line represents a spacelike geodesic anchored on the left and right (regulated) boundaries at $t_L=t_R=0$. In the absence of a shock wave, the $t=0$ slice features no wormhole behind the horizon (see Figure \ref{fig:BH_geodesics_unperturbed}), while the purple geodesic in this case does have some non-vanishing portion behind it, as illustrated in the diagram. Thus, the difference between the length of the slice anchored at $t_L=t_R=0$ with a shock wave in the bulk, and that of the $t=0$ slice in the unperturbed geometry, is a coordinate-invariant measure of the growth of the wormhole length due exclusively to the shock wave, as a function of the time $t_w$ at which it was sourced. See \cite{Shenker:2013pqa} for an explicit computation in three-dimensional gravity, whose result is given in equation \eqref{Sect_KC_holog_length_BTZ_shock}.}
    \label{fig:BH_geodesic_shockwave}
\end{figure}

A more interesting gravitational setup, which makes contact with operator growth, is that of the switchback effect. Noting that in an unperturbed black hole background the length of the Einstein-Rosen bridge behind the horizon is zero at $t=0$, we consider a protocol in which the geometry of the $t=0$ slice is disrupted by a shock wave sent from the left boundary at some time in the past $t_L = t_w >0$ (we recall that the left time increases downwards, i.e. towards the past): Always computing the length of the spacelike geodesic anchored at $t_L=t_R=0$ on the boundaries as a function of the time $t_w$ at which the shock wave was generated measures the growth of the wormhole solely due to this perturbation. Figure \ref{fig:BH_geodesic_shockwave} depicts the protocol. In words, the manifestation of the switchback effect \cite{Susskind:2018pmk} is the fact that the perturbation is essentially harmless for small values of $t_w$, until it becomes disruptive, which is captured by an initial exponential behavior. Such a behavior is indeed found in the AdS$_3$ computation that we have been alluding to throughout this discussion \cite{Shenker:2013pqa}:
\begin{equation}
    \label{Sect_KC_holog_length_BTZ_shock}
    \frac{l(t_w)}{l_{AdS}} = 2 \log\left\{ 1 + \frac{E}{8M} e^{ \frac{R}{l_{AdS}^2} t_w }\right\}~,
\end{equation}
where $M$ is the black hole mass and $E$ is the shock wave energy. The computation in \cite{Shenker:2013pqa} considered a spherically symmetric in-falling shell of matter of energy $E$, whose trajectory is time-like: If $t_w$ is small enough, the in-falling matter barely disrupts the geometry on the slice $t=0$, but for $t_w$ sufficiently big, the blueshift of its proper energy when it reaches the $t=0$ slice (as measured in the frame of the slice) is such that the matter appears to follow an almost-null trajectory, for which the backreaction on the metric can be computed using the AdS-Vaidya solution \cite{Vaidya:1943,Vaidya:1953,DRAY1985173}. We now recall the relation between the horizon radius and the mass $M$ of the BTZ black hole:
\begin{equation}
    \label{Sect_KC_holog_Rbtz_M}
    R^2 = 8G_N M l_{AdS}^2~,
\end{equation}
where $G_N$ denotes Newton's 3-dimensional constant. We also note the expression for the Bekenstein entropy \cite{Bekenstein:1973ur} of the black hole,
\begin{equation}
    \label{Sect_KC_holog_BH_entropy}
    S = \frac{A}{4 G_N}~,
\end{equation}
where $A$ is the horizon area, in this case given by $A = 2 \pi R$. With this all, we can now re-express \eqref{Sect_KC_holog_length_BTZ_shock} changing the variables $(R,M)$ by the pair $(S,\beta)$, as these will be the parameters controlling an eventual boundary computation matching the bulk result. We have:
\begin{equation}
    \label{Sect_KC_holog_length_BTZ_shock_S_beta}
    \frac{l(t_w)}{l_{AdS}^2} = 2\log\left\{ 1 + \frac{E \beta}{4S} e^{\frac{2\pi}{\beta} t_w} \right\}~.
\end{equation}
This expression features two different regimes in time, the transition between them taking place around the \textit{scrambling time}:
\begin{equation}
    \label{Sect_KC_holog_scrambling_time}
    t_{scr} := \frac{\beta}{2\pi} \log \left\{ \frac{4S}{E\beta}\right\}~.
\end{equation}
The most important feature of the scrambling time is its logarithmic growth with the number of degrees of freedom, $S$; the denominator $E\beta$ is specific to the setup considered, making the onset time $t_{scr}$ larger if the energy of the in-falling quanta is small compared to the black hole temperature, and vice-versa. Before scrambling, we have:
\begin{equation}
    \label{Sect_KC_holog_switchback_before_scrambling}
    \frac{l(t_w)}{l_{AdS}} \approx \frac{ E \beta}{2S} e^{\frac{2\pi}{\beta} t_w}~,\qquad t_w<t_{scr}~,
\end{equation}
i.e. an exponential growth with an exponent that coincides with that of the MSS bound \eqref{Sect_theoremQ_OTOC_MSS_bound}. Well after scrambling, \eqref{Sect_KC_holog_length_BTZ_shock_S_beta} behaves asymptotically linearly in time:
\begin{equation}
    \label{Sect_KC_holog_switchback_after_scrambling}
    \frac{l(t_w)}{l_{AdS}} \sim \frac{4\pi}{\beta} t_w + 2\log \left( \frac{E \beta}{4S} \right)~,\qquad t\gg t_{scr}~.
\end{equation}
Because of the behavior of the exponential function, \eqref{Sect_KC_holog_switchback_before_scrambling} does not become relevant until times of order $t_{scr}$, justifying the common qualitative description of the switchback effect, namely that the perturbation sent from the past is not very disruptive if it was sent earlier than a scrambling time in the past, effectively only kicking in if it is sent from a time in the past well greater than $t_{scr}$.

The previous discussion highlights the profile in time and the relevant time scales that one should expect from a boundary computation describing the switchback effect. Let us now argue which is the state dual to the perturbed geometry whose complexity should be computed in this scenario. The shock wave in the bulk sent from the boundary can be seen \cite{Shenker:2013pqa,Susskind:2018pmk,Witten:1998qj,Harlow:2011ke} as sourced by an operator insertion $\mathcal{O}$ in the boundary theory at the corresponding boundary time; thus, in order to prepare the microstate dual to the backreacted $t=0$ slice depicted in Figure \ref{fig:BH_geodesic_shockwave}, we need to consider the TFD state, dual to the unperturbed $t=0$ slice, evolve it to the past $t_L=t_w$, insert the operator, and evolve forward back to $t_L=0$. That is:
\begin{equation}
\label{Sect_KC_holog_TFD_operator_insertion}
    |\Psi(t_w)\rangle = e^{-it_w H_L} \mathcal{O}_L e^{it_w H_L} |TFD\rangle = \mathcal{O}_L(t_w)|TFD\rangle~,
\end{equation}
where $\mathcal{O}_L= \left(\mathcal{T}\mathcal{O}\mathcal{T}^\dagger\right)\otimes\mathds{1}$ is the one-sided operator inserted in the left copy of the Hilbert space. We recall that $e^{it_w H_L}$ evolves the left side to the past time $t_L=t_w$, and therefore $\mathcal{O}_L(t_w)$ is correctly \textit{in the past} if $t_w>0$. Any evolution involving the right Hamiltonian $H_R$ would commute with $\mathcal{O}_L$ and be eventually cancelled out by the subsequent reversed evolution, so it has been directly omitted in \eqref{Sect_KC_holog_TFD_operator_insertion}. Furthermore, noting that $|TFD\rangle$ is annihilated by the Hamiltonian $H_L-H_R$, being therefore stationary under the time evolution it generates, we can see $|\Psi(t_w)\rangle$ as a state that evolves regularly in the Schrödinger picture with such a choice of total Hamiltonian:
\begin{equation}
    \label{Sect_KC_holog_Psi_insertion_Schrodinger}
    \begin{split}
        &e^{-it_w (H_L-H_R)}|\Psi(0)\rangle = e^{-it_w(H_L-H_R)} \mathcal{O}_L |TFD\rangle \\
        &=e^{-it_w(H_L-H_R)} \mathcal{O}_L e^{it_w(H_L-H_R)}|TFD\rangle = e^{-it_w H_L} \mathcal{O}_L e^{it_w H_L} |TFD\rangle = |\Psi(t_w)\rangle~,
    \end{split}
\end{equation}
where in the second equality we have inserted a time evolution operator at will between the operator and the TFD state, using the fact that the latter is stationary, and in the third equality we have used that $[\mathcal{O}_L,H_R]=0$. Equation \eqref{Sect_KC_holog_Psi_insertion_Schrodinger} allows to give yet another interpretation to the switchback effect: The state $|TFD\rangle$ was originally stationary under the evolution generated by $H_L-H_R$; perturbing it with a localized operator insertion and then letting it evolve yields a state that remains being approximately stationary until times of order of $t_{scr}$, after which the perturbation is seen to be disruptive in the emergent bulk geometry of the microstate.

Summarizing, the switchback effect may be probed by computing the (state) complexity of the TFD state with the insertion of the time-evolved operator $\mathcal{O}_L(t_w)$, and a good candidate of holographic complexity should be expected to exhibit the profile in time described above, namely exponential growth before the scrambling time, logarithmic in the entropy, followed by a period of linear growth. If the candidate complexity is Krylov complexity, the state complexity of \eqref{Sect_KC_holog_TFD_operator_insertion} makes direct connection with operator complexity, as we shall argue now. We showed in section \ref{sect:stateKC} that the Lanczos coefficients (and hence, K-complexity) are entirely determined from the survival amplitude of the state. Using the definition of the TFD state \eqref{Sect_KC_holog_TFD} and the fact that the operator $\mathcal{O}_L(t_w)$ only has support on the left copy of the Hilbert space, it is possible to show that the survival amplitude of \eqref{Sect_KC_holog_TFD_operator_insertion} is related to the operator two-point function via
\begin{equation}
    \label{Sect_KC_holog_survivalAmplitude_TwoPt}
    S(t_w) = \langle \Psi(0) | \Psi (t_w) \rangle = \frac{1}{Z(\beta)} \text{Tr}\left[ e^{-\beta H} \mathcal{O}^\dagger \mathcal{O}(-t_w) \right] = C_{\beta} (-t_w) = C^{*}_\beta(t_w)~,
\end{equation}
where we note that the trace is over a single copy of the Hilbert space. That is, the survival amplitude equals the (conventional) thermal two-point function of the operator, whose moments, in turn, determine uniquely the operator's K-complexity as detailed in section \ref{sect:KC_op}. In particular, the Taylor expansion of the conventional thermal operator two-point function is given by
\begin{equation}
    \label{Sect_KC_holog_TwoPtTaylor_with_odd_moments}
    C_\beta (t)= \sum_{n=0}^{+\infty}\frac{(it)^n}{n!}\mu_n~,\qquad \mu_n = \left(\mathcal{O} \left|\mathcal{L}^n\right|\mathcal{O}\right)_{\beta}~,
\end{equation}
potentially admitting non-zero odd moments (which in turn imply potentially non-vanishing $a_n$ Lanczos coefficients), while the Taylor series of the state survival amplitude is written in \eqref{Sect_KCdef_survival_amplitude},
where the moments should be understood in the present context as:
\begin{equation}
    \label{Sect_KC_holog_moments_perturbed_state}
    M_n = \Big\langle \Psi (0) \Big| {\big(H_L-H_R\big)}^n\Big| \Psi(0)\Big\rangle~.
\end{equation}
Expression \eqref{Sect_KC_holog_TwoPtTaylor_with_odd_moments} differs from \eqref{Sect_KCdef_survival_amplitude} in a sign due to the difference between the Heisenberg and Schrödinger representations. Such a sign is precisely compensated by the minus sign in the argument of $C_\beta(-t_w)$ appearing in \eqref{Sect_KC_holog_survivalAmplitude_TwoPt}, allowing to conclude that the state moments of $|\Psi(t_w)\rangle$, $M_n$, are exactly equal to the operator moments of $\mathcal{O}(t_w)$, $\mu_n$, therefore yielding coincident Lanczos coefficients after plugging such quantities as an input for the recursion method (cf. section \ref{sect:Moments_to_Lanczos}). Using these Lanczos coefficients to implement the recursion \eqref{Sect_KCdef_recursion_phin_states} for the Krylov space wave function associated to the state $|\Psi(t_w)\rangle$, one will find that it differs from the corresponding recursion of the operator wave function with the same Lanczos coefficients by a minus sign on the left-hand side, again side due to the fact that $|\Psi(t_w)\rangle$ and $\mathcal{O}(t_w)$ evolve in the Schrödinger and in the Heisenberg picture, respectively, implying that the state wave function equals the complex conjugate of the operator wave function. Finally, recalling \eqref{Sect_KCdef_phin_conjugate_minust} and \eqref{Sect_KCdef_KCeven}, one concludes that the Krylov complexities of both objects coincide.
Summarizing, the K-complexity of the perturbed state $|\Psi(t_w)\rangle$ given in \eqref{Sect_KC_holog_TFD_operator_insertion} \textit{is equal} to the operator K-complexity of $\mathcal{O}(t_w)$: The switchback effect is directly probed by operator K-complexity. Let us now present an alternative proof of this fact by constructing explicitly the Krylov basis for $|\Psi(t_w)\rangle$, $\left\{|K_n\rangle\right\}_{n=0}^{K-1}$ out of the one associated to $\mathcal{O}(t_w)$, $\left\{\mathcal{O}_n\right\}_{n=0}^{K-1}$. We propose the identification:
\begin{equation}
    \label{Sect_KC_Phi_operator_Krylov_basis}
    |K_n\rangle = \mathcal{O}_{n,L}|TFD\rangle = \frac{1}{\sqrt{Z(\beta)}}\sum_{a,b=0}^{D-1} e^{-\frac{\beta E_b}{2}}\left( \mathcal{O}_n \right)_{ab}^* \left( \mathcal{T}|E_a\rangle \right)\otimes |E_b\rangle~,
\end{equation}
where $\mathcal{O}_{n,L}=\left(\mathcal{T}\mathcal{O}_n\mathcal{T}^\dagger\right)\otimes \mathds{1}$ and $\left(\mathcal{O}_n\right)_{ab}$ are the matrix elements of the operator $\mathcal{O}$ in the eigenbasis of the one-sided Hamiltonian $H$. One can prove inductively that the elements $|K_n\rangle$ satisfy the Lanczos recursion for states \eqref{Sect_KCdef_Lanczos_recursion_states}, with the Hamiltonian $H_L-H_R$ and the standard inner product in the Hilbert space of states, given that the $\mathcal{O}_n$ satisfy the Lanczos recursion for operators, with the Liouvillian $\mathcal{L}\equiv[H,\cdot]$ defined on a single copy of the Hilbert space and given the conventional thermal inner product \eqref{Sect_KCdef_thermal_inner_prod_conventional}. Since the conventional thermal inner product is a representative of the family \eqref{Sect_KCdef_thermal_inner_product_g} for which the $g(\tau)$ function is not even in the thermal circle, the Lanczos recursion for operators \eqref{Sect_KCdef_KrylovElementsRelation} in this case needs to be supplemented with $a$-coefficients, which do not vanish in general for $\beta>0$. The only non-trivial step in the proof mapping the Lanczos recursion for $|K_n\rangle$ to that of $\mathcal{O}_n$ is turning the action of the total Hamiltonian on $|K_n\rangle$ into the action of the Liouvillian (i.e. the commutator) on $\mathcal{O}_n$, which can be done, again, using the fact that the TFD state is annihilated by $H_L-H_R$:
\begin{equation}
    \label{Sect_KC_holog_Hamilt_to_commut}
     \begin{split}
         & (H_L-H_R)|K_n\rangle = (H_L-H_R) \mathcal{O}_{n,L}|TFD\rangle \\ 
         & = (H_L-H_R) \mathcal{O}_{n,L}|TFD\rangle - \mathcal{O}_{n,L} (H_L-H_R) |TFD\rangle \\
         & = [H_L,\mathcal{O}_{n,L}]|TFD\rangle- [H_R,\mathcal{O}_{n,L}] |TFD\rangle = [H_L,\mathcal{O}_{n,L}]|TFD\rangle~.
     \end{split}
\end{equation}
With this, we have that, given the $a_n$ and $b_n$ coefficients of the operator $\mathcal{O}$,
\begin{equation}
    \label{Sect_KC_holog_Kn_LanczosRec_proof}
    \begin{split}
        &(H_L-H_R - a_n)|K_n\rangle - b_n | K_{n-1}\rangle \\
        &= \Big\{ [H_L,\mathcal{O}_{n,L}] - a_n \mathcal{O}_{n,L} -b_n\mathcal{O}_{n-1,L} \Big\}|TFD\rangle \\
        &= b_{n+1} \mathcal{O}_{n+1,L}|TFD\rangle = b_{n+1}|K_{n+1}\rangle~,
    \end{split}
\end{equation}
proving that the states \eqref{Sect_KC_Phi_operator_Krylov_basis} do satisfy the Lanczos recursion with the Hamiltonian $H_L-H_R$, thus correctly being the Krylov elements of $|\Psi(t_w)\rangle$. Furthermore, the states $|K_n\rangle$ turn out to form correctly an orthonormal set provided that the elements $\mathcal{O}_n$ form such a set according to the conventional thermal inner product \eqref{Sect_KCdef_thermal_inner_prod_conventional}, thanks to the structure of the TFD state, reflected in the square root of the Boltzmann factor entering in \eqref{Sect_KC_Phi_operator_Krylov_basis}. We may even express the Krylov space wave function of $|\Psi(t_w)\rangle$,
\begin{equation}
    \label{Sect_KC_holog_wavefn_Psi}
    \phi_n^{(\Psi)}(t_w):= \langle K_n|\Psi(t_w)\rangle~, 
\end{equation}
in terms of that of $\mathcal{O}(t_w)$,
\begin{equation}
    \label{Sect_KC_holog_wavefn_Op}
    \phi_n^{(\mathcal{O})}(t_w):=\big(\mathcal{O}_n \big| \mathcal{O}(t_w)\big)_{\beta} = \frac{1}{Z(\beta)}\text{Tr}\left[ e^{-\beta H} \mathcal{O}_n^\dagger \mathcal{O}(t_w) \right]~,
\end{equation}
where we stress once again that the trace runs over a single copy of the Hilbert space. It is worth to note that, in this case, since the operator's $a$-coefficients do not necessarily vanish, its Krylov elements need not have alternating hermiticity properties and a definition of the wave function like the one in \eqref{Sect_KCdef_operator_wavefn_decomp}, where a factor of $i^n$ is pulled out, will not be useful as it will generically not make it become a real-valued function. We consider the expansion of the time-evolved operator in terms of its Krylov space wave function,
\begin{equation}
    \label{Sect_KC_holog_operator_decomp_wavefn}
    \mathcal{O}(t_w) = \sum_{n=0}^{K-1}\phi_n^{(\mathcal{O})}(t_w)~\mathcal{O}_n~.
\end{equation}
Plugging \eqref{Sect_KC_holog_operator_decomp_wavefn} into \eqref{Sect_KC_holog_TFD_operator_insertion} and recalling \eqref{Sect_KC_Phi_operator_Krylov_basis} we obtain:
\begin{equation}
    \label{Sect_KC_holog_Psi_operator_decomp_with_op_wavefn}
    |\Psi(t_w)\rangle= \sum_{n=0}^{K-1}\left[ \phi_n^{(\mathcal{O})}(t_w) \right]^{*}|K_n\rangle~,
\end{equation}
where the complex conjugation arises as a consequence of the anti-linearity of the time-reversal operator.
Finally, combining \eqref{Sect_KC_holog_Psi_operator_decomp_with_op_wavefn} with the definition \eqref{Sect_KC_holog_wavefn_Psi} yields 
\begin{equation}
    \label{Sect_KC_holog_Operator_vs_State_wavefn}
    \phi_n^{(\Psi)}(t_w) = \left[ \phi_n^{(\mathcal{O})}(t_w) \right]^{*} = \phi_n^{(\mathcal{O})}(-t_w)
\end{equation}
and, in agreement with the discussion around \eqref{Sect_KCdef_phin_conjugate_minust} and \eqref{Sect_KCdef_KCeven}, this establishes explicitly that the operator's K-complexity, $C_K^{(\mathcal{O})}(t_w)$, and the state's one, $C_K^{(\Psi)}(t_w)$, are identical:
\begin{equation}
    \label{Sect_KC_holog_complexities_state_op_switchback}
    C_K^{(\Psi)}(t_w) = \sum_{n=0}^{K-1} n ~ {\lvert\phi_n^{(\Psi)}(t_w)\rvert}^2 = \sum_{n=0}^{K-1} n ~ {\lvert\phi_n^{(\mathcal{O})}(t_w)\rvert}^2 = C_K^{(\mathcal{O})}(t_w)~.
\end{equation}
It is worth to stress that the Krylov space dimensions are equal. One is the dimension of a Krylov subspace of the operator space $\widehat{\mathcal{H}}$ of a single copy of the Hilbert space, whose dimension is $\dim(\widehat{\mathcal{H}})=D^2$, while the other is the dimension of a Krylov subspace contained in the tensor product of two copies of the Hilbert space of states, to which the state $|\Psi(t_w)\rangle$ belongs, and whose dimension is also $\dim\left( \mathcal{H}\otimes \mathcal{H} \right) = D^2$. Hence, the equality of the Krylov space dimensions is not inconsistent. Furthermore, since the Hamiltonian generating the evolution of $|\Psi(t_w)\rangle$ is $H_L-H_R$, its Krylov dimension is subject to the same upper bound as the operator's Krylov dimension, written in \eqref{Sect_KCdef_Kbound_operators}. In order to understand this, one needs to refer to the details of the proof of such a bound, which appeared in contribution \cite{I} and is contained in Chapter \ref{ch:chapter03_SYK}, but the main reason is the fact that the zero eigenvalue is $D$-times degenerate in the spectra of both the two-sided Hamiltonian $H_L-H_R$ and the single-sided Liouvillian $\mathcal{L}\equiv [H,\cdot]$, regardless of the specific features of the physical system at hand.

We have established rigorously that the Krylov complexity of a time-evolving operator in the Heisenberg picture gives the complexity of the perturbed state \eqref{Sect_KC_holog_TFD_operator_insertion} discussed in the context of the switchback effect, a proof that had not appeared in the literature so far. In order to argue whether K-complexity is suitable for holographic applications, it is left to discuss whether the profile of such an operator complexity as a function of time can be expected to display the profile featured by the gravitational result \eqref{Sect_KC_holog_length_BTZ_shock_S_beta} described previously, matching the relevant time scales. This was the goal of \cite{Barbon:2019wsy}, the first article that pointed out the potential relevance of Krylov complexity in gravity, whose main results we shall discuss now.

The aforementioned article \cite{Barbon:2019wsy} is conceived as a follow-up to the one in which K-complexity was introduced \cite{Parker:2018yvk}. The latter, where the universal operator growth hypothesis and the Q-theorem were presented, focuses on systems in the thermodynamic limit, where the number of degrees of freedom $S$ is infinite. For such infinite and chaotic systems, it was found that K-complexity grows exponentially in time, which so far is not in contradiction with gravitational expectations: In the large-$S$ limit, the scrambling time gets pushed towards infinity and only early-time dynamics, expected to be exponential in gravity as illustrated in \eqref{Sect_KC_holog_switchback_before_scrambling}, are accessible. In order to assess whether operator Krylov complexity also features a linear growth phase in chaotic systems, it is necessary to consider finite-size systems and to study their post-scrambling dynamics. For this purpose, the starting point of \cite{Barbon:2019wsy} is to consider a typical operator in a chaotic system, expected to satisfy the \textit{eigenstate thermalization hypothesis (ETH)} \cite{PhysRevA.30.504,PhysRevA.43.2046,Srednicki_1999,Srednicki:1994mfb,DAlessio:2015qtq}, which is the following Ansatz for the operator matrix elements in the energy basis:
\begin{equation}
    \label{Sect_KC_holog_ETH}
    \langle E_m | \mathcal{O} | E_n\rangle = O (\varepsilon) \delta_{mn} + e^{-\frac{S(\varepsilon)}{2}} f(\varepsilon,\omega) r_{mn}~,
\end{equation}
where for notational simplicity we have written $\varepsilon\equiv\frac{E_m+E_n}{2}$ and $\omega\equiv E_m-E_n$. $S(\varepsilon)$ denotes the microcanonical entropy, whose exponential is proportional to the density of states in a given energy shell. In this hypothesis, $O(\varepsilon)$ and $f(\varepsilon,\omega)$ are smooth functions of their arguments and $r_{mn}$ are independent (modulo hermiticity constraints) and identically distributed complex Gaussian random variables with zero mean and unit variance.

The ETH ensures \cite{Srednicki_1999,DAlessio:2015qtq} that, at sufficiently late times, the expectation value of the observable measured in a pure microstate within an energy shell is indistinguishable, up to terms suppressed by powers of $e^{-S(\varepsilon)}$, from the microcanonical expectation value in the same shell\footnote{The hypothesis itself is not specific about the time scale after which both expectation values become similar, as this depends on the particular dynamical information of the system.}. It is therefore a mechanism for thermalization of typical operators in chaotic systems. In random matrix theory \cite{DAlessio:2015qtq,stöckmann_1999} it is possible to derive an Ansatz for operator matrix elements similar to \eqref{Sect_KC_holog_ETH} but without energy dependence in the functions $S$, $O$ and $f$; intuitively, this behavior of the matrix elements can be understood as a manifestation of the lack of correlation between the eigenbases of the Hamiltonian and of the operator. One can therefore understand the ETH Ansatz as a generalization of the RMT operator Ansatz which amounts to assuming that, within each energy shell, the system effectively behaves like a random matrix theory and, in this sense, this notion of thermalization is connected to spectral chaos. The non-trivial content of the hypothesis is the Gaussian distribution for the $r_{mn}$ variables\footnote{See \cite{Rigol_XXZ} for an analysis of an integrable, non-thermalizing system in which some operators appear to satisfy an ETH-like Ansatz, with the caveat that the distribution of the off-diagonal matrix elements is not Gaussian.}, together with the suppression of the off-diagonal matrix elements and the fact that the functions $O(\varepsilon)$ and $f(\varepsilon,\omega)$ are continuous and, roughly speaking, smooth\footnote{When the ETH Ansatz applies, $O(\varepsilon)$ and $f(\varepsilon,\omega)$ give, respectively, the microcanonical one- and two-point functions of the operator. For non-Gaussian corrections to the hypothesis that give access to higher n-point functions, see the recent work \cite{Jafferis:2022uhu}.}.

The strategy of \cite{Barbon:2019wsy} is to compute two different regimes of the moments $\mu_{2n}$ for ETH operators and use them to identify the corresponding regimes in the Lanczos coefficients, by the means of the following inequality:
\begin{equation}
    \label{Sect_KC_holog_ineq_Lanczos_moments_Dyck}
    \prod_{m=1}^n b_m^2 \leq \mu_{2n} \leq C_n \prod_{m=1}^n b_m^2~,
\end{equation}
which is derived through the connection between moments and Lanczos coefficients involving Dyck paths, reviewed in section \ref{subsect:From_Lanczos_to_Moments}. The lower bound is one of the many Dyck paths contributing to $\mu_{2n}$, as each contribution is always positive; in particular, it is the path with $n$ consecutive up steps followed by $n$ consecutive descending steps. The upper bound assumes a non-decreasing Lanczos sequence, such that the contribution of all individual paths is smaller or equal than that of the aforementioned path; we recall that the Catalan number $C_n$, given in \eqref{Sect_Lanczos_Catalan_numbers}, is the total number of Dyck paths of length $2n$. If the assumption of a non-decreasing sequence cannot be made, a different upper bound, used in \cite{Parker:2018yvk}, still applies:
\begin{equation}
    \label{Sect_KC_holog_upper_bound_Dyck_Altman}
    \mu_{2n}\leq C_n \big( \max \left\{ b_m \right\}_{m=1}^{n} \big)^{2n}~.
\end{equation}

In the thermodynamic limit, given the exponentially decaying spectral function \eqref{Sect_KC_chaos_equivalence_behaviours} expected in a maximally chaotic system, 
\begin{equation}
    \label{Sect_KC_holog_spectral_funct_chaotic}
    \Phi(\omega) = e^{-\frac{\pi |\omega|}{2\alpha}}~, 
\end{equation}
the moments $\mu_{2n}$ can be obtained via \eqref{Sect_KCdef_moments_from_spectral}, which yields
\begin{equation}
    \label{Sect_KC_holog_moments_largeN}
    \mu_{2n} = {\left(\frac{2\alpha}{\pi}\right)}^{2n} (2n)!~.
\end{equation}
Assuming an asymptotically linear growth of the Lanczos coefficients, $b_n\sim \alpha n$, and plugging it into \eqref{Sect_KC_holog_ineq_Lanczos_moments_Dyck}, one finds that both the upper and the lower bound display the same asymptotic behavior at large-$n$, equal to that of \eqref{Sect_KC_holog_moments_largeN}, so the Ansatz is consistent. This is the original argument with which \cite{Parker:2018yvk} arrived at the universal operator growth hypothesis. In order to improve on this result for finite-size systems, the authors of \cite{Barbon:2019wsy} note that finite systems must necessarily have a spectral function of bounded support, since the spectrum of the system has a finite band width, which we shall assume to be generically extensive in system size, $E_{\text{max}}-E_{\text{min}}\sim \Lambda S$, where $\Lambda$ is some intensive constant with energy dimensions. This, in turn, implies \cite{ViswanathMuller} that the Lanczos coefficients asymptote to a constant $b_{\infty}$. There should therefore exist some $S$-dependent transition value $n_{*}$ at which the $b_n$ sequence transitions from initial linear growth to a plateau regime. In order order to identify $n_{*}$, we begin by noting that the integral \eqref{Sect_KCdef_moments_from_spectral} over the real line should be replaced by one over the interval $[-\Lambda S, \Lambda S]$, which is the domain of the energy differences in the spectrum:
\begin{equation}
    \label{Sect_KC_holog_moments_from_spectral_finite_support}
    \mu_{2n} = \frac{1}{2\pi} \int_{-\Lambda S}^{\Lambda S} d\omega~\omega^{2n}~\Phi(\omega)~.
\end{equation}
Assuming the behavior \eqref{Sect_KC_holog_spectral_funct_chaotic} for the spectral function, a saddle-point approximation shows that the main contribution to the integral in \eqref{Sect_KC_holog_moments_from_spectral_finite_support} comes from $\omega\sim \frac{4n\alpha}{\pi}$. Given that $\alpha$ has dimensions of energy, just like the Lanczos coefficients, and that it is intensive (since it remains finite in the thermodynamic limit), we shall pose, for the sake of this approximate argument, that $\alpha\sim \Lambda$. Hence, when $n\gg S$ we have that the saddle point sits outside of the integration domain, and there is no obvious contour deformation that would justify approximating the integral \eqref{Sect_KC_holog_moments_from_spectral_finite_support} by evaluating its integrand on the saddle point. Therefore, we have identified the transition value $n_{*}\sim S$, such that for $n>n_{*}$ one needs to compute the moments $\mu_{2n}$ out of their discrete spectral resolution. Given the infinite-temperature inner product \eqref{Sect_KCdef_inner_product_infinite_temp}, which is assumed throughout in this argument, the exact expression for the moments \eqref{Sect_KCdef_operator_moments} is:
\begin{equation}
    \label{Sect_KC_holog_moments_discrete_expression}
    \mu_{2n} = \frac{1}{D} \sum_{a,b=0}^{D-1} \omega_{ab}^{2n} |O_{ab}|^2~,
\end{equation}
where $\omega_{ab}\equiv E_a-E_b$ and $O_{ab}\equiv \langle E_a|\mathcal{O}|E_b\rangle$. For $n>0$, which is implied by $n>n_{*}$, only off-diagonal matrix elements contribute, which can be taken from the ETH estimate \eqref{Sect_KC_holog_ETH}:
\begin{equation}
    \label{Sect_KC_holog_moments_eth}
    \mu_{2n}\sim \frac{1}{D^2}\sum_{a,b} \omega_{ab}^{2n} f^2(\varepsilon,\omega)~,
\end{equation}
where the entropic suppression $e^{-S(\varepsilon)}$ has been loosely factored out of the sum as $D^{-1}$ and the random variables $|r_{mn}|^2$ inside the sum have been replaced by their average, equal to one. Accepting that the main contributions to the sum \eqref{Sect_KC_holog_moments_eth} come from the largest energy differences, and recalling that $f(\varepsilon,\omega)$ is a smoothly varying function in the ETH Ansatz, one may approximate \cite{Barbon:2019wsy} the result by the average of the power of energy differences $\omega^{2n}$ across the band $[-\Lambda S, \Lambda S]$, ignoring the details of $f(\varepsilon,\omega)$. This yields a new asymptotic behavior for the moments,
\begin{equation}
    \label{Sect_KC_holog_moments_for_plateau}
    \mu_{2n} \sim \frac{{(\Lambda S)}^{2n}}{2n}~, 
\end{equation}
which, when plugged into \eqref{Sect_KC_holog_ineq_Lanczos_moments_Dyck}, is compatible \cite{Barbon:2019wsy} with an asymptotic plateau in the Lanczos coefficients, $b_n\sim  \Lambda S \equiv b_{\infty}$. All in all, \cite{Barbon:2019wsy} predicts the following profile for the Lanczos sequence in systems with finite (but large) entropy $S$:
\begin{equation}
    \label{Sect_KC_holog_bn_profile_barbon}
    b_n \sim \left\{ \begin{aligned}
        \alpha n~,\qquad n \lesssim S~, \\
        b_{\infty}~,\qquad n \gg S~.
    \end{aligned} \right.
\end{equation}
We note that, in our order-of-magnitude estimate, $b_{\infty}\sim \Lambda S$ but also $\alpha\sim \Lambda$, and therefore one may pose $b_{\infty} \sim \alpha S$ \cite{Barbon:2019wsy}.

The profile \eqref{Sect_KC_holog_bn_profile_barbon} has interesting implications for Krylov complexity. A continuous approximation to the dynamics of the operator wave function in Krylov space, proposed in \cite{Barbon:2019wsy} and discussed in detail in Appendix \ref{Appx:Cont_approx}, results in the ballistic propagation of a point particle through a velocity field $v(n)\equiv 2 b_n$. Since the particle starts at position $n=0$, it begins probing the regime of linearly increasing Lanczos coefficients, which results in an exponential increase of its position as a function of time, captured by K-complexity: $C_K(t)\approx n(t) \sim e^{\lambda_K t}$, where we have defined $\lambda_K := 2\alpha$ for convenience. Note that this retrieves the result for infinite systems discussed in section \ref{sect:KC_chaos}. The particle will reach the region of constant Lanczos coefficients, $n>n_{*}\sim S$, at a time $t_{*}$ such that $C_K(t_{*})\sim n_{*}$, which satisfactorily gives a scrambling time scale:
\begin{equation}
    \label{Sect_KC_holog_scrambling_K}
    t_{*} \sim \frac{1}{\lambda_K} \log S~.
\end{equation}
For times larger than $t_{*}$ the particle propagates through a region of constant velocity field, $v(n)=2b_\infty \sim 2\alpha S = \lambda_K S$, implying a linear increase in K-complexity, $C_K(t)\sim \lambda_KSt$ at late times. 

Summarizing, the expected K-complexity profile for typical operators satisfying the ETH Ansatz in maximally chaotic systems is predicted by \cite{Barbon:2019wsy} to be:
\begin{equation}
    \label{Sect_KC_holog_KCprofile_Barbon}
    C_K(t)\sim \left\{ \begin{aligned}
        e^{\lambda_K t}~,\qquad t \lesssim t_{*}~, \\
        \lambda_KSt~,\qquad t \gg t_{*}~,
    \end{aligned} \right.
\end{equation}
where $\lambda_K=2\alpha$, the K-complexity exponent that bounds the Lyapunov exponent in \eqref{Sect_KC_chaos_bound_Lyapunov}, and $t_{*}=\frac{1}{\lambda_K}\log S$ is a scrambling time scale. In view of this result, the fact that operator Krylov complexity exhibits a transition from exponential to linear growth at the scrambling time led the authors of \cite{Barbon:2019wsy} to propose that it is indeed a good candidate for holographic complexity, which adds up to the considerations discussed earlier in this text on its potential suitability to describe the switchback effect.

The results on post-scrambling K-complexity dynamics that have just been reviewed are, nevertheless, incomplete to some extent: They predict an asymptotic behavior for the Lanczos coefficients at large $n$, given by a plateau, and a Krylov complexity profile that grows linearly for an indefinite amount of time after scrambling. However, for a finite system, where $S<\infty$, the Hilbert space must be finite-dimensional, just like the Krylov dimension. Therefore, the $b_n$ sequence should terminate at some point, when $n=K\sim D^2\sim e^{2S}$ if the Krylov dimension bound \eqref{Sect_KCdef_Kbound_operators} is saturated, and consequently Krylov complexity should not grow linearly indefinitely after the scrambling time, but should eventually saturate. By definition of $t_{*}$, $C_K(t_{*})\sim S$ and, since the ulterior growth is linear, K-complexity will reach the maximal value $C_K(t_K)\sim e^{2S}$ at a time scale of order
\begin{equation}
    \label{Sect_KC_holog_Heisenberg_time}
    t_K\sim \frac{1}{\lambda_K S}e^{2S}~,
\end{equation}
an exponentially large time scale known as the Heisenberg time \cite{Barbon:2003aq,Barbon:2014rma}. For times greater than the Heisenberg time, K-complexity should saturate, but the mechanism for such a saturation is not evident from the argument that has been presented, requiring further analysis. From the gravity perspective, one should also expect an eventual saturation of the expectation value of the ERB length: Considering the Bekenstein-Hawking entropy $S_{\text{BH}} = \frac{A}{4G_N}$, the sector of the gravity Hilbert space relevant to the black hole microstates has a finite dimension. During its time evolution, the microstate describing the bulk geometry explores successive orthogonal states in the Hilbert space, whose emergent geometry features increasingly larger wormhole lengths \cite{Maldacena:2013xja}, until eventually all linearly independent states have been explored by the Heisenberg time scale\footnote{The argument for claiming that all independent gravity microstates should have been explored by the Heisenberg time relies on the assumption that orthogonal states are probed at a constant rate \cite{Maldacena:2013xja}.}, after which the microstate explores nothing but linear recombinations of states previously visited, and hence the length expectation value saturates rather than growing further. This is the regime in which the geometrical description of gravity ceases to apply, and therefore such a behavior cannot be retrieved from a classical computation. Eventually, K-complexity should undergo a Poincaré recurrence at times of order $t_P\sim \exp \left(e^S\right)$, going back arbitrarily close to its initial values \cite{Barbon:2003aq}.

Probing the saturation mechanism of K-complexity in finite systems, together with the analysis of the imprints of quantum chaos or integrability in finite-size effects of Krylov space dynamics, was the main motivation behind the contributions \cite{I,II,III}, which will be presented in Chapters \ref{ch:chapter03_SYK} and \ref{ch:chapter04_Integrable}.

Finally, it is worth to close this section noting that K-complexity is also a useful candidate for the holographic protocol depicted in Figure \ref{fig:BH_geodesics_unperturbed}, i.e. the study of the growth of the wormhole length in the unperturbed geometry. As argued around \eqref{Sect_KC_holog_TFD_t}, the relevant state to be analyzed in this case is the time evolution of the TFD state generated by the two-sided Hamiltonian under which it is not stationary, $H_R+H_L$. In \cite{Balasubramanian:2022tpr} it was shown that the survival amplitude associated to such a state is given by the analytic continuation of the thermal partition function,
\begin{equation}
    \label{Sect_KC_holog_survival_tfd}
    S(t) = \langle TFD | e^{-it(H_R+H_L)}|TFD\rangle = \frac{Z(\beta + 2it)}{Z(\beta)}~,
\end{equation}
the associated probability being:
\begin{equation}
    \label{Sect_KC_holog_survivalProb_tfd}
    {|S(t)|}^2 = \frac{Z(\beta+2it)Z(\beta-2it)}{Z(\beta)^2}~,
\end{equation}
which is nothing but the \textit{spectral form factor (SFF)} \cite{Altland:2020ccq,HaakeBook}, a well studied object in the context of quantum chaos. For chaotic systems, the SFF is known to decay towards a plateau, whose value is suppressed by the Hilbert space dimension (i.e. exponentially suppressed in the number of degrees of freedom) at times exponentially large in $S$, preceded or not by a linear ramp depending on the universality class of the system. This structure of the SFF, together with the constraint of unitarity, allows the authors of \cite{Balasubramanian:2022tpr} to associate qualitatively some features of K-complexity to the SFF, as follows: Since the modulus squared of the zeroth component of the Krylov wave function entering in the computation of $C_K(t)$ in \eqref{Sect_KCdef_KC_state_def} coincides precisely with the survival probability \eqref{Sect_KC_holog_survivalProb_tfd}, $|\phi_0(t)|^2 = |S(t)|^2$, unitarity of time evolution imposes, through \eqref{Sect_KCdef_wavefn_state_normalized}, that the probabilities $|\phi_n(t)|^2$ with $n>0$ will not be allowed to grow significantly before the SFF has effectively decayed towards its plateau value at exponentially late times or, rather, that they gradually build up as the SFF progressively decays. With this, one could expect a progressive growth of state K-complexity for the state $|\Psi(t)\rangle$ in \ref{Sect_KC_holog_TFD_t} up to time scales exponentially large in $S$, after which it should saturate. This proposal is backed up by some numerical checks in random matrix theory \cite{Balasubramanian:2022tpr}. However, the absence of universal results about the \textit{early}-time features of the spectral form factor do not allow, from this perspective, to evaluate whether the state complexity profile in this case will be qualitatively similar to the gravity computation \eqref{Sect_KC_holog_length_BTZ_unperturbed}.

The last contribution of this Thesis \cite{IV}, gathered in Chapter \ref{ch:chapter05_DSSYK}, considers the state complexity of the TFD state in the double-scaled SYK model \cite{Berkooz:2018jqr,Lin:2022rbf}, a system that has been shown to be dual to two-dimensional JT gravity \cite{Jackiw:1984je,Teitelboim:1983ux}. The article presents the first explicit proof of the correspondence between Krylov complexity in the boundary theory and length expectation value in the bulk theory, by using the map between the Hilbert spaces of both dual theories to establish a correspondence between the K-complexity eigenstates and the bulk length operator eigenstates. 

\section{Research on Krylov complexity}\label{sect:reseachKC}

This section collects the main lines of research currently open in the area of Krylov complexity, together with relevant references. Inevitably, the compilation will be biased towards topics of interest in the high-energy theory community\footnote{This section contains references of works that appeared before January 1st, 2024, the date on which this Thesis was submitted for evaluation to the University of Geneva.}.

\begin{itemize}
    \item K-complexity as a probe of chaos and integrability: \cite{Parker:2018yvk,Barbon:2019wsy,Dymarsky:2021bjq,Avdoshkin:2022xuw,He:2022ryk,Du:2022ocp,Espanol:2022cqr,Bhattacharjee:2023dik,Erdmenger:2023wjg,Zhang:2023wtr,Hashimoto:2023swv,Iizuka:2023pov,Iizuka:2023fba,Bhattacharyya:2023dhp,Camargo:2023eev,Balasubramanian:2023kwd,Noh_2021,Scialchi:2023bmw}, as well as \cite{I,II,III}.
    \item Localization on the Krylov chain: \cite{II,III}, as well as \cite{Trigueros:2021rwj}.
    \item Exact (numerical) results for the Lanczos algorithm: \cite{Jian:2020qpp,Trigueros:2021rwj,Bhattacharyya:2023zda,Bhattacharjee:2022qjw,Bhattacharya:2022gbz,Liu:2022god,Kim:2021okd,He:2022ryk,Du:2022ocp,Bhattacharjee:2022lzy,Bhattacharjee:2023dik,Bhattacharya:2023zqt,Nizami:2023dkf,Hashimoto:2023swv,Camargo:2023eev} and \cite{I,II,III}.
    \item Krylov complexity in QFT and CFT: \cite{Dymarsky:2021bjq,Caputa:2021ori,Avdoshkin:2022xuw,Camargo:2022rnt,Vasli:2023syq,Kundu:2023hbk}.
    \item Analytical tools to study K-complexity, e.g. orthogonal polynomials \cite{Muck:2022xfc}, integrable structures \cite{Dymarsky:2019elm}, renormalization-group techniques \cite{Kar:2021nbm} and geometrical interpretation \cite{Caputa:2021sib,Lv:2023jbv,Aguilar-Gutierrez:2023nyk,Craps:2023ivc}.
    \item Holography and bulk dual of K-complexity: \cite{IV,Jian:2020qpp,Bhattacharjee:2022ave,Kar:2021nbm,Avdoshkin:2022xuw}.
    \item Quantum speed limits: \cite{Fan:2022mdw,Fan:2022xaa,Carabba:2022itd,Hornedal:2023xpa,Takahashi:2023nkt,Fan:2023ohh,Hornedal:2022pkc}.
    \item State Krylov complexity: \cite{Balasubramanian:2022tpr,Balasubramanian:2022dnj,Caputa:2023vyr,Bhattacharyya:2023zda,Bhattacharjee:2022qjw,Alishahiha:2022anw,Chattopadhyay:2023fob}.
    \item Krylov complexity and topological phases of matter: \cite{Caputa:2022eye,Caputa:2022yju}.
    \item K-complexity in symmetry-dominated systems, generalized coherent states: \cite{Caputa:2021sib,Patramanis:2021lkx,Balasubramanian:2022tpr,Guo:2022hui,Patramanis:2023cwz}.
    \item K-complexity in systems with saddle-dominated scrambling: \cite{Bhattacharjee:2022vlt,Huh:2023jxt}.
    \item Open and driven systems: \cite{Bhattacharya:2022gbz,Liu:2022god,Bhattacharjee:2022lzy,Bhattacharya:2023zqt,Nizami:2023dkf,Bhattacharjee:2023uwx}.
    \item K-complexity and quantum scars: \cite{Bhattacharjee:2022qjw,Nandy:2023brt}.
    \item K-complexity in tensor networks: \cite{Kim:2021okd}.
\end{itemize}

%% file: content/Chapter03.tex
\chapter{\rm\bfseries Exact results on a chaotic model: SYK}
\label{ch:chapter03_SYK}

This Chapter contains, up to minor editorial modifications, a reproduction of the publication \cite{I}, where the full sequence of Lanczos coefficients of a typical operator in a chaotic and finite (yet large) many-body system (the complex SYK$_4$ model) was computed numerically for the first time, a task for which both the Full Orthogonalization and Partial Re-Orthogonalization algorithms discussed in section \ref{sect:reorthog_algorithms} needed to be implemented in high-performance computing clusters. The plateau in the Lanczos sequence \eqref{Sect_KC_holog_bn_profile_barbon} predicted by \cite{Barbon:2019wsy} was found to receive a non-perturbative correction given by a slow descent of the Lanczos coefficients $b_n$ towards zero, eventually consistent with the finite dimensionality of Krylov space. The K-complexity profile associated to this sequence was satisfactorily found in numerical computations to transition from early exponential growth to linear growth after the scrambling time, and it furthermore depicted K-complexity saturation at times exponentially large in the number of degrees of freedom (in this case, the number of fermions). Along the way, the algebraic structure of Krylov space was explored, in developments which are at the seed of the analyses given in Chapter \ref{ch:chapter01_Lanczos}, and an upper bound on the operator Krylov space dimension was proved. The main technical achievement of this project was the successful implementation of the Lanczos algorithm for systems with Krylov spaces of dimension up to $K=63~253$.

\section{Introduction}\label{sect:intro_chapter_SYK}

The concepts of state and operator complexity lie at the intersection of quantum information, condensed matter physics, quantum field theory (QFT) and black hole physics. One is thus faced with a diversity of concepts and methods on how to precisely define and actually measure complexity. The identification of very large time scales in black hole physics and corresponding phenomena in QFT (via the AdS/CFT correspondence) \cite{Maldacena:2001kr, Barbon:2003aq, Dyson:2002pf,  Susskind:2014rva, Stanford:2014jda, Barbon:2014rma, Cotler:2016fpe, Brown:2017jil} has motivated the search for a form of complexity whose time dependence has the following distinctive characteristics: In fast-scrambling systems with finite entropy $S$, complexity should grow exponentially in time until $t_s \sim \log (S)$ (known as the scrambling time) when it reaches a value of order ${\cal O}(S)$; it then switches to a linear-in-time behavior until times of order $\exp({\cal O}(S))$ when it settles around a plateau of order $\exp({\cal O}(S))$; finally, after times of order $\exp(\exp({\cal O} (S)))$ it is expected to start performing the Poincar\'e dance. The growth of the Einstein-Rosen bridge was connected in \cite{Susskind:2014rva} with complexity growth of the quantum state in the dual CFT. A notion of complexity known as Krylov complexity, or K-complexity, was introduced in \cite{Parker:2018yvk} as a useful diagnostic of quantum chaos in the thermodynamic limit. In \cite{Barbon:2019wsy} it was argued for the first time that K-complexity of generic operators in finite-size systems exhibits the profile of complexity described above. This work reports on a complete numerical analysis of K-complexity for all time scales, including scrambling and well beyond, in concrete models demonstrating the above-described time-dependence. This is the first time that such large time scales are studied with numerical methods, and in concrete models. We uncover non-perturbative effects which were not initially anticipated. The study overcame known numerical instabilities and employed powerful algorithms executed on large computing clusters.

Roughly speaking the time evolution of complexity quantifies how quickly a reference operator grows under Heisenberg evolution in a pre-defined basis. K-complexity, denoted $C_K$, measures this growth with respect to the Krylov basis which, as we shall see below, is well-adapted to capture Heisenberg evolution efficiently. K-complexity depends on the Hamiltonian of the system and a chosen reference operator. It has an advantage over other notions of complexity: its definition does not require an arbitrary tolerance parameter. For circuit complexity\footnote{Circuit complexity of the SYK$_4$ model, the main work-horse of this project, has previously been studied in \cite{Garcia-Alvarez:2016wem,babbush2019quantum}.}, to cite but one example, such a tolerance parameter must be introduced, and its presence is crucial in establishing its boundedness \cite{Kitaev_1997, nielsen_chuang_2010, dawson2005solovaykitaev}, whereas K-complexity is naturally bounded from above\footnote{Operator size complexity \cite{Roberts:2018mnp} shares this advantage of not requiring a tolerance parameter, but differs from $C_K$ in other crucial respects. For example, it is directly bounded by the total system size, while the upper bound for K-complexity scales exponentially with system size.} \cite{Barbon:2019wsy}. It is known to be bounded by the dimension of the Hilbert space of operators; below we prove a stronger bound.

$C_K$ provides a fine-grained notion of complexity, as it manifestly distinguishes all linearly independent operators up to a given, fixed length. These features make K-complexity a natural choice for quantifying the evolution of operators at late times (see e.g. Sec. 2.1 in \cite{Altland:2020ccq} for a careful definition of chaotic time scales) which is this project’s focus.

In studying K-complexity at these time scales one should either explicitly construct, or obtain bounds on, the {\it full} sequence of Lanczos coefficients, $b_n$, which characterize the Krylov basis of an operator. These coefficients arise in the process of orthonormalizing\footnote{The process of orthonormalization requires a choice of inner product. In this work we use the standard Frobenius inner-product (see section \ref{Sec: Lanczos algorithm}). For other choices of inner product inherited from finite-temperature correlation functions see discussions in \cite{ViswanathMuller, Parker:2018yvk}.} the Krylov basis \cite{Lanczos:1950zz, ViswanathMuller}. In \cite{Parker:2018yvk} it was hypothesized that in a chaotic quantum system in the thermodynamic limit, the Lanczos coefficients grow at a linear rate. In \cite{Barbon:2019wsy} it was suggested that, in finite chaotic systems, the Lanczos sequence could exhibit plateau-like features over a certain range. The study of finite chaotic systems that is going to be presented constructs the full Lanczos sequence to reveal a non-perturbative decay of this plateau, making an initially small but eventually persistent correction to the “cliff-ended plateau" conjectured in \cite{Barbon:2019wsy}. We refer to the resulting profile as “the Descent”.

We study K-complexity in the maximally chaotic SYK$_4$ model \cite{Sachdev:1992fk, Kitaev:2015, Sachdev:2015efa, Maldacena:2016hyu}, which is by now well-established as a toy model displaying key chaotic properties of black holes, including late-time spectral chaos \cite{Garcia-Garcia:2016mno, Cotler:2016fpe}. We also study the quadratic SYK$_2$ model, which is integrable, showing Poisson statistics \cite{Garcia-Garcia:2017bkg, Garcia-Garcia:2020cdo, Haque_2019}.


\section{Time evolution of operators and physical properties of Krylov space} \label{Sec: Krylov space}

We begin by providing the definition of Krylov space and stressing its relation to the structure of the spectrum of the system under consideration (and hence to its integrable or chaotic character). Consider a Hilbert space $\mathcal{H}$, of dimension $\dim\left(\mathcal{H}\right) = D$, with associated Hilbert space of linear operators ${L}\left(\mathcal{H}\right)\equiv \widehat{\mathcal{H}}$, acting on $\mathcal{H}$ and satisfying $\dim(\widehat{\mathcal{H}})=D^2$. Now consider a system whose dynamics are described by a certain (hermitian) Hamiltonian $H\in\widehat{\mathcal{H}}$, and an observable $\mathcal{O}\in \widehat{\mathcal{H}}$.

The associated Krylov space \cite{Krylov:1931} is defined as the minimal subspace of operator space that contains the time evolution of $\mathcal{O}$ at all times. 

The time evolution of $\mathcal{O}$ is given by
\begin{equation}
    \centering
    \label{BCH-Liouvillian-def}
    \mathcal{O}(t) = e^{iHt}\mathcal{O} e^{-iHt} = e^{i\mathcal{L} t}\mathcal{O}
\end{equation}
where the Liouvillian operator is defined as $\mathcal{L}\equiv \left[H,\cdot\right]$. Thus the Krylov space is the linear span of all nested commutators of the Hamiltonian with the operator:
\begin{equation}
    \centering
    \label{Krylov-definition_maintext}
    \mathcal{H}_\mathcal{O} = \text{span}\left\{\mathcal{L}^n\mathcal{O}\right\}_{n=0}^{+\infty}= \text{span}\left\{\mathcal{O},\, [H,\mathcal{O}],\, [H,[H,\mathcal{O}]], \dots\right\}.
\end{equation}
Equivalently, $\mathcal{H}_\mathcal{O}$ is the minimal invariant subspace of the Liouvillian that contains $\mathcal{O}$. We denote $\dim\left(\mathcal{H}_\mathcal{O}\right)\equiv K$. 

$K$ is determined by studying the cardinality of the maximal set of linearly independent objects of the form $\mathcal{L}^n\mathcal{O}$, which can be found by computing the rank of
\begin{equation}
    \label{BCH-basis-rank}
    \left(\begin{matrix}
         \mathcal{O}, &
        \mathcal{LO}, &
         \mathcal{L}^2\mathcal{O},&
        \dots &
        \mathcal{L}^n\mathcal{O}, &
        \dots
    \end{matrix}\right)^T\,,
\end{equation}
in a convenient matrix representation, which we now define.
Let us choose the basis $|\omega_{ab})\equiv\ket{E_a}\bra{E_b}$, on $\widehat{\cal H}$, which is naturally induced by the eigenbasis of the Hamiltonian on ${\cal H}$. Then each nested commutator takes the form:
\begin{equation}
    \centering
    \label{Op-phases}
    \mathcal{L}^n\left|\mathcal{O}\right) =\delta_{n0} \sum_{a=1}^D O_{aa}|\omega_{aa})+\sum_{\substack{a,b = 1 \\ a\neq b}}^D  O_{ab}\,  \omega_{ab}^n \, |\omega_{ab}) 
\end{equation}
where we have defined the phases\footnote{We shall refer to energy differences as \textit{phases}, since they appear as such in the moment expansion of the two-point function of $\mathcal{O}$.} $\omega_{ab}:= E_a-E_b$, which are eigenvalues of the Liouvillian acting on $|\omega_{ab})$. Note that the $\omega$'s related to pairs of the form $(a,a)$ are zero, $\omega_{aa} = 0$ for all $a=1,...,D$. Displaying the coordinates of (\ref{Op-phases}) as rows one can construct a matrix representation of (\ref{BCH-basis-rank}), which turns out to be a Vandermonde matrix. The rank of this matrix is at most $D^2$, as it has $D^2$ columns, so we shall keep the same number of rows to make it a square Vandermonde matrix:

\begin{equation}
    \centering
    \label{Vandermonde_matrix_explicit}
    \begin{pmatrix}
    O_{11} & O_{22} & \dots & O_{DD} & O_{12} & O_{13} & \dots & O_{D-1,D} \\
    
    0 & 0 & \dots & 0 & O_{12}\; \omega_{12} & O_{13}\; \omega_{13} & \dots & O_{D-1,D}\;\omega_{D-1,D} \\
    
    0 & 0 & \dots & 0 & O_{12}\; \omega_{12}^2 & O_{13}\; \omega_{13}^2 & \dots & O_{D-1,D}\;\omega_{D-1,D}^2 \\
    \vdots & \vdots & \ddots & \vdots & \vdots & \vdots & \ddots & \vdots \\
    0 & 0 & \dots & 0 & O_{12}\; \omega_{12}^{D^2-1} & O_{13}\; \omega_{13}^{D^2-1} & \dots & O_{D-1,D}\;\omega_{D-1,D}^{D^2-1}
    \end{pmatrix} ~.
\end{equation}
To find its actual rank, we can compute its determinant, which is given by:
\begin{equation}
    \centering
    \label{Det-v2}
    \Delta\left(\left\{\omega_{ab}\right\}\right)\prod_{i,j=1}^{D} O_{ij}\,,
\end{equation}
where $\Delta\left(\left\{\omega_{ab}\right\}\right)$ is the Vandermonde determinant of the phases in the matrix. The expression (\ref{Det-v2}) will be zero if any of the phases are degenerate, and also if any of the matrix elements in the energy basis vanish, so the corresponding columns should be removed in order to be left with a matrix of maximal rank. Hence, the Krylov dimension $K$ can be estimated using the following algorithmic prescription: $K$ is equal to the number of distinct phases corresponding to the indices of non-zero matrix elements of the operator in the energy basis. The zero phase $\omega_{aa}=0$ is always at least $D$ times degenerate and therefore, since it can only be counted once, the Krylov dimension is bounded by:
\begin{equation}
    \label{Krylov-bound}
    1\leq K \leq D^2-D+1
\end{equation}
for any non-vanishing operator.

In general, if the operator $\mathcal{O}$ has a non-vanishing projection over several eigenstates of the Liouvillian that share the same degenerate eigenvalue, then this phase only contributes one dimension to the Krylov space. So a legitimate question is to wonder what particular linear combination of those eigenstates is actually contained in the Krylov space. In order to answer this, one can consider the form of the operator $\mathcal{L}^n\left|\mathcal{O}\right)$ given in (\ref{Op-phases}). Suppose that, for some set $I$ of pairs of indices, the eigenvalue is degenerate:
\begin{equation}
    \centering
    \label{Deg-phase}
    \mathcal{L}\left|\omega_{ab}\right) = \omega\left|\omega_{ab}\right)\;\;\;\forall (a,b)\in I
    \end{equation}
i.e. $\omega_{ab}=\omega$ for all $(a,b)\in I$. Assume also that $\omega_{ab}\neq\omega$ for any $(a,b)\notin I$. Inserting (\ref{Deg-phase}) in (\ref{Op-phases}) one finds:
\begin{equation}
    \centering
    \mathcal{L}^n\mathcal{O}=\omega^n \sum_{\left(a,b\right)\in I}\,O_{ab}\left|\omega_{ab}\right)+\sum_{\left(a,b\right)\notin I}O_{ab}\,\omega_{ab}^n\left|\omega_{ab}\right).
\end{equation}
It is manifest now that the direction of the $\omega$-eigenspace that contributes to the Krylov space is precisely the projection of the operator $\mathcal{O}$ over such an eigenspace:

\begin{equation}
    \centering
    \label{eigenspace-contrib}
    \left|\mathcal{K}_\omega\right):= \sum_{\left(a,b\right)\in I}\,O_{ab}\left|\omega_{ab}\right)~.
\end{equation}
Let`s call $\left|\mathcal{K}_\omega\right)$ the \textit{eigenspace representative} for the phase $\omega$. The structure of the Krylov space is now fully understood: Each eigenspace of the Liouvillian over which the operator $\mathcal{O}$ has a non-vanishing projection contributes one Krylov dimension, through a linear combination of the basis of the eigenspace of the form (\ref{eigenspace-contrib}). We can thus redefine the Krylov space as:
\begin{equation}
\centering
\label{Krylov-eigenspaces}
\mathcal{H}_\mathcal{O}=\text{span}\left\{\left|\mathcal{K}_\omega\right),\;\;\;\omega\in \sigma(\mathcal{L})\right\}
\end{equation}
where $\sigma(\mathcal{L})$ denotes the spectrum of $\mathcal{L}$. 
Finally, the Krylov dimension $K$ is simply equal to the number of non-zero eigenspace representatives $\left|\mathcal{K}_\omega\right)$.

For instance, if one considers a system whose Liouvillian has a spectrum with no degeneracies (other than the unavoidable null phases) and an operator that is dense in the energy basis, the Krylov dimension will be maximal, $K=D^2-D+1$. The only source of degeneracy in the phases will be the universal one due to diagonal phases $\omega_{aa}=0$. We can now note explicitly that the part of the operator algebra that is left out of the Krylov space belongs to the space subtended by the projectors on the energy eigenstates, since those correspond to zero phases, and the only combination of them that contributes to $\mathcal{H}_\mathcal{O}$ is the representative of the $\omega=0$ eigenspace, in this case:
\begin{equation}
    \centering
    \label{eigenspace-0-Krylov}
    \left|\mathcal{K}_0\right) = \sum_{a=1}^D O_{aa}\left|\omega_{aa}\right)=\sum_{a=1}^D O_{aa}\ket{E_a}\bra{E_a}\equiv\sum_{a=1}^DO_{aa}\mathcal{P}_a.
\end{equation}

To summarize, $K$ is determined by exploiting the advantages of the basis in operator space induced by the eigenbasis of the Hamiltonian; crucially, the determinant of a suitable matrix containing the representation of these nested commutators in the given basis reduces to a Vandermonde determinant of energy differences. The following algorithmic prescription is derived: $K$ is equal to the number of eigenspaces of the Liouvillian over which $\mathcal{O}$ has non-zero projection. The eigenvalues of the Liouvillian are precisely all possible energy differences, $\omega_{ab}=E_a-E_b$. The zero phase is always at least $D$ times degenerate, thus the Krylov dimension is bounded by (\ref{Krylov-bound}),
for any non-zero operator. The more degeneracies there are in the spectrum of the Liouvillian, the lower will be the Krylov dimension. The expectation is that the spectrum of a chaotic system will not feature degeneracies apart from those induced by the presence of extra symmetries; we therefore conjecture that typical\footnote{Along the lines of ETH \cite{PhysRevA.30.504, PhysRevA.43.2046, Srednicki_1999,DAlessio:2015qtq}, we expect typical observables in chaotic systems to be dense in the energy basis (as is the case of localized operators in local chaotic systems). Regardless of the structure of the spectrum, one can always choose a fine-tuned operator having a sparse matrix representation in the energy basis, e.g. an eigenspace projector $\ket{E}\bra{E}$; such a candidate will indeed have a very small associated Krylov space, but will not generally fulfil the requirements to be considered a \textit{typical} operator. Conserved charges are also special operators, since they commute with the Hamiltonian, and therefore their Krylov space is always one-dimensional.} operators in chaotic systems saturate the upper bound in (\ref{Krylov-bound}); for integrable systems we expect $K$ to be significantly smaller than the bound. We have confirmed both expectations by studying the chaotic SYK$_4$ model (see numerics below) as well as RMT (see Appendix \ref{Appx-RMT}), and the integrable SYK$_2$ model, see section \ref{Sec: Integrable systems} below. 

\section{\texorpdfstring{Integrable systems and SYK\textsubscript{2}}{Integrable systems and SYK2}} \label{Sec: Integrable systems}
We have conjectured that the Krylov space for the time evolution of a typical operator for integrable systems will be significantly smaller than the upper bound in (\ref{Krylov-bound}).
Before verifying this conjecture in the case of SYK$_2$, we verify it in the simpler example of the quantum harmonic oscillator $H= \omega(a^\dagger a +1/2)$ with the position operator $\hat{x}=\sqrt{\frac{1}{2\omega}}(a+a^\dagger)$, for which the Krylov space dimension is $K=2$, even though the Hilbert space is infinite\footnote{Accordingly, the Lanczos sequence contains a single element $b_1=\omega$ (see next section for the definition of the Lanczos sequence in this specific framework).}. In this case, the smallness of the Krylov space is due to the fact that the position operator has a non-vanishing projection over only two eigenspaces of the Liouvillian (those corresponding to the energy differences $\omega$ and $-\omega$).

Let us now test the conjecture in SYK$_2$, an integrable system with more structure than the harmonic oscillator.
For simplicity, let us use the Majorana (real) version of the model \cite{Kitaev:2015,Maldacena:2016hyu}:
\begin{equation}
    \centering
    \label{MajoranaSYK-q2}
    H = i \sum_{1\leq i < j \leq L}m_{ij}\,\chi_i \chi_j
\end{equation}
where the coupling strength $m_{ij}\in \mathds{R}$ is antisymmetric, $m_{ij}=-m_{ji}$, and each independent matrix element is drawn from a Gaussian distribution with zero mean and variance $\mathds{E} (m_{ij}^2) = m^2/L$.
The number of sites is even, $L=2M$, and the Majorana fermions satisfy the relations
\begin{equation}
    \label{Sect_SYK2_Majoranas_relations}
    \left\{\chi_i,\chi_j\right\}=\delta_{ij},\;\;\;\chi_i=\chi_i^\dagger~.
\end{equation}

The key point in the SYK$_2$ case is the fact that operators do not grow after commutation with the Hamiltonian, as also observed by \cite{Roberts:2018mnp, Carrega:2020mah}. Hence the subspace containing their time evolution will be, at most, that subtended by operators of fixed equal size. For example if the operator of study is a Majorana on some site with index $A$, $\mathcal{O}\equiv\chi_A$, commutation with the Hamiltonian will give $\left[\chi_i\chi_j,\chi_A\right] = \delta_{Aj}\chi_i-\delta_{Ai}\chi_j$, which is again a single-site operator (more details on the construction of the Krylov space in this case are given in Appendix \ref{Appx-SYK2}). For one-site operators, this implies an upper bound on $K$:
    \begin{equation}
        \centering
        \label{q2-KDIM-main}
        K\leq L \sim \log D \ll D^2-D+1 ~.
    \end{equation}
It should be possible to reach this conclusion by studying the degeneracy structure of the spectrum of SYK$_2$, which features Poisson level spacing statistics, and it is natural to expect other integrable systems to have a small Krylov space, due to the expected degeneracies in the spectrum of the Liouvillian and to the sparseness of simple operators in the energy basis.
In general, interacting integrable systems feature Poisson level spacing statistics, implying, at least, the existence of quasi-degenerate levels in the spectrum of the Hamiltonian, and therefore also in that of the Liouvillian. We expect these systems to behave as if the degeneracies were exact to a good degree of approximation, hence admitting a description in terms of an effectively smaller Krylov space. This argument applies, and accounts for a significant effective reduction of Krylov space, even if the operator is not sparse in the energy basis, as long as the invariant subspace of the Liouvillian over which it has a non-trivial projection contains degenerate or quasi-degenerate levels. All this leads us to suggest that integrable systems have a lower effective Krylov dimension compared with chaotic ones. 
As a first step in trying to elevate this conjecture based on a few concrete examples as well as qualitative ones, we estimate the minimal length of time for which the difference in the dimension of the explored Krylov space is large.
Since quasi-degenerate energy levels are separated by less than the mean level spacing $\Delta \sim \Gamma e^{-S}$, where $\Gamma$ is a relevant energy scale of the system such as the total spectral width, we can estimate the lowest time scale until which the Krylov dimensions of chaotic vs. integrable (with quasi-degeneracies) systems remain significantly different, to be of order of the Heisenberg time, i.e of order $e^{O(S)}$. For infinite systems this lasts forever, while for large finite systems this is way above the thermalization or scrambling time scales. The question of what ultimately generically happens between the Heisenberg time scale and the Poincar\'e time scale requires further research.
Bounds like (\ref{Krylov-bound}) and (\ref{q2-KDIM-main}) have not been noted previously in the literature.

\section{Lanczos algorithm} \label{Sec: Lanczos algorithm}

Once the Krylov space adapted to time evolution of an operator with a constant Hamiltonian is identified, one would like to construct an orthonormal basis for it, given a certain scalar product $\left(\cdot|\cdot\right)$ on operator space. This is achieved with the Lanczos algorithm, which is a particularization of the Gram-Schmidt procedure:
\begin{enumerate}
    \item set $b_0\equiv 1$ and $\left|\mathcal{O}_{-1}\right)\equiv0$
    \item $\left|\mathcal{O}_0\right) =\frac{1}{\sqrt{\left(\mathcal{O}|\mathcal{O}\right)}} \left|\mathcal{O}\right)$
    \item for $n\geq 1$: $\left|\mathcal{A}_n\right) = \mathcal{L}\left|\mathcal{O}_{n-1}\right) - b_{n-1}\left|\mathcal{O}_{n-2}\right)$
    \item set $b_n = \sqrt{\left(\mathcal{A}_n|\mathcal{A}_n\right)}$
    \item if $b_n=0$ stop; otherwise set $\left|\mathcal{O}_n\right) = \frac{1}{b_n}\left|\mathcal{A}_n\right)$ and go to step 3.
\end{enumerate}

In this work we make use of the Frobenius product $\left(A|B\right)=\frac{1}{D}\text{Tr}\left[A^\dagger B\right]$. The algorithm will construct an orthonormal set $\left\{\mathcal{O}_n\right\}_{n=0}^{K-1}$, the \textit{Krylov basis}, and the \textit{Lanczos coefficients} $\left\{b_n\right\}_{n=1}^{K-1}$.

Each Lanczos step produces an element $\left|\mathcal{A}_n\right)$ orthogonal to all previous $\left|\mathcal{O}_m\right)$ with $m<n$, so it is either zero or a new direction in Krylov space. For $n<K$, $\left|\mathcal{A}_n\right)\neq 0$ because the set that is being orthogonalized with this procedure has rank $K$ (in particular, $\left|\mathcal{A}_n\right)$ contains terms with up to $n$ nested commutators of $H$ with $\mathcal{O}$). However, $\left|\mathcal{A}_K\right)$ is orthogonal to $\left\{\mathcal{O}_n\right\}_{n=0}^{K-1}$, which is already a complete orthonormal basis of $\mathcal{H}_\mathcal{O}$, so it must therefore vanish, just as $b_K=\sqrt{\left(\mathcal{A}_K|\mathcal{A}_K\right)}=0$. We conclude that the Lanczos algorithm must terminate by hitting a zero once all directions in Krylov space have been exhausted. This is accounted for in Step 5 above.

The representation of the Liouvillian over the Krylov space in such a basis simplifies to a tridiagonal matrix $\left(\mathcal{O}_m\right|\mathcal{L}\left|\mathcal{O}_n\right)=T_{mn}$, whose entries are given by the Lanczos coefficients:
\begin{equation}
\centering
\label{L-Tridiagonal}
     T\overset{*}{=}\begin{pmatrix}
        0 & b_1 & 0 & 0 & \dots & 0\\
        b_1 & 0 & b_2 & 0 &\dots & 0 \\
        0 & b_2 & 0 & b_3 & \dots & 0 \\
        \vdots &  & \ddots & & \ddots & \vdots \\
         & & & & \ddots & b_{K-1}\\
        0 & 0 & 0 & \dots & b_{K-1} & 0
    \end{pmatrix}.
\end{equation}
The eigenvalues of this matrix are all non-degenerate, and given precisely by the phases corresponding to the eigenspaces of the Liouvillian that span the Krylov space (see section \ref{Sec: Krylov space}).

The original Lanczos algorithm described above is known to suffer from numerical instabilities which can be overcome using the re-orthogonalization algorithms FO and PRO \cite{PRO, Parlett} described in Appendix \ref{Appx-algorithms}.

\section{K-complexity and K-entropy} \label{Sec: KC and KS}
Using the sequence of Lanczos coefficients (which we will also call the $b$-sequence), one can reduce the analysis of the time-evolution of an operator $\mathcal{O}$ into the solution of a differential recurrence equation. The time-evolving operator can be expanded in the Krylov basis as:
\begin{equation}
    \label{Expansion-Krylov-basis}
    \left|\mathcal{O}(t)\right)=\sum_{n=0}^{K-1}i^n\varphi_n(t)\left|\mathcal{O}_n\right) ~,
\end{equation}
where $\varphi_n(t)$ are time-dependent coefficients which describe how the operator is distributed over the Krylov basis (they may be thought of as the ``wavefunctions" in $n$). Given the Heisenberg equation $\frac{d\mathcal{O}}{dt}=i[H,\mathcal{O}]$, they satisfy the differential recurrence equation
\begin{equation}
    \label{difrec}
    \dot{\varphi}_n(t)=b_n\varphi_{n-1}(t)-b_{n+1}\varphi_{n+1} ~.
\end{equation}
Here, $\varphi_n(0)=\delta_{n0}$, since for a normalized operator $\mathcal{O}(0)=\mathcal{O}_0$. We set $\varphi_{-1}(t)\equiv 0\equiv\varphi_K(t)$ in order to make the recurrence (\ref{difrec}) consistent with the definition (\ref{Expansion-Krylov-basis}). Also, from unitarity $\sum_{n=0}^{K-1}|\varphi_n(t)|^2=1$. The Lanczos coefficients ${\left\{ b_n \right\}}_{n=1}^{K-1}$ can be understood as hopping amplitudes for the initial operator $\mathcal{O}_0$ to explore the ``Krylov chain" and the functions $\varphi_n(t)$ can be visualized as wave-packets travelling on it \cite{Parker:2018yvk}. 

We now display two quantities which highlight broad features of the distribution $\varphi_n(t)$, viz.
\begin{itemize}
    \item \textbf{K-complexity}, which computes the average position of the distribution on the ordered Krylov basis:
    \begin{equation}
        \centering
        \label{K-Complexity}
        C_K(t)=\sum_{n=0}^{K-1}n|\varphi_n(t)|^2.
    \end{equation}
    \item \textbf{K-entropy}, which computes how randomized the distribution is:
    \begin{equation}
        \centering
        \label{K-Entropy}
        S_K(t)=-\sum_{n=0}^{K-1} |\varphi_n(t)|^2\log |\varphi_n(t)|^2.
    \end{equation}
\end{itemize}
Assuming that at very late times the operator is evenly distributed over Krylov space, there exists a saturation time $t_{sat}$ for which 
\begin{equation} \label{phi_sat}
   |\varphi_n(t\geq t_{sat})|^2\sim \frac{1}{K} ~.
\end{equation}
We can then get a rough estimate for the values of K-complexity and K-entropy at very late times by plugging (\ref{phi_sat}) into the formulas for $C_K(t)$ and $S_K(t)$:
\begin{equation}
    \label{KCsat}
    C_{K}(t\geq t_{sat})\sim\frac{1}{K} \frac{K(K-1)}{2}\sim \frac{K}{2} ,
\end{equation}

\begin{equation}
    \label{KEsat}
    S_K(t\geq t_{sat})\sim-K \frac{1}{K} \log(1/K)= \log(K).
\end{equation}
Since for chaotic systems $K\sim D^2$ and in general $D\sim e^S$, where $S$ is the entropy of the system (in the sense of ``log of the number of states"), we find that the saturation value of K-Entropy is essentially of order $S$, while the saturation value of K-complexity is of order $e^{2S}$. If $C_K(t)$ grows linearly after scrambling, the saturation time will be roughly $t_{sat}\equiv t_K\sim e^{2S}$, in agreement with the expectation in \cite{Barbon:2019wsy}. These properties are confirmed in the numerical results.

\section{Numerical results} \label{Sec: Numerical results}
In the following we present numerical\footnote{Computations were performed using the NumPy library \cite{harris2020array} for Python.} results for complex SYK$_4$, whose Hamiltonian is schematically given by:
\begin{equation}
    \centering
    \label{SYK-H}
    H = \sum_{ijkl}J_{ij;kl}c_i^\dagger c_j^\dagger c_kc_l
\end{equation}
where $i,j,k,l=1,2,...,L$ ($L$ is the system size). $J_{ij;kl}$ is a complex matrix whose independent elements follow normal distributions with:
\begin{equation}
    \centering
    \label{SYK-H-normal}
        \overline{J_{ij;kl}} = 0,\;\;\;\;\overline{{\left|J_{ij;kl}\right|}^2}=\frac{6J^2}{L^3} 
\end{equation}
where the overline denotes average over random realizations. It is worth recalling that the Hilbert space of real (Majorana) SYK with $L$ sites scales as $2^{L/2}$, while in the complex case it scales as $2^L$. However, the advantage of using this version of SYK \cite{Sonner:2017hxc} lies in the fact that the total number operator $\widehat{n}:=\sum_{i=1}^{L}c_i^\dagger c_i$ commutes with the Hamiltonian, allowing us to work in sectors of the Hilbert space with fixed occupation number, denoted by $N$ (eigenvalue of $\widehat{n}$). For the numerical computations we will take $N=\big{\lceil}{\frac{L}{2}}\big{\rceil}$.

The (hermitian) observable chosen for numerical computations is the hopping operator between sites $L$ and $L-1$:
\begin{equation}
    \centering
    \label{SYK-Hopping}
    \mathcal{O} = c_{L-1}^\dagger c_L+c_{L}^\dagger c_{L-1}\equiv h_{L-1,L} ~.
\end{equation}
Any other hopping operator $h_{ij}$ should give results equivalent to (\ref{SYK-Hopping}), due to the non-local character of the Hamiltonian (\ref{SYK-H}). In general, one could choose other non-extensive operators, such as the number operator at a particular site $n_i:=c_i^\dagger c_i$. Both $n_i$ and $h_{ij}$ have been shown numerically to satisfy ETH \cite{Sonner:2017hxc}. 

\begin{figure}[t]
    \begin{minipage}{.5\textwidth}
    \includegraphics[width=1.\linewidth]{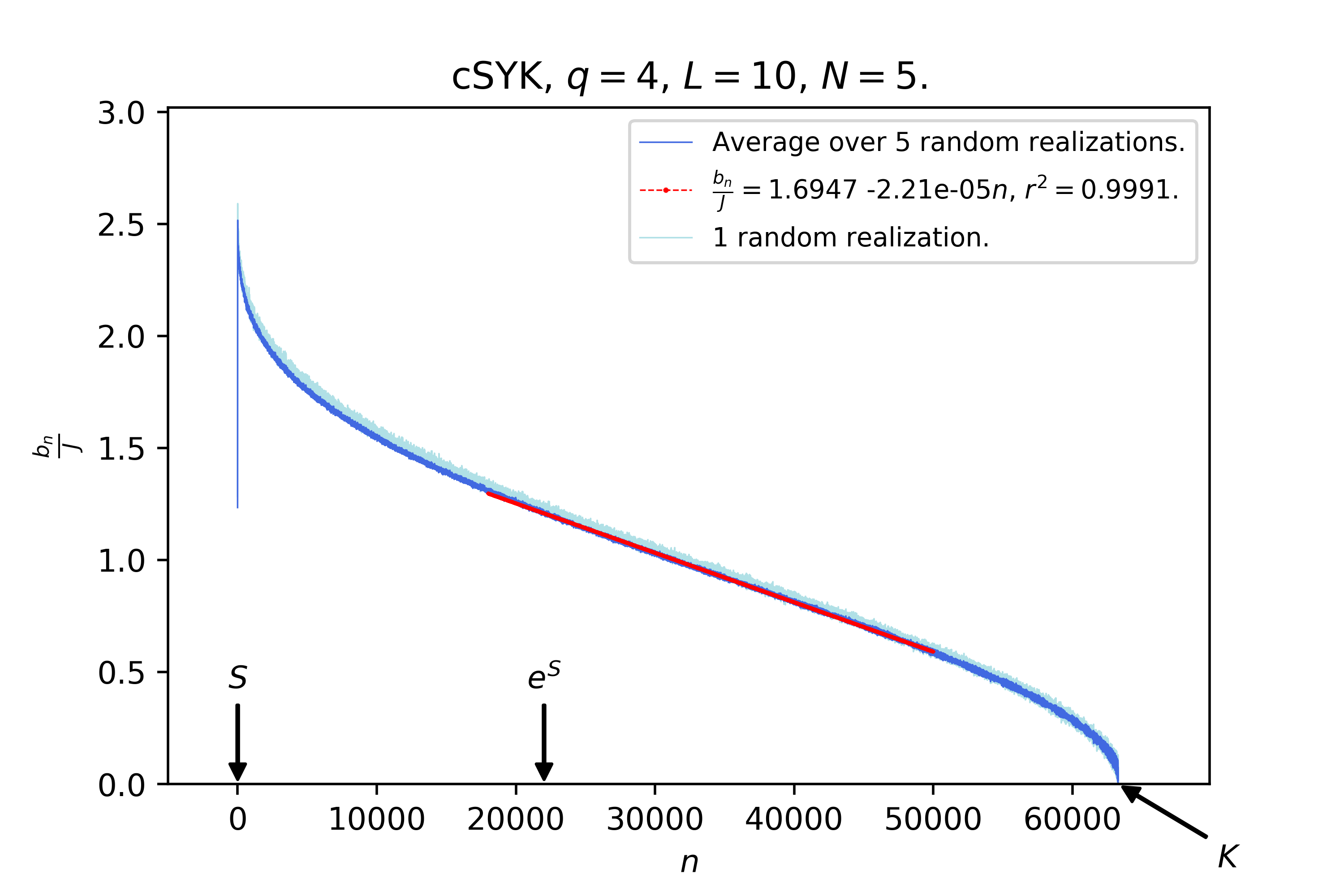}
    \end{minipage}  \quad  
    \begin{minipage}{.5\textwidth}
    \includegraphics[width=1.\linewidth]{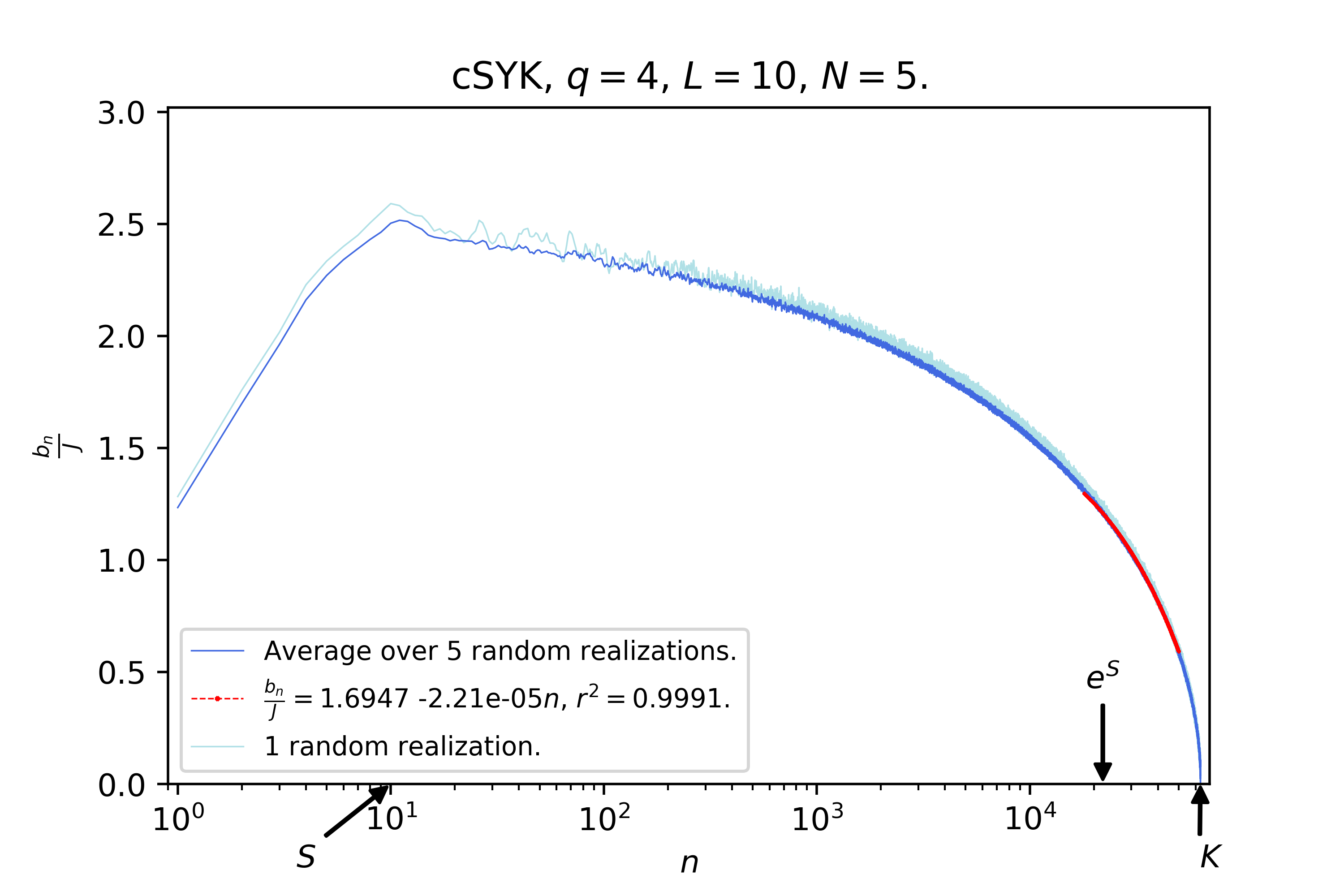}
    \end{minipage} 
    \caption{Lanczos sequence for $L=10$. \textbf{Left:} ``The Descent" depicted with linear scale along the horizontal axis. After initial growth up to $n\sim S$, a slow decrease to zero with roughly constant negative non-perturbative slope of order $\sim -\frac{1}{K} \sim - e^{-2S}$. Fitted slope (in red) of the decaying part is $-2.21\cdot 10^{-5}\approx-1.58\cdot10^{-5}=-K^{-1}$. On a 1:1 scale, the horizontal axis should be at least $843$ meters long.  \textbf{Right:} Logarithmic scale along the horizontal axis makes visible the initial ascent.}
    \label{b-L10-main}
\end{figure}

\begin{figure}[t]
    \centering
    \begin{minipage}{.45\textwidth}
	\includegraphics[width=1.\linewidth]{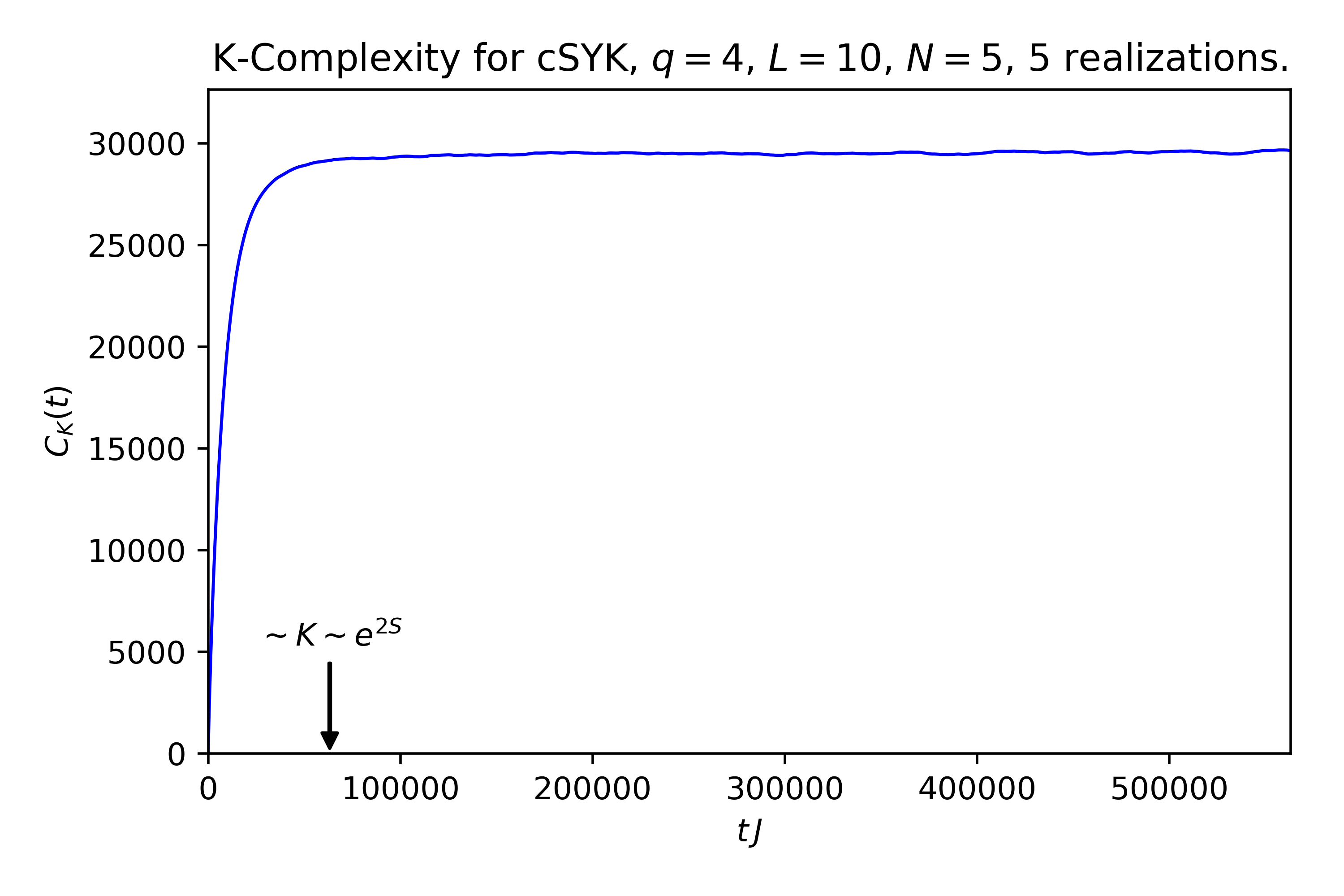}
	\end{minipage}   
	\begin{minipage}{.45\textwidth}
	\includegraphics[width=1.05\linewidth]{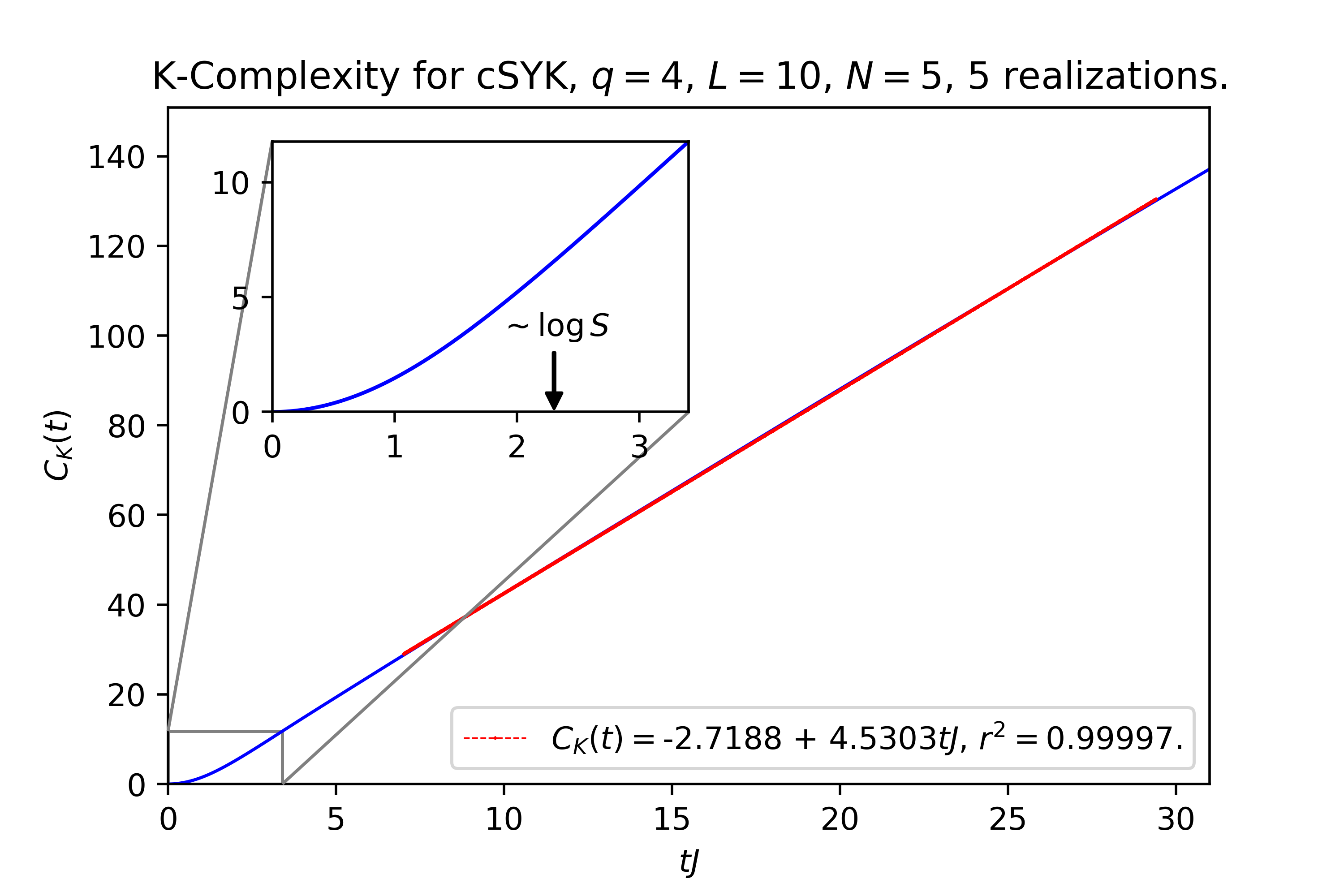}
	\end{minipage}
	\caption{K-complexity averaged over 5 random realizations with $L=10$ sites at half filling. \textbf{Left:} Full time range computed. Saturation occurs at time scales of order $t_K\sim K \sim e^{2S}$ with value near $K/2=31626.5$ (this is the time scale at which the wave-packet $\varphi_n(t)$, which propagates at roughly constant -- but slowly decreasing -- velocity, reaches the edge of the Krylov chain).  \textbf{Right:} Zoom in at early times. Note the initial non-linear growth transforming into linear growth, as verified by the linear fit (in red). Relevant time scales are indicated in the plots, although their exact location may depend on a dimensionful prefactor (the Lyapunov exponent, as discussed in \cite{Barbon:2019wsy}).}  
	\label{KC-L10-main}
\end{figure}

\begin{figure}[t]
\centering
    \begin{minipage}{.45\textwidth}
	\includegraphics[width=1.\linewidth]{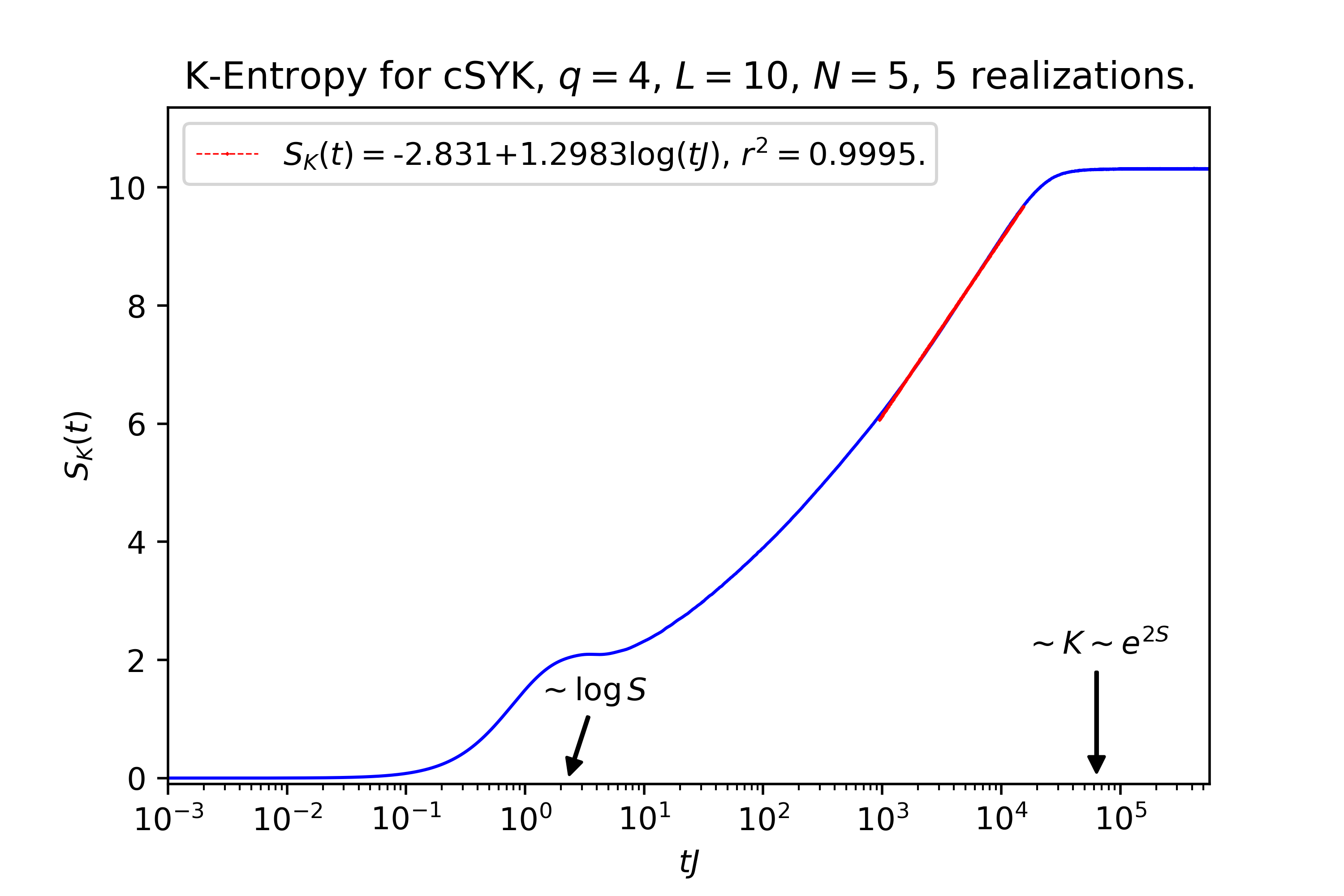}
    \end{minipage}   
    \begin{minipage}{.45\textwidth}
    \includegraphics[width=1.\linewidth]{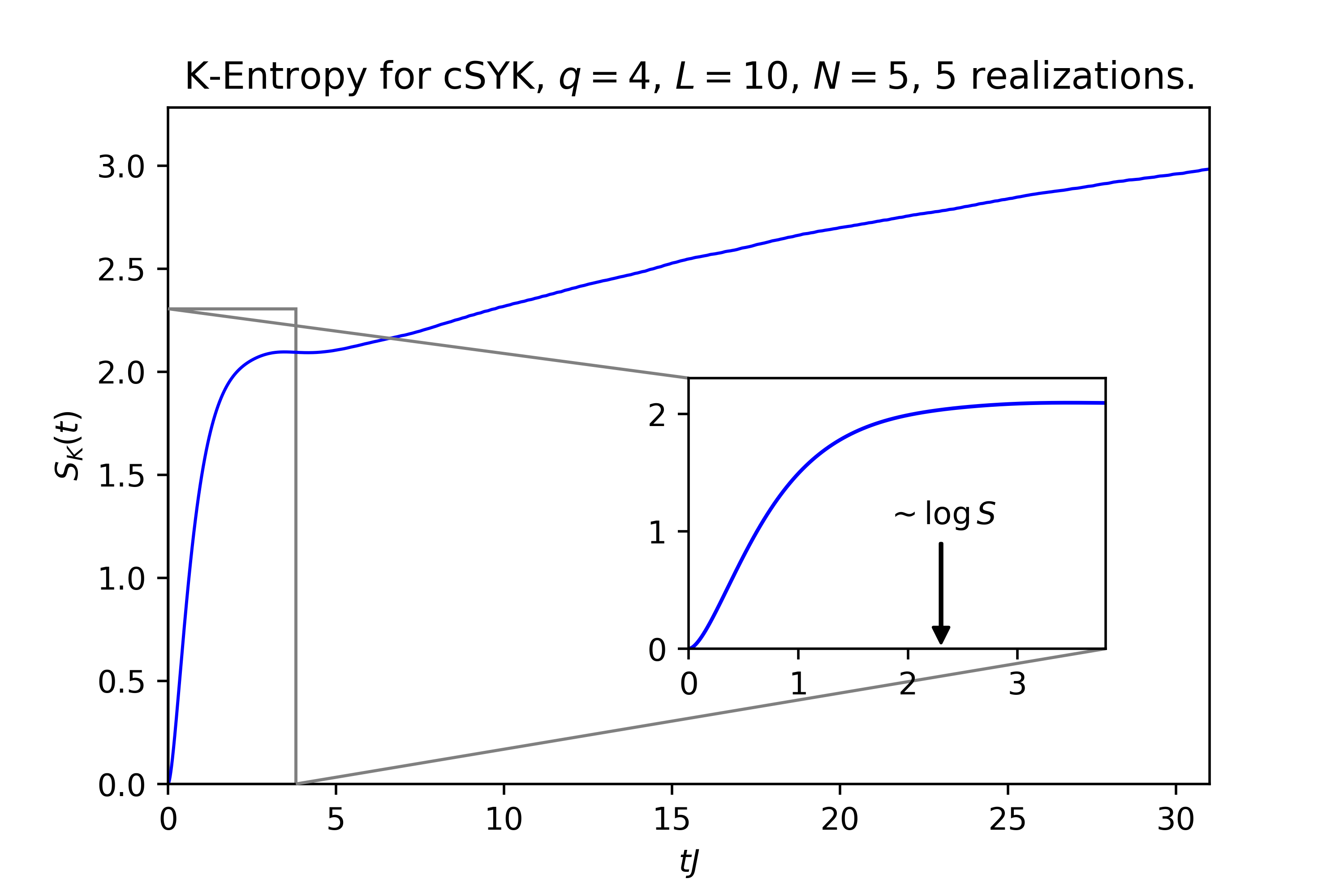}
    \end{minipage}
    \caption{K-entropy averaged over 5 random realizations with $L=10$ sites at half filling. \textbf{Left:} Full time range computed, with logarithmic scale along horizontal axis. A linear fit in the post-scrambling regime, where K-entropy is expected to grow logarithmically, is depicted in red. K-entropy grows linearly up to scrambling time, and then transitions to a logarithmic growth phase that continues until saturation around $S_K\sim L\sim S$ at exponentially late times (this is the time scale at which the wave-packet $\varphi_n(t)$ becomes fully dispersed). \textbf{Right:} K-entropy at early times with linear scale along the horizontal axis. The exact location of the time scales may depend on a dimensionful prefactor (see caption of Figure \ref{KC-L10-main}). 
}
    \label{KS-L10-main}
\end{figure}

\begin{figure}
\centering
\begin{minipage}{.35\textwidth}
    \includegraphics[width=1.\linewidth]{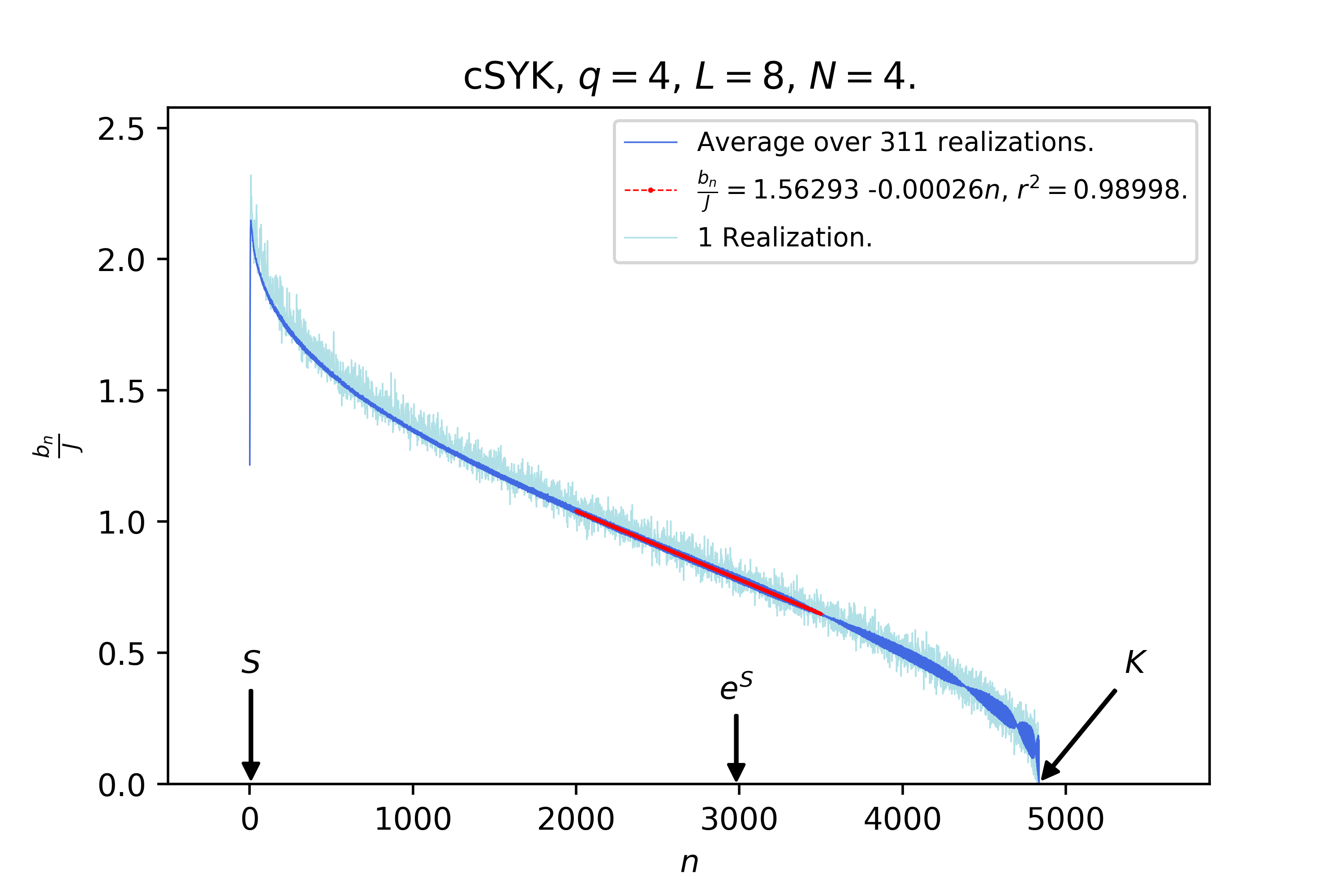}
\end{minipage} \qquad
\begin{minipage}{.35\textwidth}
    \includegraphics[width=1.\linewidth]{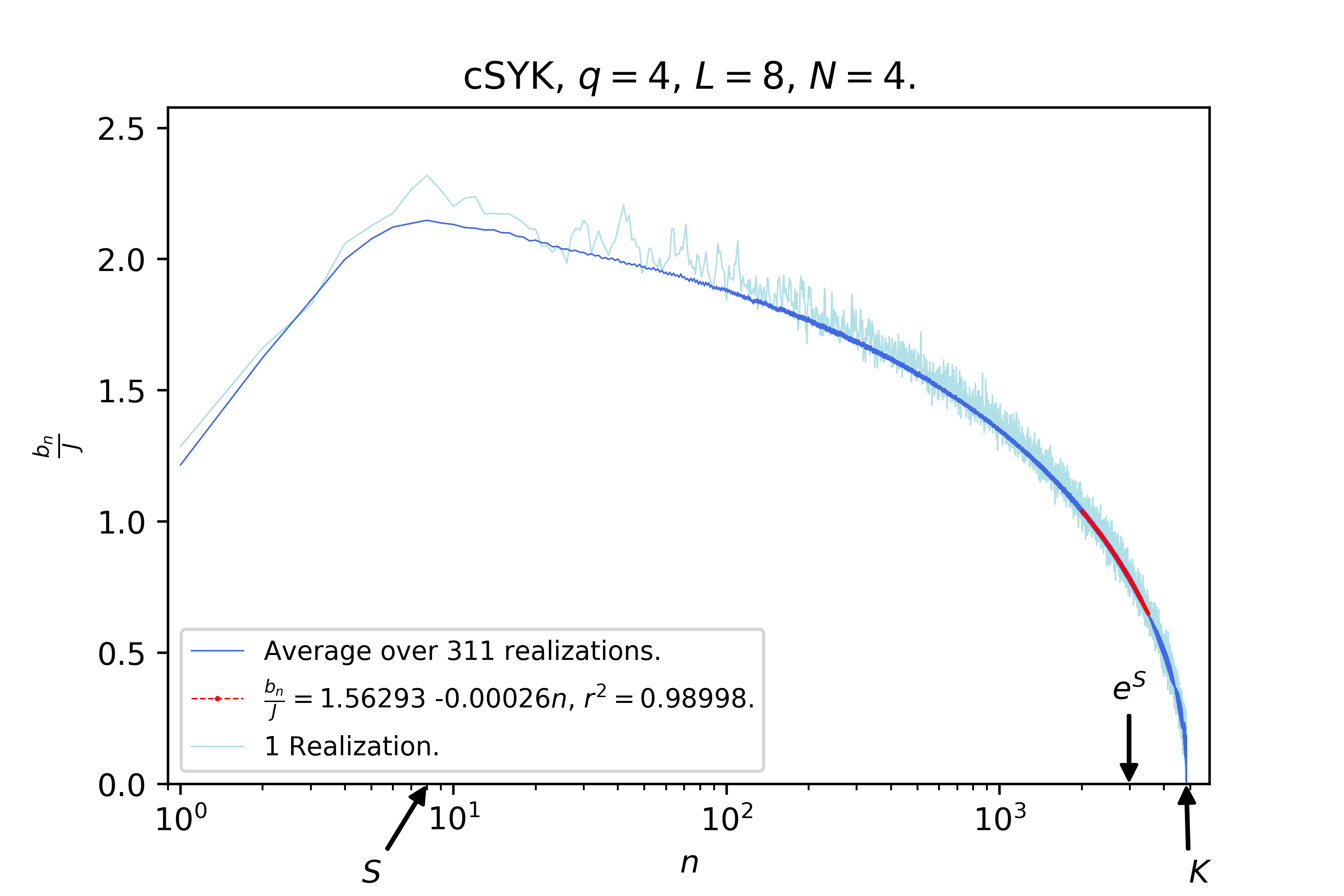}
\end{minipage} \\
\begin{minipage}{.35\textwidth}
\includegraphics[width=1.\linewidth]{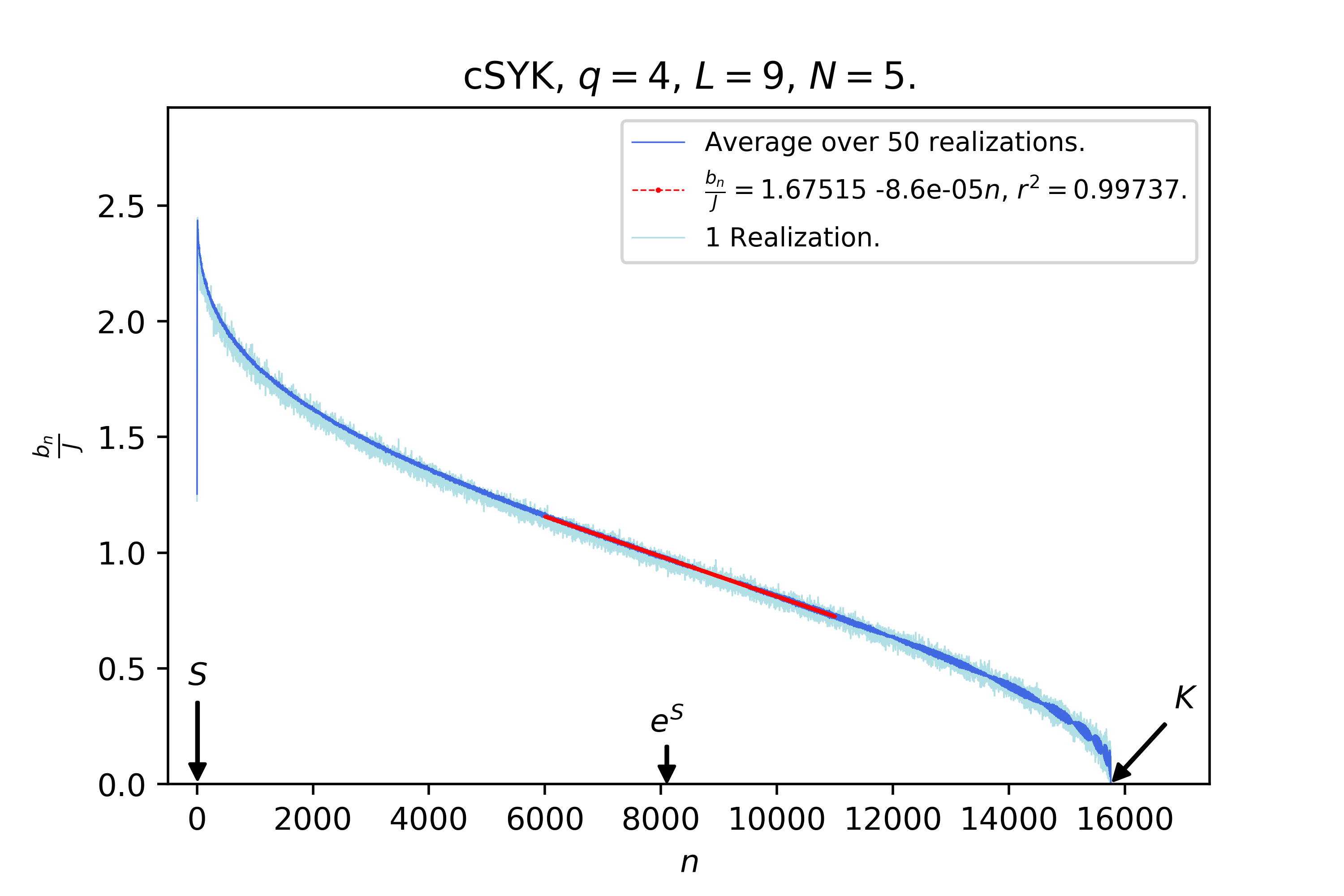}
\end{minipage}\qquad
\begin{minipage}{.35\textwidth}
\includegraphics[width=1.\linewidth]{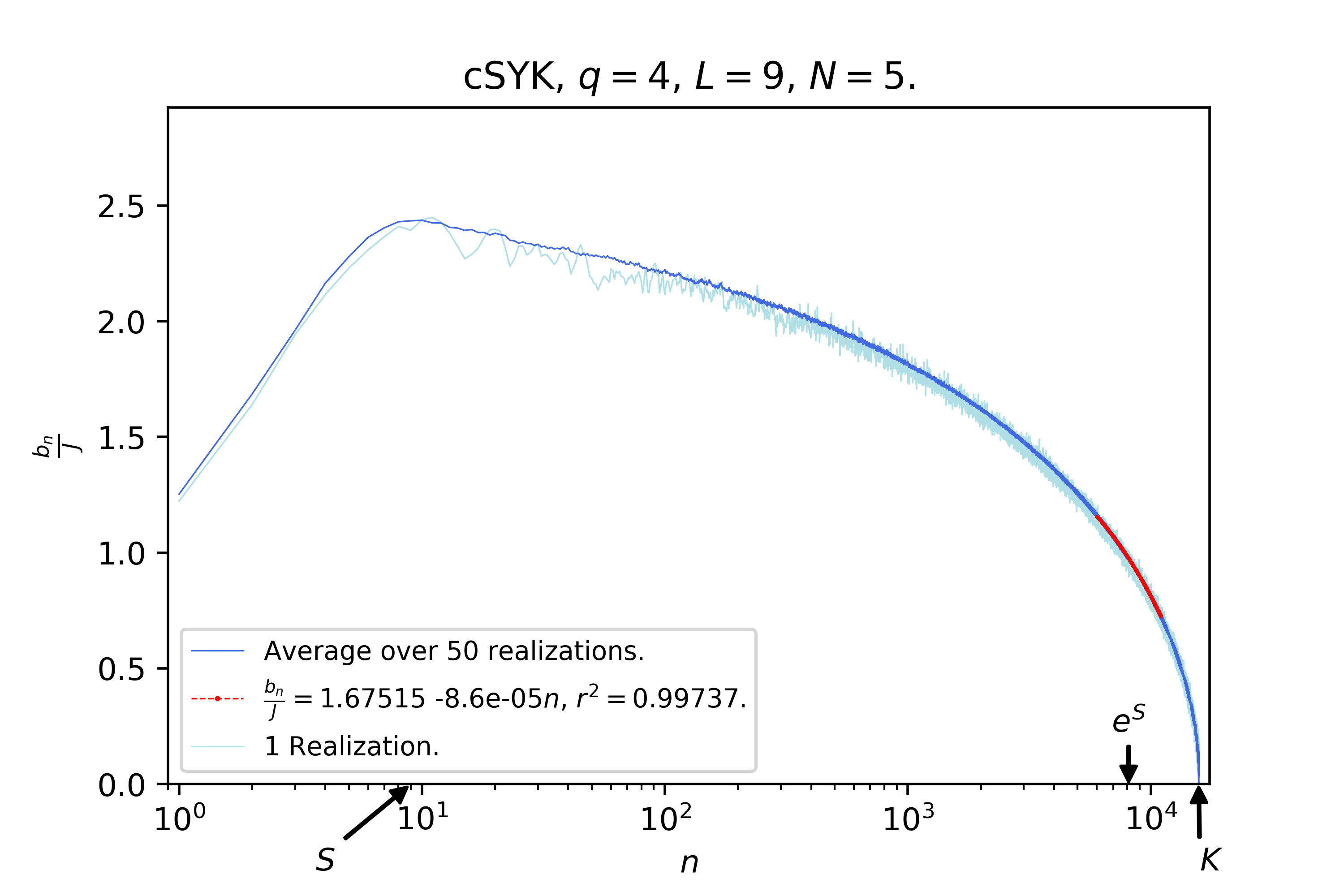}
\end{minipage} \\
\begin{minipage}{.35\textwidth}
\includegraphics[width=1.\linewidth]{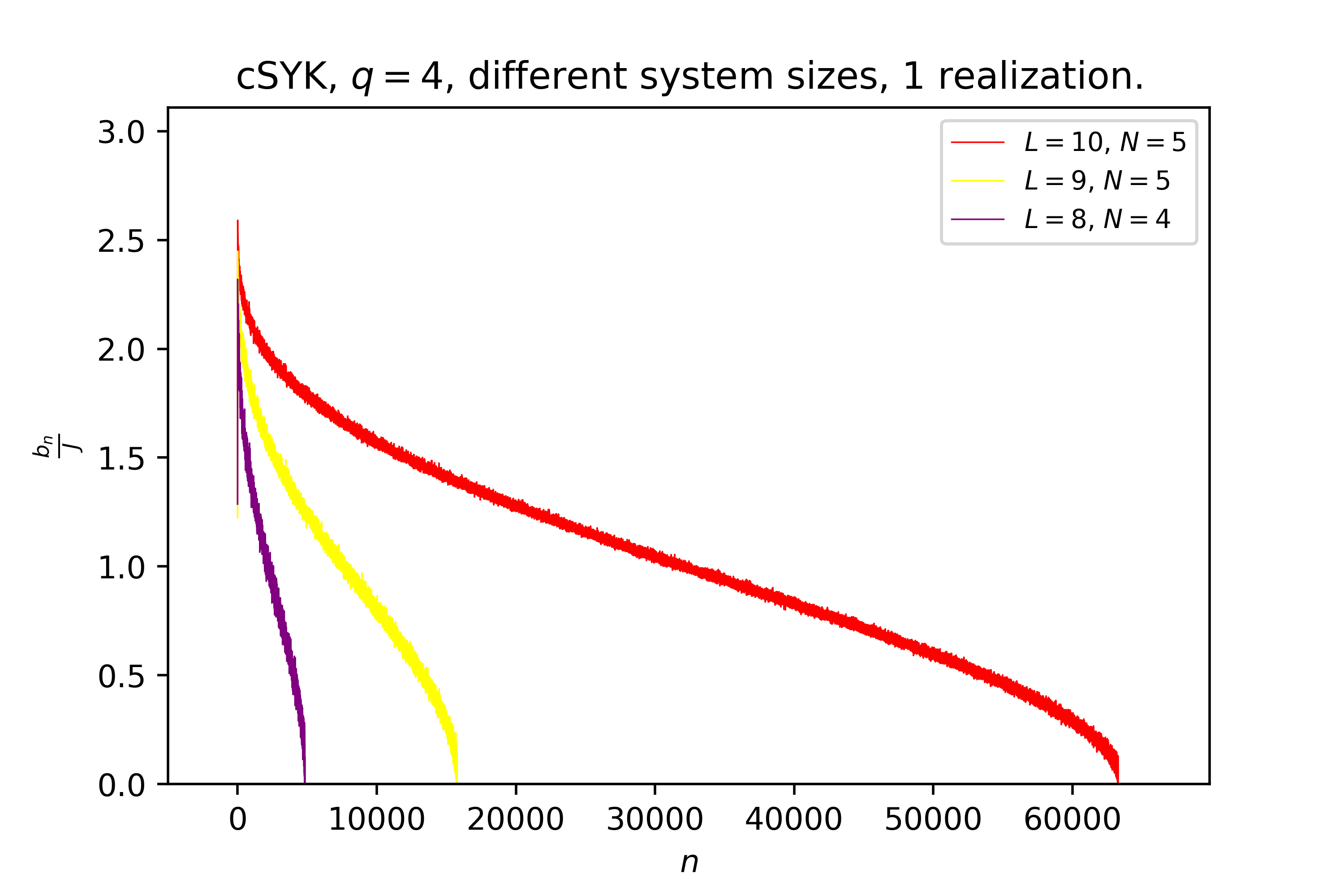}
\end{minipage}\qquad
\begin{minipage}{.35\textwidth}
\includegraphics[width=1.\linewidth]{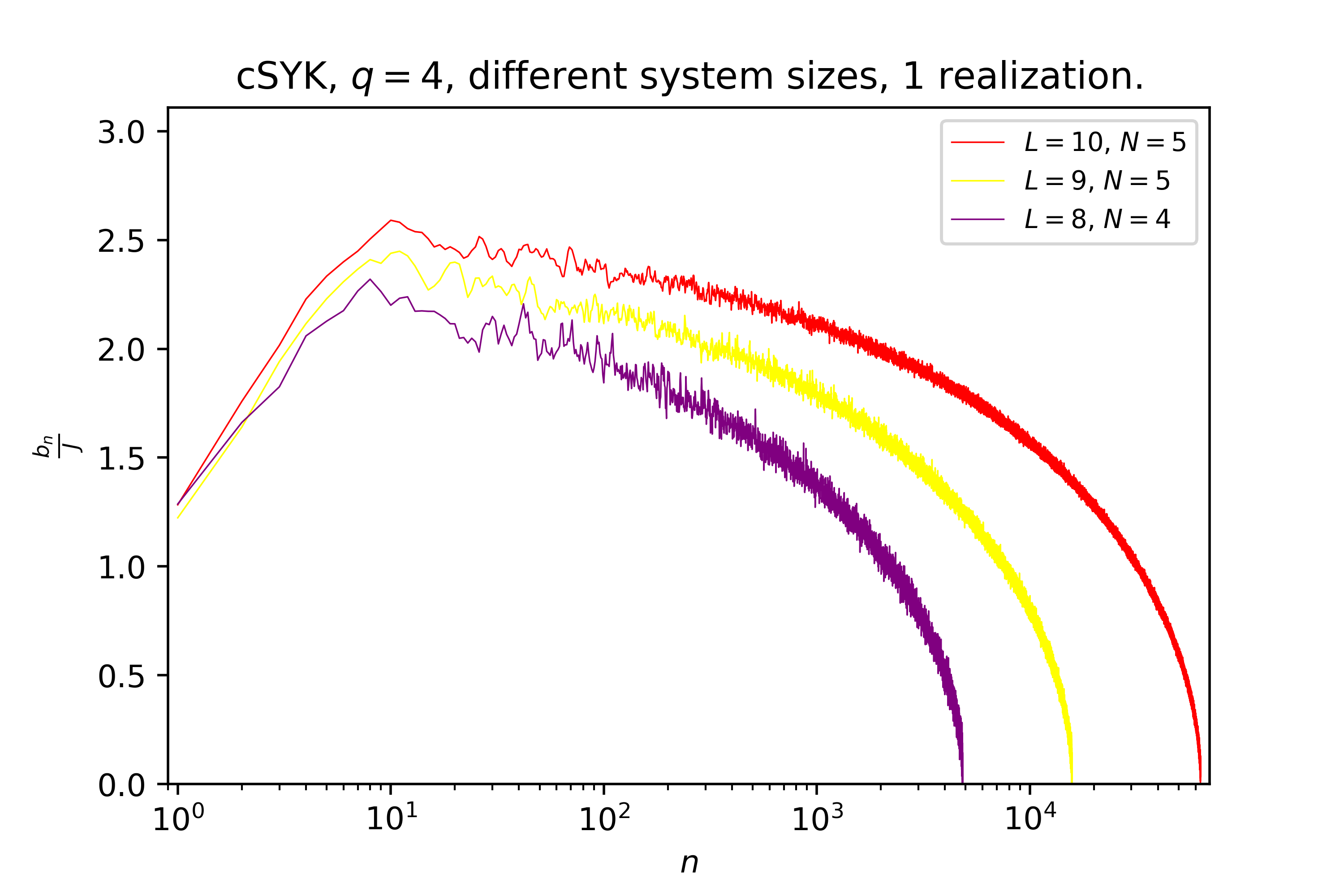}
\end{minipage}
\caption{Lanczos sequences for $L=8,\, N=4$ and $L=9,\, N=5$ and comparison of results for $L=8,9,10$. \textbf{Top row:} Results for $L=8$ in linear  (left panel) and logarithmic (right panel) scale along the horizontal axis. The plots depict both the sequence of a single random realization and the average over $311$ realizations. A linear fit is included in the decaying tail, whose slope approaches numerically the na\"ive estimate $\sim - \frac{1}{K}\approx - 0.000206$.  \textbf{Middle row:} Results for $L=9$ in linear  (left panel) and logarithmic (right panel) scale along the horizontal axis. The plots depict both the sequence of a single random realization and the average over $50$ realizations. A linear fit is included in the decaying tail, whose slope is of the order of the naive estimate $\sim-\frac{1}{K}\approx -6.3\cdot10^{-5}$.  \textbf{Bottom row:} Comparison of the Lanczos sequences for $L=8,9,10$ in linear (left panel) and logarithmic (right panel) scale along the horizontal axis.}
\label{b-sequences_8_9_Comp}
\end{figure}

\begin{figure}
\centering
\begin{minipage}{.35\textwidth}
    \includegraphics[width=1.\linewidth]{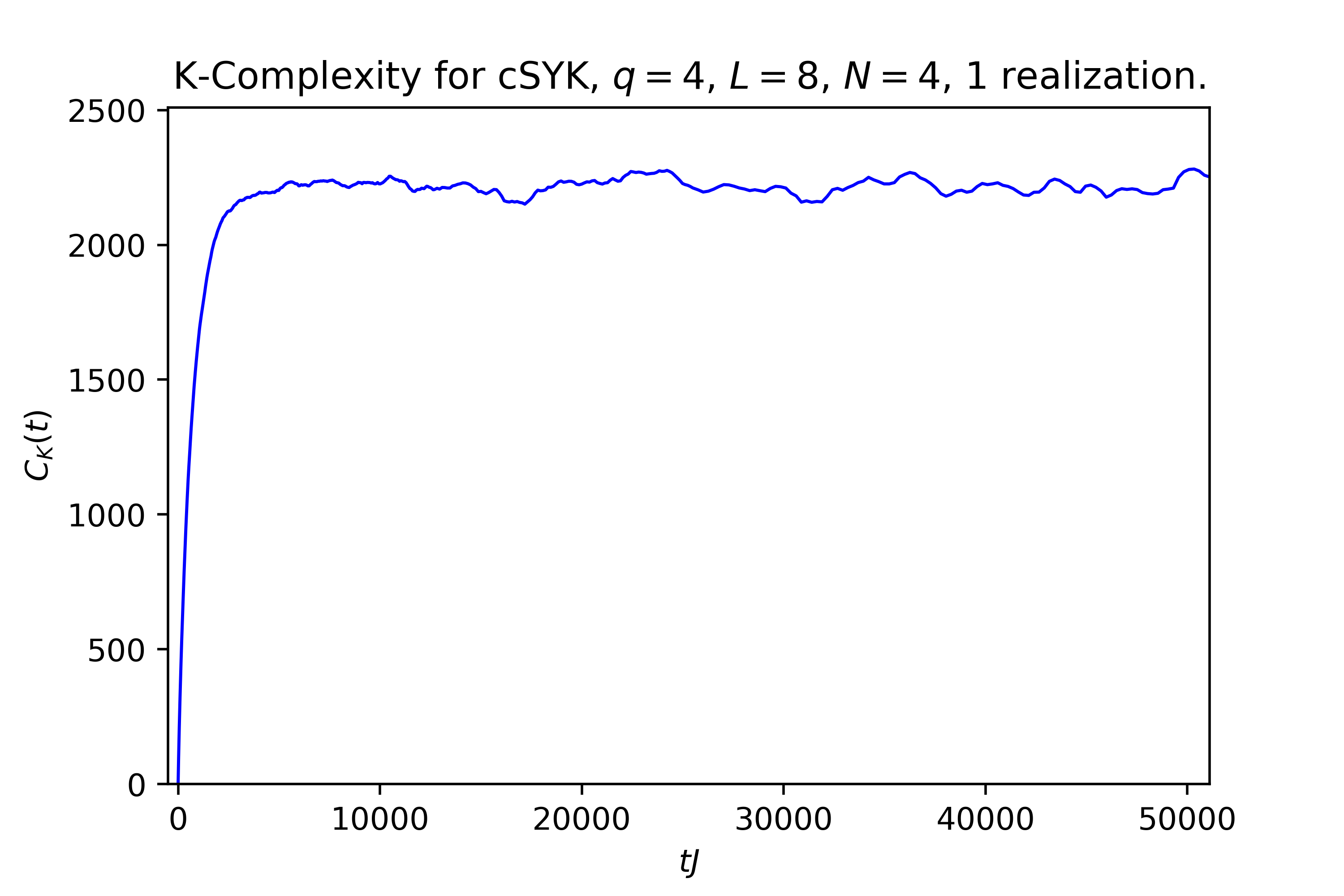}
\end{minipage} \qquad
\begin{minipage}{.35\textwidth}
    \includegraphics[width=1.\linewidth]{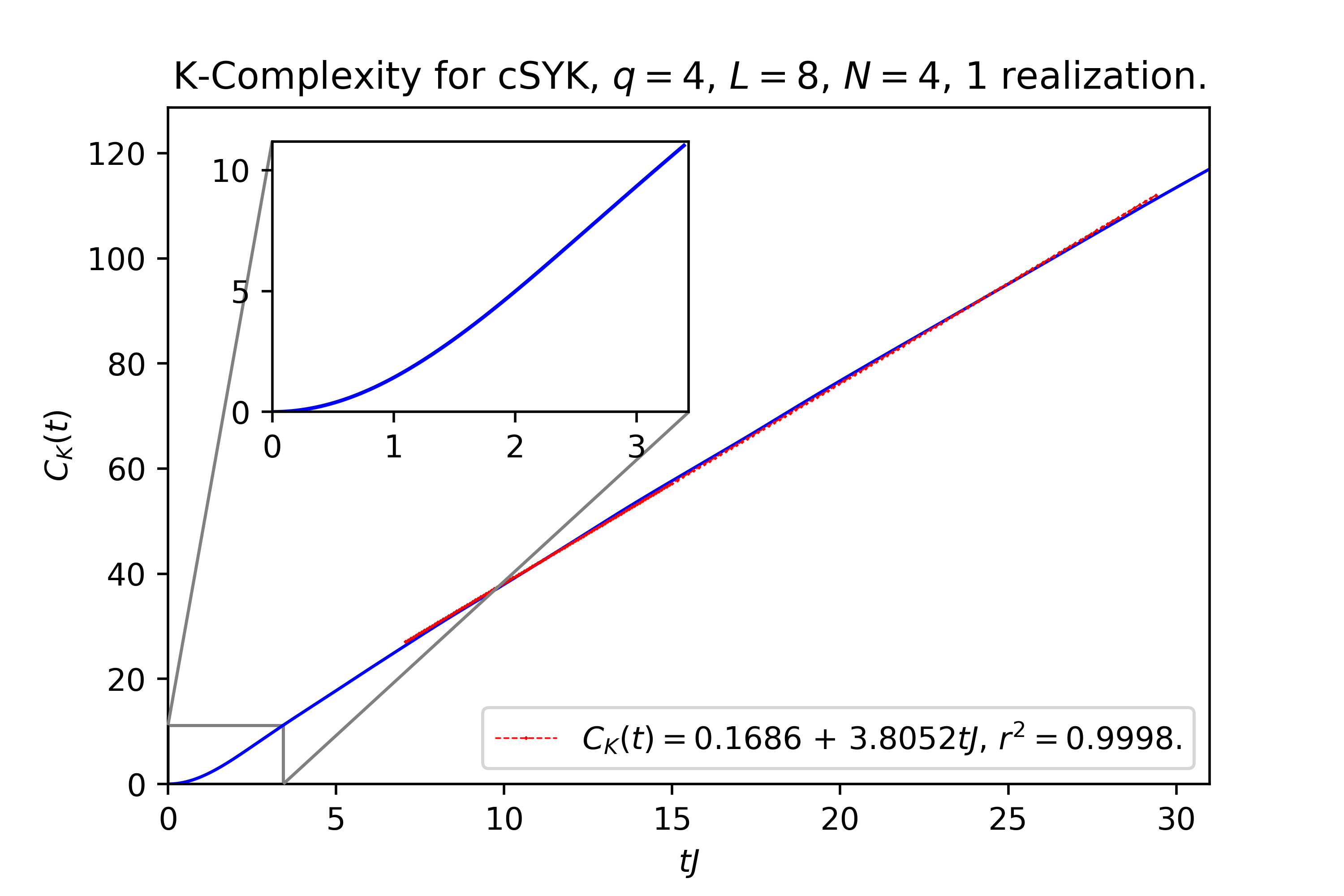}
\end{minipage} \\
\begin{minipage}{.35\textwidth}
\includegraphics[width=1.\linewidth]{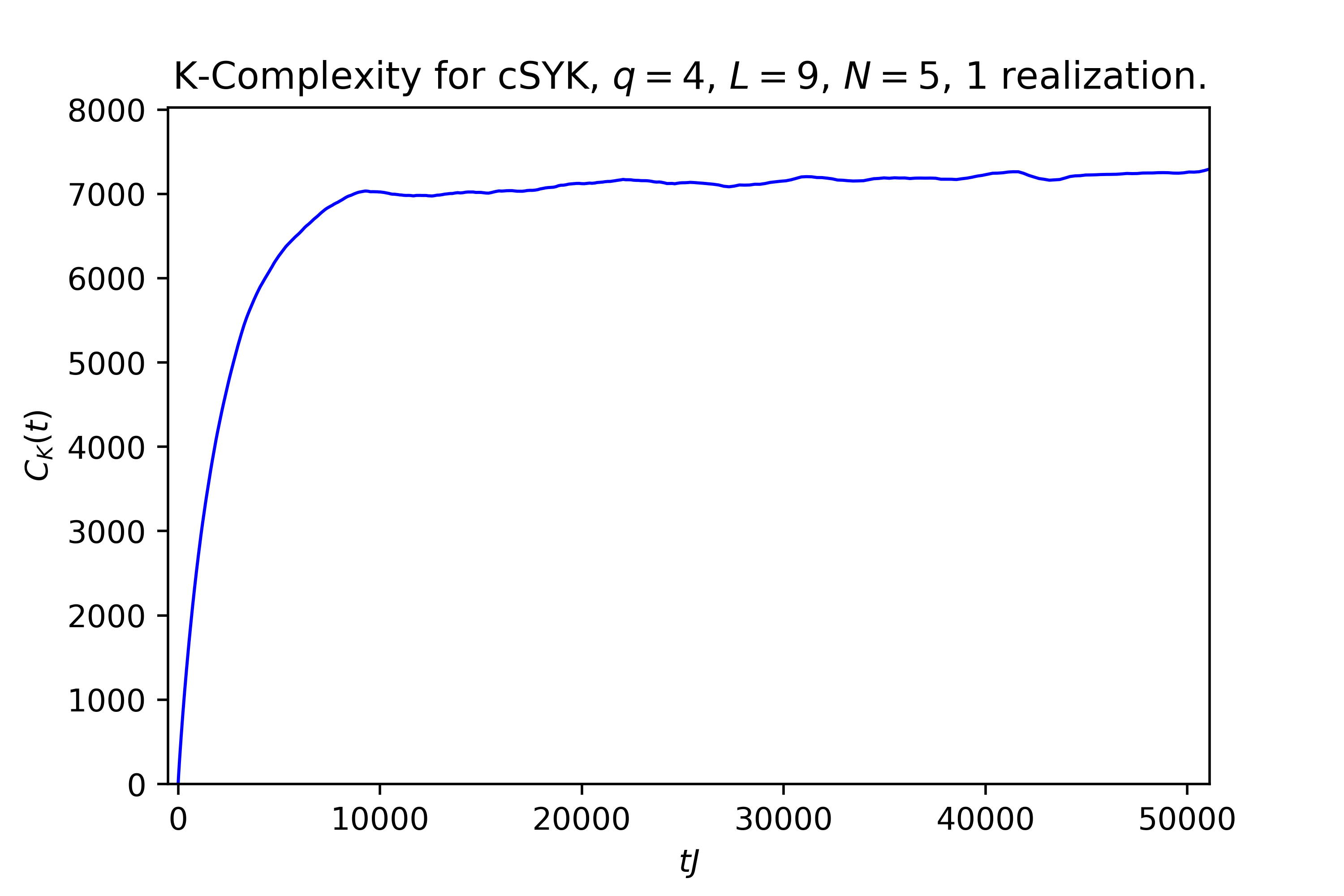}
\end{minipage}\qquad
\begin{minipage}{.35\textwidth}
\includegraphics[width=1.\linewidth]{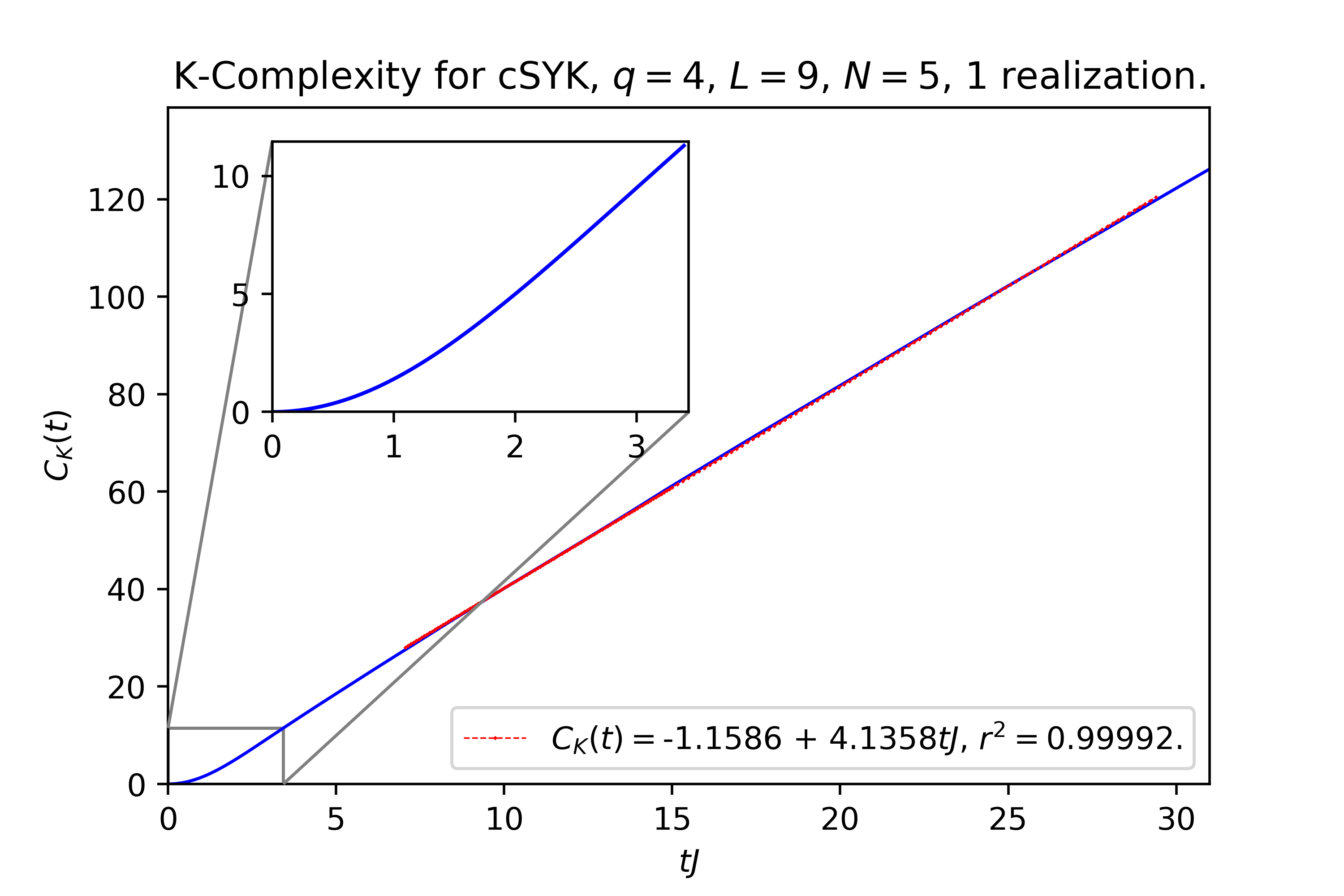}
\end{minipage} \\
\begin{minipage}{.35\textwidth}
\includegraphics[width=1.\linewidth]{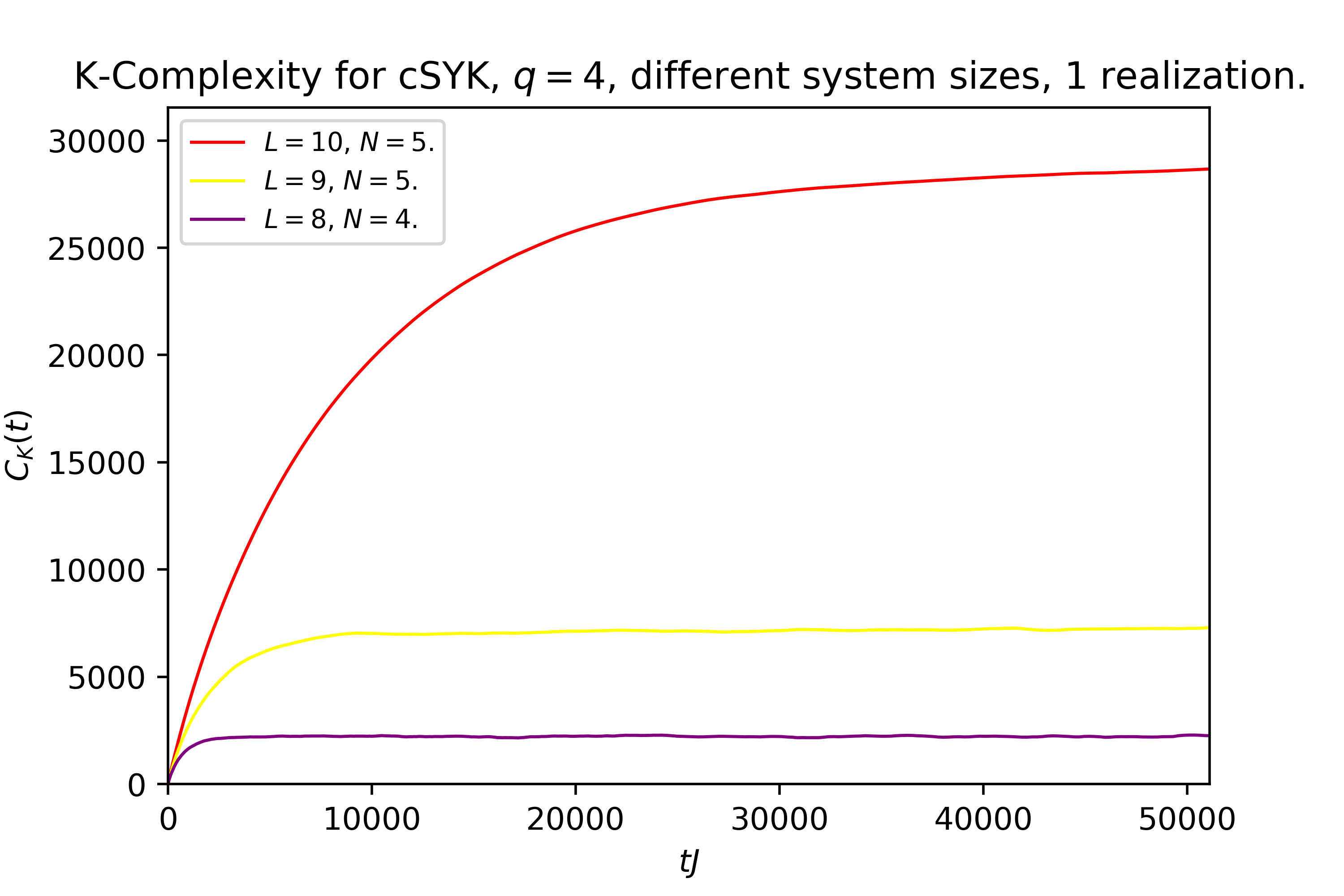}
\end{minipage}\qquad
\begin{minipage}{.35\textwidth}
\includegraphics[width=1.\linewidth]{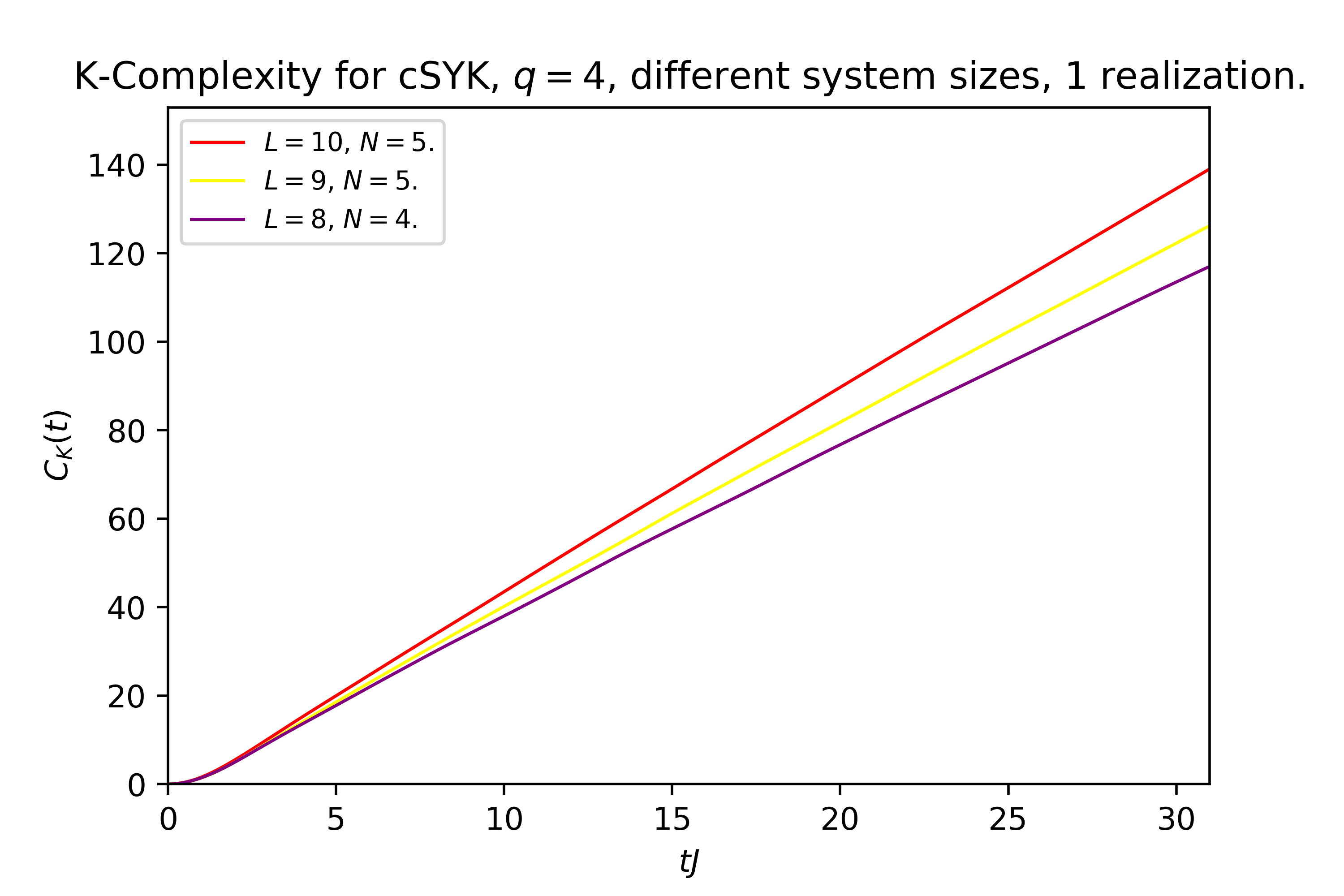}
\end{minipage}
\caption{Results for K-complexity for $L=8,\, N=4$ and $L=9,\, N=5$ and comparison of results for $L=8,9,10$. \textbf{Top row:} Results for $L=8$; for exponentially long times (left panel) and for early times (right panel). Note the change in behaviour from very early times (inset) and later times. The value at saturation is near $\sim\frac{K}{2}=2415.5$. \textbf{Middle row:} Results for $L=9$; for exponentially long times (left panel) and for early times (right panel). The value at saturation is near $\sim\frac{K}{2}=7875.5$. \textbf{Bottom row:} Comparison of results for $L=8,9,10$ for exponentially long times (left panel) and for early times (right panel).}
\label{KC_8_9_Comp}
\end{figure}

\begin{figure}
\centering
\begin{minipage}{.3\textwidth}
    \includegraphics[width=1.\linewidth]{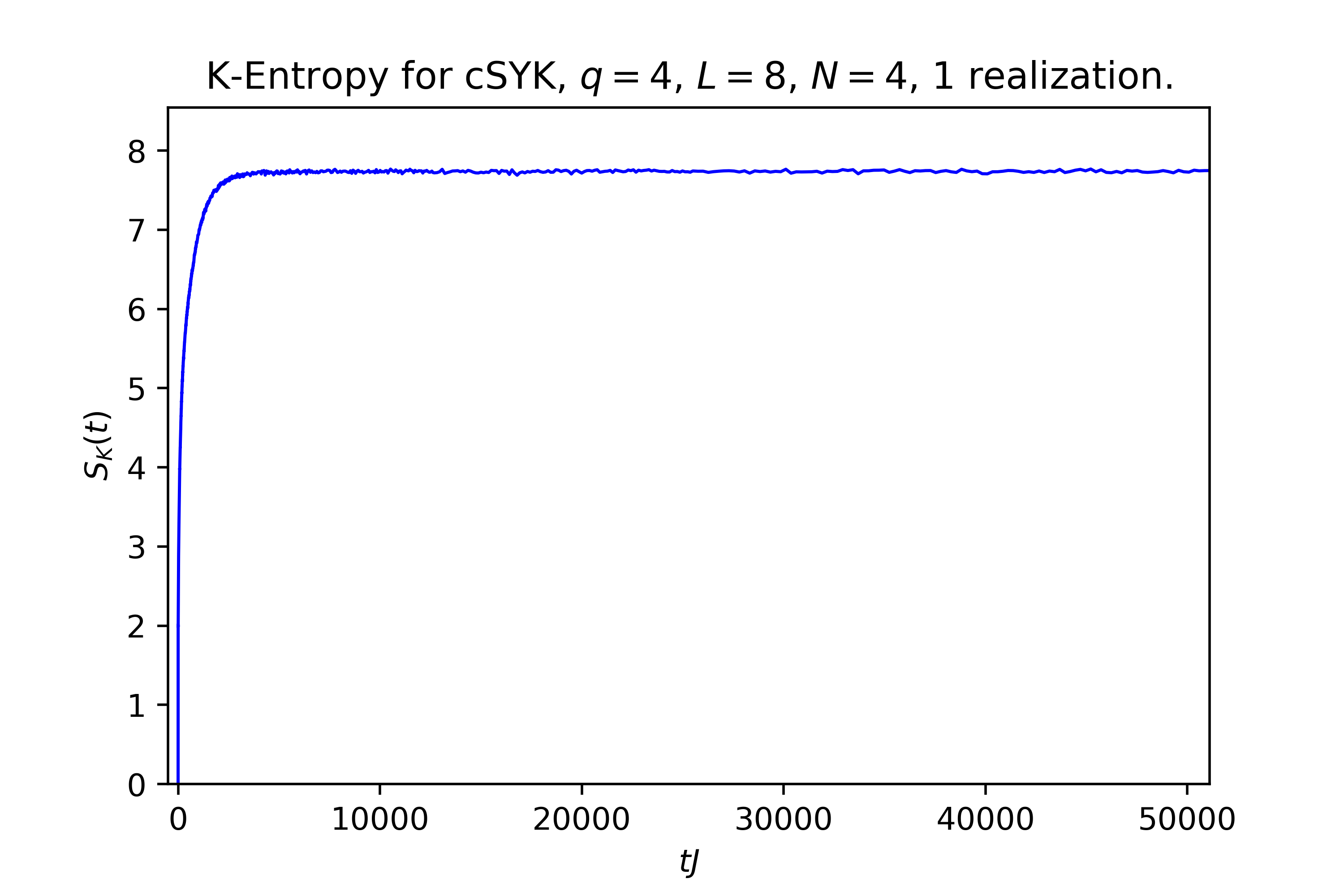}
\end{minipage} \quad
\begin{minipage}{.3\textwidth}
    \includegraphics[width=1.\linewidth]{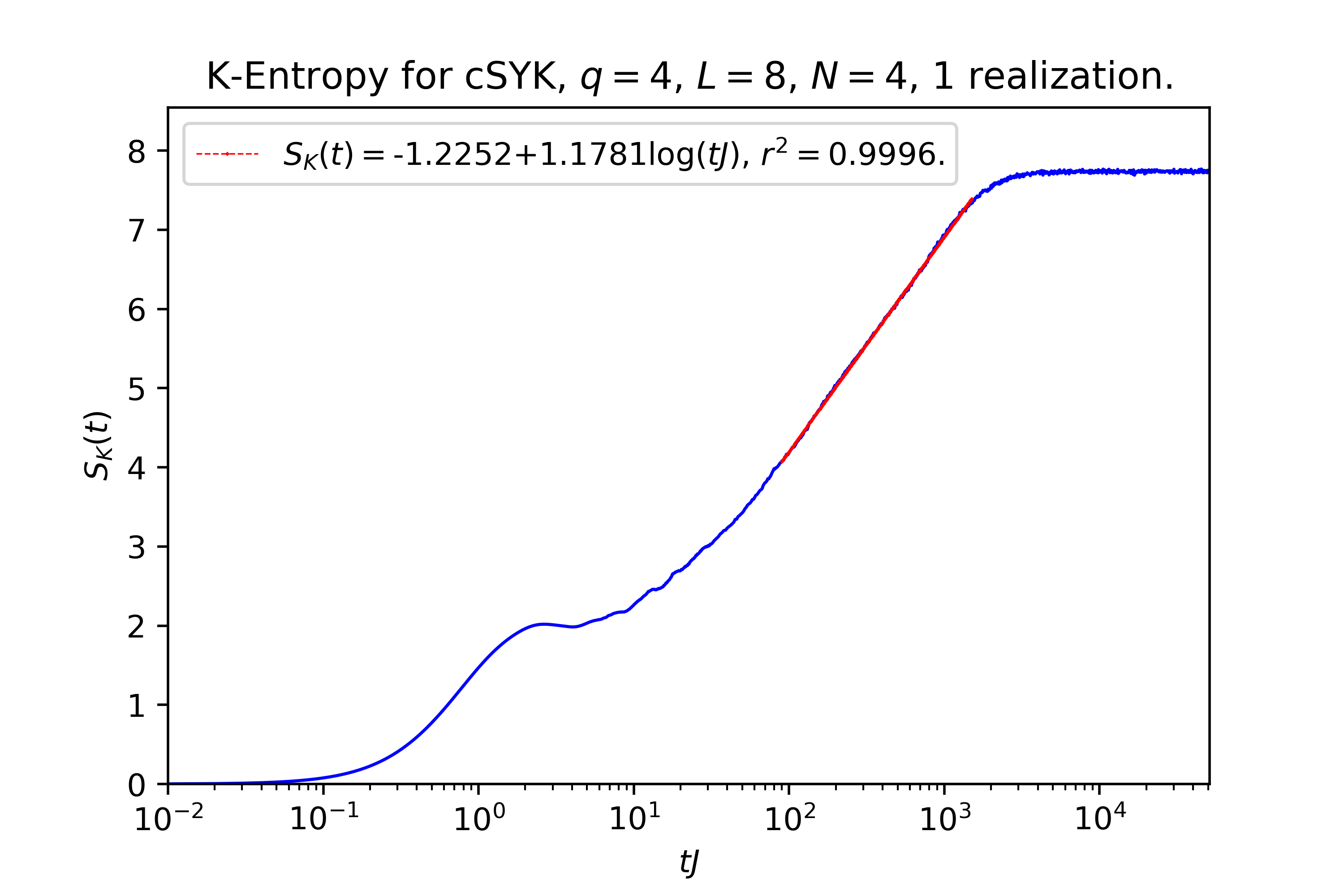}
\end{minipage} \quad
\begin{minipage}{.3\textwidth}
    \includegraphics[width=1.\linewidth]{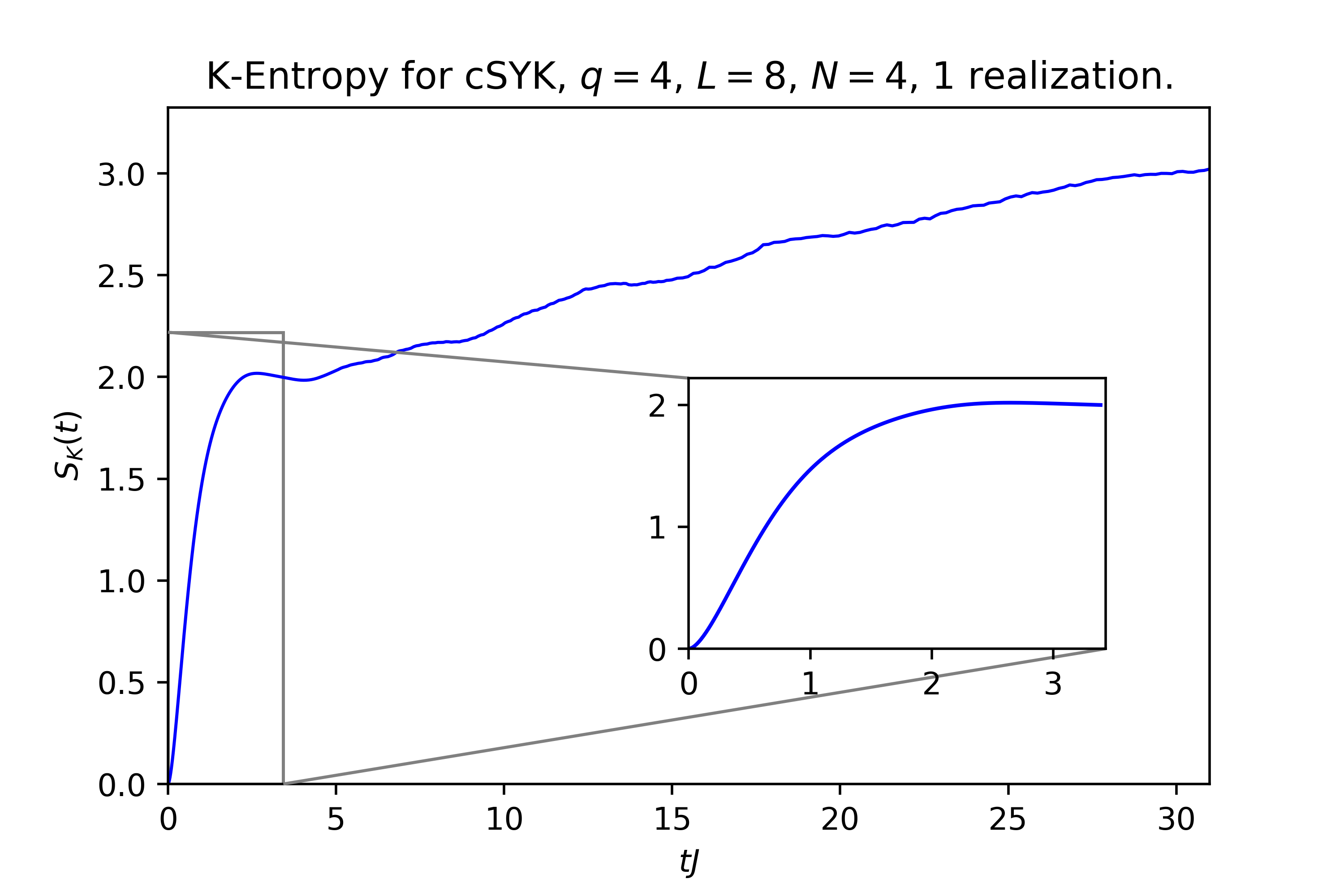}
\end{minipage}\\
\begin{minipage}{.3\textwidth}
    \includegraphics[width=1.\linewidth]{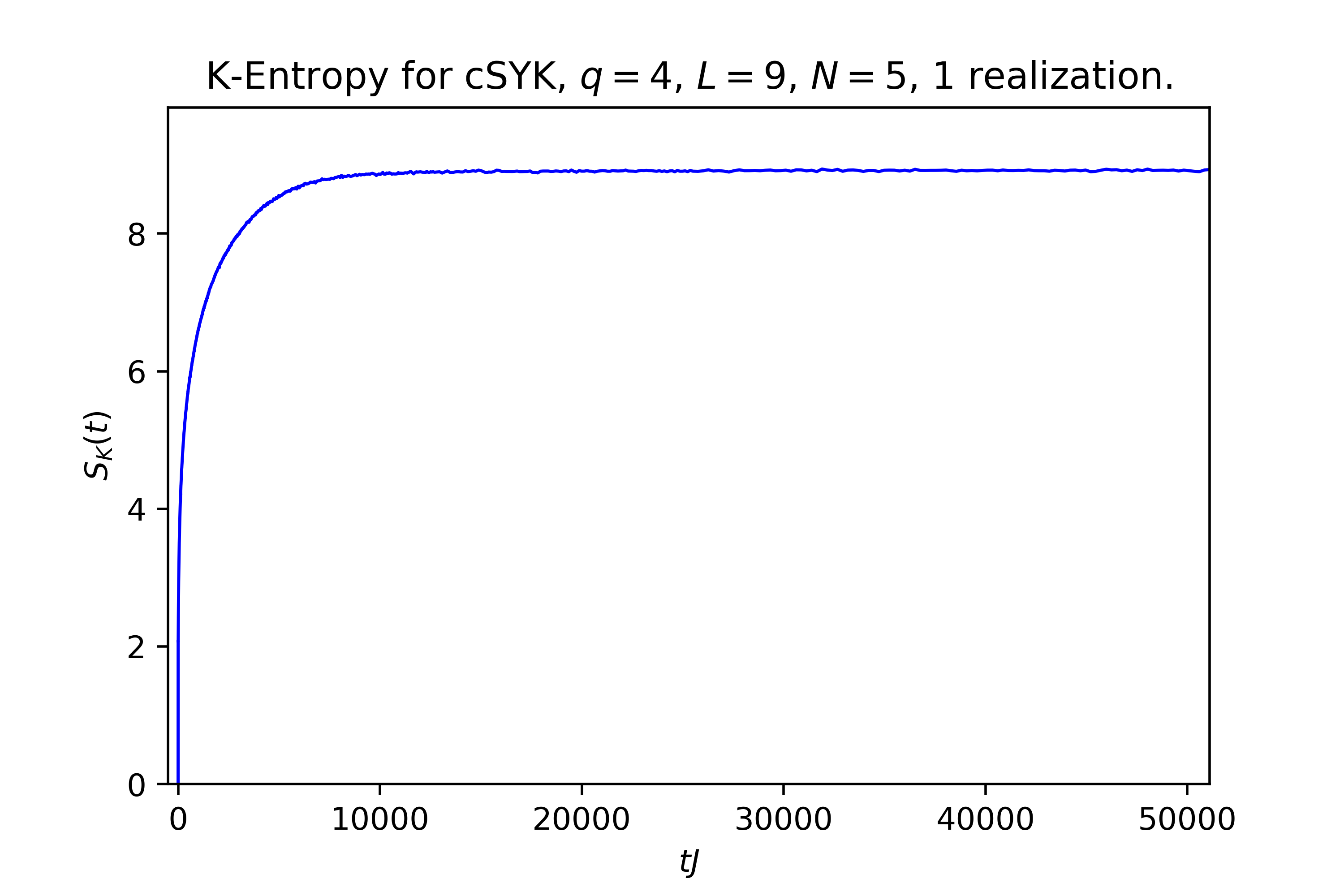}
\end{minipage} \quad
\begin{minipage}{.3\textwidth}
    \includegraphics[width=1.\linewidth]{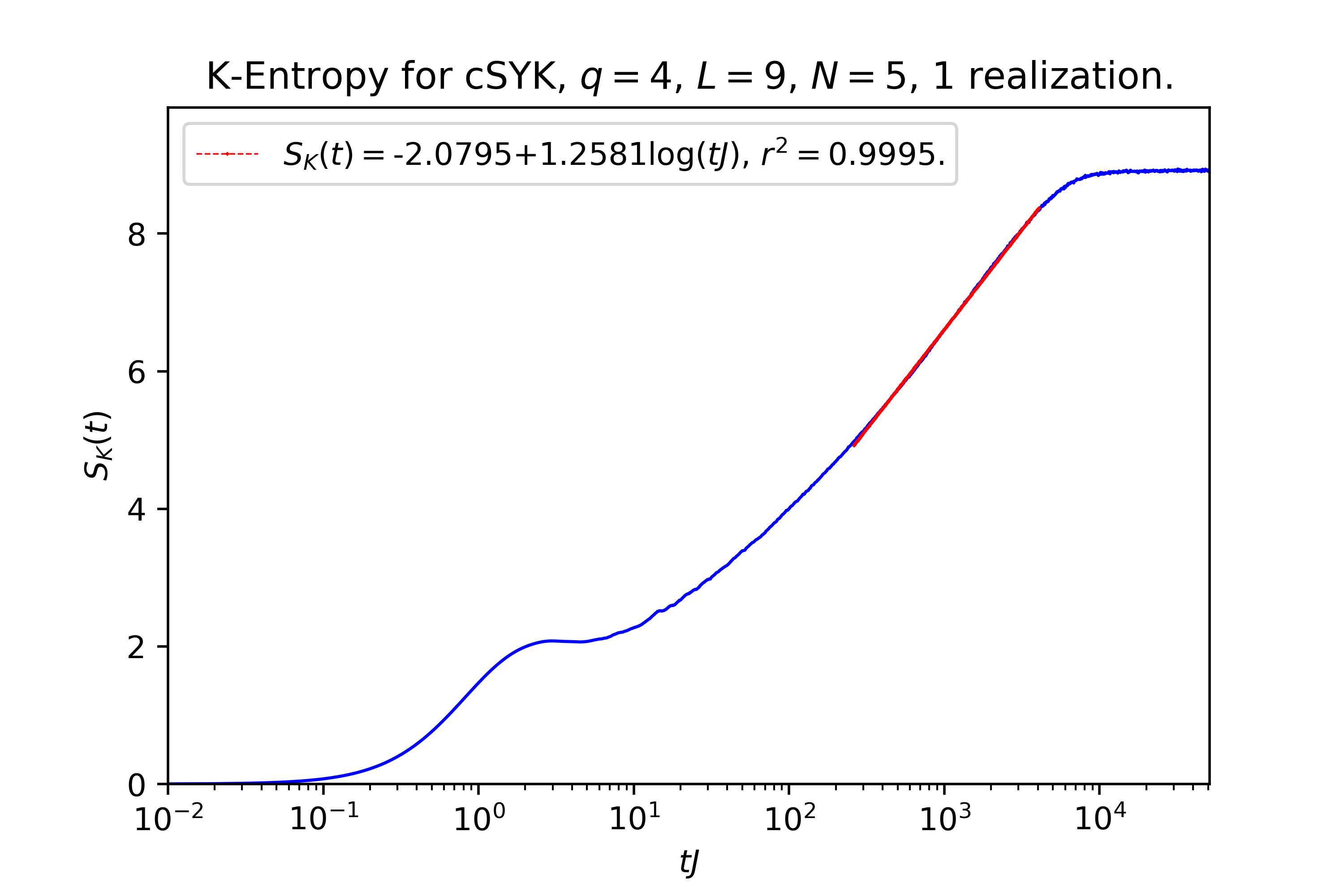}
\end{minipage} \quad
\begin{minipage}{.3\textwidth}
    \includegraphics[width=1.\linewidth]{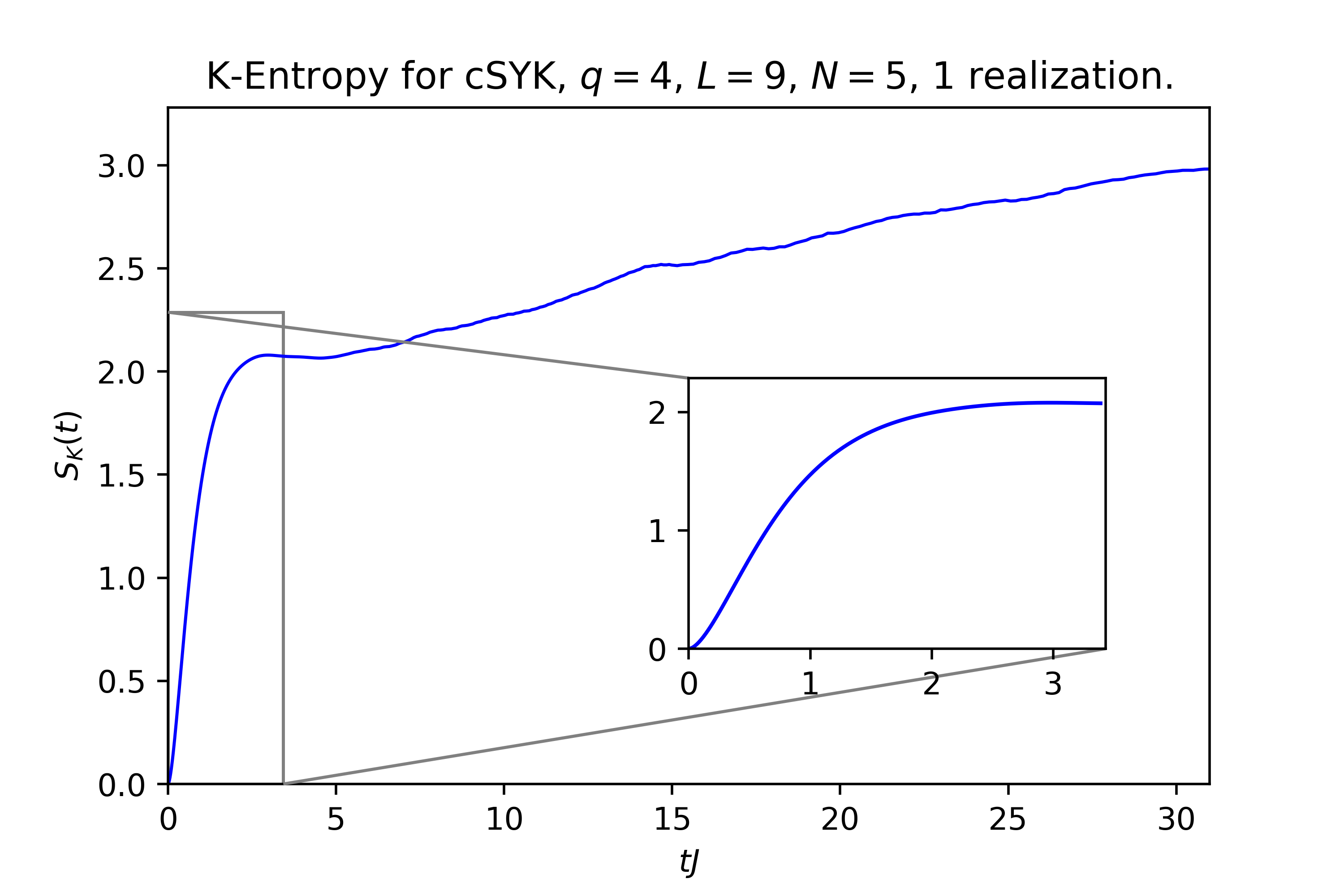}
\end{minipage}
\caption{Results for K-entropy for $L=8,\, N=4$ and $L=9,\, N=5$. \textbf{Top row:} Results for $L=8$; for exponentially long time scales in linear (left panel) and logarithmic (middle panel) scale along the horizontal axis, and for early times (right panel). The saturation value is near $L=8$.  \textbf{Bottom Row:} Results for $L=9$; for exponentially long time scales in linear (left panel) and logarithmic (middle panel) scale along the horizontal axis, and for early times (right panel). The saturation value is near $L=9$.}
\label{KS_8_9_Comp}
\end{figure}

\begin{figure}
    \centering 
    \begin{minipage}{.4\textwidth}
    \includegraphics[width=1.\linewidth]{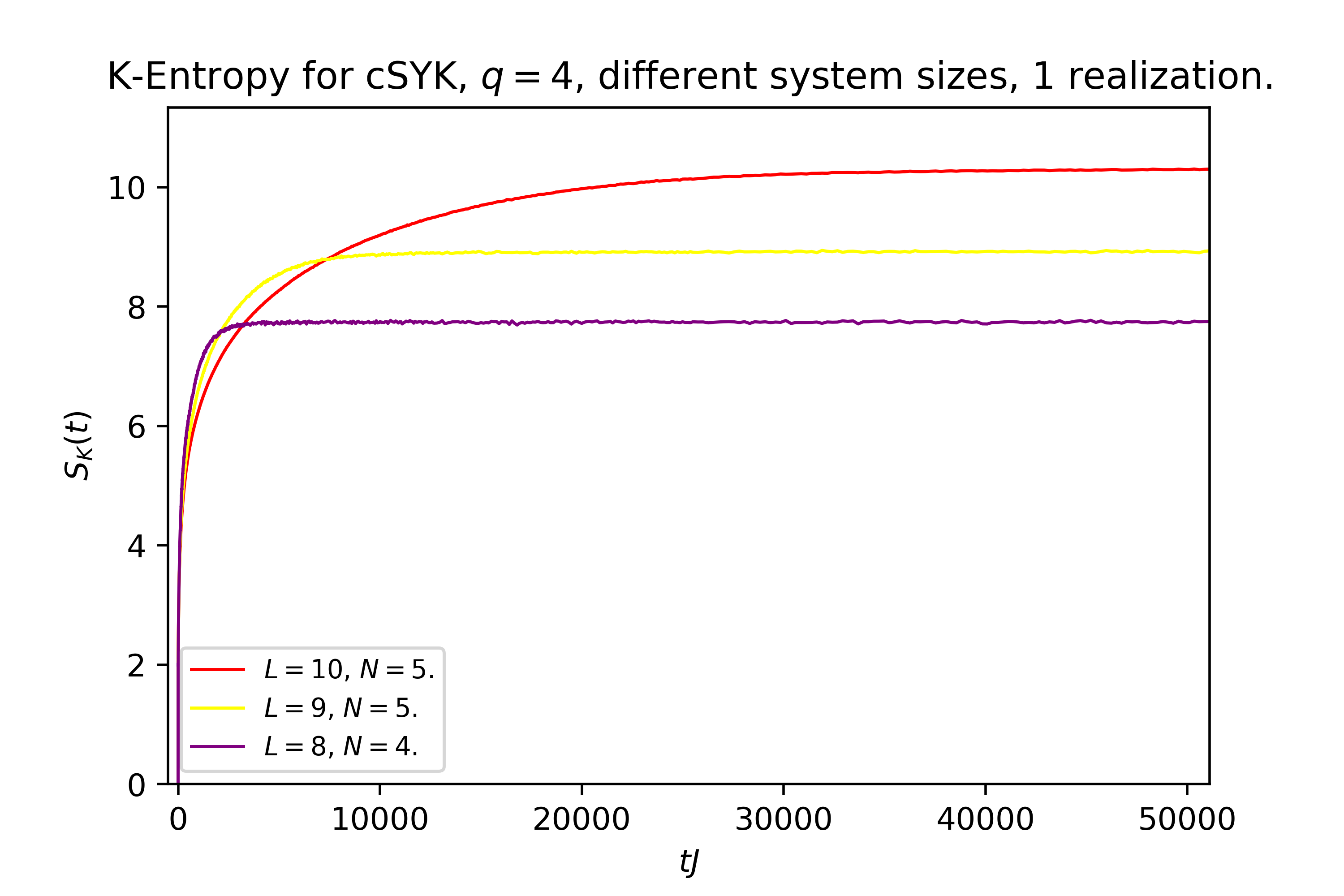}
    \end{minipage} \quad
    \begin{minipage}{.4\textwidth}
    \includegraphics[width=1.\linewidth]{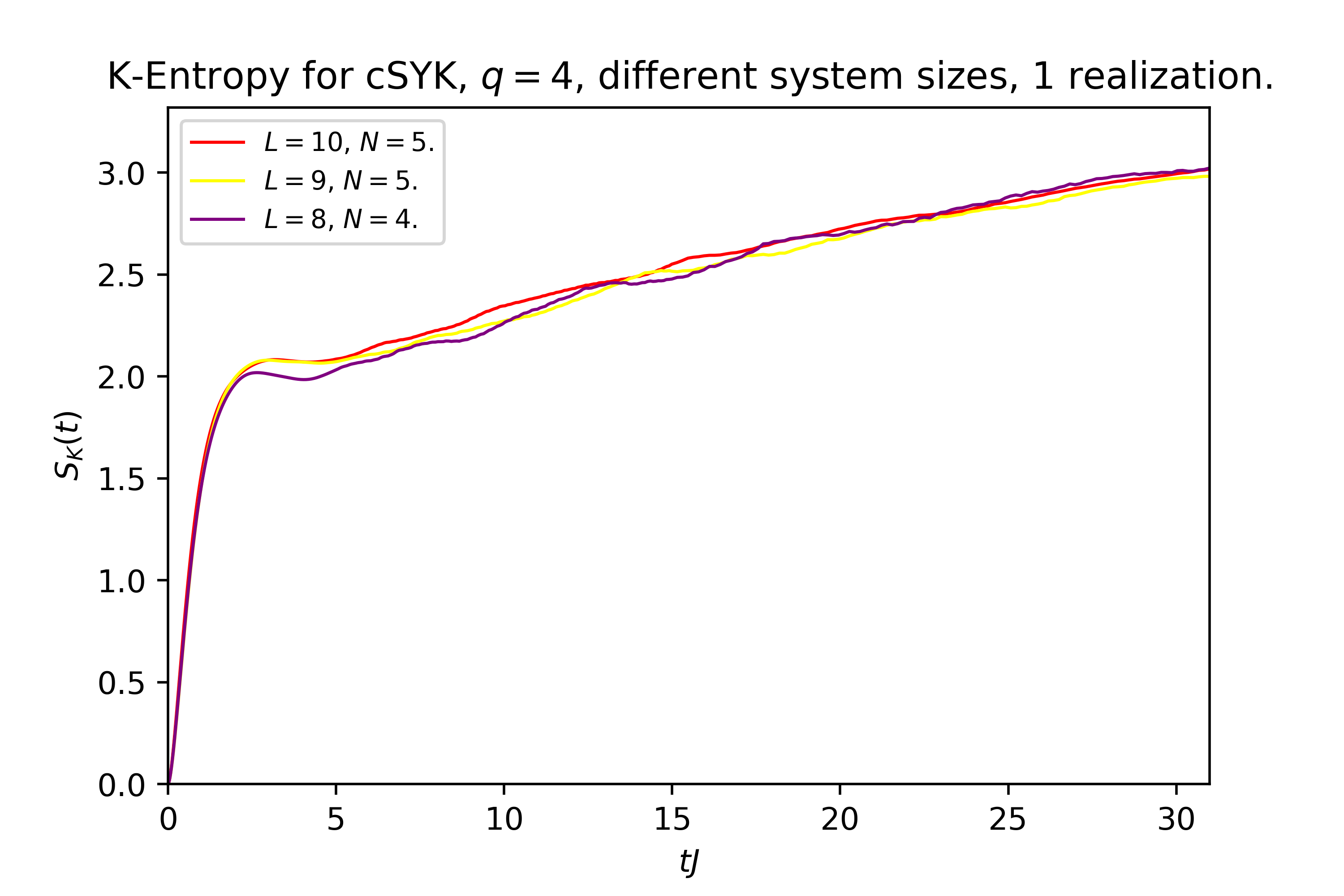}
    \end{minipage}
    \caption{Comparison of results for K-entropy for $L=8,9,10$; for long time scales (left panel), and for early times (right panel).}
    \label{KS_Size_Comp}
\end{figure}

We studied samples with $L=8,9$ and $10$ sites. The numerical results for SYK$_4$ with $L=10$ are shown in Figures \ref{b-L10-main}, \ref{KC-L10-main} and \ref{KS-L10-main}. For details on the scaling properties of the $b$-sequence, K-complexity and K-entropy with system size, see Figures \ref{b-sequences_8_9_Comp}, \ref{KC_8_9_Comp}, \ref{KS_8_9_Comp} and \ref{KS_Size_Comp}, as well as the summarizing Table \ref{tab:summary}.

The global picture emerging from these results can be summarized in the following bullet points:

\begin{itemize}
    \item \textbf{The Descent and its associated Lanczos sequence.} The length of the Lanczos sequence saturates its upper bound. It features a period of initial growth up to $n\sim S$, followed by a regime of slow decrease to zero with roughly constant negative non-perturbative slope of order $\sim - e^{-2S}$, the Descent.
    
    \item \textbf{K-complexity} features a transition from exponential growth at very early times to linear increase. At exponentially late times it saturates at half of the Krylov dimension, since by then the operator is uniformly distributed over the Krylov basis.
    \item \textbf{K-entropy} grows linearly up to scrambling time, and then transitions to a logarithmic growth phase that continues until saturation around $S_K\sim S$ at exponentially late times.
\end{itemize}

The relation between the Lanczos sequence $b_n$ and quantities like $C_K(t)$ and $S_K(t)$ is highly non-linear, which is why disorder averages need to be performed with caution. The Lanczos sequence is not really a physical observable and averaging is just used as a tool to gain knowledge about the envelope of its profile. However, in order to obtain averaged K-complexity or K-entropy as a function of time, one should compute these quantities separately for each random realization of the Hamiltonian and average over the outcomes in a final step. 
Averaging over the Lanczos sequence before computing $C_K$ and $S_K$ results in a smoothing of the $b$-sequence that stops the wave packet from randomizing efficiently before reaching the edge of the Krylov chain, being therefore reflected due to the boundary conditions. These rebounds would be reflected by large, un-physical oscillations in the profiles of $C_K$ and $S_K$, an effect that has been confirmed by numerical simulations. 



\begin{table}
    \centering
    \begin{tabular}{|c|c|c|c|c|c|c|}
        \hline
         $\log(L)$ & $L$ &  $D $ & $K/2 $ & $b_n$ slope & $C_K$ sat. & $S_K$ sat.  \\
         \hline
         2.07944 & 8 & 70 & 2415.5 & $-0.00026$ & 2215  & 7.7  \\
         2.19722 & 9 & 126 & 7875.5 & $- 8.6 \times 10^{-5}$ & 7254 & 8.9 \\
         2.30258 & 10 & 252 & 31626.5 & $-2.21 \times 10^{-5}$ & 29,618 & 10.3 \\
         \hline
    \end{tabular}
    \caption{Summary Table of the numerical phenomenology observed. Values averaged over several random realizations. In terms of $S$, $\log(L)\sim\log(S), L\sim S, D \sim e^S$ and $K\sim e^{2S}$.}
    \label{tab:summary}
\end{table}

\clearpage

\section{Discussion} \label{Sec: Discussion}
On the analytical side, we have found an algorithmic expression for the dimension of Krylov space $K$. This quantity is very sensitive to the degeneracy structure of the spectrum and is always strictly smaller than the dimension of operator space $D^2$, since it satisfies $K\leq D^2-D+1$. Typical operators in chaotic systems are expected to saturate this bound, due to the absence of degeneracies, whereas in integrable systems $K$ will typically be significantly smaller than its upper bound (at least effectively up to late times), as verified analytically for SYK$_2$. Furthermore, numerical observations in SYK$_4$ and RMT suggest that the $b_n$-profile featuring a slow decrease to zero given by a non-perturbative slope of order $e^{-2S}$ -- the Descent -- may be a generic feature of quantum chaotic systems. It is worth noting that, since K-complexity is nothing but an average position in Krylov space, the above bound on $K$ translates directly into an upper bound for $C_K(t)$. We note that although K-complexity features the profile expected from circuit complexity, there are notable differences between them: Krylov complexity does not depend on an arbitrary tolerance parameter, and it does depend on both the Hamiltonian of the system and the operator whose K-complexity is measured over time. This may have impact in situations that involve more than one operator insertion.

We end by returning to an issue we emphasized at the outset, namely that K-complexity is naturally bounded. This is a direct consequence of the finite dimensionality of the Hilbert space, and thus has the same origin as the plateau in the spectral form factor and correlation functions. In fact, such a plateau is known \cite{Altland:2020ccq,Altland:2021rqn} to be due to spectral correlations, namely level repulsion, which are probed at late times: This is qualitatively consistent with the discussion given in section \ref{sect:Lanczos_generic_behavior}, where it was argued that the last elements of the Krylov basis, which are explored by the evolving operator at late times, probe finer and finer eigenvalue differences\footnote{Note: This comment is not present in the original publication \cite{I} and has been added in the present manuscript.}. We conjecture that other late-time universal characteristics of quantum chaos, namely the dip-ramp-plateau structure \cite{Cotler:2016fpe}, also leave their imprint on K-complexity. The putative bulk realization of K-complexity should therefore be sensitive to bulk Euclidean wormholes and baby universes \cite{Saad:2019lba}, in ensemble-averaged systems, such as SYK, as well as in individual quantum chaotic systems, as has been emphasized in \cite{Altland:2020ccq}.

%% file: content/Chapter04.tex
\chapter{\rm\bfseries Integrable systems. Localization in Krylov space}
\label{ch:chapter04_Integrable}

This Chapter merges contributions \cite{II,III} in order to give an overview of the phenomenology of Krylov complexity for strongly-interacting integrable systems, mainly at late times, as opposed to chaotic ones. The works present and support the proposal of a novel Anderson localization effect operating in Krylov space that induces a smaller late-time K-complexity saturation value in integrable systems, even if their Krylov dimension is maximal. The proposal is analyzed for an XXZ spin chain with \cite{III} and without \cite{II} an integrability-breaking defect, which allows to interpolate between integrable and chaotic regimes. The current Chapter complements the discussion in the previous one on K-complexity for finite chaotic systems, and it also intends to contribute to the debate reviewed in section \ref{sect:KC_chaos} about the suitability of K-complexity as a probe of quantum chaos. Combined, the findings of \cite{I,II,III} indicate that the K-complexity saturation value is indeed sensitive to the chaotic or integrable nature of a system, which additionally makes this notion an attractive holographic complexity candidate (cf. section \ref{sect:KC_holog}). The text of this Chapter is mostly a reproduction of the contributions \cite{II,III}, which have been mildly edited in order to avoid redundancies and to make them fit within the structure of the Thesis.


\section{Introduction}\label{sect:introduction_chapter_Integrable}

At a very basic level, an appropriate notion of complexity should enable us to think of chaotic strongly correlated many-body systems, or chaotic strongly coupled quantum field theories, as being more complicated or complex than their free, or more generically interacting integrable cousins. In other words, we expect chaotic theories to time evolve simple states and operators efficiently into more complex states and/or operators, while on the other hand integrable non-chaotic theories are expected not to increase complexity as fast. 

As reproduced in the previous Chapter, the project leading to publication \cite{I} developed highly accurate and stable numerical methods, which in conjunction with analytical bounds, allowed to determine the complete $b-$sequences and resulting Krylov dynamics up to late time for a class of strongly interacting chaotic many-body systems. We found that the Krylov space in SYK$_4$ \cite{Kitaev:2015,Sachdev:2015efa,Maldacena:2016hyu} for ETH operators \cite{PhysRevA.30.504, PhysRevA.43.2046, Srednicki_1999} is maximal with a dimension that scales exponentially with system size, and that the associated Lanczos coefficients $b_n$ display an initial quasi-linear growth up to $n\sim S$ followed by a very slow nearly linear non-perturbative descent to zero at $n\sim e^S$. Note that SYK$_4$ is a rare example of a strongly interacting system, in which ETH can be shown to apply both numerically and analytically, \cite{ Sonner:2017hxc, Nayak:2019khe, Nayak:2019evx}. The aforementioned behavior of the Lanczos sequence for ETH operators then induces a K-complexity profile that starts growing exponentially up to the scrambling time $t_s\sim \log S$ and then transitions to a long period of linear growth up to eventual saturation at a value exponentially big in system size at times of the order of the Heisenberg time $t_H\sim e^S$. The qualitative shape of the profile obtained was in agreement with that of the time dependent profile expected from bulk considerations \cite{Susskind:2014rva,Brown:2015bva,Brown:2015lvg,Susskind:2018pmk}, discussed in section \ref{sect:KC_holog}. While the emphasis of that work was put on the behavior of chaotic models, it is important to contrast their behavior with that of integrable systems. This line of inquiry was initiated in \cite{I} by studying the SYK$_2$ model, and we found that its $b-$sequence terminates far below the bound $D^2  - D +1$ which in turn leads to K-complexity saturating far below the value of the chaotic SYK$_4$ model. With this motivation in mind, in \cite{II} we investigate whether under-saturation of K-complexity is a ubiquitous phenomenon in more generic integrable models, allowing thus to distinguish their complexity signature from chaotic ones. The main thrust of the argument follows from studying the $b-$sequence and its associated one-particle hopping problem in integrable theories, where we argue that the presence of near-degeneracies resulting from the Poissonian nature of the spectrum of these systems implies an erratic structure akin to that encountered in random hopping models that show wave function localization. More precisely, we will map the Krylov chain dynamics to an Anderson random hopping model with off-diagonal disorder, rather than the diagonal disorder of the original Anderson problem  \cite{Anderson_AbsDiff}. Our main finding is the observation of a disordered structure in the Lanczos sequence of a local operator in the XXZ spin chain, a canonical example of strongly-interacting integrable system solvable via the Bethe Ansatz, which induces some degree of localization that prevents maximally efficient propagation of the particle in Krylov space, resulting in a K-complexity saturation value at late times that is lower than that obtained in previous studies for SYK$_4$ \cite{I}. We confirmed that this effect was due to disorder by studying a phenomenologically-inspired model with a disordered Lanczos sequence with adjustable fluctuation strength, yielding results that reproduce closely those obtained from XXZ.


\section{Dynamics in the Krylov chain}\label{Sect_Dynamics_Krylov}

\subsection{Krylov space dynamics as a hopping problem} \label{Subsection_Krylov_hopping}

By construction, Krylov space is the minimal subspace of operator space that contains the time-evolving operator $\mathcal{O}(t)$ at all times. Thus, in order to keep track of this time evolution,  we can decompose $\Big|\mathcal{O}(t)\Big)$ in terms of the Krylov elements:

\begin{equation} \label{phi_def}
    |\mathcal{O}(t)) = e^{it\mathcal{L}} |\mathcal{O}) = \sum_{n=0}^{K-1} \phi_n(t)|\mathcal{O}_n)\,,
\end{equation}
where, for simplicity, we choose not to use the redefinition $\phi_n(t)=i^n\varphi_n(t)$ this time. The object $\phi_n(t)$ can be viewed as the wave function over the Krylov basis or, as we shall refer to it, the Krylov \textit{chain}. This terminology is justified from the fact that, as shown in (\ref{L-Tridiagonal}), $\mathcal{L}$ has a tridiagonal form in the Krylov basis, which one can write more suggestively as:

\begin{equation}
    \centering
    \label{L-tight-binding}
    \mathcal{L}=\sum_{n=0}^{K-2}b_{n+1}\Bigg( \big|\mathcal{O}_{n}\big)\big(\mathcal{O}_{n+1}\big| + \big|\mathcal{O}_{n+1}\big)\big(\mathcal{O}_{n}\big|\Bigg)~.
\end{equation}
We note that (\ref{L-tight-binding}) resembles the Hamiltonian of a one-dimensional tight-binding model on a finite chain with $K$ sites, with zero potential energies and hopping amplitudes given by the Lanczos coefficients. The statement is that the time evolution of the operator can be mapped to such a one-dimensional quantum-mechanical problem just by constructing its Krylov basis and studying its dynamics with respect to it \cite{Parker:2018yvk}, by solving the Schrödinger-like equation for the wave functions:

\begin{equation}
\label{Sect_Integr_Dynamics_Krylov_Schr_Eq}
    -i \dot{\phi}_n(t) = \sum_{m=0}^{K-1} T_{nm}\phi_m(t)=b_{n+1}\phi_{n+1}(t)+b_n\phi_{n-1}(t)\,,
\end{equation}
with the initial condition $\phi_n(0)=\delta_{n0}$ (we assume that the operator is normalized, for simplicity) and the boundary conditions $\phi_{-1}=\phi_K=0$ which ensure finiteness of the chain. 

We can thus understand the label $n$ as a position on the Krylov chain, and the Krylov elements $|\mathcal{O}_n)$ as localized states on that chain. In general, we always begin with a localized state $|\mathcal{O}_0)$ which over time can become delocalized depending on the dynamics of the Liouvillian dictated by the hopping amplitudes given by the Lanczos coefficients $b_n$.  In \cite{Dymarsky:2019elm} this perspective was used to show a relation between non-integrability and de-localization of the operator on the Krylov chain.

Two enlightening probes of the dynamics on the Krylov chain are K-complexity, $C_K(t)$ \cite{Parker:2018yvk} and K-entropy, $S_K(t)$ \cite{Barbon:2019wsy}. The former gives the average position of the propagating packet $\phi_n(t)$ over the Krylov chain,

\begin{equation}
    \centering
    \label{KC-definition}
    C_K(t)=\sum_{n=0}^{K-1}n|\phi_n(t)|^2,
\end{equation}
while the latter is nothing but another name for the Shannon entropy of the wave packet, and measures how spread or localized it is:

\begin{equation}
    \centering
    \label{KS-definition}
    S_K(t)=-\sum_{n=0}^{K-1}|\phi_n(t)|^2 \log |\phi_n(t)|^2.
\end{equation}

As we reviewed in Chapter \ref{ch:chapter02_KC}, in \cite{Parker:2018yvk} it was proposed, exploiting the relation between the asymptotics of the Lanczos coefficients and other quantities such as the two-point function, that maximally chaotic systems should display, in the thermodynamic limit, an asymptotically linear profile for the Lanczos coefficients, $b_n\sim \alpha n$, which in turn would imply an exponential growth for K-complexity $C_K(t)\sim e^{2\alpha t}$. Subsequent work \cite{Barbon:2019wsy} proposed that, for operators in finite systems satisfying the eigenstate thermalization hypothesis, ETH \cite{Srednicki:1994mfb,Srednicki_1999}, the linear growth of the Lanczos sequence should transition to a constant plateau phase around $n\sim S$, where $S$ denotes entropy or system size; this implies a transition from exponential to linear growth of K-complexity at times of order of the scrambling time $t_{s}\sim \log S$, and this linear growth should persist up to times of the order of the Heisenberg time $t_H\sim e^S$, when all the directions of the Krylov space have been explored and K-complexity would therefore saturate. 

Performing numerical simulations, in \cite{I} we were able to verify these facts for the Sachdev-Ye-Kitaev model \cite{Kitaev:2015,Sachdev:2015efa,Maldacena:2016hyu} with $4$-site interactions, encountering the additional feature of a non-perturbative descent that corrects the plateau of the Lanczos sequence making it terminate at $b_K=0$, as it should. This work also observed that typical operators\footnote{For chaotic systems, the notion of typicality amounts to satisfying the eigenstate thermalization hypothesis.} in maximally chaotic systems should saturate the Krylov dimension upper bound (\ref{Krylov-bound}) due to the absence of degeneracies in the spectrum of the Hamiltonian, whose level-spacing statistics satisfy distributions of the Wigner-Dyson type and therefore feature level repulsion \cite{BGSpaper,MehtaBook}, and to the fact that these operators are dense (in the sense of not having any zero elements) in the energy basis. In these systems, the wave packet $\phi_n(t)$ spreads in time efficiently over the Krylov chain and at times of the order of $t_H$ it tends to a uniform distribution $|\phi_n(t>t_H)|^2\sim\frac{1}{K}$, implying that K-complexity saturates at $C_K(t>t_H)\sim\frac{K}{2}$, a value wich is exponentially big in system size because $K$ saturates the bound (\ref{Krylov-bound}), and thus $K\sim D^2\sim e^{2S}$. Conversely, free or quadratic systems present in general a large number of degeneracies in the spectrum of the Hamiltonian, as well as potentially rational relations, implying degeneracies in the spectrum of the Liouvillian; this, together with abundant selection rules due to symmetries imposing the vanishing of operator matrix elements in the energy basis, reduces the Krylov dimension and therefore the upper bound for K-complexity. This was exemplified in \cite{I} taking the case of SYK with 2-site interactions, where the Krylov space of a single Majorana scales only linearly with system size, and hence so does the K-complexity upper bound.

This Chapter intends to explore the remaining middle ground: there exists a large class of strongly-interacting integrable systems which feature no exact degeneracies but still exhibit Poissonian level-spacing statistics \cite{BerryTabor} and are usually solvable by the means of the Bethe Ansatz \cite{1931ZPhy...71..205B, Samaj_bajnok_2013}.
In these systems, typical operators\footnote{In an integrable system, the notion of \textit{typicality} might be more ambiguous. For this work, we reserve this designation for local operators in the case of local systems, or non-extensive operators in the case of non-local systems with $q$-body interactions.} need not be sparse in the energy basis \cite{Rigol_XXZ}, thus admitting a Krylov space whose dimension can potentially be still exponential in system size. The difference between these systems and maximally chaotic ones should therefore appear at the dynamical level and, in this spirit, this work intends to explore whether the statistical properties of the Lanczos coefficients of the former systems may induce some form of localization preventing the effective diffusion of the wave packet through the Krylov chain, thus lowering the long-time value of the expectation value of its position, which is nothing but K-complexity.

\subsection{Disorder in Lanczos-sequences of integrable systems}
\begin{figure}[t]
    \centering
    \includegraphics[scale=0.6]{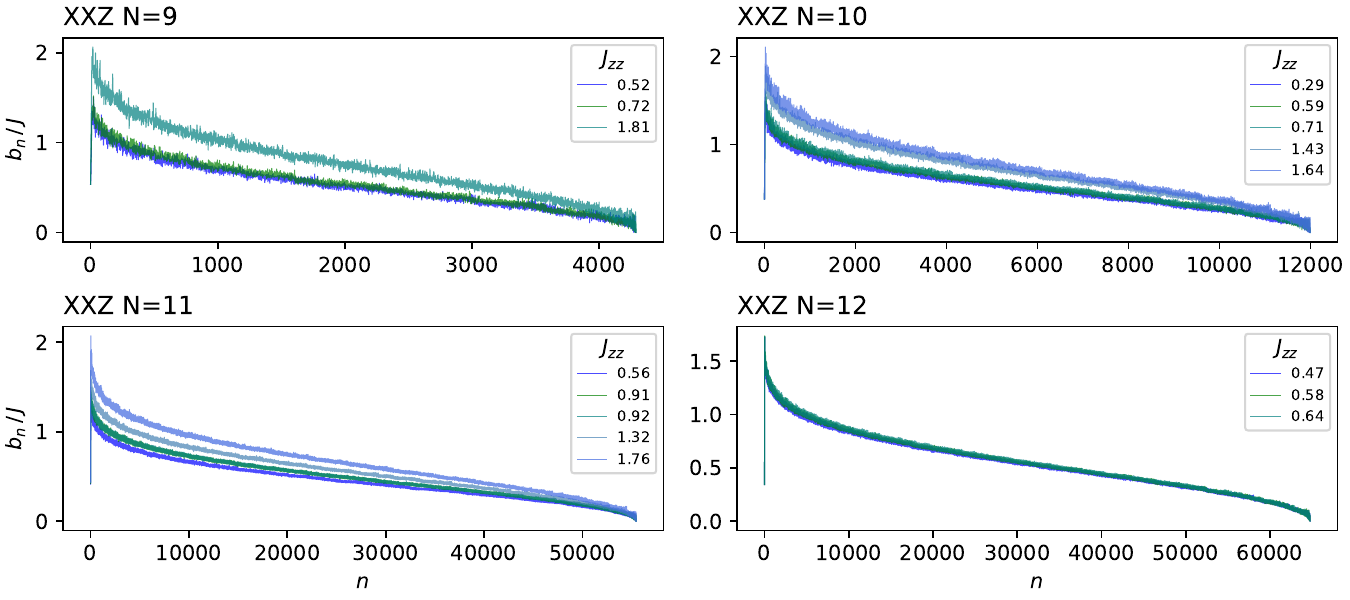}
    \caption{The Lanczos sequences for the various XXZ cases studied in this work. For the details of the XXZ model see section \ref{Section_XXZ}. For details on the one-site operator used and further numerical results, see section \ref{Sect_Numerical_Results}.}
    \label{fig:Lanczos_XXZ}
\end{figure}

\begin{figure}[t]
    \centering
    \includegraphics[scale=0.5]{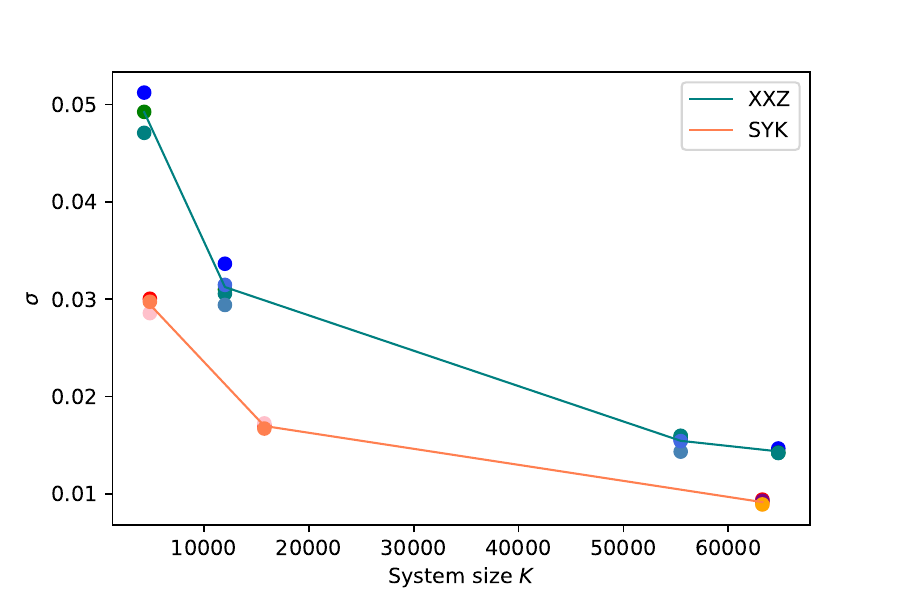}
    \caption{Using Lanczos-sequence data for various XXZ and complex SYK$_4$ systems of different Krylov dimensions, the standard deviation of the logarithm of ratios of consecutive Lanczos coefficients is represented by $\sigma$ on the $y$-axis, while the $x$-axis represents the Krylov dimension.  The smoother the Lanczos-sequence is, the closer $\sigma$ is to zero. The lines connect the average values of each set of points for a given system size and are shown to help the reader.
    For each system size, each point represents a single realization of SYK and a single $J_{zz}$ value of XXZ. In particular we present the following data.  \textbf{SYK}: 3 realizations for $L=8$ fermions, 3 realizations for $L=9$ and 5 realizations for $L=10$ (their Lanczos sequences can be found in \cite{I}); \textbf{XXZ}: 3 different $J_{zz}$ coefficients for $N=9$ spins, 5 for $N=10$, 5 for $N=11$ and 3 for $N=12$. The specific values chosen for $J_{zz}$ for each system size can be found in Figure \ref{fig:Lanczos_XXZ}.
    }
    \label{fig:sigma}
\end{figure}

As a hopping model over the Krylov chain, the properties of the Lanczos coefficients will determine the dynamics.  As we shall show below, the Lanczos coefficients depend on the Liouvillian frequencies and on the spectral decomposition of the initial operator. We will argue heuristically that the abundance of energy differences smaller than the mean level spacing $\Delta$ in integrable systems induces an erratic behavior in their Lanczos coefficients; this is a result of the Poissonian nature of their spectrum.  Conversely, chaotic systems have fewer energy differences below $\Delta$ due to level repulsion, and we will argue that one would expect a less erratic behavior in the $b$-sequence of typical operators in these systems. We have observed this effect comparing the Lanczos sequences obtained from typical operators in SYK$_4$ \cite{I} with those obtained from local operators in the XXZ spin chain, see section \ref{Subs_Anderson_loc} and figures \ref{fig:Lanczos_XXZ} and \ref{fig:sigma}.

The elements $b_n$ of the Lanczos-sequence can be written as ratios of Hankel determinants of the moments of the operator two-point function (see for example  \cite{SANCHEZDEHESA1978275}). The moments are given by: 
\begin{eqnarray} \label{moments}
    \mu_n = \sum_{i=0}^{K-1} |O_i|^2 \omega_i^n\,, 
\end{eqnarray}
where $O_i\equiv(\omega_i|\mathcal{O})$ is the spectral decomposition of the initial operator, and it can be proved (cf. section \ref{sect:Moments_to_Lanczos} and references therein) that the Lanczos coefficients satisfy the following recursion:
\begin{equation} \label{bn_Hankel}
    b_n^2 = \frac{D_{n-2} D_n}{D_{n-1}^2} \, , \quad n\geq 1\,,   
\end{equation}
where 
\begin{equation} \label{Hankel_Det}
    D_n = \begin{vmatrix} 
    \mu_0 & \mu_1 & \mu_2 & \dots & \mu_{n} \\
    \mu_1 & \mu_2 & \mu_3 & \dots & \mu_{n+1}  \\
    \vdots & \vdots & \vdots & \dots & \vdots \\
    \mu_{n} & \mu_{n+1} & \mu_{n+2} & \dots & \mu_{2n}
    \end{vmatrix} 
\end{equation}
are the so-called Hankel determinants, with $D_{-1}=1$ and $D_0=\mu_0=\sum_{i=0}^{K-1}|O_i|^2=1$ for a normalized initial operator. Note that all odd moments are zero since the operator is Hermitian and the Liouvillian spectrum is symmetric around zero. In  Appendix \ref{Appx_Hankel_Det} we show that such a Hankel determinant can be expressed directly in terms of the spectral decomposition $O_i$ and the Liouvillian frequencies $\omega_i$,
\begin{equation} \label{Dn}
    D_{n-1} = \sum_{\{i_1,i_2,\dots i_n \} \subset \{0, 1, \dots , K-1 \} } \prod_{i \in \{i_1,i_2,\dots i_n \}} |O_i|^2\prod_{a,b \in \{i_1,i_2,\dots i_n \}} (\omega_a-\omega_b)^2\,,
\end{equation}
where the sum is over all $\binom{K}{n}$ possible ways of choosing $n$ frequencies out of the $K$ frequencies in the spectrum of the Liouvillian, and the final product is the square of a Vandermonde determinant constructed out of these $n$ frequencies.  Given Poisson statistics in the level spacing of the Hamiltonian, one expects to find in the Liouvillian frequencies which are smaller than the mean level spacing, see section \ref{Section_XXZ}. With enough such frequencies, for some $n$ every choice of $n$ frequencies out of $K$ will include some very small $\omega_a-\omega_b$ in the Vandermonde determinants, reducing the value of $D_{n-1}$ compared with $D_{n-2}$ (this will also depend on the projections $O_i$ over the eigenspaces corresponding to those particular sets of frequencies). Specifically near the center of the Liouvillian spectrum, for integrable systems, we expect smaller frequencies and small frequency differences than for chaotic systems. 
Accordingly, $b_n^2$ given by the ratio (\ref{bn_Hankel}), will undergo a fluctuation, becoming either larger or smaller than the previous one, depending on whether $D_{n-1}$ is in the numerator or the denominator. Conversely in chaotic systems the smaller amount of very small frequencies, would result in a less erratic $b_n$-sequence.

\subsection{Some results from Anderson localization}\label{Subs_Anderson_loc}
As argued around equation (\ref{L-tight-binding}), dynamics on the Krylov chain constitute a one-dimensional hopping problem where the hopping amplitudes are given by the Lanczos coefficients. Thus, the erratic nature of the $b$-sequence induced by the abundance of small energy differences in integrable models can be effectively described by off-diagonal disorder in the Anderson sense. We will therefore now collect some useful results pertaining to Anderson localization that will inform our subsequent discussion. Such problems were studied as an electron localization problem (together with diagonal disorder) in \cite{Thouless_1972}; as a pure off-diagonal system in \cite{Fleishman_1977}, where it was established that the center of the band state was localized with a localization length increasing as $\sqrt{K}$, where $K$ is the chain size; also in \cite{PhysRevB.24.5698}, where localization was shown by studying the transmission coefficient; and including long-range correlations in \cite{PhysRevB.72.174207}.  For a review, see for example \cite{IZRAILEV2012125}.  

For a maximal Krylov dimension (which is attained whenever no exact degeneracies are present in the Liouvillian spectrum and the operator is fully dense in the energy basis), the Liouvillian always has a zero frequency eigenstate.  This eigenstate, which is the center-of-the-band state, can be reconstructed from the Lanczos-sequence.  In particular, expanding the Liouvillian eigenstates $|\omega)$ in the Krylov basis as

\begin{equation}
    \centering
    \label{L_eigenstate_krylov}
    |\omega ) = \sum_{n=0}^{K-1}\psi_n|\mathcal{O}_n)\,,
\end{equation}
the eigenvalue problem 
\begin{eqnarray} \label{EV_problem}
    \omega \psi_n = b_{n+1}\psi_{n+1}+b_n\psi_{n-1},\quad n=0,...,K-1
\end{eqnarray}
can be solved exactly for a given $b$-sequence for $\omega=0$.  The even elements of the eigenvector are given by:
\begin{eqnarray}
     \frac{\psi_{2n}}{\psi_0} &=& \prod_{i=1}^{n}\frac{ b_{2i-1}}{b_{2i}}, \quad n = 0,\dots, \frac{K-1}{2}
\end{eqnarray}
and all odd elements are zero.

In the absence of long-range correlations between the hopping amplitudes, it was first proposed in \cite{Fleishman_1977} that in the case of disordered $b$'s this state is localized.  The localization length was shown to be proportional to $\sqrt{K}$ and inversely proportional to $\sigma$, $l_{loc}\propto \sqrt{K}/\sigma$, where $\sigma^2$ is the variance of the logarithm of ratios of consecutive hopping amplitudes:
\begin{eqnarray} \label{disorder_strength}
    \sigma^2 = \textrm{Var}(x_i), \quad x_i \equiv \ln \Big| \frac{b_{2i-1}}{b_{2i}} \Big| ~.
\end{eqnarray}
Since $b_n$ are fluctuating, the values of $x_i$ will fluctuate around zero, and will be randomly distributed with mean 0. The variance $\sigma^2$ will depend on the details of the $b_n$ distribution. In Figure \ref{fig:Lanczos_XXZ} we depict the Lanczos sequences of a one-site operator for various instances of the XXZ spin chain (for more details, see sections \ref{Section_XXZ} and \ref{Sect_Numerical_Results}). The statistics of their fluctuations were analysed by computing (\ref{disorder_strength}) in each case\footnote{Only the first half of each sequence was used for this computation, as the statistics of the very small Lanczos coefficients towards the end of the descent are numerically less reliable due to the amplifying effect of the logarithm and the ratio in the definition of the variables $x_i$.}. The results for $\sigma$ are depicted in Figure \ref{fig:sigma}, where they are also compared to SYK$_4$ systems of similar size using data from \cite{I}. Although disorder strength is seen to decrease with the system size for both XXZ and SYK$_4$, our results indicate that indeed XXZ exhibits larger disorder in its Lanczos sequences. In particular, for comparable system sizes the disorder parameter $\sigma$ is bigger in XXZ than in SYK$_4$

\subsection{Probes of late-time behavior} \label{Sec:late-time_behavior}
In this section we introduce some of the quantities used to study and observe localization in \cite{luck:cea-01485001}, and relate them to K-complexity and its long-time-average. 

The time-dependent coefficients $\phi_n(t)$ given in (\ref{phi_def}) can be written in the basis of Liouvillian eigenvectors, 
\begin{eqnarray}
    |\mathcal{O}(t)) &=& \sum_{j=0}^{K-1} e^{i \omega_j t} |\omega_j)(\omega_j|\mathcal{O}_0) 
    =\sum_{m=0}^{K-1} \sum_{j=0}^{K-1} e^{i \omega_j t} (\omega_j|\mathcal{O}_0) (\mathcal{O}_m|\omega_j) |\mathcal{O}_m)\,,
\end{eqnarray}
where we expanded the Liouvillian eigenstates in the Krylov basis, that is 
$|\omega_j) = \sum_{m=0}^{K-1} |\mathcal{O}_m)(\mathcal{O}_m|\omega_j)$. 
From (\ref{phi_def}), using the orthonormality of the Krylov elements $(\mathcal{O}_m|\mathcal{O}_n)=\delta_{mn}$,  we recognize
\begin{eqnarray}
    \phi_n(t) = (\mathcal{O}_n|\mathcal{O}(t)) =  \sum_{j=0}^{K-1} e^{i \omega_j t} (\omega_j|\mathcal{O}_0) (\mathcal{O}_n|\omega_j)~.
\end{eqnarray}
We now give $|\phi_n(t)|^2$ the interpretation of a transition probability from $|\mathcal{O}_0)$ to $|\mathcal{O}_n)$ at time $t$:
\begin{eqnarray}
\label{P0n_def}
    P_{0n}(t) := |\phi_n(t)|^2  = \sum_{i,j=0}^{K-1} e^{i(\omega_j-\omega_i)t} (\omega_j|\mathcal{O}_0) (\mathcal{O}_n|\omega_j) (\omega_i|\mathcal{O}_n)(\mathcal{O}_0|\omega_i) ~.
\end{eqnarray}
This expression suggests that we interpret K-complexity as the average location on the Krylov chain to which the operator will transition at each time instant $t$:
\begin{eqnarray}
    C_K(t) = \sum_{n=0}^{K-1} n |\phi_n(t)|^2 = \sum_{n=0}^{K-1} n P_{0n}(t)\,.
\end{eqnarray}
To study the long-time behaviour of K-complexity, which is the main goal of this project, we perform a long-time-average over $|\phi_n(t)|^2$:
\begin{eqnarray}
\label{Q0n_longtimeavg}
    \overline{|\phi_n|^2} = \lim_{T\to \infty} \frac{1}{T}\int_0^T |\phi_n(t)|^2 dt\,.
\end{eqnarray}
By construction, the spectrum of the restriction of the Liouvillian to Krylov space has no degeneracies, and hence the phase-differences average out so that only the diagonal terms with $i=j$ in (\ref{P0n_def}) contribute to (\ref{Q0n_longtimeavg}):
\begin{eqnarray} \label{Transition_Probability}
   Q_{0n} := \overline{|\phi_n|^2} = \sum_{i=0}^{K-1}   |(\mathcal{O}_0|\omega_i)|^2 |(\mathcal{O}_n|\omega_i)|^2 ~.
\end{eqnarray}
The long-time-averaged K-complexity $\overline{C_K}$ can now be interpreted in two ways:
\begin{enumerate}
    \item The long-time-averaged expectation value of the position to which $|\mathcal{O}_0)$ can ``hop'': 
    \begin{eqnarray} \label{KC_Q0n}
        \overline{C_K} = \sum_{n=0}^{K-1} n\, Q_{0n}\,.
    \end{eqnarray}
    \item The spectral average of the K-complexities of the Liouvillian eigenstates, weighted by the overlap of such eigenstates with the initial condition:
    \begin{eqnarray} \label{KC_KCi}
        \overline{C_K} = \sum_{i=0}^{K-1} |(\mathcal{O}_0 |\omega_i)|^2   \sum_{n=0}^{K-1} n\,  |( \mathcal{O}_n |\omega_i)|^2\equiv \sum_{i=0}^{K-1}|(\mathcal{O}_0 |\omega_i)|^2\, C_K^{(i)}\,, 
    \end{eqnarray}
    where $C_K^{(i)}$ is the K-complexity of the eigenstate of phase $\omega_i$, that is, the average position on the Krylov chain of this stationary state. Eigenstates with smaller $C_K^{(i)}$ will be more localized towards the left of the chain and in general will have more significant overlap with the initial condition.
\end{enumerate}
The transition probability, and hence also $\overline{C_K}$, are influenced by localization effects in the eigenstates. The following two limiting cases are instructive:
\begin{itemize}
\item \textit{Full delocalization}: for which the eigenstates $|\omega_i)$ are fully delocalized over $|\mathcal{O}_n)$, and $|(\mathcal{O}_n |\omega_i)|^2\sim \frac{1}{K}$ for every $i,n$, implying both $Q_{0n}=\frac{1}{K}$ for all $n$ and $C_K^{(i)}=\frac{K}{2}$ for all $i$. Applying either (\ref{KC_Q0n}) or (\ref{KC_KCi}) we obtain, consistently:
    \begin{equation}
        \overline{C_K}=\frac{1}{K}\sum_{n=0}^{K-1} n \sim \frac{K}{2}\,.
    \end{equation}
\item \textit{Full localization}: for which the eigenstates $|\omega_i)$ are fully localized over $|\mathcal{O}_n)$, and $|( \mathcal{O}_n |\omega_i)|^2= \delta_{ni}$ for every $i,n$. In this case $Q_{0n}=\delta_{0n}$ for any $n$ and $C_K^{(i)}=i$ for all $i$, and we obtain:
    \begin{equation}
        \overline{C_K}= \sum_{i=0}^{K-1}\delta_{0i}\sum_{n=0}^{K-1} n\, \delta_{ni} = 0 ~.
    \end{equation}
Notice that in this extreme case $C_K(t)=0$ for all times because $|(\mathcal{O}_n | \omega_i )|^2 = \delta_{ni}$ implies that $|\omega_0)=|\mathcal{O}_0)$ (up to a phase), so that the initial condition is itself a stationary state and therefore it does not propagate.
\end{itemize}

For the case of a constant Lanczos-sequence, the eigenstates and eigenvalues of the Liouvillian as well as $\overline{C_K}$ can be computed analytically. This case is shown in detail in Appendix \ref{appx:Constant_b_analytics}.

We shall now turn to discuss the Hamiltonian of the XXZ spin chain, an interacting model whose integrable nature makes it an ideal candidate for studying the erratic structure of the Lanczos coefficients in this kind of systems and the consequent imprints that localization leaves on Krylov space dynamics.

\section{Complexity in integrable models: the XXZ spin chain}\label{Section_XXZ}

Motivated by the arguments in section \ref{Subsection_Krylov_hopping}, we intend to study operator dynamics in systems with an exponentially big Krylov dimension for typical operators but which still fall within the category of integrable systems. These conditions are fulfilled by strongly-interacting systems that are integrable via the Bethe Ansatz. In fact, in the framework of the \textit{algebraic} Bethe Ansatz it is possible to show that there exist extensively many commuting conserved charges that can be retrieved from the series expansion in the spectral parameter of the transfer matrix \cite{Samaj_bajnok_2013,slavnov2019algebraic,ReffertIntegrNotes}. However, these symmetries do not necessarily imply an extensive number of degeneracies in the spectrum of the Hamiltonian: rather, they enforce an intricate block-diagonal structure in which eigenstates carry several quantum numbers corresponding to different conserved charges (even though the underlying global symmetry might not be immediately apparent and the charges might not even be local). As a result of this, energy eigenvalues are effectively uncorrelated and therefore the nearest-neighbor level spacing statistics become Poissonian, implying that eigenvalues in the spectrum do not feature level repulsion: They can be arbitrarily close to each other, even though not exactly degenerate. Additionally, local operators in these systems can be dense in the energy basis, even though the statistical properties of their matrix elements do not follow the ETH, as exemplified in \cite{Rigol_XXZ}. These features have the effect that the Krylov space dimension for typical operators in this kind of systems is close to the upper bound (\ref{Krylov-bound}) or, at least, scales exponentially in system size as it would do for an ETH operator in a chaotic system. The distinguishing features of interacting integrable systems should be found in the impact on the actual dynamical evolution along the Krylov chain of the underlying spectral statistics and of the structure of the particular operator. The XXZ spin chain \cite{Heisenberg:1928mqa,YangXXZ_I,YangXXZ_II,Samaj_bajnok_2013} is a canonical example for this kind of models and we shall adopt it as the main workhorse of this project. Its Hamiltonian is given by:

\begin{equation}
    \centering
    \label{XXZ}
    H_{XXZ} = -\frac{J}{2} \sum_{n=1}^{N-1}\left[ \sigma_n^x \sigma_{n+1}^x+\sigma_n^y \sigma_{n+1}^y + J_{zz}\left( \sigma_n^z \sigma_{n+1}^z -1\right) \right],
\end{equation}
where $J$ is an overall dimensionful energy scale\footnote{In the numerics it was set to $1$, so it can be thought of as setting the units of the problem, as there is no other dimensionful parameter.} and $J_{zz}$ is the dimensionless coupling strength. Note that the special values $J_{zz}=0,1$ correspond to the XY and XXX models, respectively, and that the summation limits signal implicitly open boundary conditions (OBC), as opposed to periodic (PBC). Unless stated otherwise, we will use OBC in our numerical analysis.

The Hamiltonian (\ref{XXZ}) is integrable via Bethe Ansatz. In order to unveil the Poissonian nature of its spectrum, it is necessary to first mod out the discrete symmetries that do cause exact degeneracies and restrict to a Hilbert space sector whose dimension will still be exponential in system size. $H_{XXZ}$ commutes with the two following global symmetries (see for example \cite{Santos2013}):

\begin{itemize}
    \item Total spin projection along the $z$-direction (or total magnetization):
    \begin{equation}
        \centering
        \label{Total-Spin}
        S^z = \sum_{n=1}^N S_n^z\,,
    \end{equation}
    where $S_n^z=\frac{1}{2}\sigma_n^z$ denotes the spin projection along the $z$-axis at site $n$.
    \item Parity, defined as the transformation that exchanges sites as $n\longleftrightarrow{ N+1-n}$:
    \begin{equation}
        \centering
        \label{Parity}
        P = \prod_{n=1}^{\lfloor \frac{N}{2}\rfloor}\mathcal{P}_{n,N+1-n}\,,
    \end{equation}
    where $\mathcal{P}_{nm}$ is the permutation operator between sites $n$ and $m$. As defined, parity is a space inversion with respect to the chain center.
\end{itemize}

\begin{figure}
    \centering
    \includegraphics[width=10cm]{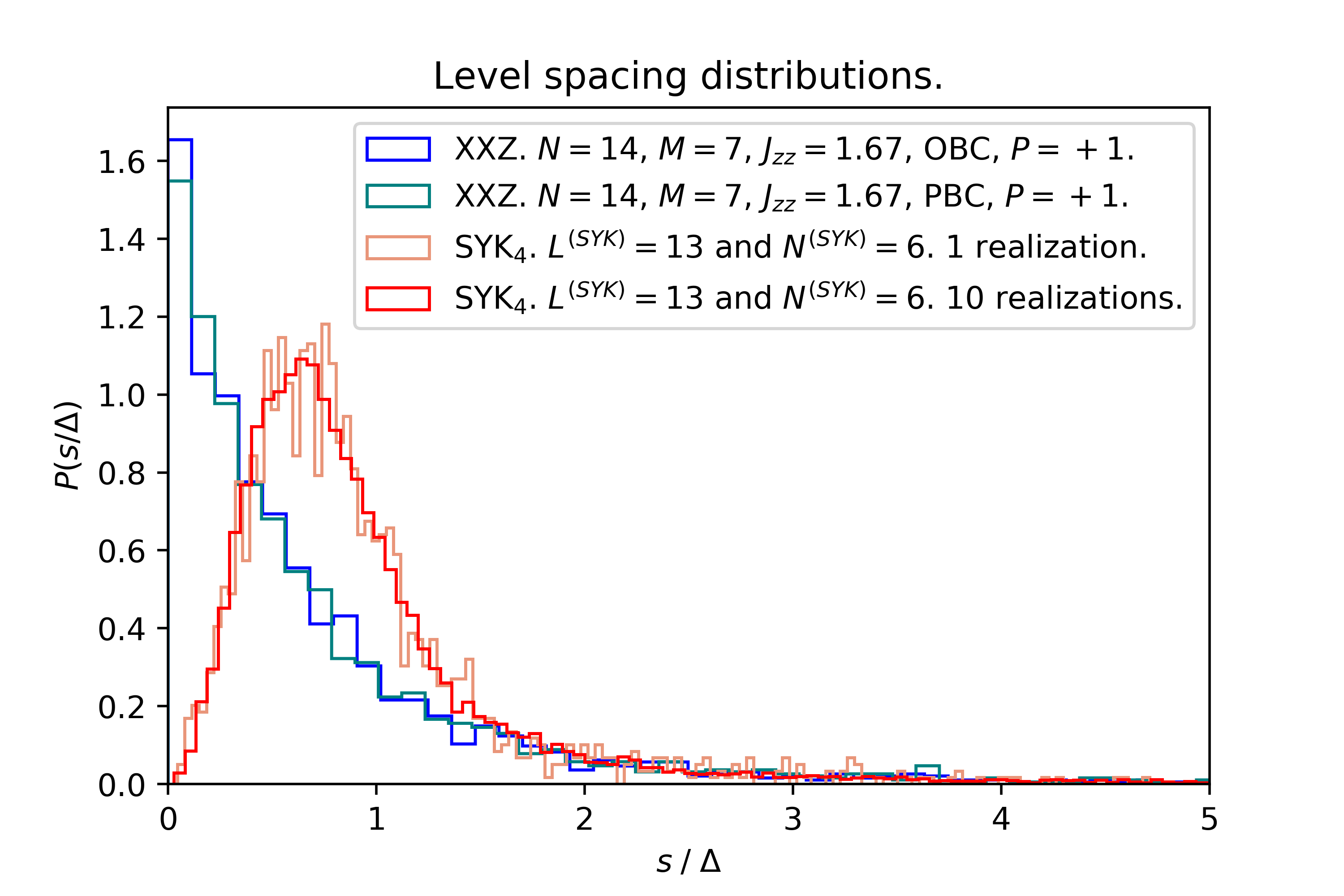}
    \caption{Nearest-neighbor level spacing distributions for XXZ with a specific $J_{zz}$ coupling with $N=14$ spins in the sector $M=7$, $P=+1$, both with open and periodic boundary conditions. For comparison, we include the corresponding distributions for complex SYK$_4$ with $L^{(SYK)}=13$ sites and occupation number $N^{(SYK)}=6$, whose Hilbert space dimension is equal to that of the positive parity sector of the above-mentioned XXZ instance. All distributions are normalized so that the area under each curve is equal to one, and all energy differences have been normalized by the mean level spacing $\Delta$ in each system.}
    \label{fig:Histogram}
\end{figure}

There are other global symmetries that can be deduced from the algebraic Bethe Ansatz, but numerical checks indicate that only parity induces exact degeneracies when OBC are chosen. For instance, we can consider \textit{charge conjugation}, which takes the form of a spin flip on each site or, up to a global phase, a rotation of angle $\pi$:

    \begin{equation}
        \centering
        \label{R-operator-charge-conj}
        R = \prod_{n=1}^{N}\sigma_n^x = i^N \exp \big( -i \pi S^x \big)~.
    \end{equation}
It is possible to show that $[H_{XXZ},R]=0=[P, R]$, however, $[R,S^z]\neq 0$. Since $\left[S^z,P\right]=0$, we choose the set of commuting observables $\left\{H_{XXZ},S^z,P \right\}$, with which it is possible to block-diagonalize $H_{XXZ}$ and restrict ourselves to a Hilbert space sector of fixed parity and total magnetization. The total Hilbert space of the system, $\mathcal{H}$, has dimension $2^N$ and, thanks to $S^z$-symmetry, it can be split as the direct sum of sectors with a fixed number $M$ of spins up (i.e. fixed magnetization sectors):

\begin{equation}
    \centering
    \label{Hilbert-M-sectors}
    \mathcal{H}=\bigoplus_{M=0}^N\mathcal{H}_M ~.
\end{equation}
In each of these sectors, whose dimension is $D_M=\binom{N}{M}$, the total magnetization operator is proportional to the identity, $S^z = M-\frac{N}{2}$ and the restriction of the Hamiltonian on each of them is closed. In future discussions, we will consider without loss of generality only sectors with $M\leq \lfloor\frac{N}{2}\rfloor$, since the $M$-sector is mapped to the $N-M$-sector by $R$ symmetry\footnote{In fact, all non-trivial physics come from \textit{antiparallel} pairs of neighbouring spins \cite{Samaj_bajnok_2013}, as can be deduced from (\ref{XXZ}).}. Furthermore, the analysis based on Bethe Ansatz shows \cite{Samaj_bajnok_2013} that, in each sector, the spectral width scales with $M$ for fixed interaction strength $J_{zz}$, since the energy of each eigenvalue can be expressed as the sum of $M$ magnon dispersion relations, which are bounded. Therefore, in order to have a spectrum whose width does not scale with system size, we choose to normalize the Hamiltonian in a fixed-magnetization sector by the number of up spins; that is, we will work with $H^{(M)}$ given by:

\begin{equation}
    \centering
    \label{XXZ-sector-normalized}
    H^{(M)} := \frac{1}{M}H_{XXZ}^{(M)},
\end{equation}
where $H_{XXZ}^{(M)}$ denotes the restriction of the XXZ Hamiltonian $H_{XXZ}$ to a fixed magnetization sector with $M$ spins up. The spectrum of $H^{(M)}$ can present a number of exact degeneracies due to parity symmetry. In fact, each magnetization sector can be further split into positive and negative parity subsectors:

\begin{equation}
    \centering
    \label{Hilbert-M-P-sectors}
    \mathcal{H}_M=\mathcal{H}_M^+\oplus \mathcal{H}_M^{-},
\end{equation}
whose dimensions $D_M^\pm$ are computed in Appendix \ref{Appx_Sectors} and summarized in Table \ref{Table_Dim_ParitySectors}. Importantly, they scale asymptotically exponentially in system size. As numerical checks indicate, once we restrict ourselves to a sector $\mathcal{H}_M^P$, the spectrum of the Hamiltonian does not feature exact degeneracies due to any other discrete symmetries, and we can extract the universal Poissonian behavior of the spectral statistics, as illustrated in Figure \ref{fig:Histogram}, where the level spacing distribution of the positive-parity sector in XXZ with both OBC and PBC are compared to that of an instance of complex SYK$_4$ with the same Hilbert space dimension. This serves to illustrate a point that could be misunderstood: the integrable character of the system reflects itself in the \textit{lack of correlation} between the energy eigenvalues, yielding a Poissonian distribution characterized by a high probability of finding energy differences smaller than the mean level spacing, as opposed to the characteristic level repulsion displayed by chaotic systems. Exact degeneracies are not necessarily a defining feature of integrable systems.

As a word of caution, we shall note that, for chains with an even number of sites $N$, there exists a zero-magnetization sector ($M=\frac{N}{2}\Longrightarrow S^z = 0$) in which accidentally $S^z$ and $R$ \textit{do} commute, since $R$ does not change the magnetization sector because there are as many spins up as spins down. Numerically (see for instance Figure \ref{fig:Histogram}) we do not observe extra degeneracies produced by this symmetry enhancement. In fact, it is conjectured in the literature \cite{Doikou:1998jh} that eigenstates within this magnetization sector are themselves eigenstates of $R$. In conclusion, for chains of even length, in the Hilbert space sector with $M=\frac{N}{2}$ and fixed $P$ we can classify eigenstates according to their $R$-eigenvalue ($\pm 1$), and the parity sectors can be split into $R$-subsectors:

\begin{equation}
    \centering
    \label{H-N/2-P-R}
    \mathcal{H}_{\frac{N}{2}}^{P}= \mathcal{H}_{\frac{N}{2}}^{P,R=+}\oplus \mathcal{H}_{\frac{N}{2}}^{P,R=-},\quad\quad\text{for }N\text{ even}
\end{equation}
whose dimensions $D_{\frac{N}{2}}^{PR}$ are also computed in Appendix \ref{Appx_Sectors} and can be written compactly as:

\begin{equation}
    \centering
    \label{R_sector_dimension}
    D_{\frac{N}{2}}^{PR} = \frac{1}{2}D_{\frac{N}{2}}^P + P\cdot R\cdot2^{\frac{N-4}{2}},\quad\quad\quad\text{for }N \text{ even.}
\end{equation}
Despite not granting exact degeneracies in the spectrum of the Hamiltonian, this symmetry enhancement can yield some selection rules for operators charged under $R$, which impose constraints on their matrix elements in the energy basis and hence can reduce the dimension of their associated Krylov space, as also discussed in Appendix \ref{Appx_Sectors}.

\begin{table}[ht]
\centering

\begin{tabular}{|c|c|c|}
\hline & & \\
       & $M$ odd & $M$ even \\ &&\\
       \hline && \\
$N$ odd  & $\begin{pmatrix} \dfrac{N-1}{2} \\[2.5ex] \dfrac{M-1}{2} \end{pmatrix}$  & $\begin{pmatrix} \dfrac{N-1}{2} \\[2.5ex] \dfrac{M}{2} \end{pmatrix}$ \\ &&      \\ \hline
&&\\
$N$ even & $0$     & $\begin{pmatrix} \dfrac{N}{2} \\[2.5ex] \dfrac{M}{2} \end{pmatrix}$ \\ && \\ \hline
\end{tabular}

\caption{The dimension of each Hilbert space sector of fixed parity and total magnetization is $D_M^{\pm}=\frac{D_M\pm A}{2}$, where $D_M=\binom{N}{M}$ and the expression for $A$, which depends on whether $N$ and $M$ are even or odd, is shown in this Table.} \label{Table_Dim_ParitySectors}
\end{table}

\section{Numerical results}\label{Sect_Numerical_Results}
In this section we present evidence that in the case of $H_{XXZ}$ discussed in the previous section and for a single-site operator which we introduce below, K-complexity indeed saturates at values below $\sim K/2$, which is the expected saturation value for chaotic systems such as SYK$_4$ studied previously in Chapter \ref{ch:chapter03_SYK}, where the Lanczos sequences featured less disorder as illustrated in Figure \ref{fig:sigma}. 

We will work with $H^{(M)}$ defined in (\ref{XXZ-sector-normalized}) in the sector of fixed magnetization and parity $\mathcal{H}_M^{+}$ and will study the time evolution of the linear operator 
\begin{equation} \label{Study_Operator}
    \mathcal{O} = \sigma^z_{m} + \sigma^z_{N-m+1}\,,
\end{equation}
where $m$ can be any site on the spin chain $1\leq m \leq N$.  This operator was chosen because it respects the sector symmetries, namely total magnetization conservation and parity, in addition to being a local one-site observable.
We choose $m$ to be near the center of the spin chain, so that boundary effects will not play a role. In general, all the matrix elements of this operator in the energy basis are non-zero and therefore, given that we work in sectors with no degeneracies, the Krylov space dimension for such an operator is expected to saturate the upper bound of (\ref{Krylov-bound}).  We tested this assertion by studying the matrix elements of this operator in the energy basis to very high precision. It would be interesting to also study other local operators; preliminary checks performed at small system sizes for linear combinations of one-site operators, two-site operators and linear combinations of two-site operators display a similar phenomenology to what will be described in this section for the operator (\ref{Study_Operator}).  Some of the Hamiltonians and operators were generated using the open-source package \cite{SciPostPhys.2.1.003}, while others were produced with programs of our own implementation, yielding coincident results.

\begin{figure}[t]
    \centering
         \begin{subfigure}[t]{0.47\textwidth}
         \centering
         \includegraphics[width=0.8\textwidth]{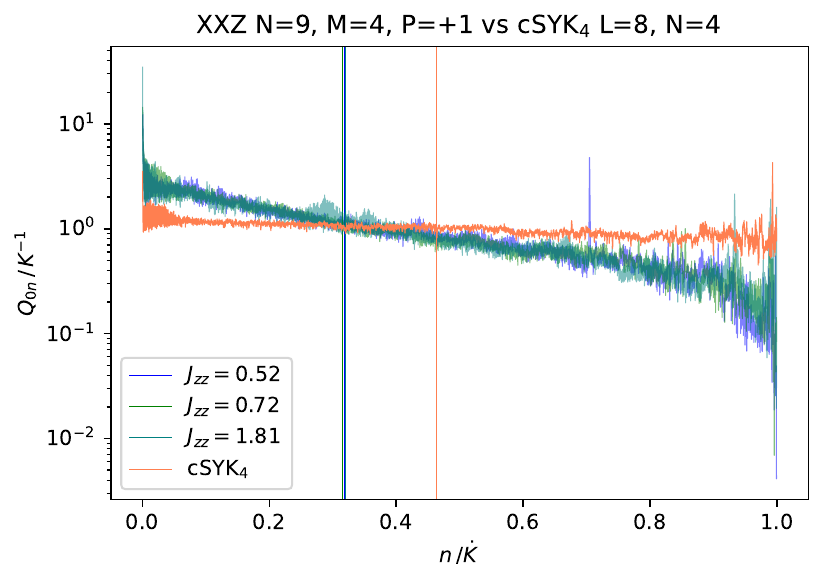}
         \caption{\footnotesize The Hilbert space dimension of XXZ with $N=9,\, M=4$, in the $P=+1$ sector is, according to Table \ref{Table_Dim_ParitySectors}, $D=66$.  For the operator $\mathcal{O}=\sigma_5^z$ the Krylov space dimension equals its upper bound $K=D^2-D+1=4291$. We find $\overline{C_K}\sim 0.318K$.  An SYK$_4$ system of Krylov space dimension $K_{SYK}=4831$ has $\overline{C_K}\approx 0.463 K_{SYK}$, much closer to the value expected for chaotic systems.}
         \label{fig:XXZ9}
        \end{subfigure}
     \hfill
     \begin{subfigure}[t]{0.47\textwidth}
         \centering
         \includegraphics[width=0.8\textwidth]{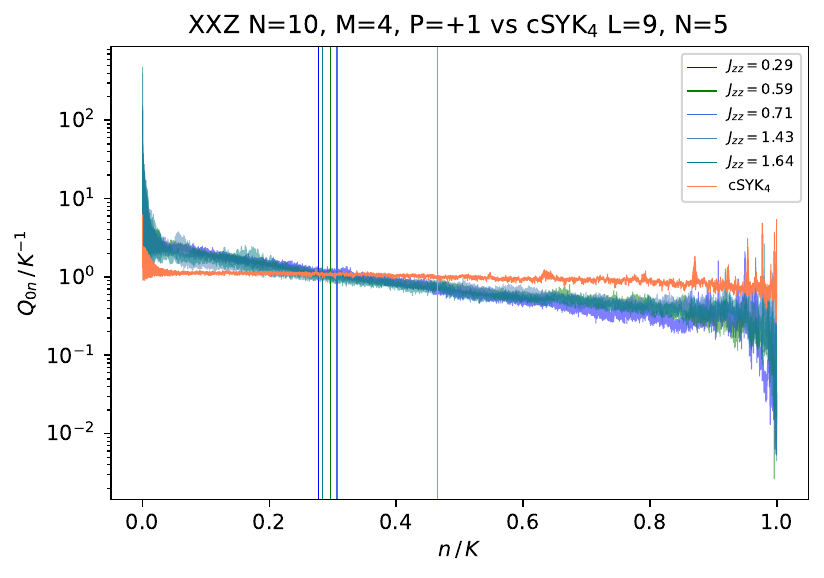}
         \caption{\footnotesize The Hilbert space dimension of XXZ with $N=10,\, M=4,\, P=+1$ is $D=110$ according to Table \ref{Table_Dim_ParitySectors}. For the operator $\mathcal{O}=\sigma_5^z+\sigma_6^z$ the Krylov space dimension saturates its upper bound and  $K=D^2-D+1=11991$. If the system were chaotic we would expect $\overline{C_K}\sim 0.5K$. We find $\overline{C_K}\approx 0.294K$.  For an SYK$_4$ system of Krylov space dimension $K_{SYK}=15751$, we find $\overline{C_K}\approx 0.466K_{SYK}$.}
         \label{fig:XXZ10}
        \end{subfigure} 
        \begin{subfigure}[t]{0.47\textwidth}
         \centering
         \includegraphics[width=0.8\textwidth]{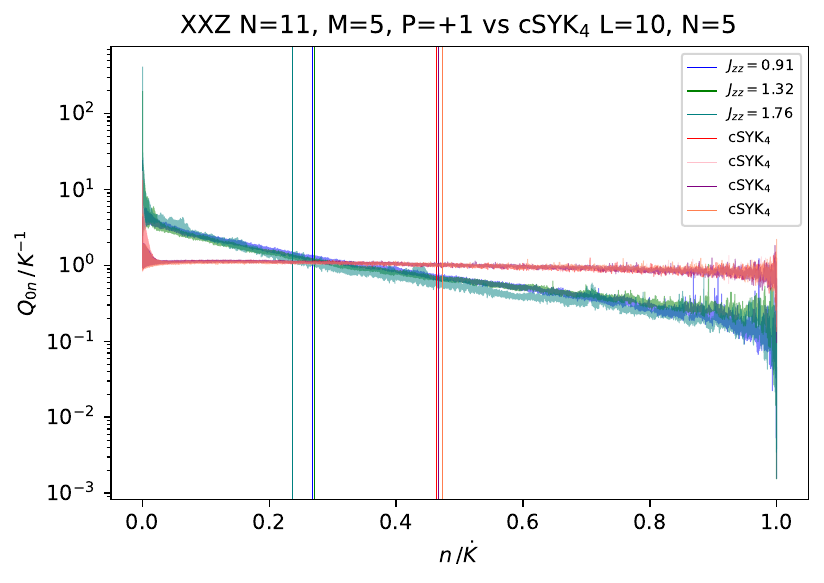}
         \caption{\footnotesize For XXZ with $N=11$ in the sector $M=5,\, P=+1$, the Hilbert space is of dimension $D=236$ (Table \ref{Table_Dim_ParitySectors}). Results for operators $\mathcal{O}=\sigma_6^z$ ($J_{zz}=0.91$) and $\mathcal{O}=\sigma_5^z+\sigma_7^z$ are shown; both saturate the upper bound on the Krylov space dimension $K=55461$. Nevertheless, the long-time average of K-complexity is below $0.5K$ and is (on average) $\overline{C_K}\sim 0.258K$. For SYK$_4$ systems of Krylov space dimension $K_{SYK}=63253$ we find (on average) $\overline{C_K}\approx 0.4675 K_{SYK}$.}
         \label{fig:XXZ11}
        \end{subfigure} 
      \hfill
     \begin{subfigure}[t]{0.47\textwidth}
         \centering
         \includegraphics[width=0.8\textwidth]{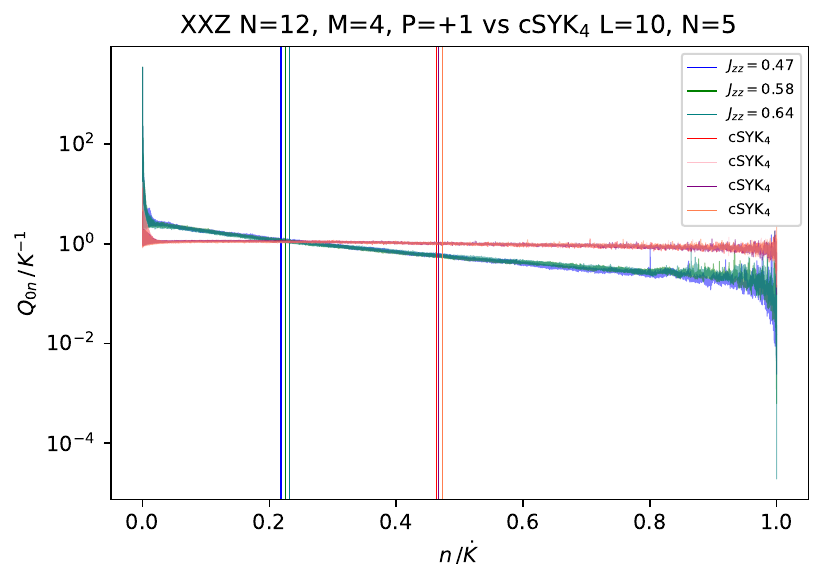}
         \caption{\footnotesize The Hilbert space dimension for XXZ with $N=12,\, M=4$ and $P=+1$ is $D=255$ (Table \ref{Table_Dim_ParitySectors}).  The operator $\mathcal{O}=\sigma_6^z+\sigma_7^z$ saturates the upper bound on the Krylov space dimension $K=D^2-D+1=64771$. The long-time average of K-complexity is below $0.5K$ and its value is (on average) $\overline{C_K}\approx 0.225 K$.  We plot again results for the same SYK$_4$ systems shown in Figure \ref{fig:XXZ11}.}
         \label{fig:XXZ12}
        \end{subfigure} 
    \caption{Numerical results for the transition probabilities $Q_{0n}$ defined in (\ref{Transition_Probability}) and $\overline{C_K}$ of XXZ systems with 9, 10, 11 and 12 spins and comparison with complex SYK$_4$ systems of 8, 9 and 10 fermions.  The vertical lines represent $\overline{C_K}$ in units of $K$ for each system.}
    \label{fig:Qn0_XXZ_vs_SYK}
\end{figure}

Using high precision arithmetic and the re-orthogonalization algorithms we developed for the study of SYK and which are described in Appendix \ref{Appx-algorithms}, we computed the Lanczos coefficients associated to operator (\ref{Study_Operator}) for various instances of XXZ, see Figure \ref{fig:Lanczos_XXZ}. Then, following the discussion in section~\ref{Sec:late-time_behavior}, we plot in Figure \ref{fig:Qn0_XXZ_vs_SYK} the time-averaged transition probabilities $Q_{0n}$ for several XXZ systems with different numbers of spins $N$, different magnetization sectors $M$ and various choices of the $J_{zz}$ coupling. The weighted average of $Q_{0n}$, which is equivalent to the long-time average of K-complexity via equation (\ref{KC_Q0n}), is plotted as a vertical line for each system. In each case, we plotted also the transition probability and $\overline{C_K}$ for a similar sized complex SYK$_4$ system studied in \cite{I}. To make the results clearer we normalized the $x$-axis with respect to the Krylov dimension $K$ for each system separately, as well as normalizing the transition probability in the $y$-axis with respect to $K^{-1}$.
We find that for XXZ the transition probability is biased towards the left side of the chain, whereas for SYK it is almost horizontal and of the order of $K^{-1}$. The left-biased transition probability for XXZ moves the saturation value of $C_K$ to a value smaller than $K/2$, while the almost flat profile of $Q_{0n}$ for SYK gives $\overline{C_K}$ a value close to $K/2$.

\begin{figure}[t]
    \centering
    \includegraphics[scale=.6]{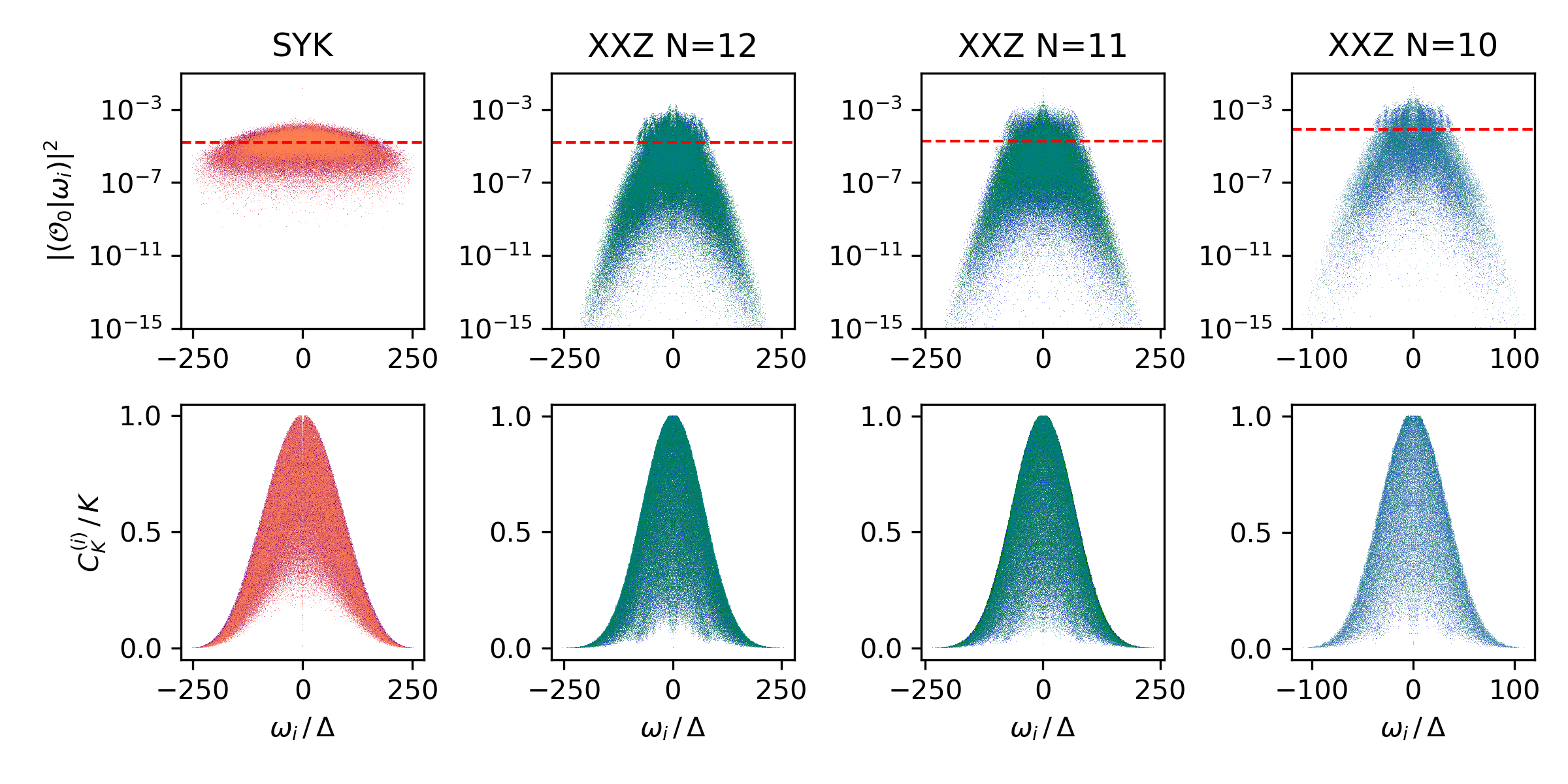}
    \caption{\footnotesize The spectral decomposition of the initial operator (top) and K-complexities of the individual Liouvillian eigenstates (bottom). Data for 4 random realization of cSYK$_4$ with 10 fermions are superimposed in the leftmost column.  Data for 3 choices of $J_{zz}$ coupling for XXZ with $N=12,\, M=4,\, P=+1$ are superimposed in the second column from the left. Data for 3 different choices of $J_{zz}$ couplings for XXZ with $N=11,\, M=5,\, P=+1$ are shown in the third column from the left; and data for 5 different choices of $J_{zz}$ couplings for XXZ with $N=10,\, M=4,\, P=+1$ are shown superimposed in the right column.  The frequencies are normalized according to $\Delta$ which is the mean level spacing computed for for each system separately. The dashed red line represents the value of $K^{-1}$ for each system. }
    \label{fig:KC_KCi}
\end{figure}

\begin{figure}[t]
    \centering
    \includegraphics[scale=.6]{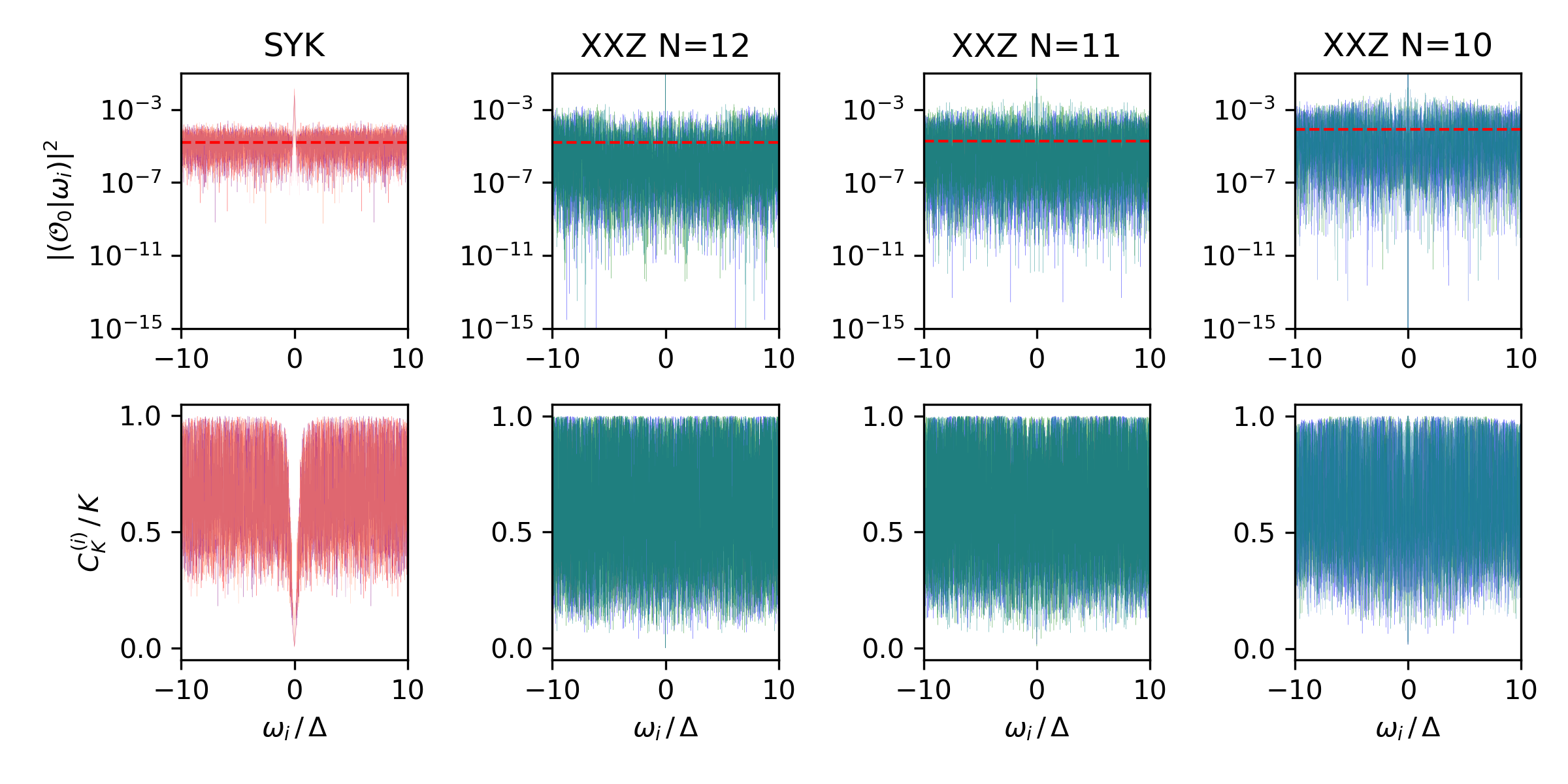}
    \caption{\footnotesize Same results as in Figure \ref{fig:KC_KCi} focused at the band-center. Note that the values of $C_K^{(i)}$ for XXZ reach smaller values than those for SYK.}
    \label{fig:KC_KCi_band_center}
\end{figure}

Equation (\ref{KC_KCi}) expresses the long-time average of K-complexity, $\overline{C_K}$, in terms the overlap of the Liouvillian eigenstates with the initial condition $|(\omega_i|\mathcal{O}_0)|^2$ and the spectral complexities $C_K^{(i)}$. The corresponding quantities computed numerically from XXZ and SYK are presented in Figures \ref{fig:KC_KCi} and \ref{fig:KC_KCi_band_center}. 
We find that two factors cooperate to decrease the value of K-complexity in XXZ compared with SYK:
(1)  In XXZ the spectral profile of the initial operator is peaked towards the band center\footnote{Here, the term \textit{band center} is taken as a synonym of the \textit{center of the spectrum}.}, while for SYK the profile is flatter and of order $K^{-1}$ as expected from ETH \cite{Srednicki_1999}.  (2) In XXZ, K-complexities of the individual eigenstates near the band center reach smaller values (Figure \ref{fig:KC_KCi_band_center}) due to enhanced localization, compared with those of SYK where most eigenstates have K-complexities of around $K/2$. These eigenstates are particularly important, since the initial condition is localized to the left of the chain and therefore will have a larger overlap with them.  In general, the eigenstates in XXZ which have a stronger overlap with the initial condition, $|(\mathcal{O}_0|\omega_i)|^2$, have smaller $C_K^{(i)}$'s than those of of SYK, thus reducing the value of $\overline{C_K}$ computed via (\ref{KC_KCi}).

\begin{figure}[t]
    \centering
    \includegraphics[scale=1]{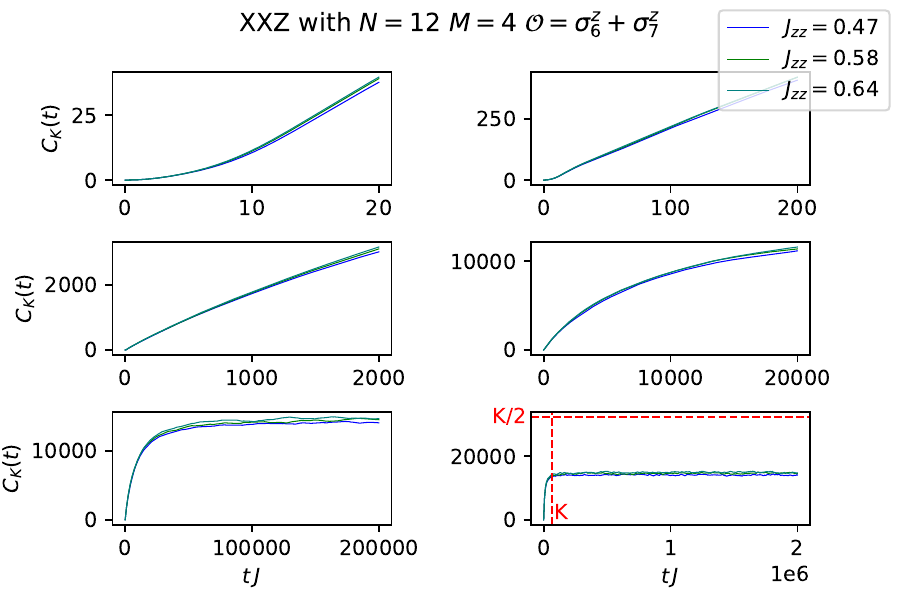}
    \caption{Time-dependent results for K-complexity. Shown here are results for XXZ with $N=12$ in the $M=4,\, P=+1$ sector at 3 different $J_{zz}$ couplings.  Each subplot shows a different time scale.  K-complexity is seen to saturate below $K/2$ at time scales of order $K$ (up to some dimensionful prefactor). }
    \label{fig:KC_time_dependent}
\end{figure}

\begin{figure}[t]
    \centering
    \includegraphics[scale=1]{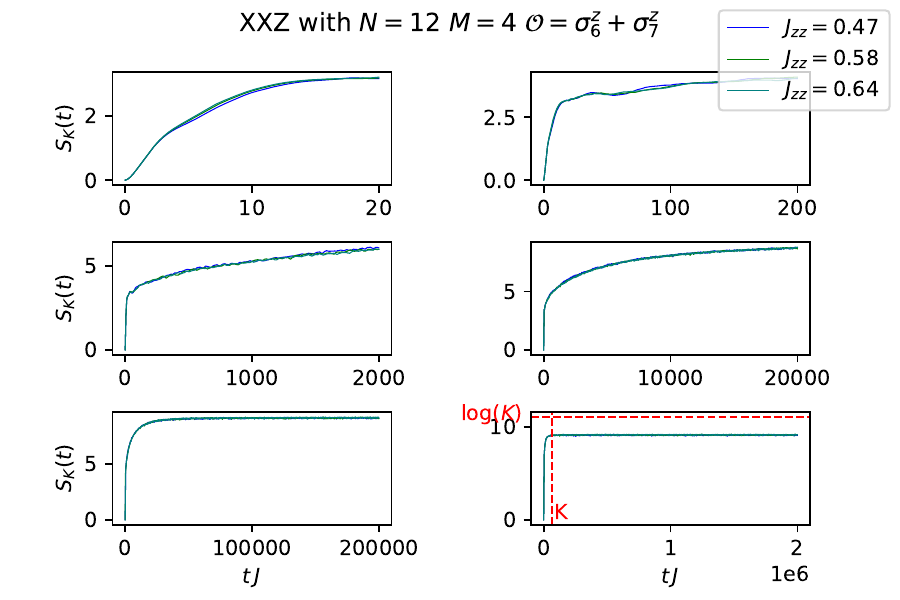}
    \caption{Time-dependent results for K-entropy.  Results for XXZ with $N=12$ in the $M=4,\, P=+1$ sector at 3 different $J_{zz}$ couplings are shown. K-entropy is seen to saturate below $\log(K)\sim N$ at time scales of order $K$. The general lower value of K-entropy is also a sign of localization.}
    \label{fig:KS_time_dependent}
\end{figure}

\begin{figure}[t]
    \centering
    \includegraphics[scale=1]{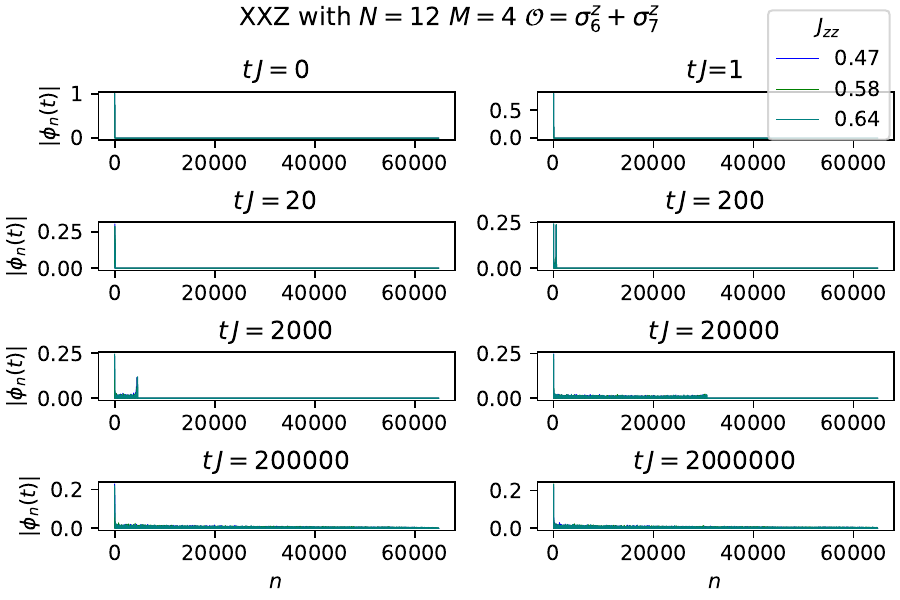}
    \caption{Snapshots of the absolute value of the wave-function at selected times for XXZ with $N=12$ in the $M=4,\, P=+1$ sector at 3 different $J_{zz}$. At $t=0$ the wave-function is a delta function on the first site on the Krylov chain. At late times the wave function  shows clear signs of localization on the first site (i.e. the initial condition), and is in general biased towards the left side of the Krylov chain. }
    \label{fig:Phi_time_dependent}
\end{figure}

Time-dependent results of K-complexity, K-entropy and snapshots of the absolute value of the wavefunction at various time scales are displayed in Figures \ref{fig:KC_time_dependent}, \ref{fig:KS_time_dependent} and \ref{fig:Phi_time_dependent}.  K-complexity is seen to saturate below $K/2$ as predicted by its late-time average. The time scale at which it saturates is of order $K$. Since $K\sim e^{2S}$ we find that the saturation \textit{time} for XXZ is similar to that observed in chaotic models such as SYK.
K-entropy saturates as well at time scales of order $K$, at a value smaller than $\log(K)\sim N$ -- the value signaling total delocalization of the operator wave function as seen in SYK in \cite{I} -- which is another indication that the wave function is more localized than in SYK. Again, the saturation \textit{time} is of order $e^{2S}$ as seen in SYK. The snapshots of the wave function at increasing time scales show that, although the wave packet is spreading towards the right of the chain, there is always a stronger support on the initial site of the Krylov chain, and in general some bias towards the left region.

\textbf{Note:} In the case of an even number of spins in the zero magnetization sector $M=N/2$, both $P$ and $R$ preserve total magnetization and in Appendix \ref{Appx_selection_rules_Krylov} we argue that $\mathcal{O}$ of the form (\ref{Study_Operator}) is charged under $R$, which creates a selection rule and reduces the Krylov space dimension. 

\clearpage

\section{Phenomenology of erratic Lanczos sequences}\label{Sect_Toys}

We have computed Lanczos sequences for various instances of XXZ and verified, in agreement with our initial expectation, that they feature disorder superimposed on the ascent-descent profile. Such Lanczos sequences play the role of disordered hopping amplitudes in the Krylov chain, and we observed that for those systems K-complexity saturated at late times at values smaller than half the Krylov space dimension, $\frac{K}{2}$, as a consequence of the long-time averaged transition probability being biased towards the left side of the chain (i.e. the side closer to the localized initial condition). In the framework of Anderson localization due to off-diagonal disorder, it is usually difficult to establish analytically features of time evolution, and customarily one focuses on properties of stationary states \cite{PhysRevB.24.5698,Fleishman_1977,IZRAILEV2012125}. For this reason, in this section we will present numerical computations with heuristically motivated Lanczos sequences whose main ingredients will be an ascent-descent profile with fluctuations on top, and we will show that they reproduce very closely the XXZ results, backing up the claim that the K-complexity phenomenology observed in this model is due to the disorder of its Lanczos coefficients. We will construct such Lanczos sequences for Krylov chains of size $K$ in successive levels of sophistication: First and for reference, we compute transition probabilities and K-complexity resulting from a flat $b$-sequence with fluctuations on top, which is nothing but the canonical setup for off-diagonal disorder; then, we will use two different Ansätze for a more realistic underlying profile displaying ascent and descent regimes. 

\subsection{Pure off-diagonal disorder}\label{Subs_Toy_FLAT}

\begin{figure}[t]
    \centering
    \includegraphics[width=0.45\textwidth]{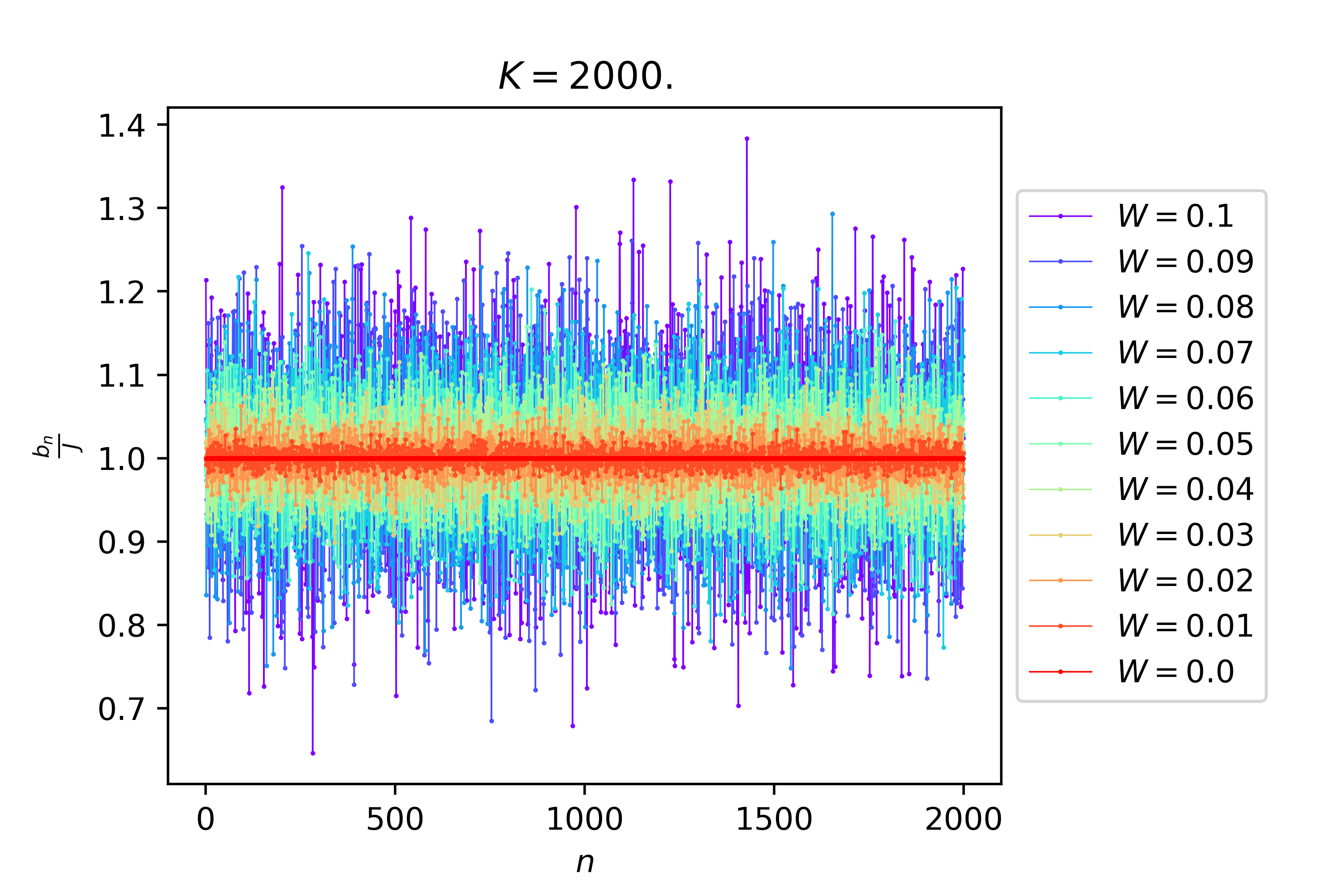} \\
    
    \includegraphics[width=0.45\textwidth]{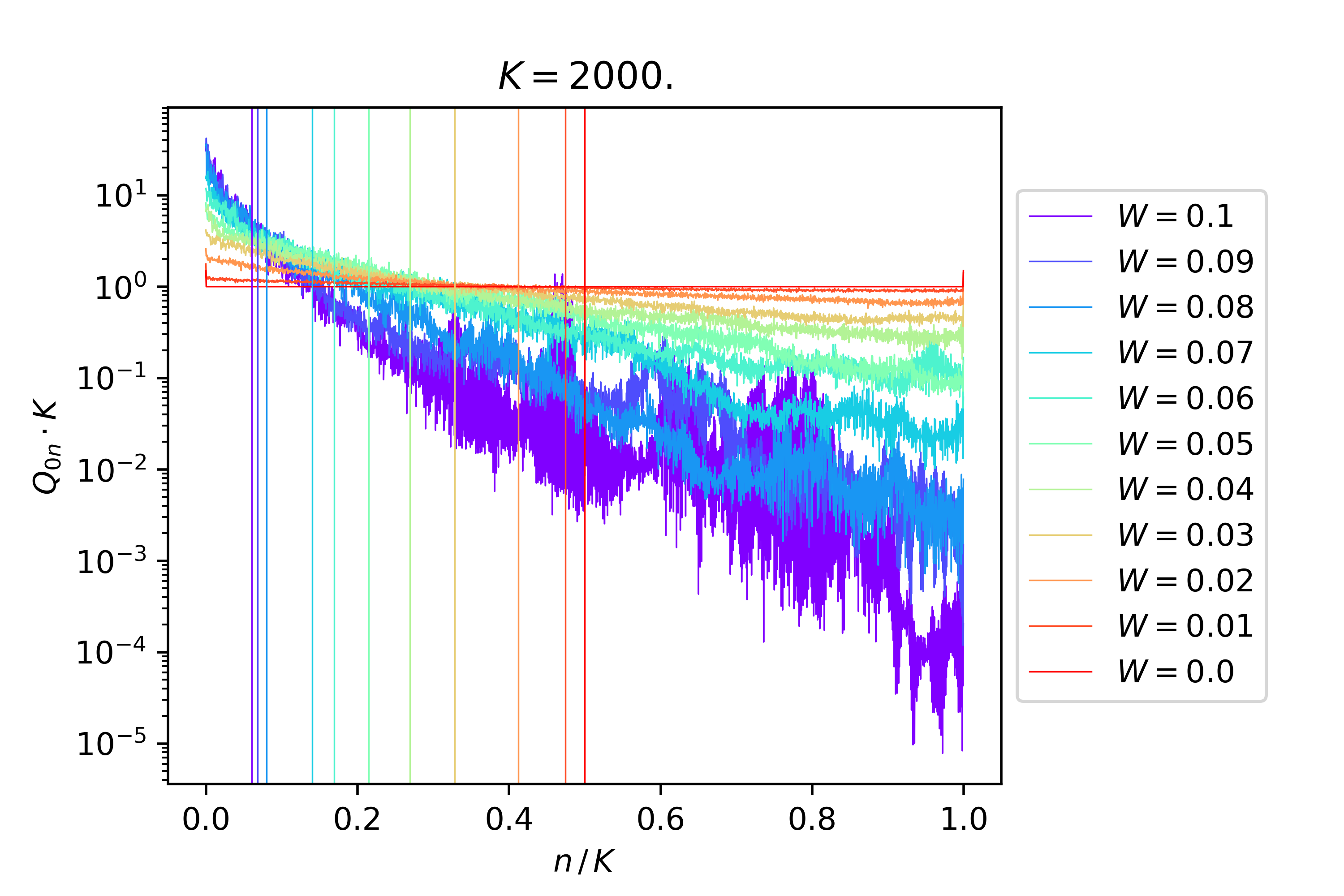} \includegraphics[width=0.45\textwidth]{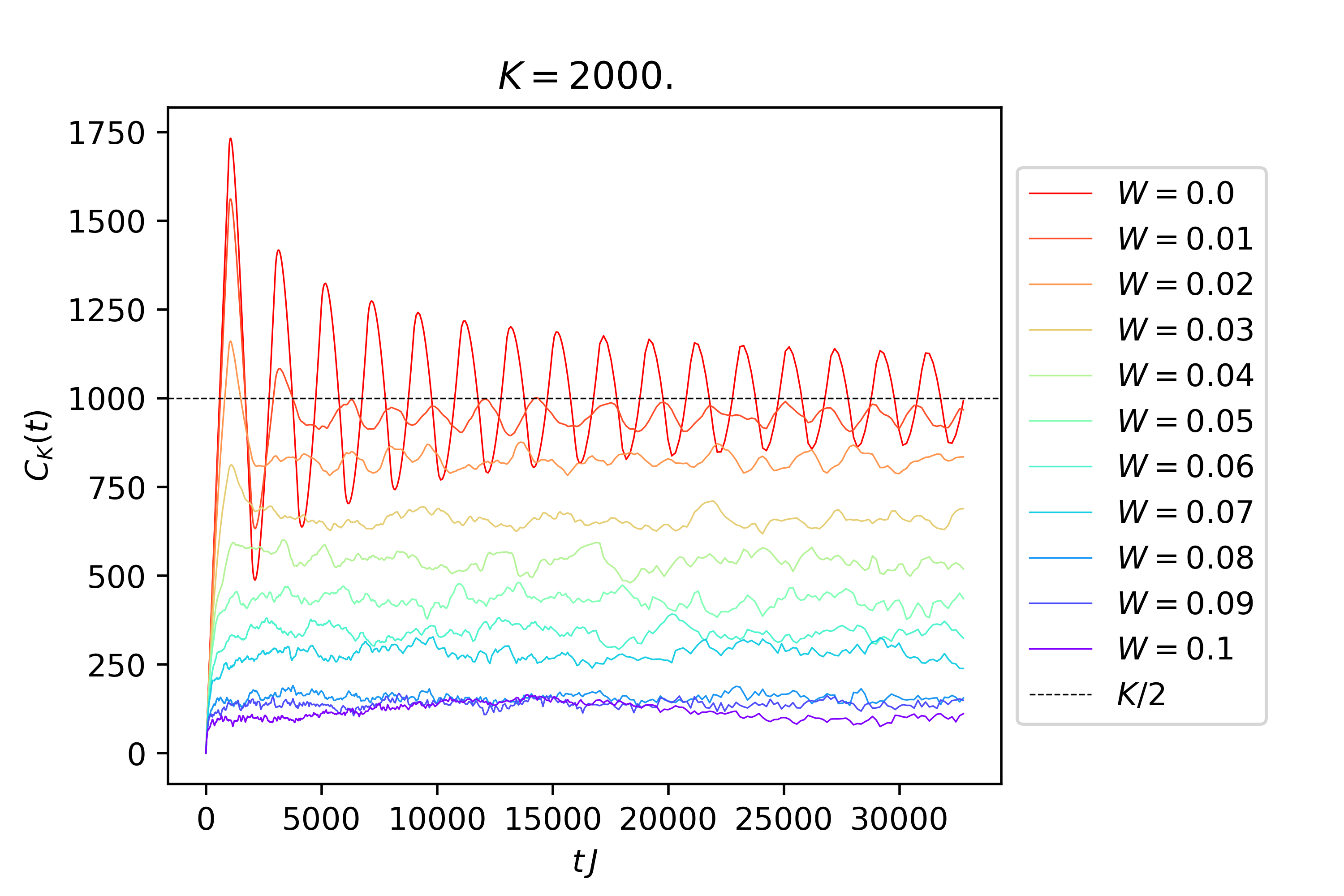}
    
    \caption{\textbf{Top:} Lanczos sequences generated by drawing each Lanczos coefficient from a normal distribution with unit mean and standard deviation $W$, following section \ref{Subs_Toy_FLAT}. Each color corresponds to a single random realization for a fixed disorder strength, as indicated in the legend. In all cases we studied a Krylov chain of length $K=2000$. \textbf{Bottom left:} Transition probabilities computed out of each Lanczos sequence. \textbf{Bottom right:} K-complexity as a function of time.}
    \label{fig:Toy_Disorder_FLAT}
\end{figure}

In this canonical model for off-diagonal disorder, we generate a random sequence of Lanczos coefficients (hopping amplitudes) $\left\{b_n\right\}_{n=1}^{K-1}$ taking all the coefficients to be independent and identically distributed (i.i.d.) Gaussian random variables with mean $J$ and standard deviation $WJ$, where $J$ is to be thought of as a dimensionful parameter setting the energy scale\footnote{Again, it was set to $1$ in the numerics, so that all dimensionful quantities computed are to be thought of as normalized by $J$.}, and $W$ is a dimensionless parameter controlling the disorder strength. Figure \ref{fig:Toy_Disorder_FLAT} depicts various random realizations of Lanczos sequences with different disorder strengths, as well as the corresponding long-time averaged transition probabilities and K-complexities as a function of time computed out of them. We observe that transition probabilities are more and more biased to the left as disorder increases, and in fact decay roughly exponentially, yielding a K-complexity profile that saturates at late times at values that decrease with disorder strength. We note that, in the disorder-free case, K-complexity shows persistent oscillations before saturating at $\frac{K}{2}$, consistent with the wave packet bouncing back and forth between the edges of the Krylov chain before diffusing efficiently into a uniform distribution; in Appendix \ref{appx:Constant_b_analytics} we show analytically that the time-averaged transition probability is indeed a constant (up to edge effects), yielding $\sim \frac{K}{2}$ as the K-complexity long-time average. The addition of disorder contributes efficiently to destroying the coherence of the propagating packet, thus washing out complexity oscillations apart from reducing its saturation value (cf. Figure \ref{fig:Toy_Disorder_FLAT} - bottom right).

\subsection{Disordered sequence with ascent and linear descent}

\begin{figure}[t]
    \centering
    \includegraphics[width=0.45\textwidth]{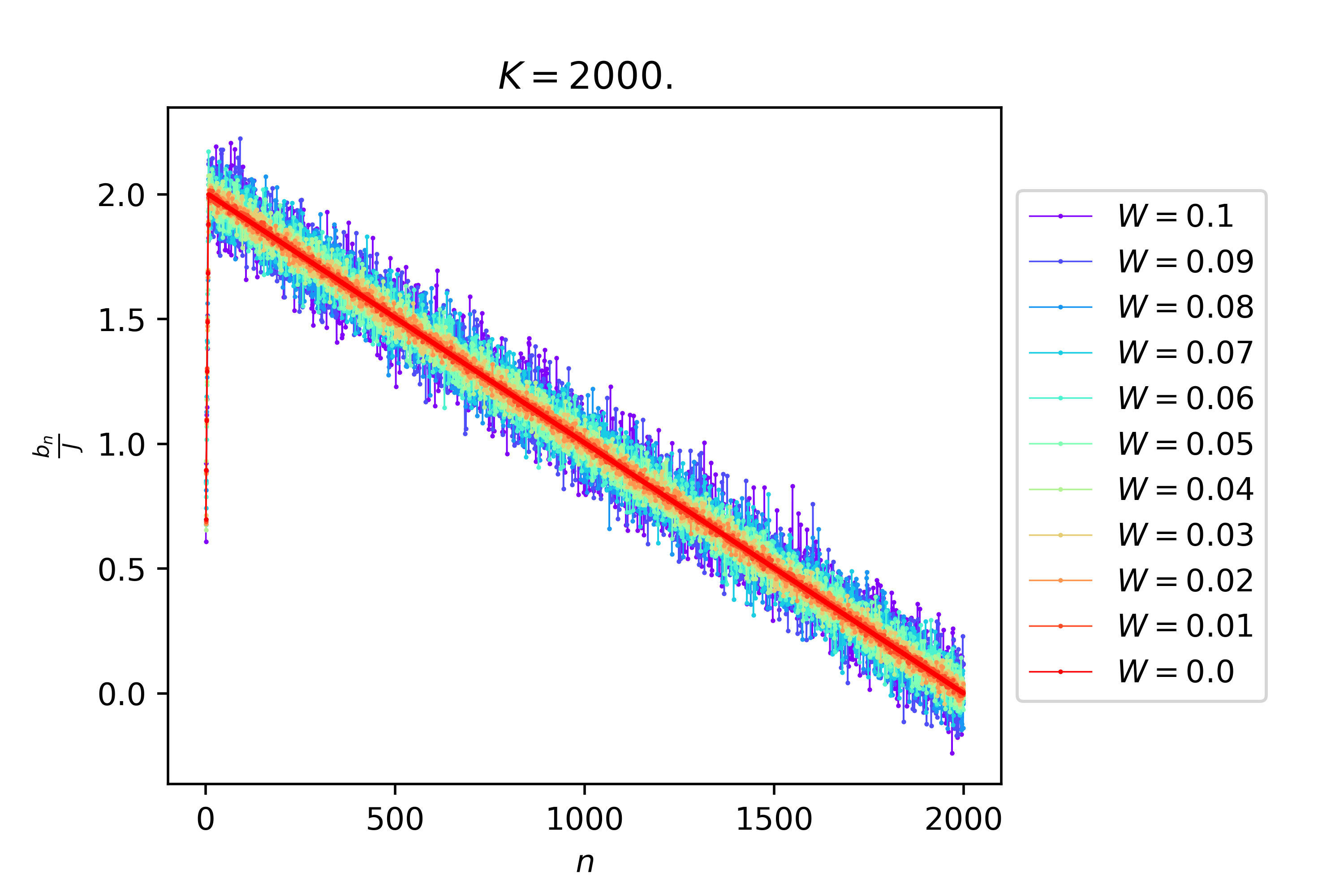} \includegraphics[width=0.45\textwidth]{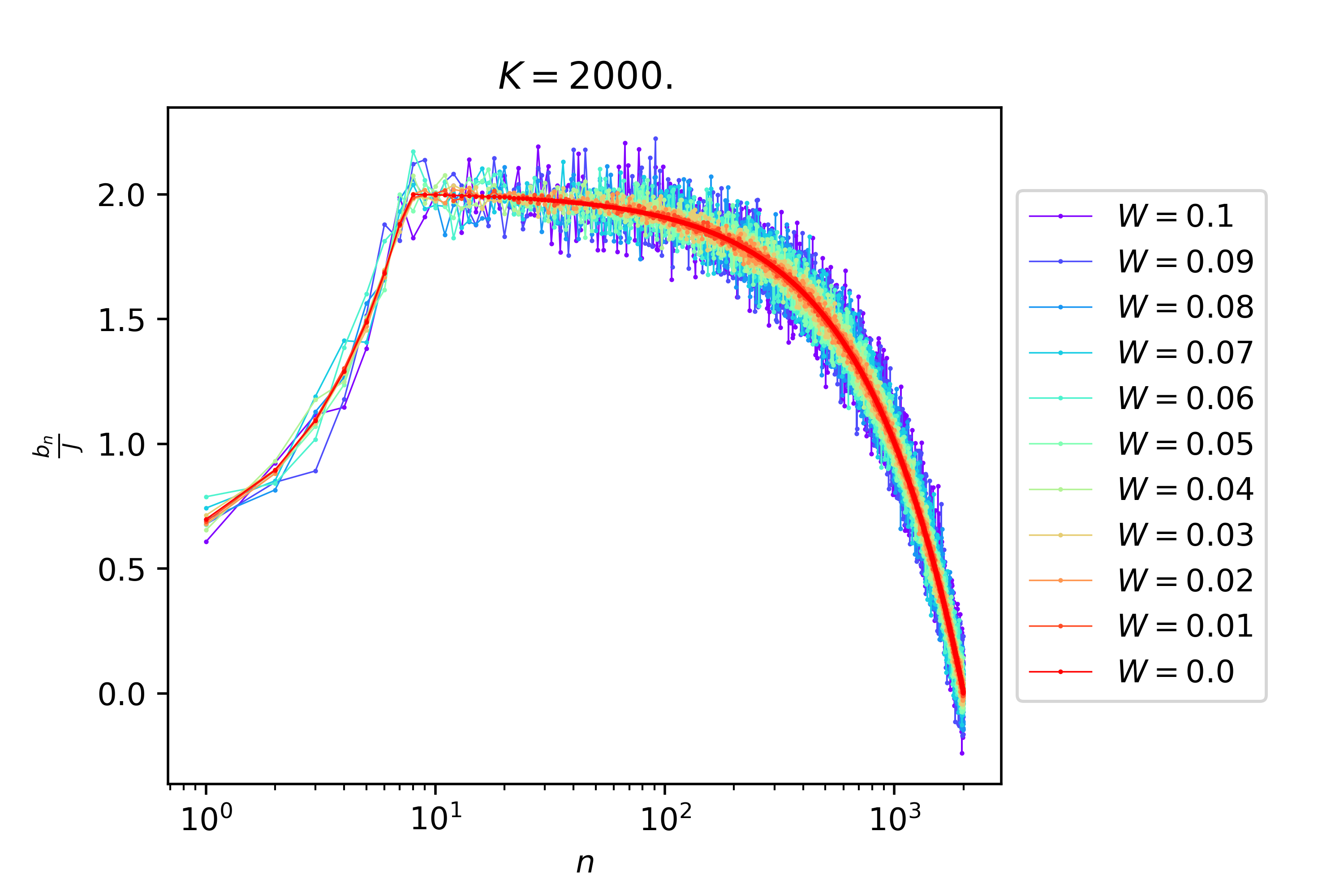} \\
    \includegraphics[width=0.45\textwidth]{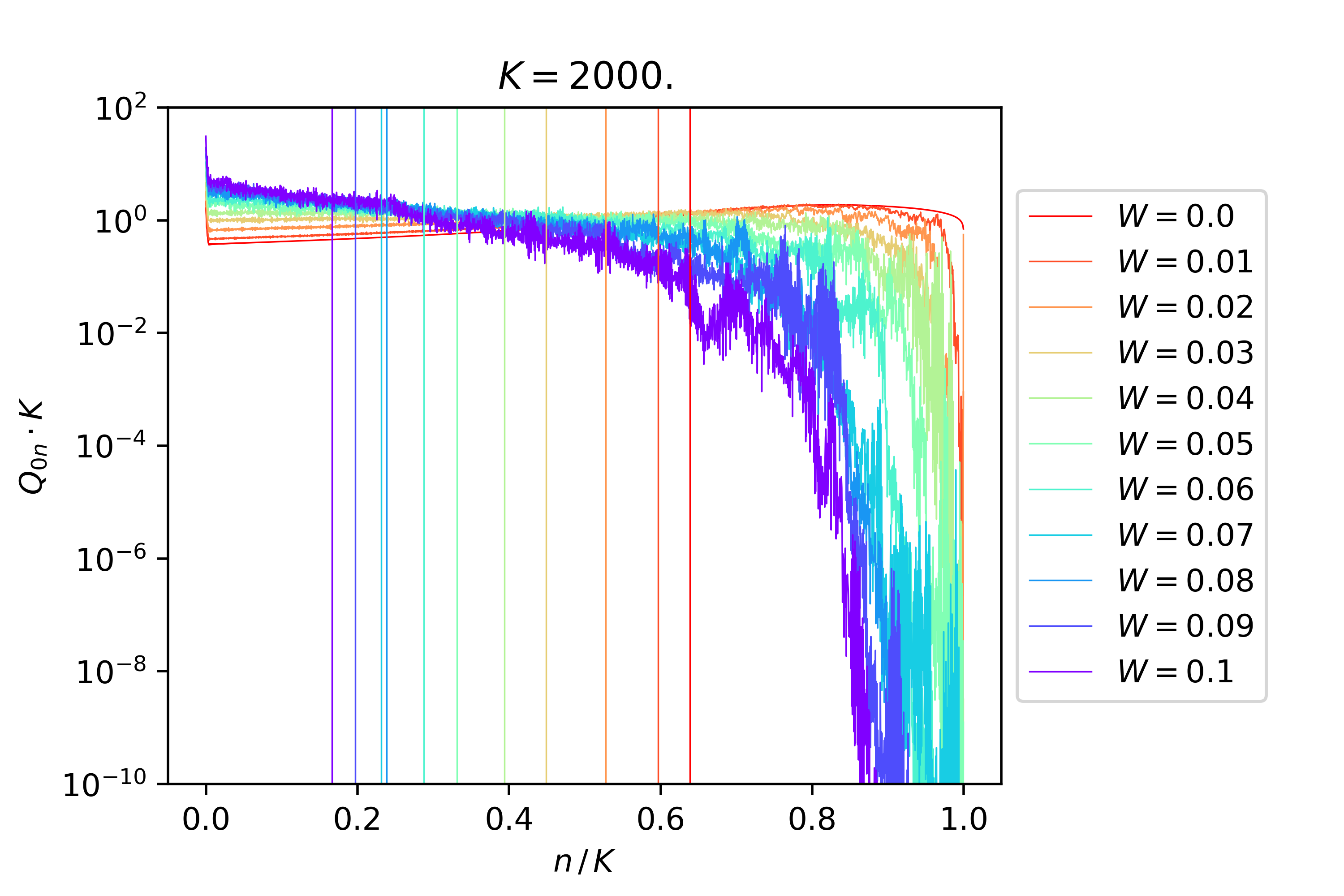} \includegraphics[width=0.45\textwidth]{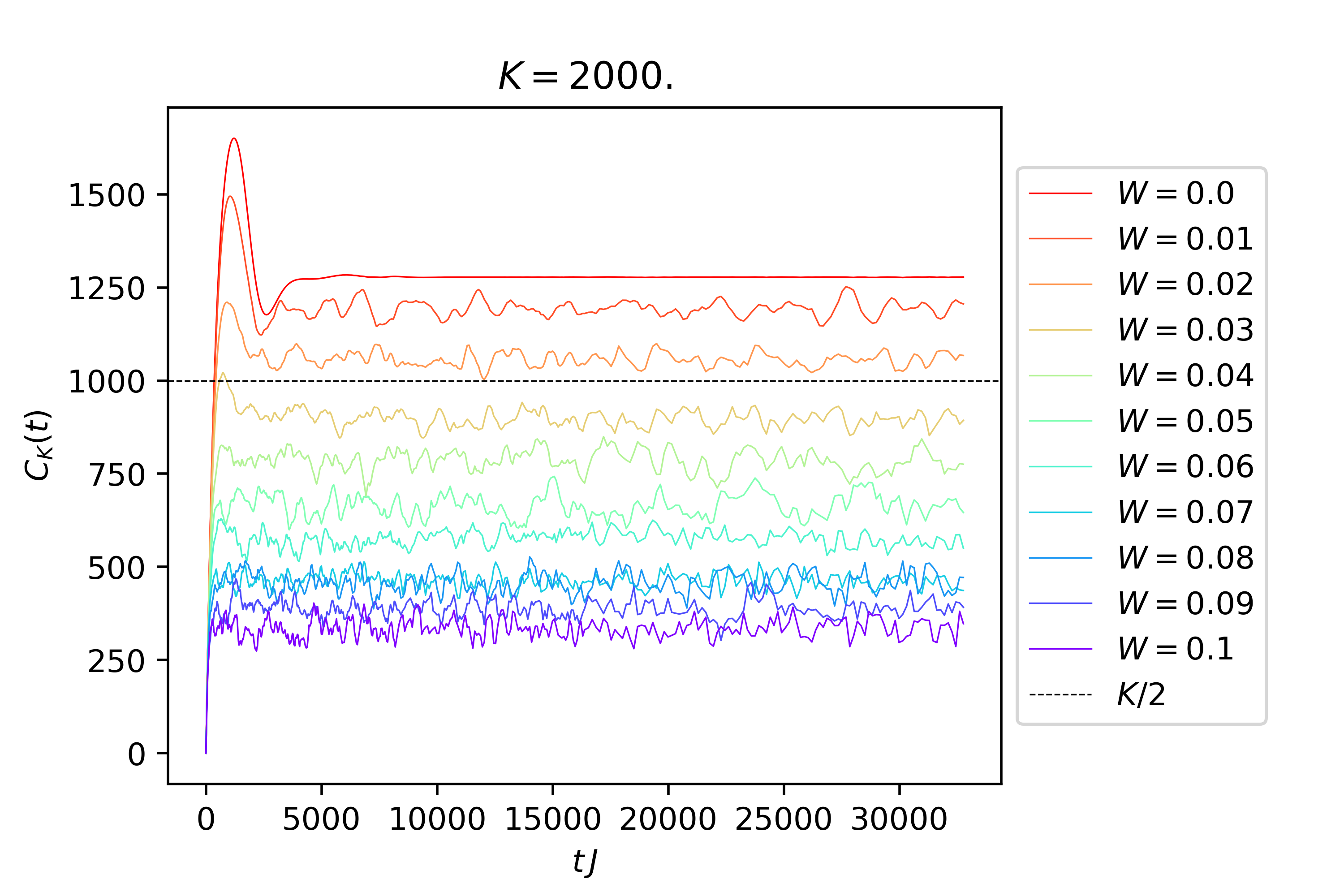}
    \caption{\textbf{Top left:} $b$-sequence for the phenomenological model (\ref{Toy_LIN}). One random realization for different values of the disorder strength is depicted. \textbf{Top right:} Same sequence, plotted with logarithmic scale along the horizontal axis. \textbf{Bottom left:} Time-averaged transition probabilities $Q_{0n}$ computed out of each sequence. As disorder increases they start to follow a profile that decreases with $n$, signaling localization. \textbf{Bottom right:} K-complexity as a function of time for each disorder strength. We observe that the late-time saturation value decreases monotonously with $W$.}
    \label{fig:Toy_LIN}
\end{figure}

This time we add i.i.d. Gaussian fluctuations on top of a Lanczos sequence $b_n$ that increases linearly up to $b_*$ at $n_*\sim \log K$ and then decays linearly to (almost) zero at $n=K-1$. That is:
\begin{equation}
    \centering
    \label{Toy_LIN}
    b_n =  \begin{cases}
        b_0 + \frac{b_*-b_0}{n_*}n + W_n\,J &  0< n < n_*\\
        b_*-\frac{b_*}{K-n_*}(n-n_*) + W_n\,J & n_*\leq n < K
    \end{cases} 
\end{equation}
where, as announced, $W_n$ are i.i.d. Gaussian random variables with zero mean and standard deviation given by the dimensionless parameter $W$. Sequences generated with different disorder strengths as well as the transition probabilities and K-complexities computed out of them are depicted in Figure \ref{fig:Toy_LIN}. The numerical values of $b_0$ and $b_*$ were chosen so that the resulting profiles reassembled those of XXZ\footnote{In fact, for adequately normalized Hamiltonians, $b_*$ should not depend strongly on system size.}.

In the absence of disorder, we observe that K-complexity saturates above $\frac{K}{2}$. This is due to the asymmetric shape of the Lanczos sequence: Since the hopping amplitudes are smaller on the right side of the chain, the packet tends to spend more time in that region on average, as reflected in the transition probability $Q_{0n}$ being slightly biased towards the right. However, as soon as disorder is turned on, the late-time saturation value of complexity drops below $\frac{K}{2}$ to some finite fraction of $K$. This localization phenomenon is also reflected in the fact that the transition probability decays with $n$ at a seemingly exponential rate (even though with a small exponent), for the disorder strengths explored.

One might naively think that under-saturation of K-complexity in the XXZ results is due to the descent in the Lanczos sequence. The model presented here contradicts that statement: We have observed that, in complete absence of disorder, the ascent-descent profile results, conversely, in ``over-saturation'' (in the sense that the late-time average is bigger than half the Krylov dimension $\frac{K}{2}$). It is thanks to the addition of disorder that the late-time complexity value can drop down, as we have explicitly demonstrated (cf. Figure \ref{fig:Toy_LIN}).

There is one slightly unsatisfactory feature of this phenomenological model: The profile of the descent on top of which disorder is added is linearly decaying, but the variance of the fluctuations does not depend on $n$. Hence, the fluctuations become stronger with respect to their mean value as $n$ increases, amplifying the decay of the transition probability. Furthermore, some coefficients towards the end of the sequence can incidentally become very close to zero, effectively ``breaking'' the chain. This all produces a big drop in the transition probability, as can be observed in Figure \ref{fig:Toy_LIN} (bottom left). Localization is, however, still operating throughout the chain, as $Q_{0n}$ decreases with $n$ even on the left side of the chain when disorder is sufficiently strong. A further refinement of the model, to be presented next, tries to get rid of the above-mentioned edge effect by adding some concavity towards the end of the descent.

\subsection{Disordered sequence with ascent and quasi-linear concave descent}

If the descent of the Lanczos sequence had a slightly concave profile, the region where the strength of the fluctuations becomes comparable to the mean value on top of which they are added would be pushed towards the right. In fact, the analytical expression for the Lanczos sequence in RMT at large size has this feature \cite{Kar:2021nbm}: the quasi-linear descent gets eventually modified by a square-root behavior. With this motivation, we use the following Ansatz for the toy Lanczos sequence:

\begin{equation}
    \centering
    \label{Toy_SQRT}
    b_n =  \begin{cases}
        b_0 + \frac{b_*-b_0}{n_*}n + W_n\,J & 0< n < n_*\\
        b_*\sqrt{1 - \frac{n-n_*}{K-n_*}} + W_n\,J & n_*\leq n < K
    \end{cases}~.
\end{equation}
For $n_*<n\ll K$ the descent is linear, but towards the edge $n\sim K$ the square-root behavior becomes dominant.

\begin{figure}[t]
    \centering
    \includegraphics[width=0.45\textwidth]{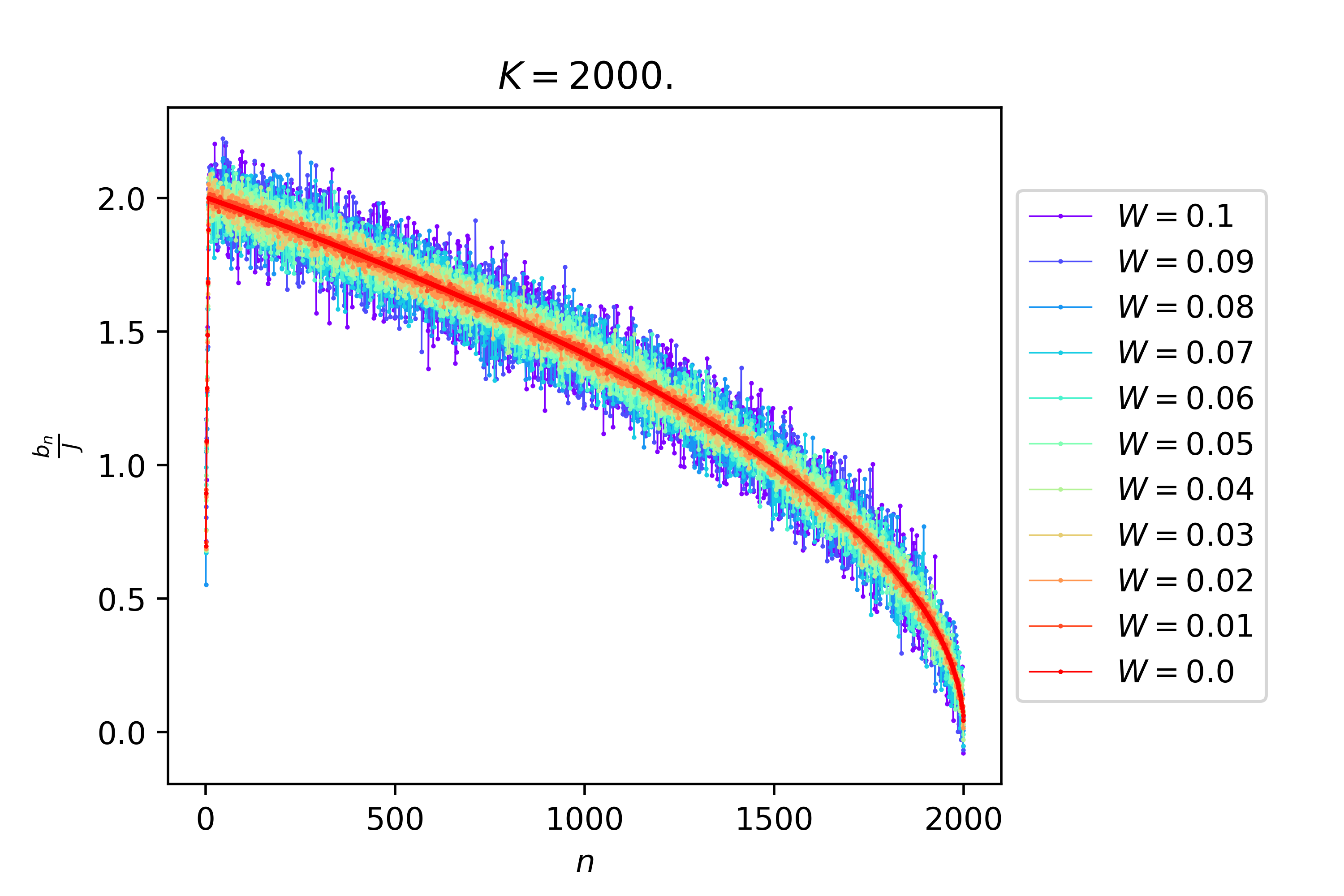} \includegraphics[width=0.45\textwidth]{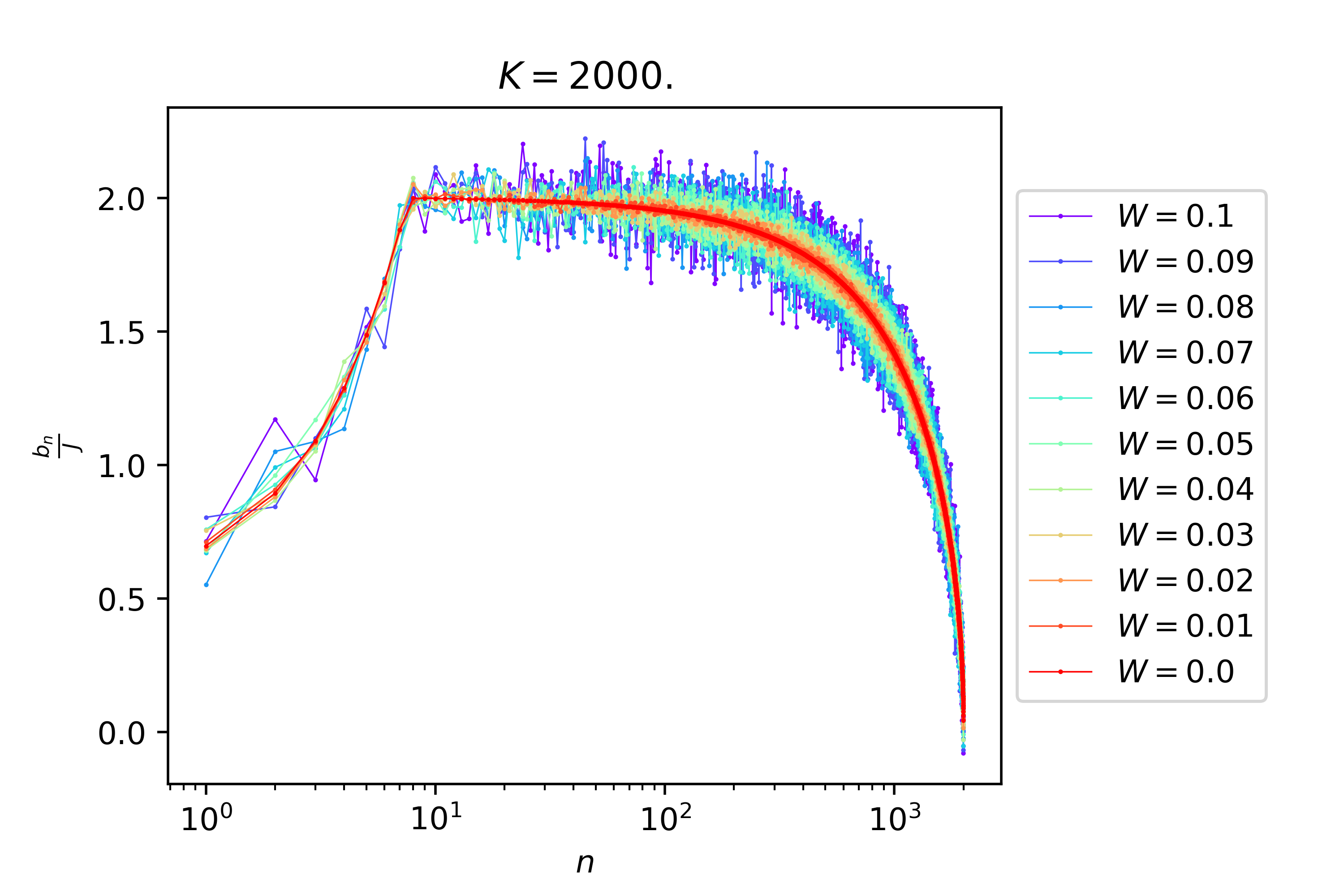} \\
    \includegraphics[width=0.45\textwidth]{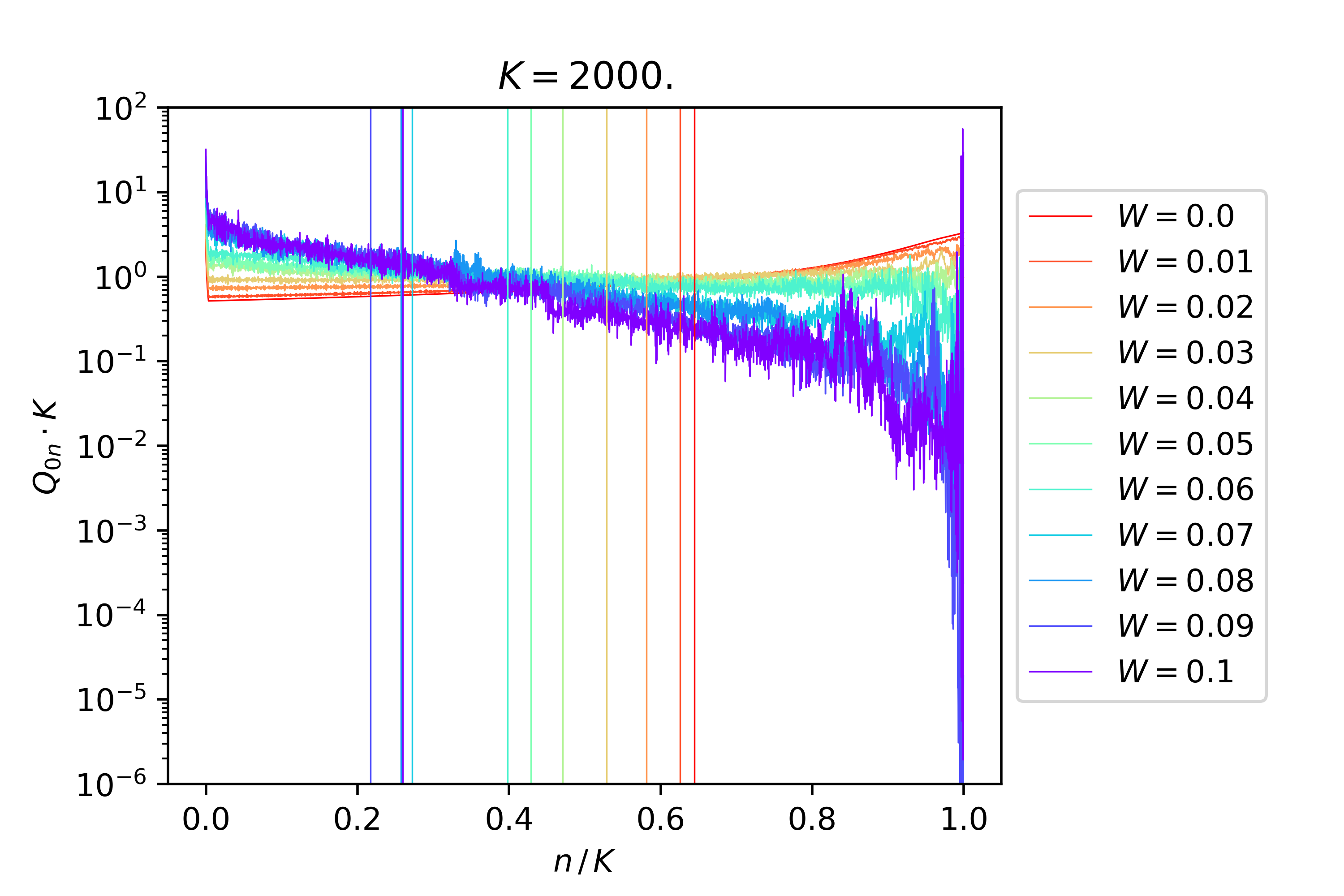} \includegraphics[width=0.45\textwidth]{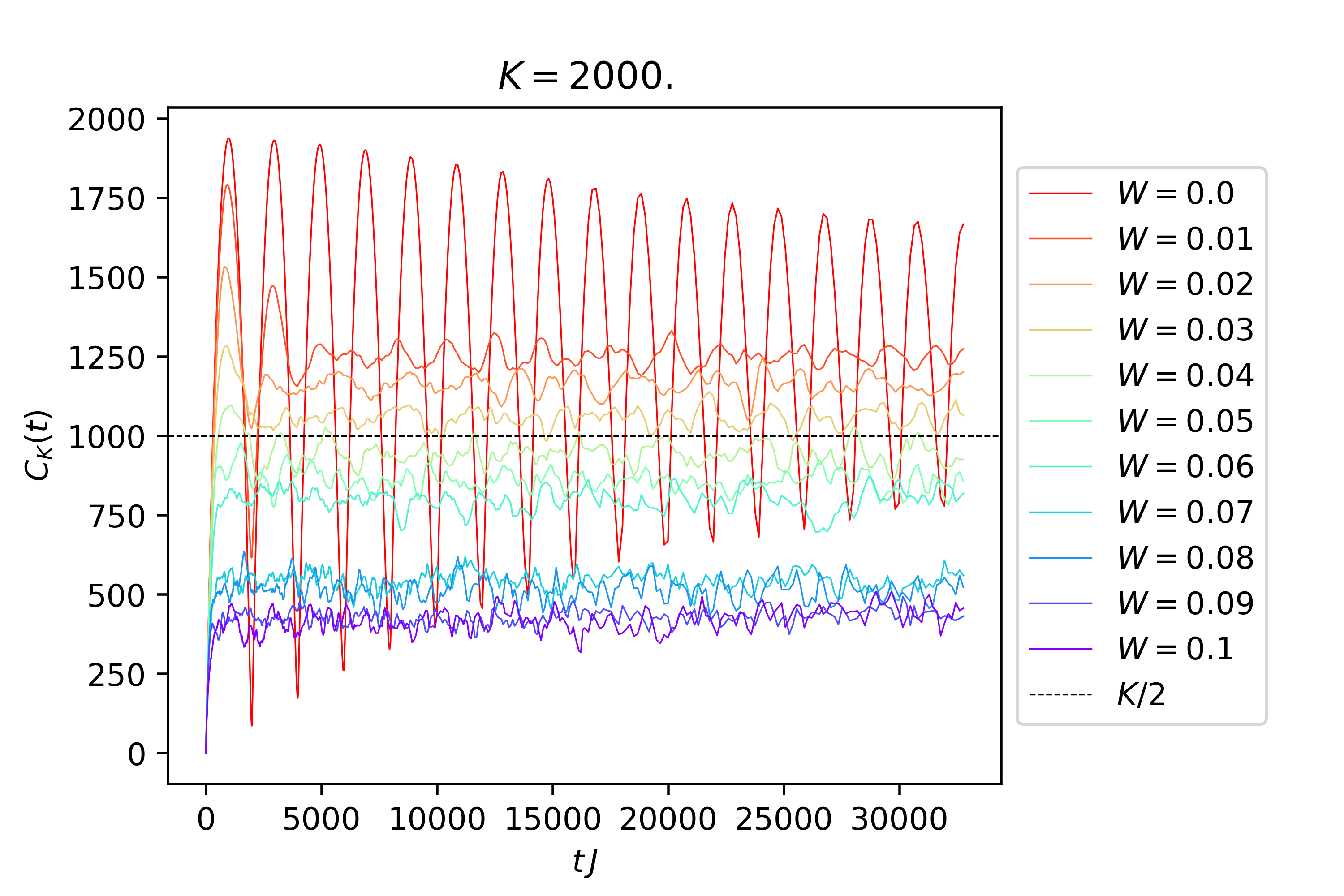}
    \caption{\textbf{Top left:} Lanczos sequences with a slightly concave shape towards the end of the descent, with fluctuations controlled by $W$. \textbf{Top right:} Idem, with logarithmic scale along the horizontal axis. \textbf{Bottom left:} Transition probabilities $Q_{0n}$ computed out of each $b$-sequence. We note the similarity to the results depicted in Figure \ref{fig:Qn0_XXZ_vs_SYK}. \textbf{Bottom right:} K-complexity as a function of time, $C_K(t)$, for each disorder strength. Large oscillations disappear when disordered is turned on, and the complexity saturation value decreases as disorder strength increases.}
    \label{fig:Toy_SQRT}
\end{figure}

Figure \ref{fig:Toy_SQRT} depicts the generated $b$-sequences as well as $Q_{0n}$ and $C_K(t)$. Their striking similarity to the XXZ results presented earlier in this Chapter backs up the claim that disorder is responsible for the obtained transition probabilities and the lower saturation value of K-complexity. A weak disorder $W=0.1$ is able to reduce the complexity saturation value from above $\frac{K}{2}$ to $\sim\frac{K}{4}$.

It is, again, interesting to note that the total absence of disorder would imply long standing coherent oscillations of complexity and a saturation value slightly above $\frac{K}{2}$ due to the asymmetric profile of the $b$-sequence that has already been discussed. This feature should remain there at larger system sizes, even though less pronounced. Hence, the fact that in a single random realization of SYK K-complexity does saturate around $\frac{K}{2}$ (in fact, even a bit below) and does not feature strong oscillations (cf. \cite{I}) indicates that even in this system the small disorder of its Lanczos sequence is operating. From this perspective, the difference between SYK and XXZ is more quantitative than qualitative: The $b$-sequence of the latter is more disordered than that of the former, but they both display localization in the Krylov chain to some extent. This is consistent with usual discussions in Anderson localization \cite{Anderson_AbsDiff,Thouless_1972}: in one spatial dimension there is no critical disorder strength, and this phenomenon is always present for any value of the variance of the fluctuations, but the localization length can vary drastically.

\section{Interlude: Summary and follow-up tasks}\label{Sect_Interlude}

So far, we have studied the behavior of K-complexity for local operators in the XXZ spin chain, a strongly-interacting integrable system solvable via the Bethe Ansatz. We have found that the associated Krylov space dimension is exponential in system size, however the disorder in the sequence of Lanczos coefficients induces a localization effect that prevents maximally efficient exploration of the Krylov chain by the operator wave function, resulting in a saturation value of K-complexity at late times below $\frac{K}{2}$. It should be noted that the starting Hamiltonian, i.e. an XXZ spin chain with constant coupling strength, was not defined through a set of disordered couplings: disorder appears in the Lanczos sequence and it is enhanced by the Poissonian statistics of its spectrum; localization operates therefore in Krylov space rather than in direct space. Results also indicate that the difference between XXZ and SYK is more quantitative than qualitative: In both cases there is some disorder in the Lanczos sequence, but it is stronger in the integrable model; since the dynamics on the Krylov chain are described by a one-dimensional hopping problem, where there is no critical value for the disorder strength, localization should be operating in both cases, to a different extent in each of them.  In turn, the enhanced disorder in the Lanczos sequences for XXZ causes a ``tilt'' in the transition probability from the initial operator $|\mathcal{O}_0)$ to $|\mathcal{O}_n)$ as can be seen both in the numerical results for XXZ (Figure \ref{fig:Qn0_XXZ_vs_SYK}) as well as in the phenomenological models in section \ref{Sect_Toys}, thus reducing the value of the long-time average of K-complexity. This is also manifest in the presence of Liouvillian eigenstates with smaller K-complexities in XXZ, especially around the band center (Figures \ref{fig:KC_KCi} and \ref{fig:KC_KCi_band_center}), which is again due to stronger localization as a result of the larger disorder in the Lanczos sequences.

More work needs to be done in order to determine how generic this phenomenon is. In particular, it is necessary to formalize more strictly the relation between Poissonian statistics (and hence, integrability) and the erratic structure of the Lanczos coefficients, in a way that allows to predict the strength of the fluctuations of the $b$-sequence from data of the initial Hamiltonian (and perhaps the operator). Eventually, it would also be enlightening to push this towards an analytical estimate for the saturation value of K-complexity, which so far can only be accessed numerically. It should be noted, however, that the latter point corresponds to a generically complicated problem in the conventional framework of Anderson localization.

Nevertheless, the most immediate check of the proposed relation between Krylov space localization and integrability that may come to mind is to test to what extent such a phenomenon applies in a system featuring both integrable and chaotic regimes. This was done for contribution \cite{III}, whose content will be reproduced in the coming sections.
In this work we explored K-complexity in a class of quantum systems that shows integrable to chaotic transitions as a function of certain control parameters. We also characterised, for the purpose of comparison, the behavior of K-complexity in random matrix theory, both with and without time reversal symmetry. Interestingly, we found agreement between the late-time behavior of K-complexity for the Hamiltonians in the chaotic regime and that of the random matrix ensembles relevant to the corresponding universality classes.

\section{XXZ and its integrability breaking} \label{sec:XXZ_rstats}
The Heisenberg XXZ spin chain is an integrable model which exhibits Poisson level-spacing statistics.  The model consists of nearest-neighbor spin interactions
\begin{equation} \label{HXXZ}
    H_{XXZ} =  \sum_{i=1}^{N-1} \left\{ J \left( S_i^x S_{i+1}^x + S_i^y S_{i+1}^y \right) + J_{zz} S_i^z S_{i+1}^z\right\}
\end{equation}
where\footnote{Note that the convention for the Hamiltonian \eqref{HXXZ} used in \cite{III} differs from \eqref{XXZ}, used in \cite{II}, by the combination of a constant shift and a rescaling by and order-one prefactor.} $S_i^\alpha =\frac{1}{2} \sigma_i^\alpha$ and $ \sigma_i^\alpha$ are the Pauli matrices with $\alpha=x,y,z$. 
In a series of papers it was shown that even the addition of a local operator such as
\begin{equation} \label{IBT_local}
    H_d =  S^z_{j}
\end{equation}
to the XXZ Hamiltonian can break its integrability \cite{Santos_2004, PhysRevE.84.016206, PhysRevB.80.125118, PhysRevB.98.235128, Rigol_XXZ, PhysRevX.10.041017} and spectral statistics will show chaotic behaviour. Another type of integrability-breaking term \cite{doi:10.1119/1.3671068} we will consider is the next-to-nearest-neighbour interaction:
\begin{equation} \label{IBT_NL}
    H_{NNN} = \sum_{i=1}^{N-2} S_i^z S_{i+2}^z ~.
\end{equation}

We will demonstrate the transition from integrability to chaos by studying the distribution of the ratios of consecutive level spacings \cite{PhysRevLett.110.084101, PhysRevB.75.155111}, and show that increasing the strength of the integrability-breaking term from zero will result in a transition in the spectral behaviour from integrable to chaotic. For an ordered set of energy eigenvalues $\{E_i\}_{i=1}^D$, consecutive level spacings are defined as $s_i=E_{i+1}-E_i$ and consecutive ratios are defined as the set $r_i = s_{i}/s_{i-1}$. 
The distribution $P(r)$ was computed in \cite{PhysRevLett.110.084101} for the random matrix ensembles GOE, GUE and GSE.
It is useful to define the quantity $\tilde{r_i} = \min\left( r_i, \frac{1}{r_i}\right)$ with distribution $P(\tilde{r})=2P(r)\theta(r-1)$ whose mean $\langle \tilde{r} \rangle$ can be used as an indicator to distinguish an integrable system from a chaotic one. For a Poissonian distribution of level-spacings $P(s)= e^{-s}$, the distribution of $r$ is given by\footnote{Making an abuse of notation, we are using the same letter $P$ for the pdf of $s$, $r$ and $\tilde{r}$.} $P(r)= (1+r)^{-2}$ \cite{PhysRevB.75.155111} and $\langle \tilde{r}\rangle = 2 \ln 2 -1 \approx 0.38629$.  For the Wigner ensembles (GOE, GUE and GSE) distinguished by their Dyson index ($\beta= 1,2$ and $4$ respectively) it was shown in \cite{PhysRevLett.110.084101} that a very good approximation for practical purposes is $P(r)=\frac{1}{Z_\beta} \frac{(r+r^2)^\beta}{(1+r+r^2)^{1+\frac{3}{2}\beta}}$, where $Z_\beta$ is a normalization constant. The $\langle \tilde{r}\rangle$ value for GOE is approximately $0.53590$.

\subsection{Choice of sector and local operator}
The XXZ Hamiltonian commutes with the operator representing the total spin in the $z$-direction, 
\begin{equation}
    M = \sum_{i=1}^N S_i^z~,
\end{equation}
and is invariant under reflection with respect to the edge of the chain, represented by the parity operator $P$ \cite{Santos2013} given in \eqref{Parity}.  To avoid degeneracies in the Hamiltonian spectrum we will work in a sector with fixed total spin and parity.  To study K-complexity we will use open boundary conditions and focus on a local operator $\mathcal{O}$ which respects these two symmetries and keeps the computation within the chosen sector:
\begin{equation} \label{Operator}
    \mathcal{O} = S_i^z + S_{N-i+1}^z~,
\end{equation}
where $i$ is chosen to be near the center of the chain.  This operator has a non-zero one-point function which, as argued in Appendix \ref{appx_Connected}, should be subtracted from it in order to probe universal features of the Krylov complexity saturation value.


When adding the integrability-breaking term (\ref{IBT_local}), we remain within the sector by using an odd-length chain and situating the impurity at the middle of the chain: 
\begin{equation} \label{IBT_Lmid}
    H_d = S_{(N+1)/2}^z ~.
\end{equation}
The integrability-breaking term (\ref{IBT_NL}) commutes both with $M$ and with $P$. 

\subsection{r-statistics for XXZ with integrability-breaking terms}
We now present results for the r-statistics of the following interpolating Hamiltonians:
\begin{equation} \label{XXZ+Hd}
    H = H_{XXZ} +\epsilon_d H_d
\end{equation}
and
\begin{equation} \label{XXZ+NNN}
    H = H_{XXZ} + J_{zz}^{(2)} H_{NNN}~,
\end{equation}
where $H_{XXZ}$ is given in (\ref{HXXZ}), $H_d$ in (\ref{IBT_Lmid}) and $H_{NNN}$ is given in (\ref{IBT_NL}).  We work with various values of $J_{zz}$ and set $J=1$ in all cases, which sets the energy units. Some of the Hamiltonians and operators used in the numerical computations were constructed using the QuSpin package \cite{SciPostPhys.2.1.003}. The Lanczos algorithm and K-complexity computations were performed using the codes developed for \cite{I} and \cite{II}.

\begin{figure}
    \centering
    \begin{subfigure}[t]{\textwidth}
    \centering
        \includegraphics[scale=0.8]{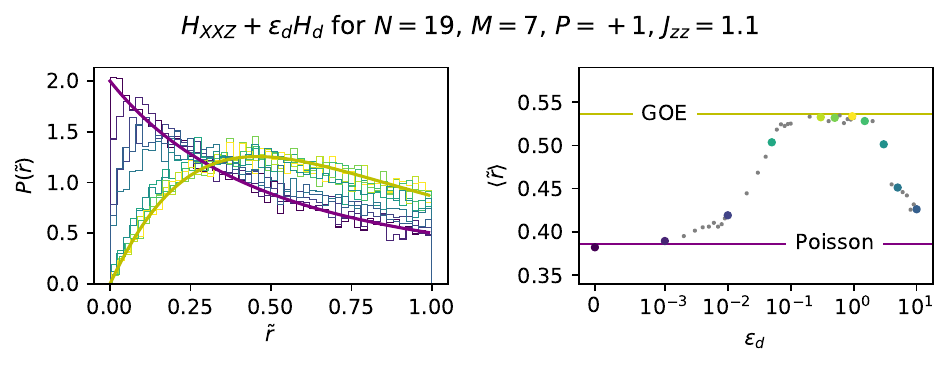}
        \caption{ $\tilde{r}$ statistics for $H_{XXZ}+\epsilon_d H_d$.}
        \label{fig:rStats_Hd}
    \end{subfigure}
    \begin{subfigure}[t]{\textwidth}
        \centering
    \includegraphics[scale=0.8]{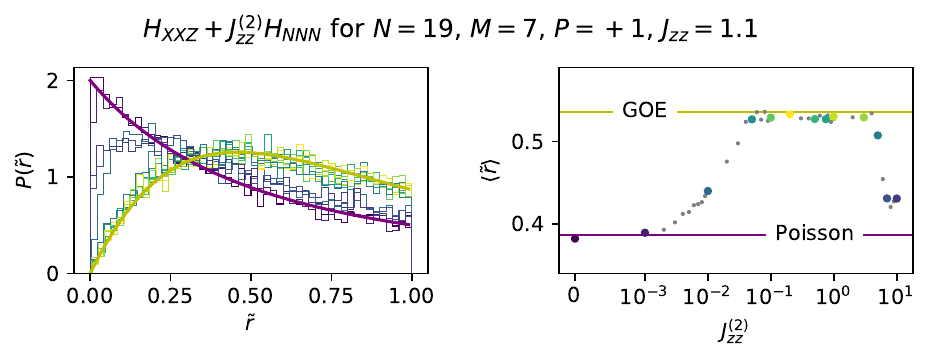}
    \caption{ $\tilde{r}$ statistics for $H_{XXZ}+J^{(2)}_{zz}H_{NNN}$.}
    \label{fig:rStats_NNN}
    \end{subfigure}
    \caption{ \textbf{Left:} Probability distribution functions for the $\tilde{r}$ statistics of  (\ref{XXZ+Hd}) (top) and (\ref{XXZ+NNN}) (bottom) with increasing value of $\epsilon_d$ and $J^{(2)}_{zz}$, respectively, computed for $N=19$ spins in the $M=7, P=+1$ sector with $J_{zz}=1.1$. The purple line represents the analytical result for $P(\tilde{r})$ in the case of Poissonian level-spacing statistics, while the yellow line represents the analytical result for the GOE ensemble. \textbf{Right:} The value of $\langle\tilde{r}\rangle$ as a function of $\epsilon_d$ (top) and $J^{(2)}_{zz}$ (bottom). Horizontal lines represent analytical values for Poisson (purple) and GOE (yellow). The colored dots represent spectra for which we plotted the $P(\tilde{r})$ distribution function in the left panel, while the gray dots represent additional data points.}
    \label{fig:rStats}
\end{figure}

Figures \ref{fig:rStats_Hd} and \ref{fig:rStats_NNN} show the $\tilde{r}$ statistics for the Hamiltonians (\ref{XXZ+Hd}) and (\ref{XXZ+NNN}) respectively.  We plot the distributions of $\tilde{r}$ for various values of the coefficient of the integrability-breaking terms, as well as the mean values $\langle\tilde{r}\rangle$. We compare the results for both $P(\tilde{r})$ and $\langle\tilde{r}\rangle$ with the analytical results for Poisson and GOE mentioned in section \ref{sec:XXZ_rstats}.  We see that increasing the strength of the integrability-breaking term makes the system transition from displaying integrable statistics to displaying chaotic statistics. Note that after the transition, increasing the value of the coefficient of the integrability-breaking term even further makes the system less chaotic, as can be seen in Fig. \ref{fig:rStats}.

\section{K-complexity and integrability-chaos transition}\label{sect_XXZ_KCsat}
\begin{figure}
    \centering
    \begin{subfigure}[t]{0.45\textwidth}
    \centering
        \includegraphics[scale=0.5]{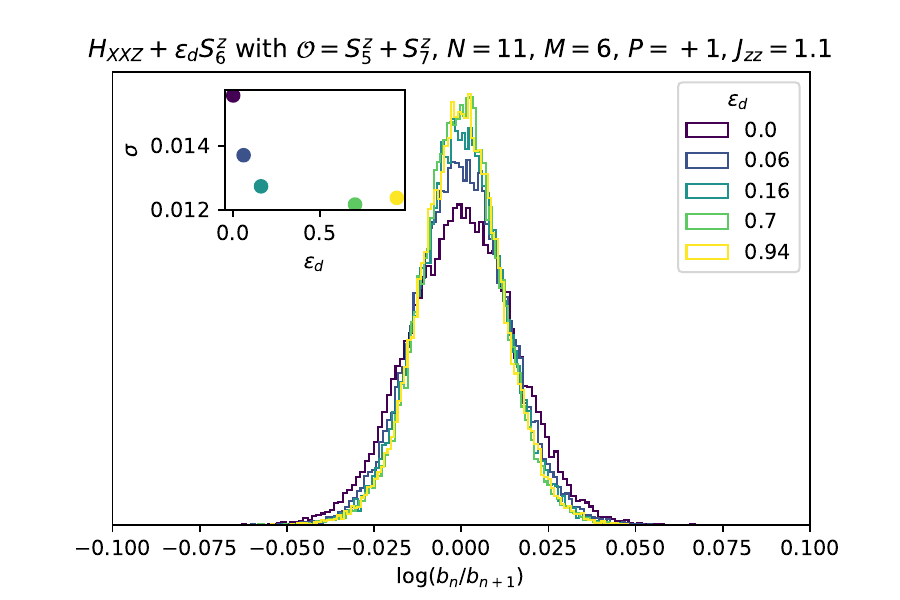}
        \caption{$H_{XXZ} +\epsilon_d H_d$ with $J_{zz}=1.1$ for $N=11$ spins in the $M=6, P=+1$ sector for the operator $\mathcal{O}=S_5^z+S_7^z$. }
    \end{subfigure}
    \hfill
    \begin{subfigure}[t]{0.45\textwidth}
    \centering
        \includegraphics[scale=0.5]{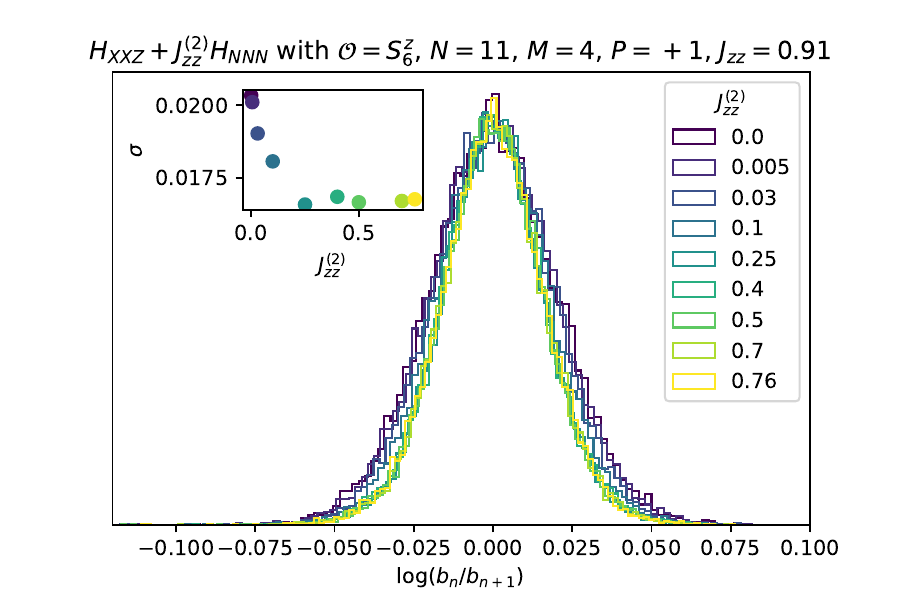}
        \caption{$H_{XXZ} +J_{zz}^{(2)} H_{NNN}$ with $J_{zz}=0.91$ for $N=11$ spins in the $M=4, P=+1$ sector for the operator $\mathcal{O}=S_6^z$.  }
    \label{}
    \end{subfigure}
    \caption{Distribution of the log of consecutive ratios of Lanczos coefficients. \textbf{Inset:} Standard-deviation $\sigma$ of this distribution as a function of the strength of the corresponding integrability-breaking term, which is generically a decreasing function.
    Comparing with the corresponding computations of the K-complexity saturation values in Figures \ref{fig:N11M6Hd} and \ref{fig:N11M4J2zz}, this is consistent with the phenomenology described in \cite{II}, namely that the saturation values of K-complexity will increase with decreasing disorder in the Lanczos coefficients.}
    \label{fig:Lanczos_stats}
\end{figure}

In \cite{II} it was shown that the saturation value of K-complexity is sensitive to the integrability/chaos of a model, by comparing results for complex SYK$_4$ systems with results for XXZ systems of similar Krylov space dimensions.
It was argued that the time evolution on the Krylov chain given by Equation (\ref{Sect_Integr_Dynamics_Krylov_Schr_Eq}) can be mapped to an Anderson problem with off-diagonal disorder. Higher disorder would imply some amount of localization for the Liouvillian eigenvectors and hence a smaller saturation value of K-complexity, while less disorder would imply less localization and higher saturation values of K-complexity.  
In this section we study the Lanczos coefficients statistics and saturation value of K-complexity for the interpolating Hamiltonians given by (\ref{XXZ+Hd}) and (\ref{XXZ+NNN}), with an operator of the type (\ref{Operator}). By increasing the value of the coefficient of the integrability-breaking term we interpolate from a fully integrable model (XXZ) to a chaotic model, as can be seen through the $\langle \tilde{r}\rangle$ transition in Figure \ref{fig:rStats}.   In Figure \ref{fig:Lanczos_stats} we plot the distribution of the log of ratios of consecutive Lanczos coefficients $\log(b_{n}/b_{n+1})$ introduced in \eqref{disorder_strength}.  The mean of this distribution is $\approx 0$ and the standard deviation generally decreases with the strength of integrability breaking, indicating less disorder in the Lanczos sequence.  

\begin{figure}[t]
    \centering
    \begin{subfigure}[t]{0.45\textwidth}
    \centering
        \includegraphics[scale=0.5]{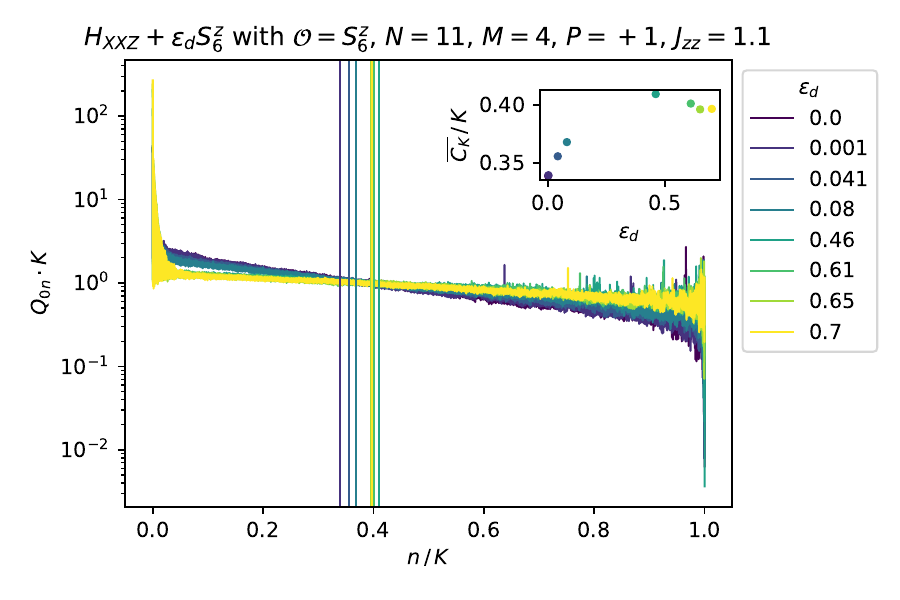}
        \caption{}
        \label{fig:N11M4Hd}
    \end{subfigure}
    \hfill
    \begin{subfigure}[t]{0.45\textwidth}
    \centering
        \includegraphics[scale=0.5]{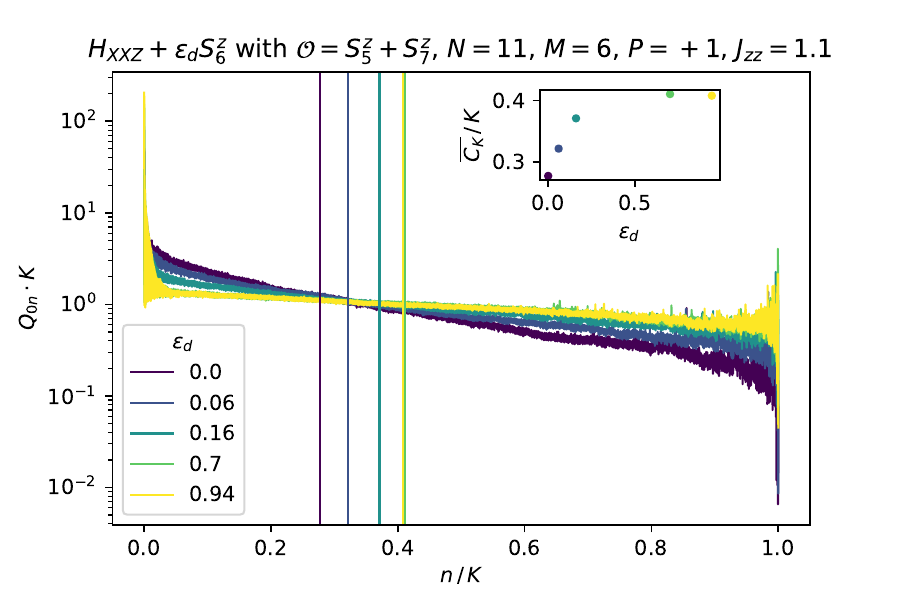}
        \caption{}
        \label{fig:N11M6Hd}
    \end{subfigure}
    \caption{Late-time transition probability results for local operator of the form (\ref{Operator}) with trace removed, for $H_{XXZ}$ with an $H_d$ integrability-breaking term. The vertical lines represent the late-time saturation value of KC as a fraction of the Krylov space dimension. \textbf{Left:} For $N=11$ spins in the sector $M=4, P=+1$ with $J_{zz}=1.1$ for the operator $\mathcal{O}=S_6^z$. The Krylov space dimension is $K=28731$. \textbf{Right:} For $N=11$ spins in the sector $M=6, P=+1$ with $J_{zz}=1.1$ for the operator $\mathcal{O}=S_5^z+S_7^z$. For this system the Krylov space dimension is $K=55461$. \textbf{Inset:} dependence of KC saturation value on the strength of the integrability-breaking term.}
    \label{fig:KC_sat_XXZ_Hd}
\end{figure}

\begin{figure}
    \centering
    \begin{subfigure}[t]{0.45\textwidth}
    \centering
        \includegraphics[scale=0.5]{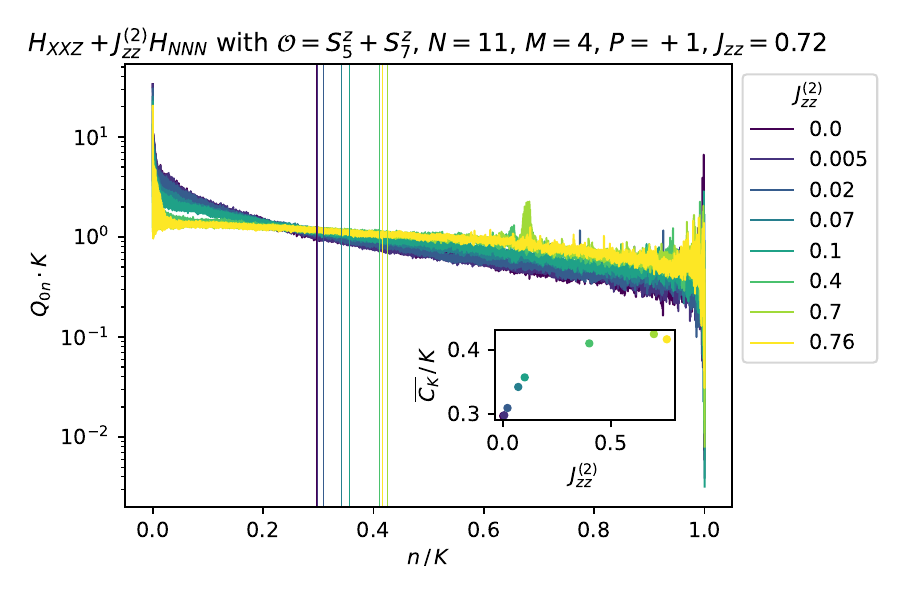}
    \end{subfigure}
    \hfill
    \begin{subfigure}[t]{0.45\textwidth}
    \centering
        \includegraphics[scale=0.5]{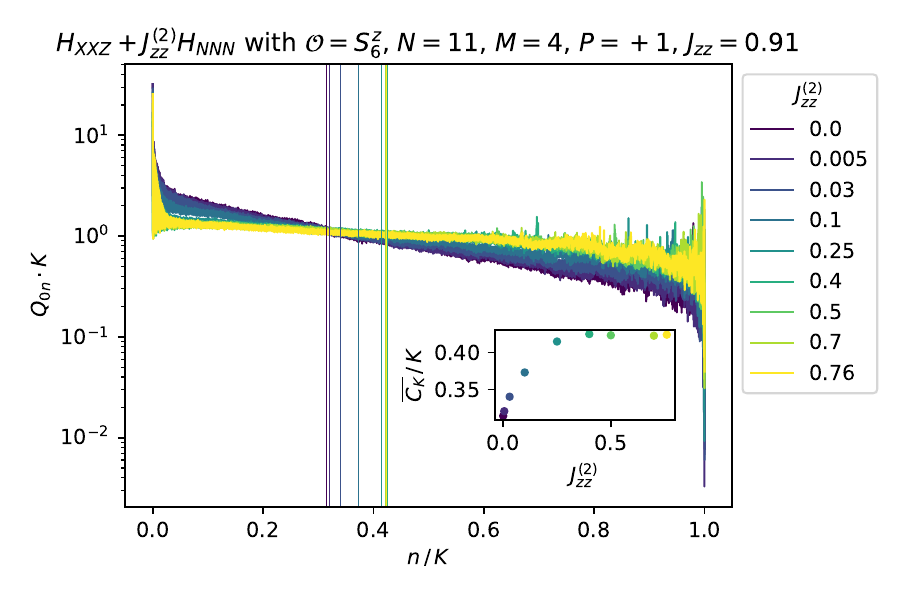}
        \caption{}
        \label{fig:N11M4J2zz}
    \end{subfigure}
    \caption{Results for the saturation value of K-complexity computed for a local operator of the form (\ref{Operator}) with trace removed, for $H_{XXZ}+J^{(2)}_{zz}\,H_{NNN}$ integrability-breaking term in the sector $N=11, M=4, P=+1$. \textbf{Left:} With $J_{zz}=0.72$, for the operator $\mathcal{O}=S_5^z+S_7^z$.   \textbf{Right:} With $J_{zz}=0.91$, for the operator $\mathcal{O}=S_6^z$. For both systems the Krylov space dimension is $K=28731$.
    \textbf{Inset:} dependence of KC saturation value on the strength of the integrability-breaking term.}
    \label{fig:KC_sat_XXZ_NNN}
\end{figure}

\begin{figure}[t]
    \centering
    \centering
    \begin{subfigure}[t]{0.3\textwidth}
    \centering
        \includegraphics[scale=0.35]{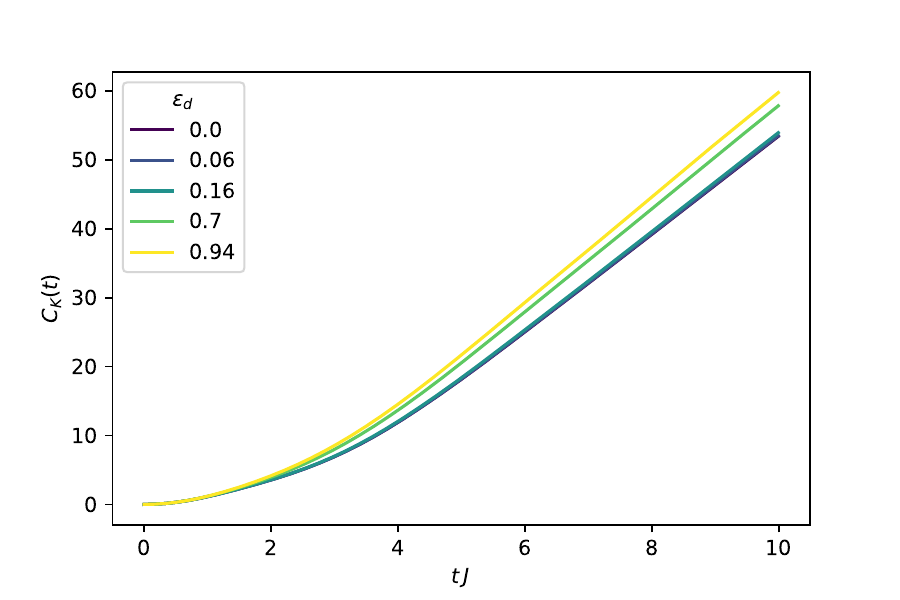}
    \end{subfigure}
    \hfill
    \begin{subfigure}[t]{0.3\textwidth}
    \centering
        \includegraphics[scale=0.35]{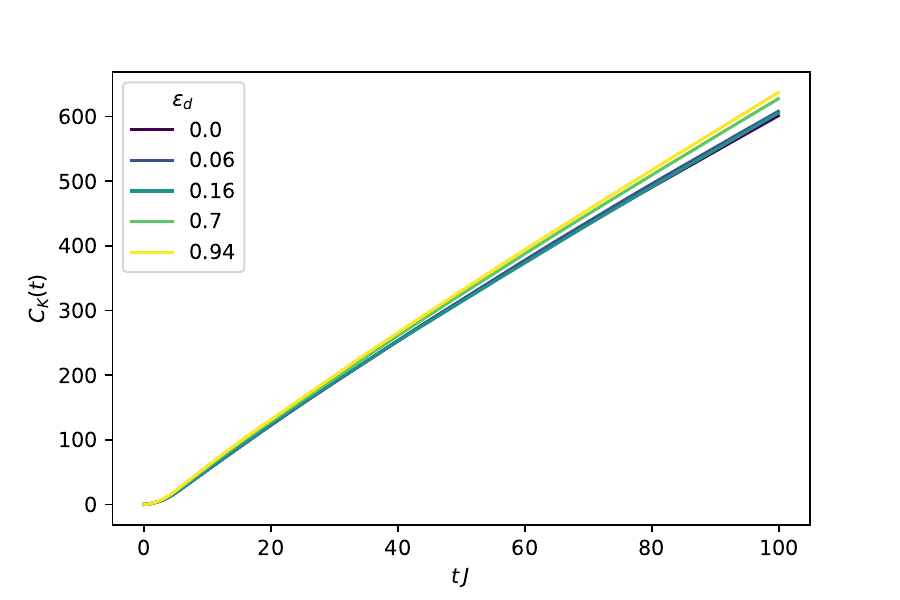}
    \end{subfigure}
    \hfill
    \begin{subfigure}[t]{0.3\textwidth}
    \centering
        \includegraphics[scale=0.35]{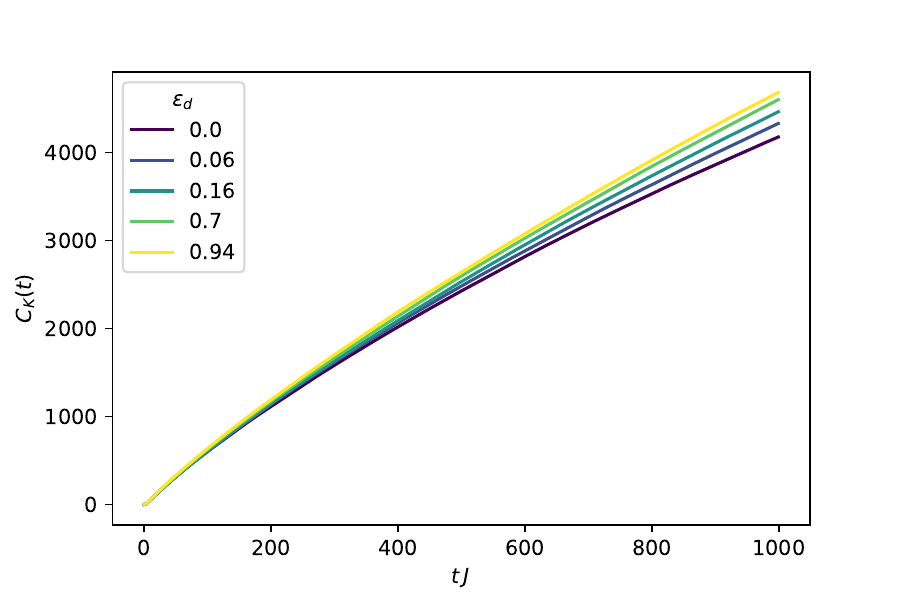}
    \end{subfigure}
    \vfill
    \begin{subfigure}[t]{0.3\textwidth}
    \centering
        \includegraphics[scale=0.35]{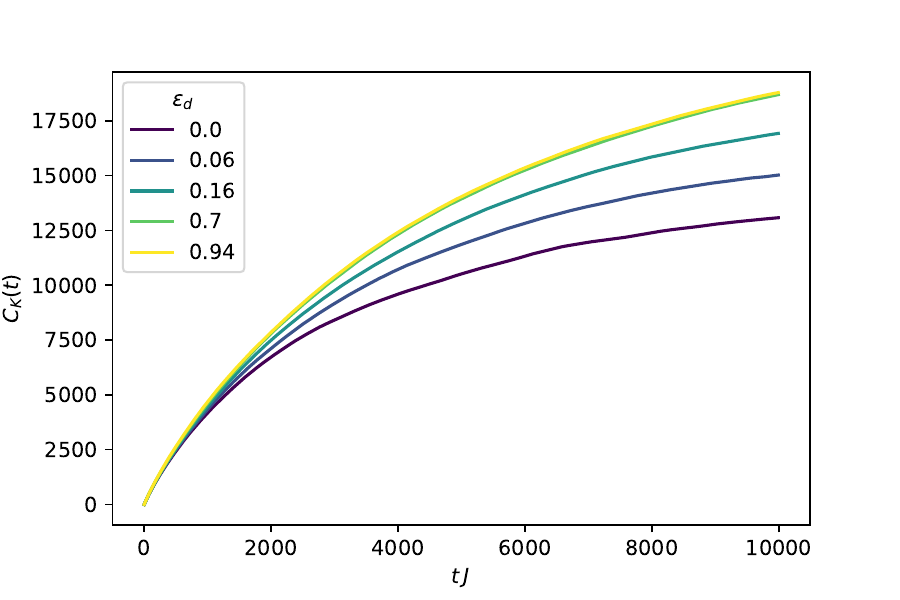}
    \end{subfigure}
    \hfill
    \begin{subfigure}[t]{0.3\textwidth}
    \centering
        \includegraphics[scale=0.35]{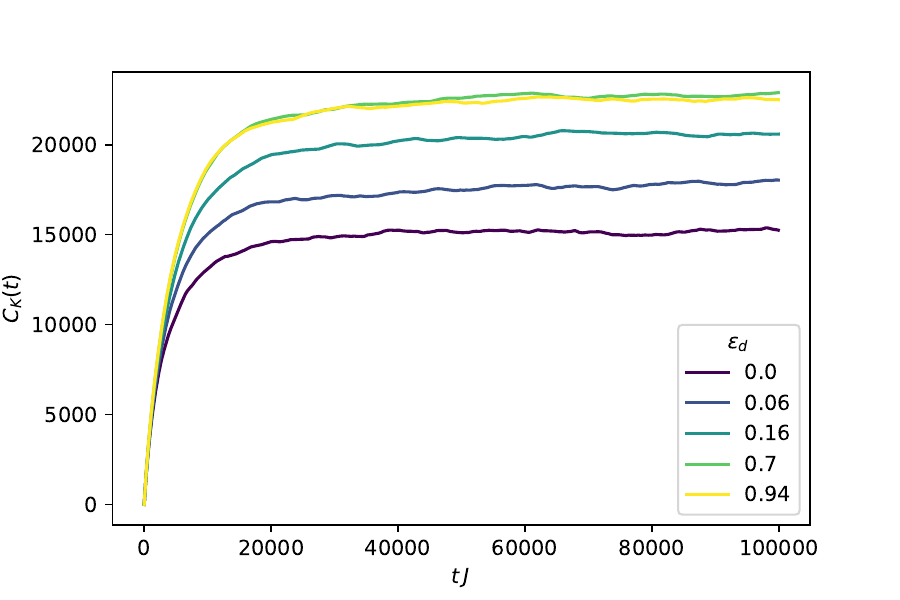}
    \end{subfigure}
    \hfill
    \begin{subfigure}[t]{0.3\textwidth}
    \centering
        \includegraphics[scale=0.35]{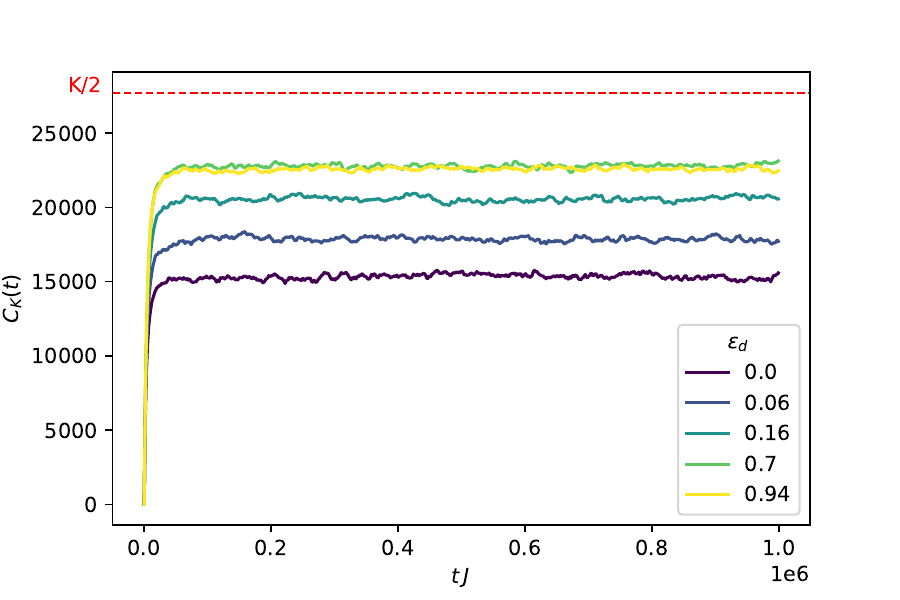}
    \end{subfigure}
    \caption{Results for the time-dependent profile of K-complexity at increasing time scales, for the same system of Fig. \ref{fig:N11M6Hd}.  The Hamiltonian is $H_{XXZ}+\epsilon_d H_d$ with $N=11$ spins in the $M=6, P=+1$ sector with $J_{zz}=1.1$, and the operator is $\mathcal{O}=S^z_{5}+S^z_{7}$ with trace removed. The Krylov space dimension for this setup is $K=55461$ which equals the upper bound for the Krylov space dimension. The final plot shows the saturation values of K-complexity, with the value of $K/2$ shown for reference.}
    \label{fig:KC_time}
\end{figure}

Indeed, we find consistently that the saturation value of K-complexity is affected by the strength of the integrability-breaking term, and generally increases with the value of the integrability-breaking coefficient, as can be seen in Figures \ref{fig:KC_sat_XXZ_Hd} and \ref{fig:KC_sat_XXZ_NNN}. The late-time saturation value of K-complexity as a fraction of the Krylov space dimension can be read off from the vertical lines in the Figures, where the $x$-axis was scaled according to the corresponding Krylov space dimension. Another interesting aspect is the time-dependent profile of K-complexity at various time scales and for different integrability breaking strength, for which results are presented in Figure \ref{fig:KC_time}.  Again we find a consistent relation between the strength of the integrability-breaking term and the late-time value of K-complexity.

\section{RMT results and dependence on universality class}\label{Sect:RMT}

This section gathers results on the saturation value of K-complexity for different operators in Random Matrix Theory, to be used as a reference to compare with the results obtained in the chaotic regime of the deformed XXZ Hamiltonian studied in sections \ref{sec:XXZ_rstats} and \ref{sect_XXZ_KCsat}.

We shall consider systems with a Hilbert space of dimension $D$, equipped with a random Hamiltonian $H$ drawn from a Gaussian ensemble, with probability measure
\begin{equation}
    \centering
    \label{RMT_Gaussian_measure}
    p(H) DH = \exp\left\{-\frac{D}{2 \sigma^2}\text{Tr}\left(H^\dagger H\right)\right\}DH\,,
\end{equation}
where $DH$ is a flat measure, the standard deviation $\sigma$ sets the energy units (and was set to $1$ in the numerics), and $H$ is a complex hermitian or a real symmetric matrix depending on whether we work with the Gaussian Unitary Ensemble (GUE) or with the Gaussian Orthogonal Ensemble (GOE), respectively\footnote{The third canonical Gaussian ensemble, which we do not study here, is the Gaussian Symplectic Ensemble (GSE). It addresses time-reversal-invariant fermionic systems displaying Kramer's degeneracy.} \cite{HaakeBook}.

\subsection{Influence of the structure of the seed operator}

A detailed numerical study reveals that the behavior of K-complexity, and in particular its late-time saturation value, is not only controlled by the statistics of the Hamiltonian spectrum, but also influenced by the structure of the operator under consideration. As an extreme illustration of this, Appendix \ref{appx_FlatOp} shows analytically that an operator that is constant in the energy basis, which is a very atypical observable in any system, features a late-time K-complexity saturation value of $\sim\frac{K}{2}$ regardless of the spectrum of the underlying Hamiltonian. In contrast, a typical operator in RMT should satisfy the \textit{RMT operator Ansatz} (see e.g. \cite{DAlessio:2015qtq}) for its matrix elements in the energy basis:
\begin{equation}
    \centering
    \label{RMT_op_Ansatz_maintext}
    \langle E_a | \mathcal{O} | E_b \rangle = O \delta_{ab} + \frac{1}{\sqrt{D}}r_{ab} ~,
\end{equation}
where all $\left\{r_{ab}\right\}$ are independent random numbers\footnote{In fact, only those $r_{ab}$ with $a\geq b$ are independent, as the rest are determined from the latter if the operator is hermitian.} drawn from a normal distribution with zero mean and unit variance; they are either real or complex depending on the universality class at hand. The one-point function term $O$ in (\ref{RMT_op_Ansatz_maintext}) will not be important for the current analysis because, as explained in Appendix \ref{appx_Connected}, we shall work with traceless operators.

\begin{figure}[t]
    \centering
    \includegraphics[width=0.45\textwidth]{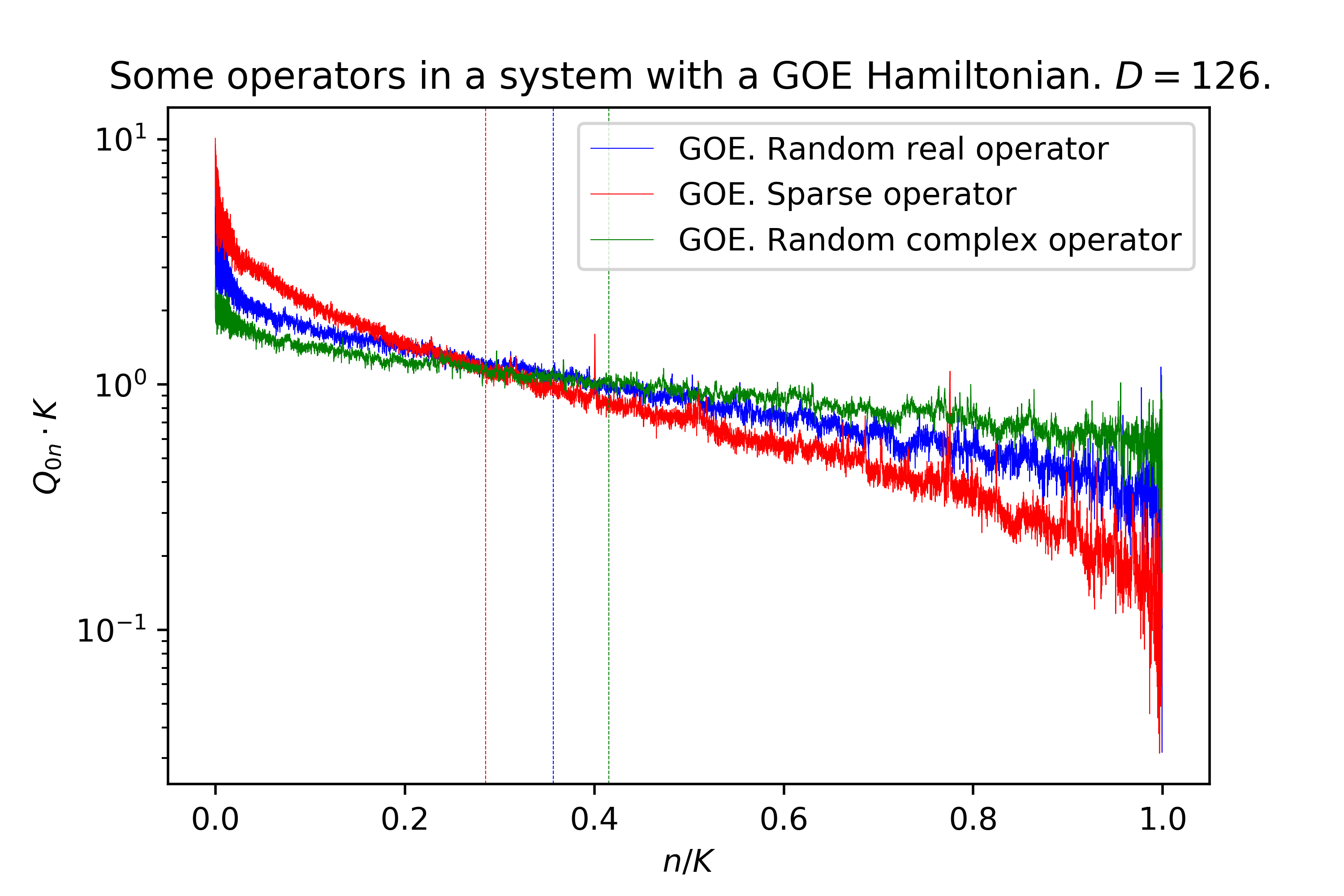} \includegraphics[width=0.45\textwidth]{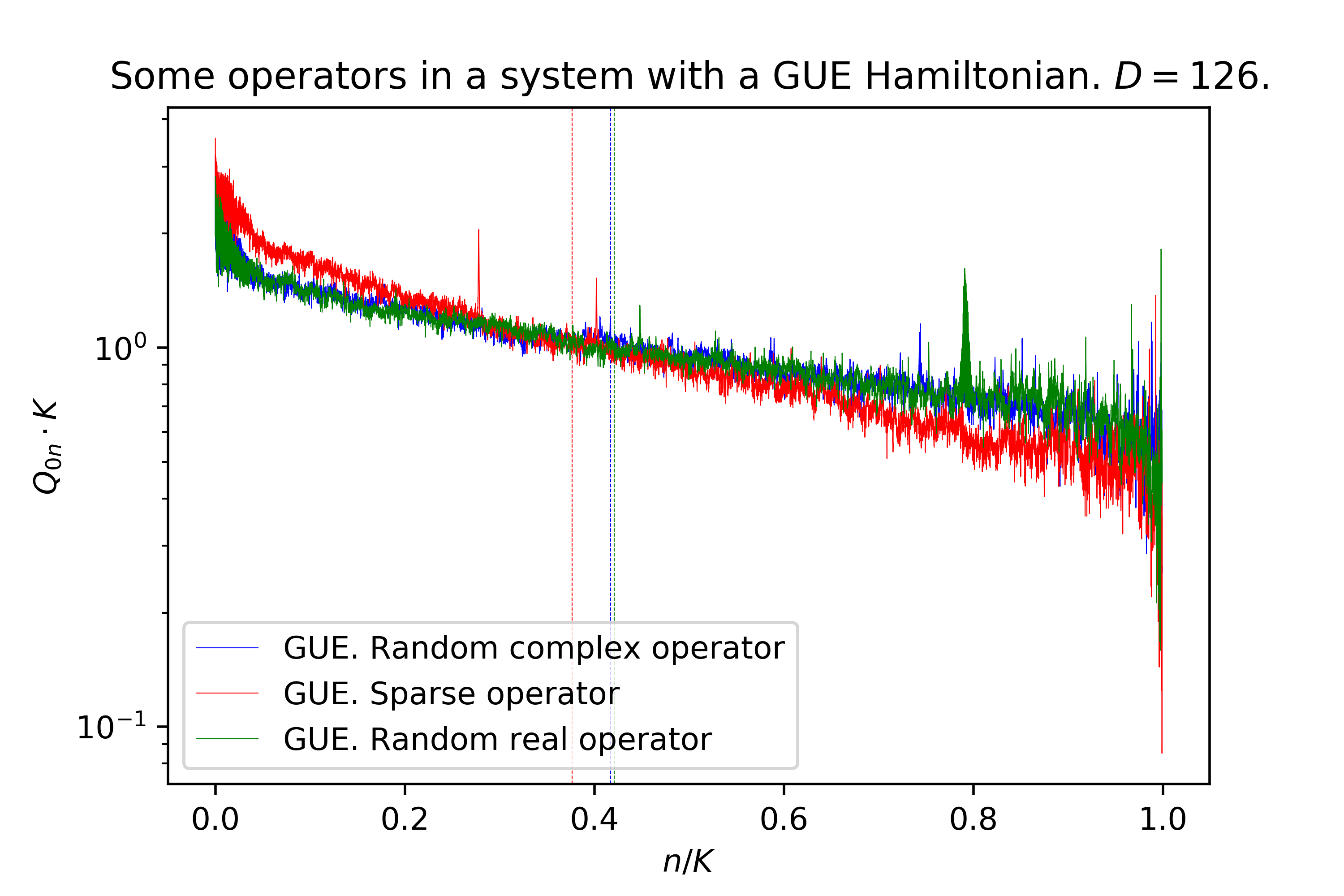}
    \caption{\small Long-time averaged operator wave packet and saturation value of K-complexity for different operator choices and Hamiltonians drawn from two RMT ensembles. K-complexity saturation values are marked by vertical lines. The Hilbert space dimension chosen was $D=126$, and the obtained Krylov dimension saturates the upper bound \cite{I}, verifying $K=D^2-D+1=15751$. \textbf{Left:} GOE Hamiltonian. Operator choices: a sparse operator in the basis in which the Hamiltonian is drawn, a random real operator drawn from a Gaussian distribution in the same basis, and a random complex operator again drawn in the same basis. We observe that the random operators saturate at higher values as compared to the sparse operator, and within them, the one that breaks time reversal (i.e. the complex one) has the highest complexity saturation value. \textbf{Right:} Same choices of operator, but for a Hamiltonian drawn from GUE. In this case time reversal is anyway broken by the Hamiltonian itself, which is why the two random operators have quantitatively very similar features, both having a complexity saturation value slightly higher than that of the sparse operator.}
    \label{fig:RMT_SomeOps}
\end{figure}

Operators satisfying the Ansatz (\ref{RMT_op_Ansatz_maintext}) can be constructed as sparse operators in the basis in which the Hamiltonian is drawn from the Gaussian ensemble, or as random matrices with independent entries. In both cases, the change-of-basis matrix that brings the operator to the energy basis is a random unitary drawn from the Haar measure and for sufficiently large $D$ they both agree with the structure (\ref{RMT_op_Ansatz_maintext}). Results on the late-time behavior of K-complexity for both operator choices in the different universality classes can be found in Figure \ref{fig:RMT_SomeOps}, which suggests that the saturation value of K-complexity is sensitive to the universality class to which the Hamiltonian belongs as well as to the choice of operator and, in particular, to whether the operator breaks time reversal or not. Note that, in general, the complexity saturation values are below $\frac{K}{2}$.

\subsection{Deviations from the RMT Ansatz: ETH operators}\label{subsect:RMT_ETH}

\begin{figure}[t]
    \centering
    \includegraphics[width=0.6\textwidth]{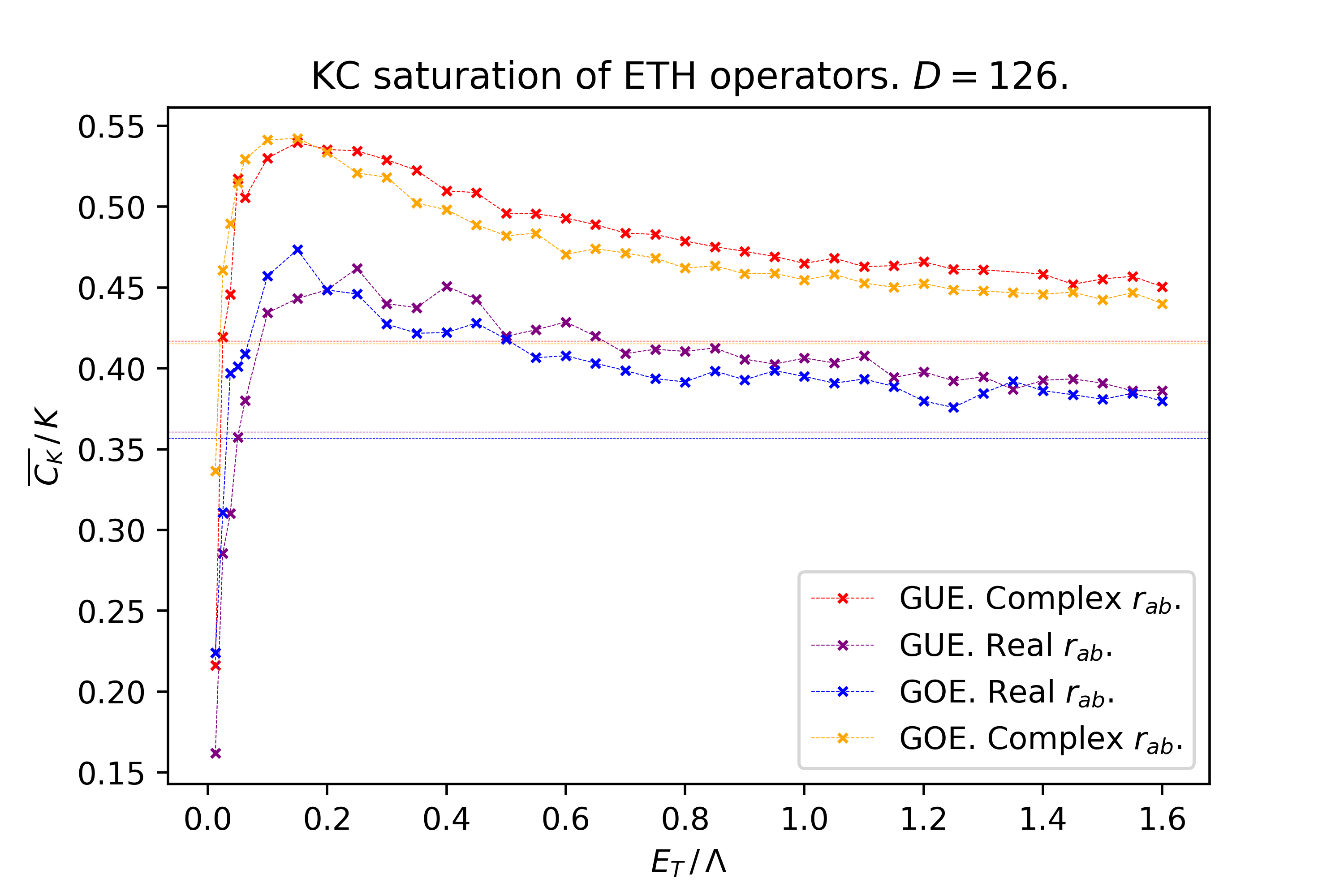}
    \caption{K-complexity saturation value as a function of the Thouless energy for different ETH operators (either real or complex in the energy basis) with Hamiltonians drawn from two different RMT ensembles (GOE and GUE). The horizontal lines mark the asymptotic value for $E_T\to\infty$, corresponding to the case of observables satisfying the pure RMT operator Ansatz. In order to mod out system-dependent scaling of the energy spectrum due to the choice of normalization of the Hamiltonian, the Thouless energy was normalized by the total bandwidth $\Lambda$, allowing for potential comparisons with other systems.}
    \label{fig:Csat_Vs_Thouless}
\end{figure}

In \cite{I} we studied complex SYK$_4$, which is a chaotic system with richer features than just RMT and displayed a complexity saturation value close to $\frac{K}{2}$.
Two features may be regarded as responsible for that behavior: the Wigner-Dyson statistics satisfied by the Hamiltonian spectrum, and the fact that the operators studied satisfy the eigenstate thermalization hypothesis (ETH) \cite{PhysRevA.43.2046,Srednicki:1994mfb,DAlessio:2015qtq,Sonner:2017hxc,Nayak:2019khe}, which is an extension of the RMT Ansatz (\ref{RMT_op_Ansatz_maintext}) that accounts for a smoothly varying density of states:

\begin{equation}
    \centering
    \label{ETH_Ansatz}
    \langle E_a | \mathcal{O} | E_b \rangle = O(\overline{E})\delta_{ab}+e^{-\frac{S(\overline{E})}{2}}\,f_{\mathcal{O}}(\overline{E},\omega)\,r_{ab}~,
\end{equation}
where $r_{ab}$ are independent (up to hermiticity), identically distributed normal random variables with zero mean and unit variance, and $\overline{E}\equiv(E_a+E_b)/2$ and $\omega \equiv E_a-E_b$ are, respectively, the average energy and the energy difference between the corresponding levels. $O(\overline{E})$ and $S(\overline{E})$ are the microcanonical one-point function and entropy, respectively, and the function $f_{\mathcal{O}}(\overline{E},\omega)$ gives the Fourier transform of the connected two-point function, sometimes denoted \textit{spectral function} \cite{Parker:2018yvk}. Disregarding the $\overline{E}$-dependence, the high-frequency tails of this function are known to be bounded from above by an exponential profile:
\begin{equation}
\centering
\label{f_bound}
    f_{\mathcal{O}}(\overline{E},\omega)\lesssim e^{-\frac{\omega}{E_T}},
\end{equation}
where $E_T$ is the Thouless energy, which is itself constrained by a system-dependent upper bound, and controls the regime of applicability of RMT. For the sake of the current analysis, we generated operators following the Ansatz (\ref{ETH_Ansatz}) where the $\overline{E}$-dependence was taken to be constant and the $\omega$-dependence was chosen to saturate the bound (\ref{f_bound}) with an adjustable Thouless energy. The (rescaled) off-diagonal elements in the energy basis $\left\{ r_{ab} \right\}$ were chosen to be either real or complex. The saturation value of K-complexity as a function of the Thouless energy for the different choices of Hamiltonian and operator are depicted in Figure \ref{fig:Csat_Vs_Thouless}. Such choices can be classified according to how they comport regarding time reversal. If we define the time reversal transformation $\mathcal{T}$ as an anti-unitary transformation that acts as complex conjugation $\mathcal{T}\overset{*}{=} K$ in the basis in which the Hamiltonian is drawn from the Gaussian ensemble, we can make the following identifications:

\begin{itemize}
    \item GOE + real $r_{ab}$: This situation matches that of a time-reversal preserving operator in a system with a Hamiltonian that preserves $\mathcal{T}$, as they both are real in the computational basis\footnote{For simplicity, here we refer to the basis in which the Hamiltonian is drawn from the corresponding ensemble as the \textit{computational basis}.}, and therefore the operator will still be real in the energy basis.
    \item GOE + complex $r_{ab}$: This case describes the situation in which the Hamiltonian is $\mathcal{T}$-invariant but the operator is not. The operator matrix elements in the computational basis will be complex and, since the change-of-basis matrix for going to the energy basis is a real orthogonal matrix, it will also have complex entries in the energy basis.
    \item GUE + complex $r_{ab}$: Since the Hamiltonian already breaks time reversal, the matrix of eigenvectors expressed in coordinates over the computational basis will be a random unitary, and hence in general the operator will have complex entries in the energy basis regardless of whether it was real or complex in the computational basis (i.e. regardless of whether it is invariant under $\mathcal{T}$ or not, respectively.)
    \item GUE + real $r_{ab}$: Along the lines of the previous point, we shall conclude that this configuration is just an atypical case, not particularly physically meaningful for discussions regarding time reversal. We have nevertheless still kept the results for this case in Figure \ref{fig:Csat_Vs_Thouless} for the sake of completeness of the analysis.
\end{itemize}

Disregarding for the discussion the seemingly unphysical GUE+real configuration, Figure \ref{fig:Csat_Vs_Thouless} illustrates the fact that in systems where time reversal is broken, either by the Hamiltonian or by the operator, depict a systematically higher K-complexity saturation value. We have also observed a continuous dependence of the saturation value on the Thouless energy that can pump up the former from the lower limiting value attained when the observable satisfies the pure RMT operator Ansatz throughout the spectrum.

\subsection{ETH in the deformed XXZ}

As the integrability-breaking defects studied in section \ref{sec:XXZ_rstats} are made stronger, the spectrum of the Hamiltonian of the deformed XXZ chain transitions from Poissonian statistics to Wigner-Dyson statistics. At the same time, it is possible to see that the seed operator under consideration transitions from a non-ETH regime when $\epsilon_d$ is small to having and ETH structure when $\epsilon_d$ attains the value that makes the spectrum of the Hamiltonian chaotic. This phenomenon was already studied in works like \cite{Rigol_XXZ,LeBlond:2019eoe,Rigol_LeBlod2020}.

\begin{figure}[t]
    \centering
    \includegraphics[width=0.4\textwidth]{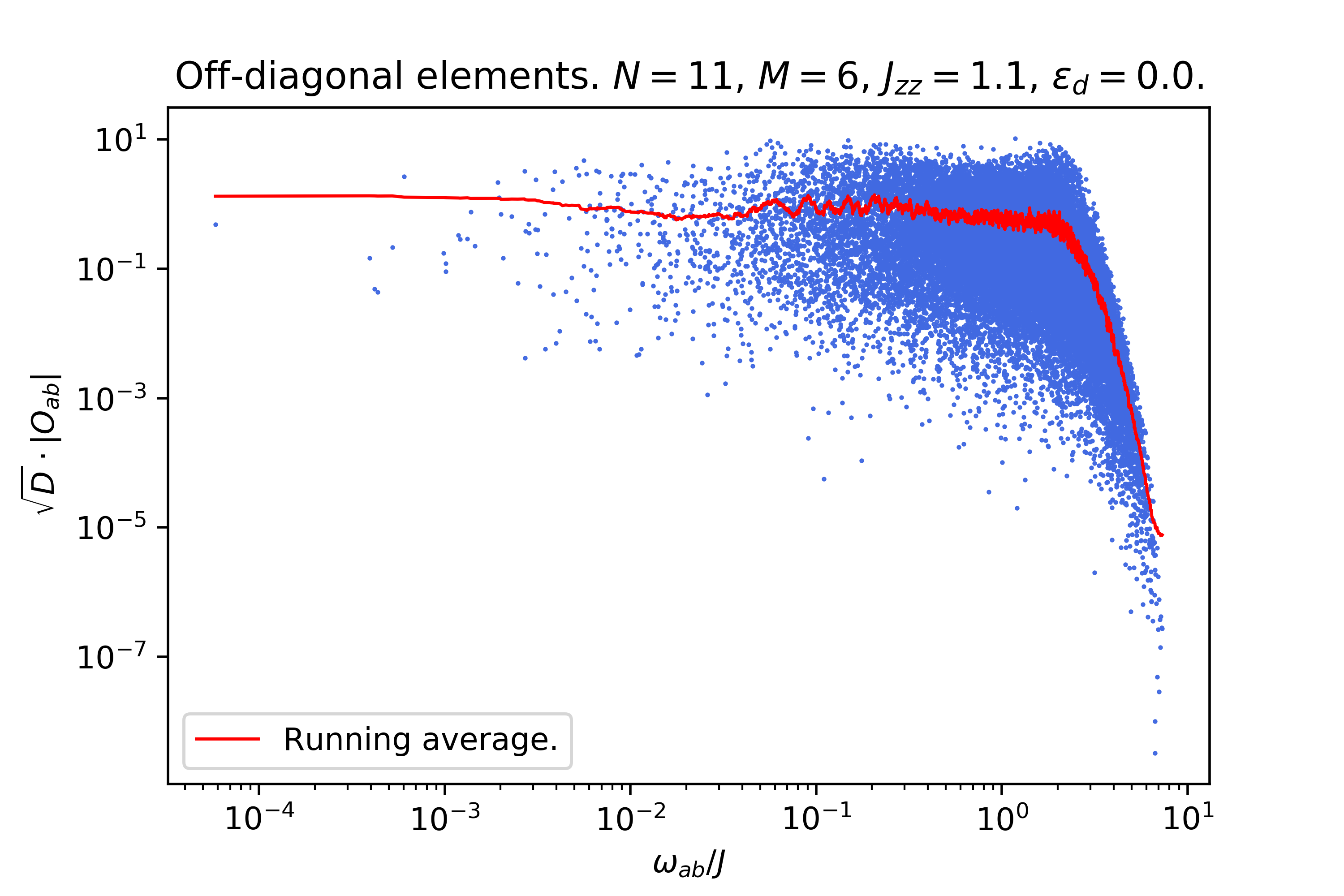} \includegraphics[width=0.4\textwidth]{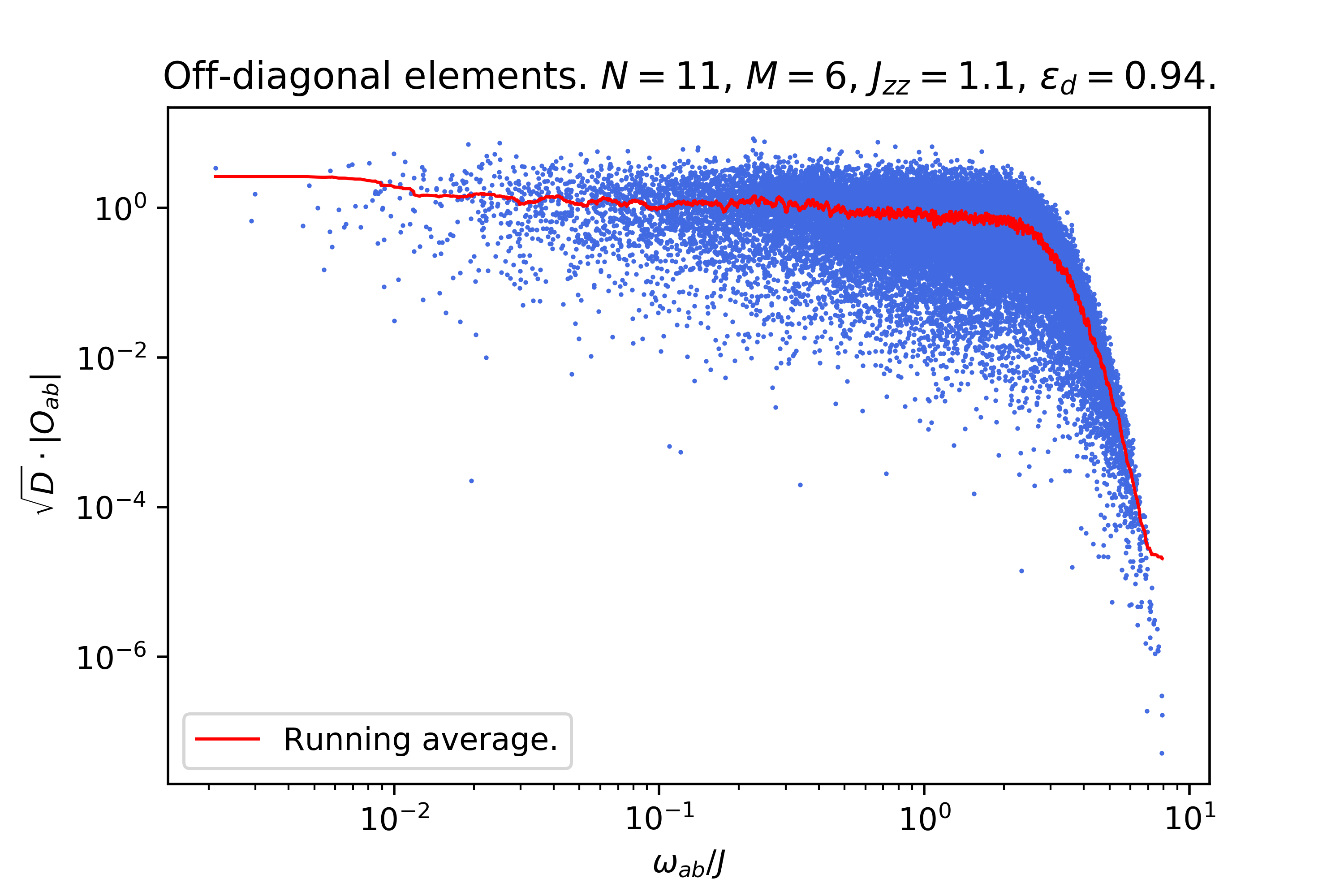} \\
    \includegraphics[width=0.4\textwidth]{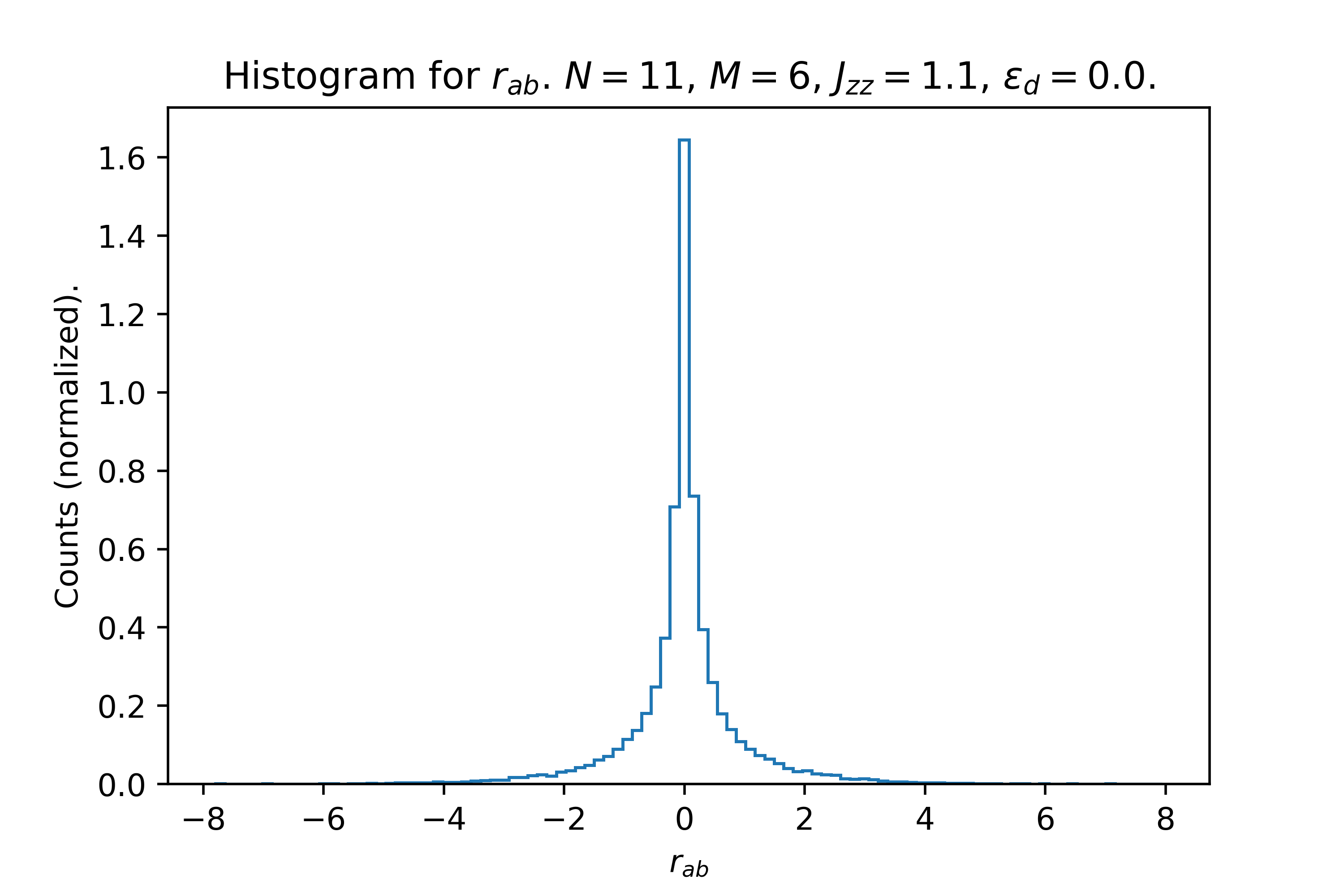} \includegraphics[width=0.4\textwidth]{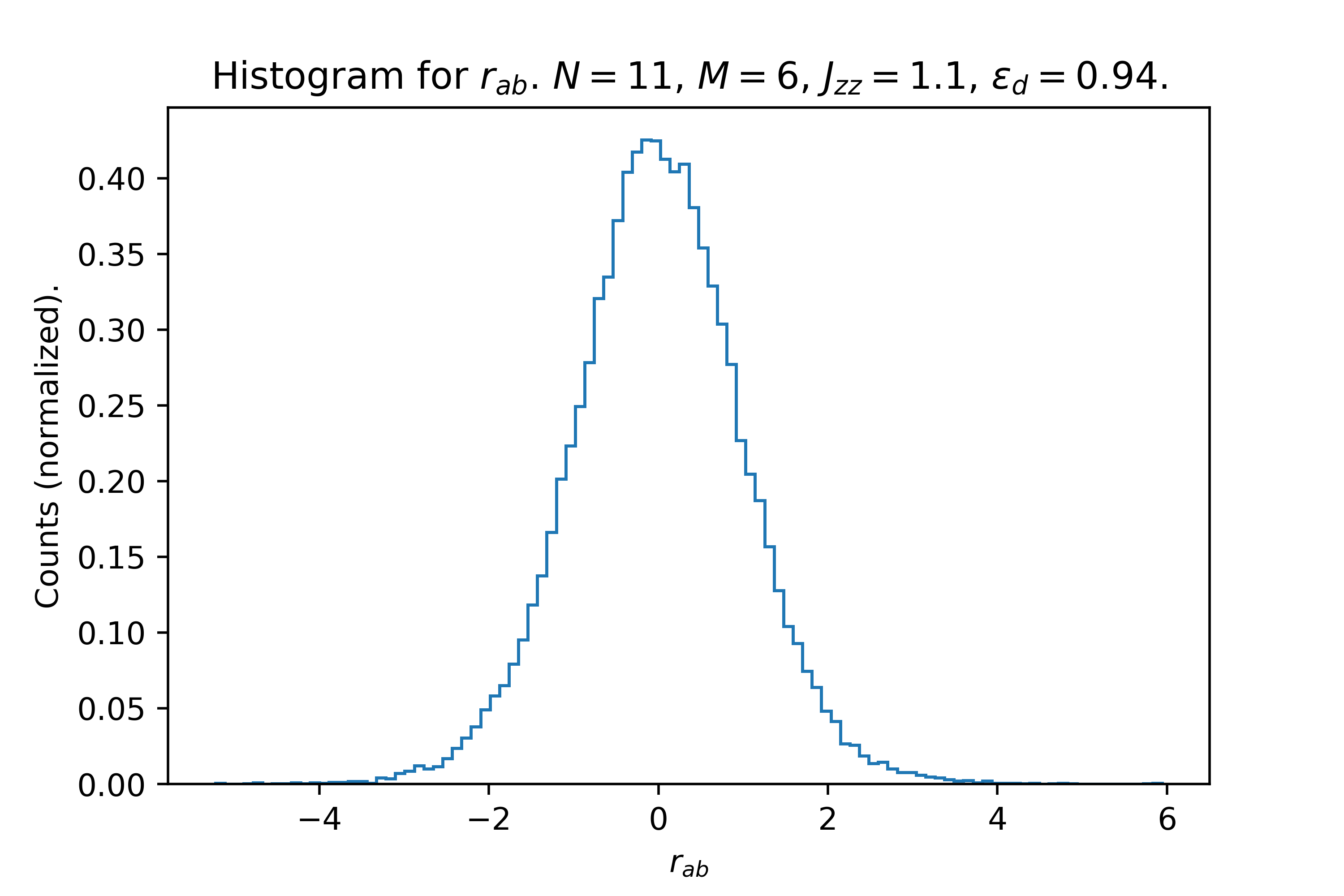} \\
    \includegraphics[width=0.4\textwidth]{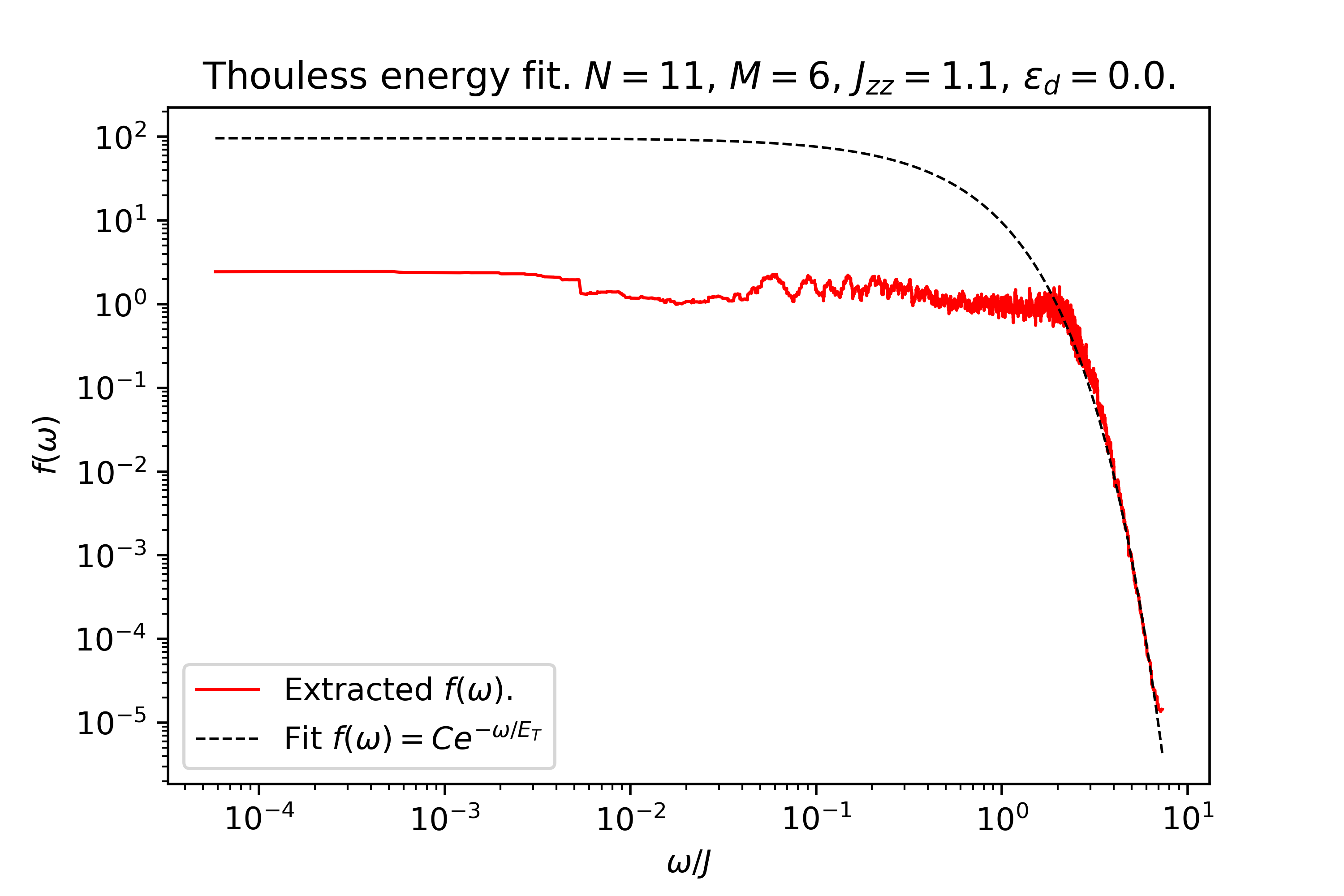} \includegraphics[width=0.4\textwidth]{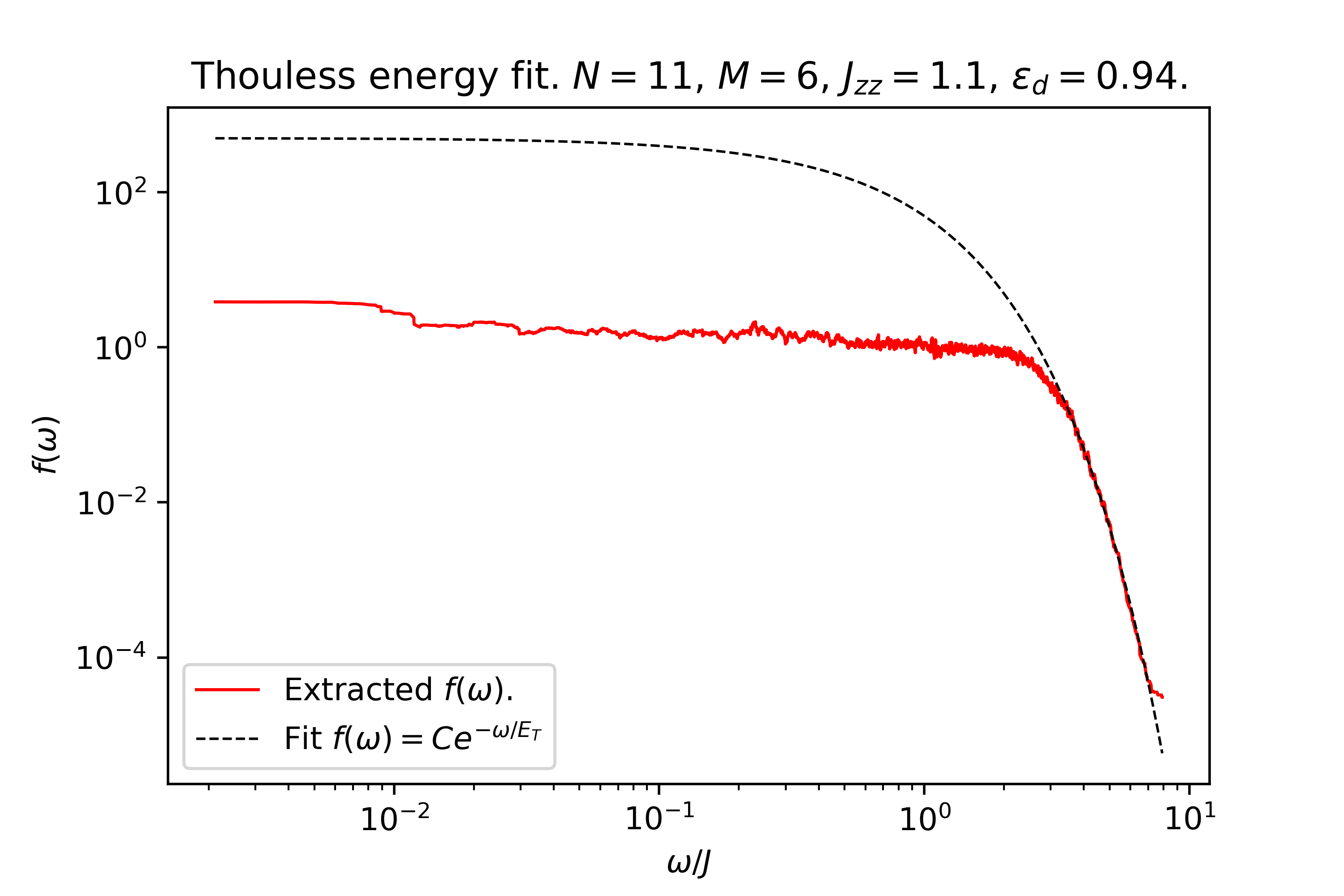}
    \caption{ \small ETH checks for the system and operator studied in Figure \ref{fig:N11M6Hd}. On the \textbf{left} we show the integrable regime $\epsilon_d=0$ while on the \textbf{right} we consider the chaotic regime with $\epsilon_d=0.94$. Since we are studying operators with a zero one-point function, the ETH check can focus on just the off-diagonal elements of the operator in the energy basis. We note, in agreement with previous works (such as \cite{Rigol_XXZ,LeBlond:2019eoe,Rigol_LeBlod2020}), that the difference between the integrable and the chaotic phase is subtle: in both cases it is possible to extract a smooth envelope $f(\omega_{ab})$ for the off-diagonal matrix elements of the operator as a function of the energy difference $\omega_{ab}\equiv E_a-E_b$, whose high-frequency tail can be fitted by an exponential form $f(\omega)=C e^{-\omega/E_T}$ yielding in both cases a very similar Thouless energy $E_T/\Lambda\sim 0.055$ (the r-value of the fit being around $0.992$), where $\Lambda$ denotes the spectral bandwidth. The difference between the integrable and the chaotic case is that in the former the fluctuations of the off-diagonal elements are not quite Gaussian, and hence they cannot be claimed to fulfill ETH, whereas they become more Gaussian in the latter, where the integrability-breaking defect is stronger. }
    \label{fig:ETH_XXZ}
\end{figure}

Here we present results on ETH checks for two extreme values of $\epsilon_d=0,\,0.94$ for the system and operator that were analyzed in Figure \ref{fig:N11M6Hd}. Figure \ref{fig:ETH_XXZ} displays the result. In the integrable regime, the operator does not fulfill the ETH Ansatz because the fluctuations are not Gaussian. In the chaotic regime ($\epsilon_d=0.94$) the operator is seen to agree with the ETH Ansatz displaying a Thouless energy normalized by the spectral bandwidth of roughly $\frac{E_T}{\Lambda}\sim 0.05$. In this chaotic regime, we have that the spectrum of the Hamiltonian is chaotic and that the operator fulfills the ETH Ansatz with a certain Thouless energy; since these are precisely the only two ingredients defining the systems studied in section \ref{subsect:RMT_ETH} and depicted in Figure \ref{fig:Csat_Vs_Thouless}, one can compare the K-complexity saturation values. The universality class at hand for our deformed XXZ is ``GOE+real'', and we note from Figure \ref{fig:Csat_Vs_Thouless} that indeed, for a Thouless energy satisfying $\frac{E_T}{\Lambda}\sim 0.05$ one expects a K-complexity saturation value around $0.4K$, consistent with what we found in Figure \ref{fig:N11M6Hd} when $\epsilon_d=0.94$.

\clearpage

\section{Discussion}\label{Sect:Disc}
The works \cite{II,III} reproduced in this Chapter have explored the behavior of the K-complexity of a strongly coupled integrable model, the XXZ spin chain, both for its original integrable version and for that with an integrability-breaking deformation allowing to interpolate between integrable and chaotic regimes. The purpose of these studies was to delineate to what extent the Krylov complexity saturation value is sensitive to integrability or chaos, as suggested in \cite{I}. We have found strong evidence for the following picture: An exponentially large K-complexity saturation value at late times is a generic feature of a quantum chaotic system, while integrable systems, even strongly coupled ones which do not show exact degeneracies of energy levels, have quantitatively lower saturation values. 
In \cite{II} it was proposed that the smaller complexity saturation value in integrable systems is due to a novel Anderson localization effect operating in Krylov space, potentially related to the Poissonian spectral statistics of the system at hand, and such a phenomenon was explored for local operators in the XXZ model, whose late-time complexity saturation value was found to be smaller than the expectation ($\sim\frac{D^2}{2}$) in chaotic systems, despite having maximal Krylov space dimensions.
Furthermore, in \cite{III} we studied K-complexity and its late-time saturation value for XXZ systems with two types of integrability-breaking terms and found that increasing the value of the coefficient of the integrability-breaking term results in an increasing value for the late-time saturation of K-complexity. We further compared the late-time saturation value in the chaotic regime to the results for RMT in the corresponding universality class, finding reasonable agreement. 


The fact that Krylov complexity values at late times are able to probe the integrable or chaotic nature of the system is a promising result for holographic applications, as argued in section \ref{sect:KC_holog}, but it also offers a plausible way out from the tension raised by \cite{Dymarsky:2021bjq} regarding the universal operator growth hypothesis, discussed in section \ref{sect:KC_chaos} and according to which the aforementioned hypothesis becomes useless at finite temperature, not probing chaotic dynamics: Their analysis was performed for systems in the thermodynamic limit, which pushes all size-dependent time scales (in particular, the scrambling time) to infinity, therefore losing all features due to finite-size effects at late times. The studies \cite{II,III} presented in this Chapter, on the contrary, were mostly agnostic about early-time dynamics and directly focused on computing late-time observables. In the realm of QFT and CFT, finite-size effects may be probed by placing the theory on a compact manifold and focusing on a microcanonical shell, which will therefore have a finite-dimensional Hilbert space. Works like \cite{Kar:2021nbm,Avdoshkin:2022xuw,Camargo:2022rnt} have started to explore this direction. 

%% file: content/Chapter05.tex
\chapter{\rm\bfseries Towards a bulk dual of Krylov complexity}
\label{ch:chapter05_DSSYK}

This Chapter will reproduce (up to minor modifications) the publication \cite{IV}, where an exact correspondence between the state Krylov complexity of the TFD state (at infinite temperature) in the low-energy regime of the double-scaled SYK model \cite{Lin:2022rbf,Berkooz:2018jqr} and bulk length in two-dimensional JT gravity \cite{Jackiw:1984je,Teitelboim:1983ux,Sarosi:2017ykf} is proved. This example, coming from the framework of low-dimensional holography, is the first instance where such a correspondence between a notion of complexity and a bulk observable is established analytically, going beyond the mere qualitative comparison of the profiles of the corresponding quantities as a function of time. The Chapter can therefore be seen as the culmination of the Thesis: After all the motivations given to study Krylov complexity in Part \ref{part:part1} and the evidence accumulated through \cite{I,II,III} (and gathered in Chapters \ref{ch:chapter03_SYK} and \ref{ch:chapter04_Integrable}) for it to be a good holographic complexity candidate, finally the project \cite{IV} succeeded in constructing its explicit holographic dual for a concrete model.

\section{Introduction}

A concrete notion of complexity which requires neither the introduction of a particular set of local unitary operations available to a quantum computer nor a choice of tolerance parameter, and under which the time evolution of complexity (for states evolving under a chaotic Hamiltonian) resembles the time-dependent profile described in section \ref{sect:KC_holog}, namely including a long period of linear growth up to exponentially late times, would be a candidate for holographic complexity. Such a candidate should be naturally defined on the quantum mechanical boundary as well as in the quantum gravity bulk, preferably relating the complexity evolution of a state on the boundary to some invariant observable's time evolution in the bulk. In view of the results of \cite{I,II,III} that have been gathered in the previous Chapters \ref{ch:chapter03_SYK} and \ref{ch:chapter04_Integrable}, we may already have a high degree of confidence about the suitability of Krylov complexity for holographic applications. It is only left to find an actual holographic instance in which K-complexity can be proved to be dual to some bulk observable, and this was precisely the goal of \cite{IV},
in which a precise one-to-one correspondence between Krylov complexity in a 1-dimensional quantum mechanical boundary and a specific bulk observable in a 2-dimensional quantum gravity theory was found.  
In particular, Krylov complexity of the thermofield double state (at infinite temperature) in the boundary theory, which is a particular limit of double-scaled SYK \cite{Lin:2022rbf,Berkooz:2018jqr},
becomes equivalent to wormhole length in the bulk theory, which is 2-sided AdS$_2$ Jackiw-Teitelboim gravity \cite{Jackiw:1984je, Teitelboim:1983ux}. 

Previous works on the subject include \cite{Jian:2020qpp} which computed operator growth in SYK via Krylov complexity and compared it with the complexity=volume formulation in JT gravity; it was shown that the two have similar qualitative behaviours in time. The work \cite{Kar:2021nbm} studied the expected behaviour of K-complexity in low-dimensional gravity theories via relationships with random matrix theory.


\section{Background} \label{Sec:Background}
In this section we review the current state of knowledge for the double-scaled Sachdev-Ye-Kitaev (DSSYK) model and chord diagrams \cite{Berkooz:2018jqr, Berkooz:2018qkz}, the phase space of JT gravity \cite{Harlow:2018tqv}, and the equivalence between DSSYK (at a specific limit) and JT gravity \cite{Lin:2022rbf}. While the material in this section largely reviews results from the literature, it exposes the main ingredients necessary to match boundary Krylov complexity to its bulk counterpart. 

\subsection{Double-scaled SYK}\label{Subsection_Background_DSSYK}

The Sachdev-Ye-Kitaev (SYK) model \cite{Sachdev_1993, Sachdev:2015efa, Kitaev:2015} is a strongly interacting, many-body quantum system described by all-to-all interactions between Majorana fermions $\psi_i$ with $1\leq i \leq N$ which satisfy $\{\psi_i,\psi_j\}=2\delta_{ij}$, grouped into $p$-body interactions $\psi_{i_1} \psi_{i_2} \dots \psi_{i_p}$ through random couplings $J_{i_1 i_2\dots i_p}$
in the form of the Hamiltonian
\begin{equation} \label{HSYK}
    H_{SYK} = i^{p/2} \sum_{1\leq i_1 < i_2 <\dots < i_p \leq N}\, J_{i_1 i_2\dots i_p} \, \psi_{i_1} \psi_{i_2} \dots \psi_{i_p} ~.
\end{equation}
The random couplings $J_{i_1 i_2\dots i_p}$ are usually taken from a Gaussian distribution with zero mean $\langle J_{i_1 i_2\dots i_p} \rangle = 0$ and an appropriate non-zero variance. The model (\ref{HSYK}) has been extensively studied for fixed $p$ and $N$, and at large $N$, as well as in the large $p$ and large $N$ limit (see e.g. \cite{Polchinski:2016xgd, Maldacena:2016hyu,  Cotler:2016fpe, Garcia-Garcia:2017pzl}).  
We will follow \cite{Berkooz:2018jqr} and take both $N$ and $p$ to infinity while keeping the ratio $2p^2/N$ fixed, a limit known as double-scaled SYK (or DSSYK), introducing the ratio parameter $\lambda$:
\begin{equation}
\label{lambda_def}
    \lambda \equiv \frac{2p^2}{N} ~,
\end{equation}
and we will take the variance $\langle J^2_{i_1 i_2\dots i_p} \rangle$ to be
\begin{equation}
\label{Variance}
    \langle J^2_{i_1 i_2\dots i_p} \rangle = \frac{1}{\lambda} \binom{N}{p}^{-1} J^2 ~.
\end{equation}
As discussed in Appendix \ref{App:coupling_variance}, double-scaled SYK with this choice of coupling variance is, for small values of $\lambda$, compatible with the results for SYK where $N$ and $p$ are large and independent respecting the hierarchy $1\ll p \ll N$, which was studied in \cite{Maldacena:2016hyu}.

\begin{figure}
    \centering
    \includegraphics[scale=0.5]{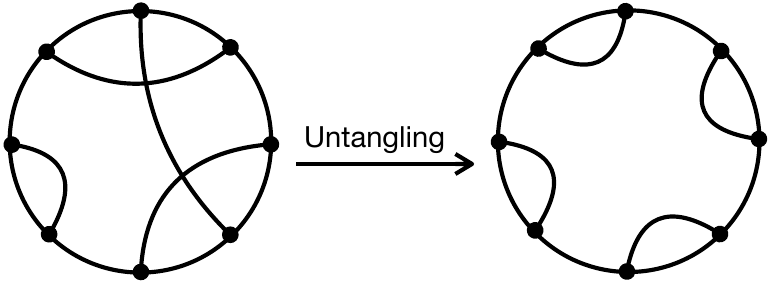}
    \caption{Untangling a chord diagram.}
    \label{fig:Chord_diagrams}
\end{figure}

\begin{figure}
    \centering
    \includegraphics[scale=0.5]{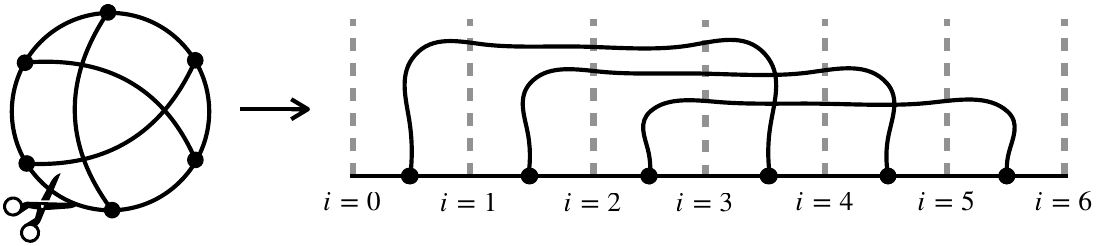}
    \caption{``Cutting open" a chord diagram \cite{Berkooz:2018jqr}. The cut-open chord diagram can be thought of as a process in which one starts with zero open chords and ends with zero open chords, while chords are created and annihilated in between.}
    \label{fig:cut_chord_diagram}
\end{figure}

In \cite{Berkooz:2018jqr} it was shown that in the limit where $N$ and $p$ go to infinity while keeping $\lambda$ fixed, the \textit{ensemble averaged} moments 
\begin{equation}
    \label{Moments_definition}
    M_{2k} := \langle \, \tr (H^{2k}) \, \rangle
\end{equation}
of the Hamiltonian (\ref{HSYK}) are given by a sum over \textit{chord diagrams}. Chord diagrams represent pairwise Wick contractions among the couplings $J_{i_1...i_p}$ of the different Hamiltonians in the product inside the trace (\ref{Moments_definition}), which make the collective indices of the corresponding monomials $\psi_I \equiv \psi_{i_1} \psi_{i_2} \dots \psi_{i_p}$ coincide. For the $2k$-th moment, there are $2k$ paired monomials in each trace. 
To find the value of such a chord diagram, it needs to be ``untangled" to a configuration where all contracted pairs are sitting one beside the other (see Figure \ref{fig:Chord_diagrams}). 
Once this is achieved, using the fact that $\psi_I^2 = i^p$ for $p$ even (it will be assumed that $p$ is even), one obtains a total phase of $i^{kp}$ coming from all the monomials squared, which exactly cancels the overall phase coming from the $i^{p/2}$ factor in the normalization of each Hamiltonian (\ref{HSYK}). Additionally, the system size dependence coming from the summation over the indices is mitigated by the binomial in the variance normalization (\ref{Variance}), yielding bounded results in the double-scaling limit. Therefore, the contribution of a given Wick contraction (i.e. a given chord diagram) boils down to the overall sign obtained after commuting the Majorana monomials through each other inside the trace until those with coincident indices are consecutive. In the double-scaling limit, it is found \cite{Berkooz:2018jqr} that the overall effect of such a commutation is captured by a multiplicative factor $q\equiv e^{-\lambda}$. The number of such commutations required to untangle a chord diagram is exactly the number of intersections of chords, and hence one is left with a simple rule for evaluating moments via chord diagrams:
\begin{equation}
     M_{2k} = \frac{J^{2k}}{\lambda^k} \sum_{\substack{\text{chord diagrams} \\ \text{ with $k$ chords}}} q^{\text{number of intersections}} ~.
\end{equation}
To evaluate the RHS, it is useful to ``cut open" each chord diagram and think of the process of constructing it as a transition from a state $|0\rangle$ with zero chords back to a state with zero chords, through a process of creating $k$ chords and eventually annihilating them all. An auxiliary Hilbert space $\{| 0 \rangle, | 1 \rangle, | 2 \rangle, \dots , | n \rangle \}$, which represents states with $0,1,2, \dots, n$ open chords, can now be introduced. An open chord is a chord that emanated from a given Hamiltonian insertion but which has not yet been closed by reaching another Hamiltonian insertion: Figure \ref{fig:cut_chord_diagram} depicts a chord diagram of six Hamiltonian insertions interpreted as a transition from zero open chords back to zero open chords through six intermediary steps (Hamiltonian insertions); if we look, for instance, at the step $i=4$, we note that there are two open chords. In the transition from $| 0 \rangle$ back to $| 0 \rangle$, chords can be ``created'' $|l\rangle \mapsto |l+1\rangle $ and ``annihilated'' $|l\rangle \mapsto |l-1\rangle $ at each step. 

All possible chord diagrams with $i$ steps (i.e. $i$ Hamiltonian insertions) are represented by a state $|\psi^{(i)}\rangle$ in the Hilbert space. We define such a state through its coordinates over the chord basis,
\begin{equation}
    \centering
    \label{State_psi_chord_basis}
    |\psi^{(i)}\rangle = \sum_{l\geq 0} \psi^{(i)}_l | l \rangle ~.
\end{equation}
The coordinates $\psi^{(i)}_l$ are defined to be the sum over all chord diagrams of $i$ steps with $l$ open chords, where each diagram is weighted by $q$ to the power of the number of intersections featured in it; that is, all diagrams with $i$ steps and $l$ open chords are represented by a state in the Hilbert space which is proportional to $|l\rangle$, with projection $\psi_{l}^{(i)}$, such that the sum of all diagrams of a fixed number of steps is represented by the linear combination $|\psi^{(i)}\rangle$ given in (\ref{State_psi_chord_basis}). The coordinates $\psi_l^{(i)}$ satisfy the following recurrence relation:
\begin{equation}\label{chord_recursion}
    \psi_l^{(i+1)} = \frac{J}{\sqrt{\lambda}}\psi_{l-1}^{(i)} +  \frac{J}{\sqrt{\lambda}}(1+q+\dots +q^{l})\psi_{l+1}^{(i)} ~,
\end{equation}
where the first term represents the possibility that a chord was created in the transition from step $i$ to step $i+1$, and the second term represents the possibility that a chord was annihilated in that transition. In the case in which the chord is annihilated at step $i+1$, there must have been $l+1$ open chords at step $i$ so that we are left with $l$ open chords at step $i+1$, and therefore the chord, in getting annihilated, can intersect up to $l$ other open chords, and in each case the diagram would get multiplied by $q$ to the number of such intersections. The term $1+q+...+q^l$ in (\ref{chord_recursion}) accounts therefore for all the possible ways to close a chord in the step $i+1$. 
This recursion (\ref{chord_recursion}) can be written concisely in a basis-independent way as

\begin{equation}
    |\psi^{(i+1)}\rangle = T |\psi^{(i)}\rangle ~,
\end{equation}
where, in coordinates over the $\{|l\rangle\}$ basis, the \textit{transfer matrix} $T$ takes the form:
\begin{equation} \label{T_non_sym}
    T \overset{*}{=} \frac{J}{\sqrt{\lambda}} \begin{pmatrix}
        0 & \frac{1-q}{1-q} & 0 & 0 & \dots  \\
        1 & 0 & \frac{1-q^2}{1-q} & 0 &  \dots  \\
        0 & 1 & 0 & \frac{1-q^3}{1-q} &  \dots \\
        0 & 0 & 1 & 0 &  \dots  \\
        \vdots & \vdots & \vdots & \vdots &  \ddots  \\
    \end{pmatrix},
\end{equation}
where the asterisk above the equality sign denotes that this is an expression in coordinates over a particular basis. 
At the first step $i=0$ there are no open chords and therefore $|\psi^{(0)}\rangle =|0\rangle$. Thus,
\begin{equation}
    \centering
    \label{psi_i_T_0}
    |\psi^{(i)}\rangle = T^i|0\rangle~.
\end{equation}
Since at each step we either create or annihilate a chord, to make a chord diagram with exactly $k$ chords we would need to apply the operator $T$ $2k$ times, creating $k$ chords and annihilating $k$ chords. The final state is $|\psi^{(2k)} \rangle= T^{2k}|0\rangle$, and thus
\begin{equation} \label{Moments_effective_Ham}
    M_{2k} = \langle 0 |\, T^{2k}\, |0\rangle = \langle 0 | \psi^{(2k)} \rangle = \psi^{(2k)}_0 ~.
\end{equation}

In view of this, $T$ can be interpreted as an effective Hamiltonian of DSSYK. In \eqref{T_non_sym} we give its expression in coordinates over the chord basis $\{ |l\rangle \}_{l=0}^\infty$, where it is represented by a non-symmetric matrix. In \cite{Berkooz:2018qkz, Berkooz:2018jqr} it was pointed out that there exists a diagonal similarity transformation that brings $T$ into a tridiagonal, symmetric form. Furthermore, in \cite{Lin:2022rbf} this transformation is explained in terms of the chord inner product defined in that article (and which was necessary to define in order to have an actual Hilbert space): according to such an inner product, the states $\{|l\rangle\}$ are orthogonal to each other, but they are not normalized (except for $|0\rangle$). The similarity transformation amounts to re-normalizing these states so that they form an orthonormal basis.
By a slight abuse of notation, from now on $|l\rangle$ will denote the \textit{normalized} fixed chord number states, i.e. $\langle l | l^\prime \rangle = \delta_{l,l^\prime}$. In this orthonormal basis, $T$ takes the form \cite{Berkooz:2018jqr,Lin:2022rbf}:
\begin{equation}
\label{T_sym}
 T    \overset{*}{=} \frac{J}{\sqrt{\lambda}} \begin{pmatrix}
        0 & \sqrt{\frac{1-q}{1-q} }& 0 & 0 & \dots  \\
        \sqrt{\frac{1-q}{1-q} } & 0 & \sqrt{\frac{1-q^2}{1-q}} & 0 &  \dots  \\
        0 & \sqrt{\frac{1-q^2}{1-q}} & 0 & \sqrt{\frac{1-q^3}{1-q}} &  \dots \\
        0 & 0 & \sqrt{\frac{1-q^3}{1-q}} & 0 &  \dots  \\
        \vdots & \vdots & \vdots & \vdots &  \ddots  
    \end{pmatrix} ~.
\end{equation} 
Algebraically, $T$ can be written as:
\begin{equation}
    \centering
    \label{T_alg}
    T = \frac{J}{\sqrt{\lambda}}\left( \alpha + \alpha^\dagger \right)~,
\end{equation}
where 
\begin{equation}
    \centering
    \label{q-annihilation_op}
    \alpha = \sum_{l\geq 0} \sqrt{[l+1]_q}\quad |l\rangle\langle l+1 |~,
\end{equation}
and where we have defined the q-number (or basic number) as:
\begin{equation}
    \centering
    \label{q_number}
    [l]_q := \frac{1-q^l}{1-q} = \sum_{k=0}^{l-1} q^k~.
\end{equation}
In fact, the ladder operators $\alpha,\,\alpha^\dagger$, together with the chord number operator 
\begin{equation}
    \centering
    \label{Chord_number_op}
    \hat{l} = \sum_{l\geq 0} l |l\rangle \langle l| ~,
\end{equation}
form the algebra of a q-deformed oscillator \cite{Arik:1976}:
\begin{equation}
    \centering
    \label{q_oscillator_algebra}
    \begin{split}
        &[\alpha,\alpha^\dagger]_q := \alpha \alpha^\dagger - q \alpha^\dagger \alpha = 1 \\
        &[\hat{l}, \alpha^\dagger] = \alpha^\dagger \\
        &[\hat{l},\alpha] = - \alpha
    \end{split}
\end{equation}
whose representation over the $|l\rangle$ basis satisfies
\begin{equation}
    \centering
    \label{q_oscillator_rep_chord_basis}
    \hat{l} |l\rangle = l |l\rangle, \qquad \alpha^\dagger |l\rangle = \sqrt{[l+1]_q} |l+1\rangle,\qquad\alpha |l\rangle = \sqrt{[l]_q} |l-1\rangle~.
\end{equation}
This algebra reduces to the usual oscillator algebra (the Heisenberg-Weyl algebra) in the limit $q\to 1$, since:
\begin{itemize}
    \item[(i)] $[A,B]_q\overset{q\to 1}{\longrightarrow} [A,B]$
    \item[(ii)] $[l]_q\overset{q\to 1}{\longrightarrow} l$.
\end{itemize}

From (\ref{T_sym}), the eigensystem equation for the components of the eigenvectors of $T$, $\psi_l(E)=\langle l|E\rangle$, is given by:
\begin{equation} \label{q-eigsys}
    E \,\psi_l(E) = \frac{J}{\sqrt{\lambda (1-q)}} \left( \sqrt{1-q^{l+1}}\,\psi_{l+1}(E) + \sqrt{1-q^{l}} \, \psi_{l-1}(E) \right)
\end{equation}
In \cite{Berkooz:2018jqr, Berkooz:2018qkz}, it was shown that the eigenvalues are a function of a continuous variable $\theta$:
\begin{equation} \label{Teigvals}
    E(\theta) = \frac{2J \, \cos \theta}{ \sqrt{\lambda(1-q)}}, \quad 0\leq \theta \leq \pi ~,
\end{equation}
and the normalized eigenvectors are given by
\begin{equation}
    \psi_l(\mu) = \sqrt{(q;q)_\infty} |(e^{2i\theta};q)_\infty| \frac{H_l(\mu|q)}{\sqrt{2\pi (q;q)_l}}, \quad \mu=\cos \theta
\end{equation}
where $(a;q)_n \equiv \prod_{k=0}^{n-1}(1-a q^k)$ is the \textit{q-Pochhammer} symbol and $H_l(\mu|q)$ are the \textit{q-Hermite polynomials}.
Since
\begin{equation} \label{T0eigenvector}
    \psi_0(\mu)=   \sqrt{\frac{(q;q)_\infty}{2\pi}} |(e^{2i\theta};q)_\infty|~,
\end{equation}
$\psi_l(\mu)$ can be written as
\begin{equation} \label{Teigenvectors}
    \psi_l(\mu)= \psi_0(\mu) \frac{H_l(\mu|q)}{\sqrt{ (q;q)_l}} ~ , \quad \mu=\cos \theta~.
\end{equation}
See Appendix \ref{Appx:EigSysDSSYK} for a detailed review of the derivation of these results.

\subsection{Phase space of JT gravity}\label{Subsect:JT}

We now turn to review some background on JT gravity and in particular its phase space, following \cite{Harlow:2018tqv, Harlow:2021dfp, Brown:2018bms}. This theory is dual to the low-energy regime of DSSYK, as we shall review in section \ref{Subsect:Bulk_Hilbert_Space}.

JT gravity is a 2-dimensional gravity theory with no propagating bulk degrees of freedom. Its action involves the metric $g_{\mu\nu}$ and the dilaton field $\Phi$ and is given by
\begin{equation} \label{S_JT}
    S_{JT} =   \int_\mathcal{M} d^2 x \sqrt{-g}\Big[\Phi_0R +  \Phi (R+2/l_{AdS}^2)\Big] +  2 \int_{\partial\mathcal{M}} dx \sqrt{\gamma} \Big[  \Phi_0 K + \Phi (K-1/l_{AdS}) \Big]
\end{equation}
where $\gamma_{\mu\nu}$ is the induced metric on the boundary, $K$ is the extrinsic curvature of the boundary and $l_{AdS}$ is a length scale of the 2-dimensional spacetime which turns out to be 2-dimensional Anti-de Sitter (AdS$_2$). This length is related to the cosmological constant. The terms involving the constant $\Phi_0$ are topological in nature and will not affect our discussion at the classical level. 
The boundary conditions
\begin{align} 
    ds^2 \big|_{\partial\mathcal{M}} &= -\frac{dt_b^2}{\epsilon^2} \label{bc_metric}\\
    \Phi \big|_{\partial\mathcal{M}} &= \frac{\phi_b}{\epsilon} \label{bc_Phi}
\end{align}
fix the induced metric $\gamma_{\mu\nu}$, with $t_b$ the time on the boundary, and fix the dilaton field $\Phi$ to a positive constant $\phi_b$ at the boundary. In the large volume limit we have $\epsilon \to 0$, i.e. $\epsilon$ plays the role of a boundary regulator.

\begin{figure}
    \centering
    \includegraphics[scale=0.5]{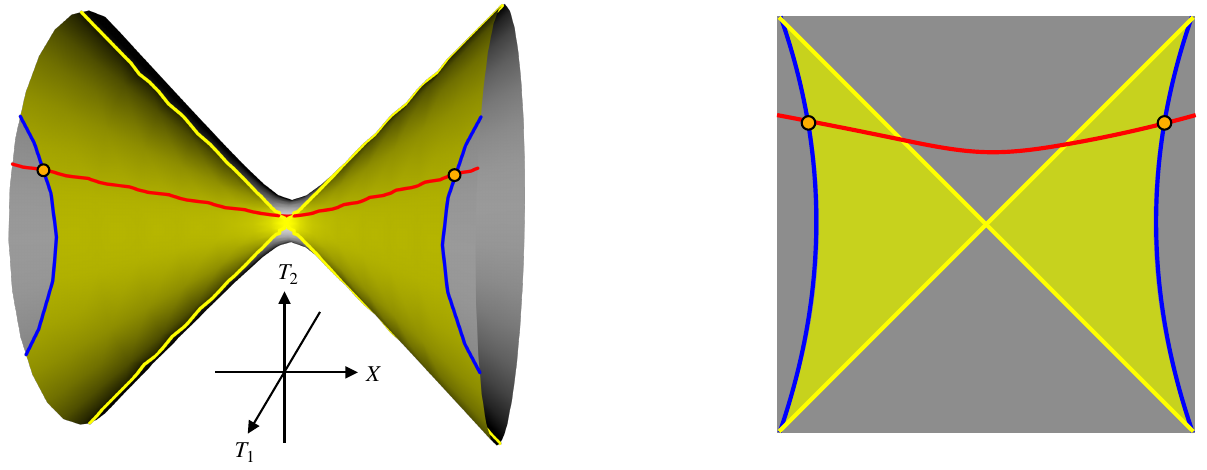}
    \caption{\textbf{Left:} The surface $-T_1^2-T_2^2+X^2=-l_{AdS}^2$ embedded in 3-dimensional space. 
    The wormhole length is represented by the red line, and the boundaries are represented by blue lines. The yellow dots represent the anchoring points at the boundary, between which the wormhole length is computed. The yellow lines represent the Schwarzschild horizon and the areas colored yellow represent the regions covered by the Schwarzschild coordinates. \textbf{Right:} Schematic Penrose diagram of the same geometry. }
    \label{fig:AdS}
\end{figure}

Variation of the action (\ref{S_JT}) gives the following equations of motion:
\begin{align}
    0 &= R+2/l_{AdS}^2 \label{metric_eq}\\
    0 &= (\nabla_\mu \nabla_\nu -g_{\mu\nu}/l_{AdS}^2)\Phi \label{Phi_eq}~.
\end{align}
The first equation of motion $R=-2/l_{AdS}^2$ tells us that the metric should have constant negative curvature, or in other words, it is described by AdS$_2$.  A metric for AdS$_2$ can be constructed via an embedding in a 3-dimensional Minkowski spacetime with two time directions and one space direction:
\begin{equation} \label{3d_metric}
    ds^2 = -dT_1^2-dT_2^2+dX^2.
\end{equation}
AdS$_2$ is the induced geometry on the hypersurface
\begin{equation} \label{AdS_embedding}
    -T_1^2-T_2^2+X^2=-l_{AdS}^2, 
\end{equation}
which has boundaries at $X\to \pm \infty$, see Figure \ref{fig:AdS}.
A set of \textit{global} coordinates which cover the whole space is:
\begin{equation} \label{global_co}
    \begin{aligned}
    T_1 &= l_{AdS}\, \sqrt{x^2+1}\cos\tau \\
    T_2 &= l_{AdS}\, \sqrt{x^2+1}\sin\tau \\
    X &= l_{AdS}\, x
    \end{aligned}
\end{equation}
with $0\leq \tau <2\pi$ and $-\infty <x <\infty$. 
Plugging these coordinates into (\ref{3d_metric}) provides an AdS$_2$ metric on the surface (\ref{AdS_embedding}):
\begin{equation} \label{metric_AdS_global}
    ds^2/l_{AdS}^2 = -(1+x^2)d\tau^2 + \frac{1}{1+x^2} dx^2 ~.
\end{equation}
Allowing $\tau\in\mathds{R}$ in \eqref{metric_AdS_global} yields the universal cover of AdS$_2$.
Turning to the second equation of motion, (\ref{Phi_eq}), its solution is given in the embedding coordinates $(T_1,T_2,X)$ by $\Phi = A T_1+ BT_2+C X$ \cite{Maldacena:2016upp, Harlow:2018tqv}. The unique solution, modulo $SO(1,2)$ rotations in embedding space, which respects the boundary condition (\ref{bc_Phi}), namely that $\Phi$ has the same (positive) value on both boundaries, can be expressed as $\Phi \propto T_1$ \cite{Harlow:2018tqv}.
That is:
\begin{equation} \label{Phi_xtau}
    \Phi(x,\tau)  
    =   \Phi_h  \sqrt{1+x^2} \, \cos\tau ~.
\end{equation}

Another set of coordinates, with a well-defined time on the boundary, is given by the \textit{Schwarzschild} coordinates
\begin{equation} \label{Sch_co}
    \begin{aligned}
    T_1 &= l_{AdS}\, r/r_s \\
    T_2 &= l_{AdS}\, \sqrt{(r/r_s)^2-1}\sinh(r_s t/l_{AdS}^2)\\
    X &= l_{AdS}\, \sqrt{(r/r_s)^2-1}\cosh(r_s t/l_{AdS}^2)~,
\end{aligned}
\end{equation}
with $r>r_s$ and $-\infty < t < \infty$. These coordinates do not cover the whole space, as shown in Figure \ref{fig:AdS}. The induced metric in Schwarzschild coordinates is given by
\begin{equation} \label{metric_AdS_Sch}
    ds^2 = -\frac{r^2-r_s^2}{l_{AdS}^2}dt^2 +\frac{ l_{AdS}^2}{r^2-r_s^2}dr^2~.
\end{equation}
The solution to (\ref{Phi_eq}) with boundary condition (\ref{bc_Phi}) acquires a simple form in terms of the Schwarzschild coordinates:
\begin{equation} \label{Phi_rt}
    \Phi(r,t)= \phi_b\, r /l_{AdS}~.
\end{equation}
Note that when $r=r_s$, the two coordinate sets (\ref{global_co}, \ref{Sch_co}) give the relationship
$1= \sqrt{1+x^2} \cos\tau$
which from (\ref{Phi_xtau}) means that at $r=r_s$ we have 
$\Phi = \Phi_h$.
On the other hand, from (\ref{Phi_rt}) we have at $r=r_s$ that
$\Phi = \phi_b r_s/l_{AdS} $.
We thus have the relation
\begin{equation} \label{rs_eq}
    r_s = l_{AdS} \frac{\Phi_h}{\phi_b}~.
\end{equation}

At the boundary, $r= l_{AdS}/\epsilon$ and so the metric (\ref{metric_AdS_Sch}) and dilaton (\ref{Phi_rt}) become
\begin{align}
    ds^2 \big|_{\text{boundary}} &= -\frac{dt^2}{\epsilon^2}(1-\epsilon^2 (r_s/l_{AdS})^2)\\
    \Phi \big|_{\text{boundary}} &= \phi_b/\epsilon
\end{align}
which from the boundary conditions (\ref{bc_metric}, \ref{bc_Phi}) shows that the Schwarzschild time $t$ becomes the boundary time when $\epsilon \to 0$.  
At the boundary, the relation between $\tau$ and $t$ can be worked out by considering the relation between the two sets of coordinates (\ref{global_co}, \ref{Sch_co}):
\begin{equation}
    \left( \frac{r}{r_s} \frac{1}{\cos \tau} \right)^2-1= \Big[\left( \frac{r}{r_s} \right)^2 -1 \Big] \cosh^2(r_s t/l_{AdS}^2)~.
\end{equation}
On the boundary $r=\frac{l_{AdS}}{\epsilon}$, and in the limit $\epsilon \to 0$ the relation between the global time $\tau$ and the boundary time $t_b$ is found to be
\begin{equation}
    \cos \tau = \frac{1}{\cosh(r_s t_b/l_{AdS}^2)} ~.
\end{equation}

\subsubsection{The length of the wormhole and its canonical conjugate}
In \cite{Harlow:2018tqv} it was shown that the phase space coordinates of JT gravity consist of: geodesic distance between the two boundaries, at a time slice defined by boundary time; and its canonical conjugate.
The geodesic distance between two points on the two asymptotic boundaries of AdS$_2$ can be identified as the length of a (Lorentzian) \textit{wormhole}, see Figure \ref{fig:AdS}. In order to compute it for a given boundary time $t_b$, we need the metric (\ref{metric_AdS_global}) and to set $\tau=\text{const}$.  We also need the value of $x$ at the boundary points labeled by time $t_b$.  Again, the relationship between the two sets of coordinates (\ref{global_co}, \ref{Sch_co}) gives
\begin{equation}
    x = \sqrt{(r/r_s)^2-1}\, \cosh(r_s t/l_{AdS}^2) ~.
\end{equation}
At the boundary $ r=l_{AdS}/\epsilon$ so setting $t=t_b$ in the above equation and taking $\epsilon \to 0$ gives
\begin{equation}
    x_b = \frac{l_{AdS}}{\epsilon \, r_s} \cosh (r_s t_b/l_{AdS}^2) ~.
\end{equation}
The wormhole length is then
\begin{align}
    l/l_{AdS} &= \int_{-x_b}^{x_b} \frac{dx}{\sqrt{1+x^2}} = 2\, \textrm{arcsinh}( x_b) = 2 \log \left(x_b+\sqrt{x_b^2+1} \right) \\
    &= 2 \log \left(  \frac{2 l_{AdS}}{\epsilon \, r_s} \cosh (r_s t_b/l_{AdS}^2)\right) + O(\epsilon) ~.
\end{align}
Using $r_s= l_{AdS} \Phi_h /\phi_b$ from (\ref{rs_eq}), the above result can be expressed as
\begin{equation}
    l/l_{AdS} = 2\log\left( \frac{2\phi_b}{ \epsilon}\right) + 2 \log \left[  \frac{1}{\Phi_h} \cosh \left(\frac{\Phi_h}{ l_{AdS}\, \phi_b}\, t_b \right)\right] + O(\epsilon) ~.
\end{equation}
The first term is infinite in the limit $\epsilon\to 0$ (since in the large volume limit the boundary of AdS is at infinity), and this result needs to be renormalized. The renormalized length of the wormhole is given by
\begin{equation}
\label{Bulk_length}
    \tilde{l}/l_{AdS}= l-2\log\left( \frac{2\phi_b}{ \epsilon}\right)  
    = 2 \log \left[\cosh \left(\frac{\Phi_h}{l_{AdS}\,\phi_b}\, t_b \right)\right] - 2 \log \Phi_h ~.
\end{equation}
In terms of the value of the Hamiltonian (or energy) on the boundary \cite{Harlow:2018tqv}
\begin{equation}\label{energy_config_bulk}
    H = \frac{2\Phi_h^2}{ l_{AdS}\,\phi_b}=E,
\end{equation}
result \eqref{Bulk_length} becomes:
\begin{equation} \label{Ren_Wormhole_length_E}
    \tilde{l}/ l_{AdS} = 2 \log \left[\cosh \left(\sqrt{\frac{E}{2  l_{AdS}\, \phi_b}}\, t_b \right)\right] -  \log \left(\frac{ l_{AdS}\,E\phi_b}{2} \right) ~.
\end{equation}
Together with the canonical conjugate of $\tilde{l}$, given by \cite{Harlow:2018tqv}
\begin{equation}
    P\,  l_{AdS} = \sqrt{2  l_{AdS}\, E\phi_b } \tanh \left( \sqrt{\frac{E}{2 l_{AdS}\,\phi_b}}t_b\right), 
\end{equation}
the Hamiltonian of JT gravity takes the form \cite{Harlow:2018tqv}
\begin{equation}\label{JT_Liouville_Ham}
    H = \frac{1}{ l_{AdS}\,\phi_b} \left( \frac{ l_{AdS}^2 \, P^2}{2} + 2 e^{-\tilde{l}/ l_{AdS}}\right) ~.
\end{equation}
This Hamiltonian is written in terms of a two-sided\footnote{ We call $\tilde{l}$ a two-sided length because it joins the two disconnected (and regularized) boundaries of AdS$_2$.} phase space variable $\tilde{l}$ and its conjugate momentum, and thus the quantum description of the system will consist of a Hilbert space spanned by the eigenfunctions of this two-sided length.  This Hilbert space is \textit{not} factorizable as a product of other Hilbert spaces describing the degrees of freedom of each side separately, as was discussed in \cite{Harlow:2018tqv}.

\subsection{Bulk Hilbert space} \label{Subsect:Bulk_Hilbert_Space}

This section reviews how the chord Hilbert space of DSSYK can be understood equivalently as the bulk Hilbert space of JT gravity. The guiding principle in \cite{Lin:2022rbf} is the following: in a given chord diagram, it is possible to identify a \textit{left} and a \textit{right} region by arbitrarily choosing two points on the circumference that will separate them; such points are identified with (Euclidean) future and past infinity, and one can define a constant (Euclidean) time slice by a line whose anchoring points are on different regions: the state on such a slice is given by the number of open chords intersecting it (defined in a unique way \cite{Lin:2022rbf}). This is nothing but a re-interpretation of the discussion around (\ref{State_psi_chord_basis}), but the picture is now very reminiscent of the non-factorizable, two-sided Hilbert space of JT gravity reviewed in the preceding section. The connection can be made even more explicit by showing that the effective DSSYK Hamiltonian becomes, in the suitable limit, the Liouville Hamiltonian of JT gravity \cite{Berkooz:2018jqr,Lin:2022rbf}, where the length operator is in fact proportional to the chord number operator. We review this below.

We will now denote the chord basis by $\{|n\rangle \}_{n=0}^\infty$. The effective Hamiltonian of the averaged theory for DSSYK, given by (\ref{T_alg}), may be written as:
\begin{equation}
    \centering
    \label{Ham_semi_inf_chain}
    T=\frac{J}{\sqrt{\lambda}}\left( \alpha + \alpha^\dagger \right) = \frac{J}{\sqrt{\lambda (1-q)}}\left( D\sqrt{1-q^{\hat{n}}} + \sqrt{1-q^{\hat{n}}} D^\dagger \right),
\end{equation}
where $\alpha,\,\alpha^\dagger$ are the ladder operators \eqref{q-annihilation_op} of the $q$-deformed harmonic oscillator and $\hat{n} | n
\rangle = n | n \rangle$ as given in \eqref{Chord_number_op}, while $D^\dagger$ is a non-normalized version of the creation operator, which therefore acts as a unit-displacement operator on the semi-infinite ordered chord basis:
\begin{equation}
    \centering
    \label{Displacement_op}
    D^\dagger |n\rangle = |n+1\rangle,\qquad D|n\rangle = |n-1\rangle,\qquad D|0\rangle = 0.
\end{equation}
On this chain, $\hat{n}$ plays the role of a position operator. Despite the inherent discreteness of the semi-infinite lattice at hand, it is possible to define a \textit{conjugate} canonical momentum $p$ for $\hat{n}$ as the generator of translations\footnote{If the discrete lattice had a finite length $K$, its Hilbert space would be finite and therefore it would be impossible to fulfill the canonical commutation relation $[\hat{n},p]=i$, but we would still call $p$ a \textit{canonical} momentum in the sense that it satisfies (\ref{Canonical_momentum_discrete}). Taking the suitable limit $K\to \infty$ one recovers the semi-infinite and discrete lattice and, restricted to a physically relevant domain, the commutator $[n,p]$ does tend to $i$, as explained in detail in \cite{Cannata:1991I,Cannata:1991II}.}:
\begin{equation}
    \centering
    \label{Canonical_momentum_discrete}
    D^\dagger \equiv e^{-ip}.
\end{equation}
Observing (\ref{Canonical_momentum_discrete}) we note that the discreteness of the lattice manifests itself in the fact that only displacements of unit length in $n$-space are allowed. A continuum limit would be achieved by allowing displacements of arbitrary length $\Delta n$, of the form $e^{-i\, \Delta n\,p}$. In this limit we would have that $p=-i\partial_n$.

Before proceeding further, it is convenient to redefine the position variable so that it acquires dimensions of length and also so that it is well-behaved in the small-$\lambda$ limit that we shall shortly take. Introducing a \textit{fundamental length scale} $l_f$ about which we do not need to be specific\footnote{It will eventually become the AdS length. It is not a parameter of the DSSYK theory, so one may equivalently say that chord number is related to bulk length \textit{normalized by AdS units.}}, we define a new length variable $l$ as:
\begin{equation}
    \centering
    \label{Length_def}
    l = l_f \lambda n,
\end{equation}
whose conjugate canonical momentum is therefore $k=\frac{p}{l_f \lambda}$, becoming $k=-i\partial_l$ in the continuum limit. With this all, the Hamiltonian becomes:
\begin{equation}
    \centering
    \label{Ham_l_k_before_inverting}
    T = \frac{J}{\sqrt{\lambda (1-q)}}\left( e^{i\lambda l_f k}\sqrt{1-e^{-\frac{l}{l_f}}} + \sqrt{1-e^{-\frac{l}{l_f}}} e^{-i\lambda l_f k} \right).
\end{equation}
The spectrum of this Hamiltonian is bounded (for $\lambda>0$) and symmetric about zero, as can be seen in (\ref{Teigvals}), so we shall do the harmless replacement $T\to -T$ that ensures that the ground state corresponds to the minimal momentum $k$ \cite{Lin:2022rbf}. This choice is arbitrary at this point but will be important later on, as it will ensure that the triple-scaled Hamiltonian is bounded from below. Let us write the explicit form of the Hamiltonian after the replacement $T\mapsto -T\equiv \tilde{T}$:
\begin{equation}
    \centering
    \label{Ham_l_k}
    \tilde{T} = -\frac{J}{\sqrt{\lambda (1-q)}}\left( e^{i\lambda l_f k}\sqrt{1-e^{-\frac{l}{l_f}}} + \sqrt{1-e^{-\frac{l}{l_f}}} e^{-i\lambda l_f k} \right).
\end{equation}

We can now take the remaining limit that brings us to the so-called \textit{triple-scaling limit}. It is a small-$\lambda$ limit in which $l$ is taken to be accordingly large, as follows\footnote{Our definition of the triple-scaling limit differs by a factor of $2$ from that in \cite{Lin:2022rbf}. See Appendix \ref{Appx:triple_scaling} for a discussion on this choice, which eventually does not imply any fundamental modification of the bulk theory.}:
\begin{equation}
    \centering
    \label{Trple-scaling-limit}
    \lambda\to 0,\qquad l\to\infty,\qquad\frac{e^{-\frac{l}{l_f}}}{(2\lambda)^2} \equiv e^{-\frac{\tilde{l}}{l_f}}\;\text{fixed.}
\end{equation}
The last condition in (\ref{Trple-scaling-limit}) can be rewritten as:
\begin{equation}
    \centering
    \label{renormalized_length}
    \frac{\tilde{l}}{l_f} = \frac{l}{l_f} -2\log\left(\frac{1}{2\lambda}\right).
\end{equation}
We shall call $\tilde{l}$ the \textit{renormalized length}\footnote{It will become the actual renormalized bulk length eventually.}. In this triple-scaled limit, the Hamiltonian (\ref{Ham_l_k}) becomes\footnote{This can be shown in the continuum limit, using $k=-i\partial_l$, it still holds for the discrete, semi-infinite chain using the commutation relation $[l,k]=i$, which is satisfied within the relevant physical domain as argued in \cite{Cannata:1991I,Cannata:1991II}.}:
\begin{equation}
    \centering
    \label{Triple_scaled_Hamiltonian}
    \tilde{T}-E_0 = 2\lambda J \left( \frac{l_f^2 k^2}{2} + 2e^{-\frac{\tilde{l}}{l_f}} \right)\;+\mathit{O}\left(\lambda^2\right),
\end{equation}
where $E_0=\frac{-2J}{\lambda}+\mathit{O}(\lambda)$ is a constant energy shift. The moral of this triple-scaling is now clear: we are zooming in near the ground state $E_0$ of the Hamiltonian (\ref{Ham_l_k}), and in this regime it takes the form of the Hamiltonian of Liouville quantum mechanics, whose spectral density is proportional to $\sinh(2\pi\sqrt{E})$ \cite{Bagrets:2016cdf}, which is in turn the Hamiltonian describing the dynamics of the single pair of phase space variables of JT gravity, as seen in (\ref{JT_Liouville_Ham}). This establishes the correspondence between \textit{triple}-scaled SYK and JT gravity: their Hilbert spaces are identified and the Hamiltonian generating dynamics is the same. Incidentally, we note that instances of DSSYK with different values of $\lambda$ and $J$ may collapse, in the triple-scaling limit, onto the same Liouville Hamiltonian if their product $\lambda J$ is equal, since this is the only parameter controlling the Liouville Hamiltonian (\ref{Triple_scaled_Hamiltonian}), and hence the gravity dual. In other words, the following parameter identification connects Hamiltonians (\ref{JT_Liouville_Ham}) and (\ref{Triple_scaled_Hamiltonian}):
\begin{equation}
    \label{Param_identification_Hamiltonians}
    l_f = l_{AdS},\qquad\qquad\qquad 2 \lambda J = \frac{1}{l_{AdS}\phi_b}.
\end{equation}
We will come back to these identifications in section \ref{Section_Gravity_matching}, where boundary K-complexity is matched to the corresponding bulk length computation.

\section{Krylov complexity and chords}\label{Sect:KC}
In this section we quickly review the definition of Krylov complexity and then provide exact results for it in DSSYK. Although the double-scaling limit of SYK is not the limit of SYK which corresponds directly to JT gravity, we study K-complexity in this limit as a warm-up and as an interesting result in itself which had not been studied before the publication of \cite{IV}. The definitions that will be provided below have already been given in Chapters \ref{ch:chapter01_Lanczos} and \ref{ch:chapter02_KC}, but they shall nevertheless be restated here in order to clearly specify the conventions and notations relevant for this Chapter.

Krylov complexity measures the spreading of states or operators in a quantum system under Hamiltonian time evolution, over a special ordered basis (the Krylov chain) adapted to the state/operator's time evolution. Here, we will focus on the definition of K-complexity for states. 
Given an initial state $|0 \rangle$ at $t=0$, its time evolution generated by the Hamiltonian $H$ in the Schrödinger picture is given by
\begin{equation}
    |\phi(t)\rangle = e^{-iHt}|0 \rangle = \sum_{n=0}^\infty \frac{(-it)^k}{k!} H^k |0\rangle 
\end{equation}
written as a linear combination over the basis $\{ |0\rangle, H|0\rangle, H^2|0\rangle,\dots \}$. The survival amplitude, which is the overlap of the time-evolving state with the state at $t=0$ (also known as \textit{fidelity}), is given by
\begin{equation} \label{survival_prob}
    \langle 0|\phi(t)\rangle = \langle 0| e^{-iHt}|0 \rangle = \sum_{n=0}^\infty \frac{(-it)^k}{k!} \langle0| H^k |0\rangle =: \sum_{n=0}^\infty \frac{(-it)^k}{k!} M_k
\end{equation}
where we define the moments $M_k := \langle0| H^k |0\rangle$.  If the survival probability $\langle 0|\phi(t)\rangle$ is an even function of $t$ then all odd moments $M_{2k+1}$ are zero.  This, in particular, is the case for the effective Hamiltonian of DSSYK with the initial state being the zero-chord state, as argued around \eqref{Moments_effective_Ham}. Recalling the discussion in section \ref{subsect:From_Lanczos_to_Moments}, this implies that the $a_n$ Lanczos coefficients vanish identically, so that the particularization of the Lanczos algorithm for the Hamiltonian and state concerned is:  

\begin{enumerate}
    \item $|A_1\rangle = H |0 \rangle$, compute $b_1 = \lVert A_1 \rVert$, if $b_1=0$ stop. Otherwise define $|1 \rangle = |A_1\rangle /b_1 $.
    \item For $n\geq 1$: $|A_{n+1}\rangle = H |n \rangle - b_{n} |n-1 \rangle$, compute $b_{n+1}=\lVert A_{n+1}\rVert$, if $b_{n+1}=0$ stop. Otherwise define $|n+1\rangle = |A_{n+1}\rangle/b_{n+1}$.
\end{enumerate}
Here it was assumed that the initial state $|0\rangle$ is normalized.
In this way the ordered Krylov basis $\{|n\rangle \}_{n=0}^{K-1}$ is constructed and the Lanczos coefficients $\{b_n\}_{n=1}^{K-1}$ are defined, where $K$ is the dimension of the Krylov space. $K$ can be infinite if the original Hilbert space has infinite dimension, and must be finite otherwise.
By construction, the Krylov basis vectors form an orthonormal basis, satisfying $\langle m |n\rangle =\delta_{mn}$. In the Krylov basis, the Hamiltonian acquires the following tridiagonal form determined by the Lanczos coefficients: 
\begin{equation} \label{HKB}
    H|n\rangle = b_{n+1}|n+1\rangle  +b_n|n-1\rangle~.
\end{equation}
The Lanczos coefficients can be determined from the moments of the survival amplitude \eqref{survival_prob}, via the iterative procedure (cf. section \ref{sect:Moments_to_Lanczos}):
\begin{align} 
    M_{2k}^{(n)} &= \frac{M_{2k}^{(n-1)}}{b_{n-1}^2}-\frac{M_{2k-2}^{(n-2)}}{b_{n-2}^2}, \quad n=1,2,\dots,L \quad k=n,n+1,\dots, L \nonumber\\
    b_n^2 &= M_{2n}^{(n)}   \label{recursion_moments} 
\end{align}
where $M_{2k}^{(0)} \equiv M_{2k}$ as well as $M_{2k}^{(-1)}=0$ and $b_{-1}=1=b_0$. Using this algorithm, the first $L$ Lanczos coefficients can be determined from the moments $M_{2k}$ with $k=0,\dots,L$. 

The time-evolving state can now be expanded over the Krylov basis:
\begin{equation}\label{phi_t_phi_n}
    |\phi(t)\rangle = \sum_{n=0}^{K-1} \phi_n(t)|n\rangle
\end{equation}
where $\phi_n(t) := \langle n|\phi(t)\rangle$ can be thought of as a ``wavefunction" spreading over the ordered Krylov basis with initial condition $\phi_n(t=0)=\langle n|0\rangle =\delta_{n0}$. From unitarity of the time evolution, the norm is preserved, i.e. $\sum_{n=0}^{K-1}|\phi_n(t)|^2=1$ for all $t$.
Note that $\phi_0(t)$ is the survival amplitude defined in (\ref{survival_prob}).

A ``position'' operator, $\hat{n}$, over the ordered Krylov basis can be defined as
\begin{eqnarray} \label{nOperator}
    \hat{n} = \sum_{n=0}^{K-1} n \, |n \rangle \langle n|,
\end{eqnarray}
and \textit{Krylov complexity} is then defined as the expectation value of $\hat{n}$ as a function of time, or equivalently, the average position of the wavefunction $|\phi(t)\rangle$ over the ordered Krylov basis:
\begin{equation} \label{KC_definition}
    C_K(t) = \langle \hat{n}(t)\rangle = \langle \phi(t)| \hat{n} | \phi(t) \rangle = \sum_{n=0}^{K-1} n \, |\phi_n(t)|^2 ~.
\end{equation}

From (\ref{HKB}) we can obtain a recurrence relation for the eigenvectors of $H$, $H|E\rangle=E|E\rangle$,
since the components of each eigenvector $|E\rangle$ over the Krylov basis, defined as $\psi_n(E)\equiv \langle n|E\rangle$, satisfy:
\begin{equation}
    E\, \psi_n(E) = b_{n+1}\psi_{n+1}(E) + b_n \psi_{n-1}(E) ~.
\end{equation}
The components of the eigenvectors of $H$ over the Krylov basis are useful in the determination of the wavefunctions $\phi_n(t)$, and if known, together with the eigenvalues of $H$, they provide the following closed-form expression for the wavefunction as a function of time:
\begin{equation} \label{Phi_EnergyB}
    \phi_n(t) \equiv \langle n|\phi(t)\rangle = \langle n|e^{-iHt} |0\rangle = \sum_E e^{-iEt}\langle n|E\rangle \langle E|0\rangle =  \sum_E e^{-iEt}\, \psi_n(E) \, \psi_0^*(E)~.
\end{equation}

\subsection{Lanczos coefficients in DSSYK}

Observing the steps leading to the expression of the effective Hamiltonian in coordinates over the basis of \textit{fixed chord number} states, given in matrix form in (\ref{T_sym}), the connection to Lanczos coefficients is argued as follows. The fixed chord number states $|n\rangle$ are an orthonormal set of linear combinations of states with up to $n$ Hamiltonian insertions $\left\{T^k|0\rangle\right\}_{k=0}^n$ and, using them as a basis, they bring the transfer matrix (or effective Hamiltonian) to a tridiagonal form. They thus turn out to be Krylov basis elements, with the matrix elements of the symmetric, tridiagonal $T$ giving the Lanczos coefficients:
\begin{equation}
    \centering
    \label{Lanczos}
    b_n = J \sqrt{\frac{[n]_q}{\lambda}}=J\sqrt{\frac{1-q^n}{\lambda(1-q)}}~.
\end{equation}
Consequently, the chord number operator $\hat{n}$, which gives the position in the chord number basis, gives equivalently the position in the Krylov basis, so it \textit{is} the K-complexity operator.

\begin{figure}
    \centering
    \includegraphics[scale=0.3]{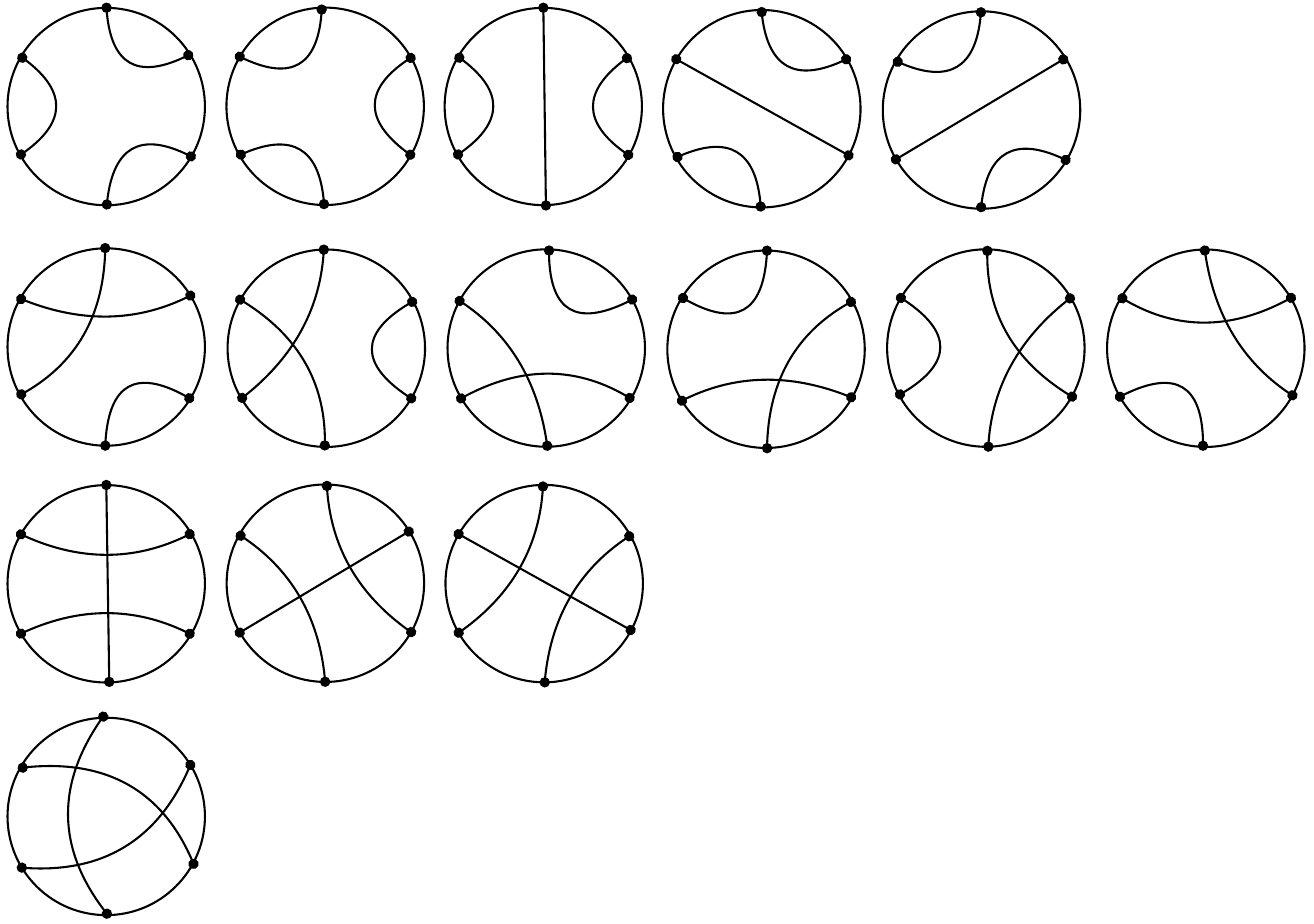}
    \caption{The 15 chord diagrams contributing to $M_6$: the \textbf{top row} shows chord diagrams with zero intersections (5 diagrams), the \textbf{second row} shows chord diagrams with 1 intersection (6 diagrams), the \textbf{third row} shows diagrams with 2 intersections (3 diagrams) and the \textbf{fourth row} shows the only chord diagram with 3 intersections. These numbers correspond to the calculation of $M_6$ in (\ref{First_three_even_moments}).}
    \label{fig:chord_diagrams_M6}
\end{figure}

There is certainly more than one orthonormal basis that can bring the effective Hamiltonian $T$ to tridiagonal form. However, given a seed state $|0\rangle$, there is a unique (up to global phases) orthonormal basis $\{|n\rangle\}$ such that the state $|n\rangle$ is a linear combination of $\{T^k |0\rangle\}_{k=0}^n$, as can be shown inductively. Such a basis can therefore be identified, up to global phases, with the Krylov basis adapted to the state $|0\rangle$ and the Hamiltonian $T$, which can be built efficiently through the Lanczos algorithm and brings the Hamiltonian to tridiagonal form with real entries. The fact that the matrix expression (\ref{T_sym}) for the effective Hamiltonian in the fixed chord number basis has positive entries indicates that such a basis coincides exactly with the Krylov basis (i.e. they do not differ even by a global phase). This argument proves that the fixed chord number states are Krylov elements and therefore that the entries of the tridiagonal version of $T$ in (\ref{T_sym}) are the Lanczos coefficients. In addition, we can perform a non-trivial check of the fact that the sequence (\ref{Lanczos}) is actually the Lanczos coefficients sequence by taking the moments $M_{2k}$ computed via chord-diagram combinatorics and showing that the usual transformation (\ref{recursion_moments}) will indeed yield a sequence of Lanczos coefficients in agreement with expression (\ref{Lanczos}). Using chord diagrams, we compute explicitly the first three even moments:
\begin{equation}
    \centering
    \label{First_three_even_moments}
    \begin{split}
        &M_2 = \frac{J^2}{\lambda} \\
        &M_4 = \frac{J^4}{\lambda^2}(2+q)\\
        &M_6=\frac{J^6}{\lambda^3}(5+6q+3q^2+q^3)
    \end{split}
\end{equation}
where $M_6$ involves evaluating the 15 chord diagrams shown in Figure \ref{fig:chord_diagrams_M6}.
With this, we can apply the recursion method (\ref{recursion_moments}):
\begin{equation}
    \centering
    \label{b_n_from_Mn_first}
    \begin{split}
        &b_1^2=M_2=\frac{J^2}{\lambda}=\frac{J^2}{\lambda}[1]_q \\
        &b_2^2 = \frac{M_4}{M_2}-M_2 = \frac{\frac{J^4}{\lambda^2}(2+q)}{\frac{J^2}{\lambda}}-\frac{J^2}{\lambda}=\frac{J^2}{\lambda}(1+q) = \frac{J^2}{\lambda} [2]_q \\
        &b_3^2 = \frac{\frac{M_6}{M_2}-M_4}{\frac{M_4}{M_2}-M_2}-\frac{M_4}{M_2}=(...)=\frac{J^2}{\lambda}(1+q+q^2) = \frac{J^2}{\lambda}[3]_q~.
    \end{split}
\end{equation}
Equations (\ref{b_n_from_Mn_first}) illustrate that the Lanczos coefficients (\ref{Lanczos}) are consistent with the Hamiltonian moments (\ref{First_three_even_moments}).

\subsubsection{State dependence of the Lanczos coefficients}\label{Subsubsect_Krylov_TFD}

An important discussion is in order: what initial state are these Lanczos coefficients being computed for? They are the Lanczos coefficients associated to the evolution under the effective Hamiltonian $T$ of the initial state $|0\rangle$, given by the state
\begin{equation}
    \centering
    \label{state_T_0}
    |\phi(t)\rangle = e^{-itT}|0\rangle,
\end{equation}
as the moments to which the coefficients (\ref{Lanczos}) are in one-to-one correspondence are the coefficients of the Taylor series of the survival amplitude of (\ref{state_T_0}):
\begin{equation}
    \centering
    \label{Survival_amplitude_zero_ket}
    \langle 0 | e^{-itT}|0\rangle =\sum_{k=0}^{+\infty} \frac{(-it)^{2k}}{(2k)!} M_{2k},
\end{equation}
where the moments $M_{2k}$ are given in (\ref{Moments_effective_Ham}).

Furthermore, the state $|0\rangle$ that seeds the evolution can be thought of as an effective version of the infinite-temperature thermofield ``double'' state in the ensemble-averaged theory, and the survival amplitude (\ref{Survival_amplitude_zero_ket}) is in fact the ensemble-averaged analytic continuation of the partition function, which is itself known to give the survival amplitude of the thermofield double state \cite{delCampo:2017bzr,Balasubramanian:2022tpr}. We shall now elaborate on this. The ensemble-averaged analytic continuation of the partition function, evaluated at infinite temperature, reads:
\begin{equation}
    \centering
    \label{Averaged_partition_fn_beta_zero}
    \left.\left\langle Z(\beta + i t) \right\rangle\right|_{\beta=0} = \left\langle \text{Tr}\left[ e^{-itH} \right] \right\rangle.
\end{equation}

Now, one can check that the trace in (\ref{Averaged_partition_fn_beta_zero}) can be rewritten as the expectation value of the evolution operator in a certain state $|\Omega\rangle$, as follows:
\begin{equation}
    \centering
    \label{Trace_omega_state}
    \text{Tr}\left[ e^{-itH} \right]=\langle \Omega | e^{-itH} |\Omega \rangle,\qquad\qquad\qquad |\Omega\rangle := \frac{1}{\sqrt{\mathcal{N}}}\sum_E |E\rangle,
\end{equation}
where $\mathcal{N}\to\infty$ denotes the Hilbert space dimension, and the factor $1/\sqrt{\mathcal{N}}$ is required for consistency with the convention $\text{Tr}[\mathds{1}]=1$, ensuring normalization of the state $|\Omega\rangle$. By inspection, we note that this state is an ``unconventional'' version of the infinite-temperature thermofield ``double'' state, in the sense that it does not belong to the tensor product of two identical Hilbert spaces as in \eqref{Sect_KC_holog_TFD} but rather to the only Hilbert space available in the problem. This feature will be consistent with the bulk interpretation of this Hilbert space as the non-factorizable two-sided Hilbert space of JT gravity \cite{Harlow:2018tqv}, as discussed in \cite{Lin:2022rbf}.

Finally, taking the ensemble average  amounts to replacing the non-averaged Hamiltonian $H$ by the effective Hamiltonian $T$, and the state $|\Omega\rangle$ by the zero-chord state $|0\rangle$~:
\begin{equation}
    \centering
    \label{Ensemble_average_survival}
      \left\langle \text{Tr}[e^{-itH}] \right\rangle = \Big\langle \langle \Omega | e^{-itH} |\Omega \rangle \Big\rangle = \langle 0 | e^{-it T} | 0 \rangle,
\end{equation}
which connects, as promised, the averaged analytic continuation of the partition function (\ref{Averaged_partition_fn_beta_zero}) to the survival amplitude (\ref{Survival_amplitude_zero_ket}) and provides the justification for considering the zero-chord state $|0\rangle$ as an effective version of the infinite-temperature thermofield ``double'' state in the averaged theory.

\subsubsection{Regimes of the Lanczos sequence}
In order to better understand the different regimes of K-complexity as a function of time, it will be useful to analyze the regimes that can be identified in the Lanczos sequence (\ref{Lanczos}).
At this point, we recall that $q=e^{-\lambda} = e^{-\frac{2p^2}{N}}$. For $\lambda>0$ we have $0<q<1$ and the Lanczos coefficients $b_n$ in (\ref{Lanczos}) are bounded. They have a horizontal asymptote at $b_\infty=\frac{J}{\sqrt{\lambda(1-q)}}>0$. In the limit $\lambda\to+\infty$ (i.e. $q\to 0$) the Lanczos sequence is constant $b_n\sim \frac{J}{\sqrt{\lambda}}\to 0$, while in the limit $\lambda\to0$ we have $q\to1$ and the Lanczos coefficients grow indefinitely as $b_n\sim J\sqrt{\frac{n}{\lambda}}$. The limit $\lambda \to 0$ is of particular importance since in \cite{Lin:2022rbf} the connection to gravity is done in a \textit{triple} scaling limit in which $\lambda \ll 1$ (i.e. $q\to 1$). 

In order to identify regimes in the Lanczos sequence, the above analysis needs to be performed more carefully. The arguments above have considered limits of $\lambda$ keeping $n$ fixed, whereas below we shall study the behavior of the $b_n$ sequence given a fixed parameter $\lambda>0$ for $n$ either sufficiently big or small compared to $\lambda$.

Inspecting (\ref{Lanczos}) we note that, given $0<q<1$, for sufficiently large $n$, the factor $q^n$ will be small and therefore we can use the approximation:
\begin{equation}
    \centering
    \label{Lanczos_large_n}
    b_n\approx \frac{J}{\sqrt{\lambda(1-q)}}\left(1-\frac{q^n}{2}\right),\qquad\text{for large }n. 
\end{equation}
Conversely, if $|n\log(q)|$ is sufficiently small, we can approximate $1-q^n=1-e^{-n \log(1/q)}\approx n\log(1/q)=\lambda n>0$ (for $n>0$), so that:
\begin{equation}
    \centering
    \label{Lanczos_small_n}
    b_n\approx J\sqrt{\frac{n}{1-q}},\qquad\text{for small }n.
\end{equation}
We can estimate the critical value $n_*(q)$ that marks the transition between the regimes (\ref{Lanczos_small_n}) and (\ref{Lanczos_large_n}) by taking the value of $n$ for which $n\log(1/q)$ becomes of order $1$, which yields:

\begin{equation}
    \centering
    \label{nstar_good_estimate}
    n_*(q) \approx \frac{1}{\log\left(\frac{1}{q}\right)}=\frac{1}{\lambda}.
\end{equation}

\begin{figure}[t]
    \centering
    \includegraphics[width=0.45\textwidth]{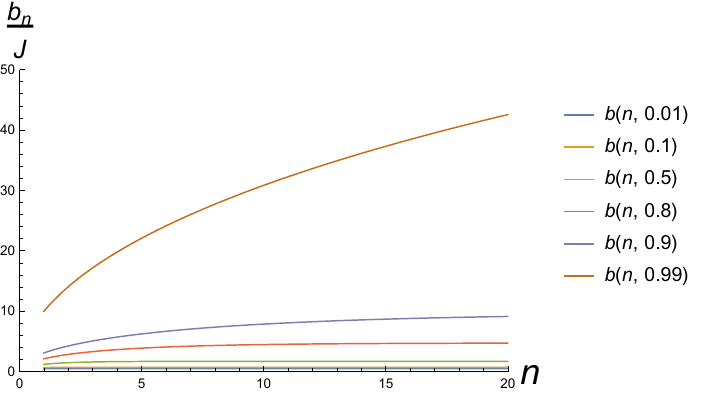} \includegraphics[width=0.45\textwidth]{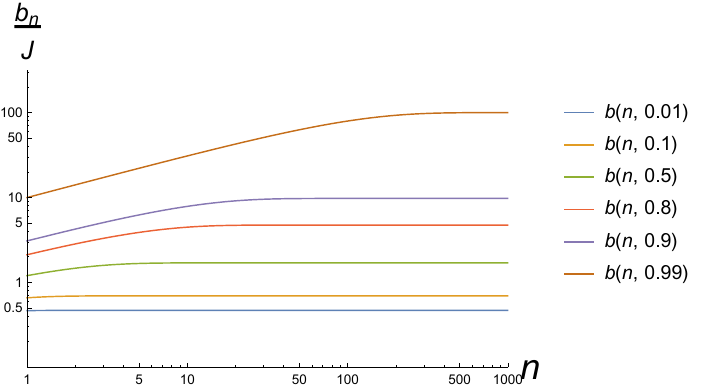}
    \caption{Lanczos coefficients in DSSYK for different values of $q$, denoted $b_n\equiv b(n,q) $. \textbf{Left:} Linear scale along both axes, focusing at small values of $n$. \textbf{Right:} Log-log plot. In this scale, the initial linear shape is compatible with a square-root behavior of $b_n$; Figure \ref{fig:Lanczos_example} illustrates this in more detail.}
    \label{fig:Lanczos_various_qvals}
\end{figure}

\begin{figure}[t]
    \centering
    \includegraphics[width=0.45\textwidth]{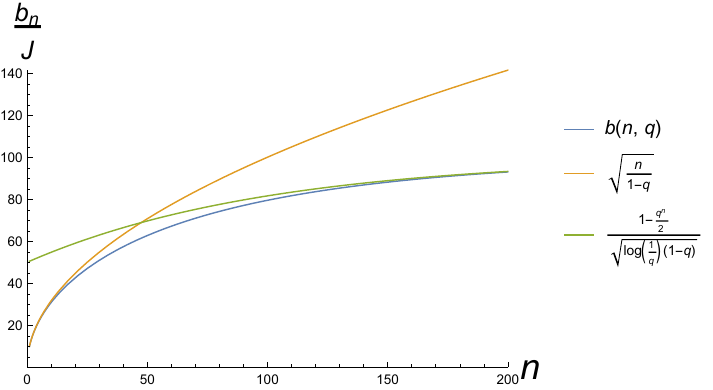} \includegraphics[width=0.45\textwidth]{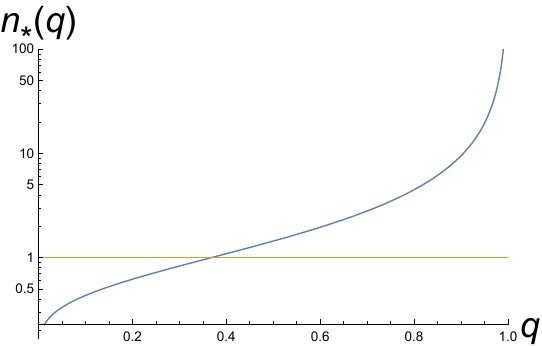}
    \caption{\textbf{Left:} Lanczos coefficients of DSSYK for $\lambda=0.01$ (i.e. $q\approx0.99005$), together with the limiting regimes for small and large $n$.
    \textbf{Right:} Transition value $n_*(q)$ as a function of $q$ (in blue). The constant $n_*=1$ has been marked in orange, for reference. Note the logarithmic scale along the vertical axis. In particular, we note that $n_*(q)$ has a vertical asymptote at $q=1$, implying that in the limit $q\to 1$, which is of interest in this work, there will be a clear separation of scales in the Lanczos sequence.}
    \label{fig:Lanczos_example}
\end{figure}

Summarizing, we have found that the Lanczos coefficients start growing as $\propto \sqrt{n}$ before $n_*(q)$, after which they saturate at a horizontal asymptote following $\propto~ 1-\frac{q^n}{2}$:
\begin{equation}
    \centering
    \label{Lanczos_regimes_summary}
    b_n =  \begin{cases}
        J\sqrt{\frac{n}{1-q}},&  n\lesssim \frac{1}{\lambda}\\
        \frac{J}{\sqrt{\lambda(1-q)}}\left(1-\frac{q^n}{2}\right),& n\gtrsim \frac{1}{\lambda}
    \end{cases} 
\end{equation}
Note that $\lim_{q\to 1^{-}} n_*(q)=\lim_{\lambda \to 0^{+}}n_*(q(\lambda))=+\infty$, consistent with the fact that when $q\to1$ we have $b_n\propto\sqrt{n}$, i.e. only the square-root behavior is featured. Additionally, inspecting (\ref{nstar_good_estimate}) we realize that for $\lambda>1$ (or, equivalently, $q<e^{-1}\approx 0.37$) the transition value of $n$ is $n_*<1$, which implies that the first regime in (\ref{Lanczos_regimes_summary}) will not be visible at all in the actual Lanczos sequence, since strictly speaking $n\in \mathds{N}$. In the cases of interest for us we will need to consider both regimes, as the connection to gravity occurs in a limit where $\lambda$ is small. Figure \ref{fig:Lanczos_various_qvals} depicts several Lanczos sequences for different values of $q$, and Figure \ref{fig:Lanczos_example} shows the different regimes in the Lanczos sequence for a particular value of $q$ close to $1$, together with the $q$-dependence of the transition value $n_*(q)$.

For reference, we can rewrite (\ref{Lanczos_regimes_summary}) for small $\lambda$ (i.e. $q$ close to $1$):
\begin{equation}
    \centering
    \label{Lanczos_regimes_summary_smallLambda}
    b_n \approx  \begin{cases}
        J\sqrt{\frac{n}{\lambda}},&  n\lesssim \frac{1}{\lambda}\\
        \frac{J}{\lambda}\left(1-\frac{e^{-n\lambda}}{2}\right),& n\gtrsim \frac{1}{\lambda}
    \end{cases} 
\end{equation}
where the corrections are subleading in an expansion in powers of $\sqrt{\lambda}$.

\subsection{K-complexity regimes in DSSYK}

The regimes of the Lanczos coefficients (\ref{Lanczos_regimes_summary}) allow us to distinguish different regimes in the growth of K-complexity. The initial condition of the wave packet is $\phi_n(0)=\delta_{n0}$, where $\phi_n(t)$ are the coordinates of the state $|\phi(t)\rangle$, defined in (\ref{state_T_0}), over the Krylov basis (i.e. the basis of fixed-chord-number states); see (\ref{Phi_EnergyB}) for the precise definition. Therefore, at sufficiently early times, the state wave packet only probes the first regime of (\ref{Lanczos_regimes_summary}), where the Lanczos coefficients exhibit a square-root behavior. Following \cite{Caputa:2021sib,Balasubramanian:2022tpr} we can recycle the results for the Heisenberg-Weyl algebra. Namely, the Hamiltonian is given by
\begin{equation}
    \centering
    \label{T_Heisenberg_Weyl}
    T = \gamma \left(a + a^\dagger\right),
\end{equation}
where $\gamma= \frac{J}{\sqrt{1-q}} = \frac{J}{\sqrt{1-e^{-\lambda}}}$ and $a,\,a^\dagger$ act over the Krylov basis as ladder operators of a simple bosonic harmonic oscillator, satisfying the usual algebra
\begin{equation}
    \centering
    \label{Heisenberg_Weyl_ladder_ops}
    [a,a^\dagger]=1,
\end{equation}
i.e. in this regime the q-deformation of the algebra (\ref{q_oscillator_algebra}) is not important as long as we restrict to the subspace of states $\{|n\rangle\;/\;n\lesssim n_*(q)\}$. Using the Baker-Campbell-Hausdorff formula and the simple commutation relation (\ref{Heisenberg_Weyl_ladder_ops}), the exact wave function $\phi_n(t)$ is computed in \cite{Caputa:2021sib}, as this is nothing but the evolution along a one-parameter family of coherent states:
\begin{equation}
    \centering
    \label{Heisenberg_Weyl_Wave_Fn}
    \phi_n(t) = e^{-\frac{\gamma^2 t^2}{2}}\frac{(-i\gamma t)^n}{\sqrt{n!}} ~.
\end{equation}
The result (\ref{Heisenberg_Weyl_Wave_Fn}) for the Krylov space wave function can equivalently be obtained using the spectral decomposition of the tridiagonal Hamiltonian. Using this method, in section~\ref{subsect:formal_KC} we shall derive a formal expression for the wave functions $\phi_n(t)$ at arbitrary $q$ and show that it reduces to (\ref{Heisenberg_Weyl_Wave_Fn}) in the $q\to 1$ limit in section \ref{Sec:qto1Limit}. Likewise, in Appendix \ref{App:Wavefunctions} we present an independent diagonalization of the tridiagonal Hamiltonian at $q=1$.

From this wave function K-complexity is given by:
\begin{equation}
    \centering
    \label{KC_early_times}
    C_K(t)=\sum_{n=0}^{+\infty}n|\phi_n(t)|^2=\gamma^2\,t^2 = \frac{(tJ)^2}{1-q}=\frac{(tJ)^2}{1-e^{-\lambda}}\,.
\end{equation}
For small $\lambda$, we can approximate (\ref{KC_early_times}) as $C_K(t)\approx\frac{(tJ)^2}{\lambda}$, up to subleading corrections in a $\lambda$-expansion.

We can use $C_K(t)$ as an estimate for the position of the peak of the coherent packet $\phi_n(t)$. Thus, the packet will start to probe the second region of (\ref{Lanczos_regimes_summary}), where the Lanczos coefficients approach a horizontal asymptote, when $C_K(t)$ becomes of the order of $n_*(q)$. We therefore define a transition time scale $t_*(q)$ by the relation
\begin{equation}
    \centering
    \label{tstar_defining_equation}
    C_K\Big(t_*(q)\Big) = n_*(q),
\end{equation}
giving
\begin{equation}
    \centering
    \label{tstar}
    t_*(q) = \frac{1}{J}\sqrt{\frac{1-q}{\log\left(\frac{1}{q}\right)}}=\frac{1}{J}\sqrt{\frac{1-e^{-\lambda}}{\lambda}},
\end{equation}
which behaves as $t_*\approx J^{-1}$ for small $\lambda$.
We note that, as $\lambda$ goes to zero, the transition value for $n$ goes to infinity, since $n_*=\frac{1}{\lambda}$. On the other hand, K-complexity at early times, $C_K(t)\sim \frac{1}{\lambda}(tJ)^2$, has an accordingly increasing growth rate, so that the $\lambda$-dependence cancels out in the expression of the transition time $t_*$, which is only controlled by $J^{-1}$.

\begin{figure}[t]
    \centering
    \includegraphics[width=7.4cm]{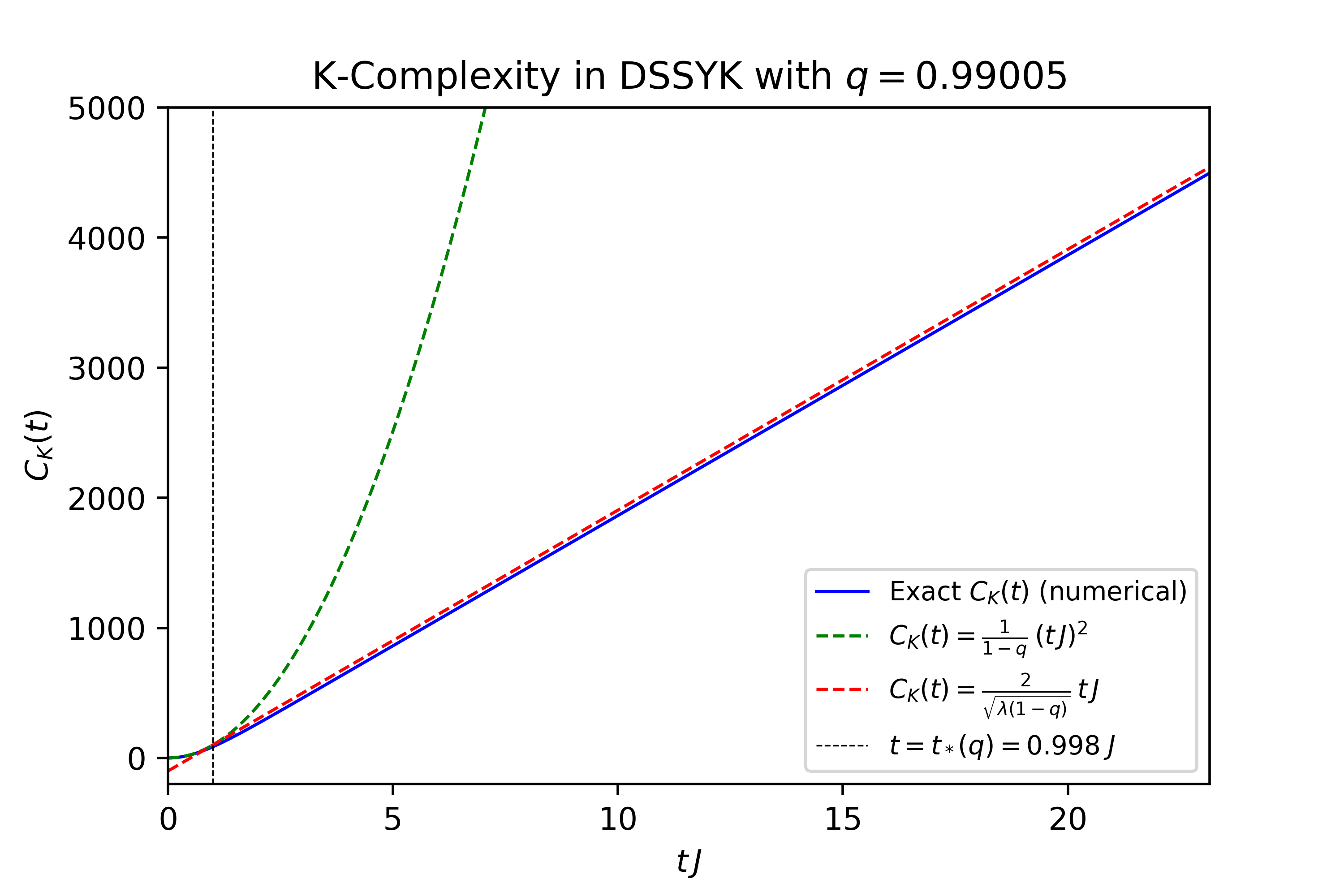} \includegraphics[width=7.4cm]{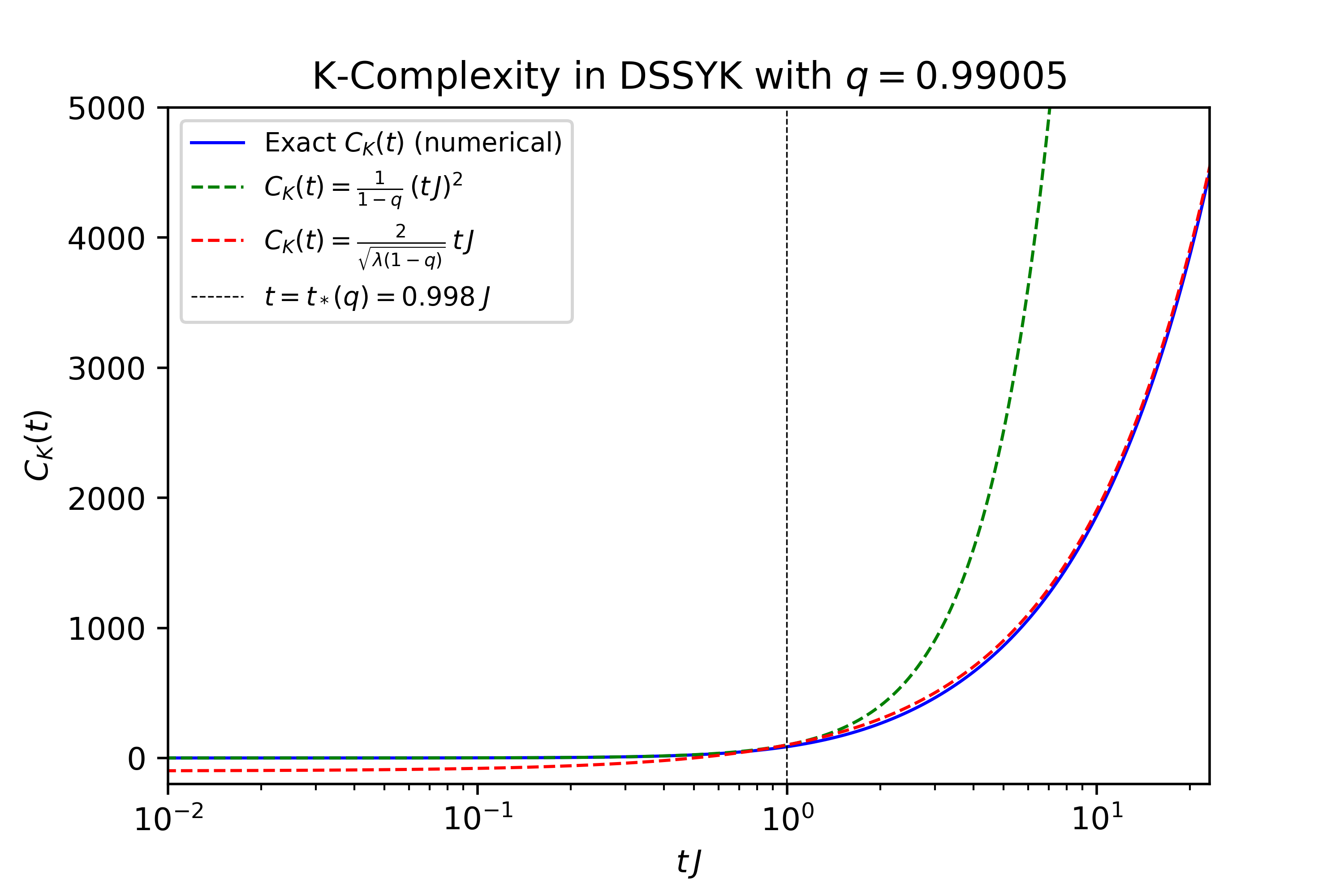}
    \caption{K-complexity regimes in DSSYK. The plots compare the exact $C_K(t)$ computed numerically (in blue) with the early (green) and late (red) time regimes described in (\ref{KC_regimes_summary}). The numerical computation imposed an artificial cutoff $K=5000$ in the length of the Lanczos sequence, but the plot only depicts a time range during which the wave packet does not yet probe the right edge of the artificially finite Krylov chain, hence not featuring spurious finite-size effects. \textbf{Left:} Linear scale on both axes. Note that the curve for the late-time approximation is parallel to the exact result but it does not exactly overlap with it; this is because of the lack of knowledge of the constant $c(\lambda)$ in (\ref{KC_regimes_summary}), which in the case of this plot was chosen to make the early- and late-time regimes match at $t=t_*(q)$. \textbf{Right:} Logarithmic scale along the horizontal axis. This allows us to note the good agreement at early times between the exact result and the corresponding approximation. Additionally, the logarithmic scale washes away the effect of the constant $c$, and the red and blue curves do overlap at late times, confirming the expected linear growth of $C_K(t)$ at times well after $t_*(q)$.}
    \label{fig:KC_numerics}
\end{figure}

Well after $t_*(q)$ the wave packet probes a region where the Lanczos coefficients are effectively constant, $b_n\approx b_{\infty}\equiv \frac{J}{\sqrt{\lambda(1-q)}}$. According to \cite{Barbon:2019wsy}, the wave functions $\phi_n(t)$ in this case are given by Bessel functions, the position of whose front-most peak evolves in time as $\sim 2 b_\infty \,t$. Using the peak position as an estimate for K-complexity, we obtain in this case:
\begin{equation}
    \centering
    \label{KC_well_after_tstar}
    C_K(t)\approx \frac{2\,tJ}{\sqrt{\lambda(1-q)}}\,. 
\end{equation}
In summary:
\begin{equation}
    \centering
    \label{KC_regimes_summary}
    C_K(t)= \begin{cases}
        \frac{(tJ)^2}{1-q}\,,& t\lesssim t_*(q) \\
        \frac{2\,tJ}{\sqrt{\lambda(1-q)}}\,+c(\lambda)\,,&  t\gg t_*(q)
    \end{cases} 
\end{equation}
where $c(\lambda)$ is some $\lambda$-dependent constant that should, strictly speaking, be there in order to ensure the matching of the two regimes.
For small $\lambda$ this becomes:
\begin{equation}
    \centering
    \label{KC_regimes_summary_small_Lambda}
    C_K(t)\approx \begin{cases}
        \frac{(tJ)^2}{\lambda}\,,& t\lesssim \frac{1}{J} \\
        \frac{2\,tJ}{\lambda}+c(\lambda)\,,& t\gg \frac{1}{J}
    \end{cases} .
\end{equation}
Figure \ref{fig:KC_numerics} depicts the different regimes of K-complexity, compared with the numerical result, for a value of $q$ close to $1$.

\subsection{Continuum approximation to K-complexity in DSSYK}\label{subsect_continuum_approx}

In this section we obtain an analytic expression for the K-complexity of the infinite-temperature thermofield ``double'' state in DSSYK (i.e. the zero-chord-number state) making use of a continuum approximation of Krylov space. Generically, the approximation developed in \cite{Barbon:2019wsy}, reviewed in Appendix \ref{Appx:Cont_approx}, consists of promoting $n$ to a continuous variable in such a way that the recursion relation describing the evolution of the wave function $\phi_n(t)$ becomes a differential equation which is, at leading order, a chiral wave equation with a velocity field given by $v(n)\equiv 2 b_n$. As a result of this equation, the wave packet propagation is ballistic and one can estimate the value of K-complexity by the position of the peak of the wave function, which behaves as a point particle propagating through the above-mentioned velocity field $v(n)$. As reviewed in Appendix \ref{Appx:Cont_approx}, for a finite Krylov space of dimension $K$, this continuum approximation is controlled by the small parameter $\varepsilon=\frac{1}{K-1}$, but it turns out that, due to some divisions by $\varepsilon$ in intermediate steps of the discussion, the applicability of this approximation in the strict thermodynamic limit $K\to\infty$ is not well under control, even if strictly speaking $\varepsilon=0$ in that limit. 

In the particular model at hand in this project, however, the parameter $\lambda$ in (\ref{lambda_def}) can be used to control the continuum approximation even in the $K\to\infty$ limit, as we shall argue here. Inspecting the expression of the Lanczos coefficients (\ref{Lanczos}) we note that we can define a dimensionless position variable as $x := \lambda n$ in such a way that its spacing goes to zero when $\lambda\to 0$, becoming a continuous variable. More precisely, a continuum limit for the Krylov space of DSSYK can be defined as the limit in which $\lambda\to 0$, $n\to\infty$ with $x=\lambda n$ fixed. In this limit, the Lanczos coefficients adopt the limiting form of a continuous function of $x$:
\begin{equation}
    \centering
    \label{Lanczos_cont_lambda}
    b_n=J\sqrt{\frac{1-e^{-\lambda n}}{\lambda(1-q)}}\longrightarrow \frac{J}{\lambda}\sqrt{1-e^{-x}}+\mathit{O}(\lambda^0)\equiv b(x).
\end{equation}
As in the argument reviewed in Appendix \ref{Appx:Cont_approx}, the starting point for studying this continuum limit is the recursion relation satisfied by the wave function. Defining a real $\varphi_n(t)$ such that $\phi_n(t)=i^{n}\varphi_n(t)$, we have that (\ref{HKB}) and (\ref{phi_t_phi_n}) imply the recursion:
\begin{equation}
    \centering
    \label{recurrence_varphi_n_maintext}
    \dot{\varphi}_n(t) = b_n\varphi_{n-1}(t)-b_{n+1}\varphi_{n+1}(t).
\end{equation}
Since in the above-mentioned $\lambda\to 0$ limit the Lanczos coefficients $b_n$ are given by a continuous function loosely denoted by $b(x)$, we can assume that the wave functions $\varphi_n(t)$ are also described by a continuous function $f(t,x)$ such that $\varphi_n(t)=f(t,n\lambda)$. With this, the recursion (\ref{recurrence_varphi_n_maintext}) reads:
\begin{equation}
    \centering
    \label{rec_cont_lambda}
    \partial_t f(t,x) = b(x)f(t,x-\lambda)-b(x+\lambda) f(t,x+\lambda).
\end{equation}
Expanding in powers of $\lambda$ we find:
\begin{equation}
\label{rec_cont_lambda_chiral}
\partial_t f(t,x) = -v(x)\partial_x f(t,x)-\frac{v^\prime(x)}{2}f(t,x)+\mathit{O}(\lambda), 
\end{equation}
which at leading order becomes a first-order wave equation for $f(t,x)$ with velocity field $v(x) \equiv 2\lambda b(x)$. Thanks to the fact that $b(x)$ scales like $\frac{1}{\lambda}$ in \eqref{Lanczos_cont_lambda}, the velocity field $v(x)$ is of order $\lambda^0$:
\begin{equation}
    v(x)=2\lambda b(x) = 2J\sqrt{1-e^{-x}}+\mathit{O}(\lambda)\overset{\lambda\to 0}{\longrightarrow} 2J\sqrt{1-e^{-x}}.
\end{equation}
In order to manipulate \eqref{rec_cont_lambda_chiral} further, we change the position variable to $y$ such that $x=0\Rightarrow y=0$ and $dy = \frac{dx}{v(x)}$. This, together with the wave function redefinition
\begin{equation}
    \centering
    \label{g_f_cont_lambda}
    g(t,y) \equiv \sqrt{v\left(x(y)\right)}f\left( t,x(y) \right), 
\end{equation}
yields the equation
\begin{equation}
    \label{g_chiral_lambda}
    \left(\partial_t+\partial_y\right)g(t,y)=0+\mathit{O}(\lambda).
\end{equation}
Equation (\ref{g_chiral_lambda}) becomes exact for $\lambda=0$. This reasoning is analogous to the analysis in Appendix \ref{Appx:Cont_approx}, where the small parameter is $\varepsilon=\frac{1}{K-1}$ instead of $\lambda$; however, in that case redefinitions like (\ref{g_f_cont_lambda}) and $dy=\frac{dx}{v(x)}$ turn out to be problematic because they imply multiplication or division by zero when $\varepsilon=0$, which does not occur in the present case because $v(x)$ remains finite in the $\lambda\to 0$ limit.

The initial condition $\phi_n(t)=\delta_{n0}$ on the Krylov chain translates, in the continuum limit, into $f(0,x)=\delta(x)$, and therefore $g(0,y)$ is also proportional to $\delta(y)$. We note that, given an initial condition $g(0,y)\equiv g_0(y)$, the solution of \eqref{g_chiral_lambda} simply propagates the initial condition as $g(t,y)=g_0(y-t)$, and hence in our case we have $g(t,y)~\propto~\delta(y-t)$. Consequently, the position expectation value of the wave packet, which gives K-complexity (up to the factor $\lambda$ used to define $x=\lambda n$ in the continuum limit) is simply given by the position of the peak $x_p(t)$, which behaves as a point particle evolving in time following the velocity field $v(x)$:
\begin{equation}
    \centering
    \label{peak_position_cont_lambda}
    t = \int_0^{y_p(t)}dy = \int_0^{x_p(t)}\frac{dx}{v(x)} ~.
\end{equation}
This approximation can therefore be thought of as a classical approximation where the evolution of the wave packet is replaced by the propagation in $x$-space of a point particle. In fact, it is possible to show that the relation $\dot{x}=v(x)=2J\sqrt{1-e^{-x}}$ is a classical solution of the equation of motion generated by a Hamiltonian with an exponential potential:
\begin{equation}
    \centering
    \label{Liouville_Ham_classical}
    H^\prime := E_0 + 2\lambda J \left( \frac{l_f^2k^2}{2} +\frac{2}{(2\lambda)^2} e^{-\frac{l}{l_f}} \right),
\end{equation}
where we have written $x$ as $\frac{l}{l_f}$.
$H^\prime$ is a Liouville Hamiltonian, but it is not quite equal to the effective Hamiltonian $\tilde{T}$ of DSSYK given in \eqref{Ham_l_k} at small $\lambda$. This is a manifestation of the fact that these two Hamiltonians are only classically equivalent. The classical limit\footnote{As discussed in \cite{Lin:2022rbf}, the classical limit is defined as the limit in which $\lambda\to 0$ and $\lambda k$ is held fixed. In this way, $[l,k]=i$ but $[l,\lambda k]=i\lambda\to 0$.} of \eqref{Ham_l_k} is:
\begin{equation}
    \centering
    \label{T_tilde_class}
    \tilde{T}_{\text{class}}=-\frac{2J}{\lambda}\cos(\lambda l_f k) \sqrt{1-e^{-\frac{l}{l_f}}}\qquad+\qquad \text{subleading} ~,
\end{equation}
and one can verify that Hamiltonians \eqref{Liouville_Ham_classical} and \eqref{T_tilde_class} yield the same Euler-Lagrange equation of motion $\ddot{x}=2J^2e^{-x}$, which produces the above-mentioned trajectory $v(x)$ upon choosing the initial conditions $x(0)=0$ and $\dot{x}(0)=0$. However, the two Hamiltonians differ at the quantum level. In order to retrieve a (quantum) Liouville Hamiltonian from DSSYK it is necessary, as described in section \ref{Subsect:Bulk_Hilbert_Space}, to take a triple-scaling limit in which $\lambda$ is taken to be small but $\frac{l}{l_f}=\lambda n$ is taken to be sufficiently large such that $\frac{e^{-l/l_f}}{(2\lambda)^2}=e^{-\tilde{l}/l_f}$ is fixed, in order to remain close to the ground state of the system. We will revisit this in section \ref{Section_Gravity_matching}, where we will compute K-complexity in the regime in which DSSYK is dual to JT gravity.

We now resume the computation of K-complexity in the continuous $\lambda\to 0$ limit of DSSYK. Since formally $x= \lambda n$ and $v(x)=2\lambda b(x)$, we can use \eqref{peak_position_cont_lambda} and give an approximation for K-complexity in DSSYK where we just promote $n$ to be a continuous variable and keep $\lambda$ fixed:

\begin{equation}
    \centering
    \label{n_preak_ballistic}
    \int_0^{n_p(t)}\frac{dn}{v(n)}=t,\qquad v(n)\equiv2b_n=2J\sqrt{\frac{1-q^n}{\lambda(1-q)}}.
\end{equation}
The integral (\ref{n_preak_ballistic}) can be performed analytically, yielding an implicit equation for $n_p(t)$:
\begin{equation}
    \centering
    \label{n_peak_implicit_eqn}
    t J = \sqrt{\frac{1-q}{\lambda}}\text{arctanh} \sqrt{1-q^{n_p(t)}},
\end{equation}
which can be solved for $n_p(t)$. We reach:
\begin{equation}
    \centering
    \label{KC_continuum_ballistic}
    C_K(t) \approx n_p(t) = \frac{2}{\lambda}\log \left\{ \cosh \left[ tJ\sqrt{\frac{\lambda}{1-q}} \right] \right\}.
\end{equation}
This approximation to K-complexity, which results from promoting $n$ (instead of $x=\lambda n$) to a continuous variable, is expected to be good at small $\lambda$, which is when the continuum limit (for the variable $x$) applies. The numerical analysis presented below confirms this expectation.

In section \ref{Section_Gravity_matching} we will recover this functional behavior of K-complexity from a gravity computation, being more specific about the bulk-boundary matching.
Furthermore, expression (\ref{KC_continuum_ballistic}) recovers exactly the early- and late-time regimes identified in (\ref{KC_regimes_summary}):

\begin{itemize}
    \item At early times, using $\cosh(x)=1+\frac{x^2}{2}+\mathit{O}(x^4)$ and $\log(1+x)=x + \mathit{O}(x^2)$, we find
    \begin{equation}
        \centering
        \label{KC_cont_ball_early}
        C_K(t)\approx \frac{2}{\lambda}\log \left\{ 1 + (tJ)^2 \frac{\lambda}{2(1-q)} \right\}\approx \frac{(tJ)^2}{1-q}\,,
    \end{equation}
    which matches perfectly the first line in (\ref{KC_regimes_summary}).
    \item At late times we just make use of the asymptotic behavior of $\cosh{x}\sim \frac{e^x}{2}$, giving
    \begin{equation}
        \centering
        \label{KC_cont_ball_late}
        C_K(t)\approx \frac{2}{\lambda}\log \left\{ \frac{1}{2} \exp \left[ tJ \sqrt{\frac{\lambda}{1-q} }\right] \right\} = \frac{2\,tJ}{\sqrt{\lambda(1-q)}}\,-\frac{2\log{2}}{\lambda},
    \end{equation}
    which also agrees with the second line of (\ref{KC_regimes_summary}) and even provides an estimate for the constant $c(\lambda)$.
\end{itemize}
Finally, the transition time between (\ref{KC_cont_ball_early}) and (\ref{KC_cont_ball_late}) can be estimated as the value of $t$ for which the argument of the $\cosh$ in (\ref{KC_continuum_ballistic}) becomes of order 1. This yields
\begin{equation}
    \centering
    \label{tstar_cont_ball}
    t_*(q) \approx \frac{1}{J}\sqrt{ \frac{1-q}{\lambda}},
\end{equation}
which agrees with our previous estimate (\ref{tstar}), and therefore it also behaves as $t_*\sim J^{-1}$ for small $\lambda$.

The fact that the regimes (\ref{KC_cont_ball_early}) and (\ref{KC_cont_ball_late}) match (\ref{KC_regimes_summary}) is a non-trivial check, since the analysis leading to (\ref{KC_regimes_summary}) did not assume the continuum approximation: it was achieved by analysing separately the different sectors of the Lanczos sequence and using known solutions for the discrete recurrence relation for such cases. We elaborate on this below:

\begin{itemize}
    \item At early times $t<t_*(q)$, both the continuum approximation and the exact discrete result derived using the Heisenberg-Weyl algebra yield $C_K(t)\approx \frac{(tJ)^2}{1-q}$. We may intuitively understand this agreement by noting that the continuum approximation can be seen as the classical propagation of a point particle in Krylov space, while the discrete result consists of the propagation of coherent states, which are known to behave semiclassically. 
    \item At late times $t\gg t_*(q)$ the exact, discrete solution for $\phi_n(t)$ consists of Bessel functions, see Appendix \ref{App:WFq0}, whose front-most peak propagates in $n$-space at a constant velocity equal to $2b_\infty = \frac{2J}{\sqrt{\lambda(1-q)}}$. Approximating K-complexity by the position of this peak yields the estimate $C_K(t)\approx \frac{2\,tJ}{\sqrt{\lambda(1-q)}}$, which agrees with the point-particle propagation from the continuum approximation at late times. However, such ballistic approximation does not always accurately apply to the discrete, exact solution, which features a wave packet that develops a tail as it propagates through Krylov space. 
\end{itemize}

\begin{figure}[t]
    \centering
    \includegraphics[width=0.45\textwidth]{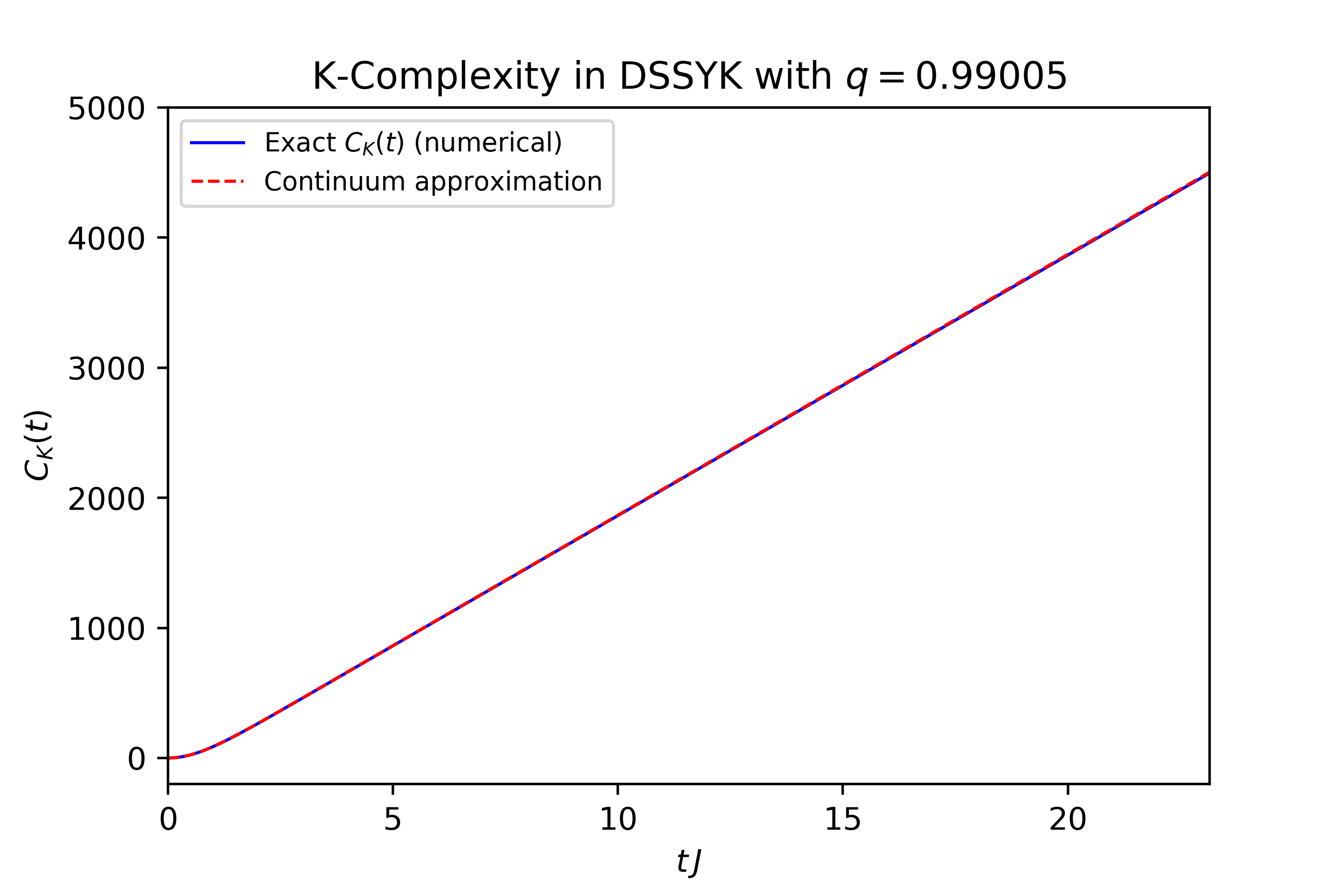} \includegraphics[width=0.45\textwidth]{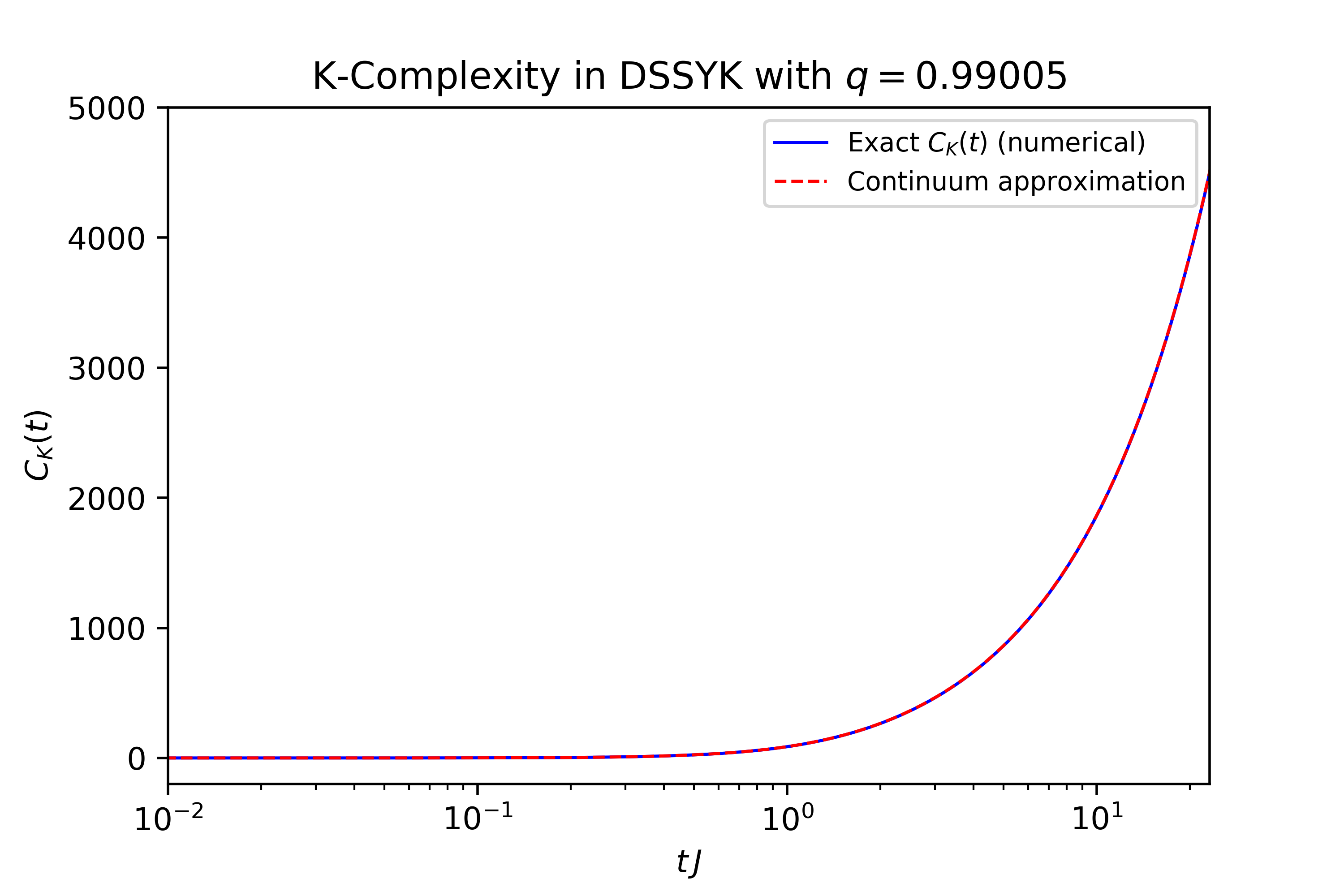} \\
    \includegraphics[width=0.45\textwidth]{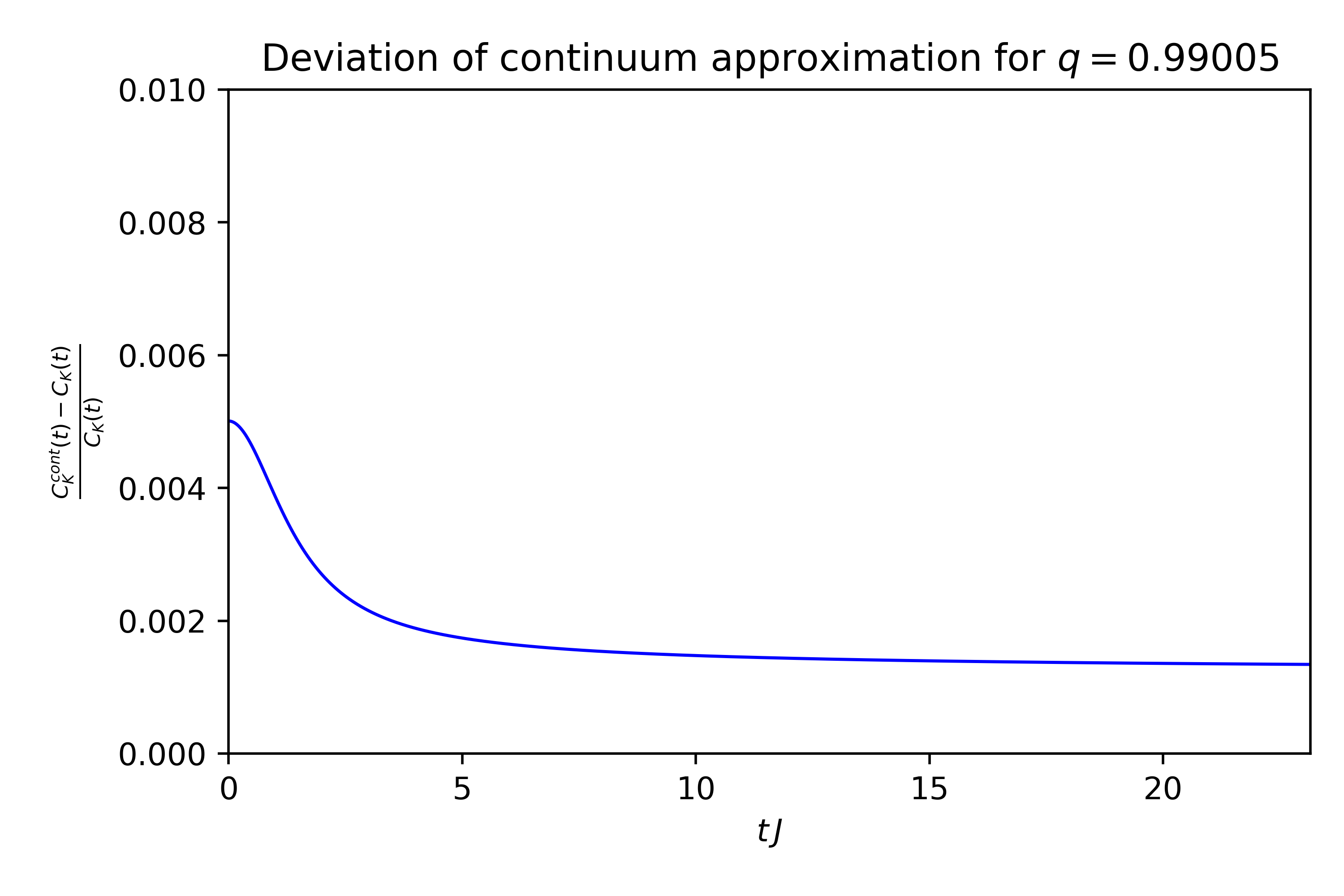}
    \caption{Comparison between the exact $C_K(t)$ computed numerically and the continuum approximation (\ref{KC_continuum_ballistic}) for DSSYK with $\lambda=0.01$ ($q\approx0.99005$), i.e. the same value as in Figure \ref{fig:KC_numerics}. The agreement is excellent. \textbf{Top left:} Linear scale along both axes. \textbf{Top right:} Logarithmic scale along the horizontal axis. \textbf{Bottom:} Normalized deviation as a function of time.}
    \label{fig:KC_exact_vs_cont_approx_q0pt9905}
\end{figure}

\begin{figure}[t]
    \centering
    \includegraphics[width=0.45\textwidth]{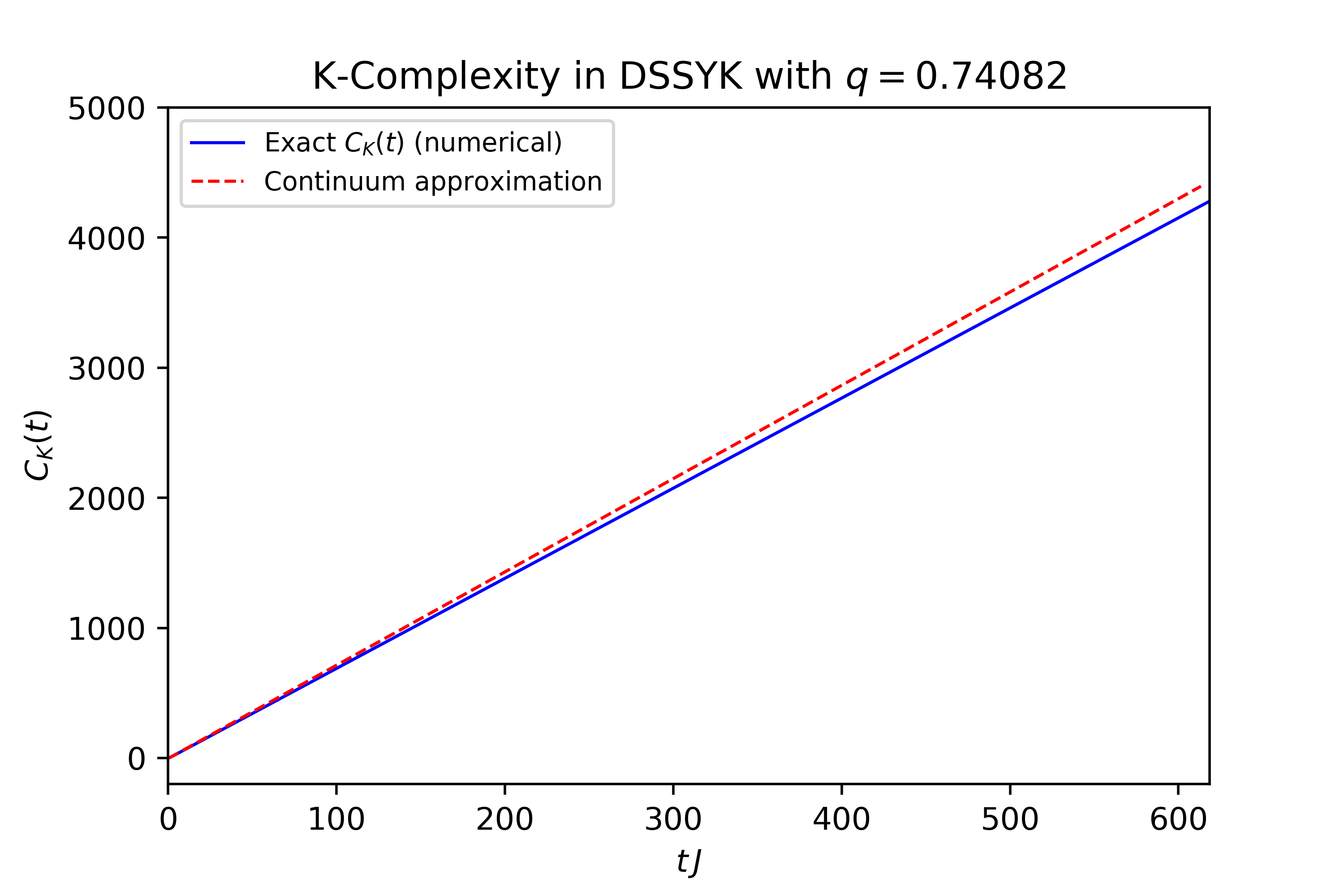} \includegraphics[width=0.45\textwidth]{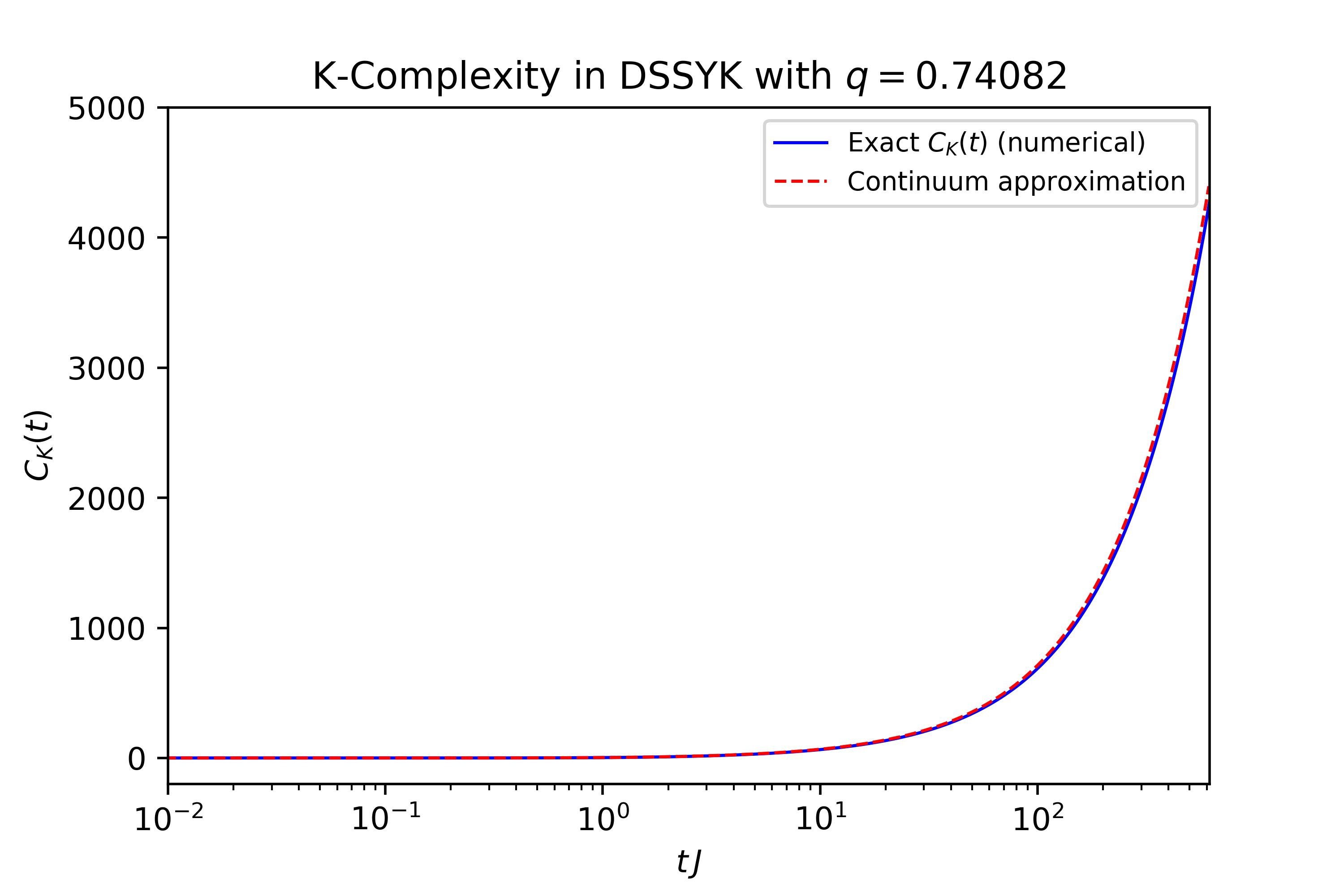} \\
    \includegraphics[width=0.45\textwidth]{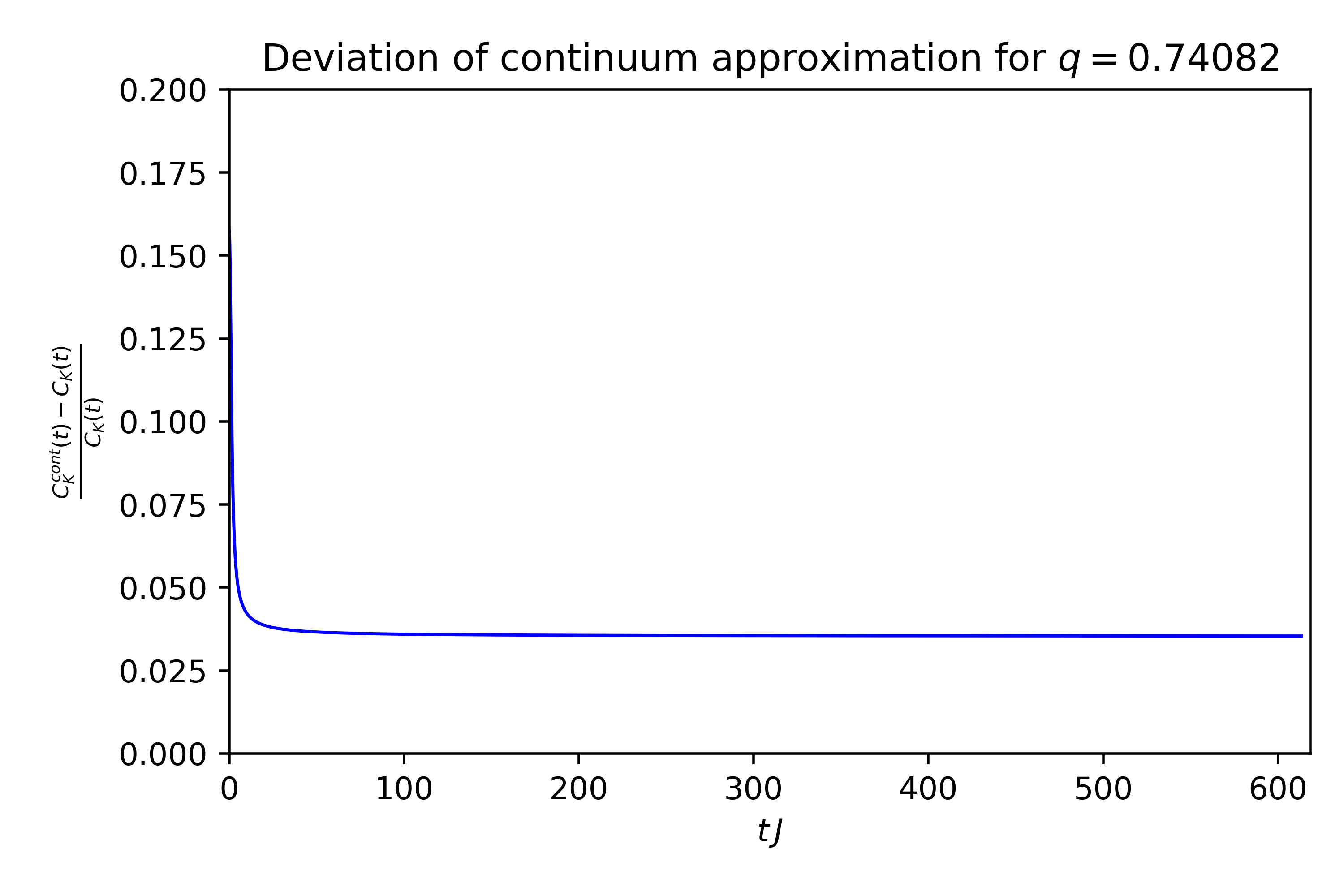}
    \caption{Continuum approximation vs exact (numerical) $C_K(t)$ for $\lambda=0.3$ (i.e. $q\approx 0.74082$).}
    \label{fig:KC_exact_vs_cont_approx_q0pt74082}
\end{figure}

\begin{figure}[t]
    \centering
    \includegraphics[width=0.45\textwidth]{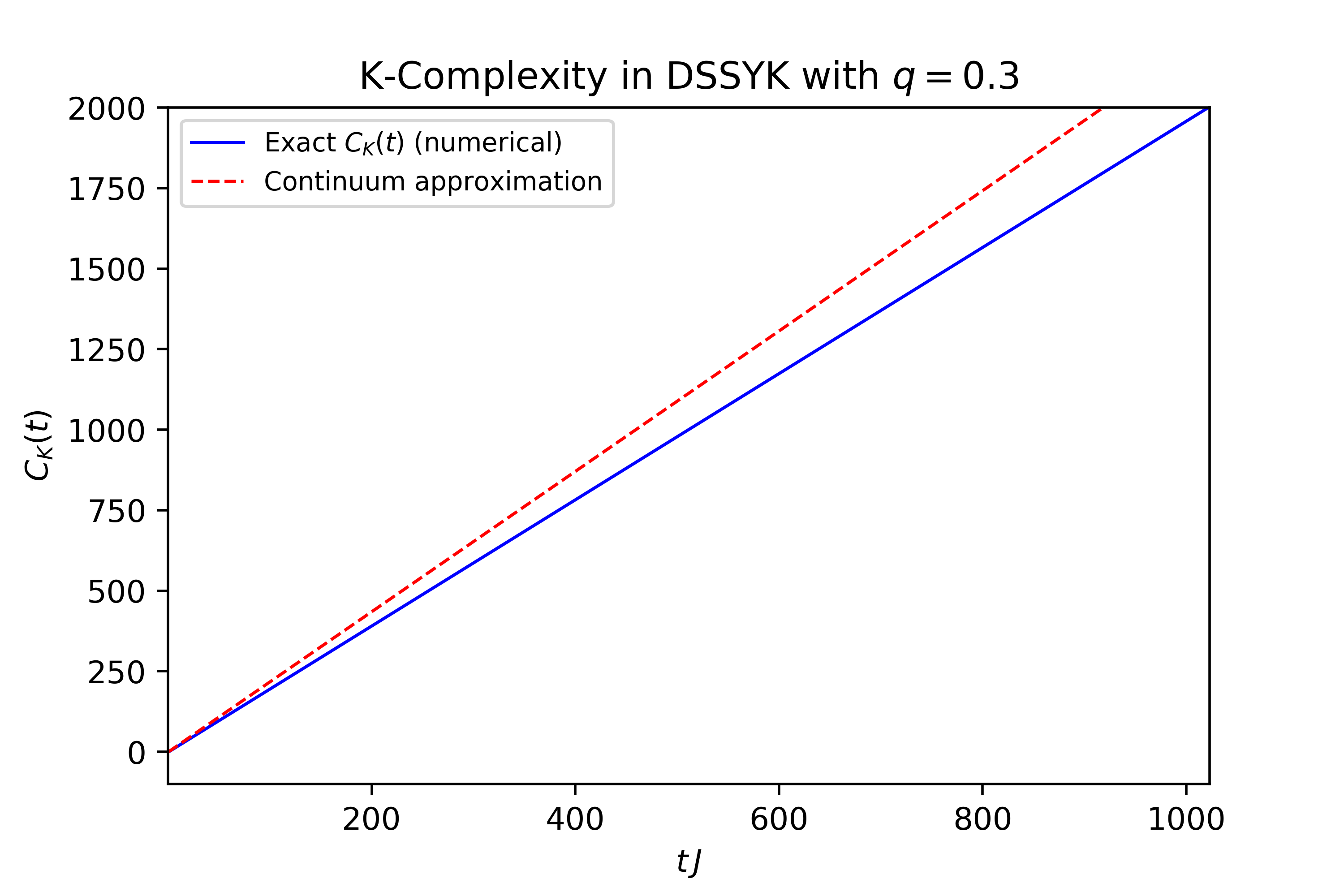} \includegraphics[width=0.45\textwidth]{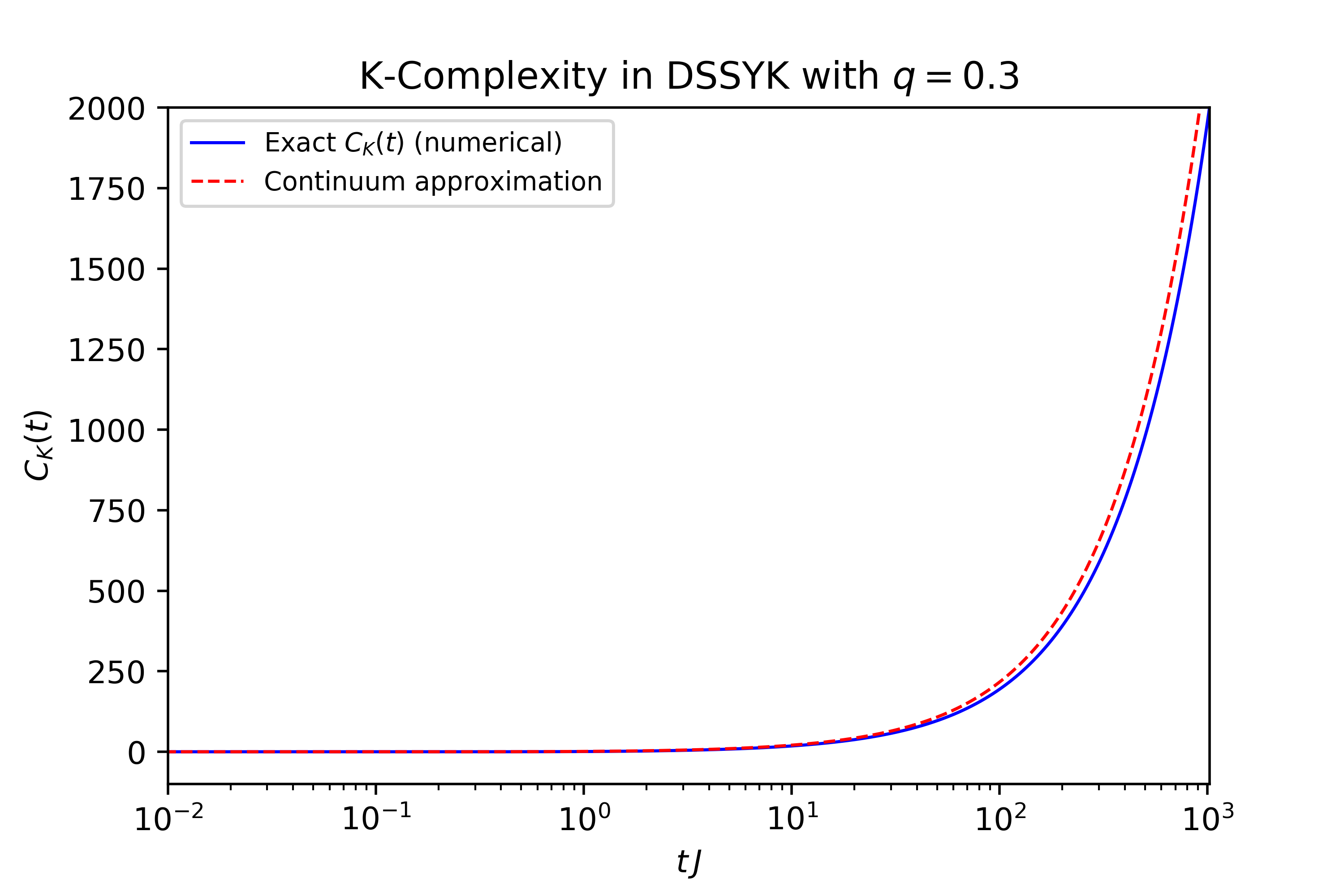} \\
    \includegraphics[width=0.45\textwidth]{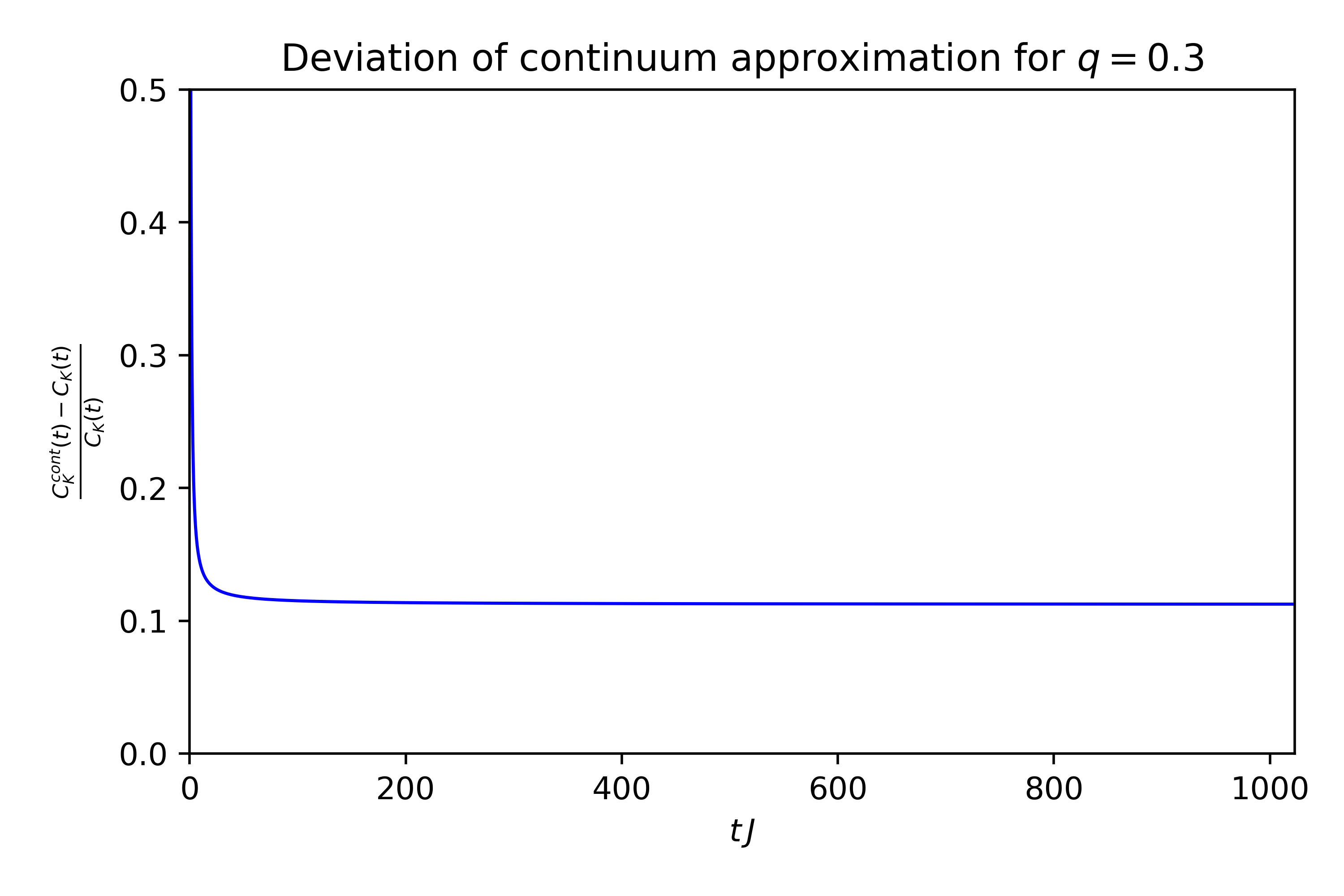}
    \caption{Continuum approximation vs exact (numerical) $C_K(t)$ for $\lambda\approx 1.20397$ (i.e. $q=0.3$).}
    \label{fig:KC_exact_vs_cont_approx_q0pt3}
\end{figure}

\begin{figure}[t]
    \centering
    \includegraphics[width=0.45\textwidth]{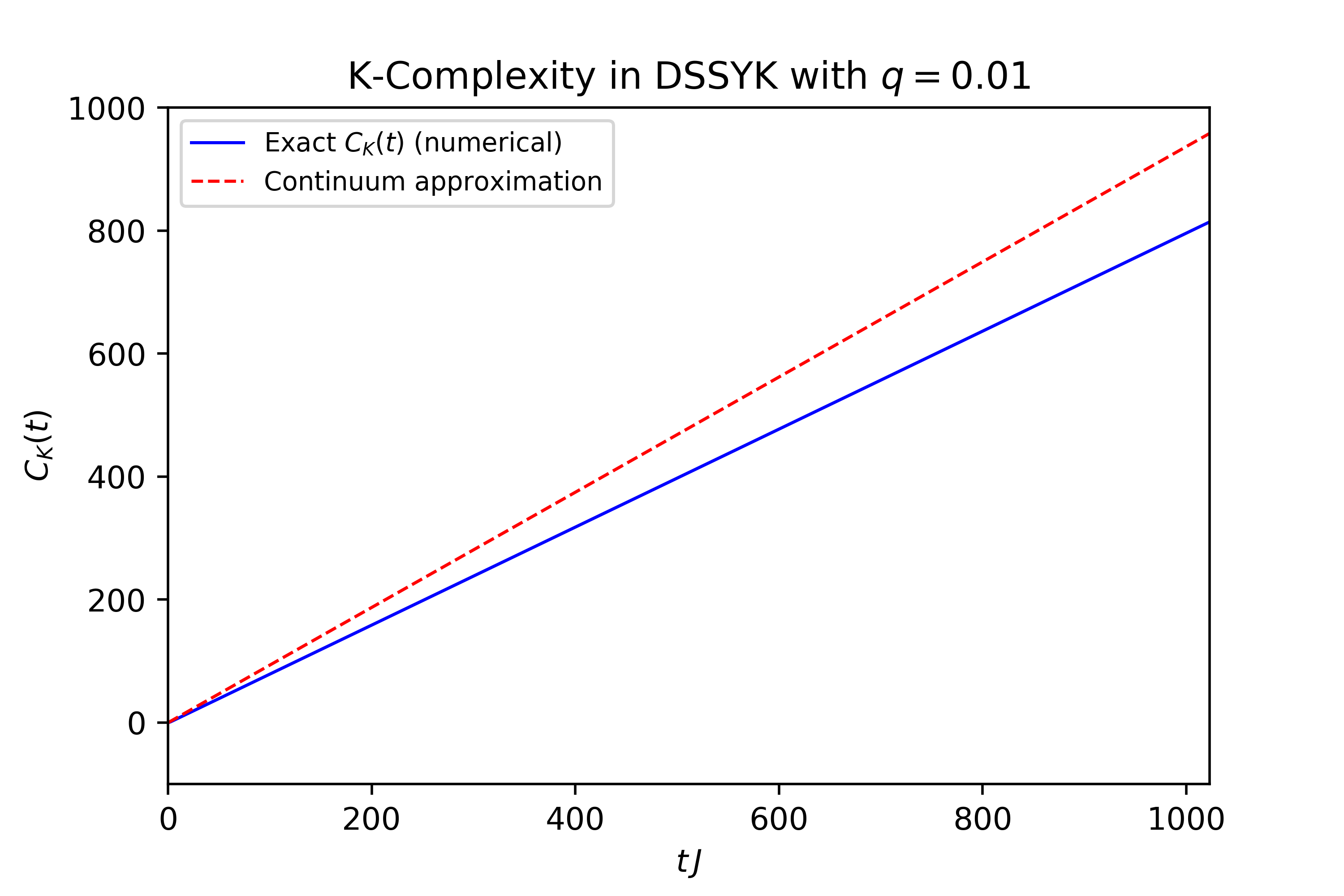} \includegraphics[width=0.45\textwidth]{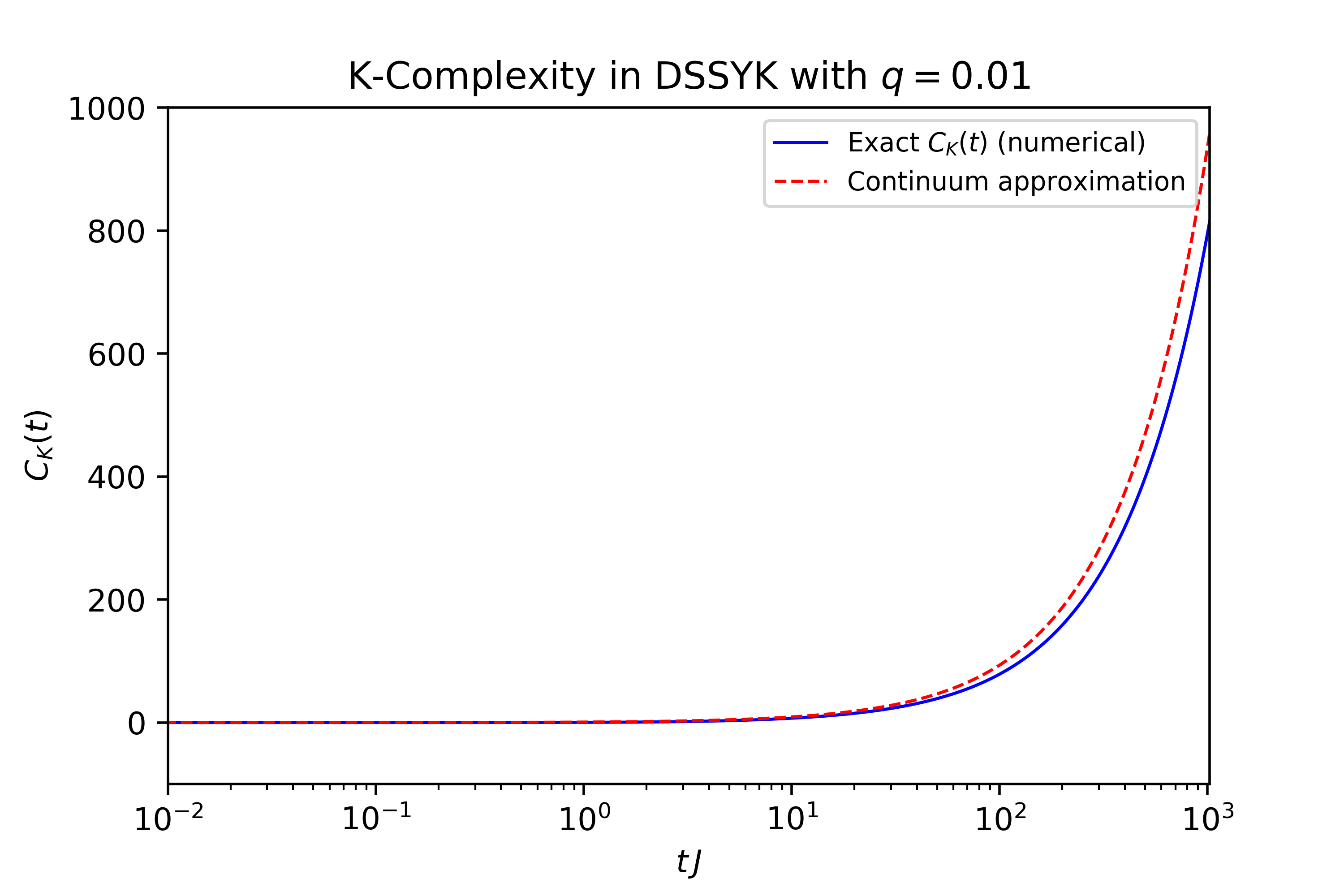}\\
    \includegraphics[width=0.45\textwidth]{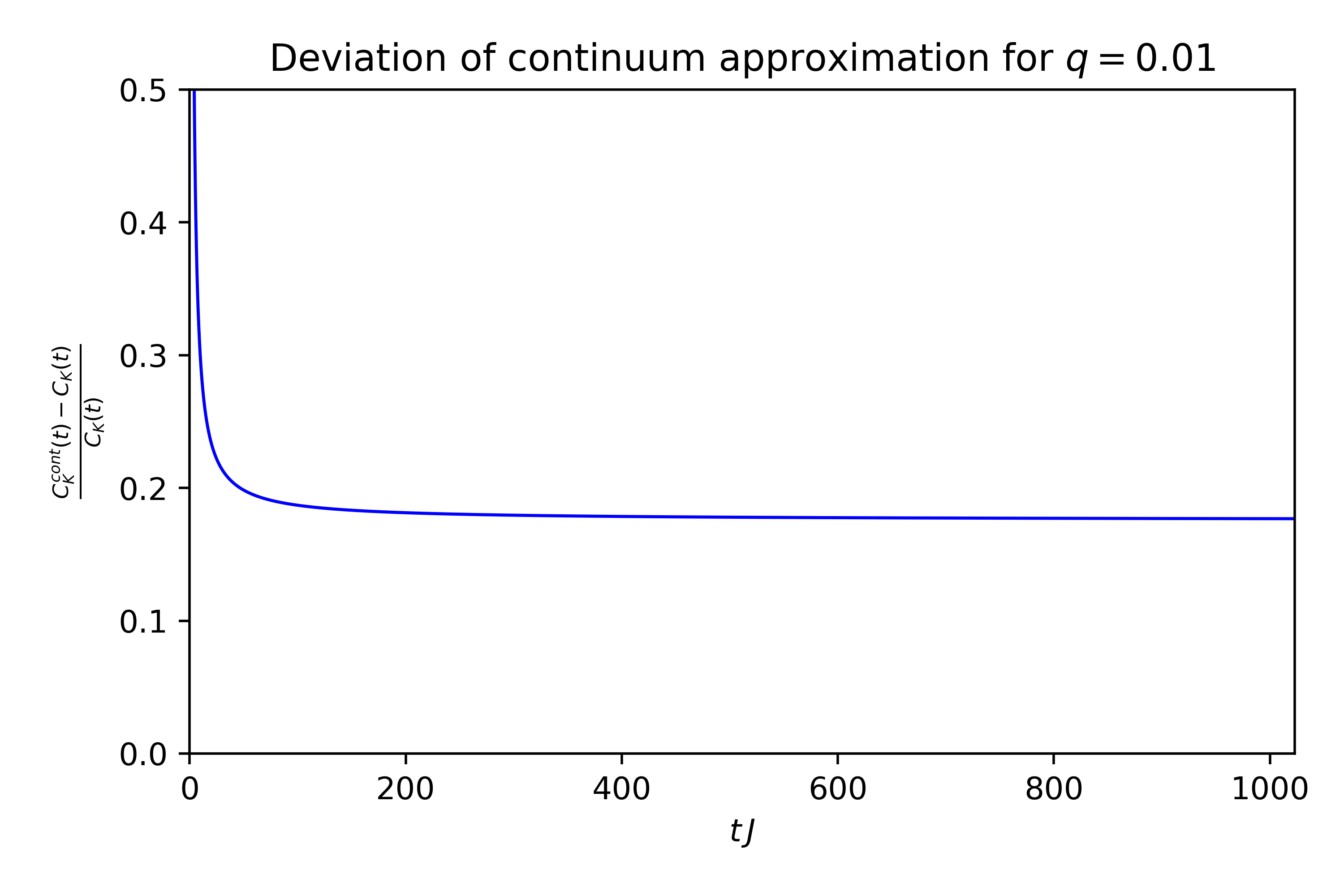}
    \caption{\textbf{Left:} Exact $C_K(t)$ computed numerically for $\lambda\approx 4.61$ (i.e. $q=0.01$), when the Lanczos sequence (\ref{Lanczos}) becomes very close to a constant $b_n\approx\frac{J}{\sqrt{\lambda}}$, vs the result of the continuum approximation. \textbf{Right:} Logarithmic scale along the horizontal axis. \textbf{Bottom:} Relative deviation. Its long-time average is $\sim 0.177$, very close to the value expected for the relative deviation of the ballistic estimate with respect to the exact result in the discrete analysis, expected to be $\sim 0.18$.}
    \label{fig:KC_exact_vs_cont_approx_q0}
\end{figure}

Numerics show that the continuum (and therefore ballistic) approximation is better the closer $q$ is to $1$, as illustrated in Figures \ref{fig:KC_exact_vs_cont_approx_q0pt9905}, \ref{fig:KC_exact_vs_cont_approx_q0pt74082} and \ref{fig:KC_exact_vs_cont_approx_q0pt3}: we studied the values $q=0.99005,\,0.74082,\,0.3$, for which the continuum approximation induces errors of around $0.1\%$, $3\%$ and $10\%$, respectively. In fact, the exact solution for the discrete problem with constant Lanczos coefficients $b_n=b$ yields $C_K(t)=\frac{16}{3\pi}bt+o(t)\approx1.7\,bt+o(t)$, as shown in (\ref{KC_linear}). An approximation estimating $C_K(t)$ from the position of the front-most peak of the wave packet would give, even in the discrete case, $C_K(t)\sim 2bt$, resulting in a relative deviation (at late times) of $\frac{2-\frac{16}{3\pi}}{\frac{16}{3\pi}}\approx 0.18$. In Figure \ref{fig:KC_exact_vs_cont_approx_q0} we observe that the relative deviation between the continuum approximation and the numerical result for K-complexity at $q=0.01$ tends precisely to this value, confirming that the error is (mainly) due to the ballistic approximation, which misses the development of a tail behind the wavefront that the exact discrete solution features. In all numerical computations we artificially truncated the Krylov chain, giving it a finite dimension $K=5000$: time-dependent results are therefore only reliable during the time interval in which the wave-packet does not yet probe this artificial edge of Krylov space, and this is the time range to which the above-mentioned plots are restricted.

For reference, we can rewrite (\ref{KC_continuum_ballistic}) for small $\lambda$ (i.e. $q$ close to $1$):
\begin{equation}
    \centering
    \label{KC_continuum_ballistic_small_lambda}
    C_K(t)\approx \frac{2}{\lambda}\log\left\{ \cosh \left(tJ\right) \right\},
\end{equation}
which in turn recovers the early- and late-time regimes listed in (\ref{KC_regimes_summary_small_Lambda}). We emphasize that this result is valid strictly in double-scaled SYK (and with small $\lambda$). In evaluating this result we have not yet taken the triple-scaling limit (\ref{Trple-scaling-limit}), for which taking $\lambda$ to be small is not enough. The derivation of K-complexity in the triple-scaling limit is deferred to section \ref{Section_Gravity_matching}.

Understood as a classical description, the results of this continuum approximation of Krylov space can be useful for the gravity interpretation: \cite{Lin:2022rbf} discusses that, since chord number (which we now understand as K-complexity) is a discrete quantity, it provides a discretized, or quantized, version of the bulk length. At this point, we propose to make the converse argument and claim that this continuum approximation of Krylov space should match gravity results obtained from a classical, geometric description. We will provide further evidence in favor of this point of view in section \ref{Section_Gravity_matching}. A further argument in favor of interpreting the continuum approximation (\ref{KC_continuum_ballistic_small_lambda}) as a classical description of Krylov space dynamics consists in noting, as we have explained, that it coincides with the classical trajectory generated by the Hamiltonian \eqref{Liouville_Ham_classical}.

\subsection{Exact formal results for K-complexity in DSSYK}\label{subsect:formal_KC}
In this section we will present the formal result for Krylov complexity in DSSYK as a function of $q$, following the prescription for the wavefunctions in (\ref{Phi_EnergyB}).  We will use the results for the eigenvectors and eigenvalues of the effective Hamiltonian $T$ (shown at the end of section \ref{Subsection_Background_DSSYK}) directly in (\ref{Phi_EnergyB}). The sum over energies will be replaced by an integral over $\theta$, as discussed in \cite{Berkooz:2018jqr, Berkooz:2018qkz} and reviewed after equation (\ref{eigvals1}) in Appendix \ref{Appx:EigSysDSSYK}. Thus:
\begin{align} 
    \phi_n(t) &= \int_0^\pi d\theta \, e^{-itE(\cos \theta)} \, \psi_n(\cos \theta)\, \psi_0(\cos \theta)\\
    & = \int_0^\pi \frac{d\theta}{2\pi} \, e^{-2iJt\frac{\cos\theta}{\sqrt{\lambda(1-q)}}} \, \frac{(q;q)_\infty}{\sqrt{(q;q)_n}}\, |(e^{2i\theta};q)_\infty|^2 \, H_n(\cos\theta|q). \label{phin_closedform}
\end{align}
In the second line we used the expressions (\ref{Teigvals}), (\ref{T0eigenvector}) and (\ref{Teigenvectors}) for the eigenvalues and eigenvectors of $T$. The small $\lambda$ limit, $q\to 1$, of this expression is discussed in section \ref{Sec:qto1Limit}.
Exact results for the wavefunctions for $q=0$ and for $q=1$ are given in Appendix \ref{App:Wavefunctions}.  

It can be checked that the wavefunction is normalized, i.e. $\sum_{n=0}^\infty |\phi_n(t)|^2=1$, as follows:
\begin{align*}
    \sum_{n=0}^\infty |\phi_n(t)|^2 &= [(q;q)_\infty]^2 \int_0^\pi \frac{d\theta}{2\pi} \int_0^\pi \frac{d\phi}{2\pi} e^{-\frac{2iJt}{\sqrt{\lambda(1-q)}}(\cos\theta-\cos\phi)}|(e^{2i\theta};q)_\infty|^2 |(e^{2i\phi};q)_\infty|^2 \nonumber\\ 
    &\times \sum_{n=0}^\infty \frac{1}{(q;q)_n} 
      H_n(\cos \theta|q)H_n(\cos \phi|q) =\frac{(q;q)_\infty}{2\pi} \int_0^\pi d\theta \, |(e^{2i\theta};q)_\infty|^2 = 1
\end{align*}
where (\ref{qH_orthogonality}) was used in the second equality and (\ref{exp_qH_identity}) with $n=m=0$ was used in the final equality.

We are now ready to write down a formal expression for K-complexity in double-scaled SYK as a function of $q=e^{-\lambda}$:
\begin{align}
    C_K(t) =\sum_{n=0}^\infty n |\phi_n(t)|^2 =& [(q;q)_\infty]^2 \int_0^\pi \frac{d\theta}{2\pi} \int_0^\pi \frac{d\phi}{2\pi} e^{-\frac{2iJt}{\sqrt{\lambda(1-q)}}(\cos\theta-\cos\phi)}|(e^{2i\theta};q)_\infty|^2 |(e^{2i\phi};q)_\infty|^2 \nonumber\\ 
    &\times \sum_{n=0}^\infty \frac{n}{(q;q)_n} 
      H_n(\cos \theta|q)H_n(\cos \phi|q) ~.
\end{align}

\subsubsection{The $q\to 1$ limit} \label{Sec:qto1Limit}

In the $q\to 1$ limit we can reproduce the result (\ref{Heisenberg_Weyl_Wave_Fn}) by carefully taking the $\lambda\to 0$ limit in the general expression for the wavefunction (\ref{phin_closedform}). We begin by finding expressions \cite{weisstein} for $(x;q)_\infty$ and $(q;q)_\infty$ for $q=e^{-\lambda}$ in the limit $\lambda \to 0$, as well as for $H(\cos \theta|q)$ and $(q;q)_n$:
\begin{align}
    (x;q)_\infty &\approx \exp\Big[ -\frac{1}{\lambda} \mathrm{Li}_2(x)+\frac{1}{2}\log(1-x)+O(\lambda) \Big]\\
    (q;q)_\infty &\approx \sqrt{\frac{2\pi}{\lambda}} \exp \Big[ -\frac{\pi^2}{6\lambda}+\frac{\lambda}{24} +O(\lambda^2) \Big] \\
    (q;q)_n &\approx n! \lambda^n \\
    H(\cos \theta; q) &\approx 2^n \cos^n(\theta) ~.
\end{align}
From the first identity above we find that  for $q=e^{-\lambda}$ in the limit $\lambda \to 0$,
\begin{align}
    |(e^{2i\theta};q)_\infty|^2 &=(e^{2i\theta};q)_\infty(e^{-2i\theta};q)_\infty \approx  \sqrt{1-e^{2 i \theta}}\sqrt{1-e^{-2 i \theta}}\; e^ {-\frac{1}{\lambda}[\mathrm{Li}_2(e^{2i\theta})+\mathrm{Li}_2(e^{-2i\theta})]} \nonumber\\
    &=2|\sin \theta| \; e^ {-\frac{1}{\lambda}[\mathrm{Li}_2(e^{2i\theta})+\mathrm{Li}_2(e^{-2i\theta})]}~.
\end{align}

Plugging these expansions into (\ref{phin_closedform}) while keeping only first order terms in $\lambda$ and expanding $\sqrt{1-q}\approx \sqrt{\lambda}$, we have:
\begin{align}
    \phi_n(t) & \approx  2\sqrt{\frac{2\pi}{\lambda}} \frac{2^n}{\sqrt{n! \lambda^n}} \int_0^\pi \frac{d\theta}{2\pi} e^{-\frac{1}{\lambda}\left[\frac{\pi^2}{6} + \mathrm{Li}_2(e^{2i\theta})+\mathrm{Li}_2(e^{-2i\theta}) +2itJ\cos \theta \right] } |\sin \theta| \cos^n(\theta).
\end{align}
To proceed we will use the identity\footnote{See for example \url{https://dlmf.nist.gov/25.12} (last consulted on 01/01/2024).}
\begin{equation}
    \mathrm{Li}_2(e^{2i\theta})+\mathrm{Li}_2(e^{-2i\theta}) = \frac{\pi^2}{3}-2\pi \theta +2\theta^2 
\end{equation}
which gives us $-\frac{1}{\lambda}[\pi^2/2 -2\pi \theta +2\theta^2] =-\frac{2}{\lambda}(\theta-\pi/2)^2$ in the exponential term. We change variables to $x = \theta -\pi/2$ to get
\begin{align}
    \phi_n(t) \approx 2\sqrt{\frac{1}{2\pi\lambda}} \frac{2^n}{\sqrt{n! \lambda^n}}\int_{-\pi/2}^{\pi/2} dx \,  e^{-\frac{2}{\lambda}(x^2-itJ\sin x)} |\cos x| (-\sin x)^n ~.
\end{align}

For $\lambda \to 0$, and $tJ\ll 1$, the main contribution to this integral is for $|x|\ll 1$, hence we expand $\sin(x) \approx x$ and $ \cos x \approx 1$:
\begin{align}
    \phi_n(t) \approx \sqrt{\frac{2}{\pi\lambda}} \frac{2^n}{\sqrt{n! \lambda^n}}\int_{-\pi/2}^{\pi/2} dx \, e^{-\frac{2}{\lambda}(x^2-2itJ x)} (-x)^n ~.
\end{align}
Completing the square, we find another Gaussian integral:
\begin{align}
    \phi_n(t) \approx \sqrt{\frac{2}{\pi\lambda}} \frac{(-2)^n}{\sqrt{n! \lambda^n}} e^{-\frac{(tJ)^2}{2\lambda}}\int_{-\pi/2}^{\pi/2} dx \, e^{-\frac{2}{\lambda}\left(x-\frac{itJ}{2}\right)^2} x^n ~.
\end{align}
Using a saddle-point approximation for $\lambda \to 0$, we set $x\approx \frac{itJ}{2}$ and perform the Gaussian integral:
\begin{equation}
     \phi_n(t) \approx \sqrt{\frac{2}{\pi\lambda}} \frac{(-2)^n}{\sqrt{n! \lambda^n}} e^{-\frac{(tJ)^2}{2\lambda}} \sqrt{\frac{\pi \lambda}{2}}\left( \frac{itJ}{2}\right)^n = e^{-\frac{(tJ)^2}{2\lambda}} \frac{(-itJ/\sqrt{\lambda})^n}{\sqrt{n!}} ~,
\end{equation}
arriving at the result (\ref{Heisenberg_Weyl_Wave_Fn}) which is indeed expected for $\lambda \to 0 $ and $tJ\ll 1$. Interestingly, this analysis agrees with our previous estimate (\ref{tstar_cont_ball}) and tells us that in the limit $q\to 1$ the transition time remains finite, $\lim_{q \to 1^{-}} t_*(q)=J^{-1}$, and hence the Heisenberg-Weyl result still only applies at early times.

\section{Bulk wormhole length and Krylov complexity} \label{Section_Gravity_matching}
We have now assembled all the ingredients necessary to carry out the main task we set ourselves in this work: giving a precise match of bulk and boundary Krylov complexity in the context of DSSYK. Let us recall the two main building blocks for the argument:

\begin{itemize}
    \item In section \ref{Subsect:Bulk_Hilbert_Space} we reviewed the identification between the bulk JT Hilbert space and the boundary Hilbert space of double-scaled SYK, whose dynamics in the triple-scaling limit are generated by the Liouville Hamiltonian. From this identification one concludes that fixed chord-number states are bulk length eigenstates.
    \item In section \ref{Sect:KC} we showed that fixed chord-number states are equal to the Krylov basis elements associated to the zero chord-number state (or the infinite-temperature thermofield ``double'' state). 
\end{itemize}

Putting these two items together it follows that \textit{the Krylov basis elements are bulk length eigenstates. Therefore, the position expectation value on the Krylov chain (i.e., K-complexity, as expressed in (\ref{KC_definition}) via the position operator (\ref{nOperator})) gives the length expectation value. This establishes the correspondence between K-complexity and two-sided bulk length}.

\begin{figure}
    \centering
    \includegraphics[scale=0.6]{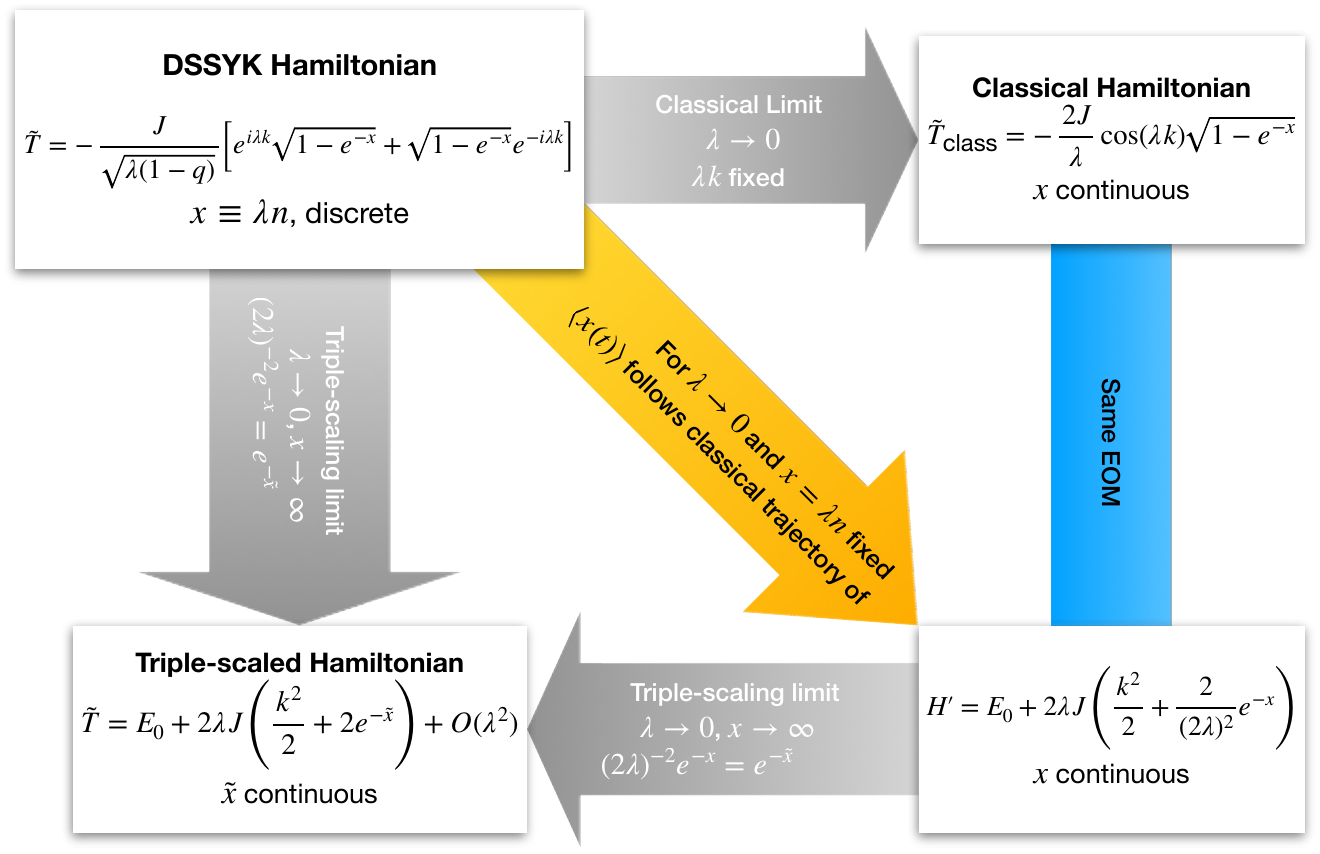}
    \caption{Diagram relating the various limiting forms of the Hamiltonian studied in this work. The \textbf{top left} expression gives the effective Hamiltonian for the averaged theory of DSSYK, written in \eqref{Ham_l_k}. Its classical limit is the \textbf{top right} equation, stated in \eqref{T_tilde_class}. As explained in section \ref{subsect_continuum_approx}, this Hamiltonian yields the same Euler-Lagrange equation of motion as the Liouville Hamiltonian depicted in the \textbf{bottom right} block and presented in \eqref{Liouville_Ham_classical}. As indicated by the yellow arrow, this Hamiltonian is relevant because the continuous analysis of the DSSYK Hamiltonian presented in \ref{subsect_continuum_approx} yields a position expectation value that follows a classical trajectory generated by this bottom-right Liouville Hamiltonian. Finally, the \textbf{bottom left} box contains the triple-scaled Hamiltonian \eqref{Triple_scaled_Hamiltonian}: a Liouville Hamiltonian to which the full DSSYK Hamiltonian reduces in the triple-scaling limit. This limit has the effect of zooming in near the ground state, and therefore this is an effective low-energy Hamiltonian. This is the regime where DSSYK is dual to JT gravity. More technically, the introduction of the regularized position variable is required in order to obtain potential and kinetic terms of the same order in a $\lambda$-expansion, as outlined by the horizontal arrow at the bottom of the diagram. For simplicity and despite the notational abuse, in this Figure $k$ denotes the (dimensionless) conjugate momentum of $x$.}
    \label{fig:Hamiltonians}
\end{figure}

In order to be more specific, 
we need to go to the regime of DSSYK in which its Hamiltonian becomes that of JT gravity. We argued in section \ref{subsect_continuum_approx} that only taking the continuous $\lambda\to 0$ limit is not sufficient for this, because in that case the DSSYK Hamiltonian is only classically equivalent to a Liouville Hamiltonian.
As reviewed in section \ref{Subsect:Bulk_Hilbert_Space}, in order to find a quantum Liouville Hamiltonian as some limiting form of the full DSSYK Hamiltonian, it is necessary to perform a triple-scaling limit \eqref{Trple-scaling-limit} that effectively zooms in near the ground state of the spectrum. In terms of the regularized length $\tilde{l}$ defined in this limit, the resulting Hamiltonian is
\eqref{Triple_scaled_Hamiltonian}, which we restate here:
\begin{equation}
    \centering
    \label{Triple_scaled_Hamiltonian_restatement_gravity_discussion}
    \tilde{T} = E_0 + 2\lambda J \left( \frac{l_f^2 k^2}{2} + 2e^{-\frac{\tilde{l}}{l_f}} \right)\;+\mathit{O}\left(\lambda^2\right).
\end{equation}
Incidentally, we note that \eqref{Triple_scaled_Hamiltonian_restatement_gravity_discussion} can be reached as the result of applying the triple-scaling limit to (\ref{Liouville_Ham_classical}). In this procedure, the regularized length introduced through the triple-scaling limit becomes necessary in order to get a well-behaved limiting Hamiltonian in which kinetic and potential terms appear at the same order. For the sake of clarity, Figure~\ref{fig:Hamiltonians} depicts a diagram relating the various limiting forms of the Hamiltonian discussed in this work.

In short, in this section we will perform a K-complexity calculation analogous to that in section \ref{subsect_continuum_approx}, specialized to the regime of DSSYK in which it becomes dual to JT gravity. For this, let us first argue that it is still possible to relate the evolution of \textit{regularized} length to K-complexity in the triple-scaled system.
Eigenstates of $\tilde{l}$ can be regarded as the continuum version of position eigenstates on a lattice with labels $\tilde{n}$ such that $\frac{\tilde{l}}{l_f}=\lambda\tilde{n}$. We note that $\tilde{n}$ differs from $n$ in a $\lambda$-dependent shift, necessary to zoom in near the ground state of the original $n$-lattice. Via an argument analogous to that in section \ref{Subsubsect_Krylov_TFD} relating the infinite-temperature thermofield ``double'' state of the full Hamiltonian to the $|n=0\rangle$ state in the averaged theory, we can associate the infinite-temperature thermofield ``double'' state of the low-energy Hamiltonian to the $|\tilde{n}=0\rangle$ state in the averaged theory.
Finally, from the fact that $\frac{\widehat{\tilde{l}}}{l_f}=\lambda \widehat{\tilde{n}}$ holds as an operator identity, taking expectation value on both sides yields the relation $\frac{\tilde{l}(t)}{l_f}=\lambda \widetilde{C_K}(t)$, between bulk regularized length in JT and the K-complexity $\widetilde{C_K}(t)$ of the infinite-temperature thermofield double state of triple-scaled SYK in the averaged theory. 

We can elaborate further on K-complexity from the point of view of the $\tilde{n}$-lattice. Its Lanczos coefficients are obtained by performing the triple-scaling protocol on (\ref{Lanczos}):
\begin{equation}
    \centering
    \label{Triple-scaled-Lanczos}
    b_{\tilde{n}}=\frac{J}{\lambda}\sqrt{1-(2\lambda)^2 q^{\tilde{n}}}+\mathit{O}(\lambda^0)=b -2\lambda J q^{\tilde{n}}+\mathit{O}(\lambda^2),
\end{equation}
where the constant $b = \frac{J}{\lambda}+\mathit{O}(\lambda^0)$ is related to the ground-state energy. We note that $q^{\tilde{n}}=e^{-\lambda \tilde{n}}$ is fixed in the triple-scaling limit in such a way that $\lambda \tilde{n}\equiv \frac{\tilde{l}}{l_f}$ does not scale with $\lambda$. In this limit, the variable $\tilde{l}$ becomes continuous; however, the analysis of the recurrence equation in Krylov space in this continuum limit is slightly different from the case in section~\ref{subsect_continuum_approx}: here, one does not reach the conclusion that $2\lambda b(\tilde{x})$ (where $\tilde{x}\equiv\frac{\tilde{l}}{l_f}$) plays the role of a velocity, because the orders of the $\lambda$-expansion of $b(\tilde{x})$ get mixed with those coming from the displacement operator. Instead, it can be checked that the differential equation that this continuous analysis yields is nothing but the Schrödinger equation dictated by the triple-scaled Hamiltonian (\ref{Triple_scaled_Hamiltonian_restatement_gravity_discussion}), consistently. Furthermore, the small-$\lambda$ prefactor of this Hamiltonian allows for a classical approximation in which the expectation value of $\tilde{l}$ is given by the solution to the equation of motion dictated by the Liouville Hamiltonian \eqref{Triple_scaled_Hamiltonian_restatement_gravity_discussion} derived on the boundary. We perform such a classical evaluation using the initial conditions $\frac{\widetilde{l}(0)}{l_{AdS}}=\widetilde{x}_0$, $\dot{\widetilde{l}}(0)=0$ and, setting $l_f=l_{AdS}$ as announced in \eqref{Param_identification_Hamiltonians}, we reach:
\begin{equation}
    \centering
    \label{Length-KC-tilde}
    \lambda \widetilde{C_K}(t)=\frac{\widetilde{l}(t)}{l_{AdS}}=\widetilde{x}_0+2\log\left\{ \cosh \left( 2\lambda J e^{-\widetilde{x}_0/2}\, t \right) \right\},
\end{equation}
which is a solution with energy\footnote{For the sake of notational simplicity, we define the energy $E$ as the difference between the total energy and the ground-state energy, i.e. $E\equiv \langle \tilde{T} \rangle-E_0$, where $\tilde{T}$ is given in \eqref{Triple_scaled_Hamiltonian_restatement_gravity_discussion}.} 
\begin{equation}
    \centering
    \label{Energy_solution_init_cond}
    E=4\lambda J e^{-\tilde{x}_0}.
\end{equation}
For fixed $\widetilde{x}_0$, this is consistent with the fact that the Hamiltonian \eqref{Triple_scaled_Hamiltonian_restatement_gravity_discussion} describes configurations that constitute excitations of energies of order $\lambda$, i.e. close to the ground state. This allows us to rewrite \eqref{Length-KC-tilde} as:
\begin{equation}
    \centering
    \label{Length-KC-tilde_E}
    \lambda \widetilde{C_K}(t)=\frac{\widetilde{l}(t)}{l_{AdS}}=2\log\left\{ \cosh \left( t\, \sqrt{E \lambda J} \right) \right\} -\log\left(\frac{E}{4\lambda J}\right).
\end{equation}
Recalling that the identification between the bulk JT Hamiltonian (\ref{JT_Liouville_Ham}) and the triple-scaled boundary Hamiltonian \eqref{Triple_scaled_Hamiltonian_restatement_gravity_discussion} reviewed in section \ref{Subsect:Bulk_Hilbert_Space} implies $2\lambda J = \frac{1}{l_{AdS} \phi_b}$, as stated in \eqref{Param_identification_Hamiltonians}, we have that \eqref{Length-KC-tilde_E} reads:
\begin{equation}
    \centering
    \label{Length-KC-tilde_E_grav}
    \lambda \widetilde{C_K}(t)=\frac{\widetilde{l}(t)}{l_{AdS}}=2\log\left\{ \cosh \left( t\, \sqrt{\frac{E}{2l_{AdS}\phi_b}} \right) \right\} -\log\left(\frac{l_{AdS}E\phi_b}{2}\right),
\end{equation}
which coincides exactly with the gravity computation of the regularized length \eqref{Ren_Wormhole_length_E}. 

To summarize, we have succeeded in demonstrating a direct match between the calculated bulk length (\ref{Ren_Wormhole_length_E}) and the boundary K-complexity in the triple-scaled limit \eqref{Length-KC-tilde_E}.  This precise match is expressed in (\ref{Length-KC-tilde_E_grav}) after making the appropriate identification between bulk and boundary parameters.

\subsection{Some remarks on the bulk-boundary matching}
We can elaborate further on the role of the initial condition $\tilde{x}_0$ in the bulk.
Comparing the expression for the configuration energy \eqref{Energy_solution_init_cond} with that in the gravity computation \eqref{energy_config_bulk}, we find that:
\begin{equation}
    \centering
    \label{init_cond_is_phi_h}
    \widetilde{x}_0=-2\log\Phi_h~.
\end{equation}
That is: from the bulk point of view, the choice of $\Phi_h$
amounts to the choice of a coordinate patch adapted to an accelerated (Rindler) observer in the boundary \cite{Spradlin:1999bn,Maldacena:2016upp} that sees the bulk as a black hole whose horizon lies at the locus where the dilaton takes the value $\Phi_h$. From the perspective of the triple-scaled Hamiltonian derived from the boundary theory, this is equivalent to choosing the initial condition for the solution of the equation of motion. The relation between such an initial condition and the dilaton field at the horizon is precisely \eqref{init_cond_is_phi_h}. In particular, the K-complexity of the infinite-temperature thermofield ``double'' state starts at zero when $t=0$, corresponding to the choice $\Phi_h = 1$.

To summarize, in this section we have revisited the need to perform a triple-scaling limit in order to find a (quantum) Liouville Hamiltonian near the ground state of DSSYK, hence retrieving the regime in which the bulk theory is described by JT gravity. This limit defines a notion of regularized length in terms of which the Krylov problem can be posed in a manner analogous to that in section \ref{Sect:KC}, where we studied DSSYK away from the triple-scaling limit. Making use of the bulk/boundary Hilbert space identification reviewed in section \ref{Subsect:Bulk_Hilbert_Space}, we were able to identify the K-complexity of the infinite-temperature thermofield ``double'' state of the triple-scaled Hamiltonian with the expectation value of bulk regularized length, which we could evaluate classically thanks to the smallness of $\lambda$, which plays the role of an $\hbar$ parameter controlling the semiclassical expansion. This correspondence is summarized in \eqref{Length-KC-tilde_E}, where the excitation energy is given by
\begin{equation}
    \centering
    \label{energy_bulk_bdry}
    E = \frac{2\Phi_h^2}{l_{AdS}\phi_b} = 4\lambda J e^{-\tilde{x}_0}.
\end{equation}
Plugging this result into either \eqref{Length-KC-tilde_E} or \eqref{Length-KC-tilde_E_grav} one recovers \eqref{Bulk_length}.

For completeness, we may now elaborate on the features of the emergent bulk described by the triple-scaled SYK Hamiltonian, together with the accelerated observer patch specified by the initial condition $\tilde{x}_0$. This discussion is based on the parameter identifications \eqref{Param_identification_Hamiltonians} and \eqref{init_cond_is_phi_h}.
The bulk consists of a black hole with a horizon radius given by\footnote{Since $l_f=l_{AdS}$, we shall avoid notational cluttering by simply denoting this length scale by $L$.} \eqref{rs_eq}:
\begin{equation}
    \centering
    \label{rs_J}
    r_s = \frac{\Phi_h}{\phi_b}L = 2\lambda JL^2e^{-\tilde{x}_0/2}~.
\end{equation}
Similarly, the black hole temperature can be computed through \cite{Brown:2018bms}:
\begin{equation}
    \centering
    \label{BH_temp_J}
    T_{\text{BH}}=\frac{r_s}{2\pi L^2} = \frac{\lambda J}{\pi}e^{-\tilde{x}_0/2}~.
\end{equation}
For the particular case of the infinite-temperature thermofield ``double'' state of triple-scaled SYK, we have $\tilde{x}_0=0$ and therefore $T_{BH}=\frac{\lambda J}{\pi}$, that is, a temperature of order $\lambda$. This is consistent with the fact that it is the infinite temperature thermofield ``double'' state of the low-energy Hamiltonian, rather than that of the full DSSYK Hamiltonian, with respect to which it is actually a low-temperature state because it zooms in near the ground state of the system.

As we have stressed, neither $r_s$ nor $T_{\text{BH}}$ are a property of global AdS$_2$, but of the coordinates adapted to an observer that sees a horizon in the same way that a Rindler observer in flat space would perceive a horizon due to its acceleration\footnote{They may be understood as an actual black hole radius and temperature whenever the 2-dimensional gravity setup comes from the near-horizon limit of a higher dimensional near-extremal black hole \cite{Brown:2018bms,Sarosi:2017ykf}. In such a case, $\Phi_h$ can also be understood as a measure of the black hole entropy.}. Accordingly, the bulk theory (\ref{Triple_scaled_Hamiltonian}) is solely controlled by the parameters $L$ and $2 \lambda J$, while the choice of a patch (specified by $\Phi_h$) is equivalent to the choice of the initial condition $\widetilde{x}_0$ for the equation of motion of \eqref{Triple_scaled_Hamiltonian_restatement_gravity_discussion}.

\section{Summary and discussion} \label{Sec:Discussion}
The main result of this work is a precise match between the (renormalized) wormhole length in JT gravity \eqref{Ren_Wormhole_length_E} and Krylov complexity in the triple-scaling limit of SYK \eqref{Length-KC-tilde_E}, for the infinite temperature TFD state. To achieve this we related the fixed-chord-number states in DSSYK to Krylov basis elements and used the fact that Krylov complexity is the expectation value of the position operator on the Krylov chain.  We then used the result of \cite{Lin:2022rbf} that in the triple-scaled limit of DSSYK, fixed-chord-number states are fixed-wormhole-length states in JT gravity.  Together with the identification of fixed-chord-number states as Krylov basis elements, this provides a direct match between K-complexity -- or the expectation value of position -- in the triple-scaling limit, and the wormhole length in JT gravity.   We now provide a more detailed summary and some discussion of our conclusions.

\subsection{Results for DSSYK}
We began by establishing that the fixed chord-number states, in terms of which the Hilbert space of DSSYK can be constructed, are nothing but the Krylov basis elements for the effective Hamiltonian of the averaged theory and for the infinite-temperature thermofield ``double'' state, understood as the linear combination with equal weights of all the eigenstates of the theory. The associated sequence of Lanczos coefficients was read off the matrix elements of the effective Hamiltonian expressed in coordinates over the chord basis, and they were used to compute the K-complexity profile as a function of time following various procedures involving increasingly higher levels of sophistication and detail, which we summarize below.

In a first approach, we used the different regimes (in $n$-space) of the Lanczos coefficients in order to estimate the K-complexity profile in its different time regimes. These regimes are defined as the intervals of time during which the wave packet defined over Krylov space remains mostly contained within the corresponding region where the Lanczos coefficients take a particular limiting form. We found that the $b_n$ sequence of the infinite-temperature thermofield ``double'' state transitions from a square-root growth, $b_n \sim \sqrt{n}$, to saturation at a plateau in $n$-space, which implied a K-complexity profile that transitions from quadratic growth at early times to linear growth at times greater than $J^{-1}$, the inverse of the DSSYK coupling strength. We did not obtain a saturation of K-complexity at late times as, due to the fact that we are studying the double-scaled SYK, the system is already in the thermodynamic limit, where the Hilbert space has infinite-dimension: K-complexity saturation is a non-perturbative effect that appears at finite late times that are exponentially large in the number of degrees of freedom \cite{Barbon:2019wsy, I}, and therefore this time scale gets pushed away to infinity in the limit under consideration. Note that in this limit the spectrum of the effective Hamiltonian is continuous. Even for finite systems with a continuous spectrum non-perturbative behaviors require a case-by-case consideration: in some cases the behavior associated with the discreteness of the spectrum survives (e.g. in averaged but finite SYK the plateau behavior persists in the spectral form factor), while calculating a two-point function in a black hole background \cite{Maldacena:2001kr, Barbon:2003aq} yields a result that tends asymptotically to zero as a function of time. 

A further refinement consists in using a continuum approximation to Krylov space dynamics in order to obtain a smooth function $C_K(t)$ for the K-complexity profile. As a result of this approximation, the evolution of the wave packet in Krylov space was reduced to the propagation of a point particle driven by a space-dependent velocity field whose profile is given by the (continuous limit of the) Lanczos coefficients. We found that $C_K(t)$ grows in time following a log-cosh functional form that recovers the quadratic- and linear-growth behaviors in the corresponding regimes, consistent with the previous analysis. The accuracy of this continuum approximation is determined by $\lambda$, the `` 't Hooft coupling'' in DSSYK, which was found to play the role of an $\hbar$-like parameter controlling the semiclassical expansion.

We also provided a formal expression of K-complexity obtained by introducing a resolution of the identity in the definition of the Krylov space wave packet in terms of the spectral decomposition of the effective Hamiltonian of DSSYK. (The spectral decomposition of DSSYK was worked out in \cite{Berkooz:2018jqr,Berkooz:2018qkz}.) The resulting expression involves integrals and sums of q-deformed special functions. We analyzed the exact analytical K-complexity profile for specific limits of the 't Hooft coupling of DSSYK, $\lambda\to\infty$ and $\lambda\to 0$, verifying that it reduces to the expected K-complexity profiles for a flat Lanczos sequence and for a sequence dictated by the Heisenberg-Weyl algebra, respectively.

\subsection{Boundary-bulk correspondence}
As argued in \cite{Lin:2022rbf}, taking a controlled small-$\lambda$ limit in double-scaled SYK (the so-called triple-scaling limit), one zooms in near the ground state of the system. In this limit, where one expects dynamics to be governed by the Schwarzian action \cite{Maldacena:2016hyu}, one finds, consistently, that the Hamiltonian takes the form of a Liouville Hamiltonian, which is the ADM Hamiltonian of JT gravity \cite{Harlow:2018tqv}. \cite{Lin:2022rbf} exploited this further and noted that the Hilbert spaces of JT and DSSYK can be identified in such a way that fixed chord-number states are bulk length eigenstates. We combined this with our result on chord states being Krylov basis elements associated to the infinite-temperature thermofield ``double'' (TFD) state, and concluded that Krylov elements are, in turn, bulk length eigenstates. Therefore, K-complexity agrees quantitatively with bulk length through the relation (\ref{Length-KC-tilde_E_grav}), which matches the K-complexity of the TFD state of the triple-scaled Hamiltonian to the time evolution of bulk regularized length, previously computed in \cite{Harlow:2018tqv}. This matching was established upon a consistent identification of parameters in the bulk and boundary theories given in (\ref{Param_identification_Hamiltonians}) and \eqref{init_cond_is_phi_h}. The log-cosh behavior previously obtained for $C_K(t)$ is responsible for the well-known log-cosh growth of bulk length in the unperturbed geometry, which we matched including numerical pre-factors. From the bulk point of view, it is natural to argue that the lack of late-time saturation of the length as a function of time is due to the absence of (doubly) non-perturbative contributions of higher-genus, which lead to the plateau behaviour of the spectral form factor \cite{Altland:2020ccq,Altland:2022xqx} and correlation functions \cite{Altland:2021rqn}, as we are only considering the disk topology\footnote{See for example \cite{Iliesiu:2021ari} for a bulk computation of a notion of bulk volume associated to complexity which does saturate at late times due to contributions of non-trivial topology.}; this is consistent with the fact that the boundary theory is considered in the thermodynamic limit. For finite-size systems, the K-complexity late-time saturation value for the TFD state can be related to the late-time plateau of the spectral form factor, as proposed in \cite{Balasubramanian:2022tpr} and explored further in \cite{Erdmenger:2023wjg}. Finally, let us note that the growth of the bulk length in a back-reacted geometry in the presence of a matter shock wave may be studied by computing the K-complexity of an operator insertion in DSSYK \footnote{This possibility was suggested by E. Witten in conversations prior to the publication of \cite{IV}, and it makes connection with the discussion on the switchback effect given in section \ref{sect:KC_holog}.}.

\subsection{Discussion}
The agreement between the DSSYK Hamiltonian and JT gravity occurs in the triple-scaling limit (which implies a small value of $\lambda$) because this is the regime in which the SYK Hamiltonian takes a Liouville form. In this regime, the theories are dual to each other, both classically and quantum-mechanically. In this work we were able to explore this duality by a semiclassical computation of both K-complexity on the boundary and the geometric description in the bulk. We expect that quantum corrections will still match by construction, since both the Hilbert spaces and the Hamiltonians are identified \cite{Berkooz:2018jqr,Lin:2022rbf}. On top of this, going away from the small-$\lambda$ limit in DSSYK would amplify the corrections at higher orders in $\lambda$ in the effective Hamiltonian (\ref{Triple_scaled_Hamiltonian}). Hence, the gravity theory gradually deviates from JT towards the arbitrary-$\lambda$ Hamiltonian (\ref{Ham_l_k}). 

In this work we have focused on Krylov complexity for the infinite temperature TFD state. An important future direction is to explore other states, in DSSYK as well as in the triple-scaling limit. A class of such states are the thermal states.

Finally, the calculations presented in this Chapter provide an explicit matching between K-complexity and bulk length in an instance of lower-dimensional holography. It is legitimate to wonder how generic this is, and what lessons may be extracted for higher dimensions. We advocate for an optimistic point of view and claim that this low-dimensional model may be reductionist for some aspects but not for the ones of interest: we investigated the relation between K-complexity and bulk length, using a model of holography where the gravity theory possesses a phase space consisting exclusively of the two-sided length variable (and its conjugate momentum); in this sense, the model is the simplest it can be while still preserving the ingredients relevant for the question at hand. Furthermore, some higher-dimensional scenarios involving near-extremal black holes may be reduced to a two-dimensional gravity problem when analysed in the near-horizon regime (see e.g. \cite{Sarosi:2017ykf} and references therein), which would make direct contact with JT gravity and therefore with our results. It would be interesting to work this out in detail. Another avenue to explore K-complexity in higher dimensional scenarios may consist of studying black hole microstates in microcanonical windows at fixed energy, whose Hilbert space dimension will be finite if the theory is defined on a compact manifold; the finiteness of such a Hilbert subspace should allow us to probe non-perturbative effects such as the saturation of complexity \cite{Kar:2021nbm}.

%% file: content/Discussion.tex
\chapter*{\rm\bfseries Discussion and conclusions}\label{ch:Conclusion}
\addcontentsline{toc}{part}{\protect\numberline{}Discussion and conclusions}%

As in every area of open research, the conclusions can only be partial. Krylov complexity is a notion whose definition is, at least in appearance, not immediately related to other well-known and established instances of quantum complexity. For instance, it constructs a basis which is specifically adapted to the time evolution of the given initial condition, not being immediately suited for the evolution of any other seed state or operator, even if the latter is qualitatively similar to the original seed (e.g. typical, local operators in the case of operator Krylov complexity in chaotic $k$-local systems). Additionally, it has some seemingly counter-intuitive properties, such as the invariance of the K-complexity profile as a function of time under a time translation of the seed, which can be deduced from the symmetry considerations in section \ref{sect:Lanczos_alg_syms} and which has been recently discussed in \cite{Aguilar-Gutierrez:2023nyk}. Nevertheless, the theorems due to \cite{Parker:2018yvk} and \cite{Balasubramanian:2022tpr}, analyzed in section \ref{sect:Two_theorems}, come at rescue, formally putting Krylov complexity in relation to generalized notions of complexity that include operator size and out-of-time-ordered correlators, or in general any other complexity that may be expressed as the position expectation value with respect to some complexity eigenbasis, and showing that it stands out from such a class of complexities because it is in a sense optimal for tracking time evolution, by construction of the Krylov basis. Krylov complexity might be thought of as some generalization of circuit complexity where the only fundamental gate is the time evolution operator itself: A state (or an operator) is understood as \textit{K-complex} if the time evolution of a certain initial condition will take long time to develop a significant overlap with it.

Furthermore, as has been emphasized on multiple occasions throughout this manuscript, Krylov complexity is uniquely specified once an initial condition and a time evolution generator are given within a Hilbert space with a well-defined inner product. The lack of arbitrariness in its definition makes it a promising candidate for becoming the complexity item in the holographic dictionary, as argued in Chapter \ref{ch:chapter02_KC}. This possibility is further supported by its sensitivity to the chaotic or integrable nature of systems, proposed in the seminal article \cite{Parker:2018yvk} and tested further by \cite{Barbon:2019wsy} and by the publications \cite{I,II,III} contributing to this Thesis. In addition, the growth of the Einstein-Rosen bridge length in gravity is a manifestation of the fact that the time evolution of the black hole microstate probes successively orthogonal microstates of the Hilbert space whose associated emergent geometry features a progressively bigger wormhole length \cite{Maldacena:2013xja}: The Krylov basis, which is the output of the Gram-Schmidt process applied to the nested powers of the Hamiltonian (resp. Liouvillian) acting on the initial state (resp. operator), is by construction adapted to such a description.

The publications on which this Thesis is based have provided significant insights on the aspects discussed above:

\begin{itemize}
    \item The article \cite{I} was the first to depict the full profile of Krylov complexity in a finite-size chaotic many-body system with an holographic bulk dual, up to exponentially late time scales including the complexity saturation regime, observing its compatibility with holographic expectations. This, together with the implementation of re-orthogonalization algorithms that allow to run successfully the Lanczos algorithm for large system sizes with floating-point precision and avoiding the loss of orthogonality of the Krylov basis, are the reasons why this work has had a notable impact in the field.
    \item Articles \cite{II,III} explored the behavior of Krylov complexity in integrable systems and proposed the existence of a localization effect in Krylov space, which is enhanced for integrable systems as opposed to chaotic ones, yielding a smaller late-time complexity saturation value for the former. This mechanism, due to the statistics of the Lanczos sequence, which are in turn related to the spectral statistics, is an interesting proposal which nevertheless requires further investigation.
    \item Finally, as emphasized in Chapter \ref{ch:chapter05_DSSYK}, the publication \cite{IV} may be regarded as the culmination of this Thesis, establishing an explicit correspondence between Krylov complexity and bulk length in an instance of low-dimensional holography. This result surely motivates further research on the holographic applications of K-complexity\footnote{The article \cite{IV} was presented at the international Strings 2023 conference \cite{StringsRuth}.}, in a context in which its suitability for such a purpose is not widely accepted yet (see e.g. \cite{Avdoshkin:2022xuw}).
\end{itemize}

All the above are arguments in favour of pursuing a deeper study on Krylov complexity in the realms of quantum chaos and holography, justifying the outburst of works in the area that has taken place in recent years (cf. section \ref{sect:reseachKC}). There remain many open questions to be settled, of which we may list some below:

\begin{itemize}
    \item The Krylov localization phenomenon still needs to be understood better. Publication \cite{III} gives strong evidence for it being a probe of integrability because it shows that its effect is reduced in the regime in which the system perturbed by an integrability-breaking defect is more chaotic. Nevertheless, a better understanding of the relation between the statistics of the Lanczos coefficients controlling this phenomenon and the spectral statistics is lacking, together with precise estimates of the late-time saturation value of Krylov complexity based on this. Furthermore, as already hinted in \cite{II,III} and \cite{Espanol:2022cqr}, these results also have an important dependence on the initial condition, for which a systematic analysis is still missing.
    \item To present day, it is still not obvious how to generalize the notion of Krylov complexity to systems with time-dependent Hamiltonians, a context with potential applications in holography \cite{Nozaki:2013wia}. The Lanczos algorithm appears to be intrinsically well-suited for time-evolution operators given by the exponential of a constant generator, and generalizing it to time-ordered exponentials without the introduction of unwanted extrinsic tolerance parameters remains an unsolved challenge.
    \item On a similar note, in section \ref{sect:KryPolCont} we mentioned that a putative extension of the Lanczos algorithm to Hilbert spaces of uncountably infinite dimension is also unexplored. This constitutes an intriguing direction of thought with potential applications to quantum field theory and gravity \cite{Witten:2021jzq}.
    \item The most immediate question in view of the holographic results in \cite{IV} is the establishment of the correspondence between operator Krylov complexity in double-scaled SYK and bulk length in JT gravity\footnote{At the moment of the submission of this Thesis, this constitutes work in progress by the authors of \cite{I,II,III,IV}, together with M. Ambrosini.}.
    \item The question on the suitability of Krylov complexity for holography inevitably passes by the study of higher-dimensional scenarios, which remain unexplored in this framework. Two approaches might be taken in this case: On the one hand, an analysis of finite-size effects in a higher-dimensional instance of holography, in order to elucidate whether the Krylov complexity profile takes the expected form, may be achieved by considering microcanonical spectral windows of the boundary theory, potentially placed in a compact manifold in order to make the Hilbert space dimension effectively finite \cite{Kar:2021nbm}; on the other hand, one might make the reassuring point that the near-horizon geometry of extremal black holes in higher dimensions is described by two-dimensional JT gravity \cite{Sarosi:2017ykf}, making direct contact with the results in \cite{IV}.
\end{itemize}

All in all, the motivations to study Krylov complexity, as well as the open questions remaining to be answered, are abundant. An arbitrarily long Thesis could be written about this area of research, combining past experience on applications of Krylov methods and the Lanczos algorithm to condensed matter and high energy physics with a discussion on the current debates and future lines of investigation to be pursued. But one needs to graduate at some point.

%% file: content/AppxCh01.tex
\chapter{\rm\bfseries Appendices to Chapter \ref{ch:chapter01_Lanczos}}
\label{ch:AppxCh01}

\section{Numerical results for the toy models}\label{appx:Lanczos_toys_numerics}

This Appendix gathers all the figures pertaining to the numerical results obtained analysing the toy models presented in section \ref{sect:Lanczos_Numerics}. The defining features of each model are:
\begin{itemize}
    \item \textbf{Model 1:} A spectrum with an exponential hierarchy \eqref{Sect_Lanczos_eigenvalues_parametric_separation} and a seed vector \eqref{Sect_Lanczos_Numerics_seed_exp} which is itself exponentially peaked near the largest eigenvalue.
    \item  \textbf{Model 2:} Same spectrum as model 1, this time with a probe \eqref{Sect_Lanczos_Numerics_seed_flat} whose profile in the eigenbasis is flat.
    \item \textbf{Model 3:} A flat probe and a spectrum \eqref{Sect_Lanczos_Numerics_spectrum_quasideg} with an exponential hierarchy and a quasi-degeneracy near the largest eigenvalue.
\end{itemize}

In all cases, the parameter controlling the spectral scale was set to $\lambda=2$ and the Krylov dimension was chosen to be $K=20$. The numerical computations were run using the software Mathematica \cite{Mathematica} at a working precision of $100$ decimal digits.


\begin{figure}
    \centering
    \includegraphics[width = 6.cm]{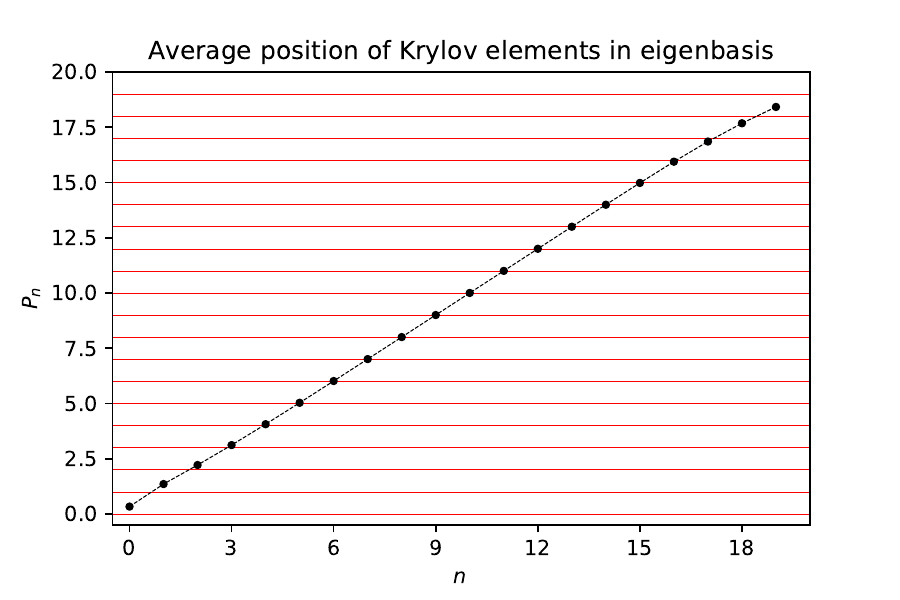} \includegraphics[width=6.cm]{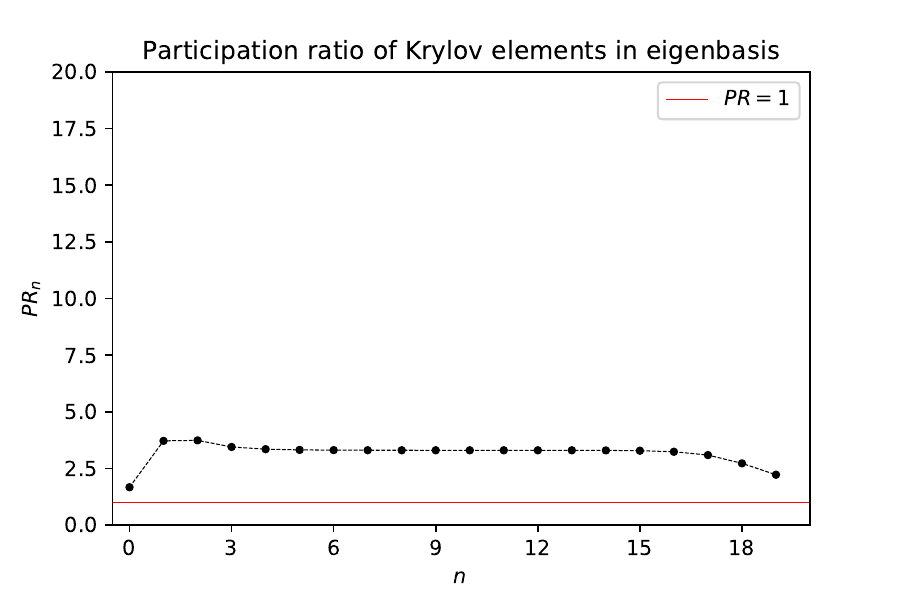}
    \caption{\textbf{Model 1.} 
    \textbf{Left:} Average position of the Krylov elements on the eigenbasis, $P(|K_n\rangle)\equiv P_n$. Horizontal red lines indicate integer values, corresponding to the position of each eigenvector, as can be seen from the definition of \eqref{Sect_Lanczos_Numerics_position_ebasis_def}. \textbf{Right:} Participation ratio of each Krylov basis element, $PR(|K_n\rangle)\equiv PR_n$. Note that the range of the vertical axes is $[0,K]$. Combining the information from both plots, we can conclude that the $n$-th Krylov element is heavily localized near the $n$-th eigenspace. 
    }
    \label{fig:Sect_Lanczos_Numerics_exp_seed_Krylov_elements}
\end{figure}

\begin{figure}
    \centering
    \includegraphics[width=6cm]{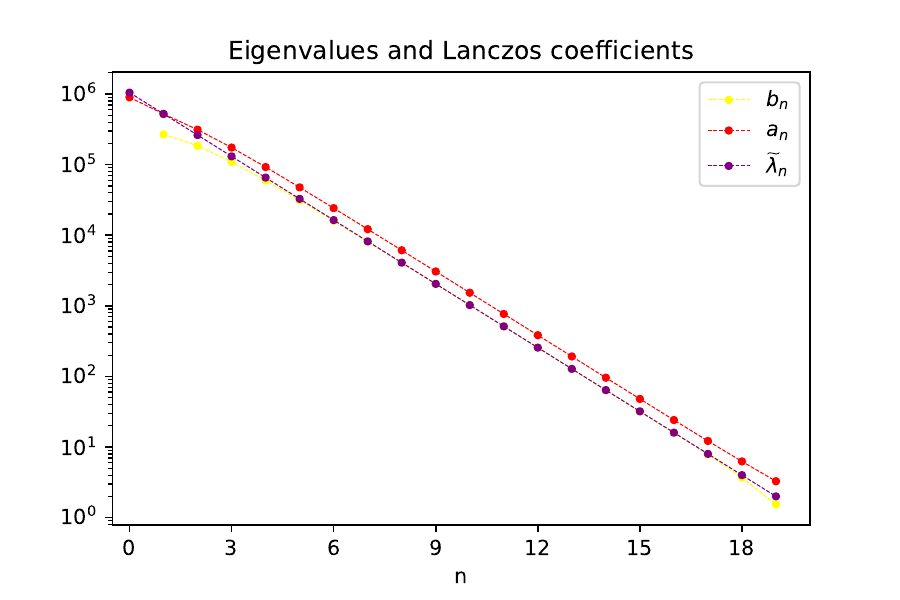}
    \caption{\textbf{Model 1.} 
    Lanczos coefficients and eigenvalues of $H$. The plot illustrates how these coefficients are roughly of the same order of magnitude as the eigenvalues of $H$, in qualitative agreement with the fact that $H$ takes a tridiagonal form \eqref{Sect_Lanczos_Tridiag_H_matrix} when written in coordinates over the Krylov basis, where the items of such a matrix are given by the $a$- and $b$-coefficients. For the model at hand, it is natural to plot $\widetilde{\lambda}_n$ together with $a_n$ and $b_n$, sharing the same horizontal axis, in the sense that the corresponding numerical approximations to the eigenvalues in a given Krylov space truncation of dimension $n_{*}$, denoted $\left\{\widetilde{\lambda}_m^{(n_{*})}\right\}_{m=0}^{n_{*}-1}$, are given by the eigenvalues of the tridiagonal matrix built out of $\left\{a_m\right\}_{m=0}^{n_{*}-1}$ and $\left\{b_m\right\}_{m=1}^{n_{*}-1}$.}
    \label{fig:Sect_Lanczos_Numerics_exp_seed_Lanczos_coeffs_vs_evals}
\end{figure}

\begin{figure}
    \centering
    \includegraphics[width=6cm]{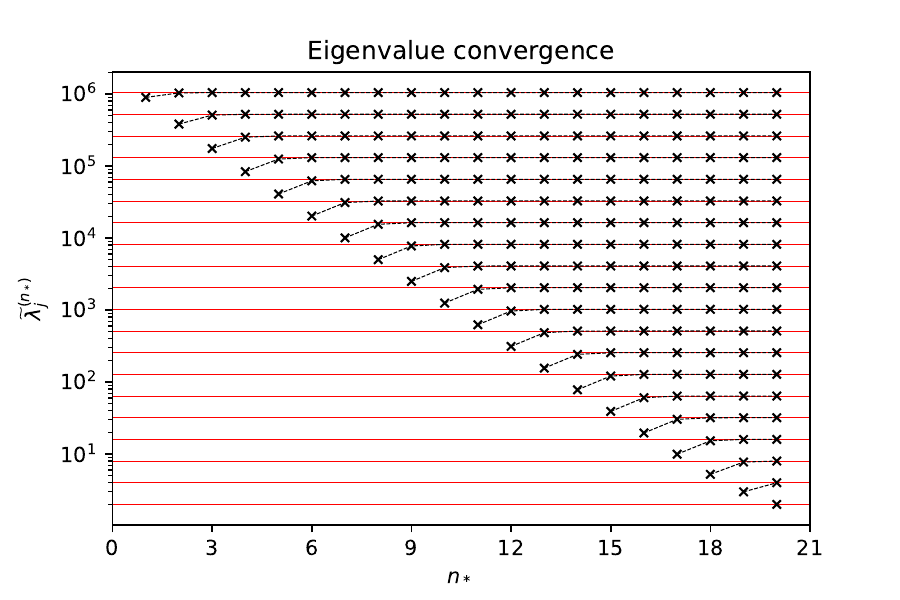} \\
    \includegraphics[width=6cm]{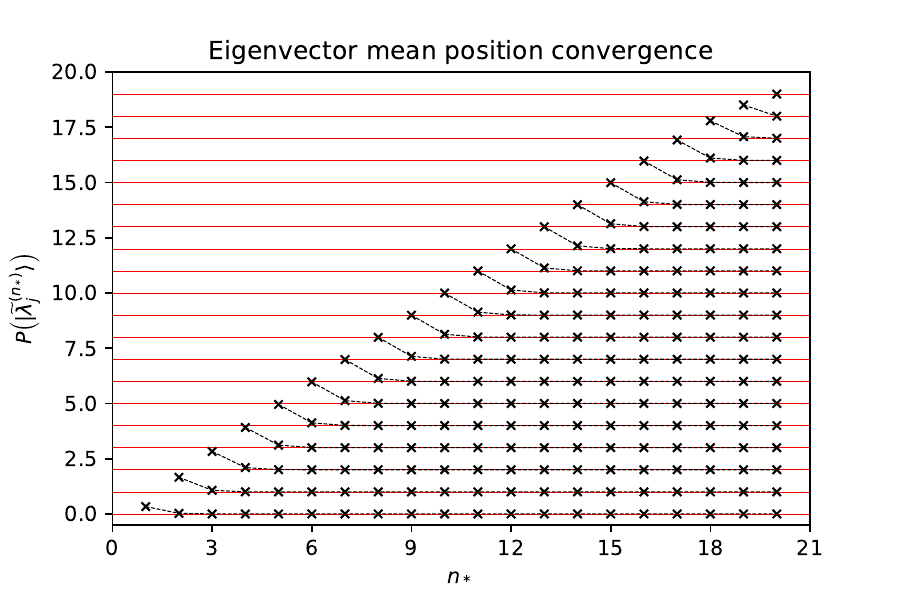} \includegraphics[width=6cm]{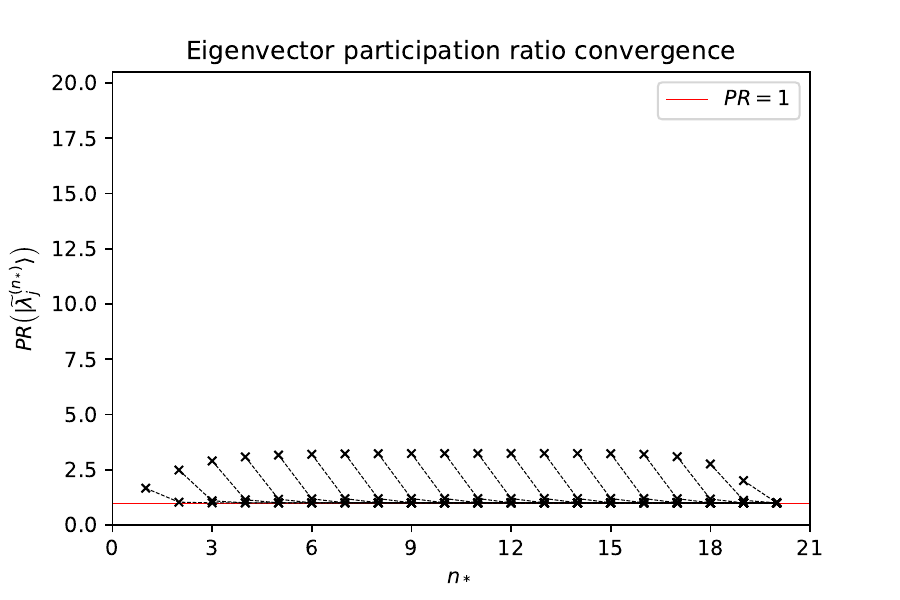}
    \caption{\textbf{Model 1.} 
    Convergence of approximate eigenvalues and eigenvectors in the successive truncations of Krylov space, as a function of the dimension of the truncation $n_{*}$. \textbf{Top:} Eigenvalue approximations. For each value $n_{*}$ of the truncated Krylov subspace dimension, markers indicate the $n_{*}$ approximate eigenvalues obtained. Strictly speaking, since $n_{*}$ cannot be continuously varied, in principle there is no reason to identify the approximate eigenvalues of different truncations with each other; however, for the model at hand, since each time $n_{*}$ increases by one a new eigenvalue smaller than the rest is generated, we chose a numbering criterion for $\widetilde{\lambda}_j^{(n_{*})}$ such that $\widetilde{\lambda}_0^{(n_{*})} > \widetilde{\lambda}_1^{(n_{*})} > \dots >\widetilde{\lambda}_{n_{*}-1}^{(n_{*})}$ for every fixed $n_{*}$, and joined the spectra of different truncations accordingly with dashed lines. The background red lines give the exact eigenvalues, to which the approximate ones converge when $n_{*}=K$. \textbf{Bottom left:} Mean position on the eigenbasis of the approximate eigenvectors at each truncation. These vectors are computed, via \eqref{Sect_Lanczos_evec_approx_truncated}, out of the same Krylov polynomials that yield the approximate eigenvalues, and hence the pattern to join with dashed lines the results of different truncations is inherited from the criterion for the eigenvalues. Red background lines indicate the position of the exact eigenvectors (i.e. integer positions over the eigenbasis). \textbf{Bottom right:} Participation ratio of the approximate eigenvectors. Since they tend to $1$ in the limit $n_{*}=K$, we confirm that they become Kronecker deltas (and hence, exact eigenvectors) in such a limit.}
    \label{fig:Sect_Lanczos_Numerics_exp_seed_EvalEvec_Convergence}
\end{figure}

\begin{figure}
    \centering
    \includegraphics[width=6cm]{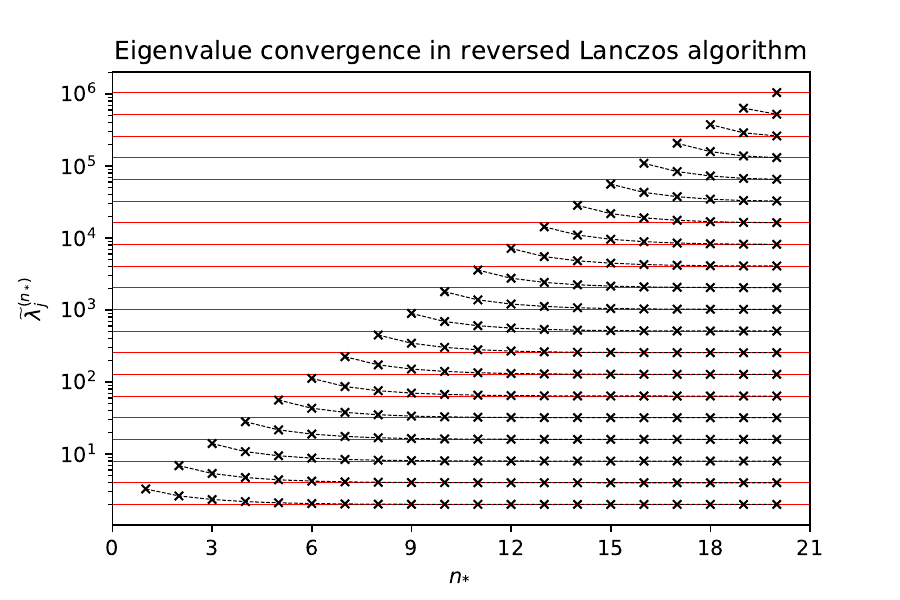} \\
    \includegraphics[width=6cm]{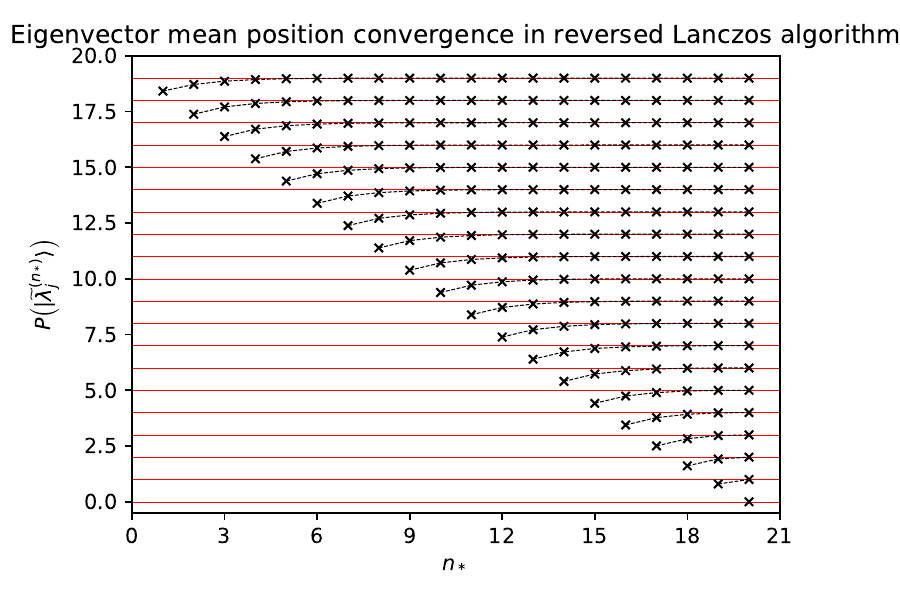} \includegraphics[width=6cm]{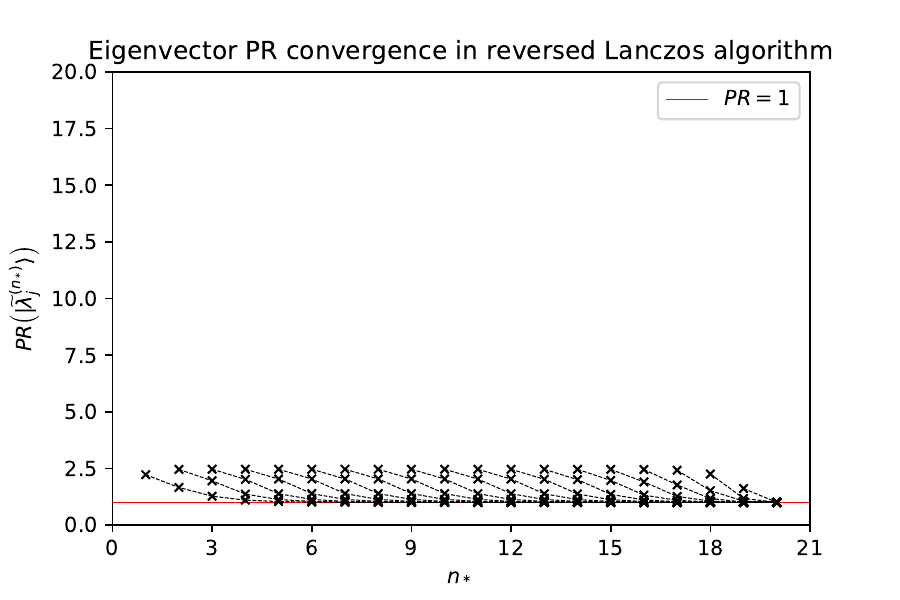}
    \caption{\textbf{Model 1.} Convergence of approximate eigenvalues and eigenvectors obtained from the reversed Lanczos algorithm. \textbf{Top:} Approximate eigenvalues. Each time the dimension of the Krylov truncation increases, a new, bigger eigenvalue appears, and eigenvalues across different truncations are joined with dashed lines accordingly. Background red lines give the exact eigenvalues of $H$, reproduced when $n_{*}=K$. As announced, this time the smaller eigenvalues are reproduced first and the largest eigenvalue is not reproduced until $n_{*}=K$. \textbf{Bottom left:} Average position in the eigenbasis of the approximate eigenvectors at each truncation. Background lines give the position of the exact eigenvectors. \textbf{Bottom right:} Participation ratio of the approximate eigenvectors, which correctly tends to $1$ when $n_{*}=K$.}
    \label{fig:Sect_Lanczos_Numerics_exp_seed_EvalEvec_Convergence_Reversed}
\end{figure}

\clearpage

\begin{figure}
    \centering
    \includegraphics[width=6cm]{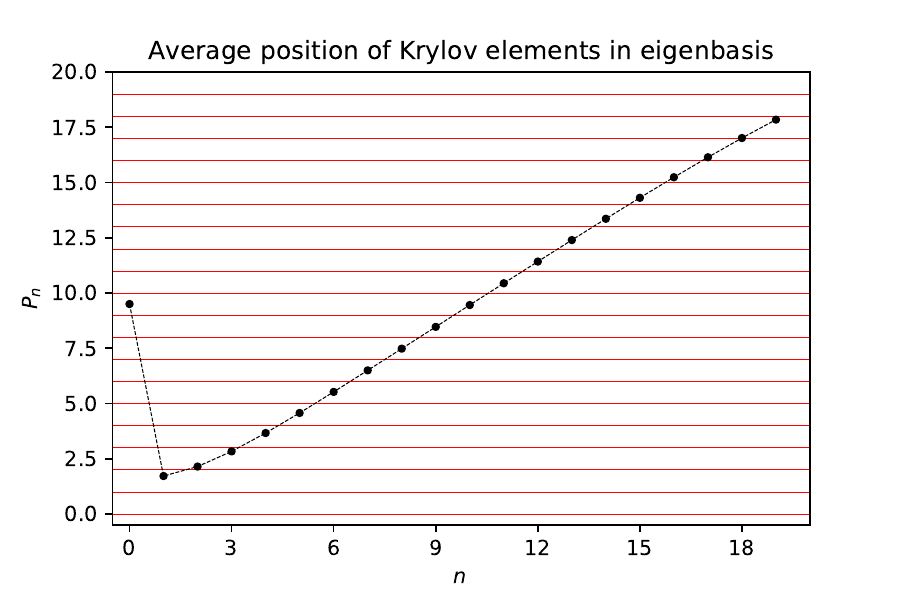} \includegraphics[width=6cm]{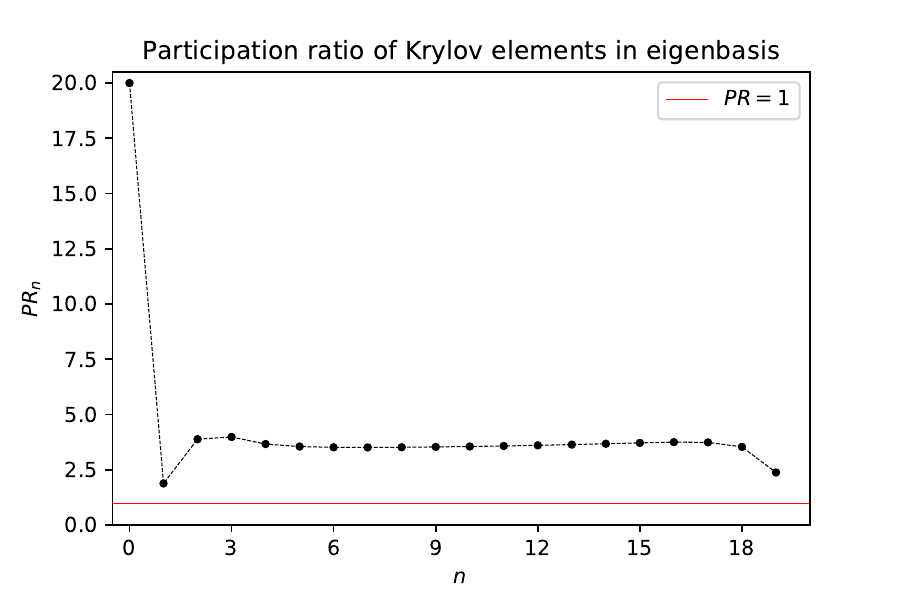}
    \caption{\textbf{Model 2.} \textbf{Left:} Average position of the Krylov elements in the eigenbasis. We observe that $|K_0\rangle = |\Omega\rangle$ has an average position of approximately half the Krylov dimension, since it is a linear combination of all eigenvectors with equal coefficients, by construction. Subsequently, each Krylov element $|K_n\rangle$ with $n>0$ is located near $|\widetilde{\lambda}_n\rangle$. \textbf{Right:} Participation ratio of the Krylov elements in the eigenbasis. The plot manifests that $|K_0\rangle$ indeed receives contributions from all eigenvectors, while the rest of the Krylov elements effectively receive contributions from a a few eigenvectors, signaling that they are peaked near their average position.}
    \label{fig:Sect_Lanczos_Numerics_flat_seed_Krylov_elements}
\end{figure}

\begin{figure}
    \centering
    \includegraphics[width=6cm]{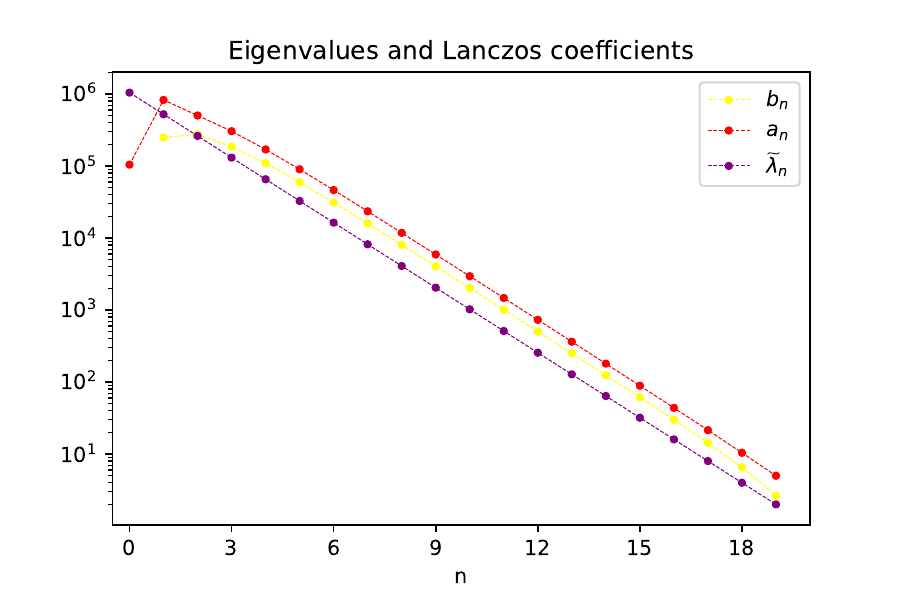}
    \caption{\textbf{Model 2.} Lanczos coefficients $a_n$, $b_n$, and eigenvalues $\widetilde{\lambda}_n$, plotted against a common horizontal axis for the index $n$. We observe that $a_0$ is not of the same order as $\widetilde{\lambda}_0$, as expected, since it is given by the arithmetic average of the spectrum, according to \eqref{Sect_Lanczos_a0_explicit}. Since Krylov elements $|K_n\rangle$ with $n>0$ are significantly peaked near exact eigenvectors, we observe that the rest of the Lanczos coefficients are, in order, numerically comparable to the corresponding eigenvalues.}
    \label{fig:Sect_Lanczos_Numerics_flat_seed_Lanczos_coeffs_vs_evals}
\end{figure}

\begin{figure}
    \centering
    \includegraphics[width=6cm]{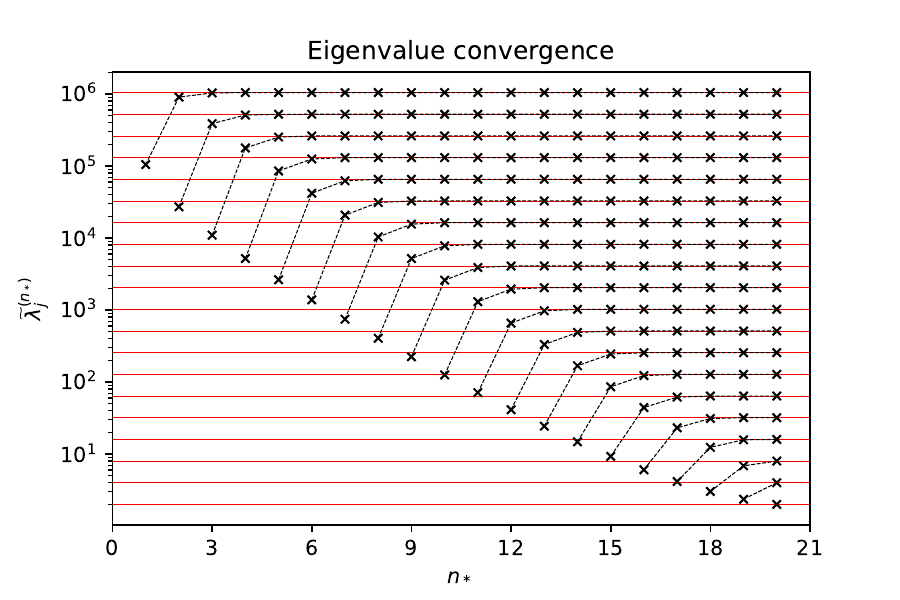} \\
    \includegraphics[width=6cm]{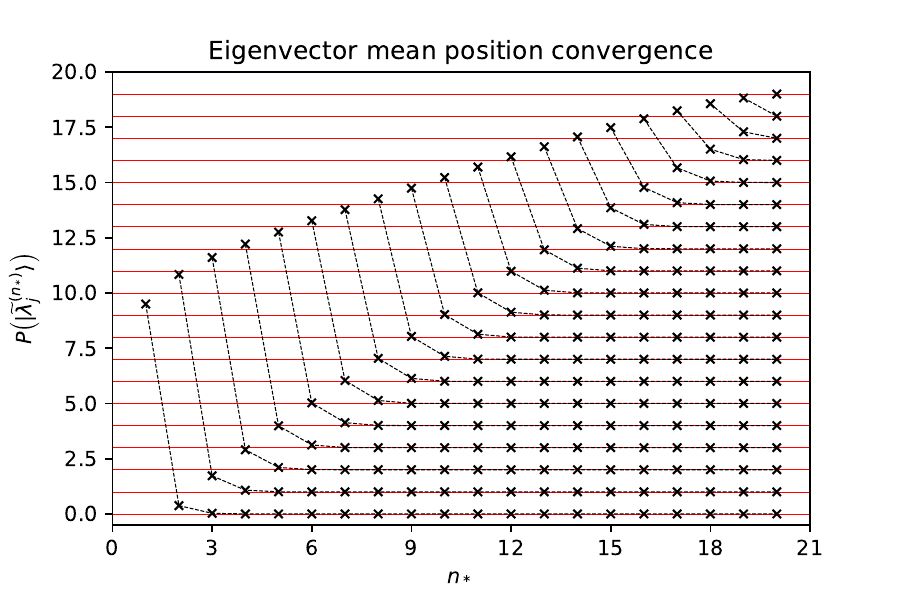} \includegraphics[width=6cm]{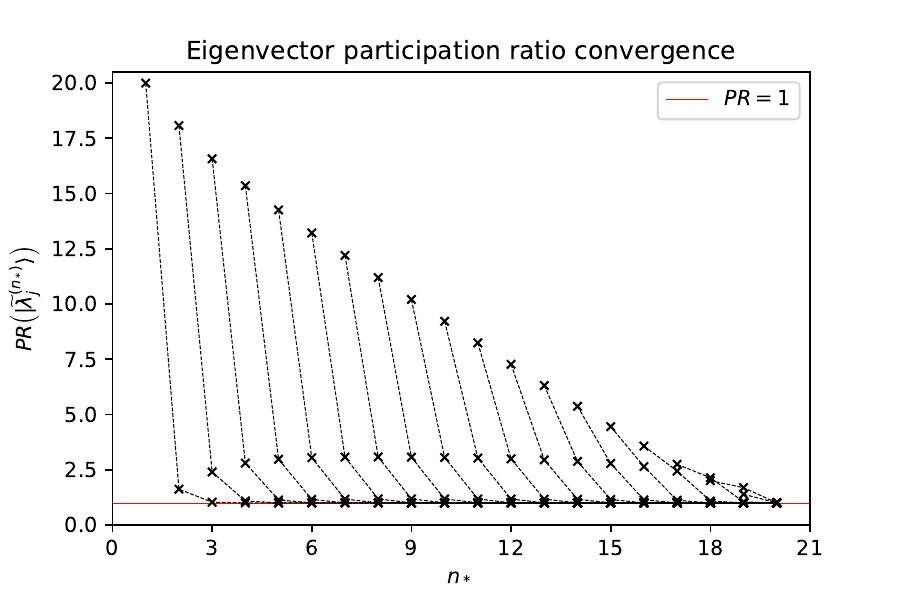}
    \caption{\textbf{Model 2.} \textbf{Top:} Eigenvalue convergence as a function of the dimension $n_{*}$ of the truncated Krylov subspace. Eigenvalues obtained in different truncations are joined by dashed lines with the same criterion as in Figure \ref{fig:Sect_Lanczos_Numerics_exp_seed_EvalEvec_Convergence} for Model 1. \textbf{Bottom left:} Convergence of the position in the eigenbasis of the approximate eigenvectors associated to the previous eigenvalues. We notice that in each truncation, the position of the new approximate eigenvector obtained is roughly a central position in the complement of the truncated Krylov subspace. \textbf{Left:} Participation ratio of the approximate eigenvectors, again as a function of the truncated Krylov dimension $n_{*}$. The $PR$ of the new approximate eigenvector obtained in each truncation is indeed similar to the number of directions excluded from the truncated Krylov subspace, confirming that this vector is a linear combination with similar coefficients of all the exact eigenvectors effectively not probed by the truncation. All participation ratios still tend to $1$ when $n_{*}=K$, confirming that the exact eigenvectors are obtained, as expected.}
    \label{fig:Sect_Lanczos_Numerics_flat_seed_EvalEvec_Convergence}
\end{figure}

\begin{figure}
    \centering
    \includegraphics[width=6cm]{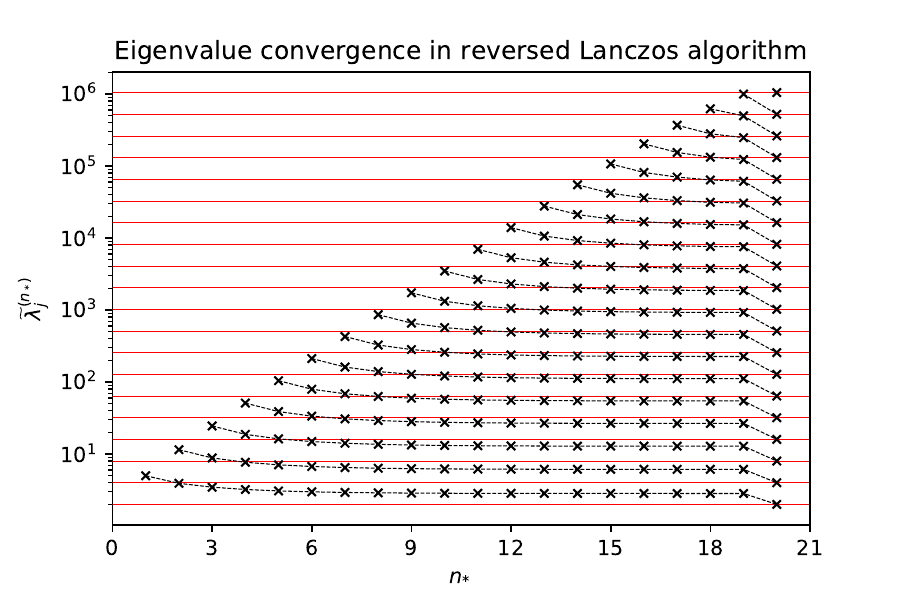} \\
    \includegraphics[width=6cm]{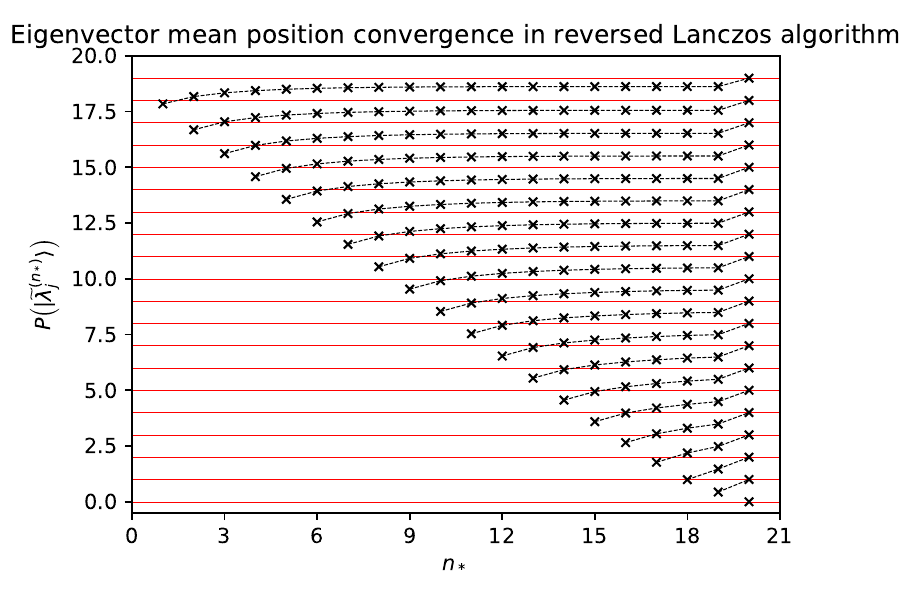} \includegraphics[width=6cm]{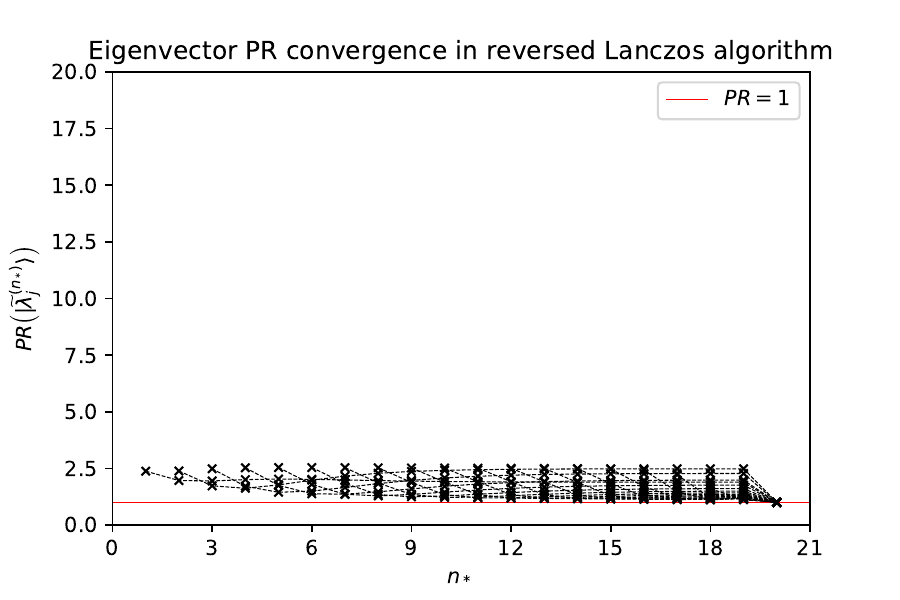}
    \caption{\textbf{Model 2.} \textbf{Top:} Eigenvalue convergence in the reversed Lanczos algorithm, as a function of the truncated Krylov space dimension $n_{*}$. Correctly, the lower eigenvalues are reproduced first. They are joined across the different truncations by dashed lines following the same criterion as in Figure \ref{fig:Sect_Lanczos_Numerics_exp_seed_EvalEvec_Convergence_Reversed}. \textbf{Bottom left:} Convergence of the position of the associated approximate eigenvectors. They are seen to tend to the position of the corresponding exact eigenvectors when $n_{*}=K$. \textbf{Bottom right:} Participation ratio of the approximate eigenvectors as a function of $n_{*}$. Again, when $n_{*}=K$, all participation ratios are equal to one.}
    \label{fig:Sect_Lanczos_Numerics_flat_seed_EvalEvec_Convergence_Reversed}
\end{figure}

\clearpage

\begin{figure}
    \centering
    \includegraphics[width=6cm]{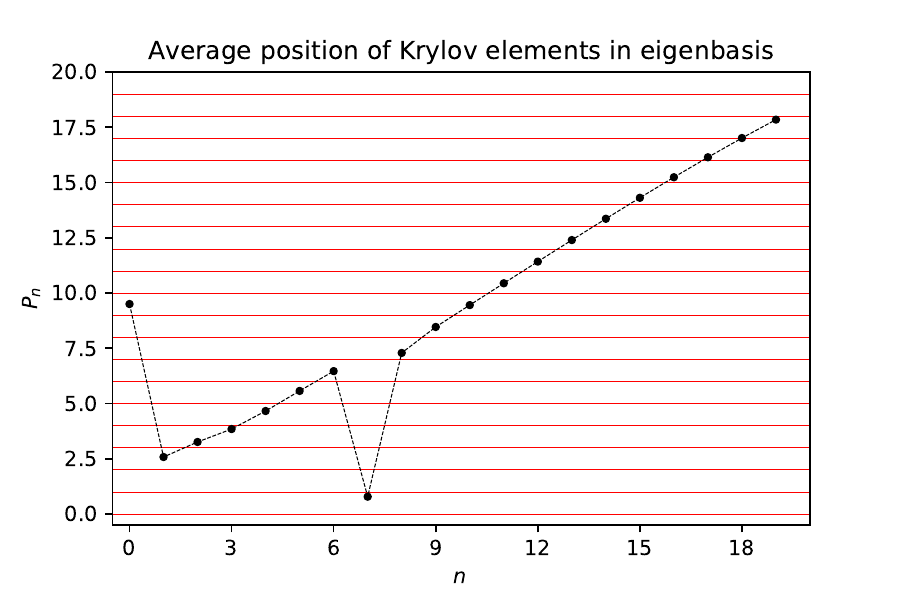} \includegraphics[width=6cm]{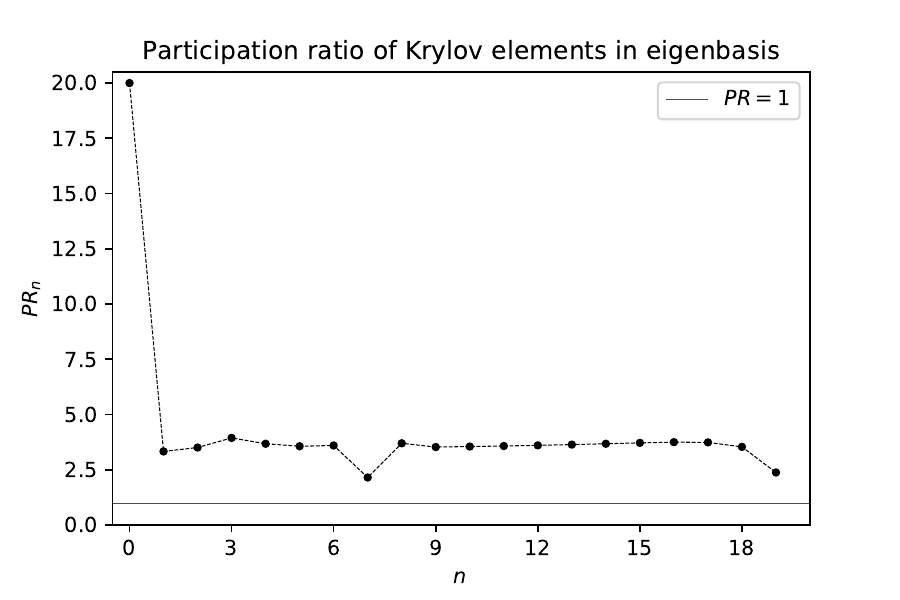}
    \caption{\textbf{Model 3.} \textbf{Left:} Average position of Krylov elements in the eigenbasis of $H$. The relevant features are the following: $|K_0\rangle = |\Omega\rangle$ is a flat linear combination of all eigenvectors, so its average position is close to $\frac{K}{2}$; $|K_1\rangle$ and $|K_7\rangle$ are located near eigenvectors of large eigenvalues, while the rest of the Krylov elements probe progressively eigenvectors of smaller eigenvalues. \textbf{Right:} Participation ratio of each Krylov element, confirming that they all receive contributions from a few exact eigenvectors, except for $|K_0\rangle$ which, by construction, has $PR_0=K$. }
    \label{fig:Sect_Lanczos_Numerics_flat_seed_quasideg_Krylov_elements}
\end{figure}

\begin{figure}
    \centering
    \includegraphics[width=6cm]{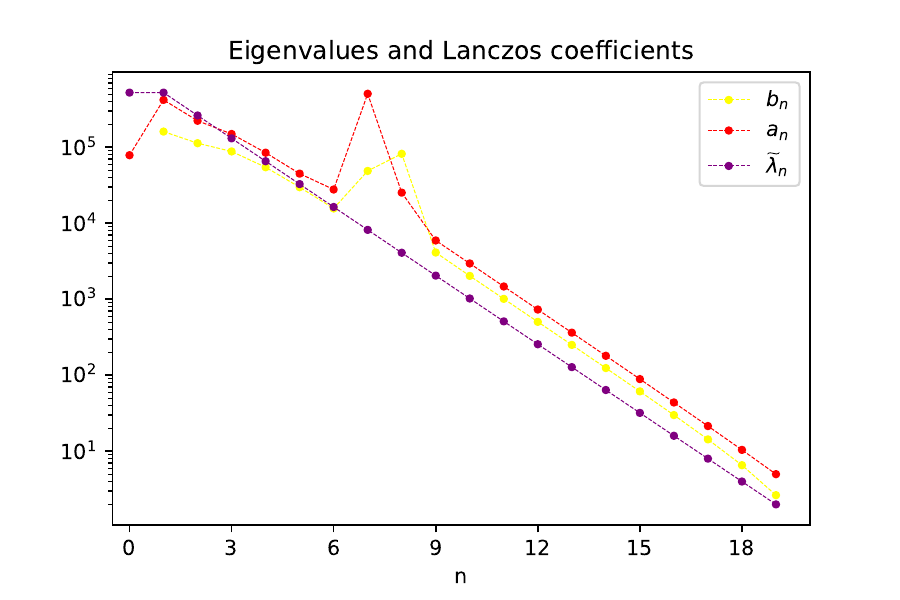}
    \caption{\textbf{Model 3.} Lanczos coefficients $a_n$, $b_n$ and exact eigenvalues $\widetilde{\lambda}_n$ plotted against a common horizontal axis for the index $n$. This time, $a_7$ is not seen to be comparable to $\widetilde{\lambda}_7$, because $|K_7\rangle$ is not peaked near $|\widetilde{\lambda}_7\rangle$, as it probes $|\widetilde{\lambda}_0\rangle$ and $|\widetilde{\lambda}_1\rangle$ instead, $a_7$ being therefore of the order of the largest eigenvalues of the spectrum. The rest of the plot is qualitatively similar to those in Figures \ref{fig:Sect_Lanczos_Numerics_exp_seed_Lanczos_coeffs_vs_evals} and \ref{fig:Sect_Lanczos_Numerics_flat_seed_Lanczos_coeffs_vs_evals}.}
    \label{fig:Sect_Lanczos_Numerics_flat_seed_quasideg_Lanczos_coeffs_vs_evals}
\end{figure}

\begin{figure}
    \centering
    \includegraphics[width=6cm]{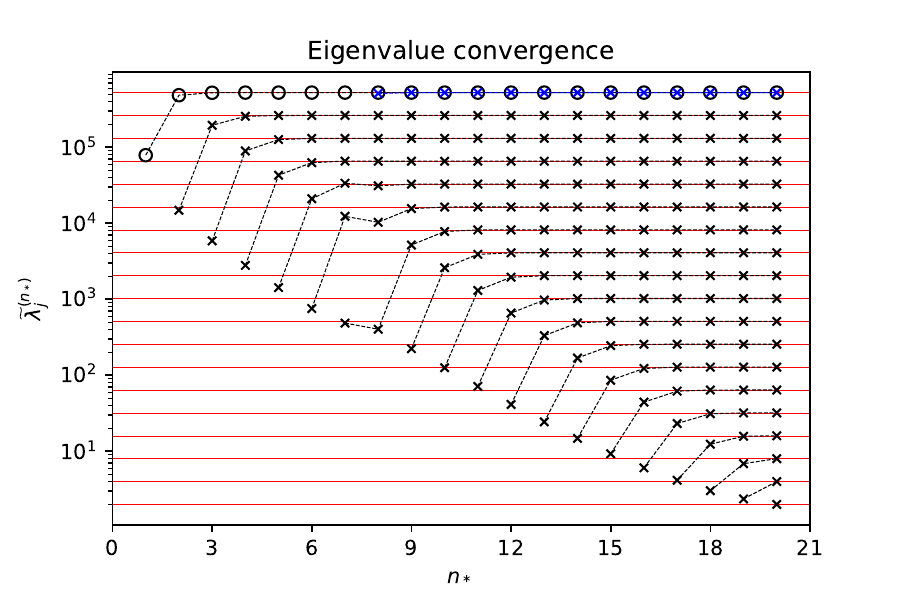} \\
    \includegraphics[width=6cm]{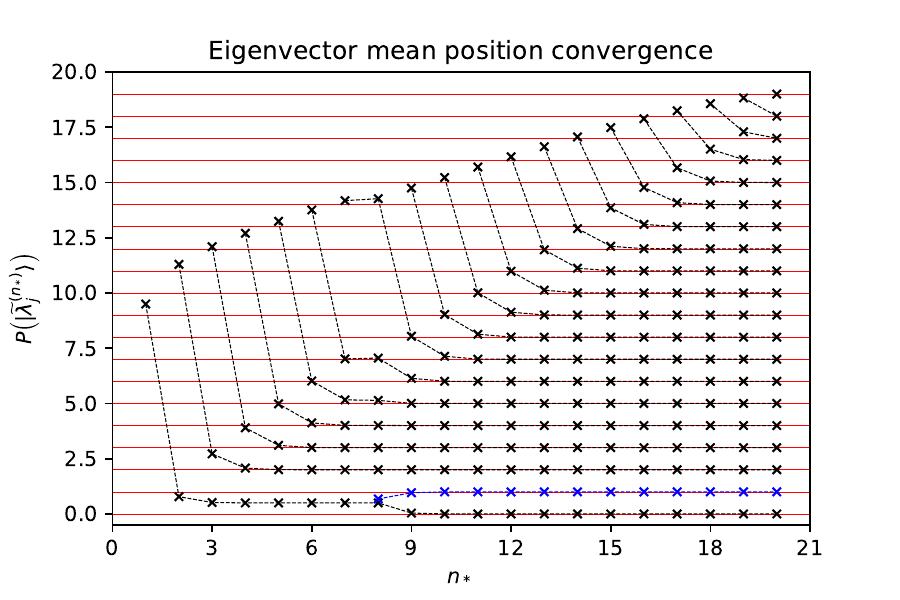} \includegraphics[width=6cm]{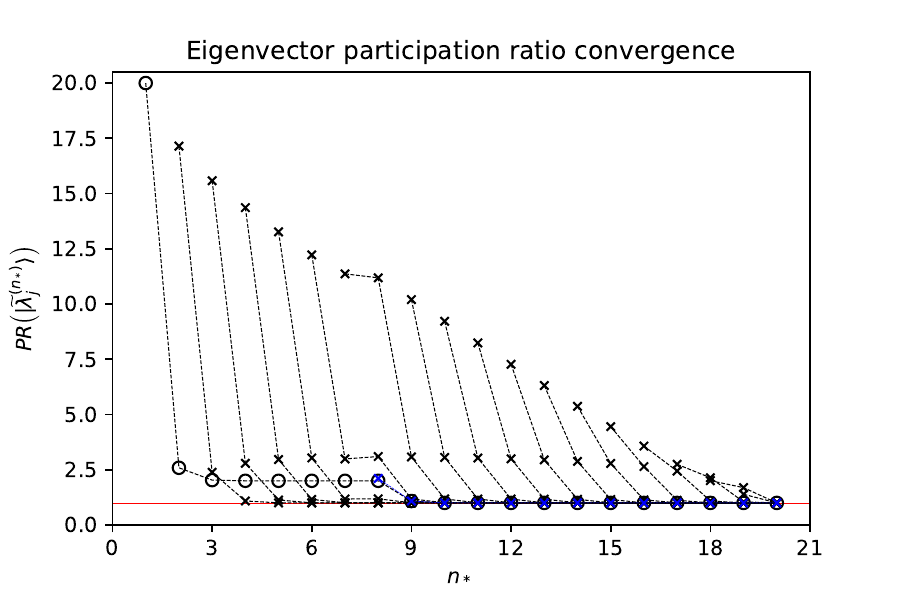}
    \caption{\textbf{Model 3.} \textbf{Top:} Eigenvalue convergence across the various Krylov space truncations, as a function of the subspace dimension $n_{*}$. For $n_{*}<8$, there is only one approximate eigenvalue numerically comparable to $\widetilde{\lambda}_0$ and $\widetilde{\lambda}_1$. For $n_{*}\geq 8$, when both $|K_0\rangle$ and $|K_7\rangle$ are included in the truncated Krylov subspace, the quasi-degeneracy is resolved and a second eigenvalue (marked in blue) appears near the upper edge of the spectrum. \textbf{Bottom left:} Approximate eigenvector position convergence as a function of $n_{*}$. We note that for $ n_{*}< 8$ the approximate eigenvector of the highest approximate eigenvalue is located between positions $j=0$ and $j=1$, confirming that it is some linear combination of $|\widetilde{\lambda}_0\rangle$ and $|\widetilde{\lambda}_1\rangle$. When $n_{*}=8$, the new approximate eigenvector (in blue) that appears is another orthogonal linear combination of such vectors, hence resolving the degeneracy. \textbf{Bottom right:} Participation ratio of approximate eigenvectors as a function of $n_{*}$. Note that, just like in the results of Model 2, the approximate eigenvector corresponding to the smallest eigenvalue in each truncation receives contributions from all the exact eigenvectors effectively left out of the Krylov subspace. The approximate eigenvector associated to the largest approximate eigenvalue retains a $PR$ of approximately two until the vector (in blue) resolving the quasi-degeneracy appears at $n_{*}=8$, with a similar $PR$, after which both vectors feature participation ratios much closer to unity.}
    \label{fig:Sect_Lanczos_Numerics_flat_seed_quasideg_EvalEvec_Convergence}
\end{figure}

\begin{figure}
    \centering
    \includegraphics[width=6cm]{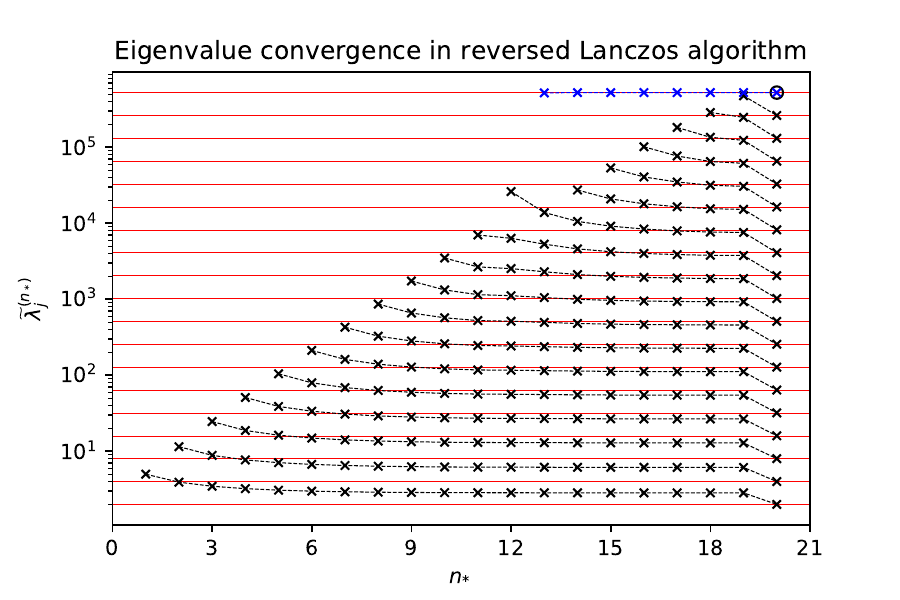} \\
    \includegraphics[width = 6cm]{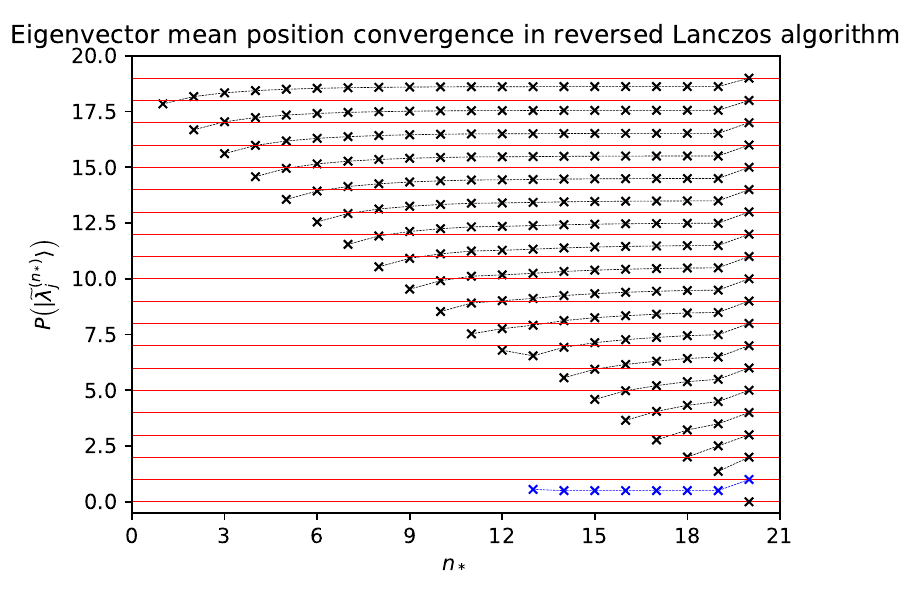} \includegraphics[width=6cm]{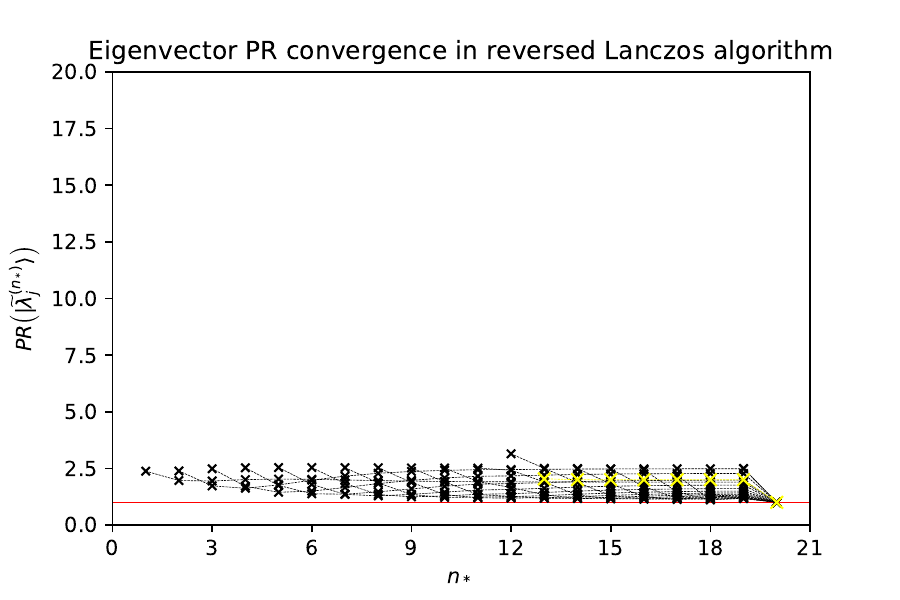}
    \caption{\textbf{Model 3.} Convergence of approximate eigenvalues (\textbf{top}), approximate eigenvector position (\textbf{bottom left}) and participation ratio (\textbf{bottom right}) in the reversed Lanczos algorithm, as a function of the truncated Krylov subspace dimension $n_{*}$. Eigenvalues and eigenvectors are progressively approximated starting from the bottom of the spectrum, until a large eigenvalue appears at $n_{*}=13$. This is because the last Krylov element contained in this truncation is, according to \eqref{Sect_Lanczos_reversed_vs_original}, $|\overline{K}_{12}\rangle = |K_7\rangle$, which is peaked near the eigenvectors of largest eigenvalues, as confirmed by Figure \ref{fig:Sect_Lanczos_Numerics_flat_seed_quasideg_Krylov_elements}. The other large eigenvalue is only generated in the last truncation $n_{*}=K=20$, which reaches $|\overline{K}_{19}\rangle = |K_0\rangle$, also peaked in the same region of the spectrum. The color of the new approximate eigenvalue and eigenvector generated at $n_{*}=13$ has been chosen to be either blue or yellow in each plot, for the sake of contrast.}
    \label{fig:Sect_Lanczos_Numerics_flat_seed_quasideg_EvalEvec_Convergence_Reversed}
\end{figure}

%% file: content/AppxCh03.tex
\chapter{\rm\bfseries Appendices to Chapter \ref{ch:chapter03_SYK}}
\label{ch:AppxCh03}

\section{Lanczos sequence in RMT}\label{Appx-RMT}
\begin{figure}
    \begin{minipage}{.44\textwidth}
    \includegraphics[width=1.\linewidth]{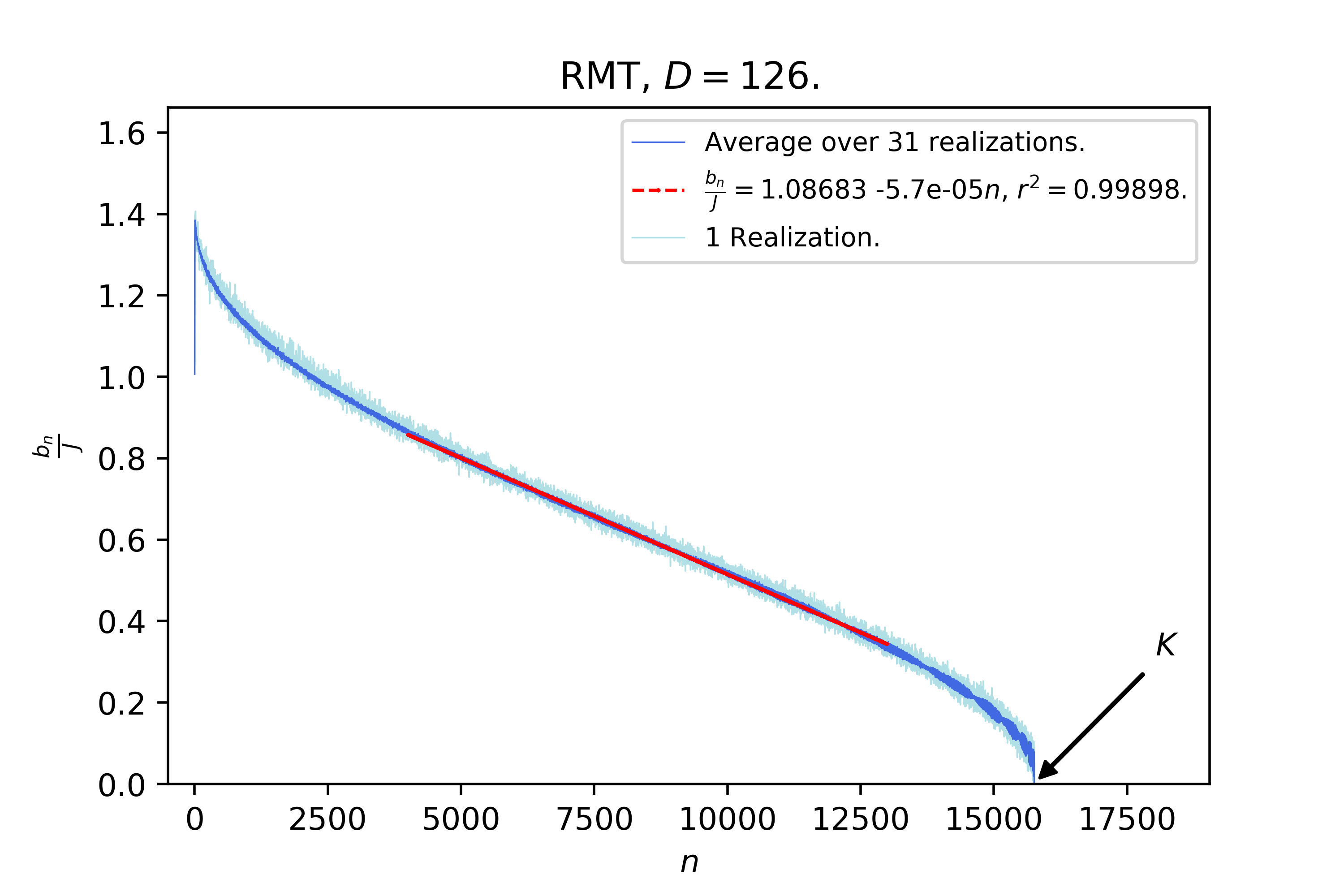}
    \end{minipage} \quad
    \begin{minipage}{.44\textwidth}
    \includegraphics[width=1.\linewidth]{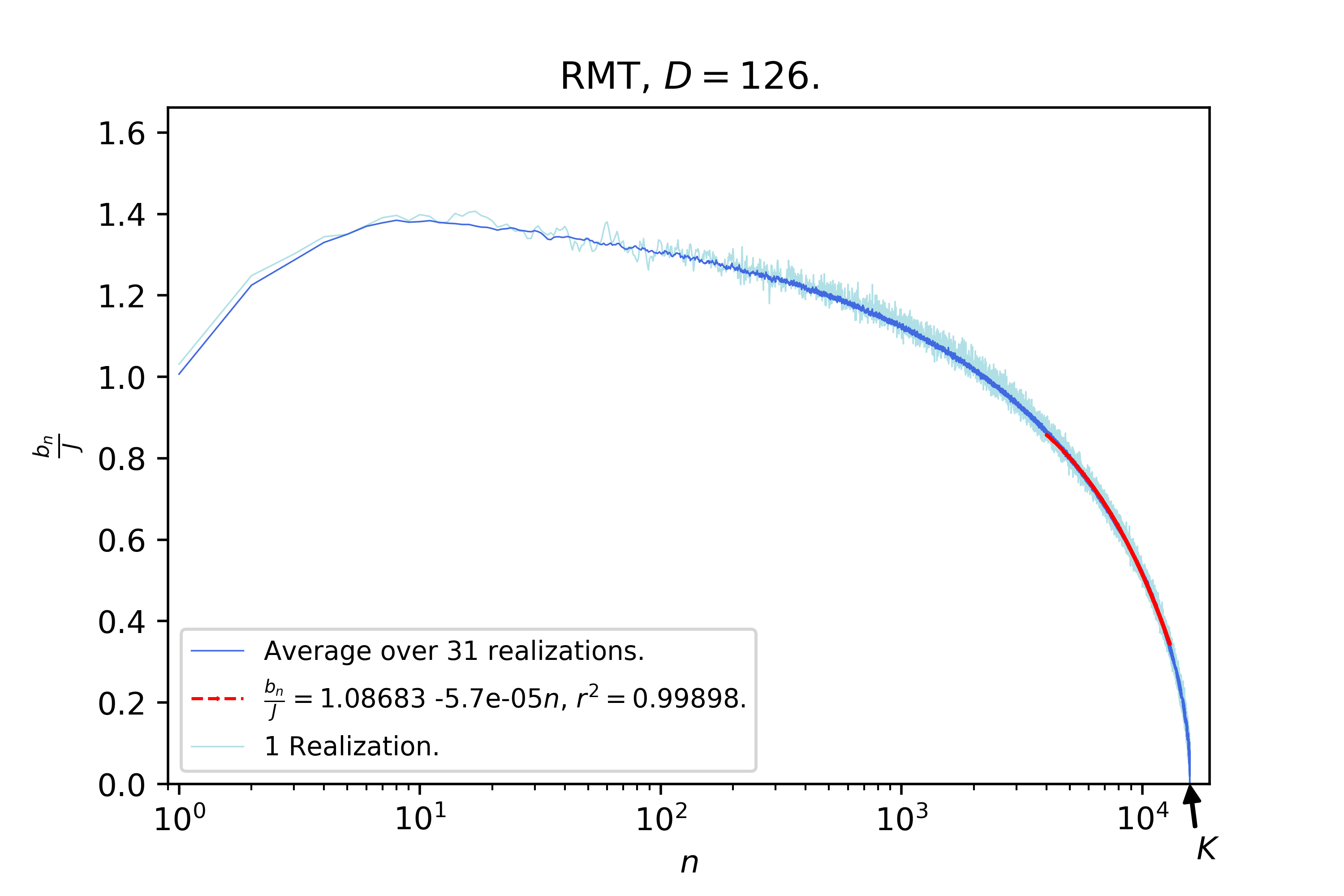}
    \end{minipage}
    \caption{Lanczos sequence for RMT drawn from the GUE with potential (\ref{RMT-GUE}). The Hilbert space dimension was chosen to match that of cSYK with $L=9$ and $N=5$. The right panel features the same data, with logarithmic scale along horizontal axis, allowing more resolution on the initial part of the sequence.}
    \label{b-RMT-D126}
\end{figure}

Some preliminary numerical checks have indicated that RMT reproduces qualitatively the features observed in complex SYK with $q=4$, that is: saturation of the upper bound for Krylov space, $K=D^2-D+1$, and slow decrease of the $b$-sequence to zero after initial growth, with a non-perturbative slope of order $\sim-\frac{1}{K}\sim-e^{-2S}$, $S$ being the entropy (system size). Figure \ref{b-RMT-D126} depicts the Lanczos sequence of a system whose Hamiltonian $H$ is drawn from a GUE with potential:
\begin{equation}
\centering
\label{RMT-GUE}
V(H) = \frac{D}{2J^2}\text{Tr}\left(H^2\right)
\end{equation}
where $D$ is the Hilbert space dimension and $J$ is the coupling strength that sets energy scales. These preliminary checks make it natural to conjecture that the discussed features are universal in chaotic systems.

\section{Krylov space for \texorpdfstring{SYK\textsubscript{2}}{SYK2}}\label{Appx-SYK2}
Given the SYK$_2$ model described in \ref{MajoranaSYK-q2}, we can proceed to apply the Lanczos algorithm analytically to the operator $\mathcal{O}=\chi_A$.

We will use the Frobenius scalar product defined in the main text, recalling that the dimension of the Hilbert space of states in this case is $D=2^M$. It is not difficult to prove that:
\begin{equation}
    \centering
    \label{Majoranas-overlap}
    \left(\chi_i|\chi_j\right) = \frac{1}{2}\delta_{ij} ~.
\end{equation}

\begin{itemize}
    \item $\mathcal{O}_0$:
    Making use of (\ref{Majoranas-overlap}) one finds that our starting operator is not normalized, since $\left(\mathcal{O}|\mathcal{O}\right)=\frac{1}{2}$, so:
    \begin{equation}
        \centering
        \label{q2-O0}
        \mathcal{O}_0 = \frac{\mathcal{O}}{\sqrt{\left(\mathcal{O}|\mathcal{O}\right)}} = \sqrt{2}\mathcal{O}=\sqrt{2}\,\chi_A ~.
    \end{equation}
    
    \item $\mathcal{O}_1$:
    One first needs to compute $\mathcal{A}_1$:
    \begin{equation}
        \centering
        \label{q2-A1}
        \mathcal{A}_1 = \left[H,\mathcal{O}_0\right] = i\frac{\sqrt{2}}{2}\sum_{i,j=1}^{L=2M}m_{ij}\left[\chi_i\chi_j,\chi_A\right]
    \end{equation}
    We make use of the commutator of the fermionic bilinear with the single Majorana:
    \begin{equation}
        \centering
        \label{q2-Commut}
        \left[\chi_i\chi_j,\chi_A\right] = \delta_{Aj}\chi_i-\delta_{Ai}\chi_j
    \end{equation} 
    so that finally
    \begin{equation}
        \centering
        \label{q2-A1-final}
        \mathcal{A}_1=\left[H,\mathcal{O}_0\right] = i\sqrt{2}\sum_{\substack{i = 1 \\ i\neq A}}^{L=2M}m_{iA}\chi_i ~.
    \end{equation}
    
    The first Lanczos coefficient is:
    \begin{equation}
        \centering
        \label{q2-b1}
        b_1 = \sqrt{\left(\mathcal{A}_1|\mathcal{A}_1\right)}=\sqrt{2\sum_{\substack{i,j=1 \\ i\neq A \\ j\neq A}}^{L=2M}m_{iA}m_{jA}\left(\chi_i|\chi_j\right)}
    \end{equation}
    and recalling (\ref{Majoranas-overlap}) one finds:
    \begin{equation}
        \centering
        \label{q2-b1-final}
        b_1 = \sqrt{\sum_{\substack{i=1 \\ i\neq A}}^{L=2M}m_{iA}^2} ~.
    \end{equation}
    Thus, the next Krylov element is:
    \begin{equation}
        \centering
        \label{q2-O1}
        \mathcal{O}_1 = \frac{1}{b_1}\mathcal{A}_1 = \frac{i\sqrt{2}}{\sqrt{\mathlarger{\sum}_{\substack{i=1 \\ i\neq A}}^{L=2M}m_{iA}^2}} \sum_{\substack{i=1 \\ i\neq A}}^{L=2M}m_{iA}\,\chi_i ~.
    \end{equation}
    The form of (\ref{q2-b1-final}) allows the computation of the ensemble average of the first Lanczos coefficient, since the calculation will turn out to be simple if one makes use of spherical coordinates in sample space:
    \begin{equation}
        \centering
        \label{q2-Eb1}
        \mathds{E}\left(b_1\right) = \int d^{L-1} x\, b(\mathbf{x}) P(\mathbf{x})
    \end{equation}
    where we defined $b(\mathbf{x})\equiv\sqrt{\mathbf{x}^2}$, and note that there are $L-1$ random variables because (\ref{q2-b1-final}) only involves $L-1$ independent coupling strengths. The probability distribution is a product of Gaussians:
    \begin{equation}
        \centering
        \label{q2-b1-pdf}
        P(\mathbf{x}) = \prod _{i=1}^{L-1}\rho(x_i),\;\;\;\;\rho(x_i)=\frac{1}{\sigma \sqrt{2\pi}}e^{-\frac{x_i^2}{2\sigma^2}}
    \end{equation}
    where, in agreement with $ \mathds{E} (m_{ij}^2) = m^2/L$, the standard deviation is given by $\sigma = \frac{m}{\sqrt{L}}$.
    As promised, the use of spherical coordinates greatly simplifies the computation and the result is:
    \begin{equation}
        \centering
        \label{q2-Eb1-final}
        \mathds{E}\left(b_1\right) = \frac{\Gamma\left(\frac{L}{2}\right)}{\Gamma\left(\frac{L-1}{2}\right)}\sqrt{\frac{2}{L}}m\overset{L\to +\infty}{\longrightarrow} m ~.
    \end{equation}
    
    \item \textbf{More elements:} It can be an interesting task to compute in closed form the rest of the Lanczos sequence, but we can already anticipate what its length will be: As shown in (\ref{q2-Commut}), the commutator of a bilinear with a single-site operator will still yield a linear combination of Majoranas on a single site, so $\mathcal{A}_2=\left[H,\mathcal{O}_1\right]-b_1 \mathcal{O}_0$ will still be a one-site operator, and so will the rest of the Krylov elements. Hence the maximum possible number of linearly independent Krylov elements will be given by the number of sites of the system, that is to say, in this integrable system the Krylov dimension $K$ is bounded by:
    \begin{equation}
        \centering
        \label{q2-KDIM}
        K\leq L \sim \log D \ll D^2-D+1
    \end{equation}
    i.e. the Krylov space scales at most linearly with entropy (system size), instead of with the operator space dimension (exponential in entropy).
\end{itemize}

\section{Numerical algorithms}\label{Appx-algorithms}

It is known that the original Lanczos algorithm, as presented in section \ref{Sec: Lanczos algorithm}, features an important numerical instability \cite{PRO,SO,Parlett}: the construction of each Krylov element makes use of the two previous ones, so errors due to finite-precision arithmetic accumulate dramatically and orthogonality of the Krylov basis is soon lost in numerical computations. Residual overlaps between Krylov elements grow exponentially (or even faster) with the iteration number $n$, which makes the Lanczos coefficients unreliable after a few iterations. In particular, the original Lanczos algorithm does not feature the termination of the sequence by hitting a zero at $n=K$ when run with finite precision, and instead it generally outputs a Lanczos sequence that oscillates wildly around some constant value, whose disorder average yields a completely flat sequence (after initial growth) in complex SYK; this is purely a product of numerical precision errors and needs to be corrected in order to shed light on the structure of the $b$-sequence along the full length of the Krylov chain. Some numerical algorithms need to be implemented in order to cure this instability, allowing observation of the slow decay to zero of the Lanczos coefficients after the initial growth in complex SYK. Such algorithms are described below.

\subsection{Full Orthogonalization (FO)}

This algorithm \cite{Parlett} performs a brute-force re-orthogonalization of the newly constructed Krylov element with respect to the previous ones at every iteration of the Lanczos algorithm, ensuring orthonormality of the Krylov basis up to machine precision $\varepsilon_M$. The algorithm is not very efficient time-wise, and is also costly in terms of memory, since the whole Krylov basis needs to be stored and is used in every iteration, it is therefore used mainly when one wants to compute only part of the Lanczos-sequence (see for example the work \cite{Yates_2020}). However, for small samples, FO can be used safely and one can check that the results yielded agree with the theoretical predictions described previously in this article (length of the Lanczos sequence, termination by hitting a zero, eigenvalues of the tri-diagonal matrix matching those corresponding to the eigenspace representatives that span the Krylov space). 

The FO algorithm amounts to performing explicit Gram-Schmidt at every iteration in the Lanczos algorithm to ensure orthogonality (up to machine precision). For numerical purposes, it is usually optimal to perform Gram-Schmidt \textit{twice} every time:

\begin{enumerate}
    
    \item $\left|\mathcal{O}_0\right) = \frac{1}{\sqrt{\left(\mathcal{O}|\mathcal{O}\right)}}\left|\mathcal{O}\right)$.
    \item For $n\geq 1$: Compute $\left|\mathcal{A}_n\right) = \mathcal{L}\left|\mathcal{O}_{n-1}\right)$.
    \item Re-orthogonalize $\left|\mathcal{A}_n\right)$ explicitly with respect to all previous Krylov elements: \\
    $\left|\mathcal{A}_n\right)\longmapsto \left|\mathcal{A}_n\right)-\sum_{m=0}^{n-1}\left|\mathcal{O}_m\right)\left(\mathcal{O}_m|\mathcal{A}_n\right)$.
    \item Repeat step 3.
    \item Set $b_n = \sqrt{\left(\mathcal{A}_n|\mathcal{A}_n\right)}$.
    \item If $b_n=0$ stop; otherwise set $\left|\mathcal{O}_n\right) = \frac{1}{b_n}\left|\mathcal{A}_n\right)$ and go to step 2.
\end{enumerate}

\subsection{Partial Re-Orthogonalization (PRO)}

This algorithm \cite{PRO} allows the residual overlaps between Krylov elements to grow up to a certain threshold, and re-orthogonalization is only performed when the threshold is crossed. For a machine precision $\varepsilon_M$, the threshold is typically taken to be $\sqrt{\varepsilon_M}$.

In our notation, the recursion relation for the Krylov basis is, including finite-precision errors:
\begin{equation}
\centering
\label{Lanczos-recursion}
b_{n}\left| \mathcal{O}_n \right) = \mathcal{L}\left| \mathcal{O}_{n-1} \right)-b_{n-1}\left| \mathcal{O}_{n-2} \right)+\left| \xi_{n-1} \right)
\end{equation}
where $\left| \xi_{n-1} \right)$ accounts for some spurious vector generated by accumulated numerical errors. All the objects denoted above represent the \textit{quantities numerically computed}, rather than the actual (analytically exact) Lanczos coefficients and Krylov elements. Acting with $\left( \mathcal{O}_{k} \right|$ from the left:
\begin{equation}
\centering
\label{Act-from-left}
b_{n}\left( \mathcal{O}_{k} \right.\left| \mathcal{O}_n \right) = \left( \mathcal{O}_{k} \right|\mathcal{L}\left| \mathcal{O}_{n-1} \right)-b_{n-1}\left( \mathcal{O}_{k} \right.\left| \mathcal{O}_{n-2} \right)+\left( \mathcal{O}_{k} \right.\left| \xi_{n-1} \right)
\end{equation}
where we recall that $\left( \mathcal{O}_{k} \right|\mathcal{L}\left| \mathcal{O}_{n-1} \right) = T_{k,n-1}$, $T$ being the tri-diagonal symmetric matrix built out of the Lanczos sequence, see (\ref{L-Tridiagonal}). We now define the matrix $W$, with items $W_{kn}=\left( \mathcal{O}_{k} \right.\left| \mathcal{O}_n \right)$, that is:
\begin{equation}
\label{w-items}
\centering
{\displaystyle
\left(W_{kn}\right)=\begin{pmatrix}
W_{00}=\left( \mathcal{O}_{0} \right.\left| \mathcal{O}_0 \right)&W_{01}=\left( \mathcal{O}_{0} \right.\left| \mathcal{O}_1 \right)&W_{02}=\left( \mathcal{O}_{0} \right.\left| \mathcal{O}_2 \right)&\dots\\
\cdot&W_{11}=\left( \mathcal{O}_{1} \right.\left| \mathcal{O}_1 \right)&W_{12}=\left( \mathcal{O}_{1} \right.\left| \mathcal{O}_2 \right)&\dots\\
\cdot&\cdot&W_{22}=\left( \mathcal{O}_{2} \right.\left| \mathcal{O}_2 \right)&\dots\\
\vdots&\vdots&\vdots&\ddots
\end{pmatrix}}
\end{equation}

Even though we do not write all the entries in the matrix, it is hermitian by definition of the scalar product $\left(\cdot | \cdot \right)$, so $W=W^\dagger\Longleftrightarrow W_{kn}=W_{nk}^{*}$. In the PRO algorithm, however, we construct iteratively the entries written explicitly in (\ref{w-items}): For a given $n$ we will want to estimate $W_{kn}$, for $k\leq n$. We do so by noticing that (\ref{Act-from-left}) is nothing but:
\begin{equation}
\centering
\label{Act-from-left-w}
b_{n} W_{kn} = T_{k,n-1} -b_{n-1} W_{k,n-2} + \left( \mathcal{O}_{k} \right.\left| \xi_{n-1} \right).
\end{equation}
Renaming indices $k\leftrightarrow n-1$ (i.e. $k\mapsto n-1$ and $n\mapsto k+1$):
\begin{equation}
\centering
\label{Act-from-left-w-rename-ind}
b_{k+1} W_{n-1,k+1} = T_{n-1,k} -b_{k} W_{n-1,k-1} + \left( \mathcal{O}_{n-1} \right.\left| \xi_{k} \right).
\end{equation}

Computing $(\ref{Act-from-left-w})-(\ref{Act-from-left-w-rename-ind})$, recalling that $T$ is symmetric and solving for $W_{kn}$:
\begin{equation}
\centering
\label{Recursion-w}
\begin{split}
W_{kn} & =\frac{1}{b_n}\left[b_{k+1} W_{k+1,n-1}^{*} + b_k W_{k-1,n-1}^{*}-b_{n-1} W_{k,n-2}\right. \\ 
& +\left. \left( \mathcal{O}_{k} \right.\left| \xi_{n-1} \right)- \left( \mathcal{O}_{n-1} \right.\left| \xi_{k} \right)\right] .
\end{split}
\end{equation}

We want to use (\ref{Recursion-w}) to determine, for a fixed $n$, all $W_{kn}$ with $k\leq n$ given that we know all $\left\{ W_{ij},\;i=0,...,j,\;\forall j=0,...,n-1 \right\}$ (according to what we depicted in (\ref{w-items}): we compute iteratively each upper-diagonal column making use of the previous ones). 

We note that $W_{nn}$ is not determined by (\ref{Recursion-w}) in terms of previous upper-diagonal columns, but we can set it to $W_{nn}=1$ because in the $n$-th Lanczos step, $\left| \mathcal{O}_n \right)$ is explicitly normalized to unity. Likewise, $W_{n-1,n}$ is not determined from (\ref{Recursion-w}) in terms of previous columns, and the Lanczos recursion (\ref{Lanczos-recursion}) does not explicitly orthogonalize $\left| \mathcal{O}_n \right)$ against $\left| \mathcal{O}_{n-1} \right)$, so we will need to orthogonalize them explicitly, and then set $W_{n-1,n}=\mathit{O}\left(\varepsilon_M\right)$ (i.e. zero up to machine precision).

Additionally, an estimate for $\left( \mathcal{O}_{k} \right.\left| \xi_{n-1} \right)- \left( \mathcal{O}_{n-1} \right.\left| \xi_{k} \right)$ is needed. One can take something of order of the machine precision times the norm of the Liouvillian:
\begin{equation}
\centering
\label{estimate}
\Big{|}\left( \mathcal{O}_{k} \right.\left| \xi_{n-1} \right)- \left( \mathcal{O}_{n-1} \right.\left| \xi_{k} \right)\Big{|} \sim 2\varepsilon_M\|\mathcal{L}\|
\end{equation}
where $\big{|}\big{|}\mathcal{L}\big{|}\big{|}$ should be the norm of the Liouvillian induced by the scalar product $\left(\cdot | \cdot\right)$ in the Hilbert space of operators. For practical applications, this contribution can just be ignored, since in any case each iteration of the algorithm will already generate spurious errors of order $\varepsilon_M$.

All in all, the LanPRO algorithm reads:
\begin{itemize}
	\item Compute $\left|\mathcal{O}_{0}\right)=\frac{1}{\sqrt{\left(\mathcal{O}\right.\left|\mathcal{O}\right)}}\left|\mathcal{O}\right)$.
	\begin{itemize}
		\item Set $W_{00}=1$.
	\end{itemize}
	\item Compute $\left|\mathcal{A}_{1}\right) = \mathcal{L}\left|\mathcal{O}_{0}\right)$. 
		
		\begin{itemize}
		
		\item Orthogonalize it explicitly with respect to $\left|\mathcal{O}_{0}\right)$.
		\item Compute $	b_1=\sqrt{\left(\mathcal{A}_1|\mathcal{A}_1\right)}$. \textbf{If} $b_1<\sqrt{\varepsilon_M}$ \textbf{stop.} Otherwise compute $\left| \mathcal{O}_1 \right)=\frac{1}{b_1}\left|\mathcal{A}_{1}\right) $.
		\item Set $W_{01}=\varepsilon_M$ and $W_{11}=1$.
		\end{itemize}
	
	\item \textbf{Loop} for $n\geq 2$, and for every $n$ \textbf{do}:
	\begin{itemize}
		\item Compute $\left|\mathcal{A}_{n}\right) = \mathcal{L}\left|\mathcal{O}_{n-1}\right)-b_{n-1}\left|\mathcal{O}_{n-2}\right)$.
		\item Compute the \textit{a-priori Lanczos coefficient:} \\ $b_n=\sqrt{\left(\mathcal{A}_n|\mathcal{A}_n\right)}$.
		\item \textbf{If} $b_n<\sqrt{\varepsilon_M}$ \textbf{break}, otherwise continue...

		\item Orthogonalize explicitly $\left| \mathcal{A}_n \right)$ with respect to $\left| \mathcal{O}_{n-1}\right)$.
		\item Set $W_{n,n}=1$ and $W_{n-1,n}=\varepsilon_M$.
		\item \textbf{Loop} for all $k=0,...,n-2$, determine $W_{kn}$ \textbf{doing}:
		\begin{equation}
		\centering
		\label{Assign-w}
		\begin{split}
		&\widetilde{W} = b_{k+1} W_{k+1,n-1}^{*} + b_k W_{k-1,n-1}^{*}-b_{n-1} W_{k,n-2}  \\
		&W_{kn}=\frac{1}{b_n}\left[\widetilde{W}+\frac{\widetilde{W}}{\big|\widetilde{W}\big|}\cdot 2\varepsilon_M\|\mathcal{L}\| \right]
		\end{split}	
		\end{equation}
		
		\item\textbf{If} there is some $k\leq n-2$ such that $W_{kn}>\sqrt{\varepsilon_M}$,  \textbf{do}: 
		\begin{itemize}
			\item Re-orthogonalize explicitly $\left| \mathcal{A}_n \right)$ and $\left| \mathcal{A}_{n-1} \right)$ with respect to all previous Krylov elements.
			\item From the new $\left|\mathcal{A}_{n-1}\right)$, re-compute $b_{n-1}$. \textbf{Break} if $b_{n-1}<\sqrt{\varepsilon_M}$, \textbf{otherwise} re-compute $\left|\mathcal{O}_{n-1}\right)$.
			\item From the new $\left|\mathcal{A}_{n}\right)$, re-compute $b_{n}$. \textbf{Break} if $b_{n}<\sqrt{\varepsilon_M}$, \textbf{otherwise} compute $\left|\mathcal{O}_{n}\right)=\frac{1}{b_n}\left|\mathcal{A}_n\right)$.
			\item Set $W_{a,n-1}=\delta_{a,n-1}$+$\left(1-\delta_{a,n-1}\right)\varepsilon_M$, for all $a=0,...,n-1$.
			\item Set $W_{a,n}=\delta_{a,n}$+$\left(1-\delta_{a,n}\right)\varepsilon_M$, for all $a=0,...,n$.
		\end{itemize}
		\item \textbf{Otherwise} compute  $\left|\mathcal{O}_{n}\right)=\frac{1}{b_n}\left|\mathcal{A}_n\right)$.
		
	\end{itemize}
 
\end{itemize}

(End of algorithm).

Some comments are in order:

\begin{itemize}
	\item In every iteration $n\geq 2$, only the two previous upper-diagonal columns $\left\{W_{k,n-1}\right\}_{k=0}^{n-1}$ and $\left\{W_{k,n-2}\right\}_{k=0}^{n-2}$ of the matrix $W$ are required. This is why, whenever re-orthogonalization is required, one only carries it out for $\left|\mathcal{A}_{n-1}\right)$ and $\left|\mathcal{A}_{n}\right)$.
	\item Whenever re-orthogonalization is needed, it is optimal to perform it twice, in the same way Gram-Schmidt is applied twice at every step of the FO algorithm.
\end{itemize}

In the case of complex SYK$_4$, for the system sizes studied in this article and a typical floating point precision of $\varepsilon_M\sim 10^{-15}$, PRO reduces the number of re-orthogonalizations required by approximately a factor of $10$, as compared to FO.

%% file: content/AppxCh04.tex
\chapter{\rm\bfseries Appendices to Chapter \ref{ch:chapter04_Integrable}}
\label{ch:AppxCh04}

\section{Moments and Hankel determinants} \label{Appx_Hankel_Det}
In this Appendix we show how to arrive at (\ref{Dn}) starting with (\ref{Hankel_Det}).  Firstly, let us look at 
\begin{eqnarray}
    D_{K-1} &=& \begin{vmatrix} 
    \mu_0 & \mu_1 & \mu_2 & \dots & \mu_{K-1} \\
    \mu_1 & \mu_2 & \mu_3 & \dots & \mu_{K}  \\
    \vdots & \vdots & \vdots & \dots & \vdots \\
    \mu_{K-1} & \mu_K & \mu_{K+1} & \dots & \mu_{2K-2}
    \end{vmatrix} ~.
\end{eqnarray}
Given (\ref{moments}), the matrix in the determinant can be written as
\begin{eqnarray} \label{DK-1}
    \begin{pmatrix} 
    1 & \sum_i |O_i|^2 \omega_i & \sum_i |O_i|^2 \omega_i^2 & \dots & \sum_i |O_i|^2 \omega_i^{K-1} \\
    \sum_i |O_i|^2 \omega_i & \sum_i |O_i|^2 \omega_i^2 & \sum_i |O_i|^2 \omega_i^3 & \dots & \sum_i |O_i|^2 \omega_i^K  \\
    \vdots & \vdots & \vdots & \dots & \vdots \\
    \sum_i |O_i|^2 \omega_i^{K-1} & \sum_i |O_i|^2 \omega_i^K & \sum_i |O_i|^2 \omega_i^{K+1} & \dots & \sum_i |O_i|^2 \omega_i^{2K-2}
    \end{pmatrix}\,,
\end{eqnarray}
where we have assumed the operator is normalized $\sum_{i=0}^{K-1}|O_i|^2=1$. (\ref{DK-1}) can be decomposed as
\small{\begin{eqnarray}
    \begin{pmatrix} 
    1 & 1  &  1 & \dots & 1 \\
    \omega_0 &  \omega_1 & \omega_2 & \dots & \omega_{K-1}  \\
    \vdots & \vdots & \vdots & \dots & \vdots \\
     \omega_0^{K-1} &  \omega_1^{K-1} &   \omega_2^{K-1} & \dots & \omega_{K-1}^{K-1}
    \end{pmatrix}
    \begin{pmatrix}
    |O_0|^2 & & & \\
     & |O_1|^2 & & \\
     & & \ddots & \\
     & & & |O_{K-1}|^2
    \end{pmatrix}
    \begin{pmatrix} 
    1 & \omega_0  &  \omega_0^2 & \dots & \omega_0^{K-1} \\
    1 &  \omega_1 &  \omega_1^2 & \dots &  \omega_1^{K-1} \\
    \vdots & \vdots & \vdots & \dots & \vdots \\
     1 & \omega_{K-1} &  \omega_{K-1}^2 & \dots &  \omega_{K-1}^{K-1}
    \end{pmatrix} \nonumber\,.
\end{eqnarray}}
In this form the determinant is immediate 
\begin{equation}
    D_{K-1} = \prod_{i=0}^{K-1} |O_i|^2 \prod_{0\leq i < j \leq K-1} (\omega_j-\omega_i)^2\,,
\end{equation} 
where the final product is the square of the Vandermonde determinant of the Liouvillian frequencies $\{\omega_i\}_{i=0}^{K-1}$.

Hankel determinants $D_n$ with $n<K-1$ are a bit more complicated since the Vandermonde matrices in this case are not square. The matrix in the determinant of $D_n$, is the product of matrices with sizes $(n+1\times K)$ and $(K \times n+1)$, namely
\small{\begin{eqnarray}
    \begin{pmatrix} 
    |O_0|^2 & |O_1|^2  &  |O_2|^2 & \dots & |O_{K-1}|^2 \\
    |O_0|^2\omega_0 &  |O_1|^2\omega_1 & |O_2|^2\omega_2 & \dots & |O_{K-1}|^2\omega_{K-1}  \\
    \vdots & \vdots & \vdots & \dots & \vdots \\
     |O_0|^2\omega_0^{n} &  |O_1|^2\omega_1^n &   |O_2|^2\omega_2^{n} & \dots & |O_{K-1}|^2\omega_{K-1}^{n}
    \end{pmatrix}
    \begin{pmatrix} 
    1 & \omega_0  &  \omega_0^2 & \dots & \omega_0^{n} \\
    1 &  \omega_1 &  \omega_1^2 & \dots &  \omega_1^{n} \\
    \vdots & \vdots & \vdots & \dots & \vdots \\
     1 & \omega_{K-1} &  \omega_{K-1}^2 & \dots &  \omega_{K-1}^{n}
    \end{pmatrix}\,.
\end{eqnarray}}
The determinant of a matrix which is a product of two rectangular matrices, can be reduced via the Cauchy-Binet formula (see for example \cite{enwiki:1030584213}) to the form (\ref{Dn}).

\section{Hilbert space dimension of XXZ sectors}\label{Appx_Sectors}

This Appendix presents the details of the computation of the dimensions of Hilbert space sectors of XXZ with fixed magnetization, parity and $R$-charge (whenever the latter is also a conserved quantum number within a given sector).

\subsection{Parity sectors at fixed magnetization}\label{Appx_Sectors_Parity}

We start by considering the sector $\mathcal{H}_M$ of fixed magnetization with $M$ spins up, whose dimension is given by:

\begin{equation}
    \centering
    \label{Dim-M}
    D_M = \binom{N}{M}\,,
\end{equation}
and we shall proceed to split it in sectors of fixed parity, $\mathcal{H}_M=\mathcal{H}_M^+\oplus \mathcal{H}_M^{-}$, whose dimensions we denote by $D_M^+$ and $D_M^{-}$, respectively. In order to study these parity sectors, it is convenient to use the natural tensor product basis of the spins in the XXZ chain. For the sake of language economics, we shall refer to it as the ``computational basis'', identifying ones with spins up and zeroes with spins down. If we examine individually each element $\ket{\psi}$ of this basis, we find the two mutually exclusive possibilities:

\begin{itemize}
    \item[a)] $P\ket{\psi}=\ket{\psi}$, i.e. $\ket{\psi}$ is already invariant under parity, hence it already belongs to $\mathcal{H}_M^+$.
    \item[b)] $\bra{\psi}P \ket{\psi}=0$, i.e. $\ket{\psi}$ is not invariant, so the action of parity gives a different element in the computational basis, and therefore their overlap vanishes due to orthogonality. For the states in this class, it immediately follows that $\ket{\psi}+P\ket{\psi}$ has positive parity while $\ket{\psi}-P \ket{\psi}$ has negative parity.
\end{itemize}

We denote $A,B$ as the number of states in the computational basis that satisfy (a), (b), respectively. $B$ needs to be always even because if $\ket{\psi}$ belongs to (b) then so does $P \ket{\psi}$; the fact that $P^2=\mathds{1}$ groups the states in the class (b) pairwise and would lead to a contradiction if $B$ is odd. We note that the $\mathcal{H}_M^+$ sector receives $A$ states from class (a) and $\frac{B}{2}$ states from class (b), where the factor of $\frac{1}{2}$ comes from the fact that half of the independent linear combinations of pairs constructed in (b) have positive parity and half have negative parity. To summarize:

\begin{equation}
    \centering
    \label{Parity-dims-1}
    \begin{split}
        & D_M^+ = A+\frac{B}{2} \\
        &D_M^{-} = \frac{B}{2} ~.
    \end{split} 
\end{equation}
Since (a) and (b) are mutually exclusive, we have that:

\begin{equation}
    \centering
    \label{a-b-exclusive}
    A+B = D_M\,,
\end{equation}
and therefore:

\begin{equation}
    \centering
    \label{DplusDminus_A}
    \begin{split}
        &D_M^+ = \frac{D_M+A}{2}\\
        &D_M^{-}=\frac{D_M-A}{2}\,.
    \end{split}
\end{equation}
So all that is left is to determine $A$, i.e. the number of states in the computational basis that are invariant under parity. Since parity performs a mirror transformation with respect to the chain center, the general philosophy is to count the number of states in (a) by finding the number of ways to arrange half of the ones in the left half of the chain, as the requirement of invariance under parity will determine the position of the remaining ones in the other half. However, depending on whether $N$ and $M$ are even or odd, some subtleties need to be taken into account.

\begin{itemize}
    \item[1.] \textbf{$N$ odd:} In this case there is always a site at the chain center that is fixed under parity transformations. This resource shall be conveniently exploited.
    \begin{itemize}
        \item[1.1.] \textbf{$M$ odd:} We need to put a one on the chain center, and then find all possible ways to arrange $\frac{M-1}{2}$ ones on the $\frac{N-1}{2}$ positions of the left side of the chain. That is:
        
        \begin{equation}
            \centering
            \label{A-1-1}
            A = \binom{\frac{N-1}{2}}{\frac{M-1}{2}}\,,
        \end{equation}
        
        and using (\ref{DplusDminus_A}) this yields:
        
        \begin{equation}
            \centering
            \label{DplusDminus_1_1}
            D_M^{\pm} = \frac{1}{2}\binom{N}{M}\pm\frac{1}{2}\binom{\frac{N-1}{2}}{\frac{M-1}{2}}\,.            \end{equation}
        \item[1.2.] \textbf{$M$ even:} In this case we cannot put a one on the chain center. We thus find all possibilities to arrange $\frac{M}{2}$ ones on $\frac{N-1}{2}$ positions:
        
        \begin{equation}
            \centering
            \label{A-1-2}
            A = \binom{\frac{N-1}{2}}{\frac{M}{2}}\,,
        \end{equation}
        
        which implies:
        
        \begin{equation}
            \centering
            \label{DplusDminus_1_2}
            D_M^{\pm}=\frac{1}{2}\binom{N}{M}\pm \frac{1}{2}\binom{\frac{N-1}{2}}{\frac{M}{2}} ~.
        \end{equation}
    \end{itemize}
    \item[2.] \textbf{$N$ even:} This time there is no fixed point under parity, which constrains the possibility of having states in class (a) at all.
    \begin{itemize}
        \item[2.1.] \textbf{$M$ odd:} In this case there is no way to arrange half of the ones in half of the chain, so simply:
        
        \begin{equation}
            \centering
            \label{A-2-1}
            A=0\,,
        \end{equation}
        and thus:
        \begin{equation}
            \centering
            \label{DplusDminus_2_1}
            D_M^+=D_M^{-}=\frac{1}{2}\binom{N}{M} ~.       \end{equation}
            \item[2.2.] \textbf{$M$ even:} We just need to find all possible ways to arange $\frac{M}{2}$ ones on $\frac{N}{2}$ sites:
            \begin{equation}
                \centering
                \label{A_2_2}
                A = \binom{\frac{N}{2}}{\frac{M}{2}}\,,
            \end{equation}
            so that:
            \begin{equation} \label{DplusDminus_2_2}
                \centering
                D_M^{\pm}=\frac{1}{2}\binom{N}{M}\pm\frac{1}{2}\binom{\frac{N}{2}}{\frac{M}{2}}\,.
                \end{equation}
    \end{itemize}
\end{itemize}
We have therefore succeeded in computing the dimension of all Hilbert space sectors of fixed magnetization and parity. The results are summarized in Table \ref{Table_Dim_ParitySectors}.

\subsection{\texorpdfstring{$R$}{R}-subsectors}\label{Appx_Sectors_R}

For XXZ chains with an even number of sites $N$ there exists a zero-magnetization sector with $M=\frac{N}{2}$ in which both $P$ and $R$ are conserved quantum numbers due to the symmetry enhancement discussed in Section \ref{Section_XXZ}. Therefore, we can further split the parity sectors into subsectors of fixed $R=\pm 1$, $\mathcal{H}_{\frac{N}{2}}^P=\mathcal{H}_{\frac{N}{2}}^{P,+}\oplus \mathcal{H}_{\frac{N}{2}}^{P,-}$. Their dimensions $D_{\frac{N}{2}}^{PR}$ can again be computed making use of the computational basis for the starting sector $\mathcal{H}_{\frac{N}{2}}$. Again splitting the elements of this basis into classes (a) and (b) defined earlier in \ref{Appx_Sectors_Parity}, we can make the following observations:

\begin{itemize}
    \item No state $\ket{\psi}$ in the computational basis is invariant under $R$. In particular, no state in (a) is invariant under $R$, even though they are all parity eigenstates with positive eigenvalue.
    \item There is a subset of the computational basis states $\ket{\psi}$ in (b) that satisfy:
    
    \begin{equation}
        \centering
        \label{States_P_R}
        P\ket{\psi} = R\ket{\psi}.
    \end{equation}
    
    For a state fulfilling this property, we can build two orthonormal combinations $\ket{\psi_+}$ and $\ket{\psi_-}$ as:
    
    \begin{equation}
        \centering
        \label{R-combinations}
        \ket{\psi_\pm}=\frac{1}{\sqrt{2}}\big(\ket{\psi}\pm P\ket{\psi}\big)=\frac{1}{\sqrt{2}}\big(\ket{\psi}\pm R\ket{\psi}\big). 
    \end{equation}
    
    And we note that $\ket{\psi_{\pm}}$ verifies simultaneously $P=R=\pm 1$. The states of the form of $\ket{\psi_+}$ introduce a bias towards $R=+1$ in the $P=+1$ sector, while those of the form $\ket{\psi_-}$ introduce a bias towards $R=-1$ in the $P=-1$ sector.
    
\end{itemize}

We denote the number of states fulfilling (\ref{States_P_R}) as $B^\prime<B$. In order to find this number, we need to count the total number of possible configurations of the left half of the spin chain: Once half the chain is determined, the other half is fixed by reflecting and ``flipping''. Note that the requirement of $M=\frac{N}{2}$ is baked in, since if there are $m$ 1's in one half of the chain, there will be $\frac{N}{2}-m$ 1's in the other half. This yields $B^\prime = 2^{\frac{N}{2}}$, and with this we are in conditions to proceed with the counting of the dimension of the $R$-subsectors within each parity sector:
    
    \begin{itemize}
        \item $\mathbf{P=+1.}$ We can enumerate the following contributions:
        
        \begin{itemize}
            \item[1.] $\frac{B^\prime}{2}$ out of the $\frac{B}{2}$ positive-parity states coming from the class (b) are of the form $\ket{\psi_+}$ in (\ref{R-combinations}) and thus have directly $R=+1$.
            \item[2.] The $\frac{B-B^\prime}{2}$ remaining positive-parity states coming from class (b) are not invariant under $R$.
            \item[3.] The $A$ states in class (a), which are all positive-parity states, are not invariant under $R$.
        \end{itemize}
        
        Hence, points $2$ and $3$ above give $A+\frac{B-B^\prime}{2}=D_{\frac{N}{2}}^+-\frac{B^\prime}{2}$ positive-parity states that are not invariant under $R$. Since $R^2=1$, we proceed analogously as we did in section \ref{Appx_Sectors_Parity} in order to construct parity eigenstates out of states in (b): We group them pairwise in linear combinations that give as many states with $R+1$ as states with $R=-1$. This yields:
        
        \begin{equation}
            \centering
            \label{Sectors_Pplus_R}
            \begin{split}
                &D_{\frac{N}{2}}^{++} = \frac{B^\prime}{2}+\frac{1}{2}\left(D_{\frac{N}{2}}^+-\frac{B^\prime}{2}\right) = \frac{1}{2}D_{\frac{N}{2}}^++2^{\frac{N-4}{2}} \\
                &D_{\frac{N}{2}}^{+-} = \frac{1}{2}\left(D_{\frac{N}{2}}^+-\frac{B^\prime}{2}\right) =\frac{1}{2}D_{\frac{N}{2}}^+-2^{\frac{N-4}{2}}\,.
            \end{split}
        \end{equation}
        
        \item $\mathbf{P=-1.}$ In this case we have the contributions:
        
        \begin{itemize}
            \item[1.] $\frac{B^\prime}{2}$ out of the $\frac{B}{2}$ negative-parity states coming from the class (b) are of the form $\ket{\psi_-}$ in (\ref{R-combinations}) and thus have directly $R=-1$.
            \item[2.] The remaining $\frac{B-B^\prime}{2}=D_{\frac{N}{2}}^--\frac{B^\prime}{2}$ negative-parity states coming from (b) are not $R$-invariant, and thus have to be grouped pairwise in linear combinations giving as many $R=+1$ as $R=-1$ eigenstates.
        \end{itemize}
        
        This yields the expressions:
        
        \begin{equation}
            \centering
            \label{Sectors_Pminus_R}
            \begin{split}
                & D_{\frac{N}{2}}^{-+} = \frac{1}{2}\left(D_{\frac{N}{2}}^--\frac{B^\prime}{2}\right)=\frac{1}{2}D_{\frac{N}{2}}^--2^{\frac{N-4}{2}} \\
                & D_{\frac{N}{2}}^{--}=\frac{B^\prime}{2}+\frac{1}{2}\left(D_{\frac{N}{2}}^--\frac{B^\prime}{2}\right)=\frac{1}{2}D_{\frac{N}{2}}^-+2^{\frac{N-4}{2}}\,. 
            \end{split}
        \end{equation}
        
    \end{itemize}
    Happily, equations (\ref{Sectors_Pplus_R}) and (\ref{Sectors_Pminus_R}) can be merged together into a single expression:
    
    \begin{equation}
        \centering
        \label{Sectors_P_R_appx}
        D_{\frac{N}{2}}^{PR}=\frac{1}{2}D_{\frac{N}{2}}^P+P\cdot R \cdot 2^{\frac{N-4}{2}},\quad\quad\quad\text{for }N \text{ even.}
    \end{equation}

\subsubsection{Selection rules and Krylov dimension for operators charged under \texorpdfstring{$R$}{R}}\label{Appx_selection_rules_Krylov}
Because of the $R$-symmetry enhancement, operators charged under $R$ are constrained by selection rules that enforce nullity of a part of their matrix elements in the energy basis, which has the effect of reducing the dimension of their associated Krylov space. The reason why these selection rules exist is that eigenstates of the Hamiltonian are themselves $R$-eigenstates, rather than being degenerate, as proposed in \cite{Doikou:1998jh} and as we have confirmed performing numerical checks. We shall now prove explicitly those selection rules and give the expression for the Krylov dimension that they imply in each case.

\begin{itemize}
    \item \textbf{Negatively charged operator.} We consider an observable $\mathcal{O}$ acting on the sector $\mathcal{H}_{\frac{N}{2}}^P$ that verifies:
    
    \begin{equation}
        \centering
        \label{Op_R_neg_charged}
        R \mathcal{O} R^{\dagger} = -\mathcal{O}\quad\Longleftrightarrow \quad\left[R,\mathcal{O}\right]=-2\mathcal{O}R  ~.
    \end{equation}
    
    Since eigenstates in the sector of zero magnetization and fixed parity $P$ of this chain of even length are also eigenstates of $R$ with eigenvalue $\pm 1$, it follows that the action of $\mathcal{O}$ on these states has the effect of changing the sign of the $R$-eigenvalue. To show this, we shall label eigenstates in the sector explicitly with their energy and $R$-eigenvalues, $\ket{E,r}$ with $r=\pm 1$. Using (\ref{Op_R_neg_charged}) it is possible to show that:

\begin{equation}
    \centering
    \label{R-flip}
    R \mathcal{O}\ket{E,r}=-r\,\mathcal{O}\ket{E,r} ~.
\end{equation}
That is: While $\ket{E,r}$ is an eigenstate of $R$ with eigenvalue $r$, $\mathcal{O}\ket{E,r}$ is an eigenstate of $R$ with eigenvalue $-r$. This in turn implies the selection rule:

\begin{equation}
    \centering
    \label{Selection-rule}
    \langle E^\prime,r^\prime | \mathcal{O}|E,r\rangle=0 \quad\text{if}\quad r^\prime = r ~.
\end{equation}
But (\ref{Selection-rule}) are the matrix elements of the observable in the energy basis, so they are direcly related to the Krylov dimension, and the fact that many of them vanish identically because of the selection rule has the effect of lowering the Krylov dimension with respect to its upper bound $D^2-D+1$ (where we use $D$ instead of $D_{\frac{N}{2}}^P$ for the sake of notational simplicity). To give the upper bound for the Krylov dimension in this case, let us assume that all matrix elements that are not forced to vanish by the selection rule are non-zero. Then, the number of non-zero matrix elements $\langle E^\prime,r^\prime | \mathcal{O}|E,r\rangle$ (and hence the Krylov space dimension, as there are no degeneracies in the spectrum of the Hamiltonian in this sector) is given by:

\begin{equation}
    \centering
    \label{K-selection}
    K = 2R^+R^-\,,
\end{equation}
where, again for notational simplicity, $R^\pm \equiv D_{\frac{N}{2}}^{P,R=\pm}$ denote schematically the number of eigenstates with $R$-eigenvalues $r=\pm1$ in the sector $\mathcal{H}_{\frac{N}{2}}^P$.

\item $\mathbf{R}$\textbf{-invariant operator.} In this case, we consider an observable $\mathcal{O}$ acting on the sector $\mathcal{H}_{\frac{N}{2}}^P$ that verifies:

    \begin{equation}
        \centering
        \label{Op_R_pos_charged}
        R \mathcal{O} R^{\dagger} = \mathcal{O}\quad\Longleftrightarrow \quad\left[R,\mathcal{O}\right]=0 ~.
    \end{equation}
    
    The selection rule in this case is: 
    
    \begin{equation}
    \centering
    \label{TwoSiteCenter_Selection}
    \langle E^\prime,r^\prime|\mathcal{O}|E,r\rangle = 0 \quad \text{if}\quad r^\prime = -r ~.
\end{equation}
Again, assuming that all matrix elements unconstrained by (\ref{TwoSiteCenter_Selection}) are non-zero, and noting that some of them are diagonal elements, we reach the prediction:

\begin{eqnarray}
    \text{Number of non zero matrix elements:} \quad (R^+)^2+(R^-)^2 \\
    K = R^+(R^+-1)+R^-(R^--1)+1=(R^+)^2+(R^-)^2-D+1\,,
    \label{K_R_inv}
\end{eqnarray}
where again $D\equiv D_{\frac{N}{2}}^P$ and $R^\pm \equiv D_{\frac{N}{2}}^{P,R=\pm}$ for simplicity. In fact, since $\mathcal{O}$ is invariant under $R$, we could have just considered separately its restrictions on $\mathcal{H}_{\frac{N}{2}}^{P,+}$ and $\mathcal{H}_{\frac{N}{2}}^{P,-}$. It is apparent from this perspective that (\ref{K_R_inv}) is nothing but the combination of the upper bounds of the form (\ref{Krylov-bound}) applied separately to each $R$-sector, where the operator is dense in the energy basis.
\end{itemize}

\begin{figure}[ht]
     \centering
     \begin{subfigure}[b]{0.4\textwidth}
         \centering
         \includegraphics[width=\textwidth]{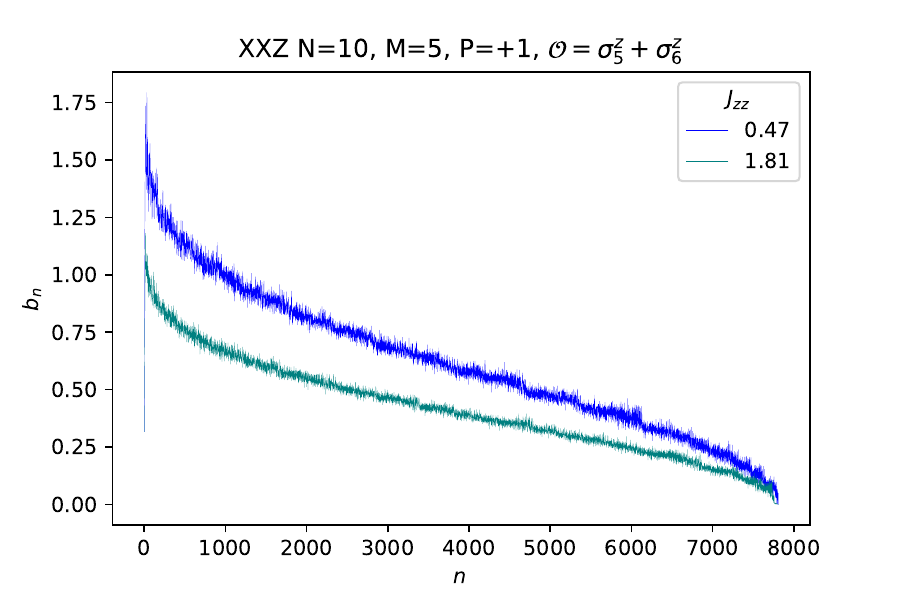}
         \caption{Lanczos sequences}
         \label{fig:Lanczos_N10}
     \end{subfigure}
     \begin{subfigure}[b]{0.25\textwidth}
         \centering
         \includegraphics[width=\textwidth]{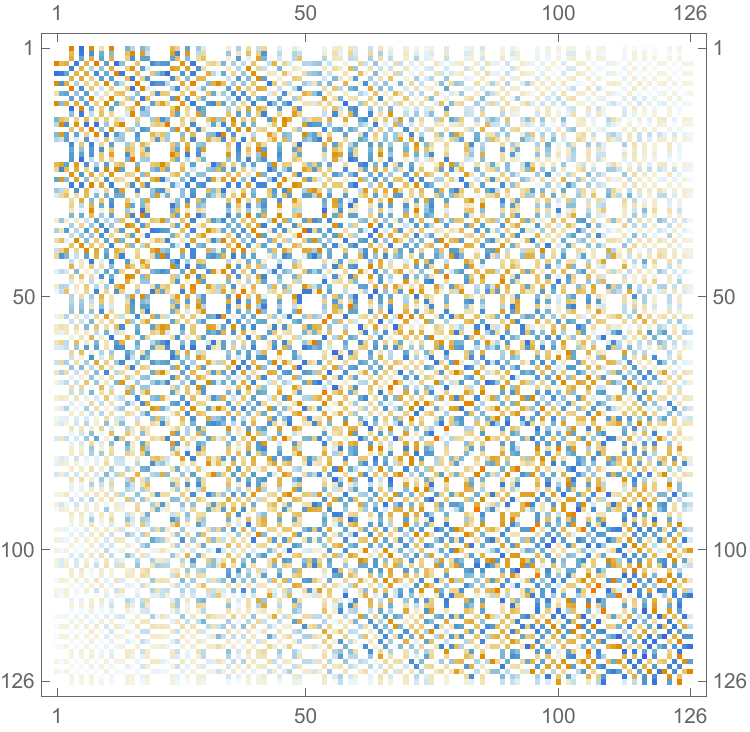}
         \caption{$J_{zz}=0.47$}
         \label{fig:Jzz047}
     \end{subfigure}
     \begin{subfigure}[b]{0.25\textwidth}
         \centering
         \includegraphics[width=\textwidth]{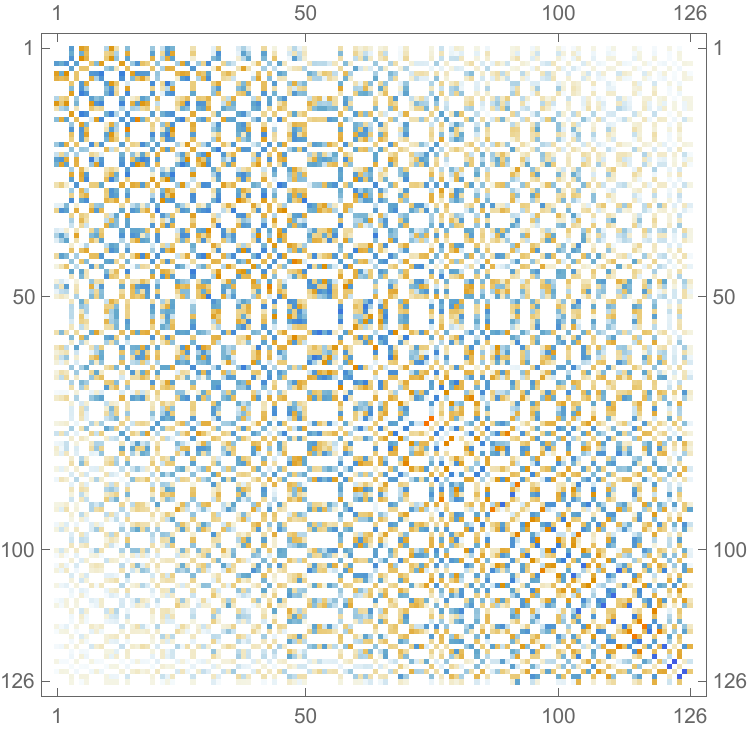}
         \caption{$J_{zz}=1.81$}
         \label{fig:Jzz181}
     \end{subfigure}
        \caption{For XXZ with $N=10$ in the sector $M=5,\, P=+1$ for which the Hilbert space dimension is $D=126$ according to Table \ref{Table_Dim_ParitySectors}, the operator $\mathcal{O}=\sigma_5^z+\sigma_6^z$ is not dense in the energy basis.  The left panel shows the Lanczos sequences which terminate at $K=7810$, and the middle and right panels show matrix plots of the operator in the energy basis for the two $J_{zz}$ couplings for which the Lanczos sequences were computed.}
        \label{fig:XXZ_N10_M5}
\end{figure}

Consider the operator we study, $\mathcal{O}=\sigma_m^z+\sigma_{N-m+1}^z$  in an XXZ system with an even number of spins $N$ in the zero magnetization sector $M=N/2$ and $P=+1$. As discussed in Section \ref{Section_XXZ}, in this sector both $P$ and $R$ are symmetries of the Hamiltonian. It can be shown, using the property $\sigma^j\sigma^k = \delta^{jk}+i\varepsilon^{jkl}\sigma^l$ of the Pauli matrices at a fixed site, that the operator $\mathcal{O}$ chosen here satisfies (\ref{Op_R_neg_charged}), thus being negatively charged under $R$. We therefore expect it to span (in the absence of other degeneracies in the spectrum) a Krylov space of dimension (\ref{K-selection}). We confirm this expectation for the case of XXZ with $N=10, M=5$ in the $P=+1$ sector with the operator $\mathcal{O}=\sigma^z_5+\sigma^z_6$, where we find the Krylov space dimension to be $K=7810$, as indeed predicted by (\ref{K-selection}), once we work with high enough accuracy.  In Figure \ref{fig:XXZ_N10_M5} we show the Lanczos sequences as well as matrix plots of the operator in the energy basis, for two different $J_{zz}$ coefficients.

\section{Analytical results for a constant \texorpdfstring{$b$}{b}-sequence}\label{appx:Constant_b_analytics}
To contrast with the disordered systems we study in the main text, we discuss here a flat, smooth $b$-sequence with no disorder. Such systems were considered in the context of Anderson localization, for example in \cite{luck:cea-01485001}.  

Consider the eigenvalue problem (\ref{EV_problem}) with $b_n=1$ (in some units) for all $n$, i.e.
\begin{equation}
    \omega \psi_n = \psi_{n+1}+\psi_{n-1}, \quad n=0,1,\dots, K-1\,,
\end{equation}
where we want to find $\psi_n$ and $\omega$ which solve this equation. 
Assuming boundary conditions $\psi_{-1}=\psi_{K}=0$, it can be solved via the transfer matrix recurrence equation:
\begin{equation}
    \begin{pmatrix}
        \psi_{n+1}\\ \psi_{n} 
    \end{pmatrix} = 
    \begin{pmatrix}
        \omega & -1 \\ 1 & 0
    \end{pmatrix}
    \begin{pmatrix}
        \psi_n\\ \psi_{n-1}
    \end{pmatrix}~.
\end{equation}
The solution in terms of the transfer matrix and boundary conditions is
\begin{equation}
    \begin{pmatrix}
        \psi_{n}\\ \psi_{n-1} 
    \end{pmatrix} = 
    \begin{pmatrix}
        \omega & -1 \\ 1 & 0
    \end{pmatrix}^n
    \begin{pmatrix}
        \psi_0\\ 0
    \end{pmatrix}~.
\end{equation}
Finding the eigenvalues and eigenvectors of the transfer matrix, it can be written as
\begin{equation}
    \begin{pmatrix}
        \omega & -1 \\ 1 & 0
    \end{pmatrix} = \frac{1}{\lambda_1-\lambda_2}\begin{pmatrix}
        \lambda_1 & \lambda_2 \\ 1 & 1
    \end{pmatrix} \begin{pmatrix}
        \lambda_1 & 0 \\ 0 & \lambda_2
    \end{pmatrix}\begin{pmatrix}
        1 & -\lambda_2 \\ -1 & \lambda_1
    \end{pmatrix}\,,
\end{equation}
where $\lambda_{1,2}=\frac{1}{2}\left(\omega\pm \sqrt{\omega^2-4} \right)$.  This allows us to write the solution for any $n$ in terms of $\psi_0$ and $\lambda_{1,2}$, as follows:
\begin{equation} \label{eigvec_eq}
    \begin{pmatrix}
        \psi_{n}\\ \psi_{n-1} 
    \end{pmatrix} = 
    \frac{1}{\lambda_1-\lambda_2}\begin{pmatrix}
        \lambda_1 & \lambda_2 \\ 1 & 1
    \end{pmatrix} \begin{pmatrix}
        \lambda_1^n & 0 \\ 0 & \lambda_2^n
    \end{pmatrix}\begin{pmatrix}
        1 & -\lambda_2 \\ -1 & \lambda_1
    \end{pmatrix}\begin{pmatrix}
        \psi_0\\ 0
    \end{pmatrix} = \frac{\psi_0}{\lambda_1-\lambda_2} \begin{pmatrix}
        \lambda_1^{n+1}-\lambda_2^{n+1}\\ \lambda_1^n -\lambda_2^n
    \end{pmatrix}~.
\end{equation}
Now we can use the other boundary condition, namely $\psi_{K}=0$:
\begin{equation}
    \begin{pmatrix}
        0\\ \psi_{K-1} 
    \end{pmatrix} = \frac{\psi_0}{\lambda_1-\lambda_2} \begin{pmatrix}
        \lambda_1^{K+1}-\lambda_2^{K+1}\\ \lambda_1^K -\lambda_2^K
    \end{pmatrix}\,,
\end{equation}
from which we extract the condition $\lambda_1^{K+1}-\lambda_2^{K+1}=0$, which in terms of $\omega$ reads
\begin{equation}\label{E_condition}
    \left(\omega+\sqrt{\omega^2-4}\right)^{K+1} - \left(\omega-\sqrt{\omega^2-4}\right)^{K+1} = 0 ~.
\end{equation}
A family of solutions is
\begin{equation}\label{E_no_disorder}
    \omega_i = 2 \cos\left( \frac{\pi (i+1)}{K+1}\right), \quad i=0,1,2,\dots, K-1 ~.
\end{equation}
It is also clear now from (\ref{eigvec_eq}), that the eigenvector element $\psi_{ni}$ for eigenvalue $\omega_i$ is:
\begin{equation}
    \centering
    \label{eigenvector-disorder-free}
    \psi_{ni} = \psi_{0i}\frac{\sin \left(\frac{(n+1)(i+1)}{K+1}\pi\right)}{\sin \left(\frac{i+1}{K+1}\pi\right)}\equiv C_i \sin \left(\frac{(n+1)(i+1)}{K+1}\pi\right),\;\;n,i=0,1,2,...,K-1\,,
\end{equation}
where we have absorbed the denominator in the normalization constant because it does not depend on $n$ (recall that the label $i$ designates the eigenstate). $C_i$ can now be fixed from normalization, and the normalized eigenstates are
\begin{eqnarray}
   \begin{pmatrix}
       \psi_{0i}\\
       \psi_{1i}\\
       \vdots \\
       \psi_{ni}\\
       \vdots \\
       \psi_{K-1,i}
   \end{pmatrix} =
   \sqrt{\frac{2}{K+1}} \begin{pmatrix}
        \sin \frac{ (i+1) \pi}{K+1} \\
        \sin \frac{2 (i+1) \pi}{K+1} \\
        \vdots \\
        \sin \frac{(n+1) (i+1) \pi}{K+1} \\
        \vdots \\
        \sin \frac{K (i+1) \pi}{K+1}
   \end{pmatrix}, \quad i = 0,1,2, \dots, K-1~.
\end{eqnarray}
In terms of Krylov elements, we found that
\begin{eqnarray}
    ( \mathcal{O}_n |\omega_i) = \psi_{ni}= \sqrt{\frac{2}{K+1}} \sin \frac{(n+1) (i+1) \pi}{K+1} \quad  i,n=0,1,\dots, K-1\,, 
\end{eqnarray}
from which we can compute K-complexity for the pure no-disorder system:
\begin{eqnarray}
    \overline{C_K} = \sum_{i=0}^{K-1} |( \mathcal{O}_0 |\omega_i)|^2   \sum_{n=0}^{K-1} n\,  |(\mathcal{O}_n |\omega_i)|^2 = \sum_{n=0}^{K-1} n \sum_{i=0}^{K-1} \psi_{0i}^2 \psi_{ni}^2 ~.
\end{eqnarray}
The transition probability (\ref{Transition_Probability}) is\footnote{To arrive at this result we used $
    \sum_{n=0}^{N}e^\frac{2i\pi m n}{N+1} = \frac{1-e^{2i\pi m}}{1-e^{\frac{2i\pi m }{N+1}}} = \begin{cases}
        0, & (N+1) \text{ not } |m\\
        N+1, & (N+1) | m 
    \end{cases}$. See \cite{luck:cea-01485001}.}
\begin{eqnarray}
    Q_{0n} &=&
    \left(\frac{2}{K+1}\right)^2 \sum_{i=0}^{K-1} \sin^2\frac{ (i+1)\pi}{K+1} \sin^2 \frac{(n+1) (i+1) \pi}{K+1} \nonumber\\
    &=& \frac{1}{K+1}\left(1+\frac{1}{2} \delta_{n0}+\frac{1}{2}\delta_{n,K-1} \right)
\end{eqnarray}
and the time-averaged K-complexity is given by,
\begin{eqnarray}
    \overline{C_K} &=&\frac{1}{K+1} \sum_{n=0}^{K-1} n \left(1+\frac{1}{2} \delta_{n0}+\frac{1}{2}\delta_{n,K-1} \right)\nonumber\\ &=&\frac{K(K-1)}{2(K+1)} +0+  \frac{K-1}{2(K+1)}   = \frac{K}{2}\times O(1) +O(1)\sim \frac{K}{2} ~.
\end{eqnarray}
This analytical result shows that for a flat and smooth Lanczos-sequence K-complexity will saturate at $K/2$.


\section{Connected part of autocorrelation function and saturation value of K-complexity}\label{appx_Connected}
This Appendix analyses the impact of the operator one-point function on the saturation value of K-complexity at late times. 
Let us first remind how the operator one-point function dominates the two-point function plateau. We shall do so by assuming that the operator satisfies the Eigenstate Thermalization Hypothesis (ETH).
Consider a hermitian normalized operator whose elements in the energy basis are given by
\begin{eqnarray}
\label{Op_spectral_decomp}
    \mathcal{O}&=&\sum_{a,b=1}^D  O_{ab}|E_a\rangle \langle E_b|
\end{eqnarray}
where $D$ is the Hilbert space dimension. Note that in (\ref{Op_spectral_decomp}) the operator matrix elements in the energy basis are defined such that $O_{ab}=\langle E_a | \mathcal{O} | E_b \rangle$.
With this convention, the ETH Anstatz takes the usual form; suppressing energy dependence in the matrix elements of the Ansatz (since we are interested in order-of-magnitude estimates), it boils down to the RMT operator Ansatz:
\begin{equation}
    \centering
    \label{RMT_op_Ansatz}
    O_{ab} = O\delta_{ab}+\frac{1}{\sqrt{D}}r_{ab},
\end{equation}
where $O$ gives (up to non-perturbative corrections) the one-point function of the operator and is taken not to scale with $D$, and the matrix $(r_{ab})$ is drawn from a Gaussian ensemble with unit variance (and hence the elements $r_{ab}$ are also of order $D^0$). Note that the operator (\ref{Op_spectral_decomp}) with the matrix elements given by (\ref{RMT_op_Ansatz}) is normalized\footnote{By this, we mean that the norm of the operator whose matrix elements satisfy (\ref{RMT_op_Ansatz}) does not scale with $D$.} according to the operator inner product
\begin{eqnarray}
\label{Inner_prod_norm_one}
    \|\mathcal{O}\|^2 = \frac{1}{D} \mathrm{Tr}\Big[\mathcal{O}^\dagger \mathcal{O} \Big] = \frac{1}{D}\sum_{a,b=1}^D |O_{ab}|^2 = 1.
\end{eqnarray}
The autocorrelation function is given by
\begin{eqnarray}
\label{Two-pt-function}
    \phi_0(t)&=&\big\langle \mathcal{O}^\dagger \mathcal{O}(t)\big\rangle=\frac{1}{D} \mathrm{Tr}\Big[\mathcal{O}^\dagger \mathcal{O}(t) \Big] = \frac{1}{D} \sum_{a,b=1}^D |O_{ab}|^2 e^{i(E_a-E_b)t} ~.
\end{eqnarray}
Due to normalization of the operator (\ref{Inner_prod_norm_one}), the two-point function (\ref{Two-pt-function}) starts at $1$, i.e. $\phi_0(0)=1$. The Ansatz (\ref{RMT_op_Ansatz}) has some implications on the late-time behavior of $\phi_0(t)$, which we can study by performing a long-time average:

\begin{equation}
    \centering
    \label{Two-Pt-LongTAvg}
    \overline{\phi_0}:=\lim_{T\to+\infty}\frac{1}{T}\int_0^T dt\, \phi_0(t).
\end{equation}
We can now use that, for $\omega\neq 0$:
\begin{equation}
    \centering
    \label{Phases_average_out}
    \lim_{T\to +\infty}\frac{1}{T}\int_{0}^T dt\, e^{i\omega t}=\lim_{T\to +\infty}\frac{1}{T}\left. \left[ \frac{e^{i\omega t}}{i\omega} \right] \right|_{t=0}^T=\frac{1}{i\omega}\lim_{T\to+\infty}\frac{e^{i\omega T}-1}{T}=0.
\end{equation}
With this, assuming no exact degeneracies in the energy spectrum, the long-time average of (\ref{Two-pt-function}) eliminates the contribution of the off-diagonal matrix elements and yields:
\begin{equation}
    \centering
    \label{Two-Pt_LongT_Diagonal}
    \overline{\phi_0}=\frac{1}{D}\sum_{a=1}^D|O_{aa}|^2=\frac{1}{D}\sum_{a=1}^D \left( O + \frac{r_{aa}}{\sqrt{D}} \right)^2= O^2 + \frac{2O}{D^{3/2}}\sum_{a=1}^D r_{aa} + \frac{1}{D^2}\sum_{a=1}^D r_{aa}^2,
\end{equation}
where in the second equality we have used the Ansatz (\ref{RMT_op_Ansatz}). We thus conclude that the long-time average of the two-point function is dominated by the square of the one-point function. This fact is also in qualitative agreement with large-N factorization, i.e. in the thermodynamic limit the two-point function becomes disconnected at late times. 

In order to probe spectral correlations we can choose to subtract explicitly the one-point function squared from the auto-correlation function (\ref{Two-pt-function}), which defines the so-called connected two-point function:

\begin{equation}
    \centering
    \label{Two-pt-connected}
    \phi_0^{(c)}(t):= \Bigg\langle \bigg( \mathcal{O} - \langle \mathcal{O} \rangle\bigg)\bigg(\mathcal{O}(t)-\langle\mathcal{O}\rangle\bigg) \Bigg\rangle = \big\langle \mathcal{O}\mathcal{O}(t) \big\rangle-\big\langle \mathcal{O} \big\rangle^2,
\end{equation}
where, to alleviate notational crowding, we have implicitly assumed that $\mathcal{O}$ is hermitian, and we have made use of the fact that the one-point function is time-independent, $\big\langle \mathcal{O}(t) \big\rangle = \big\langle \mathcal{O} \big\rangle$. Again, making use of the expression of the operator $\mathcal{O}$ in the energy basis, we can write (\ref{Two-pt-connected}) as:
\begin{equation}
    \centering
    \label{Two_Pt_Conn_Ebasis}
    \phi_0^{(c)}(t) = \frac{1}{D}\sum_{a,b=1}^D|O_{ab}|^2 e^{it(E_a-E_b)}\,-\,\frac{1}{D^2}\sum_{a,b=1}^DO_{aa}O_{bb}, 
\end{equation}
where we have used that $\langle \mathcal O \rangle = \frac{1}{D}\text{Tr}[\mathcal O]$ is the infinite-temperature one-point function of the operator $\mathcal{O}$. As defined in (\ref{Two-pt-connected}), the connected two-point function is not normalized so that its value at $t=0$ is exactly one, but this is not important because we can still prove that $\phi_0^{(c)}(t=0)$ is of order one, i.e. its value does not scale with $D$:

\begin{equation}
    \centering
    \label{Two_Pt_Conn_t0}
    \phi_0^{(c)}(t=0)=\frac{1}{D}\sum_{a,b=1}^D |O_{ab}|^2 - \frac{1}{D^2}\sum_{a,b=2}^D O_{aa}O_{bb} = \frac{1}{D^2}\sum_{a,b=1}^D |r_{ab}|^2-\frac{1}{D}\left\{ \frac{1}{D^2}\sum_{a,b=1}^Dr_{aa}r_{bb} \right\}, 
\end{equation}
as can be seen plugging in the Ansatz (\ref{RMT_op_Ansatz}). We note that the leading term in (\ref{Two_Pt_Conn_t0}) is the first term of the last expression, consisting of a sum of $D^2$ numbers of order one, divided by a $D^2$ factor, and hence $\phi_0^{(c)}(t=0)$ is a number of order $D^0$.

Now, we can estimate the height of the late-time plateau by computing the long-time average of the connected two-point function:
\begin{equation}
    \centering
    \label{Two_Pt_Conn_LongT_Avg}
    \overline{\phi_0^{(c)}}= \lim_{T\to \infty} \frac{1}{T}\int_0^T dt \, \phi_0^{(c)}(t) = \frac{1}{D}\sum_{a=1}^D O_{aa}^2 - \frac{1}{D^2}\sum_{a,b=1}^D O_{aa} O_{bb}.
\end{equation}
Plugging the Ansatz (\ref{RMT_op_Ansatz}) in (\ref{Two_Pt_Conn_LongT_Avg}) we again find that the terms involving the order-one quantity $O$ cancel out, yielding:
\begin{equation}
    \centering
    \label{Two_Pt_Conn_LongT_Avg_result}
    \overline{\phi_0^{(c)}}= \frac{1}{D}\left\{ \frac{1}{D}\sum_{a=1}^D r_{aa}^2 + \frac{1}{D^2}\sum_{a,b=1}^D r_{aa}r_{bb} \right\}.
\end{equation}
The quantity inside the braces is of order $D^{0}$. We thus conclude that the connected two-point function has a long-time average of order $\frac{1}{D}$, and that this is deduced from the ETH-like Ansatz (\ref{RMT_op_Ansatz}). To argue that $\phi_0^{(c)}(t)$ actually plateaus at $\frac{1}{D}$, one should prove that its long time variance is (exponentially) suppressed\footnote{We shall refer to quantities of order $\frac{1}{D}$ or smaller as \textit{exponentially suppressed} because the Hilbert space dimension is typically exponential in the number of degrees of freedom $S$ of the system, i.e. $D\sim e^{S}$.}, so that the function remains close to its long-time average at late times. We shall do that later, when studying the long time average of the square of the two-point function. But before that, we can note that there was a simpler way to derive the previous results, by defining a new operator $\widetilde{\mathcal{O}}$ obtained by subtracting the one-point function from the initial operator $\mathcal{O}$:
\begin{equation}
    \centering
    \label{Connected_Op}
    \widetilde{\mathcal{O}}=\mathcal{O}-\mathds{1}\langle \mathcal{O} \rangle = \mathcal{O}-\frac{1}{D}\text{Tr}[\mathcal{O}]\mathds{1}.
\end{equation}
Using the operator Ansatz for $\mathcal{O}$ given in (\ref{RMT_op_Ansatz}), we note that the matrix elements of $\widetilde{\mathcal{O}}$ in the energy basis are given by:
\begin{equation}
    \centering
    \label{Matrix_elements_otilde}
    \widetilde{O}_{ab} = \frac{1}{\sqrt{D}}\widetilde{r}_{ab},
\end{equation}
where all $\widetilde{r}_{ab}$ are of order one and the matrix $\widetilde{R}\equiv \left( \widetilde{r}_{ab} \right)$ is related to the matrix $R\equiv \left( r_{ab} \right)$ through:
\begin{equation}
    \centering
    \label{Rtilde_vs_R}
    \widetilde{R} = R - \mathds{1} \langle R \rangle.
\end{equation}
In particular:
\begin{equation}
    \centering
    \label{rtilde_vs_r}
    \widetilde{r}_{ab} = r_{ab}-\frac{\delta_{ab}}{D}\sum_{c=1}^D r_{cc},
\end{equation}
from where it is apparent that all $\widetilde{r}_{ab}$ are of order one and that they follow the exact constraint $\text{Tr}[\widetilde{R}]=\sum_{a=1}^D\widetilde{r}_{aa}=0$.

This operator redefinition is useful because we immediately note that the connected two-point function of $\mathcal{O}$ is identically equal to the full two-point function of $\widetilde{\mathcal{O}}$:

\begin{equation}
    \centering
    \label{Two_Pt_Conn_Tilde}
    \phi_0^{(c)}(t) = \big\langle \widetilde{\mathcal{O}} \widetilde{\mathcal{O}}(t) \big\rangle.
\end{equation}
And thus, using the expression of $\widetilde{O}$ in the energy basis (\ref{Matrix_elements_otilde}) it is straightforward to see that:
\begin{equation}
    \centering
    \label{Two_Pt_Otilde}
    \phi_0^{(c)}(t) = \frac{1}{D}\sum_{a,b=1}^D |\widetilde{O}_{ab}|^2 e^{it(E_a-E_b)} = \frac{1}{D^2}\sum_{a,b=1}^D |\widetilde{r}_{ab}|^2 e^{it(E_a-E_b)},
\end{equation}
from where it is immediate that $\phi_0^{(c)}(t=0)\sim D^0$ and that $\overline{\phi_0^{(c)}}\sim \frac{1}{D}$.

A similar argument can now be applied to transition probabilities on the Krylov chain.
The long-time average of the transition probability is given by (\ref{Transition_Probability})
\begin{eqnarray}
    Q_{0n} := \overline{|\phi_n|^2} = \lim_{T\to \infty} \frac{1}{T} \int_0^T |\phi_n(t)|^2 dt~.
\end{eqnarray}
which for $n=0$ takes the form 
\begin{eqnarray}
    \label{appx_Q00}
    Q_{00} = \frac{1}{D^2} \Big[ \sum_{a,b=1}^D  |O_{aa}|^2 |O_{bb}|^2 + \sum_{a\neq b=1}^D  |O_{ab}|^4  \Big]~.
\end{eqnarray}
From the Ansatz (\ref{RMT_op_Ansatz}) we find that:
\begin{eqnarray}
     \frac{1}{D^2}\sum_{a,b=1}^D |O_{aa}|^2 |O_{bb}|^2 \sim O^4 + \mathit{O}\left( \frac{1}{D} \right).
\end{eqnarray}
\begin{eqnarray}
    \frac{1}{D^2}\sum_{a\neq b=1}^D  |O_{ab}|^4 \sim \mathit{O}\left(\frac{1}{D^2}\right).
\end{eqnarray}
We thus find that the long-time-average of the square of the autocorrelation function behaves like
\begin{eqnarray} \label{large_phi0}
     \overline{|\phi_0|^2} &\sim& O(1)+ O\left(\frac{1}{D}\right).
\end{eqnarray}
The time averaged transition probability $Q_{00}$ takes an order-one value controlled by the one-point function. In order to see an exponentially suppressed plateau, we again need to work with the connected two-point function $\phi_0^{(c)}(t)$, and the associated probability $P_{00}^{(c)}(t):=\phi_0^{(c)}(t)^2$ (note that the two-point function is always real provided that the operator is hermitian, which we assume), whose long-time average we shall denote $Q_{00}^{(c)}$. As we showed in (\ref{Two_Pt_Conn_t0}), $\phi_0^{(c)}(t=0)\sim 1$, and therefore $P_{00}^{(c)}(t=0)\sim 1$. Likewise, $Q_{00}^{(c)}$ can be expressed in terms of the matrix elements of the traceless operator $\widetilde{\mathcal{O}}$:
\begin{equation}
    \centering
    \label{Q00_Conn_otilde}
    Q_{00}^{(c)}=\frac{1}{D^2}\sum_{a,b=1}^D\left[ \widetilde{O}_{aa}^2\widetilde{O}_{bb}^2 + |\widetilde{O}_{ab}|^4 \right] = \frac{1}{D^4}\sum_{a,b=1}^D \left[ \widetilde{r}_{aa}^2\widetilde{r}_{bb}^2 + |\widetilde{r}_{ab}|^4 \right] \sim \frac{D^2}{D^4}=\frac{1}{D^2}.
\end{equation}
And hence $P_{00}^{(c)}(t)$ plateaus\footnote{Actually, in order to show that $P_{00}^{(c)}(t)$ plateaus at the long-time average $Q_{00}^{(c)}$, we should also show that the long-time variance of $P_{00}^{(c)}(t)$ is suppressed, otherwise a strongly oscillating function could still be compatible with the long-time average prediction. This proof is doable (even though cumbersome), but $P_{00}^{(c)}(t)\sim e^{-2S}$ at late times seems to hold for ETH operators in chaotic systems like cSYK$_4$ according to numerical checks.} at late times at $\frac{1}{D^2}\sim e^{-2S}$. Incidentally, note that $Q_{00}^{(c)}$ gives the long-time variance of $\phi_0^{(c)}(t)$, and hence showing that (\ref{Q00_Conn_otilde}) is suppressed concludes the proof that the connected two-point function is close to the plateau value at late times, as anticipated above. 

\subsection{Example: Hopping vs number operator in \texorpdfstring{cSYK$_4$}{cSYK4}} \label{app:SYK}

The work on cSYK$_4$ \cite{I}, studied the K-complexity of hopping operators, $h_{ij}=c_i^\dagger c_j + h.c$. These operators have a zero one-point function, and hence their two-point function is connected. However, non-universal effects due to a non-zero one-point function can be probed if we consider, for example, an on-site number operator $n_i=c_i^\dagger c_i$. Indeed, since in cSYK$_4$ we work in fixed occupation sectors, the one-point functions of the on-site number operators are constrained by the relation:

\begin{equation}
    \centering
    \label{Number_ev}
    N=\sum_{i=1}^L n_i\;\Longrightarrow\;\langle N \rangle = \sum_{i=1}^L \langle n_i \rangle, 
\end{equation}
where $N$ is the total number operator and $L$ is the number of sites (or rather, the number of complex fermions). Hence, (\ref{Number_ev}) together with the fact that $\langle n_i \rangle \geq 0$ implies that at least one on-site number operator needs to have a non-zero expectation value whenever $\langle N \rangle > 0$. In fact, the chaotic character of cSYK$_4$ seems to distribute equally the expectation value accross all the $n_i$, and given a fixed occupation sector we can thus estimate $\langle n_i \rangle \sim \frac{\langle N \rangle}{L}$ for all $i=1,...,L$, i.e. the one-point function equals the filling ratio.

\begin{figure}[t]
    \centering
    \includegraphics[width=0.5\textwidth]{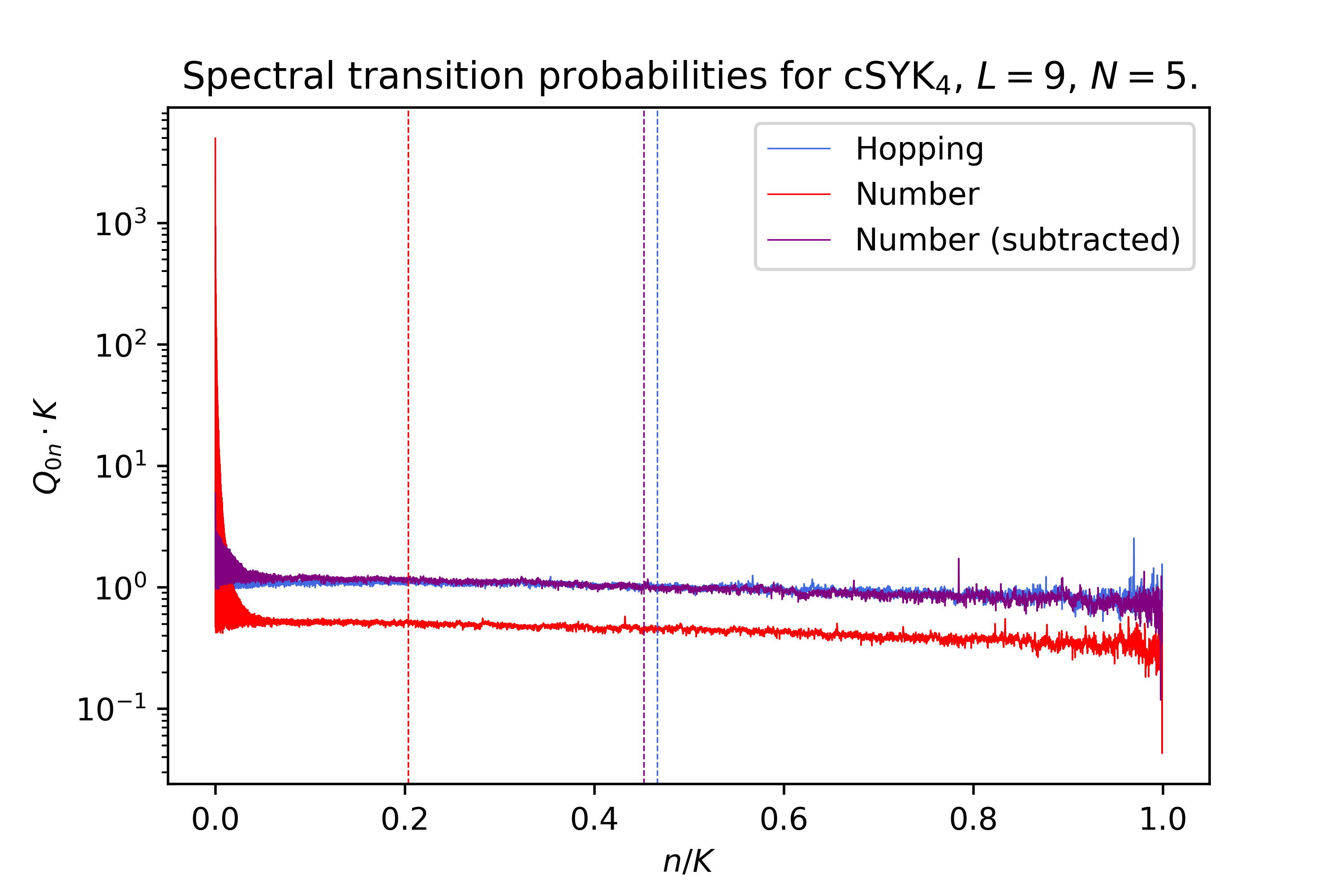}
    \caption{Averaged transition probabilities $Q_{0n}$ in cSYK$_4$ for the hopping operator and both the full and the subtracted version of the number operator. Vertical lines mark the estimated K-complexity saturation value. For this system size, $K=15751\sim 10^4$, and we observe that $Q_{00}\sim 1$ for the full number operator, signaling that indeed the one-point function dominates the late-time regime of the transition probability. The rest of the $Q_{0n}$ with $n>1$ for the full number operator seem to be rather uniformly distributed, but with a lower value due to the constraint $\sum_{n=0}^{K-1}Q_{0n}=1$. This eventually enforces a complexity saturation value of $\sim 0.2K$, much below the naively expected $\sim\frac{K}{2}$, which the hopping operator does display. Conversely, both for the hopping operator and for the subtracted number operator, which have a zero one-point function by construction, all the transition probabilities are rather uniformly distributed around $\frac{1}{K}$, yielding a K-complexity saturation value much closer to $\frac{K}{2}$.}
    \label{fig:Transition_Prob_SYK_Ops}
\end{figure}

This non-zero one-point function controls the averaged transition probability $Q_{00}$, which becomes of order one in system size and has the effect of lowering the K-complexity saturation value in (\ref{KC_Q0n}). Figure \ref{fig:Transition_Prob_SYK_Ops} illustrates this claim.

In order to avoid this, we can seed the Lanczos algorithm with the subtracted version of the operator, $\widetilde{\mathcal{O}}$, which by construction has a zero one-point function. Following the usual arguments \cite{ViswanathMuller, Parker:2018yvk}, we find that the Lanczos coefficients $\widetilde{b}_n$ of $\widetilde{\mathcal{O}}$ are in one-to-one correspondence with the moments $\widetilde{\mu}_n$ of the connected two-point function of $\mathcal{O}$, $\phi_0^{(c)}(t)$, and that the K-complexity long-time average is given by:

\begin{equation}
    \centering
    \label{KC-longT-otilde}
    \overline{C_K}=\sum_{n=0}^{K-1} n \widetilde{Q_{0n}},
\end{equation}
where $\widetilde{Q_{00}}=Q_{00}^{(c)}$. Therefore, for an ETH operator this last quantity will be exponentially suppressed, hence not competing with the other uniformly distributed $\widetilde{Q_{0n}}$ with $n>0$ and allowing for a K-complexity saturation value closer to $K/2$. This is illustrated in Figure \ref{fig:Transition_Prob_SYK_Ops}.

\subsection{Role of the one-point function in XXZ} \label{app:XXZ}
\begin{figure}[t]
    \centering
    \includegraphics[width=0.5\textwidth]{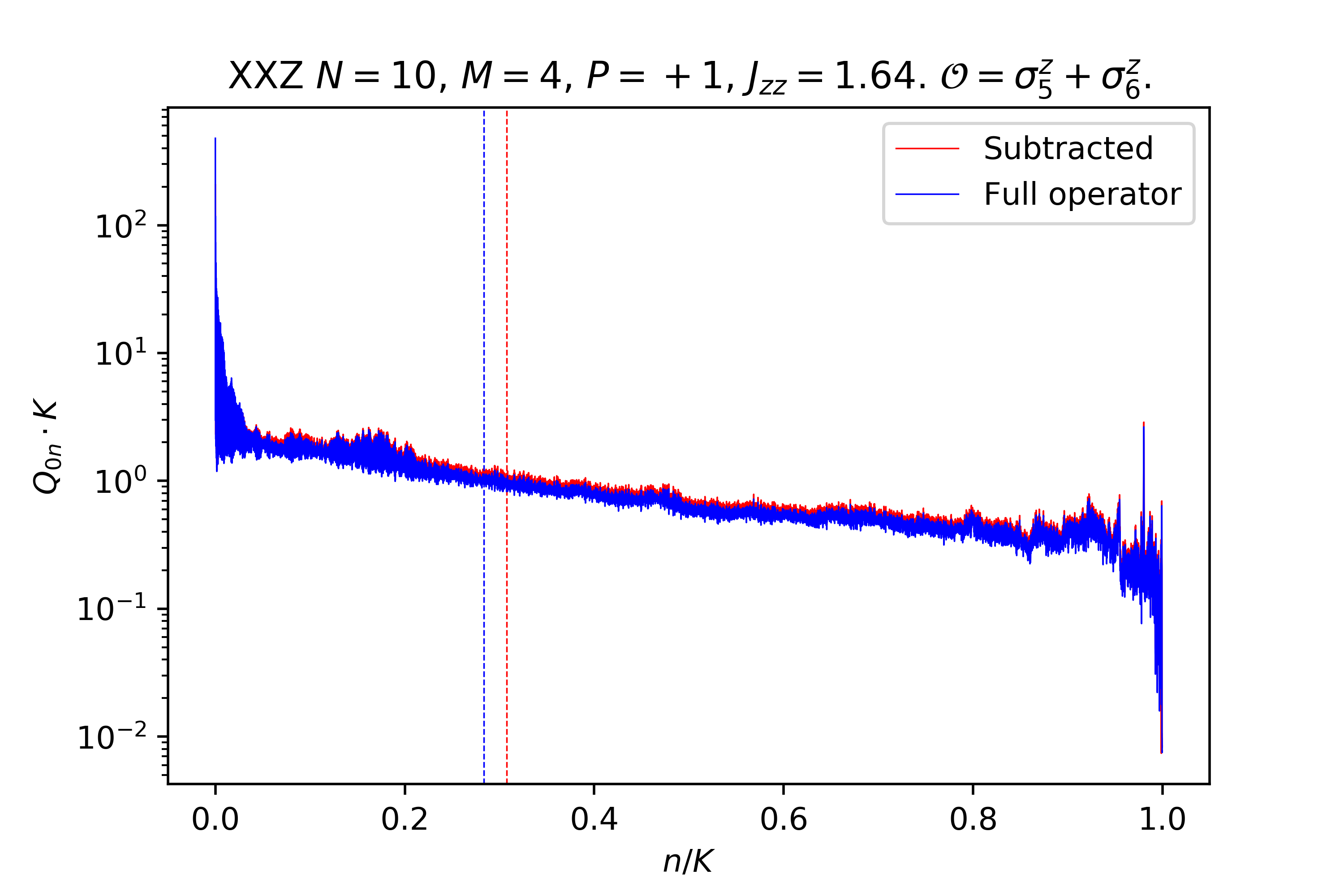}
    \caption{Long-time averaged transition probabilities for an instance of XXZ studied in \cite{II}, this time also considering a version of the operator where the non-vanishing one-point function has been subtracted. In contrast to the cSYK$_4$ case, this time the subtraction of the one-point function does not alter drastically the K-complexity saturation value, since in this case undersaturation is due to the monotonously decaying profile of $Q_{0n}$ that we associated to Anderson localization on the Krylov chain in \cite{II}, together with the fact that even the connected part of the two-point function is itself not exponentially suppressed at late times due to the integrable nature of the system.}
    \label{fig:XXZ_Connected_VS_Full}
\end{figure}

This may raise some concern regarding previous work in XXZ \cite{II}, as the operators used in that case were on-site Pauli sigma matrices, whose one-point function in fixed-magnetization Hilbert space sectors are also constrained by the value of the total magnetization in the given sector, through:

\begin{equation}
    \centering
    \label{Total_Mag_ev}
    S^z = \frac{1}{2}\sum_{n=1}^N \sigma_n^z\;\longrightarrow\;\langle S^z \rangle = \frac{1}{2}\sum_{n=1}^N\langle \sigma_n^z \rangle ,
\end{equation}
where, for XXZ, $N$ denotes the number of chain sites. Since XXZ is integrable, we do not necessarily assume that all $\langle \sigma_n^z \rangle$ are similar, but it is anyway clear from (\ref{Total_Mag_ev}) that in general they need not be zero. This might make one think that the XXZ calculations should be re-made taking as an input the subtracted version of on-site Pauli matrices. Figure \ref{fig:XXZ_Connected_VS_Full} illustrates that, in this case, subtracting the one-point function does not alter qualitatively the results because the connected part of the two-point function in this integrable system is already not exponentially suppressed, and hence the late-time value of the two-point function does not change drastically if one subtracts the disconnected part from it.

\section{Profile of transition probability for flat operator}\label{appx_FlatOp}
Consider a dense operator with constant matrix elements $O_{ab}=1$ for all $a,b=1,\dots,D$. In the Krylov basis such an operator has the following profile
\begin{eqnarray} \label{const_operator}
    |\mathcal{O}_0)=\Big(\underbrace{\frac{1}{D},\dots, \frac{1}{D}}_{\text{$\frac{D(D-1)}{2}$ terms} },\sqrt{\frac{1}{D}}, \underbrace{\frac{1}{D},\dots, \frac{1}{D}}_{\text{$\frac{D(D-1)}{2}$ terms} } \Big)
\end{eqnarray}
such that $\sum_{i=0}^{K-1}|O_i|^2=1$.  We will call such an operator a ``flat'' operator.

We recall from \cite{II} that for an odd number of elements in the Krylov basis, which is  the case when no degeneracies are present and the operator has a non-zero projection over all Liouvillian frequencies (in such a case $K=D^2-D+1$ which is an odd number), the  Liouvillian eigenvector at the middle of its spectrum has zero eigenvalue 
\begin{eqnarray}
    \mathcal{L}|\omega_{\text{middle}}) = \omega_{\text{middle}}|\omega_{\text{middle}})=0 ~.
\end{eqnarray}
In general, the Liovillian eigenvectors can be expanded in the Krylov basis:
\begin{eqnarray}\label{omega_eigvec}
    |\omega_i) = \sum_{n=0}^{K-1} \psi_{ni}|\mathcal{O}_n)~.
\end{eqnarray}
For $|\omega_{\text{middle}})$ the coefficients $\psi_{n}$ satisfy
\begin{eqnarray}
    \psi_{2n} &=& (\mathcal{O}_{2n}|\omega_{\text{middle}}) = \psi_0 \prod_{i=1}^n \frac{b_{2i-1}}{b_{2i}} \equiv \psi_0 X_n \label{psi_even}\\
    \psi_{2n+1} &=& (\mathcal{O}_{2n+1}|\omega_{\text{middle}})  = 0~, \label{psi_odd}
\end{eqnarray}
where in (\ref{psi_even}) we defined $\prod_{i=1}^n \frac{b_{2i-1}}{b_{2i}} \equiv X_n$ for later convenience. 
Note that $\psi_0$ is determined by the middle element of (\ref{const_operator}), i.e. 
\begin{eqnarray} \label{psi0}
    \psi_0=\sqrt{\frac{1}{D}}
\end{eqnarray}
since in $\mathcal{L}$'s eigenvector matrix, the middle element of $|\mathcal{O}_0)$ is the first element in $|\omega_{\text{middle}}) $ as can be seen from (\ref{omega_eigvec}). 

\begin{figure}[t]
    \centering
    \begin{subfigure}[t]{0.45\textwidth}
    \centering
        \includegraphics[scale=0.45]{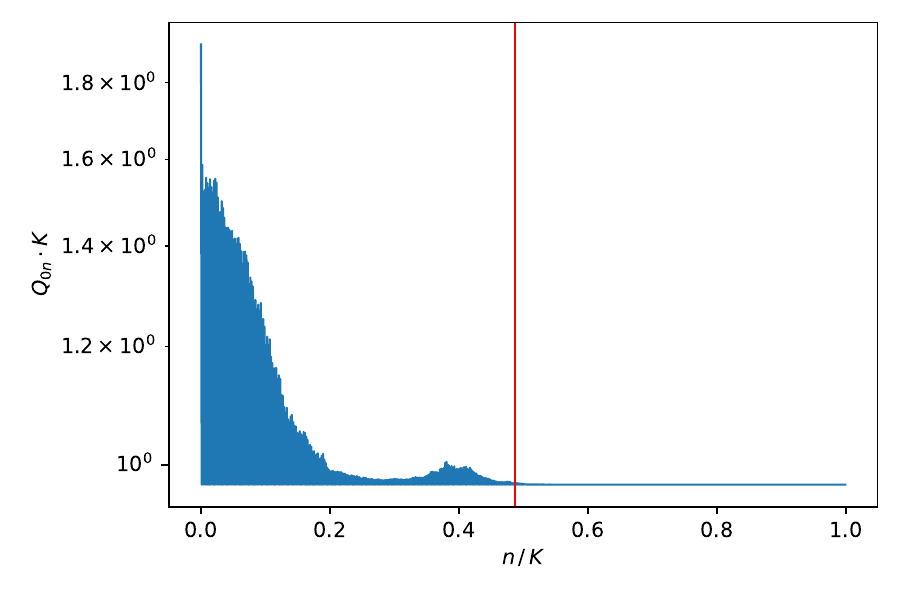}
    \end{subfigure}
    \hfill
    \begin{subfigure}[t]{0.45\textwidth}
    \centering
        \includegraphics[scale=0.45]{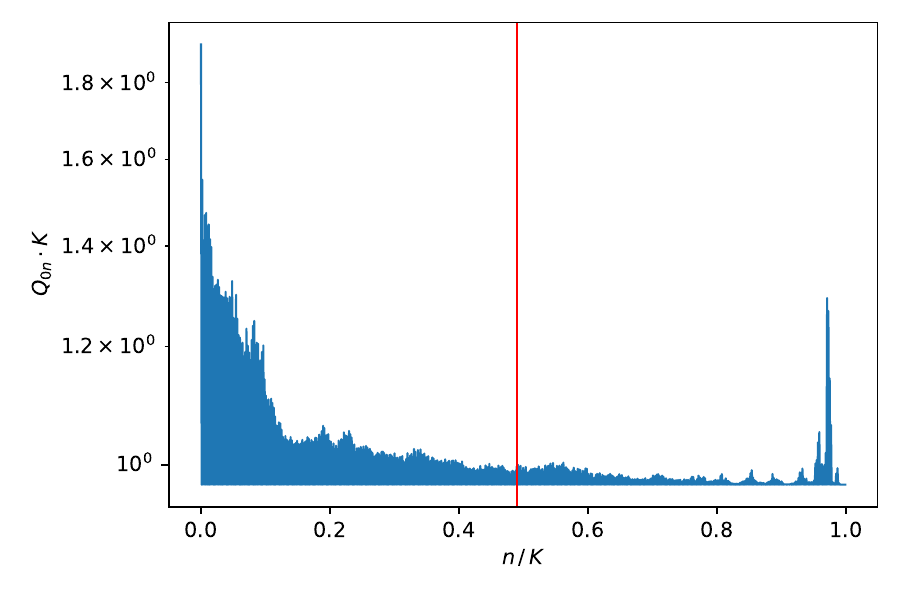}
    \end{subfigure}
    \caption{K-complexity saturation value for constant operator evolving under Hamiltonian taken from a GOE ensemble (left) and Hamiltonian with Poissonian statistics (right), both computed at $D=32$.  Note that both cases exhibit saturation value close to $K/2$.}
    \label{fig:GOE_vs_Poisson}
\end{figure}

With the information from (\ref{const_operator}), (\ref{psi_even}, \ref{psi_odd}) and  (\ref{psi0}) we can compute $Q_{0n}$ in terms of the Lanczos coefficients.
Starting with the definition (\ref{Transition_Probability})
\begin{eqnarray}
    Q_{0n}=\sum_{i=0}^{K-1} |(\mathcal{O}_0|\omega_i)|^2 |(\mathcal{O}_n|\omega_i)|^2 
\end{eqnarray}
the first element is given by
\begin{eqnarray}
    Q_{00}=\sum_{i=0}^{K-1} |(\mathcal{O}_0|\omega_i)|^4 = D(D-1) \frac{1}{D^4} + \frac{1}{D^2} = \frac{2D-1}{D^3}~.
\end{eqnarray}
where we used (\ref{const_operator}) directly. The next element is
\begin{eqnarray}
    Q_{01}=\sum_{i=0}^{K-1} |(\mathcal{O}_0|\omega_i)|^2 |(\mathcal{O}_1|\omega_i)|^2 = \frac{1}{D^2} \sum_{i\neq \text{middle}} |(\mathcal{O}_1|\omega_i)|^2 + \frac{1}{D}\cdot 0 = \frac{1}{D^2}
\end{eqnarray}
where in the second equality we used the fact that the Krylov elements are normalized, hence $1=\sum_{i\neq \text{middle}} |(\mathcal{O}_1|\omega_i)|^2+|(\mathcal{O}_1|\omega_{\text{middle}})|^2$ and since from (\ref{psi_odd}), $(\mathcal{O}_1|\omega_{\text{middle}})=0$ we deduce that $\sum_{i\neq \text{middle}} |(\mathcal{O}_1|\omega_i)|^2=1$.

\begin{eqnarray}
    Q_{02} &=&\sum_{i=0}^{K-1} |(\mathcal{O}_0|\omega_i)|^2 |(\mathcal{O}_2|\omega_i)|^2 = \frac{1}{D^2} \sum_{i\neq \text{middle}} |(\mathcal{O}_2|\omega_i)|^2 + \frac{1}{D}  |(\mathcal{O}_2|\omega_{\text{middle}})|^2 \nonumber\\
    &=& \frac{1}{D^2}\left(1-\frac{1}{D}X_1^2\right) +\frac{1}{D}\left(\frac{1}{D}X_1^2\right) = \frac{1}{D^2} + \frac{X_1^2}{D^2}\left( 1-\frac{1}{D} \right) ~.
\end{eqnarray}
The rest of $Q_{0n}$ can be computed in a similar manner, and we conclude that for $n\geq 1$
\begin{eqnarray}
    Q_{0,2n} &=& \frac{1}{D^2} + \frac{X_n^2}{D^2}\left( 1-\frac{1}{D} \right) \\
    Q_{0,2n+1} &=& \frac{1}{D^2} ~.
\end{eqnarray}
One can check that this result is normalized
\begin{eqnarray}
    \sum_{n=0}^{K-1} Q_{0n} = \frac{2D-1}{D^3} +\left(\frac{K-1}{2}\right)\frac{1}{D^2}+\left(\frac{K-1}{2}\right)\frac{1}{D^2} +  \frac{D-1}{D^2} \sum_{n=1}^{\frac{K-1}{2}}\frac{X_n^2}{D} = 1
\end{eqnarray}
where we used $K=D^2-D+1$ and from normalization of $|\omega_{\text{middle}})$ we know that $\frac{1}{D}+\sum_{n=1}^{\frac{K-1}{2}} \frac{X_n^2}{D}=1$.
The value of K-complexity can then be estimated as follows:
\begin{eqnarray}
    \overline{C_K} &=& \sum_{n=0}^{K-1} n Q_{0n} = \frac{1}{D^2}\sum_{n=1}^{K-1} n + \frac{D-1}{D^2}  \sum_{n=1}^{\frac{K-1}{2}} 2n \frac{X_n^2}{D} \nonumber \\
    &=&\frac{1}{2}K(K-1)\frac{1}{D^2} + \frac{D-1}{D^2} {C_K}_{\text{middle}}
\end{eqnarray}
where ${C_K}_{\text{middle}}$ is the K-complexity of the eigenvector $|\omega_{\text{middle}})$\footnote{K-complexity for individual eigenvectors of the Liouvillian was defined in \cite{II}.} and by definition $0\leq {C_K}_{\text{middle}}\leq K$. Hence it is found that for a flat operator
\begin{eqnarray}
    \frac{D^2}{2}-D+1-\frac{1}{2D} \leq \overline{C_K} \leq \frac{D^2}{2}-1+\frac{3}{2D}-\frac{1}{D^2}
\end{eqnarray}
which for large enough $D$ indicates that 
\begin{eqnarray}
    \overline{C_K} \sim \frac{D^2}{2} \sim \frac{K}{2}
\end{eqnarray}
independently of the spectrum or Lanczos coefficients data.  We show numerically that this is indeed the case by studying flat operators with Hamiltonians of GOE statistics and Poissonian statistics in Figure \ref{fig:GOE_vs_Poisson}.

%% file: content/AppxCh05.tex
\chapter{\rm\bfseries Appendices to Chapter \ref{ch:chapter05_DSSYK}}
\label{ch:AppxCh05}

\section{Coupling variance choices} \label{App:coupling_variance}

For reference, this Appendix discusses the relation between the various choices of SYK coupling variance in the literature. Different choices are suitable depending on the specific limit in which the system is studied, as they will produce bounded results for observables such as expectation values of operators and correlation functions.

In \cite{Maldacena:2016hyu}, the SYK model is studied analytically at large $N$, first at fixed $p$ and later on at also large $p$. The Hamiltonian in that article is:

\begin{equation}
    \label{SYKHam_Malda_fixed_q}
    H^{(1)} = i^{p/2}\sum_{1\leq i_1<...<i_p\leq N} J^{(1)}_{i_1...i_p} \chi_{i_1}...\chi_{i_p},
\end{equation}
where the Majoranas are normalized such that they square to $\frac{1}{2}\mathds{1}$, that is:
\begin{equation}
    \label{Majoranas_anticom_equals_1}
    \left\{\chi_i,\chi_j\right\}=\delta_{ij}.
\end{equation}
For the sake of studying the model in the large-$N$ limit at fixed $p$, the coupling variance used in \cite{Maldacena:2016hyu} is
\begin{equation}
    \centering
    \label{Variance_Malda_large_N_fixed_q}
    \left\langle \left( J_{i_1...i_p}^{(1)} \right)^2 \right\rangle = \frac{(p-1)! J^2}{N^{p-1}}.
\end{equation}

When \cite{Maldacena:2016hyu} turns to studying SYK at large $N$ and large $p$, the Hamiltonian keeps the same structure,
\begin{equation}
    \label{SYKHam_Malda_large_q}
    H^{(2)} = i^{p/2}\sum_{1\leq i_1<...<i_p\leq N} J^{(2)}_{i_1...i_p} \chi_{i_1}...\chi_{i_p},
\end{equation}
but the coupling variance is chosen to have an extra $p$-dependent scaling to get a more uniform limit:
\begin{equation}
    \label{Variance_Malda_large_N_large_q}
    \left\langle \left( J_{i_1...i_p}^{(2)} \right)^2 \right\rangle = \frac{2^{p-1}(p-1)!}{p N^{p-1}}J^2.
\end{equation}
The origin of the $2^{p-1}$ factor is related to the normalization of the Majoranas (\ref{Majoranas_anticom_equals_1}): Since they square to $\frac{1}{2}\mathds{1}$, they will give an additional suppression of $2^{-p}$ every time a Wick contraction pairs two Hamiltonians inside of a trace, which needs to be compensated by the coupling variance in order to obtain finite results in the large-$p$ limit. More explicitly, redefining the Majoranas as
\begin{equation}
    \label{Majoranas_redef}
    \chi_i = \frac{1}{\sqrt{2}} \psi_i,
\end{equation}
we obtain operators $\psi_i$ that are normalized in a way such that they square to one:
\begin{equation}
    \label{Majoranas_anticom_equals_2}
    \left\{ \psi_i,\psi_j \right\}=2\delta_{ij}.
\end{equation}
In terms of these operators, the Hamiltonian (\ref{SYKHam_Malda_large_q}) reads:
\begin{equation}
    \label{SYKHam_Malda_large_q_normalized_Majoranas}
    H^{(2)} = i^{p/2}\sum_{1\leq i_1<...<i_p\leq N} J^{(2)}_{i_1...i_p} 2^{-\frac{p}{2}}\psi_{i_1}...\psi_{i_p}\equiv i^{p/2}\sum_{1\leq i_1<...<i_p\leq N} \widetilde{J^{(2)}}_{i_1...i_p} \psi_{i_1}...\psi_{i_p},
\end{equation}
so that the effective coupling strength variance gets modified to
\begin{equation}
    \label{Variance_Malda_large_q_effective}
    \left\langle \left( \widetilde{J^{(2)}}_{i_1...i_p} \right)^2 \right\rangle=\left\langle \left( J^{(2)}_{i_1...i_p} \right)^2 \right\rangle 2^{-p} = \frac{(p-1)!}{2pN^{p-1}}J^2.
\end{equation}

When turning to double-scaled SYK, extensively studied in \cite{Berkooz:2018jqr}, it is more enlightening to work with normalized Majoranas $\psi_i$ satisfying (\ref{Majoranas_anticom_equals_2}) since, as mentioned in section \ref{Subsection_Background_DSSYK}, the monomials $\psi_I$, where $I$ denotes a collective index $i_1...i_p$, square to a phase, and therefore the combinatorial factor coming from the sum in the Hamiltonian can be directly compensated by the normalization of the coupling strength in order to yield bounded moments of the Hamiltonian. The Hamiltonian in \cite{Berkooz:2018jqr} is:
\begin{equation}
    \label{SYKHam_DSSYK_berkooz}
    H^{(3)}=i^{p/2}\sum_{1\leq i_1<...<i_p\leq N}J^{(3)}_{i_1...i_p}\psi_{i_1}...\psi_{i_p},
\end{equation}
and the coupling strength variance is given by
\begin{equation}
    \label{Variance_berkooz}
    \left\langle \left( J^{(3)}_{i_1...i_p} \right)^2 \right\rangle = {\binom{N}{p}}^{-1}J^2.
\end{equation}
As an example, this normalization ensures that $\left\langle \text{Tr}\left[ \left(H^{(3)}\right)^2 \right] \right\rangle = J^2$, even at finite and independent $N$ and $p$.

For the developments in Chapter \ref{ch:chapter05_DSSYK}, we are interested in double-scaled SYK but we do not use the coupling strength variance (\ref{Variance_berkooz}). Instead, we follow \cite{Lin:2022rbf,Goel:2023svz} and use a variance which, when taken in the double-scaling limit, has an asymptotic behavior that makes contact with (\ref{Variance_Malda_large_q_effective}) in a way about which we shall be precise below. Our Hamiltonian is:
\begin{equation}
    \label{SYKHam_Ours}
    H = i^{p/2}\sum_{1\leq i_1 <...<i_p\leq N} J_{i_1...i_p}\psi_{i_1}...\psi_{i_p}.
\end{equation}
We study this system in the double-scaling limit, taking $N$ and $p$ to infinity keeping the ratio
\begin{equation}
    \label{DS_thooft_coupling}
    \lambda=\frac{2p^2}{N}
\end{equation}
fixed. Given $\lambda$, the coupling strength variance is chosen to be:
\begin{equation}
    \label{Variance_ours}
    \left\langle J_{i_1...i_p}^2 \right\rangle=\frac{1}{\lambda}{\binom{N}{p}}^{-1}\,J^2.
\end{equation}
To compare with (\ref{Variance_Malda_large_q_effective}), we can perform an asymptotic analysis already within the double-scaling limit. That is, we take (\ref{Variance_ours}) and implicitly assume everywhere that $N\equiv N(p) = \frac{2p^2}{\lambda}$. Then, for large $p$ we find the asymptotic behavior:
\begin{equation}
    \centering
    \label{Variance_ours_asymp}
    \left\langle J_{i_1...i_p}^2 \right\rangle=\frac{1}{\lambda}{\binom{N(p)}{p}}^{-1}\,J^2\sim \frac{e^{\lambda/4}}{2p}\frac{(p-1)!}{{N(p)}^{p-1}}J^2 = \frac{(p-1)!}{2p {N(p)}^{p-1}} J^2 + \mathit{O}(\lambda).
\end{equation}
Where $f(p)\sim g(p)\Leftrightarrow \lim_{p\to+\infty} \frac{f(p)}{g(p)}=1$.
This means that, in the double-scaling  limit, the variance (\ref{Variance_ours}) is equivalent to the choice (\ref{Variance_Malda_large_q_effective}) upon the additional assumption of small $\lambda$, which is precisely the regime on which papers like \cite{Lin:2022rbf,Goel:2023svz} focus, since it zooms in near the ground state of SYK and yields the Liouville Hamiltonian which makes contact with Schwarzian dynamics and hence with an AdS bulk dual.

In view of this, the small-$\lambda$ regime of DSSYK can be seen as a controlled way to approach large-$N$ and large-$p$ SYK with the hierarchy $1\ll p \ll N$, as also discussed in \cite{Mukhametzhanov:2023tcg}.

\section{A note on the triple-scaling limit}\label{Appx:triple_scaling}

A priori, there is a whole family of triple scaling limits that one can take which, applied to the Hamiltonian (\ref{Ham_l_k}), yield Liouville-like Hamiltonians with a different relative weight between the kinetic term and the potential term. The family of limits is parametrized by some $a\in\mathds{R}$ as follows:
\begin{equation}
    \centering
    \label{triple-scaling-family}
    \lambda\to 0,\qquad l\to\infty,\qquad(a\lambda)^{-2}e^{-\frac{l}{l_f}}=e^{-\frac{\tilde{l}}{l_f}}\,\text{fixed}.
\end{equation}
Applied to (\ref{Ham_l_k}), this limit yields:
\begin{equation}
    \centering
    \label{Ham_triple_scaled_family}
    \tilde{T} - E_0 = 2\lambda J\left( \frac{l_f^2 k^2}{2} + \frac{a^2}{2}e^{-\frac{\tilde{l}}{l_f}} \right)\;+\mathit{O}\left(\lambda^2\right).
\end{equation}
However, all the Hamiltonians in the family (\ref{Ham_triple_scaled_family}) are equivalent in the sense that they only differ by a finite translation of the length variable. Their density of states is therefore the same and the eigenfunctions of one Hamiltonian in the family can be obtained from those of any other by just a shift. In our work, we choose $a=2$ (instead of the $a=1$ convention in \cite{Lin:2022rbf}) because in this case the resulting triple-scaled Hamiltonian makes contact with the form of the Liouville Hamiltonian for JT gravity written explicitly in \cite{Harlow:2018tqv}, restated in \eqref{JT_Liouville_Ham}. Nevertheless, we stress that, since all the Hamiltonians in the family (\ref{Ham_triple_scaled_family}) are equivalent in the sense that we have just explained, changing $a$ in the definition of the triple-scaling limit (\ref{triple-scaling-family}) will not modify the gravity theory fundamentally. This ambiguity in the definition of $\tilde{l}$ may be seen as related to the choice of regularization scheme for the boundary divergence from the bulk perspective.

\section{Eigenvalues and eigenvectors of the effective Hamiltonian}\label{Appx:EigSysDSSYK}
The discussion in this Appendix closely follows \cite{Berkooz:2018jqr, Berkooz:2018qkz} and is presented for the sake of completeness.

The symmetric version of $T$ given in (\ref{T_sym}) satisfies, in the chord basis,
\begin{equation} \label{T_chordB}
    T\,|l\rangle = \frac{J}{\sqrt{\lambda (1-q)}} \left( \sqrt{1-q^{l+1}}\,|l+1\rangle + \sqrt{1-q^{l}} \, |l-1\rangle \right) ~.
\end{equation}
To find the eigenvalues of $T$ we write down the eigensystem equation for the components of the eigenvectors of $T$, $\psi_l(E)=\langle l|E\rangle$:
\begin{equation} \label{q-eigsys1}
    E \,\psi_l(E) = \frac{J}{\sqrt{\lambda (1-q)}} \left( \sqrt{1-q^{l+1}}\,\psi_{l+1}(E) + \sqrt{1-q^{l}} \, \psi_{l-1}(E) \right)
\end{equation}
 We note that $E$ is required to be independent of the position label $l$ as (\ref{q-eigsys1}) is an eigenvalue problem; therefore, assuming there are no bound states, we can take the large $l$ limit of the above relation where $T$ becomes a tridiagonal Toeplitz\footnote{A Toeplitz matrix is a matrix with equal entries along the diagonals.} matrix with eigensystem equation
\begin{equation}
     E \,\psi_l(E) = \frac{J}{\sqrt{\lambda (1-q)}} \left( \psi_{l+1}(E) + \psi_{l-1}(E) \right)~.
\end{equation}
The eigenvalues of such an $L$-dimensional  tridiagonal Toeplitz matrix are given by:
\begin{equation}
    E_k = \frac{2J}{ \sqrt{\lambda(1-q)}}\cos \left(\frac{k\pi}{L+1} \right), \quad k=1,\dots, L ~.
\end{equation}
When $L$ is large the argument of the cosine, $\theta =\frac{k\pi}{L+1}$, becomes a continuous variable between $0$ and $\pi$ and the spectrum has values between $E_{min}=-\frac{2J}{\lambda \sqrt{1-q}}$ and $E_{max}=\frac{2J}{\lambda \sqrt{1-q}}$. The eigenvalues are then a function of the continuous variable\footnote{When $l$ becomes continuous, $\theta$ can be interpreted as momentum.} $\theta$:
\begin{equation} \label{eigvals1}
    E(\theta) = \frac{2J \, \cos \theta}{ \sqrt{\lambda(1-q)}}, \quad 0\leq \theta \leq \pi ~.
\end{equation}
Note that as $L\to \infty$, the variable $\theta$ becomes continuous in a uniform way and thus summing over eigenvalues $\sum_{k=1}^{L}$ becomes $\int_0^\pi d\theta$ with uniform density $\rho(\theta)=1$. This fact will be important for later results. 

To find the eigenvectors, we go back to (\ref{q-eigsys1}) using the result (\ref{eigvals1}) and defining for convenience $\mu=\cos \theta$ (we will use $v_l$ instead of $\psi_l$, reserving the latter to the final normalized result):
\begin{equation} \label{eigsys1}
    2\mu\,v_l(\mu) = \sqrt{1-q^{l+1}}\,v_{l+1}(\mu)+ \sqrt{1-q^l}\, v_{l-1}(\mu) ~.
\end{equation}
This equation can be identified once $v_l(\mu)$ is redefined as follows
\begin{equation}
    v_l(\mu)=\frac{\tilde{v}_l(\mu)}{\sqrt{(q;q)_l}}
\end{equation}
where $(a;q)_n \equiv \prod_{k=0}^{n-1}(1-a q^k)$ is the $q$-Pochhammer symbol. Plugging this redefinition into (\ref{eigsys1}) and multiplying both sides of the equation by $\sqrt{(q;q)_l}$ gives:
\begin{equation}
    2\mu \tilde{v}_l(\mu) = (1-q^l)\tilde{v}_{l-1}(\mu)+\tilde{v}_{l+1}(\mu)
\end{equation}
which now can be identified as the recurrence relation for the \textit{q-Hermite polynomials} (see Appendix \ref{Appx:qHermite}), hence the solution to (\ref{eigsys1}) is
\begin{equation}
    v_l(\mu)=\frac{H_l(\mu | q)}{\sqrt{(q;q)_l}}, \quad \mu=\cos(\theta)~.
\end{equation}
Using the identity (\ref{qH_orthogonality}) and the fact that $0\leq \theta\leq \pi$, it can be shown that the norm squared of these eigenvectors is given by
\begin{equation}
     \sum_{l=0}^\infty |v_l(\mu)|^2 = \sum_{l=0}^\infty \frac{H_l(\mu | q)H_l(\mu|q)}{(q;q)^l}=\frac{2\pi }{|(e^{2i\theta};q)_\infty|^2(q;q)_\infty}~,
\end{equation}
and therefore the normalized eigenvector is given by
\begin{equation}
    \psi_l(\mu) = \sqrt{(q;q)_\infty} |(e^{2i\theta};q)_\infty| \frac{H_l(\mu|q)}{\sqrt{2\pi (q;q)_l}}, \quad \mu=\cos \theta~.
\end{equation}
We note that 
\begin{equation} 
    \psi_0(\mu)=  \sqrt{(q;q)_\infty} |(e^{2i\theta};q)_\infty| \frac{H_0(\mu|q)}{\sqrt{2\pi (q;q)_0}} =  \sqrt{\frac{(q;q)_\infty}{2\pi}} |(e^{2i\theta};q)_\infty|
\end{equation}
where we used $H_0(x|q)=1$ and $(q;q)_0=1$.  We thus can write
\begin{equation} 
\label{Teigenvectors1}
    \psi_l(\mu)= \psi_0(\mu) \frac{H_l(\mu|q)}{\sqrt{ (q;q)_l}} , \quad \mu=\cos \theta.
\end{equation}
The orthogonality relations for these eigenvectors are shown in Appendix \ref{App:Orthogonality}.

\subsection{q-Hermite Polynomials} \label{Appx:qHermite}
The q-Hermite polynomial is defined as \cite{weisstein}:
\begin{equation}
    H_n(\cos \theta|q)= e^{in\theta} {}_2 \phi_0\Big[ \begin{matrix}
        q^{-n}, 0 \\ - 
    \end{matrix};\; q,q^ne^{-2i\theta} \Big]
\end{equation}
where ${}_2 \phi_0$ is a q-hypergeometric function, defined by \cite{weisstein}:
\begin{equation}
    {}_j \phi_k\Big[ \begin{matrix}
        a_1, a_2, \dots, a_j \\ b_1, b_2, \dots, b_k
    \end{matrix};\; q,z \Big] = \sum_{n=0}^\infty \frac{(a_1;q)_n\, (a_2;q)_n\, \dots (a_j;q)_n}{(b_1;q)_n\, (b_2;q)_n\, \dots (b_k;q)_n} \Big[ (-1)^n q^{n(n-1)/2}\Big]^{1+k-j} \frac{z^n}{(q;q)_n} ~.
\end{equation}
Using this definition we have
\begin{equation}
    H_n(\cos \theta|q)=  \sum_{k=0}^\infty \frac{(q^{-n};q)_k}{(q;q)_k} (-1)^k q^{nk-\frac{k(k-1)}{2}} e^{i(n-2k)\theta}
\end{equation}
Now note that, $(q^{-n};q)_k=(1-q^{-n})(1-q^{-n+1})(1-q^{-n+2})\dots (1-q^{-n+k-1}) $ so for $k=n+1$ we have $(q^{-n};q)_{n+1}=(1-q^{-n})(1-q^{-n+1})(1-q^{-n+2})\dots (1-1) =0$. We thus find that for $(q^{-n};q)_{k>n+2}=0$ and the sum becomes truncated at $k=n$:
\begin{equation}
    H_n(\cos \theta|q)=  \sum_{k=0}^n \frac{(q^{-n};q)_k}{(q;q)_k} (-1)^k q^{nk-\frac{k(k-1)}{2}} e^{i(n-2k)\theta}~.
\end{equation}
From this expression we can get another useful expression for $H_n(\cos \theta|q)$, as follows:
\begin{align*}
    (q^{-n};q)_k &= (1-q^{-n})(1-q^{-n+1})(1-q^{-n+2})\dots (1-q^{-n+k-1}) \\
    &= (-1)^k (q^{-n}-1)(q^{-n+1}-1)(q^{-n+2}-1)\dots (q^{-n+k-1}-1) \\
    &= (-1)^k\, q^{-n}(1-q^n)\,q^{-n+1}(1-q^{n+1}))\dots q^{-n+k-1}(1-q^{n-k+1}) \\
    &= (-1)^k\, q^{-nk+\frac{k(k-1)}{2}} \frac{(q;q)_n}{(q;q)_{n-k}}~,
\end{align*}
and thus, another expression for the q-Hermite polynomial is
\begin{equation} \label{qHermite_exp2}
    H_n(\cos \theta|q)=  \sum_{k=0}^n \frac{(q;q)_n}{(q;q)_k\,(q;q)_{n-k}} e^{i(n-2k)\theta}~.
\end{equation}
\subsection{Orthogonality relations}
A useful identity for q-Hermite polynomials with $x=\cos \theta, y=\cos \phi$ (see for example \cite{qHermite_formula}) is:
\begin{equation}
    \sum_{l=0}^\infty H_l(x|q)H_l(y|q) \frac{t^l}{(q;q)_l} = \frac{(t^2;q)_\infty}{(te^{i(\theta+\phi)};q)_\infty (te^{i(\theta-\phi)};q)_\infty(te^{-i(\theta-\phi)};q)_\infty (te^{-i(\theta+\phi)};q)_\infty},
\end{equation}
which for $t=1$ and $x=y$ is exactly the needed identity.
Noting that 
\begin{equation*}
    (te^{i(\theta+\phi)};q)_\infty (te^{-i(\theta+\phi)};q)_\infty = \prod_{k=0}^\infty (1-te^{i(\theta+\phi)}q^k)(1-te^{-i(\theta+\phi)}q^k)=|(te^{i(\theta+\phi)};q)_\infty|^2~,
\end{equation*}
and similarly for $e^{i(\theta-\phi)}$, we have:
\begin{equation*}
    \frac{(t^2;q)_\infty}{|(te^{i(\theta+\phi)};q)_\infty|^2 |(te^{i(\theta-\phi)};q)_\infty|^2 } = \frac{(1-t^2)(qt^2;q)_\infty}{|1-te^{i(\theta+\phi)}|^2|1-te^{i(\theta-\phi)}|^2|(qte^{i(\theta+\phi)};q)_\infty|^2 |(qte^{i(\theta-\phi)};q)_\infty|^2},
\end{equation*}
where it is made clear that for $t=1$ the expression vanishes for all $\theta, \phi$ except for $\theta=\pm \phi$. Taking  $t=1-\epsilon$, in the limit $\epsilon \to 0$ $\theta \to \phi$ we have
\begin{align*}
    |1-te^{i(\theta-\phi)}|^2 &= 1-t[e^{i(\theta-\phi)}+ e^{-i(\theta-\phi)}] +t^2 \\
    &= 1-t[1+i(\theta-\phi)-\frac{(\theta-\phi)^2}{2}+ 1-i(\theta-\phi)-\frac{(\theta-\phi)^2}{2}+\dots] +t^2\\
    &= 1-(1-\epsilon)[2-(\theta-\phi)^2+O((\theta-\phi)^4)] +(1-\epsilon)^2 \\
    &= \epsilon^2 + (\theta-\phi)^2 + O(\epsilon (\theta-\phi)^2)+\dots
\end{align*}
and thus
\begin{align*}
    &\frac{(1-t^2)(qt^2;q)_\infty}{|1-te^{i(\theta+\phi)}|^2|1-te^{i(\theta-\phi)}|^2|(qte^{i(\theta+\phi)};q)_\infty|^2 |(qte^{i(\theta-\phi)};q)_\infty|^2} \\
    &\to \frac{2\epsilon}{\epsilon^2 + (\theta-\phi)^2} \frac{(q;q)_\infty}{|(e^{2i\theta};q)_\infty|^2|(q;q)_\infty|^2}~.
\end{align*}
In the limit $\epsilon \to 0$, the first term is recognized as a delta function $\delta(t)=\lim_{\epsilon\to 0} \frac{1}{\pi} \frac{\epsilon}{\epsilon^2+t^2}$ and taking into account also the possibility $\theta \to -\phi$ it is found that
\begin{equation} \label{qH_orthogonality}
    \sum_{l=0}^\infty \frac{H_l(x|q)H_l(y|q)}{(q;q)_l} = \frac{2\pi [\delta(\theta-\phi)+\delta(\theta+\phi)]}{|(e^{2i\theta};q)_\infty|^2(q;q)_\infty}~.
\end{equation}

\subsection{Orthogonality of the eigenvectors of DSSYK} \label{App:Orthogonality} 
In this Appendix we show that the eigenvectors (\ref{Teigenvectors1}) satisfy orthogonality relations in energy and in chord number.  Firstly we check orthogonality in energy:
\begin{equation}
    \sum_{l=0}^\infty \psi_l(\cos \theta) \psi_l(\cos \phi) = \frac{(q;q)_\infty}{2\pi} |(e^{2i\theta};q)_\infty|^2 \sum_{l=0}^\infty   \frac{H_l(\cos \theta|q)H_l(\cos \phi|q)}{ (q;q)_l} = \delta(\theta-\phi)~,
\end{equation}
where we used the result (\ref{qH_orthogonality}) together with the knowledge that both $\theta$ and $\phi$ get values between $0$ and $\pi$. 
Secondly, we check orthogonality in chord number:
\begin{equation}
    \int_0^\pi d\theta \, \psi_n(\cos \theta) \psi_m(\cos \theta) = \frac{(q;q)_\infty}{\sqrt{ (q;q)_n (q;q)_m}} \int_0^\pi  \frac{d\theta}{2\pi} |(e^{2i\theta};q)_\infty|^2 H_n(\cos \theta|q)H_m(\cos \theta|q)~.
\end{equation}
To compute this we need the following identity:
\begin{equation} \label{exp_qH_identity}
    \int_0^\pi  \frac{d\theta}{2\pi} |(e^{2i\theta};q)_\infty|^2 H_n(\cos \theta|q)H_m(\cos \theta|q) = \frac{(q;q)_n}{(q;q)_\infty}\delta_{nm}~,
\end{equation}
which gives immediately
\begin{equation}
    \int_0^\pi d\theta \, \psi_n(\cos \theta) \psi_m(\cos \theta) = \delta_{nm} ~.
\end{equation}

\section{Continuum approximation of Krylov space}\label{Appx:Cont_approx}

This section provides a derivation of the continuum approximation of Krylov space developed in \cite{Barbon:2019wsy} in order to understand better its applicability and its relation to ballistic propagation.

In the case in which there are no diagonal Lanczos coefficients, $a_n=0$, it is useful to redefine the wavefunctions as
\begin{equation}
    \centering
    \label{Phi_redefinition_i_n}
    \phi_n(t)=i^n\varphi_n(t),
\end{equation}
so that $\varphi_n(t)\in\mathds{R}$, satisfying the recurrence relation:
\begin{equation}
    \centering
    \label{recurrence_varphi_n}
    \dot{\varphi}_n(t) = b_n\varphi_{n-1}(t)-b_{n+1}\varphi_{n+1}(t),
\end{equation}
for all $n=0,...,K-1$ and with initial condition $\varphi_n(0)=\delta_{n,0}$. Note that, for the sake of rigor, we start by considering a finite Krylov space of dimension $K$; we shall eventually consider the limit of large $K$. For consistency, $b_0\equiv 1$ and $\varphi_{-1}(t)\equiv 0$.

We will start by performing some exact manipulations that amount to rewriting (\ref{recurrence_varphi_n}) in a more suggestive form. Since Krylov space is discrete, position is labelled by $n=0,...,K-1$ and the lattice spacing is constant and equal to $1$. Thus, strictly speaking, it does not go to zero in any sense. In order to make a continuum limit possible, we use a change of variables $\xi = \frac{n}{K-1}\equiv \varepsilon n$, where we identify the small parameter $\varepsilon=\frac{1}{K-1}$. As a result, position in Krylov space is denoted by a rescaled variable $\xi$ which takes discrete values contained in the interval $[0,1]$:
\begin{equation}
    \centering
    \label{xi_domain}
    \xi \in \left\{ \frac{n}{K-1},\; n = 0,1,...,K-1 \right\} = \left\{ \varepsilon n,\;n=0,1,...,\frac{1}{\varepsilon} \right\}.
\end{equation}
The spacing between the allowed values of $\xi$ is $\varepsilon=\frac{1}{K-1}$, and we note that in the limit $K\to+\infty$ they accumulate to $\xi\in [0,1]$, hence yielding a continuum limit. But let us still keep $K$ fixed.

An important technicality has to be considered before passing to the continuum limit. We need to assume that the Lanczos coefficients $b_n$ are given by the evaluation at discrete values of the domain of a certain smooth function that we loosely denote $b(n)$. This has to be assumed, since analytic continuation cannot be invoked because for it to apply one needs to have a function defined over at least an interval, not a discrete domain. Our Lanczos coefficients for DSSYK (\ref{Lanczos}) certainly satisfy this assumption, but there are examples in the literature of Lanczos sequences that do not fulfill it: the works \cite{Dymarsky:2021bjq,Avdoshkin:2022xuw,Camargo:2022rnt} have studied systems where the Lanczos sequence shows staggering, the even and odd coefficients following different profiles. For such systems, the Lanczos sequence does not have a well-defined continuum limit.

Upon the previous variable rescaling, we have:
\begin{equation}
    \centering
    \label{b_n_xi}
    b_n = b(n) = b\big( (K-1)\xi \big) =: w(\xi),
\end{equation}
where we have introduced the function $w$, defined over the rescaled compact domain (\ref{xi_domain}). 

Having assumed that the Lanczos coefficients $b_n$ are given by evaluating a smooth function $b(x)$ at discrete values $x=n$, it is now justified to make the same assumption for $\varphi_n(t) = \varphi(t,n)$. Then again, upon the variable rescaling:
\begin{equation}
    \centering
    \label{varphi_n_xi}
    \varphi_n(t) = \varphi(t,n) = \varphi\big(t,(K-1)\xi\big) =: f(t,\xi),
\end{equation}
where the domain for the second variable of $f$ is (\ref{xi_domain}). We can now rewrite (\ref{recurrence_varphi_n}) as:
\begin{equation}
    \centering
    \label{recurrence_f_xi}
    \partial_t f(t,\xi) = w(\xi) f(t,\xi - \varepsilon) -w(\xi + \varepsilon) f(t,\xi + \varepsilon).
\end{equation}
We emphasize that (\ref{recurrence_f_xi}) is an exact rewriting of the recurrence (\ref{recurrence_varphi_n}), with $\varepsilon=\frac{1}{K-1}$. A solution $f(t,\xi)$ of (\ref{recurrence_f_xi}) gives a solution for (\ref{recurrence_varphi_n}) by evaluating $\varphi_n(t)=f(t,\frac{n}{K-1})$.

Expanding (\ref{recurrence_f_xi}), which we can do since we have assumed that $w(\xi)$ is smooth and we have argued that $f(t,\xi)$ can also be taken to be so, we arrive at
\begin{equation}
    \centering
    \label{Chiral_wave_eqn_approx}
    \partial_t f(t,\xi) + v(\xi) \partial_\xi f(t,\xi) + \frac{1}{2}v^\prime (\xi) f(t,\xi) = \mathit{O}(\varepsilon^2),
\end{equation}
where we have defined $v(\xi):=2\varepsilon w(\xi)$. The leading term of (\ref{Chiral_wave_eqn_approx}) is a chiral wave equation with velocity field $v(\xi)$ and mass term $\frac{1}{2}v^\prime (\xi) f(t,\xi)$. This equation can be solved using a change of variables $\xi \mapsto \chi$ such that $\chi=0$ when $\xi = 0$ and $\frac{d\xi}{v(\xi)}=d\chi$. Since $v(\xi)$ is of order $\varepsilon$, this change of variables unwraps the compact domain of $\xi$ and has again a non-compact domain, but the discussion about the compact domain was technically necessary in order to explain in what sense there is a continuum limit, and why $\varepsilon=\frac{1}{K-1}$ controls the deviation from it. Introducing additionally the function 
\begin{equation}
    \centering
    \label{g_chi_def}
    g(t,\chi):= \sqrt{v(\xi(\chi))} f(t,\xi(\chi)),
\end{equation}
equation (\ref{recurrence_f_xi}) simplifies further into:
\begin{equation}
    \centering
    \label{wave_eqn_g_chi}
    \left(\partial_t + \partial_\chi \right)g(t,\chi) = \mathit{O}(\varepsilon^{5/2}),
\end{equation}
where we remind that $\varepsilon=\frac{1}{K-1}$. In the limit $K\to\infty$ equation (\ref{wave_eqn_g_chi}) becomes exactly a chiral wave equation with unit velocity for the wave function $g$ with coordinates $t,\,\chi$. However, this does not imply that the continuum approximation of Krylov space becomes exact and indistinguishable from the discrete problem in the large-$K$ limit: The intermediate steps of rescaling $n$ to a compact domain $\xi = \varepsilon n$ fail strictly speaking when $\varepsilon=0$, and the continuum approximation need not apply in general. It may be applicable for propagating wave packets whose typical length scale is larger than the lattice spacing, but in our cases of interest the initial condition for the discrete wave packet is $\varphi_n(t=0)=\delta_{n0}$, which therefore probes the discreteness of the Krylov chain at early times.

Interestingly, studying (\ref{wave_eqn_g_chi}) we note that this continuum approximation behaves as a classical limit, because it results in ballistic propagation, as we shall explain. Given an initial condition $g(0,\chi)=g_0(\chi)$, which in our case will be proportional to a delta function centered at $\chi=0$, the generic solution of (\ref{wave_eqn_g_chi}) simply propagates this packet to the right, without spreading it:
\begin{equation}
    \centering
    \label{g_solution}
    g(t,\chi)=g_0(\chi - t).
\end{equation}
Since the packet $g(t,\chi)$ is a propagating delta function, K-complexity coincides with the position of its peak, which propagates at velocity $1$ in $\chi$-space. Undoing the change of variables:
\begin{equation}
    \centering
    \label{packet_peak_chi_xi}
    t = \int_0^{\chi(t)}d\chi = \int_{0}^{\xi(t)}\frac{d\xi}{v(\xi)},
\end{equation}
and using $v(\xi)=2\varepsilon w(\xi)= 2 \varepsilon b(\xi / \varepsilon)$, together with $\xi = \varepsilon n$, we reach the equation\footnote{Jos\'e Barb\'on pointed out this expression in discussions prior to the publication of \cite{IV}.}: 
\begin{equation}
    \centering
    \label{n_peak_cont_estimate}
    t = \int_0^{n(t)}\frac{dn}{2b(n)},
\end{equation}
which can be used as an estimate for the position of the peak $n(t)$ in $n$-space, in an approximation in which $n$ is directly promoted to be a continuous variable.

We remark that the continuum approximation gives ballistic propagation as an output, rather than it being an additional, independent assumption. In this sense, this approximation can be thought of as a classical approximation, because the propagation of the wave packet has turned into the problem of the propagation of a localized particle with well-defined position that travels driven by a velocity field $2 b(n)$ in $n$-space.

\section{Wavefunctions and Krylov complexity for specific values of q} \label{App:Wavefunctions}
In this Appendix we evaluate the exact wavefunctions, $\phi_n(t)$, for $q=0$ and for $q=1$. For $q=0$ we can use the result (\ref{phin_closedform}) by plugging $q=0$ directly into the equation.  For $q=1$ we  start from the eigensystem equation with Lanczos coefficients $b_n \sim \sqrt{n}$ and find the eigenvectors in order to compute $\phi_n(t)$.  The analysis of $q\to 1$ is done in the main text.
\subsection{Exact wavefunctions and K-complexity for q=0} \label{App:WFq0}

For $q\to 0$ the Lanczos coefficients are given by $b_n\sim\frac{J}{\sqrt{\lambda}}\equiv b$ ($\to 0$) for all $n=0,1,2,\dots$. When the tridiagonal matrix with these Lanczos coefficients above and below the diagonal is a Liouvillian, the result for $\phi_n(t)$ was found in \cite{Barbon:2019wsy} to be
\begin{equation} \label{phin_q0}
    \phi_n(t) = i^n \varphi_n(t)= i^n\frac{(n+1)}{bt} J_{n+1}(2bt) ~,
\end{equation}
whereas in our case (dynamics in the Schrödinger picture) we need to replace $t\to-t$.
We can arrive at this result from (\ref{phin_closedform}) by setting $q=0$:
\begin{align}
    (0;0)_\infty &= \prod_{k=0}^\infty (1-0\cdot 0^k)=1 = (0;0)_n \\
    (e^{2i\theta};0)_\infty &= 1-e^{2i\theta} \Rightarrow |(e^{2i\theta};0)_\infty|^2 = 4 \sin^2\theta \\
    H_n(\cos\theta|0) &= \sum_{k=0}^n \frac{(0;0)_n}{(0;0)_k(0;0)_{n-k}} e^{i(n-2k)\theta} = \sum_{k=0}^n e^{i(n-2k)\theta} = \frac{\sin[(1+n)\theta]}{\sin \theta}~,
\end{align}
where we used (\ref{qHermite_exp2}) to evaluate $H_n(\cos \theta|0)$. Plugging all this into (\ref{phin_closedform}) we have
\begin{align}
    \phi_n(t) &= \frac{2}{\pi} \int_0^\pi d\theta e^{-2ibt \cos \theta} \sin \theta \sin [(1+n)\theta] \nonumber\\
    &= \frac{1}{\pi} \int_0^\pi d\theta\, e^{-2ibt \cos \theta} \cos(n\,\theta) - \frac{1}{\pi} \int_0^\pi d\theta\, e^{-2ibt \cos \theta} \cos[(n+2)\,\theta] \nonumber\\
    &= i^n J_n(-2bt) + i^{n+2} J_{n+2}(-2bt) = i^n\Big[ J_n(-2bt)+J_{n+2}(-2bt)\Big] \nonumber\\
    &= i^n \frac{(n+1)}{(-bt)}J_{n+1}(-2bt) ~,\label{Bessel_func}
\end{align}
where in the second line we used the trigonometric identity $\sin \alpha \sin \beta = \frac{1}{2} [\cos(\alpha-\beta)-\cos(\alpha+\beta)]$, in the third line we used one of the integral representations of the Bessel function of the first kind\footnote{\url{https://dlmf.nist.gov/10.9} (last consulted on 01/01/2024).} $J_n(z) = \frac{{i^{-n}}}{\pi} \int_0^\pi d\theta \, e^{i z \cos \theta} \cos(n\theta)$, and in the fourth line we used the Bessel function identity \cite{weisstein} $J_n(z)+J_{n+2}(z)=\frac{2(n+1)}{z}J_{n+1}(z)$.  We managed to recover the result (\ref{phin_q0}) with $t\to -t$ as required.

With the wavefunction given by (\ref{Bessel_func}) the result for Krylov complexity can be evaluated exactly as follows:
\begin{align}
    C_K(t) &= \sum_{n=0}^\infty n |\phi_n(t)|^2 = \frac{1}{(bt)^2} \sum_{n=0}^\infty n (n+1)^2 J^2_{n+1}(-2bt)=\frac{1}{(bt)^2} \sum_{n=1}^\infty (n-1)n^2J_n^2(2bt) \nonumber\\
    &=\frac{16}{3}(bt)^2[J_0^2(2bt)+J_1^2(2bt)]+J_0^2(2bt)+\frac{1}{3}J_1^2(2bt)-\frac{8}{3}btJ_0(2bt)J_1(2bt)-1.
\end{align}
Using the asymptotic forms of $J_0$ and $J_1$,
\begin{eqnarray}  
J_0(x)&\sim& \sqrt{\frac{2}{\pi x}}\cos(x-1/4\pi)~,\\
J_1(x)&\sim& \sqrt{\frac{2}{\pi x}}\sin(x-1/4\pi)~,
\end{eqnarray}
gives
\begin{eqnarray}\label{KC_linear}
    C_K(t) &\sim& \frac{16}{3\pi}bt+\frac{4}{3\pi}\cos(4bt)-1+O(1/t)~,
\end{eqnarray}
which is linear in $t$ for large $t$.

\subsection{Exact wavefunctions for q=1}

For $q\to 1$ the Lanczos coefficients are given by $b_n\sim J \sqrt{\frac{n}{\lambda}}$, and the eigensystem equation is $T|v(E)\rangle=E |v(E)\rangle$. Re-defining the energy variable as $E=\frac{J}{\sqrt{\lambda}}\varepsilon$, the equation can be written component-wise as
\begin{equation}
    \varepsilon\,v_l(\varepsilon) = \sqrt{l} \, v_{l-1}(\varepsilon)+\sqrt{l+1} \, v_{l+1}(\varepsilon)~.
\end{equation}
In a similar manner to the general $q$ case, we redefine the eigenvector components as
\begin{equation}
v_l(\varepsilon)=\frac{\tilde{v}_l(\varepsilon)}{\sqrt{l!}}~,
\end{equation}
which leads to the equation
\begin{equation}
    \varepsilon\, \tilde{v}_l(\varepsilon) = l\, \tilde{v}_{l-1}(\varepsilon) + \tilde{v}_{l+1}(\varepsilon) ~.
\end{equation}
This equation is solved by the (regular) Hermite polynomials given by $H_{e_n}(x)=2^{-n/2} H_n(x/\sqrt{2})$. We thus have 
\begin{equation}
    v_l(\varepsilon)= \frac{H_{e_l}(\varepsilon)}{\sqrt{l!}}= \frac{1}{2^{l/2}\sqrt{l!}}H_l\left(\frac{\varepsilon}{\sqrt{2}}\right) ~.
\end{equation}
As in the q-Hermite case, we will use Hermite identities to normalize the eigenvectors.
\begin{equation}
   \sum_{l=0}^\infty v_l(\varepsilon_1) v_l(\varepsilon_2) = \sum_{l=0}^\infty \frac{H_l(\varepsilon_1/\sqrt{2})H_l(\varepsilon_2/\sqrt{2})}{2^l \, l!} = \sqrt{2\pi} e^{\frac{\varepsilon_1^2}{2}}\delta(\varepsilon_1-\varepsilon_2),
\end{equation}
where we used the identity \cite{weisstein}: $\sum_{n=0}^\infty \frac{H_n(x)H_n(y)}{2^n \, n!}=\sqrt{\pi} e^{\frac{1}{2}(x^2+y^2)}\delta(x-y)$ with $x=y=\varepsilon/\sqrt{2}$ (using $\delta((\varepsilon_1-\varepsilon_2)/\sqrt{2})=\sqrt{2}\delta(\varepsilon_1-\varepsilon_2)$). The normalized eigenvectors are then given by
\begin{equation}\label{evecs_q_1}
    \psi_l(\varepsilon)= \frac{1}{(2\pi)^{1/4}2^{l/2}\sqrt{l!}}\, e^{-\varepsilon^2/4}\, H_l\left(\frac{\varepsilon}{\sqrt{2}}\right)~.
\end{equation}
This result can be used directly to find the result (\ref{Heisenberg_Weyl_Wave_Fn}) for the wavefunction in the case $q=1$:

\begin{align}
     \phi_l(t) &= \int_{-\infty}^\infty d\varepsilon\, e^{-i\frac{J}{\sqrt{\lambda}}t\varepsilon} \psi_l(\varepsilon)\psi_0(\varepsilon) = \frac{1}{\sqrt{2\pi}\, 2^{l/2} \sqrt{l!}}  \int_{-\infty}^\infty d\varepsilon\, e^{-i\frac{J}{\sqrt{\lambda}}t\varepsilon} \, e^{-\frac{\varepsilon^2}{2}} \, H_l\left(\frac{\varepsilon}{\sqrt{2}}\right) \\
     &= e^{-\frac{(\gamma t)^2}{2}} \frac{(-i\,\gamma t)^l}{\sqrt{l!}},
\end{align}
where $\gamma = \frac{J}{\sqrt{\lambda}}$. Here, the identity \cite{weisstein} $\int_{-\infty}^\infty dx \, e^{-(z-x)^2}H_n(x)=2^n \sqrt{\pi} z^n$, was used. We have also assumed that the density of states is flat in $\varepsilon$; this has not been proved here, but it can be checked a posteriori by noting that the orthonormal eigenvectors (\ref{evecs_q_1}) indeed satisfy the completeness relation with respect to the flat measure $d\varepsilon$.

%% file: references.bib
@article{IV,
    author = "Rabinovici, E. and S\'anchez-Garrido, A. and Shir, R. and Sonner, J.",
    title = "{A bulk manifestation of Krylov complexity}",
    eprint = "2305.04355",
    archivePrefix = "arXiv",
    primaryClass = "hep-th",
    doi = "10.1007/JHEP08(2023)213",
    journal = "JHEP",
    volume = "08",
    pages = "213",
    year = "2023"
}

@article{III,
    author = "Rabinovici, E. and S\'anchez-Garrido, A. and Shir, R. and Sonner, J.",
    title = "{Krylov complexity from integrability to chaos}",
    eprint = "2207.07701",
    archivePrefix = "arXiv",
    primaryClass = "hep-th",
    doi = "10.1007/JHEP07(2022)151",
    journal = "JHEP",
    volume = "07",
    pages = "151",
    year = "2022"
}

@article{II,
    author = "Rabinovici, E. and S\'anchez-Garrido, A. and Shir, R. and Sonner, J.",
    title = "{Krylov localization and suppression of complexity}",
    eprint = "2112.12128",
    archivePrefix = "arXiv",
    primaryClass = "hep-th",
    doi = "10.1007/JHEP03(2022)211",
    journal = "JHEP",
    volume = "03",
    pages = "211",
    year = "2022"
}

@article{I,
    author = "Rabinovici, E. and S\'anchez-Garrido, A. and Shir, R. and Sonner, J.",
    title = "{Operator complexity: a journey to the edge of Krylov space}",
    eprint = "2009.01862",
    archivePrefix = "arXiv",
    primaryClass = "hep-th",
    doi = "10.1007/JHEP06(2021)062",
    journal = "JHEP",
    volume = "06",
    pages = "062",
    year = "2021"
}

@misc{StringsRuth,
    title={Research Talk 23 - A holographic dual for Krylov complexity}, 
    author={R. Shir},
    year={2023},
    url = {https://events.perimeterinstitute.ca/event/29/contributions/364/},
    note = {Contributed talk for Strings 2023 at the Perimeter Institute of Theoretical Physics, Waterloo, Canada (Online, last consulted on 01/01/2024)}
}

@article{Lanczos:1950zz,
    author = "Lanczos, Cornelius",
    title = "{An iteration method for the solution of the eigenvalue problem of linear differential and integral operators}",
    doi = "10.6028/jres.045.026",
    journal = "J. Res. Natl. Bur. Stand. B",
    volume = "45",
    pages = "255--282",
    year = "1950"
}

@article{Krylov:1931,
    author = "A. N. Krylov",
    title = "{On the numerical solution of the equation determining the frequencies of small oscillations of material systems in applied mechanics questions.}",
    journal = "Proceedings of the Academy of Sciences of the USSR. VII series. Department of Mathematical and Natural Sciences",
    volume = "4",
    pages = "491--539",
    year = "1931",
    note = "(In Russian. Zentralblatt MATH number: 2561215, JFM: 57.1454.02)",
    zbMATH = {2561215},
    JFM = {57.1454.02}
}

@Article{Luzin:1931,
 Author = {Luzin, N.},
 Title = {Sur la m{\'e}thode de {M}. {A}. {Krylov} de composition de l'{\'e}quation s{\'e}culaire.},
 FJournal = {Bulletin de l'Acad{\'e}mie des Sciences de l'URSS. VII. S{\'e}rie},
 Journal = {Bull. Acad. Sci. URSS},
 Volume = {7},
 Pages = {903--958},
 Year = {1931},
 note = "(In Russian. Zentralblatt MATH number: 2561216. JFM: 57.1455.01)",
 zbMATH = {2561216},
 JFM = {57.1455.01}
}

@article{Parker:2018yvk,
    author = "Parker, Daniel E. and Cao, Xiangyu and Avdoshkin, Alexander and Scaffidi, Thomas and Altman, Ehud",
    title = "{A Universal Operator Growth Hypothesis}",
    eprint = "1812.08657",
    archivePrefix = "arXiv",
    primaryClass = "cond-mat.stat-mech",
    doi = "10.1103/PhysRevX.9.041017",
    journal = "Phys. Rev. X",
    volume = "9",
    number = "4",
    pages = "041017",
    year = "2019"
}

@article{Barbon:2019wsy,
    author = "Barb\'on, J. L. F. and Rabinovici, E. and Shir, R. and Sinha, R.",
    title = "{On The Evolution Of Operator Complexity Beyond Scrambling}",
    eprint = "1907.05393",
    archivePrefix = "arXiv",
    primaryClass = "hep-th",
    reportNumber = "IFT-UAM/CSIC-19-98",
    doi = "10.1007/JHEP10(2019)264",
    journal = "JHEP",
    volume = "10",
    pages = "264",
    year = "2019"
}

@article{Jian:2020qpp,
    author = "Jian, Shao-Kai and Swingle, Brian and Xian, Zhuo-Yu",
    title = "{Complexity growth of operators in the SYK model and in JT gravity}",
    eprint = "2008.12274",
    archivePrefix = "arXiv",
    primaryClass = "hep-th",
    doi = "10.1007/JHEP03(2021)014",
    journal = "JHEP",
    volume = "03",
    pages = "014",
    year = "2021"
}

@article{Dymarsky:2019elm,
    author = "Dymarsky, Anatoly and Gorsky, Alexander",
    title = "{Quantum chaos as delocalization in Krylov space}",
    eprint = "1912.12227",
    archivePrefix = "arXiv",
    primaryClass = "cond-mat.stat-mech",
    doi = "10.1103/PhysRevB.102.085137",
    journal = "Phys. Rev. B",
    volume = "102",
    number = "8",
    pages = "085137",
    year = "2020"
}

@article{Dymarsky:2021bjq,
    author = "Dymarsky, Anatoly and Smolkin, Michael",
    title = "{Krylov complexity in conformal field theory}",
    eprint = "2104.09514",
    archivePrefix = "arXiv",
    primaryClass = "hep-th",
    doi = "10.1103/PhysRevD.104.L081702",
    journal = "Phys. Rev. D",
    volume = "104",
    number = "8",
    pages = "L081702",
    year = "2021"
}

@article{Avdoshkin:2022xuw,
    author = "Avdoshkin, Alexander and Dymarsky, Anatoly and Smolkin, Michael",
    title = "{Krylov complexity in quantum field theory, and beyond}",
    eprint = "2212.14429",
    archivePrefix = "arXiv",
    primaryClass = "hep-th",
    month = "12",
    year = "2022"
}

@article{Kundu:2023hbk,
    author = "Kundu, Arnab and Malvimat, Vinay and Sinha, Ritam",
    title = "{State Dependence of Krylov Complexity in $2d$ CFTs}",
    eprint = "2303.03426",
    archivePrefix = "arXiv",
    primaryClass = "hep-th",
    month = "3",
    year = "2023"
}

@article{Caputa:2021sib,
    author = "Caputa, Pawel and Magan, Javier M. and Patramanis, Dimitrios",
    title = "{Geometry of Krylov complexity}",
    eprint = "2109.03824",
    archivePrefix = "arXiv",
    primaryClass = "hep-th",
    doi = "10.1103/PhysRevResearch.4.013041",
    journal = "Phys. Rev. Res.",
    volume = "4",
    number = "1",
    pages = "013041",
    year = "2022"
}

@article{Patramanis:2021lkx,
    author = "Patramanis, Dimitrios",
    title = "{Probing the entanglement of operator growth}",
    eprint = "2111.03424",
    archivePrefix = "arXiv",
    primaryClass = "hep-th",
    doi = "10.1093/ptep/ptac081",
    journal = "PTEP",
    volume = "2022",
    number = "6",
    pages = "063A01",
    year = "2022"
}

@article{Caputa:2021ori,
    author = "Caputa, Pawel and Datta, Shouvik",
    title = "{Operator growth in 2d CFT}",
    eprint = "2110.10519",
    archivePrefix = "arXiv",
    primaryClass = "hep-th",
    reportNumber = "CERN-TH-2021-151",
    doi = "10.1007/JHEP12(2021)188",
    journal = "JHEP",
    volume = "12",
    pages = "188",
    year = "2021",
    note = "[Erratum: JHEP 09, 113 (2022)]"
}

@article{Balasubramanian:2022tpr,
    author = "Balasubramanian, Vijay and Caputa, Pawel and Magan, Javier M. and Wu, Qingyue",
    title = "{Quantum chaos and the complexity of spread of states}",
    eprint = "2202.06957",
    archivePrefix = "arXiv",
    primaryClass = "hep-th",
    doi = "10.1103/PhysRevD.106.046007",
    journal = "Phys. Rev. D",
    volume = "106",
    number = "4",
    pages = "046007",
    year = "2022"
}

@article{Caputa:2022eye,
    author = "Caputa, Pawel and Liu, Sinong",
    title = "{Quantum complexity and topological phases of matter}",
    eprint = "2205.05688",
    archivePrefix = "arXiv",
    primaryClass = "hep-th",
    doi = "10.1103/PhysRevB.106.195125",
    journal = "Phys. Rev. B",
    volume = "106",
    number = "19",
    pages = "195125",
    year = "2022"
}

@article{Caputa:2022yju,
    author = "Caputa, Pawel and Gupta, Nitin and Haque, S. Shajidul and Liu, Sinong and Murugan, Jeff and Van Zyl, Hendrik J. R.",
    title = "{Spread complexity and topological transitions in the Kitaev chain}",
    eprint = "2208.06311",
    archivePrefix = "arXiv",
    primaryClass = "hep-th",
    doi = "10.1007/JHEP01(2023)120",
    journal = "JHEP",
    volume = "01",
    pages = "120",
    year = "2023"
}

@article{Balasubramanian:2022dnj,
    author = "Balasubramanian, Vijay and Magan, Javier M. and Wu, Qingyue",
    title = "{Tridiagonalizing random matrices}",
    eprint = "2208.08452",
    archivePrefix = "arXiv",
    primaryClass = "hep-th",
    doi = "10.1103/PhysRevD.107.126001",
    journal = "Phys. Rev. D",
    volume = "107",
    number = "12",
    pages = "126001",
    year = "2023"
}

@article{Caputa:2023vyr,
    author = "Caputa, Pawel and Magan, Javier M. and Patramanis, Dimitrios and Tonni, Erik",
    title = "{Krylov complexity of modular Hamiltonian evolution}",
    eprint = "2306.14732",
    archivePrefix = "arXiv",
    primaryClass = "hep-th",
    month = "6",
    year = "2023"
}

@article{Patramanis:2023cwz,
    author = "Patramanis, Dimitrios and Sybesma, Watse",
    title = {{Krylov complexity in a natural basis for the Schr\"odinger algebra}},
    eprint = "2306.03133",
    archivePrefix = "arXiv",
    primaryClass = "quant-ph",
    month = "6",
    year = "2023"
}

@article{Balasubramanian:2023kwd,
    author = "Balasubramanian, Vijay and Magan, Javier M. and Wu, Qingyue",
    title = "{Quantum chaos, integrability, and late times in the Krylov basis}",
    eprint = "2312.03848",
    archivePrefix = "arXiv",
    primaryClass = "hep-th",
    month = "12",
    year = "2023"
}

@article{Kar:2021nbm,
    author = "Kar, Arjun and Lamprou, Lampros and Rozali, Moshe and Sully, James",
    title = "{Random matrix theory for complexity growth and black hole interiors}",
    eprint = "2106.02046",
    archivePrefix = "arXiv",
    primaryClass = "hep-th",
    doi = "10.1007/JHEP01(2022)016",
    journal = "JHEP",
    volume = "01",
    pages = "016",
    year = "2022"
}

@article{Trigueros:2021rwj,
    author = "Trigueros, Fabian Ballar and Lin, Cheng-Ju",
    title = "{Krylov complexity of many-body localization: Operator localization in Krylov basis}",
    eprint = "2112.04722",
    archivePrefix = "arXiv",
    primaryClass = "cond-mat.dis-nn",
    doi = "10.21468/SciPostPhys.13.2.037",
    journal = "SciPost Phys.",
    volume = "13",
    number = "2",
    pages = "037",
    year = "2022"
}

@article{Bhattacharjee:2022vlt,
    author = "Bhattacharjee, Budhaditya and Cao, Xiangyu and Nandy, Pratik and Pathak, Tanay",
    title = "{Krylov complexity in saddle-dominated scrambling}",
    eprint = "2203.03534",
    archivePrefix = "arXiv",
    primaryClass = "quant-ph",
    doi = "10.1007/JHEP05(2022)174",
    journal = "JHEP",
    volume = "05",
    pages = "174",
    year = "2022"
}

@article{Muck:2022xfc,
    author = {M\"uck, Wolfgang and Yang, Yi},
    title = "{Krylov complexity and orthogonal polynomials}",
    eprint = "2205.12815",
    archivePrefix = "arXiv",
    primaryClass = "hep-th",
    doi = "10.1016/j.nuclphysb.2022.115948",
    journal = "Nucl. Phys. B",
    volume = "984",
    pages = "115948",
    year = "2022"
}

@article{Carabba:2022itd,
    author = {Carabba, Nicoletta and H\"ornedal, Niklas and del Campo, Adolfo},
    title = "{Quantum speed limits on operator flows and correlation functions}",
    eprint = "2207.05769",
    archivePrefix = "arXiv",
    primaryClass = "quant-ph",
    doi = "10.22331/q-2022-12-22-884",
    journal = "Quantum",
    volume = "6",
    pages = "884",
    year = "2022"
}

@article{Takahashi:2023nkt,
    author = "Takahashi, Kazutaka and del Campo, Adolfo",
    title = "{Shortcuts to Adiabaticity in Krylov Space}",
    eprint = "2302.05460",
    archivePrefix = "arXiv",
    primaryClass = "quant-ph",
    month = "2",
    year = "2023"
}

@article{Bhattacharya:2022gbz,
    author = "Bhattacharya, Aranya and Nandy, Pratik and Nath, Pingal Pratyush and Sahu, Himanshu",
    title = "{Operator growth and Krylov construction in dissipative open quantum systems}",
    eprint = "2207.05347",
    archivePrefix = "arXiv",
    primaryClass = "quant-ph",
    doi = "10.1007/JHEP12(2022)081",
    journal = "JHEP",
    volume = "12",
    pages = "081",
    year = "2022"
}

@article{Liu:2022god,
    author = "Liu, Chang and Tang, Haifeng and Zhai, Hui",
    title = "{Krylov complexity in open quantum systems}",
    eprint = "2207.13603",
    archivePrefix = "arXiv",
    primaryClass = "cond-mat.str-el",
    doi = "10.1103/PhysRevResearch.5.033085",
    journal = "Phys. Rev. Res.",
    volume = "5",
    number = "3",
    pages = "033085",
    year = "2023"
}

@article{Bhattacharjee:2022lzy,
    author = "Bhattacharjee, Budhaditya and Cao, Xiangyu and Nandy, Pratik and Pathak, Tanay",
    title = "{Operator growth in open quantum systems: lessons from the dissipative SYK}",
    eprint = "2212.06180",
    archivePrefix = "arXiv",
    primaryClass = "quant-ph",
    reportNumber = "YITP-22-152",
    doi = "10.1007/JHEP03(2023)054",
    journal = "JHEP",
    volume = "03",
    pages = "054",
    year = "2023"
}

@article{Bhattacharya:2023zqt,
    author = "Bhattacharya, Aranya and Nandy, Pratik and Nath, Pingal Pratyush and Sahu, Himanshu",
    title = "{On Krylov complexity in open systems: an approach via bi-Lanczos algorithm}",
    eprint = "2303.04175",
    archivePrefix = "arXiv",
    primaryClass = "quant-ph",
    reportNumber = "YITP-23-25",
    month = "3",
    year = "2023"
}

@article{Bhattacharjee:2023uwx,
    author = "Bhattacharjee, Budhaditya and Nandy, Pratik and Pathak, Tanay",
    title = "{Operator dynamics in Lindbladian SYK: a Krylov complexity perspective}",
    eprint = "2311.00753",
    archivePrefix = "arXiv",
    primaryClass = "quant-ph",
    reportNumber = "YITP-23-133, RIKEN-iTHEMS-Report-23",
    month = "11",
    year = "2023"
}

@article{Nizami:2023dkf,
    author = "Nizami, Amin A. and Shrestha, Ankit W.",
    title = "{Krylov construction and complexity for driven quantum systems}",
    eprint = "2305.00256",
    archivePrefix = "arXiv",
    primaryClass = "quant-ph",
    month = "4",
    year = "2023"
}

@article{Bhattacharjee:2022qjw,
    author = "Bhattacharjee, Budhaditya and Sur, Samudra and Nandy, Pratik",
    title = "{Probing quantum scars and weak ergodicity breaking through quantum complexity}",
    eprint = "2208.05503",
    archivePrefix = "arXiv",
    primaryClass = "quant-ph",
    doi = "10.1103/PhysRevB.106.205150",
    journal = "Phys. Rev. B",
    volume = "106",
    number = "20",
    pages = "205150",
    year = "2022"
}

@article{Nandy:2023brt,
    author = "Nandy, Sourav and Mukherjee, Bhaskar and Bhattacharyya, Arpan and Banerjee, Aritra",
    title = "{Quantum state complexity meets many-body scars}",
    eprint = "2305.13322",
    archivePrefix = "arXiv",
    primaryClass = "quant-ph",
    month = "5",
    year = "2023"
}

@article{Bhattacharjee:2022ave,
    author = "Bhattacharjee, Budhaditya and Nandy, Pratik and Pathak, Tanay",
    title = "{Krylov complexity in large-$q$ and double-scaled SYK model}",
    eprint = "2210.02474",
    archivePrefix = "arXiv",
    primaryClass = "hep-th",
    reportNumber = "YITP-22-106",
    month = "10",
    year = "2022"
}

@article{Erdmenger:2023wjg,
    author = "Erdmenger, Johanna and Jian, Shao-Kai and Xian, Zhuo-Yu",
    title = "{Universal chaotic dynamics from Krylov space}",
    eprint = "2303.12151",
    archivePrefix = "arXiv",
    primaryClass = "hep-th",
    doi = "10.1007/JHEP08(2023)176",
    journal = "JHEP",
    volume = "08",
    pages = "176",
    year = "2023"
}

@article{Bhattacharyya:2023zda,
    author = "Bhattacharyya, Arpan and Haque, S. Shajidul and Jafari, Ghadir and Murugan, Jeff and Rapotu, Dimakatso",
    title = "{Krylov complexity and spectral form factor for noisy random matrix models}",
    eprint = "2307.15495",
    archivePrefix = "arXiv",
    primaryClass = "hep-th",
    doi = "10.1007/JHEP10(2023)157",
    journal = "JHEP",
    volume = "10",
    pages = "157",
    year = "2023"
}

@article{Kim:2021okd,
    author = "Kim, Joonho and Murugan, Jeff and Olle, Jan and Rosa, Dario",
    title = "{Operator delocalization in quantum networks}",
    eprint = "2109.05301",
    archivePrefix = "arXiv",
    primaryClass = "quant-ph",
    doi = "10.1103/PhysRevA.105.L010201",
    journal = "Phys. Rev. A",
    volume = "105",
    number = "1",
    pages = "L010201",
    year = "2022"
}

@article{Fan:2022xaa,
    author = "Fan, Zhong-Ying",
    title = "{Universal relation for operator complexity}",
    eprint = "2202.07220",
    archivePrefix = "arXiv",
    primaryClass = "quant-ph",
    doi = "10.1103/PhysRevA.105.062210",
    journal = "Phys. Rev. A",
    volume = "105",
    number = "6",
    pages = "062210",
    year = "2022"
}

@article{Fan:2022mdw,
    author = "Fan, Zhong-Ying",
    title = "{The growth of operator entropy in operator growth}",
    eprint = "2206.00855",
    archivePrefix = "arXiv",
    primaryClass = "hep-th",
    doi = "10.1007/JHEP08(2022)232",
    journal = "JHEP",
    volume = "08",
    pages = "232",
    year = "2022"
}

@article{Guo:2022hui,
    author = "Guo, Shiyong",
    title = "{Operator growth in SU(2) Yang-Mills theory}",
    eprint = "2208.13362",
    archivePrefix = "arXiv",
    primaryClass = "hep-th",
    month = "8",
    year = "2022"
}

@article{He:2022ryk,
    author = "He, Song and Lau, Pak Hang Chris and Xian, Zhuo-Yu and Zhao, Long",
    title = "{Quantum chaos, scrambling and operator growth in $ T\overline{T} $ deformed SYK models}",
    eprint = "2209.14936",
    archivePrefix = "arXiv",
    primaryClass = "hep-th",
    doi = "10.1007/JHEP12(2022)070",
    journal = "JHEP",
    volume = "12",
    pages = "070",
    year = "2022"
}

@article{Du:2022ocp,
    author = "Du, Bao-ning and Huang, Min-xin",
    title = "{Krylov Complexity in Calabi-Yau Quantum Mechanics}",
    eprint = "2212.02926",
    archivePrefix = "arXiv",
    primaryClass = "hep-th",
    reportNumber = "USTC-ICTS/PCFT-22-32",
    month = "12",
    year = "2022"
}

@article{Espanol:2022cqr,
    author = "Espa\~nol, Bernardo L. and Wisniacki, Diego A.",
    title = "{Assessing the saturation of Krylov complexity as a measure of chaos}",
    eprint = "2212.06619",
    archivePrefix = "arXiv",
    primaryClass = "quant-ph",
    doi = "10.1103/PhysRevE.107.024217",
    journal = "Phys. Rev. E",
    volume = "107",
    number = "2",
    pages = "024217",
    year = "2023"
}

@article{Alishahiha:2022anw,
    author = "Alishahiha, Mohsen and Banerjee, Souvik",
    title = "{A universal approach to Krylov State and Operator complexities}",
    eprint = "2212.10583",
    archivePrefix = "arXiv",
    primaryClass = "hep-th",
    month = "12",
    year = "2022"
}

@article{Camargo:2022rnt,
    author = "Camargo, Hugo A. and Jahnke, Viktor and Kim, Keun-Young and Nishida, Mitsuhiro",
    title = "{Krylov complexity in free and interacting scalar field theories with bounded power spectrum}",
    eprint = "2212.14702",
    archivePrefix = "arXiv",
    primaryClass = "hep-th",
    doi = "10.1007/JHEP05(2023)226",
    journal = "JHEP",
    volume = "05",
    pages = "226",
    year = "2023"
}

@article{Hornedal:2023xpa,
    author = {H\"ornedal, Niklas and Carabba, Nicoletta and Takahashi, Kazutaka and del Campo, Adolfo},
    title = "{Geometric Operator Quantum Speed Limit, Wegner Hamiltonian Flow and Operator Growth}",
    eprint = "2301.04372",
    archivePrefix = "arXiv",
    primaryClass = "quant-ph",
    doi = "10.22331/q-2023-07-11-1055",
    journal = "Quantum",
    volume = "7",
    pages = "1055",
    year = "2023"
}

@article{Bhattacharjee:2023dik,
    author = "Bhattacharjee, Budhaditya",
    title = "{A Lanczos approach to the Adiabatic Gauge Potential}",
    eprint = "2302.07228",
    archivePrefix = "arXiv",
    primaryClass = "quant-ph",
    month = "2",
    year = "2023"
}

@article{Chattopadhyay:2023fob,
    author = "Chattopadhyay, Arghya and Mitra, Arpita and van Zyl, Hendrik J. R.",
    title = "{Spread complexity as classical dilaton solutions}",
    eprint = "2302.10489",
    archivePrefix = "arXiv",
    primaryClass = "hep-th",
    doi = "10.1103/PhysRevD.108.025013",
    journal = "Phys. Rev. D",
    volume = "108",
    number = "2",
    pages = "025013",
    year = "2023"
}

@article{Lv:2023jbv,
    author = "Lv, Chenwei and Zhang, Ren and Zhou, Qi",
    title = "{Building Krylov complexity from circuit complexity}",
    eprint = "2303.07343",
    archivePrefix = "arXiv",
    primaryClass = "quant-ph",
    month = "3",
    year = "2023"
}

@article{Zhang:2023wtr,
    author = "Zhang, Ren and Zhai, Hui",
    title = "{Universal Hypothesis of Autocorrelation Function from Krylov Complexity}",
    eprint = "2305.02356",
    archivePrefix = "arXiv",
    primaryClass = "cond-mat.stat-mech",
    month = "5",
    year = "2023"
}

@article{Hashimoto:2023swv,
    author = "Hashimoto, Koji and Murata, Keiju and Tanahashi, Norihiro and Watanabe, Ryota",
    title = "{Krylov complexity and chaos in quantum mechanics}",
    eprint = "2305.16669",
    archivePrefix = "arXiv",
    primaryClass = "hep-th",
    reportNumber = "KUNS-2967",
    month = "5",
    year = "2023"
}

@article{Iizuka:2023pov,
    author = "Iizuka, Norihiro and Nishida, Mitsuhiro",
    title = "{Krylov complexity in the IP matrix model}",
    eprint = "2306.04805",
    archivePrefix = "arXiv",
    primaryClass = "hep-th",
    reportNumber = "OU-HET-1189",
    doi = "10.1007/JHEP11(2023)065",
    journal = "JHEP",
    volume = "11",
    pages = "065",
    year = "2023"
}

@article{Iizuka:2023fba,
    author = "Iizuka, Norihiro and Nishida, Mitsuhiro",
    title = "{Krylov complexity in the IP matrix model. Part II}",
    eprint = "2308.07567",
    archivePrefix = "arXiv",
    primaryClass = "hep-th",
    doi = "10.1007/JHEP11(2023)096",
    journal = "JHEP",
    volume = "11",
    pages = "096",
    year = "2023"
}

@article{Bhattacharyya:2023dhp,
    author = "Bhattacharyya, Arpan and Ghosh, Debodirna and Nandi, Poulami",
    title = "{Operator growth and Krylov Complexity in Bose-Hubbard Model}",
    eprint = "2306.05542",
    archivePrefix = "arXiv",
    primaryClass = "hep-th",
    month = "6",
    year = "2023"
}

@article{Camargo:2023eev,
    author = "Camargo, Hugo A. and Jahnke, Viktor and Jeong, Hyun-Sik and Kim, Keun-Young and Nishida, Mitsuhiro",
    title = "{Spectral and Krylov Complexity in Billiard Systems}",
    eprint = "2306.11632",
    archivePrefix = "arXiv",
    primaryClass = "hep-th",
    reportNumber = "IFT-UAM/CSIC-23-61",
    month = "6",
    year = "2023"
}

@article{Fan:2023ohh,
    author = "Fan, Zhong-Ying",
    title = "{Generalised Krylov complexity}",
    eprint = "2306.16118",
    archivePrefix = "arXiv",
    primaryClass = "hep-th",
    month = "6",
    year = "2023"
}

@article{Vasli:2023syq,
    author = "Vasli, M. J. and Babaei Velni, K. and Mohammadi Mozaffar, M. R. and Mollabashi, A. and Alishahiha, M.",
    title = "{Krylov Complexity in Lifshitz-type Scalar Field Theories}",
    eprint = "2307.08307",
    archivePrefix = "arXiv",
    primaryClass = "hep-th",
    month = "7",
    year = "2023"
}

@article{Aguilar-Gutierrez:2023nyk,
    author = "Aguilar-Gutierrez, Sergio E. and Rolph, Andrew",
    title = "{Krylov complexity is not a measure of distance between states or operators}",
    eprint = "2311.04093",
    archivePrefix = "arXiv",
    primaryClass = "hep-th",
    doi = "10.1103/PhysRevD.109.L081701",
    journal = "Phys. Rev. D",
    volume = "109",
    number = "8",
    pages = "L081701",
    year = "2024"
}

@article{Huh:2023jxt,
    author = "Huh, Kyoung-Bum and Jeong, Hyun-Sik and Pedraza, Juan F.",
    title = "{Spread complexity in saddle-dominated scrambling}",
    eprint = "2312.12593",
    archivePrefix = "arXiv",
    primaryClass = "hep-th",
    month = "12",
    year = "2023"
}

@article{Scialchi:2023bmw,
    author = "Scialchi, Gast\'on F. and Roncaglia, Augusto J. and Wisniacki, Diego A.",
    title = "{Integrability to chaos transition through Krylov approach for state evolution}",
    eprint = "2309.13427",
    archivePrefix = "arXiv",
    primaryClass = "quant-ph",
    month = "9",
    year = "2023"
}

@book{nielsen_chuang_2010, place={Cambridge}, title={Quantum Computation and Quantum Information: 10th Anniversary Edition}, DOI={10.1017/CBO9780511976667}, publisher={Cambridge University Press}, author={Nielsen, Michael A. and Chuang, Isaac L.}, year={2010}}

@book{HaakeBook, 
title={Quantum Signatures of Chaos},
DOI={https://doi.org/10.1007/978-3-319-97580-1},
publisher={Springer Nature Switzerland }, 
author={Fritz Haake, Sven Gnutzmann, Marek Kuś}, 
year={2018}}

@book{stöckmann_1999, place={Cambridge}, title={Quantum Chaos: An Introduction}, DOI={10.1017/CBO9780511524622}, publisher={Cambridge University Press}, author={Stöckmann, Hans-Jürgen}, year={1999}}

@book{MehtaBook, 
title={Random Matrices},
DOI={https://doi.org/10.1016/C2009-0-22297-5},
publisher={Elsevier Inc.}, 
author={MADAN LAL MEHTA}, 
year={1990}}

@book{efetov_1996, place={Cambridge}, title={Supersymmetry in Disorder and Chaos}, DOI={10.1017/CBO9780511573057}, publisher={Cambridge University Press}, author={Efetov, Konstantin}, year={1996}}

@book{ViswanathMuller, 
title={The recursion method. Application to many-body dynamics},
publisher={Springer-Verlag Berlin Heidelberg}, 
author={V.S. Viswanath and G. Mueller}, 
year={1994}}

@InProceedings{Magnus,
author="Magnus, Alphonse",
editor="Pettifor, D. G.
and Weaire, D. L.",
title="Asymptotic Behaviour of Continued Fraction Coefficients Related to Singularities of the Weight Function",
booktitle="The Recursion Method and Its Applications",
year="1987",
publisher="Springer Berlin Heidelberg",
address="Berlin, Heidelberg",
pages="22--45"
}

@book{Parlett, 
title={The Symmetric Eigenvalue Problem},
publisher={Society for Industrial and Applied Mathematics}, 
author={Beresford N. Parlett}, 
year={1998}}

@book{DE_Caesa, 
title={Differential Equations: An Operational Approach},
publisher={Rinton Press}, 
author={H. Moya-Cessa and F. Soto-Eguibar}, 
year={2011}}

@book{DE_Teschl, 
title={Ordinary Differential Equations and Dynamical Systems},
publisher={American Mathematical Society}, 
author={G. Teschl}, 
year={2012}}

@book{GalindoI,
    author = {A. Galindo and P. Pascual},
    title = {Quantum Mechanics I},
    publisher = {Springer Berlin, Heidelberg},
    year = {2012}
}

@book{GalindoII,
    author = {A. Galindo and P. Pascual},
    title = {Quantum Mechanics II},
    publisher = {Springer Berlin, Heidelberg},
    year = {2012}
}

@book{Arnold:1989who,
    author = "Arnold, V. I.",
    title = "{Mathematical Methods of Classical Mechanics}",
    doi = "10.1007/978-1-4757-2063-1",
    publisher = "Springer",
    series = "Graduate Texts in Mathematics",
    year = "1989"
}

@book{ClassicalChaosBook,
    author = {A. J. Lichtenberg and M. A. Lieberman},
    title = {Regular and Chaotic Dynamics},
    publisher = {Springer New York, NY},
    year = {2010},
    url = {https://doi.org/10.1007/978-1-4757-2184-3}
}

@book{KrylovBook_Vorst,
    author = {Henk A. van der Vorst},
    title = {Iterative Krylov Methods for Large Linear Systems},
    publisher = {Cambridge University Press},
    year = {2009}
}

@book{KrylovBook_Liesen,
    author = {J. Liesen and Z. Strakos},
    title = "{Krylov Subspace Methods: Principles and Analysis}",
    publisher = {Oxford University Press},
    year = {2012},
}

@book{KrylovBiography_Shtraykh,
    author = {S. Y. Shtraykh},
    title = {Alekséi Nikoláyevich Krylov, His Life and Work},
    publisher = {Gosudarstv. Izdat. Tehn.-Teor. Lit., Moscow-Leningrad},
    year = {1950}
}

@book{KrylovWorks_Smirnov,
    author = {V. I. Smirnov (ed.)},
    title = {The manuscript legacy of academician Aleksei Nikolaevic Krylov. Scientific description},
    publisher = {Izdat. 'Nauka', Leningrad. Otdel, Leningrad},
    year = {1969}
}

@book{MartianBook_Marx,
    author = {György Marx},
    title = {The Voice of the Martians. Hungarian Scientists who Shaped the 20th Century in the West},
    publisher = {Akadémiai Kiadó},
    year = {2001}
}

@book{MartianBook_Hargittai,
    author = {István Hargittai},
    title = {Martians of Science. Five Physicists Who Changed the Twentieth Century},
    publisher = { Oxford University Press},
    year = {2009}
}

@book{MartianBook_Marton,
    author = {Kati Marton} ,
    title = {The Great Escape. Nine Jews Who Fled Hitler and Changed the World},
    publisher = {Simon \& Schuster} ,
    year = {2006}
}

@book{LanczosBiography_Gellai,
    author = {Barbara Gellai},
    title = {The Intrinsic Nature of Things. The Life and Science of Cornelius Lanczos},
    publisher = {American Mathematical Society},
    year = {2010}
}

@book{DiFrancesco:1997nk,
    author = "Di Francesco, P. and Mathieu, P. and Senechal, D.",
    title = "{Conformal Field Theory}",
    doi = "10.1007/978-1-4612-2256-9",
    publisher = "Springer-Verlag",
    address = "New York",
    series = "Graduate Texts in Contemporary Physics",
    year = "1997"
}

@article{Hull:1998vg,
    author = "Hull, C. M.",
    title = "{Timelike T duality, de Sitter space, large N gauge theories and topological field theory}",
    eprint = "hep-th/9806146",
    archivePrefix = "arXiv",
    reportNumber = "QMW-PH-98-28",
    doi = "10.1088/1126-6708/1998/07/021",
    journal = "JHEP",
    volume = "07",
    pages = "021",
    year = "1998"
}

@article{Strominger:2001pn,
    author = "Strominger, Andrew",
    title = "{The dS / CFT correspondence}",
    eprint = "hep-th/0106113",
    archivePrefix = "arXiv",
    doi = "10.1088/1126-6708/2001/10/034",
    journal = "JHEP",
    volume = "10",
    pages = "034",
    year = "2001"
}

@inproceedings{Witten:2001kn,
    author = "Witten, Edward",
    title = "{Quantum gravity in de Sitter space}",
    booktitle = "{Strings 2001: International Conference}",
    eprint = "hep-th/0106109",
    archivePrefix = "arXiv",
    month = "6",
    year = "2001"
}

@article{Klemm:2001ea,
    author = "Klemm, Dietmar",
    title = "{Some aspects of the de Sitter / CFT correspondence}",
    eprint = "hep-th/0106247",
    archivePrefix = "arXiv",
    reportNumber = "IFUM-690-FT",
    doi = "10.1016/S0550-3213(02)00007-X",
    journal = "Nucl. Phys. B",
    volume = "625",
    pages = "295--311",
    year = "2002"
}

@article{Balasubramanian:2002zh,
    author = "Balasubramanian, Vijay and de Boer, Jan and Minic, Djordje",
    editor = "de Wit, B. and Vandoren, S.",
    title = "{Notes on de Sitter space and holography}",
    eprint = "hep-th/0207245",
    archivePrefix = "arXiv",
    reportNumber = "VPI-IPPAP-02-05, UPR-1008-T, IFTA-2002-26",
    doi = "10.1016/S0003-4916(02)00020-9",
    journal = "Class. Quant. Grav.",
    volume = "19",
    pages = "5655--5700",
    year = "2002"
}

@article{Anninos:2017hhn,
    author = "Anninos, Dionysios and Hofman, Diego M.",
    title = "{Infrared Realization of dS$_2$ in AdS$_2$}",
    eprint = "1703.04622",
    archivePrefix = "arXiv",
    primaryClass = "hep-th",
    doi = "10.1088/1361-6382/aab143",
    journal = "Class. Quant. Grav.",
    volume = "35",
    number = "8",
    pages = "085003",
    year = "2018"
}

@article{Anninos:2018svg,
    author = "Anninos, Dionysios and Galante, Dami\'an A. and Hofman, Diego M.",
    title = "{De Sitter horizons \& holographic liquids}",
    eprint = "1811.08153",
    archivePrefix = "arXiv",
    primaryClass = "hep-th",
    doi = "10.1007/JHEP07(2019)038",
    journal = "JHEP",
    volume = "07",
    pages = "038",
    year = "2019"
}

@article{Anninos:2021ihe,
    author = "Anninos, Dionysios and Harris, Eleanor",
    title = "{Three-dimensional de Sitter horizon thermodynamics}",
    eprint = "2106.13832",
    archivePrefix = "arXiv",
    primaryClass = "hep-th",
    doi = "10.1007/JHEP10(2021)091",
    journal = "JHEP",
    volume = "10",
    pages = "091",
    year = "2021"
}

@article{Anninos:2022ujl,
    author = {Anninos, Dionysios and Galante, Dami\'an A. and M\"uhlmann, Beatrix},
    title = "{Finite features of quantum de Sitter space}",
    eprint = "2206.14146",
    archivePrefix = "arXiv",
    primaryClass = "hep-th",
    doi = "10.1088/1361-6382/acaba5",
    journal = "Class. Quant. Grav.",
    volume = "40",
    number = "2",
    pages = "025009",
    year = "2023"
}

@article{Galante:2022nhj,
    author = "Galante, Damian",
    title = "{Geodesics, complexity and holography in (A)dS$_2$}",
    doi = "10.22323/1.406.0359",
    journal = "PoS",
    volume = "CORFU2021",
    pages = "359",
    year = "2022"
}

@article{Galante:2023uyf,
    author = "Galante, Damian A.",
    title = "{Modave lectures on de Sitter space \& holography}",
    eprint = "2306.10141",
    archivePrefix = "arXiv",
    primaryClass = "hep-th",
    doi = "10.22323/1.435.0003",
    journal = "PoS",
    volume = "Modave2022",
    pages = "003",
    year = "2023"
}

@book{Strominger:2017zoo,
    author = "Strominger, Andrew",
    title = "{Lectures on the Infrared Structure of Gravity and Gauge Theory}",
    eprint = "1703.05448",
    archivePrefix = "arXiv",
    primaryClass = "hep-th",
    isbn = "978-0-691-17973-5",
    month = "3",
    year = "2017"
}

@article{Donnay:2020guq,
    author = "Donnay, Laura and Pasterski, Sabrina and Puhm, Andrea",
    title = "{Asymptotic Symmetries and Celestial CFT}",
    eprint = "2005.08990",
    archivePrefix = "arXiv",
    primaryClass = "hep-th",
    reportNumber = "CPHT-RR022.042020",
    doi = "10.1007/JHEP09(2020)176",
    journal = "JHEP",
    volume = "09",
    pages = "176",
    year = "2020"
}

@article{Pasterski:2021rjz,
    author = "Pasterski, Sabrina",
    title = "{Lectures on celestial amplitudes}",
    eprint = "2108.04801",
    archivePrefix = "arXiv",
    primaryClass = "hep-th",
    doi = "10.1140/epjc/s10052-021-09846-7",
    journal = "Eur. Phys. J. C",
    volume = "81",
    number = "12",
    pages = "1062",
    year = "2021"
}

@inproceedings{Pasterski:2021raf,
    author = "Pasterski, Sabrina and Pate, Monica and Raclariu, Ana-Maria",
    title = "{Celestial Holography}",
    booktitle = "{Snowmass 2021}",
    eprint = "2111.11392",
    archivePrefix = "arXiv",
    primaryClass = "hep-th",
    month = "11",
    year = "2021"
}

@article{Mizera:2022sln,
    author = "Mizera, Sebastian and Pasterski, Sabrina",
    title = "{Celestial geometry}",
    eprint = "2204.02505",
    archivePrefix = "arXiv",
    primaryClass = "hep-th",
    doi = "10.1007/JHEP09(2022)045",
    journal = "JHEP",
    volume = "09",
    pages = "045",
    year = "2022"
}

@article{Gonzo:2022tjm,
    author = "Gonzo, Riccardo and McLoughlin, Tristan and Puhm, Andrea",
    title = "{Celestial holography on Kerr-Schild backgrounds}",
    eprint = "2207.13719",
    archivePrefix = "arXiv",
    primaryClass = "hep-th",
    reportNumber = "CPHT-RR046.062022, HU-EP-22/26, SAGEX-22-24-E, TCD 22-04",
    doi = "10.1007/JHEP10(2022)073",
    journal = "JHEP",
    volume = "10",
    pages = "073",
    year = "2022"
}

@article{Sonner:2018rmt,
    author = "Sonner, Julian and Withers, Benjamin",
    title = "{Linear gravity from conformal symmetry}",
    eprint = "1810.12923",
    archivePrefix = "arXiv",
    primaryClass = "hep-th",
    doi = "10.1088/1361-6382/ab0d3f",
    month = "10",
    year = "2018"
}

@article{Susskind:2014rva,
    author = "Susskind, Leonard",
    title = "{Computational Complexity and Black Hole Horizons}",
    eprint = "1403.5695",
    archivePrefix = "arXiv",
    primaryClass = "hep-th",
    doi = "10.1002/prop.201500092",
    journal = "Fortsch. Phys.",
    volume = "64",
    pages = "24--43",
    year = "2016",
    note = "[Addendum: Fortsch.Phys. 64, 44--48 (2016)]"
}

@article{Stanford:2014jda,
    author = "Stanford, Douglas and Susskind, Leonard",
    title = "{Complexity and Shock Wave Geometries}",
    eprint = "1406.2678",
    archivePrefix = "arXiv",
    primaryClass = "hep-th",
    doi = "10.1103/PhysRevD.90.126007",
    journal = "Phys. Rev. D",
    volume = "90",
    number = "12",
    pages = "126007",
    year = "2014"
}

@article{Brown:2015bva,
    author = "Brown, Adam R. and Roberts, Daniel A. and Susskind, Leonard and Swingle, Brian and Zhao, Ying",
    title = "{Holographic Complexity Equals Bulk Action?}",
    eprint = "1509.07876",
    archivePrefix = "arXiv",
    primaryClass = "hep-th",
    doi = "10.1103/PhysRevLett.116.191301",
    journal = "Phys. Rev. Lett.",
    volume = "116",
    number = "19",
    pages = "191301",
    year = "2016"
}

@article{Brown:2015lvg,
    author = "Brown, Adam R. and Roberts, Daniel A. and Susskind, Leonard and Swingle, Brian and Zhao, Ying",
    title = "{Complexity, action, and black holes}",
    eprint = "1512.04993",
    archivePrefix = "arXiv",
    primaryClass = "hep-th",
    doi = "10.1103/PhysRevD.93.086006",
    journal = "Phys. Rev. D",
    volume = "93",
    number = "8",
    pages = "086006",
    year = "2016"
}

@article{Ben-Ami:2016qex,
    author = "Ben-Ami, Omer and Carmi, Dean",
    title = "{On Volumes of Subregions in Holography and Complexity}",
    eprint = "1609.02514",
    archivePrefix = "arXiv",
    primaryClass = "hep-th",
    doi = "10.1007/JHEP11(2016)129",
    journal = "JHEP",
    volume = "11",
    pages = "129",
    year = "2016"
}

@article{Chapman:2016hwi,
    author = "Chapman, Shira and Marrochio, Hugo and Myers, Robert C.",
    title = "{Complexity of Formation in Holography}",
    eprint = "1610.08063",
    archivePrefix = "arXiv",
    primaryClass = "hep-th",
    doi = "10.1007/JHEP01(2017)062",
    journal = "JHEP",
    volume = "01",
    pages = "062",
    year = "2017"
}

@article{Carmi:2017jqz,
    author = "Carmi, Dean and Chapman, Shira and Marrochio, Hugo and Myers, Robert C. and Sugishita, Sotaro",
    title = "{On the Time Dependence of Holographic Complexity}",
    eprint = "1709.10184",
    archivePrefix = "arXiv",
    primaryClass = "hep-th",
    doi = "10.1007/JHEP11(2017)188",
    journal = "JHEP",
    volume = "11",
    pages = "188",
    year = "2017"
}

@article{Belin:2021bga,
    author = "Belin, Alexandre and Myers, Robert C. and Ruan, Shan-Ming and S\'arosi, G\'abor and Speranza, Antony J.",
    title = "{Does Complexity Equal Anything?}",
    eprint = "2111.02429",
    archivePrefix = "arXiv",
    primaryClass = "hep-th",
    reportNumber = "CERN-TH-2021-181, YITP-22-02",
    doi = "10.1103/PhysRevLett.128.081602",
    journal = "Phys. Rev. Lett.",
    volume = "128",
    number = "8",
    pages = "081602",
    year = "2022"
}

@article{Belin:2022xmt,
    author = "Belin, Alexandre and Myers, Robert C. and Ruan, Shan-Ming and S\'arosi, G\'abor and Speranza, Antony J.",
    title = "{Complexity equals anything II}",
    eprint = "2210.09647",
    archivePrefix = "arXiv",
    primaryClass = "hep-th",
    reportNumber = "CERN-TH-2022-159; YITP-22-101",
    doi = "10.1007/JHEP01(2023)154",
    journal = "JHEP",
    volume = "01",
    pages = "154",
    year = "2023"
}

@article{BGSpaper,
  title = {Characterization of Chaotic Quantum Spectra and Universality of Level Fluctuation Laws},
  author = {Bohigas, O. and Giannoni, M. J. and Schmit, C.},
  journal = {Phys. Rev. Lett.},
  volume = {52},
  issue = {1},
  pages = {1--4},
  numpages = {0},
  year = {1984},
  month = {1},
  publisher = {American Physical Society},
  doi = {10.1103/PhysRevLett.52.1},
  url = {https://link.aps.org/doi/10.1103/PhysRevLett.52.1}
}

@InProceedings{Bohigas_survey_proceedings,
author="Bohigas, O.
and Haq, R. U.
and Pandey, A.",
editor="B{\"o}ckhoff, K. H.",
title="Fluctuation Properties of Nuclear Energy Levels and Widths : Comparison of Theory with Experiment",
booktitle="Nuclear Data for Science and Technology",
year="1983",
publisher="Springer Netherlands",
address="Dordrecht",
pages="809--813"
}

@article{Bohigas_survey_paper,
  title = {Fluctuation Properties of Nuclear Energy Levels: Do Theory and Experiment Agree?},
  author = {Haq, R. U. and Pandey, A. and Bohigas, O.},
  journal = {Phys. Rev. Lett.},
  volume = {48},
  issue = {16},
  pages = {1086--1089},
  numpages = {0},
  year = {1982},
  month = {4},
  publisher = {American Physical Society},
  doi = {10.1103/PhysRevLett.48.1086},
  url = {https://link.aps.org/doi/10.1103/PhysRevLett.48.1086}
}

@article{BerryTabor,
 ISSN = {00804630},
 URL = {http://www.jstor.org/stable/79349},
 author = {M. V. Berry and M. Tabor},
 journal = {Proceedings of the Royal Society of London. Series A, Mathematical and Physical Sciences},
 number = {1686},
 pages = {375--394},
 publisher = {The Royal Society},
 title = {Level Clustering in the Regular Spectrum},
 urldate = {2023-08-15},
 volume = {356},
 year = {1977}
}

@article{Altland:2020ccq,
    author = "Altland, Alexander and Sonner, Julian",
    title = "{Late time physics of holographic quantum chaos}",
    eprint = "2008.02271",
    archivePrefix = "arXiv",
    primaryClass = "hep-th",
    doi = "10.21468/SciPostPhys.11.2.034",
    journal = "SciPost Phys.",
    volume = "11",
    pages = "034",
    year = "2021"
}

@article{Altland:2021rqn,
    author = "Altland, Alexander and Bagrets, Dmitry and Nayak, Pranjal and Sonner, Julian and Vielma, Manuel",
    title = "{From operator statistics to wormholes}",
    eprint = "2105.12129",
    archivePrefix = "arXiv",
    primaryClass = "hep-th",
    doi = "10.1103/PhysRevResearch.3.033259",
    journal = "Phys. Rev. Res.",
    volume = "3",
    number = "3",
    pages = "033259",
    year = "2021"
}

@article{Altland:2022xqx,
    author = "Altland, Alexander and Post, Boris and Sonner, Julian and van der Heijden, Jeremy and Verlinde, Erik",
    title = "{Quantum chaos in 2D gravity}",
    eprint = "2204.07583",
    archivePrefix = "arXiv",
    primaryClass = "hep-th",
    doi = "10.21468/SciPostPhys.15.2.064",
    journal = "SciPost Phys.",
    volume = "15",
    pages = "064",
    year = "2023"
}

@misc{slavnov2019algebraic,
      title={Algebraic Bethe ansatz}, 
      author={N. A. Slavnov},
      year={2019},
      eprint={1804.07350},
      archivePrefix={arXiv},
      primaryClass={math-ph}
}

@misc{ReffertIntegrNotes,
      title={Introduction to Integrable Models [Online; last accessed 28-December-2023]}, 
      author={S. Reffert},
      year={2014},
      url = {http://www.reffert.itp.unibe.ch/Lectures.pdf}
}

@article{Maldacena:2015waa,
    author = "Maldacena, Juan and Shenker, Stephen H. and Stanford, Douglas",
    title = "{A bound on chaos}",
    eprint = "1503.01409",
    archivePrefix = "arXiv",
    primaryClass = "hep-th",
    doi = "10.1007/JHEP08(2016)106",
    journal = "JHEP",
    volume = "08",
    pages = "106",
    year = "2016"
}

@article{Roberts:2014isa,
    author = "Roberts, Daniel A. and Stanford, Douglas and Susskind, Leonard",
    title = "{Localized shocks}",
    eprint = "1409.8180",
    archivePrefix = "arXiv",
    primaryClass = "hep-th",
    reportNumber = "MIT-CTP-4594, SU-ITP-14-20",
    doi = "10.1007/JHEP03(2015)051",
    journal = "JHEP",
    volume = "03",
    pages = "051",
    year = "2015"
}

@article{Shenker:2013pqa,
    author = "Shenker, Stephen H. and Stanford, Douglas",
    title = "{Black holes and the butterfly effect}",
    eprint = "1306.0622",
    archivePrefix = "arXiv",
    primaryClass = "hep-th",
    reportNumber = "SU-ITP-13-08",
    doi = "10.1007/JHEP03(2014)067",
    journal = "JHEP",
    volume = "03",
    pages = "067",
    year = "2014"
}

@article{Srednicki:1994mfb,
    author = "Srednicki, Mark",
    title = "{Chaos and Quantum Thermalization}",
    eprint = "cond-mat/9403051",
    archivePrefix = "arXiv",
    doi = "10.1103/PhysRevE.50.888",
    month = "3",
    year = "1994"
}

@article{Schlenker:2022dyo,
    author = "Schlenker, Jean-Marc and Witten, Edward",
    title = "{No ensemble averaging below the black hole threshold}",
    eprint = "2202.01372",
    archivePrefix = "arXiv",
    primaryClass = "hep-th",
    doi = "10.1007/JHEP07(2022)143",
    journal = "JHEP",
    volume = "07",
    pages = "143",
    year = "2022"
}

@article{Berkooz:2018jqr,
    author = "Berkooz, Micha and Isachenkov, Mikhail and Narovlansky, Vladimir and Torrents, Genis",
    title = "{Towards a full solution of the large N double-scaled SYK model}",
    eprint = "1811.02584",
    archivePrefix = "arXiv",
    primaryClass = "hep-th",
    doi = "10.1007/JHEP03(2019)079",
    journal = "JHEP",
    volume = "03",
    pages = "079",
    year = "2019"
}

@article{Lin:2022rbf,
    author = "Lin, Henry W.",
    title = "{The bulk Hilbert space of double scaled SYK}",
    eprint = "2208.07032",
    archivePrefix = "arXiv",
    primaryClass = "hep-th",
    doi = "10.1007/JHEP11(2022)060",
    journal = "JHEP",
    volume = "11",
    pages = "060",
    year = "2022"
}

@article{Barbon:2003aq,
    author = "Barbon, J. L. F. and Rabinovici, E.",
    title = "{Very long time scales and black hole thermal equilibrium}",
    eprint = "hep-th/0308063",
    archivePrefix = "arXiv",
    reportNumber = "CERN-TH-2003-167, RI-03-07-007",
    doi = "10.1088/1126-6708/2003/11/047",
    journal = "JHEP",
    volume = "11",
    pages = "047",
    year = "2003"
}

@article{Barbon:2014rma,
    author = "Barbon, Jose L. F. and Rabinovici, Eliezer",
    title = "{Geometry And Quantum Noise}",
    eprint = "1404.7085",
    archivePrefix = "arXiv",
    primaryClass = "hep-th",
    reportNumber = "IFT-UAM-CSIC-14-034",
    doi = "10.1002/prop.201400044",
    journal = "Fortsch. Phys.",
    volume = "62",
    pages = "626--646",
    year = "2014"
}

@article{Maldacena:1997re,
    author = "Maldacena, Juan Martin",
    title = "{The Large N limit of superconformal field theories and supergravity}",
    eprint = "hep-th/9711200",
    archivePrefix = "arXiv",
    reportNumber = "HUTP-97-A097, HUTP-98-A097",
    doi = "10.4310/ATMP.1998.v2.n2.a1",
    journal = "Adv. Theor. Math. Phys.",
    volume = "2",
    pages = "231--252",
    year = "1998"
}

@article{Schwarzschild:1916uq,
    author = "Schwarzschild, Karl",
    title = "{On the gravitational field of a mass point according to Einstein's theory}",
    eprint = "physics/9905030",
    archivePrefix = "arXiv",
    journal = "Sitzungsber. Preuss. Akad. Wiss. Berlin (Math. Phys. )",
    volume = "1916",
    pages = "189--196",
    year = "1916"
}

@article{Schwarzschild:1916ae,
    author = "Schwarzschild, Karl",
    title = "{On the gravitational field of a sphere of incompressible fluid according to Einstein's theory}",
    eprint = "physics/9912033",
    archivePrefix = "arXiv",
    journal = "Sitzungsber. Preuss. Akad. Wiss. Berlin (Math. Phys. )",
    volume = "1916",
    pages = "424--434",
    year = "1916"
}

@article{Einstein:1916vd,
    author = "Einstein, Albert",
    editor = "Hsu, Jong-Ping and Fine, D.",
    title = "{The foundation of the general theory of relativity.}",
    doi = "10.1002/andp.19163540702",
    journal = "Annalen Phys.",
    volume = "49",
    number = "7",
    pages = "769--822",
    year = "1916"
}

@article{Lanczos:1924zz,
    author = "Lanczos, Cornelius",
    title = "{Uber eine stationare kosmologie im sinne der einsteinschen gravitationstheorie}",
    doi = "10.1007/BF01328251",
    journal = "Z. Phys.",
    volume = "21",
    pages = "73--110",
    year = "1924"
}

@article{Lanczos:1932zz,
    author = "Lanczos, Cornelius",
    title = "{Electricity as a natural property of Riemannian geometry}",
    doi = "10.1103/RevModPhys.39.716",
    journal = "Rev. Mod. Phys.",
    volume = "39",
    pages = "716--736",
    year = "1932"
}

@article{Lanczos:1938sf,
    author = "Lanczos, Cornelius",
    title = "{A Remarkable property of the Riemann-Christoffel tensor in four dimensions}",
    doi = "10.2307/1968467",
    journal = "Annals Math.",
    volume = "39",
    pages = "842--850",
    year = "1938"
}

@article{Lanczos:1942zz,
    author = "Lanczos, Cornelius",
    title = "{Matter waves and electricity}",
    doi = "10.1103/PhysRev.61.713",
    journal = "Phys. Rev.",
    volume = "61",
    pages = "713--720",
    year = "1942"
}

@article{Lanczos:1949zz,
    author = "Lanczos, Cornelius",
    title = "{Lagrangian multiplier and Riemannian spaces}",
    doi = "10.1103/RevModPhys.21.497",
    journal = "Rev. Mod. Phys.",
    volume = "21",
    pages = "497--502",
    year = "1949"
}

@article{Lanczos:1957zz,
    author = "Lanczos, Cornelius",
    title = "{Electricity and general relativity}",
    doi = "10.1103/RevModPhys.29.337",
    journal = "Rev. Mod. Phys.",
    volume = "29",
    pages = "337--350",
    year = "1957"
}

@article{Lanczos:1962zz,
    author = "Lanczos, Cornelius",
    title = "{The splitting of the Riemann tensor}",
    doi = "10.1103/RevModPhys.34.379",
    journal = "Rev. Mod. Phys.",
    volume = "34",
    pages = "379--389",
    year = "1962"
}

@article{Lanczos:1963zz,
    author = "Lanczos, Cornelius",
    title = "{Undulatory Riemannian spaces}",
    doi = "10.1063/1.1704021",
    journal = "J. Math. Phys.",
    volume = "4",
    pages = "951--959",
    year = "1963"
}

@article{Lanczos:1975su,
    author = "Lanczos, Cornelius",
    title = "{Gravitation and Riemannian Space}",
    doi = "10.1007/BF01100310",
    journal = "Found. Phys.",
    volume = "5",
    pages = "9--18",
    year = "1975"
}

@article{Witten:1998qj,
    author = "Witten, Edward",
    title = "{Anti-de Sitter space and holography}",
    eprint = "hep-th/9802150",
    archivePrefix = "arXiv",
    reportNumber = "IASSNS-HEP-98-15",
    doi = "10.4310/ATMP.1998.v2.n2.a2",
    journal = "Adv. Theor. Math. Phys.",
    volume = "2",
    pages = "253--291",
    year = "1998"
}

@article{Anous:2019yku,
    author = "Anous, Tarek and Sonner, Julian",
    title = "{Phases of scrambling in eigenstates}",
    eprint = "1903.03143",
    archivePrefix = "arXiv",
    primaryClass = "hep-th",
    doi = "10.21468/SciPostPhys.7.1.003",
    journal = "SciPost Phys.",
    volume = "7",
    pages = "003",
    year = "2019"
}

@article{Sekino:2008he,
    author = "Sekino, Yasuhiro and Susskind, Leonard",
    title = "{Fast Scramblers}",
    eprint = "0808.2096",
    archivePrefix = "arXiv",
    primaryClass = "hep-th",
    reportNumber = "SU-ITP-08-18, OIQP-08-08",
    doi = "10.1088/1126-6708/2008/10/065",
    journal = "JHEP",
    volume = "10",
    pages = "065",
    year = "2008"
}

@article{Gubser:1998bc,
    author = "Gubser, S. S. and Klebanov, Igor R. and Polyakov, Alexander M.",
    title = "{Gauge theory correlators from noncritical string theory}",
    eprint = "hep-th/9802109",
    archivePrefix = "arXiv",
    reportNumber = "PUPT-1767",
    doi = "10.1016/S0370-2693(98)00377-3",
    journal = "Phys. Lett. B",
    volume = "428",
    pages = "105--114",
    year = "1998"
}

@article{Heemskerk:2009pn,
    author = "Heemskerk, Idse and Penedones, Joao and Polchinski, Joseph and Sully, James",
    title = "{Holography from Conformal Field Theory}",
    eprint = "0907.0151",
    archivePrefix = "arXiv",
    primaryClass = "hep-th",
    reportNumber = "NSF-KITP-09-110",
    doi = "10.1088/1126-6708/2009/10/079",
    journal = "JHEP",
    volume = "10",
    pages = "079",
    year = "2009"
}

@article{Penedones:2010ue,
    author = "Penedones, Joao",
    title = "{Writing CFT correlation functions as AdS scattering amplitudes}",
    eprint = "1011.1485",
    archivePrefix = "arXiv",
    primaryClass = "hep-th",
    doi = "10.1007/JHEP03(2011)025",
    journal = "JHEP",
    volume = "03",
    pages = "025",
    year = "2011"
}

@article{Fitzpatrick:2011ia,
    author = "Fitzpatrick, A. Liam and Kaplan, Jared and Penedones, Joao and Raju, Suvrat and van Rees, Balt C.",
    title = "{A Natural Language for AdS/CFT Correlators}",
    eprint = "1107.1499",
    archivePrefix = "arXiv",
    primaryClass = "hep-th",
    reportNumber = "SLAC-PUB-14506, HRI-ST-1107",
    doi = "10.1007/JHEP11(2011)095",
    journal = "JHEP",
    volume = "11",
    pages = "095",
    year = "2011"
}

@article{Harlow:2011ke,
    author = "Harlow, Daniel and Stanford, Douglas",
    title = "{Operator Dictionaries and Wave Functions in AdS/CFT and dS/CFT}",
    eprint = "1104.2621",
    archivePrefix = "arXiv",
    primaryClass = "hep-th",
    reportNumber = "SU-ITP-11-22",
    month = "4",
    year = "2011"
}

@article{Brown:1986nw,
    author = "Brown, J. David and Henneaux, M.",
    title = "{Central Charges in the Canonical Realization of Asymptotic Symmetries: An Example from Three-Dimensional Gravity}",
    doi = "10.1007/BF01211590",
    journal = "Commun. Math. Phys.",
    volume = "104",
    pages = "207--226",
    year = "1986"
}

@article{Aharony:1999ti,
    author = "Aharony, Ofer and Gubser, Steven S. and Maldacena, Juan Martin and Ooguri, Hirosi and Oz, Yaron",
    title = "{Large N field theories, string theory and gravity}",
    eprint = "hep-th/9905111",
    archivePrefix = "arXiv",
    reportNumber = "CERN-TH-99-122, HUTP-99-A027, LBNL-43113, RU-99-18, UCB-PTH-99-16, LBL-43113",
    doi = "10.1016/S0370-1573(99)00083-6",
    journal = "Phys. Rept.",
    volume = "323",
    pages = "183--386",
    year = "2000"
}

@article{Henningson:1998gx,
    author = "Henningson, M. and Skenderis, K.",
    title = "{The Holographic Weyl anomaly}",
    eprint = "hep-th/9806087",
    archivePrefix = "arXiv",
    reportNumber = "CERN-TH-98-188, KUL-TF-98-21",
    doi = "10.1088/1126-6708/1998/07/023",
    journal = "JHEP",
    volume = "07",
    pages = "023",
    year = "1998"
}

@article{deBoer:1999tgo,
    author = "de Boer, Jan and Verlinde, Erik P. and Verlinde, Herman L.",
    title = "{On the holographic renormalization group}",
    eprint = "hep-th/9912012",
    archivePrefix = "arXiv",
    reportNumber = "PUPT-1898, ITFA-99-39, SPIN-1999-29",
    doi = "10.1088/1126-6708/2000/08/003",
    journal = "JHEP",
    volume = "08",
    pages = "003",
    year = "2000"
}

@article{Skenderis:2002wp,
    author = "Skenderis, Kostas",
    editor = "de Wit, B. and Vandoren, S.",
    title = "{Lecture notes on holographic renormalization}",
    eprint = "hep-th/0209067",
    archivePrefix = "arXiv",
    reportNumber = "PUTP-2047",
    doi = "10.1088/0264-9381/19/22/306",
    journal = "Class. Quant. Grav.",
    volume = "19",
    pages = "5849--5876",
    year = "2002"
}

@article{Einstein:1935tc,
    author = "Einstein, Albert and Rosen, N.",
    title = "{The Particle Problem in the General Theory of Relativity}",
    doi = "10.1103/PhysRev.48.73",
    journal = "Phys. Rev.",
    volume = "48",
    pages = "73--77",
    year = "1935"
}

@article{Bekenstein:1973ur,
    author = "Bekenstein, Jacob D.",
    title = "{Black holes and entropy}",
    doi = "10.1103/PhysRevD.7.2333",
    journal = "Phys. Rev. D",
    volume = "7",
    pages = "2333--2346",
    year = "1973"
}

@article{Hilbert:1928,
    author = {D. Hilbert, J. v. Neumann and L. Nordheim},
    title = {Über die Grundlagen der Quantenmechanik},
    journal = {Mathematische Annalen},
    volume = {98},
    pages = {1--30},
    year = {1928},
    month = {3},
    doi= {10.1007/BF01451579}
}

@article{Neumann:1930,
    author = {v. Neumann, J.},
    title = {Allgemeine Eigenwerttheorie Hermitescher Funktionaloperatoren},
    journal = { Mathematische Annalen },
    volume = {102},
    pages = {49--131},
    year = {1930},
    month = {12},
    doi= {10.1007/BF01782338}
}

@article{Jordan:1933vh,
    author = "Jordan, Pascual and von Neumann, J. and Wigner, Eugene P.",
    title = "{On an Algebraic generalization of the quantum mechanical formalism}",
    doi = "10.2307/1968117",
    journal = "Annals Math.",
    volume = "35",
    pages = "29--64",
    year = "1934"
}

@article{Kramer_1993,
doi = {10.1088/0034-4885/56/12/001},
url = {https://dx.doi.org/10.1088/0034-4885/56/12/001},
year = {1993},
month = {12},
publisher = {},
volume = {56},
number = {12},
pages = {1469},
author = {B Kramer and  A MacKinnon},
title = {Localization: theory and experiment},
journal = {Reports on Progress in Physics}
}

@misc{Mathematica,
  author = {Wolfram Research{,} Inc.},
  title = {Mathematica, {V}ersion 13.3},
  url = {https://www.wolfram.com/mathematica},
  note = {Champaign, IL, 2023}
}

@misc{weisstein,
    author   = {Weisstein, Eric W.},
    title    = {Tree. {From MathWorld---A Wolfram Web Resource}},
    url      = {\url{http://mathworld.wolfram.com/Tree.html}},
    note     = {Last visited on 13/4/2012},
}

@Article{         harris2020array,
 title         = {Array programming with {NumPy}},
 author        = {Charles R. Harris and K. Jarrod Millman and St{\'{e}}fan J.
                 van der Walt and Ralf Gommers and Pauli Virtanen and David
                 Cournapeau and Eric Wieser and Julian Taylor and Sebastian
                 Berg and Nathaniel J. Smith and Robert Kern and Matti Picus
                 and Stephan Hoyer and Marten H. van Kerkwijk and Matthew
                 Brett and Allan Haldane and Jaime Fern{\'{a}}ndez del
                 R{\'{i}}o and Mark Wiebe and Pearu Peterson and Pierre
                 G{\'{e}}rard-Marchant and Kevin Sheppard and Tyler Reddy and
                 Warren Weckesser and Hameer Abbasi and Christoph Gohlke and
                 Travis E. Oliphant},
 year          = {2020},
 month         = {9},
 journal       = {Nature},
 volume        = {585},
 number        = {7825},
 pages         = {357--362},
 doi           = {10.1038/s41586-020-2649-2},
 publisher     = {Springer Science and Business Media {LLC}},
 url           = {https://doi.org/10.1038/s41586-020-2649-2}
}

@article{MotzkinRef,
  title        = {Motzkin paths, Motzkin polynomials and recurrence relations},
  author       = {Oste, Roy and Van der Jeugt, Joris},
  journal      = {ELECTRONIC JOURNAL OF COMBINATORICS},
  volume       = {22},
  year         = {2015},
  url = {https://doi.org/10.37236/4781}
}

@article{MotzkinNumbers,
    author = {Aigner, M.},
    title = {Motzkin Numbers} ,
    journal = {European Journal of Combinatorics} ,
    year = {1998},
    volume = {19},
    pages = {663--675},
    url = {http://dx.doi.org/10.1006/eujc.1998.0235}
}

@article{BERNHART199973,
title = {Catalan, Motzkin, and Riordan numbers},
journal = {Discrete Mathematics},
volume = {204},
number = {1},
pages = {73-112},
year = {1999},
note = {Selected papers in honor of Henry W. Gould},
issn = {0012-365X},
doi = {https://doi.org/10.1016/S0012-365X(99)00054-0},
url = {https://www.sciencedirect.com/science/article/pii/S0012365X99000540},
author = {Frank R. Bernhart}
}

@book{Stanley_2015, place={Cambridge}, title={Catalan Numbers}, publisher={Cambridge University Press}, author={Stanley, Richard P.}, year={2015}}

@book{Whittaker_Watson_1996, place={Cambridge}, edition={4}, series={Cambridge Mathematical Library}, title={A Course of Modern Analysis}, publisher={Cambridge University Press}, author={Whittaker, E. T. and Watson, G. N.}, year={1996}, collection={Cambridge Mathematical Library}}

@article{Witten:2021jzq,
    author = "Witten, Edward",
    title = "{Why Does Quantum Field Theory In Curved Spacetime Make Sense? And What Happens To The Algebra of Observables In The Thermodynamic Limit?}",
    eprint = "2112.11614",
    archivePrefix = "arXiv",
    primaryClass = "hep-th",
    month = "12",
    year = "2021"
}

@article{Alday:2020eua,
    author = "Alday, Luis F. and Kologlu, Murat and Zhiboedov, Alexander",
    title = "{Holographic correlators at finite temperature}",
    eprint = "2009.10062",
    archivePrefix = "arXiv",
    primaryClass = "hep-th",
    reportNumber = "CERN-TH-2020-155",
    doi = "10.1007/JHEP06(2021)082",
    journal = "JHEP",
    volume = "06",
    pages = "082",
    year = "2021"
}

@article{Hawking:1982dh,
    author = "Hawking, S. W. and Page, Don N.",
    title = "{Thermodynamics of Black Holes in anti-De Sitter Space}",
    reportNumber = "PRINT-83-0019 (CAMBRIDGE)",
    doi = "10.1007/BF01208266",
    journal = "Commun. Math. Phys.",
    volume = "87",
    pages = "577",
    year = "1983"
}

@article{DeBoer:2019yoe,
    author = "De Boer, Jan and Van Breukelen, Rik and Lokhande, Sagar F. and Papadodimas, Kyriakos and Verlinde, Erik",
    title = "{Probing typical black hole microstates}",
    eprint = "1901.08527",
    archivePrefix = "arXiv",
    primaryClass = "hep-th",
    doi = "10.1007/JHEP01(2020)062",
    journal = "JHEP",
    volume = "01",
    pages = "062",
    year = "2020"
}

@article{Banados:1992wn,
    author = "Banados, Maximo and Teitelboim, Claudio and Zanelli, Jorge",
    title = "{The Black hole in three-dimensional space-time}",
    eprint = "hep-th/9204099",
    archivePrefix = "arXiv",
    reportNumber = "PRINT-92-0151 (CHILE), IASSNS-HEP-92-29",
    doi = "10.1103/PhysRevLett.69.1849",
    journal = "Phys. Rev. Lett.",
    volume = "69",
    pages = "1849--1851",
    year = "1992"
}

@article{Maldacena:2013xja,
    author = "Maldacena, Juan and Susskind, Leonard",
    title = "{Cool horizons for entangled black holes}",
    eprint = "1306.0533",
    archivePrefix = "arXiv",
    primaryClass = "hep-th",
    doi = "10.1002/prop.201300020",
    journal = "Fortsch. Phys.",
    volume = "61",
    pages = "781--811",
    year = "2013"
}

@article{Vaidya:1943,
    author = {Vaidya, P. C.},
    title = {The External Field of a Radiating Star in General Relativity},
    journal = {General Relativity and Gravitation},
    year = {1999},
    volume = {31},
    pages = {119-120},
    url = {https://doi.org/10.1023/A:1018871522880}
}

@article{Vaidya:1953,
    author = {Vaidya, P. C.},
    title = {Newtonian Time in General Relativity},
    journal = {Nature} ,
    year = {1953},
    volume= {171},
    pages = {60--61},
    url = {https://doi.org/10.1038/171260a0}
}

@article{DRAY1985173,
title = {The gravitational shock wave of a massless particle},
journal = {Nuclear Physics B},
volume = {253},
pages = {173-188},
year = {1985},
issn = {0550-3213},
doi = {https://doi.org/10.1016/0550-3213(85)90525-5},
url = {https://www.sciencedirect.com/science/article/pii/0550321385905255},
author = {Tevian Dray and Gerard {'t Hooft}}
}

@article{Jafferis:2022uhu,
    author = "Jafferis, Daniel Louis and Kolchmeyer, David K. and Mukhametzhanov, Baur and Sonner, Julian",
    title = "{Matrix Models for Eigenstate Thermalization}",
    eprint = "2209.02130",
    archivePrefix = "arXiv",
    primaryClass = "hep-th",
    doi = "10.1103/PhysRevX.13.031033",
    journal = "Phys. Rev. X",
    volume = "13",
    number = "3",
    pages = "031033",
    year = "2023"
}

@article{Nozaki:2013wia,
    author = "Nozaki, Masahiro and Numasawa, Tokiro and Takayanagi, Tadashi",
    title = "{Holographic Local Quenches and Entanglement Density}",
    eprint = "1302.5703",
    archivePrefix = "arXiv",
    primaryClass = "hep-th",
    reportNumber = "YITP-13-14, IPMU-13-0045",
    doi = "10.1007/JHEP05(2013)080",
    journal = "JHEP",
    volume = "05",
    pages = "080",
    year = "2013"
}

@article{Sonner:2017hxc,
    author = "Sonner, Julian and Vielma, Manuel",
    title = "{Eigenstate thermalization in the Sachdev-Ye-Kitaev model}",
    eprint = "1707.08013",
    archivePrefix = "arXiv",
    primaryClass = "hep-th",
    doi = "10.1007/JHEP11(2017)149",
    journal = "JHEP",
    volume = "11",
    pages = "149",
    year = "2017"
}

@article{PRO,
 ISSN = {00255718, 10886842},
 URL = {http://www.jstor.org/stable/2007563},
 author = {Horst D. Simon},
 journal = {Mathematics of Computation},
 number = {165},
 pages = {115--142},
 publisher = {American Mathematical Society},
 title = {The Lanczos Algorithm With Partial Reorthogonalization},
 volume = {42},
 year = {1984}
}

@article{SO,
 ISSN = {00255718, 10886842},
 URL = {http://www.jstor.org/stable/2006037},
 author = {B. N. Parlett and D. S. Scott},
 journal = {Mathematics of Computation},
 number = {145},
 pages = {217--238},
 publisher = {American Mathematical Society},
 title = {The Lanczos Algorithm with Selective Orthogonalization},
 volume = {33},
 year = {1979}
}

@article{Roberts:2018mnp,
    author = "Roberts, Daniel A. and Stanford, Douglas and Streicher, Alexandre",
    title = "{Operator growth in the SYK model}",
    eprint = "1802.02633",
    archivePrefix = "arXiv",
    primaryClass = "hep-th",
    doi = "10.1007/JHEP06(2018)122",
    journal = "JHEP",
    volume = "06",
    pages = "122",
    year = "2018"
}

@article{Kitaev_1997,
	doi = {10.1070/rm1997v052n06abeh002155},
	url = {https://doi.org/10.1070%2Frm1997v052n06abeh002155},
	year = 1997,
	month = {12},
	publisher = {{IOP} Publishing},
	volume = {52},
	number = {6},
	pages = {1191--1249},
	author = {A Yu Kitaev},
	title = {Quantum computations: algorithms and error correction},
	journal = {Russian Mathematical Surveys}
}

@misc{dawson2005solovaykitaev,
    title={The Solovay-Kitaev algorithm},
    author={Christopher M. Dawson and Michael A. Nielsen},
    year={2005},
    eprint={quant-ph/0505030},
    archivePrefix={arXiv},
    primaryClass={quant-ph}
}

@article{Garcia-Garcia:2017bkg,
    author = "Garc\'{i}a-Garc\'{i}a, Antonio M. and Loureiro, Bruno and Romero-Berm\'{u}dez, Aurelio and Tezuka, Masaki",
    title = "{Chaotic-Integrable Transition in the Sachdev-Ye-Kitaev Model}",
    eprint = "1707.02197",
    archivePrefix = "arXiv",
    primaryClass = "hep-th",
    doi = "10.1103/PhysRevLett.120.241603",
    journal = "Phys. Rev. Lett.",
    volume = "120",
    number = "24",
    pages = "241603",
    year = "2018"
}

@article{Haque_2019,
   title={Eigenstate thermalization scaling in Majorana clusters: From chaotic to integrable Sachdev-Ye-Kitaev models},
   volume={100},
   ISSN={2469-9969},
   url={http://dx.doi.org/10.1103/PhysRevB.100.115122},
   DOI={10.1103/physrevb.100.115122},
   number={11},
   journal={Physical Review B},
   publisher={American Physical Society (APS)},
   author={Haque, Masudul and McClarty, P. A.},
   year={2019},
   month={9}
}

@article{Garcia-Garcia:2020cdo,
    author = "Garc\'{i}a-Garc\'{i}a, Antonio M. and Jia, Yiyang and Rosa, Dario and Verbaarschot, Jacobus J.M.",
    title = "{Sparse Sachdev-Ye-Kitaev model, quantum chaos and gravity duals}",
    eprint = "2007.13837",
    archivePrefix = "arXiv",
    primaryClass = "hep-th",
    month = "7",
    year = "2020"
}

@article{Maldacena:2001kr,
    author = "Maldacena, Juan Martin",
    title = "{Eternal black holes in anti-de Sitter}",
    eprint = "hep-th/0106112",
    archivePrefix = "arXiv",
    reportNumber = "NSF-ITP-01-59",
    doi = "10.1088/1126-6708/2003/04/021",
    journal = "JHEP",
    volume = "04",
    pages = "021",
    year = "2003"
}

@article{Dyson:2002pf,
    author = "Dyson, Lisa and Kleban, Matthew and Susskind, Leonard",
    title = "{Disturbing implications of a cosmological constant}",
    eprint = "hep-th/0208013",
    archivePrefix = "arXiv",
    reportNumber = "SU-ITP-02-25, MIT-CTP-3295",
    doi = "10.1088/1126-6708/2002/10/011",
    journal = "JHEP",
    volume = "10",
    pages = "011",
    year = "2002"
}

@article{Cotler:2016fpe,
    author = "Cotler, Jordan S. and Gur-Ari, Guy and Hanada, Masanori and Polchinski, Joseph and Saad, Phil and Shenker, Stephen H. and Stanford, Douglas and Streicher, Alexandre and Tezuka, Masaki",
    title = "{Black Holes and Random Matrices}",
    eprint = "1611.04650",
    archivePrefix = "arXiv",
    primaryClass = "hep-th",
    reportNumber = "SU-ITP-16-19, SU-ITP-16/19, YITP-16-124",
    doi = "10.1007/JHEP05(2017)118",
    journal = "JHEP",
    volume = "05",
    pages = "118",
    year = "2017",
    note = "[Erratum: JHEP 09, 002 (2018)]"
}

@article{Brown:2017jil,
    author = "Brown, Adam R. and Susskind, Leonard",
    title = "{Second law of quantum complexity}",
    eprint = "1701.01107",
    archivePrefix = "arXiv",
    primaryClass = "hep-th",
    doi = "10.1103/PhysRevD.97.086015",
    journal = "Phys. Rev. D",
    volume = "97",
    number = "8",
    pages = "086015",
    year = "2018"
}

@article{PhysRevA.30.504,
  title = {Ergodicity and mixing in quantum theory. I},
  author = {Peres, Asher},
  journal = {Phys. Rev. A},
  volume = {30},
  issue = {1},
  pages = {504--508},
  numpages = {0},
  year = {1984},
  month = {7},
  publisher = {American Physical Society},
  doi = {10.1103/PhysRevA.30.504},
  url = {https://link.aps.org/doi/10.1103/PhysRevA.30.504}
}

@article{PhysRevA.43.2046,
  title = {Quantum statistical mechanics in a closed system},
  author = {Deutsch, J. M.},
  journal = {Phys. Rev. A},
  volume = {43},
  issue = {4},
  pages = {2046--2049},
  numpages = {0},
  year = {1991},
  month = {2},
  publisher = {American Physical Society},
  doi = {10.1103/PhysRevA.43.2046},
  url = {https://link.aps.org/doi/10.1103/PhysRevA.43.2046}
}

@article{Srednicki_1999,
   title={The approach to thermal equilibrium in quantized chaotic systems},
   volume={32},
   ISSN={1361-6447},
   url={http://dx.doi.org/10.1088/0305-4470/32/7/007},
   DOI={10.1088/0305-4470/32/7/007},
   number={7},
   journal={Journal of Physics A: Mathematical and General},
   publisher={IOP Publishing},
   author={Srednicki, Mark},
   year={1999},
   month={1},
   pages={1163–1175}
}

@article{DAlessio:2015qtq,
    author = "D'Alessio, Luca and Kafri, Yariv and Polkovnikov, Anatoli and Rigol, Marcos",
    title = "{From quantum chaos and eigenstate thermalization to statistical mechanics and thermodynamics}",
    eprint = "1509.06411",
    archivePrefix = "arXiv",
    primaryClass = "cond-mat.stat-mech",
    doi = "10.1080/00018732.2016.1198134",
    journal = "Adv. Phys.",
    volume = "65",
    number = "3",
    pages = "239--362",
    year = "2016"
}

@article{Maldacena:2016hyu,
    author = "Maldacena, Juan and Stanford, Douglas",
    title = "{Remarks on the Sachdev-Ye-Kitaev model}",
    eprint = "1604.07818",
    archivePrefix = "arXiv",
    primaryClass = "hep-th",
    doi = "10.1103/PhysRevD.94.106002",
    journal = "Phys. Rev. D",
    volume = "94",
    number = "10",
    pages = "106002",
    year = "2016"
}

@article{Sachdev:1992fk,
    author = "Sachdev, Subir and Ye, Jinwu",
    title = "{Gapless spin fluid ground state in a random, quantum Heisenberg magnet}",
    eprint = "cond-mat/9212030",
    archivePrefix = "arXiv",
    reportNumber = "PRINT-93-0077",
    doi = "10.1103/PhysRevLett.70.3339",
    journal = "Phys. Rev. Lett.",
    volume = "70",
    pages = "3339",
    year = "1993"
}

@article{Sachdev:2015efa,
    author = "Sachdev, Subir",
    title = "{Bekenstein-Hawking Entropy and Strange Metals}",
    eprint = "1506.05111",
    archivePrefix = "arXiv",
    primaryClass = "hep-th",
    doi = "10.1103/PhysRevX.5.041025",
    journal = "Phys. Rev. X",
    volume = "5",
    number = "4",
    pages = "041025",
    year = "2015"
}

@article{Kitaev:2015,
    author = "Kitaev, Alexei",
    title = "{A simple model of quantum holography}",
    journal = "KITP talks",
    year = "2015"
}

@article{Saad:2019lba,
    author = "Saad, Phil and Shenker, Stephen H. and Stanford, Douglas",
    title = "{JT gravity as a matrix integral}",
    eprint = "1903.11115",
    archivePrefix = "arXiv",
    primaryClass = "hep-th",
    month = "3",
    year = "2019"
}

@article{Garcia-Alvarez:2016wem,
    author = "Garc\'{i}a-\'{A}lvarez, L. and Egusquiza, I.L. and Lamata, L. and del Campo, A. and Sonner, J. and Solano, E.",
    title = "{Digital Quantum Simulation of Minimal AdS/CFT}",
    eprint = "1607.08560",
    archivePrefix = "arXiv",
    primaryClass = "quant-ph",
    doi = "10.1103/PhysRevLett.119.040501",
    journal = "Phys. Rev. Lett.",
    volume = "119",
    number = "4",
    pages = "040501",
    year = "2017"
}

@article{babbush2019quantum,
  title={Quantum simulation of the Sachdev-Ye-Kitaev model by asymmetric qubitization},
  author={Babbush, Ryan and Berry, Dominic W and Neven, Hartmut},
  journal={Physical Review A},
  volume={99},
  number={4},
  pages={040301},
  year={2019},
  publisher={APS}
}

@article{Garcia-Garcia:2016mno,
    author = "Garc\'ia-Garc\'ia, Antonio M. and Verbaarschot, Jacobus J. M.",
    title = "{Spectral and thermodynamic properties of the Sachdev-Ye-Kitaev model}",
    eprint = "1610.03816",
    archivePrefix = "arXiv",
    primaryClass = "hep-th",
    doi = "10.1103/PhysRevD.94.126010",
    journal = "Phys. Rev. D",
    volume = "94",
    number = "12",
    pages = "126010",
    year = "2016"
}

@article{Carrega:2020mah,
    author = "Carrega, Matteo and Kim, Joonho and Rosa, Dario",
    title = "{Unveiling operator growth in SYK quench dynamics}",
    eprint = "2007.03551",
    archivePrefix = "arXiv",
    primaryClass = "quant-ph",
    reportNumber = "KIAS-P20033",
    month = "7",
    year = "2020"
}

@article{Yates_2020,
   title={Dynamics of almost strong edge modes in spin chains away from integrability},
   volume={102},
   ISSN={2469-9969},
   url={http://dx.doi.org/10.1103/PhysRevB.102.195419},
   DOI={10.1103/physrevb.102.195419},
   number={19},
   journal={Physical Review B},
   publisher={American Physical Society (APS)},
   author={Yates, Daniel J. and Abanov, Alexander G. and Mitra, Aditi},
   year={2020},
   month={11}
}

@article{PhysRevLett.115.256803,
  title = {Exponentially Slow Heating in Periodically Driven Many-Body Systems},
  author = {Abanin, Dmitry A. and De Roeck, Wojciech and Huveneers, Fran\ifmmode \mbox{\c{c}}\else \c{c}\fi{}ois},
  journal = {Phys. Rev. Lett.},
  volume = {115},
  issue = {25},
  pages = {256803},
  numpages = {5},
  year = {2015},
  month = {12},
  publisher = {American Physical Society},
  doi = {10.1103/PhysRevLett.115.256803},
  url = {https://link.aps.org/doi/10.1103/PhysRevLett.115.256803}
}

@inproceedings{Susskind:2018pmk,
    author = "Susskind, Leonard",
    title = "{Three Lectures on Complexity and Black Holes}",
    eprint = "1810.11563",
    archivePrefix = "arXiv",
    primaryClass = "hep-th",
    doi = "10.1007/978-3-030-45109-7",
    publisher = "Springer",
    series = "SpringerBriefs in Physics",
    month = "10",
    year = "2018"
}

@article{Iliesiu:2021ari,
    author = "Iliesiu, Luca V. and Mezei, M\'ark and S\'arosi, G\'abor",
    title = "{The volume of the black hole interior at late times}",
    eprint = "2107.06286",
    archivePrefix = "arXiv",
    primaryClass = "hep-th",
    reportNumber = "CERN-TH-2021-102",
    month = "7",
    year = "2021"
}

@article{luck:cea-01485001,
  TITLE = {{An investigation of equilibration in small quantum systems: the example of a particle in a 1D random potential}},
  AUTHOR = {Luck, J M},
  URL = {https://hal-cea.archives-ouvertes.fr/cea-01485001},
  NOTE = {19 pages, 7 figures, 1 table},
  JOURNAL = {{Journal of Physics A: Mathematical and Theoretical}},
  HAL_LOCAL_REFERENCE = {t15/182 },
  PUBLISHER = {{IOP Publishing}},
  VOLUME = {49},
  NUMBER = {11503},
  YEAR = {2016},
  MONTH = Mar,
  DOI = {10.1088/1751-8113/49/11/115303},
  HAL_ID = {cea-01485001},
  HAL_VERSION = {v1},
}

@article{Santos2013,
   title={An introduction to the spectrum, symmetries, and dynamics of spin-1/2 Heisenberg chains},
   volume={81},
   ISSN={1943-2909},
   url={http://dx.doi.org/10.1119/1.4798343},
   DOI={10.1119/1.4798343},
   number={6},
   journal={American Journal of Physics},
   publisher={American Association of Physics Teachers (AAPT)},
   author={Joel, Kira and Kollmar, Davida and Santos, Lea F.},
   year={2013},
   month={6},
   pages={450–457}
}

@article{Thouless_1972,
	doi = {10.1088/0022-3719/5/1/010},
	url = {https://doi.org/10.1088/0022-3719/5/1/010},
	year = 1972,
	month = {1},
	publisher = {{IOP} Publishing},
	volume = {5},
	number = {1},
	pages = {77--81},
	author = {D J Thouless},
	title = {A relation between the density of states and range of localization for one dimensional random systems},
	journal = {Journal of Physics C: Solid State Physics}
}

@article{Fleishman_1977,
	doi = {10.1088/0022-3719/10/6/003},
	url = {https://doi.org/10.1088/0022-3719/10/6/003},
	year = 1977,
	month = {3},
	publisher = {{IOP} Publishing},
	volume = {10},
	number = {6},
	pages = {L125--L126},
	author = {L Fleishman and D C Licciardello},
	title = {Fluctuations and localization in one dimension},
	journal = {Journal of Physics C: Solid State Physics}
}

@article{PhysRevB.24.5698,
  title = {Off-diagonal disorder in one-dimensional systems},
  author = {Soukoulis, C. M. and Economou, E. N.},
  journal = {Phys. Rev. B},
  volume = {24},
  issue = {10},
  pages = {5698--5702},
  numpages = {0},
  year = {1981},
  month = {11},
  publisher = {American Physical Society},
  doi = {10.1103/PhysRevB.24.5698},
  url = {https://link.aps.org/doi/10.1103/PhysRevB.24.5698}
}

@article{IZRAILEV2012125,
title = {Anomalous localization in low-dimensional systems with correlated disorder},
journal = {Physics Reports},
volume = {512},
number = {3},
pages = {125-254},
year = {2012},
note = {Anomalous localization in low-dimensional systems with correlated disorder},
issn = {0370-1573},
doi = {https://doi.org/10.1016/j.physrep.2011.11.002},
url = {https://www.sciencedirect.com/science/article/pii/S0370157311002936},
author = {F.M. Izrailev and A.A. Krokhin and N.M. Makarov}
}

@article{PhysRevB.72.174207,
  title = {Localization-delocalization transition in a one one-dimensional system with long-range correlated off-diagonal disorder},
  author = {Cheraghchi, H. and Fazeli, S. M. and Esfarjani, K.},
  journal = {Phys. Rev. B},
  volume = {72},
  issue = {17},
  pages = {174207},
  numpages = {8},
  year = {2005},
  month = {11},
  publisher = {American Physical Society},
  doi = {10.1103/PhysRevB.72.174207},
  url = {https://link.aps.org/doi/10.1103/PhysRevB.72.174207}
}

@article{SANCHEZDEHESA1978275,
title = {The spectrum of Jacobi matrices in terms of its associated weight function},
journal = {Journal of Computational and Applied Mathematics},
volume = {4},
number = {4},
pages = {275-283},
year = {1978},
issn = {0377-0427},
doi = {https://doi.org/10.1016/0771-050X(78)90026-8},
url = {https://www.sciencedirect.com/science/article/pii/0771050X78900268},
author = {Jesus Sanchez-Dehesa}
}

@misc{ enwiki:1030584213,
    author = "{Wikipedia contributors}",
    title = "{Cauchy–Binet formula --- Wikipedia, The Free Encyclopedia [Online; last accessed 19-December-2021]}",
    URL = {https://en.wikipedia.org/wiki/Cauchy-Binet\_formula}
  }

@article{Nayak:2019khe,
    author = "Nayak, Pranjal and Sonner, Julian and Vielma, Manuel",
    title = "{Eigenstate Thermalisation in the conformal Sachdev-Ye-Kitaev model: an analytic approach}",
    eprint = "1903.00478",
    archivePrefix = "arXiv",
    primaryClass = "hep-th",
    doi = "10.1007/JHEP10(2019)019",
    journal = "JHEP",
    volume = "10",
    pages = "019",
    year = "2019"
}

@article{Nayak:2019evx,
    author = "Nayak, Pranjal and Sonner, Julian and Vielma, Manuel",
    title = "{Extended Eigenstate Thermalization and the role of FZZT branes in the Schwarzian theory}",
    eprint = "1907.10061",
    archivePrefix = "arXiv",
    primaryClass = "hep-th",
    doi = "10.1007/JHEP03(2020)168",
    journal = "JHEP",
    volume = "03",
    pages = "168",
    year = "2020"
}

@book{Samaj_bajnok_2013, 
place={Cambridge}, 
title={Introduction to the Statistical Physics of Integrable Many-body Systems}, DOI={10.1017/CBO9781139343480}, 
publisher={Cambridge University Press},
author={Šamaj, Ladislav and Bajnok, Zoltán}, 
year={2013}
}

@article{Rigol_XXZ,
  title = {Low-frequency behavior of off-diagonal matrix elements in the integrable XXZ chain and in a locally perturbed quantum-chaotic XXZ chain},
  author = {Brenes, Marlon and Goold, John and Rigol, Marcos},
  journal = {Phys. Rev. B},
  volume = {102},
  issue = {7},
  pages = {075127},
  numpages = {7},
  year = {2020},
  month = {8},
  publisher = {American Physical Society},
  doi = {10.1103/PhysRevB.102.075127},
  url = {https://link.aps.org/doi/10.1103/PhysRevB.102.075127}
}

@article{doi:10.1126/science.1121541,
author = {Michael A. Nielsen  and Mark R. Dowling  and Mile Gu  and Andrew C. Doherty },
title = {Quantum Computation as Geometry},
journal = {Science},
volume = {311},
number = {5764},
pages = {1133-1135},
year = {2006},
doi = {10.1126/science.1121541}
}

@ARTICLE{1931ZPhy...71..205B,
       author = {{Bethe}, H.},
        title = "{Zur Theorie der Metalle}",
      journal = {Zeitschrift fur Physik},
         year = 1931,
        month = mar,
       volume = {71},
       number = {3-4},
        pages = {205-226},
          doi = {10.1007/BF01341708},
       adsurl = {https://ui.adsabs.harvard.edu/abs/1931ZPhy...71..205B}
}

@article{Anderson_AbsDiff,
  title = {Absence of Diffusion in Certain Random Lattices},
  author = {Anderson, P. W.},
  journal = {Phys. Rev.},
  volume = {109},
  issue = {5},
  pages = {1492--1505},
  numpages = {0},
  year = {1958},
  month = {3},
  publisher = {American Physical Society},
  doi = {10.1103/PhysRev.109.1492},
  url = {https://link.aps.org/doi/10.1103/PhysRev.109.1492}
}

@article{Heisenberg:1928mqa,
    author = "Heisenberg, W.",
    title = "{Zur Theorie des Ferromagnetismus}",
    doi = "10.1007/BF01328601",
    journal = "Z. Phys.",
    volume = "49",
    number = "9-10",
    pages = "619--636",
    year = "1928"
}

@article{YangXXZ_I,
  title = {One-Dimensional Chain of Anisotropic Spin-Spin Interactions. I. Proof of Bethe's Hypothesis for Ground State in a Finite System},
  author = {Yang, C. N. and Yang, C. P.},
  journal = {Phys. Rev.},
  volume = {150},
  issue = {1},
  pages = {321--327},
  numpages = {0},
  year = {1966},
  month = {10},
  publisher = {American Physical Society},
  doi = {10.1103/PhysRev.150.321},
  url = {https://link.aps.org/doi/10.1103/PhysRev.150.321}
}

@article{YangXXZ_II,
  title = {One-Dimensional Chain of Anisotropic Spin-Spin Interactions. II. Properties of the Ground-State Energy Per Lattice Site for an Infinite System},
  author = {Yang, C. N. and Yang, C. P.},
  journal = {Phys. Rev.},
  volume = {150},
  issue = {1},
  pages = {327--339},
  numpages = {0},
  year = {1966},
  month = {10},
  publisher = {American Physical Society},
  doi = {10.1103/PhysRev.150.327},
  url = {https://link.aps.org/doi/10.1103/PhysRev.150.327}
}

@article{Doikou:1998jh,
    author = "Doikou, Anastasia and Nepomechie, Rafael I.",
    title = "{Parity and charge conjugation symmetries and S matrix of the XXZ chain}",
    eprint = "hep-th/9810034",
    archivePrefix = "arXiv",
    reportNumber = "UMTG-215",
    month = "10",
    year = "1998"
}

@Article{SciPostPhys.2.1.003,
	title={{QuSpin: a Python Package for Dynamics and Exact Diagonalisation of  Quantum Many Body Systems part I: spin chains}},
	author={Phillip Weinberg and Marin Bukov},
	journal={SciPost Phys.},
	volume={2},
	issue={1},
	pages={003},
	year={2017},
	publisher={SciPost},
	doi={10.21468/SciPostPhys.2.1.003},
	url={https://scipost.org/10.21468/SciPostPhys.2.1.003},
}

@article{Santos_2004,
	doi = {10.1088/0305-4470/37/17/004},
	url = {https://doi.org/10.1088/0305-4470/37/17/004},
	year = 2004,
	month = {4},
	publisher = {{IOP} Publishing},
	volume = {37},
	number = {17},
	pages = {4723--4729},
	author = {L F Santos},
	title = {Integrability of a disordered Heisenberg spin-1/2 chain},
	journal = {Journal of Physics A: Mathematical and General}
}

@article{PhysRevE.84.016206,
  title = {Domain wall dynamics in integrable and chaotic spin-1$/$2 chains},
  author = {Santos, Lea F. and Mitra, Aditi},
  journal = {Phys. Rev. E},
  volume = {84},
  issue = {1},
  pages = {016206},
  numpages = {8},
  year = {2011},
  month = {7},
  publisher = {American Physical Society},
  doi = {10.1103/PhysRevE.84.016206},
  url = {https://link.aps.org/doi/10.1103/PhysRevE.84.016206}
}

@article{PhysRevB.80.125118,
  title = {Incoherent transport induced by a single static impurity in a Heisenberg chain},
  author = {Bari\ifmmode \check{s}\else \v{s}\fi{}i\ifmmode \acute{c}\else \'{c}\fi{}, O. S. and Prelov\ifmmode \check{s}\else \v{s}\fi{}ek, P. and Metavitsiadis, A. and Zotos, X.},
  journal = {Phys. Rev. B},
  volume = {80},
  issue = {12},
  pages = {125118},
  numpages = {4},
  year = {2009},
  month = {9},
  publisher = {American Physical Society},
  doi = {10.1103/PhysRevB.80.125118},
  url = {https://link.aps.org/doi/10.1103/PhysRevB.80.125118}
}

@article{PhysRevB.98.235128,
  title = {High-temperature coherent transport in the XXZ chain in the presence of an impurity},
  author = {Brenes, Marlon and Mascarenhas, Eduardo and Rigol, Marcos and Goold, John},
  journal = {Phys. Rev. B},
  volume = {98},
  issue = {23},
  pages = {235128},
  numpages = {14},
  year = {2018},
  month = {12},
  publisher = {American Physical Society},
  doi = {10.1103/PhysRevB.98.235128},
  url = {https://link.aps.org/doi/10.1103/PhysRevB.98.235128}
}

@article{PhysRevX.10.041017,
  title = {Adiabatic Eigenstate Deformations as a Sensitive Probe for Quantum Chaos},
  author = {Pandey, Mohit and Claeys, Pieter W. and Campbell, David K. and Polkovnikov, Anatoli and Sels, Dries},
  journal = {Phys. Rev. X},
  volume = {10},
  issue = {4},
  pages = {041017},
  numpages = {16},
  year = {2020},
  month = {10},
  publisher = {American Physical Society},
  doi = {10.1103/PhysRevX.10.041017},
  url = {https://link.aps.org/doi/10.1103/PhysRevX.10.041017}
}

@article{doi:10.1119/1.3671068,
author = {Gubin,Aviva  and F. Santos,Lea },
title = {Quantum chaos: An introduction via chains of interacting spins 1/2},
journal = {American Journal of Physics},
volume = {80},
number = {3},
pages = {246-251},
year = {2012},
doi = {10.1119/1.3671068},

URL = { 
        https://doi.org/10.1119/1.3671068
    
},
eprint = { 
        https://doi.org/10.1119/1.3671068
    
}

}

@article{PhysRevLett.110.084101,
  title = {Distribution of the Ratio of Consecutive Level Spacings in Random Matrix Ensembles},
  author = {Atas, Y. Y. and Bogomolny, E. and Giraud, O. and Roux, G.},
  journal = {Phys. Rev. Lett.},
  volume = {110},
  issue = {8},
  pages = {084101},
  numpages = {5},
  year = {2013},
  month = {2},
  publisher = {American Physical Society},
  doi = {10.1103/PhysRevLett.110.084101},
  url = {https://link.aps.org/doi/10.1103/PhysRevLett.110.084101}
}

@article{PhysRevB.75.155111,
  title = {Localization of interacting fermions at high temperature},
  author = {Oganesyan, Vadim and Huse, David A.},
  journal = {Phys. Rev. B},
  volume = {75},
  issue = {15},
  pages = {155111},
  numpages = {5},
  year = {2007},
  month = {4},
  publisher = {American Physical Society},
  doi = {10.1103/PhysRevB.75.155111},
  url = {https://link.aps.org/doi/10.1103/PhysRevB.75.155111}
}

@article{Hornedal:2022pkc,
    author = {H\"ornedal, Niklas and Carabba, Nicoletta and Matsoukas-Roubeas, Apollonas S. and del Campo, Adolfo},
    title = "{Ultimate Physical Limits to the Growth of Operator Complexity}",
    eprint = "2202.05006",
    archivePrefix = "arXiv",
    primaryClass = "quant-ph",
    month = "2",
    year = "2022"
}

@article{LeBlond:2019eoe,
    author = "LeBlond, Tyler and Mallayya, Krishnanand and Vidmar, Lev and Rigol, Marcos",
    title = "{Entanglement and matrix elements of observables in interacting integrable systems}",
    eprint = "1909.09654",
    archivePrefix = "arXiv",
    primaryClass = "cond-mat.stat-mech",
    doi = "10.1103/PhysRevE.100.062134",
    journal = "Phys. Rev. E",
    volume = "100",
    number = "6",
    pages = "062134",
    year = "2019"
}

@article{Rigol_LeBlod2020,
  title = {Eigenstate Thermalization in a Locally Perturbed Integrable System},
  author = {Brenes, Marlon and LeBlond, Tyler and Goold, John and Rigol, Marcos},
  journal = {Phys. Rev. Lett.},
  volume = {125},
  issue = {7},
  pages = {070605},
  numpages = {6},
  year = {2020},
  month = {8},
  publisher = {American Physical Society},
  doi = {10.1103/PhysRevLett.125.070605},
  url = {https://link.aps.org/doi/10.1103/PhysRevLett.125.070605}
}

@article{Noh_2021,
	doi = {10.1103/physreve.104.034112},
	url = {https://doi.org/10.1103%2Fphysreve.104.034112},
	year = 2021,
	month = {9},
	publisher = {American Physical Society ({APS})},
	volume = {104},
	number = {3},
	author = {Jae Dong Noh},
	title = {Operator growth in the transverse-field Ising spin chain with integrability-breaking longitudinal field},
	journal = {Physical Review E}
}

@article{Maldacena:2016upp,
    author = "Maldacena, Juan and Stanford, Douglas and Yang, Zhenbin",
    title = "{Conformal symmetry and its breaking in two dimensional Nearly Anti-de-Sitter space}",
    eprint = "1606.01857",
    archivePrefix = "arXiv",
    primaryClass = "hep-th",
    doi = "10.1093/ptep/ptw124",
    journal = "PTEP",
    volume = "2016",
    number = "12",
    pages = "12C104",
    year = "2016"
}

@article{Spradlin:1999bn,
    author = "Spradlin, Marcus and Strominger, Andrew",
    title = "{Vacuum states for AdS(2) black holes}",
    eprint = "hep-th/9904143",
    archivePrefix = "arXiv",
    reportNumber = "HUTP-99-A014",
    doi = "10.1088/1126-6708/1999/11/021",
    journal = "JHEP",
    volume = "11",
    pages = "021",
    year = "1999"
}

@article{Jackiw:1984je,
    author = "Jackiw, R.",
    editor = "Baier, R. and Satz, H.",
    title = "{Lower Dimensional Gravity}",
    reportNumber = "MIT-CTP-1203",
    doi = "10.1016/0550-3213(85)90448-1",
    journal = "Nucl. Phys. B",
    volume = "252",
    pages = "343--356",
    year = "1985"
}

@article{Teitelboim:1983ux,
    author = "Teitelboim, C.",
    title = "{Gravitation and Hamiltonian Structure in Two Space-Time Dimensions}",
    doi = "10.1016/0370-2693(83)90012-6",
    journal = "Phys. Lett. B",
    volume = "126",
    pages = "41--45",
    year = "1983"
}

@article{Garcia-Garcia:2017pzl,
    author = "Garc\'\i{}a-Garc\'\i{}a, Antonio M. and Verbaarschot, Jacobus J. M.",
    title = "{Analytical Spectral Density of the Sachdev-Ye-Kitaev Model at finite N}",
    eprint = "1701.06593",
    archivePrefix = "arXiv",
    primaryClass = "hep-th",
    doi = "10.1103/PhysRevD.96.066012",
    journal = "Phys. Rev. D",
    volume = "96",
    number = "6",
    pages = "066012",
    year = "2017"
}

@article{Polchinski:2016xgd,
    author = "Polchinski, Joseph and Rosenhaus, Vladimir",
    title = "{The Spectrum in the Sachdev-Ye-Kitaev Model}",
    eprint = "1601.06768",
    archivePrefix = "arXiv",
    primaryClass = "hep-th",
    doi = "10.1007/JHEP04(2016)001",
    journal = "JHEP",
    volume = "04",
    pages = "001",
    year = "2016"
}

@article{Berkooz:2018qkz,
    author = "Berkooz, Micha and Narayan, Prithvi and Simon, Joan",
    title = "{Chord diagrams, exact correlators in spin glasses and black hole bulk reconstruction}",
    eprint = "1806.04380",
    archivePrefix = "arXiv",
    primaryClass = "hep-th",
    doi = "10.1007/JHEP08(2018)192",
    journal = "JHEP",
    volume = "08",
    pages = "192",
    year = "2018"
}

@article{Goel:2023svz,
    author = "Goel, Akash and Narovlansky, Vladimir and Verlinde, Herman",
    title = "{Semiclassical geometry in double-scaled SYK}",
    eprint = "2301.05732",
    archivePrefix = "arXiv",
    primaryClass = "hep-th",
    month = "1",
    year = "2023"
}

@article{Harlow:2018tqv,
    author = "Harlow, Daniel and Jafferis, Daniel",
    title = "{The Factorization Problem in Jackiw-Teitelboim Gravity}",
    eprint = "1804.01081",
    archivePrefix = "arXiv",
    primaryClass = "hep-th",
    doi = "10.1007/JHEP02(2020)177",
    journal = "JHEP",
    volume = "02",
    pages = "177",
    year = "2020"
}

@article{Cannata:1991I,
    author = "F. Cannata and L. Ferrari",
    title = "{Canonical conjugate momentum of discrete label operators in quantum mechanics I: Formalism}",
    doi = "10.1007/BF00689891",
    journal = "Foundations of Physics Letters",
    volume = "04",
    pages = "557",
    year = "1991"
}

@article{Cannata:1991II,
    author = "F. Cannata and L. Ferrari",
    title = "{Canonical conjugate momentum of discrete label operators in quantum mechanics II: Formalism}",
    doi = "10.1007/BF00689892",
    journal = "Foundations of Physics Letters",
    volume = "04",
    pages = "569",
    year = "1991"
}

@article{Brown:2018bms,
    author = "Brown, Adam R. and Gharibyan, Hrant and Lin, Henry W. and Susskind, Leonard and Thorlacius, L\'arus and Zhao, Ying",
    title = "{Complexity of Jackiw-Teitelboim gravity}",
    eprint = "1810.08741",
    archivePrefix = "arXiv",
    primaryClass = "hep-th",
    doi = "10.1103/PhysRevD.99.046016",
    journal = "Phys. Rev. D",
    volume = "99",
    number = "4",
    pages = "046016",
    year = "2019"
}

@article{Harlow:2021dfp,
    author = "Harlow, Daniel and Wu, Jie-qiang",
    title = "{Algebra of diffeomorphism-invariant observables in Jackiw-Teitelboim gravity}",
    eprint = "2108.04841",
    archivePrefix = "arXiv",
    primaryClass = "hep-th",
    doi = "10.1007/JHEP05(2022)097",
    journal = "JHEP",
    volume = "05",
    pages = "097",
    year = "2022"
}

@article{Sachdev_1993,
	doi = {10.1103/physrevlett.70.3339}, 
	url = {https://doi.org/10.1103\%2Fphysrevlett.70.3339},  
	year = 1993,
	month = {5}, 
	publisher = {American Physical Society ({APS})},  
	volume = {70},  
	number = {21},  
	pages = {3339--3342},  
	author = {Subir Sachdev and Jinwu Ye},  
	title = {Gapless spin-fluid ground state in a random quantum Heisenberg magnet}, 
	journal = {Physical Review Letters}
}

@article{Arik:1976,
author = {Arik,M.  and Coon,D. D. },
title = {Hilbert spaces of analytic functions and generalized coherent states},
journal = {Journal of Mathematical Physics},
volume = {17},
number = {4},
pages = {524-527},
year = {1976},
doi = {10.1063/1.522937},

URL = { 
        https://aip.scitation.org/doi/abs/10.1063/1.522937
    
},
eprint = { 
        https://aip.scitation.org/doi/pdf/10.1063/1.522937
    
}

}

@article{Mukhametzhanov:2023tcg,
    author = "Mukhametzhanov, Baur",
    title = "{Large p SYK from chord diagrams}",
    eprint = "2303.03474",
    archivePrefix = "arXiv",
    primaryClass = "hep-th",
    month = "3",
    year = "2023"
}

@article{Bagrets:2016cdf,
    author = "Bagrets, Dmitry and Altland, Alexander and Kamenev, Alex",
    editor = "Unno, Yoshinobu and Ohsugi, Takashi and Hou, Suen and Sadrozinski, Hartmut F. -W. and Lou, Xinchou and Zhu, Hongbo and Ouyang, Qun",
    title = "{Sachdev\textendash{}Ye\textendash{}Kitaev model as Liouville quantum mechanics}",
    eprint = "1607.00694",
    archivePrefix = "arXiv",
    primaryClass = "cond-mat.str-el",
    doi = "10.1016/j.nuclphysb.2016.08.002",
    journal = "Nucl. Phys. B",
    volume = "911",
    pages = "191--205",
    year = "2016"
}

@article{delCampo:2017bzr,
    author = "del Campo, A. and Molina-Vilaplana, J. and Sonner, J.",
    title = "{Scrambling the spectral form factor: unitarity constraints and exact results}",
    eprint = "1702.04350",
    archivePrefix = "arXiv",
    primaryClass = "hep-th",
    doi = "10.1103/PhysRevD.95.126008",
    journal = "Phys. Rev. D",
    volume = "95",
    number = "12",
    pages = "126008",
    year = "2017"
}

@article{qHermite_formula,
 ISSN = {00222518, 19435258},
 URL = {http://www.jstor.org/stable/24892881},
 author = {D. M. Bressoud},
 journal = {Indiana University Mathematics Journal},
 number = {4},
 pages = {577--580},
 publisher = {Indiana University Mathematics Department},
 title = {A Simple Proof of Mehler's Formula for q-Hermite Polynomials},
 note = {[Online; last accessed on 9-April-2023]},
 volume = {29},
 year = {1980}
}

@article{Sarosi:2017ykf,
    author = "S\'arosi, G\'abor",
    title = "{AdS$_{2}$ holography and the SYK model}",
    eprint = "1711.08482",
    archivePrefix = "arXiv",
    primaryClass = "hep-th",
    doi = "10.22323/1.323.0001",
    journal = "PoS",
    volume = "Modave2017",
    pages = "001",
    year = "2018"
}

@article{Cottrell:2018ash,
    author = "Cottrell, William and Freivogel, Ben and Hofman, Diego M. and Lokhande, Sagar F.",
    title = "{How to Build the Thermofield Double State}",
    eprint = "1811.11528",
    archivePrefix = "arXiv",
    primaryClass = "hep-th",
    doi = "10.1007/JHEP02(2019)058",
    journal = "JHEP",
    volume = "02",
    pages = "058",
    year = "2019"
}

@article{Craps:2023ivc,
    author = "Craps, Ben and Evnin, Oleg and Pascuzzi, Gabriele",
    title = "{A Relation between Krylov and Nielsen Complexity}",
    eprint = "2311.18401",
    archivePrefix = "arXiv",
    primaryClass = "quant-ph",
    doi = "10.1103/PhysRevLett.132.160402",
    journal = "Phys. Rev. Lett.",
    volume = "132",
    number = "16",
    pages = "160402",
    year = "2024"
}
